\documentclass[aps,prl,reprint,longbibliography]{revtex4-2}
\usepackage{pdfpages}
\usepackage{appendix}
\usepackage{graphicx}
\usepackage{longtable}
\usepackage{hyperref}
\hypersetup{colorlinks=true,linkcolor=blue,filecolor=gray,urlcolor=blue,citecolor=blue}
\usepackage{chemformula} 
\usepackage{easyReview}

\makeatletter
\AtBeginDocument{\let\LS@rot\@undefined}
\makeatother

\begin{document}

\title{Hydride Units Filled B--C Clathrate: A New Pathway for High-Temperature Superconductivity at Ambient Pressure}

\author{Ying Sun}
\affiliation{Department of Physics, Rutgers University, Newark, New Jersey 07102, United States}
\author{Li Zhu}
\email{li.zhu@rutgers.edu}
\affiliation{Department of Physics, Rutgers University, Newark, New Jersey 07102, United States}


\begin{abstract}
	The pursuit of room-temperature superconductors has recently advanced with the discovery of high-temperature superconductivity in compressed hydrides, although sustaining the superconductivity of hydrides at ambient pressure remains challenging. In parallel, $sp^3$--bonded frameworks comprising lightweight elements have emerged as another avenue for developing ambient-pressure superconductors. However, despite their stability at low pressures, the critical temperature ($T_\text{c}$) values observed in these materials have not yet reached the impressive benchmarks set by hydride-based superconductors. Here we propose a novel design strategy for achieving high-temperature superconductivity at ambient pressure by integrating hydride units into B--C clathrate structures. This approach exploits the beneficial properties of hydrogen, the lightest element, to enhance the superconductivity beyond that of the parent compounds. For instance, our computational predictions indicate that doping \ch{SrB3C3} with ammonium (\ch{NH4}) yields a \ch{SrNH4B6C6} compound with an estimated $T_\text{c}$ of 85 K at ambient pressure---more than double that of its precursor (31 K). Extensive substitution across the periodic table results in a family of \ch{MNH4B6C6} superconductors that are predicted to be superconducting at ambient pressure, with the highest predicted $T_\text{c}$ of 115 K in \ch{PbNH4B6C6}. Our findings present a promising strategy for discovering high-$T_\text{c}$ superconductors at ambient pressure, potentially revolutionizing technologies reliant on superconducting materials.
\end{abstract}

\maketitle

\section*{Introduction}
\label{introduction}

The quest for room-temperature superconductivity has been a long-standing aspiration and a central focus of research in the field of condensed matter physics. 
In recent years, significant progress has been made in the realm of high-temperature superconducting compressed hydrides \cite{sun2023,Lilia2022,Lv2020}. Theory-oriented discoveries of \ch{H3S}  (203 K at 200 GPa) \cite{li2014,duan2014,drozdov2015} and a family of clathrate superhydrides (e.g., 215 K at 172 GPa in \ch{CaH6} \cite{wang2012b,li2022,ma2022}, 243 K at 201 GPa in \ch{YH9} \cite{peng2017,kong2021,snider2021,wang2022}, and 260 K at 180 GPa in \ch{LaH10} \cite{peng2017,liu2017,drozdov2019,somayazulu2019,errea2020}) have continually broken records in measured $T_\text{c}$.   
These materials have demonstrated that the dream of room-temperature superconductivity is not just a theoretical possibility but could be within our reach. 

However, a significant challenge lies in preserving these high-temperature superconducting compressed hydrides at ambient pressure: the lowest superconducting pressure record among the experimentally confirmed hydrogen-rich high-temperature superconductors (exhibit superconductivity above liquid nitrogen temperature) is \ch{LaBeH8}, which has a measured $T_\text{c}$ of 110 K at 80 GPa \cite{song2023stoichiometric}.
Nevertheless, efforts to stabilize high-temperature superconducting hydrides at ambient pressure persist.

For example, \ch{LaBeH8} has been predicted to remain stable down to 20 GPa \cite{zhang2022design} and \ch{BaSiH8}, which is isomorphic to \ch{LaBeH8}, was predicted to remain dynamically stable down to 3 GPa with a predicted $T_\text{c}$ of 71 K \cite{lucrezi2022silico}.
Two research groups have recently predicted the dynamic stability of \ch{Mg2IrH6} at ambient pressure through high-throughput computational methods \cite{sanna2023,dolui2023}. However, a divergence exists in the predicted $T_\text{c}$ for this compound, indicating the complexities in predicting the properties of potential ambient-pressure superconducting hydrides.

Yet, the realm of conventional high-temperature superconductivity is not limited to hydrides. Other light-element materials, including boron/carbon-based compounds, also exhibit superconductivity. The robust covalent bonds intrinsic to these materials contribute to their stability, even at 1 atmosphere (\textit{atm.}). 
A number of light-element materials have been theoretically proposed to be superconductive at ambient or near-ambient pressure, including but not limited to heavily boron-doped diamond (55 K at 1 \textit{atm.}) \cite{Moussa2008}, \ch{BC5} (41.5 K at 1 \textit{atm.}) \cite{Li2010}, \ch{BaB8} (62 K at 1 \textit{atm.}) \cite{Ma2023}, \ch{HC8} (117.7 K at 5 GPa) \cite{Ding2023}, and a family of light-element clathrate superconductors represented by their first member \ch{SrB3C3} (31 K at 1 \textit{atm.}) \cite{zhu2020,zhu2023}.

\ch{SrB3C3} \cite{zhu2020}, a boron-carbide analogue of the clathrate hexahydrides \ch{CaH6} \cite{wang2012b}, has been predicted via the CALYPSO method \cite{Wang2010,Wang2012,gao2019} to dynamically stabilize in the bipartite sodalite (Type-VII clathrate) structure. The estimated ambient-pressure $T_\text{c}$ of 31 K is attributed to the strong coupling between the $sp^3$ $\sigma$-bonding bands and boron-associated $E_g$ modes \cite{zhu2023}.
Subsequently, \ch{SrB3C3} was synthesized at a pressure of 57 GPa and quenched to ambient-pressure conditions in an inert atmosphere \cite{zhu2020}. Recent observations \cite{zhu2023} have provided evidence for a superconducting transition in this material, marking a significant milestone in the field.
Following the successful synthesis of \ch{SrB3C3} \cite{zhu2020}, an isotypic carbon-boron clathrate phase filled with lanthanum, \ch{LaB3C3}, was also realized \cite{Strobel2020}. Predicted to possess an indirect gap of approximately 1.3 eV, this compound was synthesized under more moderate pressures (38 GPa) and similarly quenched to ambient conditions \cite{Strobel2020}, indicating the prospect for future synthesis of similar clathrate materials under even less extreme conditions.

Inspired by the synthesis of \ch{SrB3C3} \cite{zhu2020} and \ch{LaB3C3} \cite{Strobel2020}, there has been a surge in theoretical investigations into similar materials. 
Numerous ternary and quaternary isotypic boron-carbon clathrate structures, with different guest metal atoms, have been thoroughly studied \cite{DiCataldo2022,Geng2023}. A comprehensive analysis of the first 57 elements from the periodic table revealed that only five elements, namely Ca, Sr, Y, Ba, and La, form stable \ch{XB3C3} compounds at ambient or moderate pressures, including already synthesized \ch{SrB3C3} and \ch{LaB3C3}. \ch{CaB3C3} and \ch{BaB3C3} were predicted to exhibit $T_\text{c}$ values of 48 K at 1 \textit{atm.} and 50 K at 30 GPa, respectively \cite{DiCataldo2022}.
For binary-guest compounds, over twenty \ch{XYB6C6} variants (with X and Y representing electropositive metal atoms) are predicted to be dynamically stable at 1 \textit{atm.}, with expected ambient-pressure $T_\text{c}$ values up to 88 K in \ch{KPbB6C6} \cite{Geng2023}.
Besides, theoretical research on analogous clathrate cages with inequivalent C:B ratios \cite{Cui2022} identified \ch{CaB2C4} and \ch{SrB4C2} as superconductors with $T_\text{c}$ values of 2 and 19 K, respectively. 
Metal-doped clathrate materials with pure carbon frameworks or boron-nitrogen frameworks are also predicted to exhibit high-$T_\text{c}$ superconductivity at ambient pressure, such as \ch{NaC6} (116 K) \cite{Hai2023}, \ch{CsC10} (74 K) \cite{Hai2023}, \ch{Al2(BN)6} (72 K) \cite{Hai2022} and \ch{La(BN)5} (69 K) \cite{Ding2022}.

Despite significant advancements in light-element superconductivity, the values of $T_\text{c}$ observed in these systems still lag behind those achieved in hydride-based superconductors. This gap underscores a pivotal challenge in the field: the integration of the inherent stability of B--C superconductors at ambient pressure with the superior $T_\text{c}$ characteristic of hydrides. The combination of these properties has the potential to propel the field forward by integrating the advantages intrinsic to both classes of materials. In this work, we propose a novel approach to meet this challenge by introducing hydride units into B--C clathrate superconductors. This strategy aims to leverage the structural stability of B--C clathrate materials while harnessing the high $T_\text{c}$ found in hydrides, presenting a promising pathway toward high-temperature superconductivity at ambient pressure. To test our hypothesis, we conducted a series of high-throughput calculations, leading to the identification of several \ch{NH4}-doped B--C clathrate superconductors that demonstrate stability with enhanced superconductivity. 
For instance, when ammonium (\ch{NH4}) is doped into \ch{SrB3C3}, the predicted $T_\text{c}$ for \ch{SrNH4B6C6} soars to 85 K at ambient pressure, more than doubling that of the original \ch{SrB3C3} (31 K \cite{zhu2023}), proving the feasibility of this new approach.
Furthermore, we expanded the scope of our study to include other metal atoms, resulting in the identification of 55 \ch{MNH4B6C6} compounds that are dynamically stable at ambient pressure, with \ch{PbNH4B6C6} exhibiting the highest predicted $T_\text{c}$ of 115 K, exceeding the boiling point (77 K) of liquid nitrogen.
Our results provide a viable strategy for enhancing $T_\text{c}$ and designing new high-$T_\text{c}$ superconductors at ambient pressure.

\section*{Results and Discussion}
\label{results_discussion}

\begin{figure}[t]
	\includegraphics[width=8.5cm,angle=0]{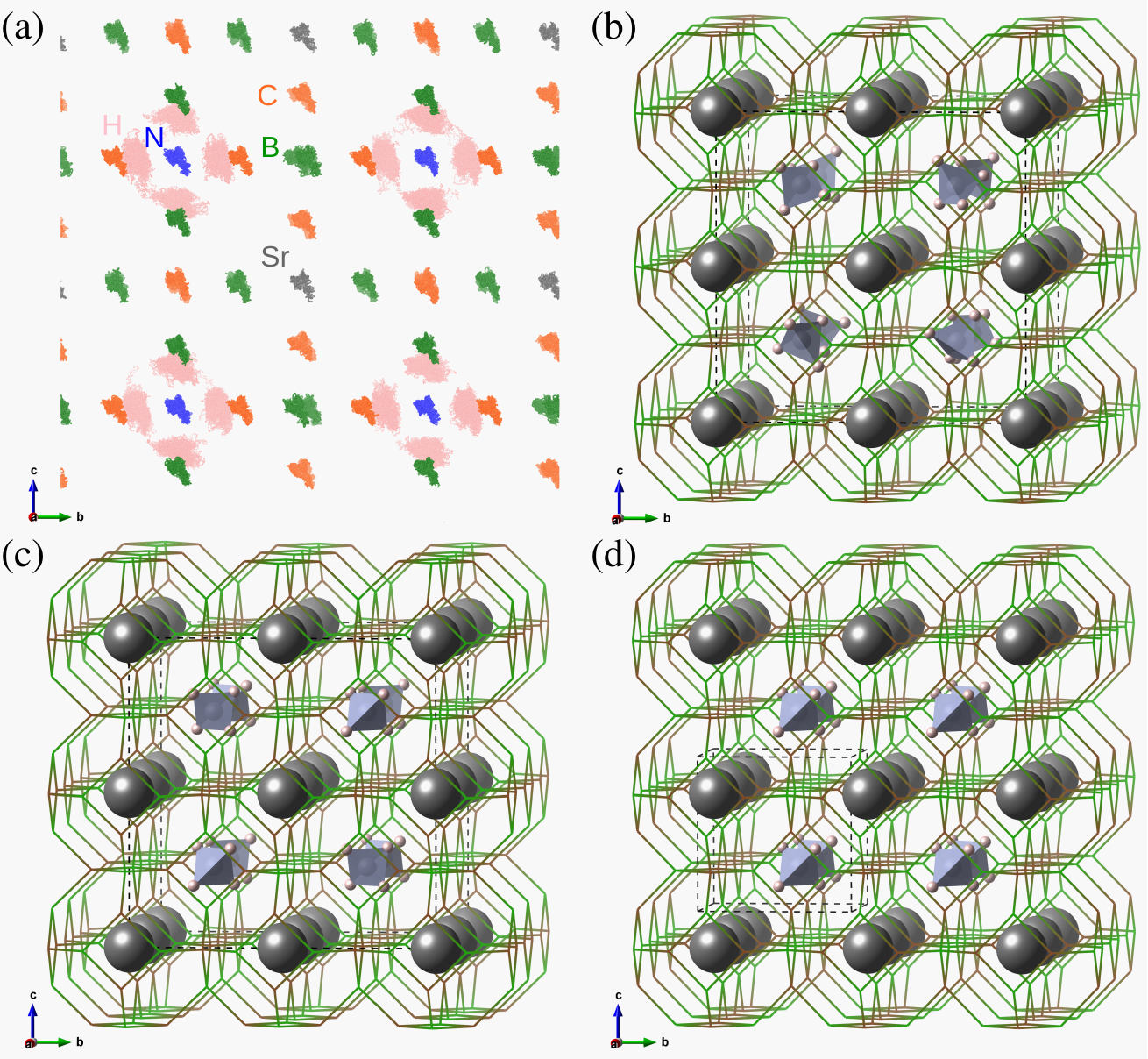}
	\caption{(a) The atomic trajectory of \textit{I}23--\ch{SrNH4B6C6} during the 30--40 \textit{ps} period in the molecular dynamics simulation at 100 K and ambient pressure, where the guest ammonium (\ch{NH4}) unit rotates at a small angle in the equilibrium position. Supercells of (b) \textit{I}23, (c) \textit{Fmmm}, and (d) \textit{P}1 phase of \ch{SrNH4B6C6}. The large, medium, and small spheres represent the Sr, N, and H atoms, respectively, the translucent regular tetrahedrons represent the \ch{NH4} units, and the gradient line represents the B--C framework. 
	\label{fig1}}
\end{figure}

\begin{figure}[t]
	\includegraphics[width=8.5cm,angle=0]{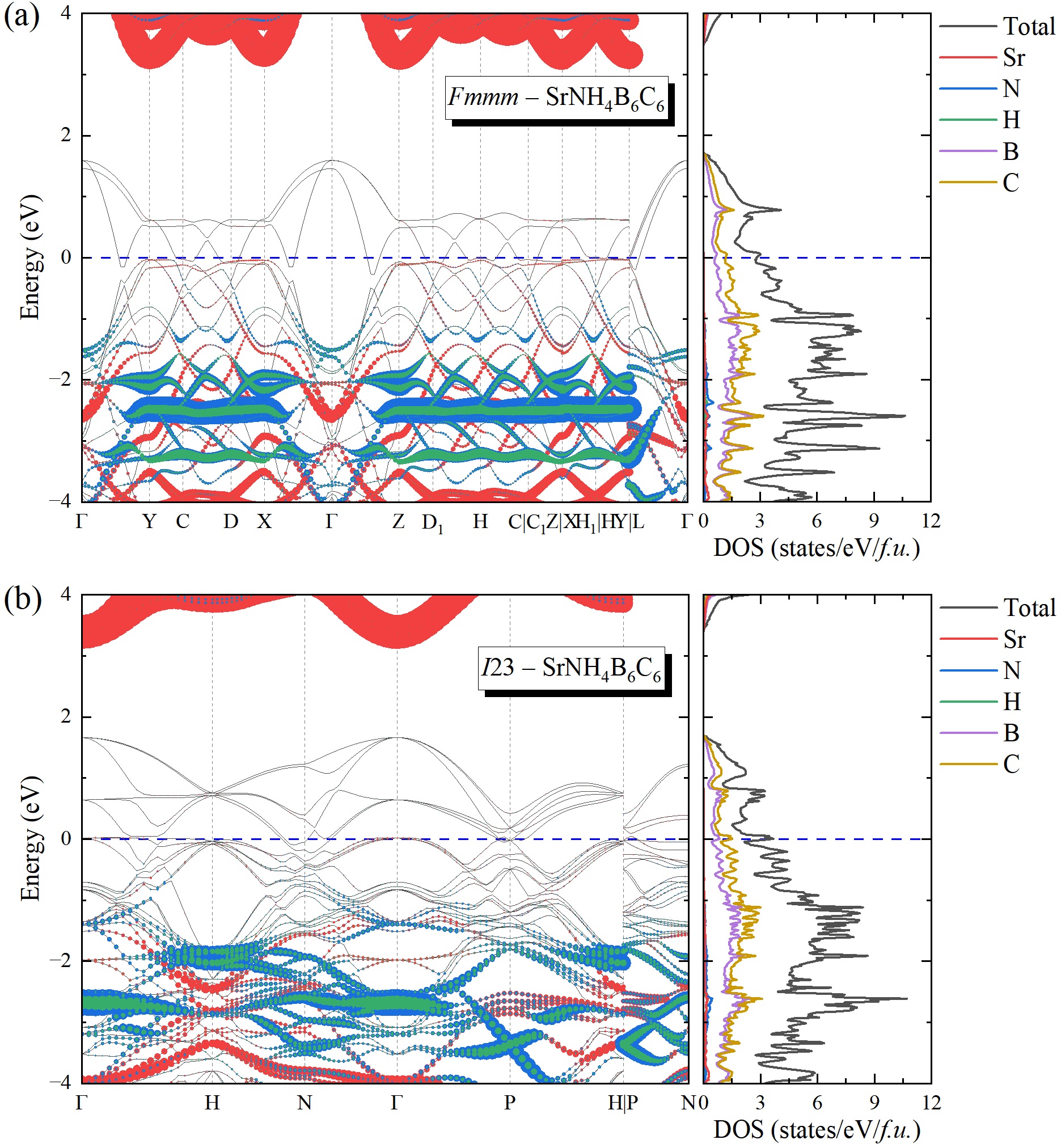}
	\caption{Electronic band structure (left panel) and projected density of states (DOS, right panel) for (a) \textit{Fmmm} and (b) \textit{I}23 phase of \ch{SrNH4B6C6} at ambient pressure. The bands projected onto the Sr, N, and H species are represented by red, blue, and green points, respectively. The size of each point is proportional to the weight of the corresponding orbital character. The blue dashed line indicates the Fermi energy. \label{fig2}}
\end{figure}

\begin{figure*}[t]
	\includegraphics[width=17cm,angle=0]{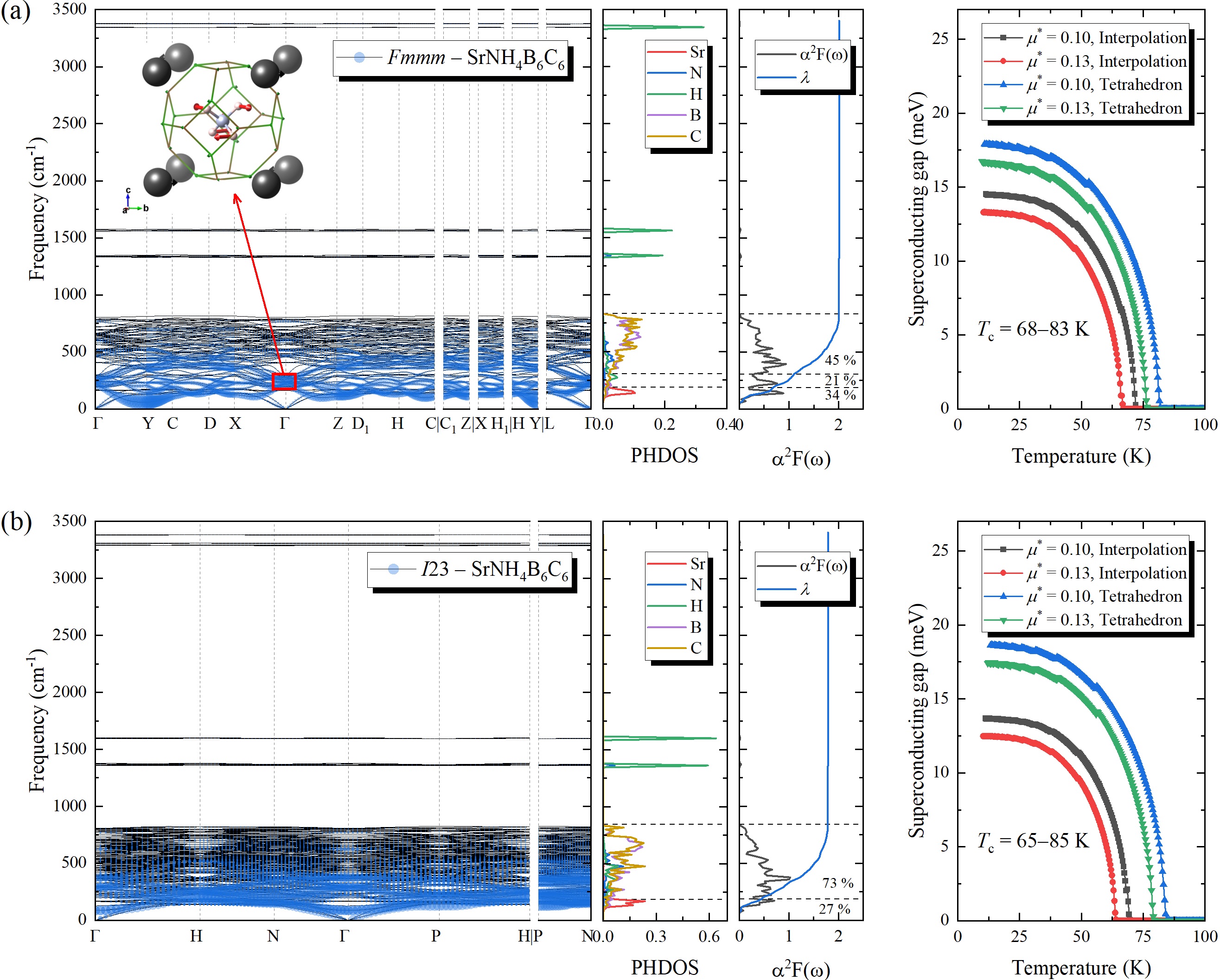}
	\caption{Phonon dispersion, phonon density of states (PHDOS), phonon spectroscopic function $\alpha^2F(\omega)$, electron-phonon integral $\lambda$, and calculated isotropic superconducting gap as a function of temperature for the (a) \textit{Fmmm} and (b) \textit{I}23 phase of \ch{SrNH4B6C6} at ambient pressure. The size of blue circles in the phonon dispersion curves is proportional to the electron-phonon coupling strength. The highest temperature at which the superconducting gap value is nonzero defines the $T_\text{c}$. \label{fig3}}
\end{figure*}

\begin{figure*}[t]
	\includegraphics[width=17cm,angle=0]{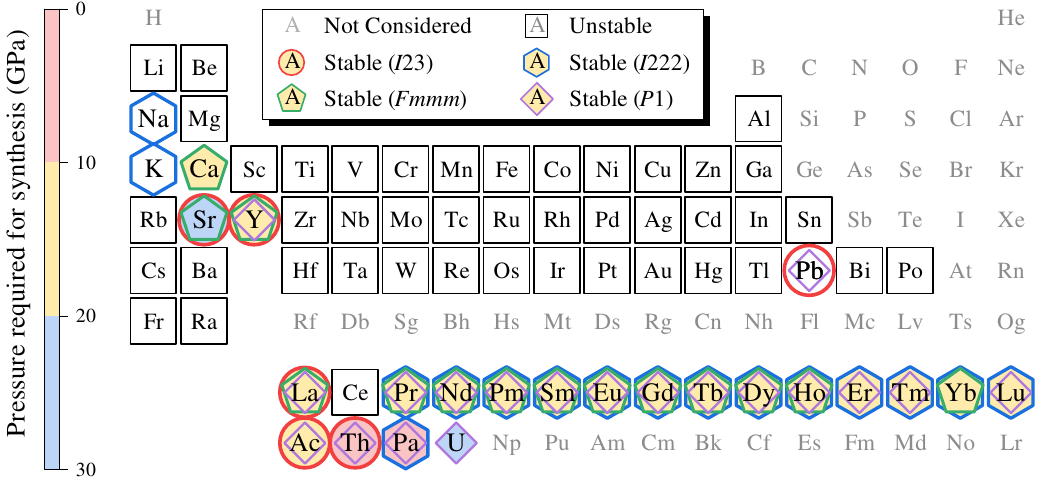}
	\caption{The stability periodic table of \ch{MNH4B6C6} in different symmetries at ambient pressure is presented. Phases that are dynamically unstable from phonon calculations are marked with a black square border, while dynamically stable phases are highlighted with colored borders. The red circle, blue hexagon, green pentagon, and purple diamond borders represent that the corresponding compound can be dynamically stable in $I$23, $I$222, $Fmmm$, and $P$1 phases, respectively. The colored block filling with pink, yellow, or blue indicates that the lowest thermally stable pressure (with respect to the ground-state elemental structures) of the corresponding compound falls within the 0--10, 10--20, and 20--30 GPa range, respectively.\label{fig4}}
\end{figure*}

\begin{figure*}[t]
	\includegraphics[width=15cm,angle=0]{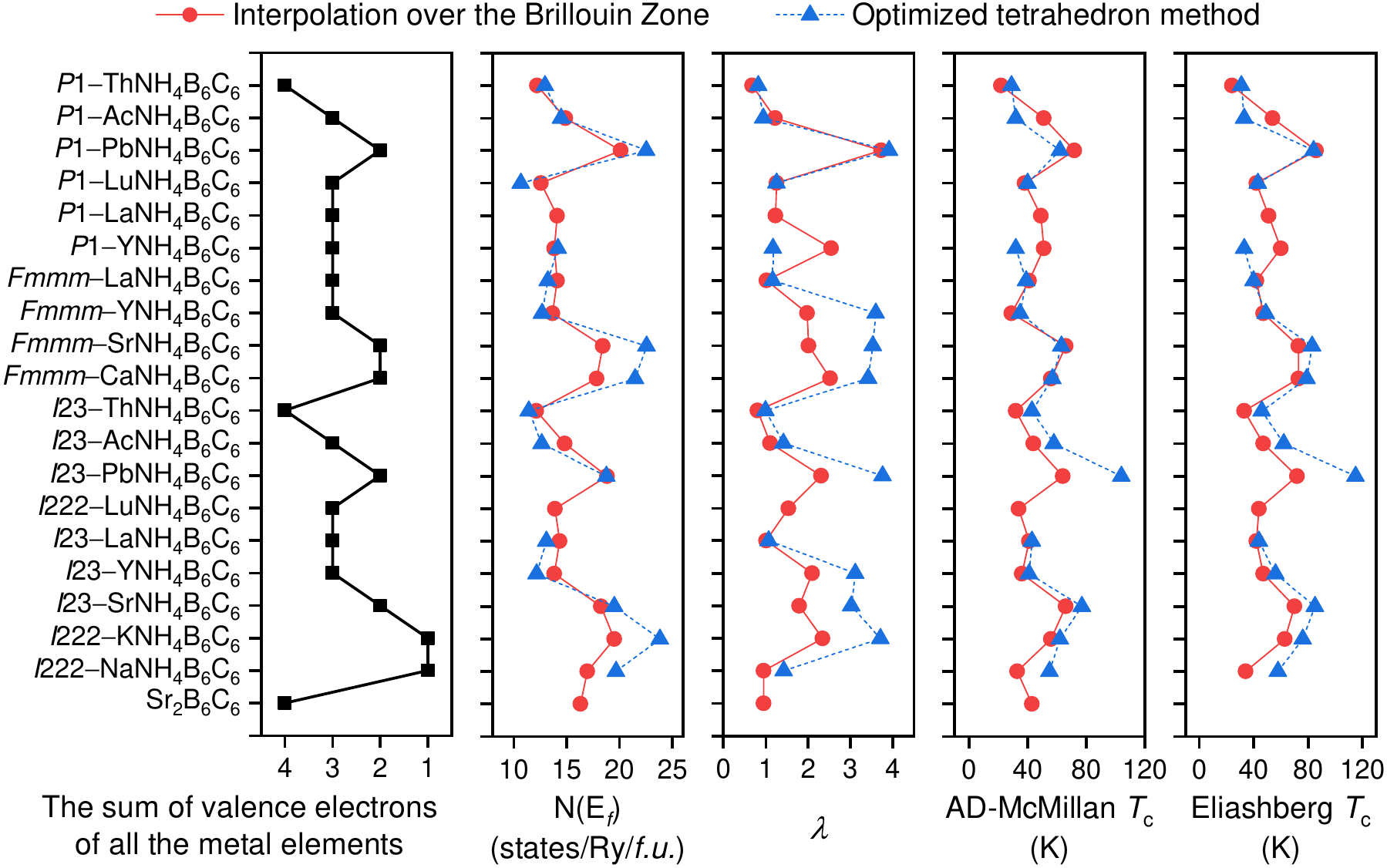}
	\caption{Superconducting mechanism of \ch{MNH4B6C6} clathrate structures: comparison of the total number of valence electrons of the metal elements in the B--C clathrate compounds, against the electronic density of states at the Fermi level ($N(E_f)$ in states/Ry/\textit{f.u.}), electron-phonon coupling constant ($\lambda$), and the critical temperature ($T_\text{c}$ in K) values estimated using Coulomb pseudopotential values of $\mu$$^\star$ = 0.10 by solving the AD-McMillan equation as well as the Eliashberg equation at ambient pressure. All the electron-phonon interaction properties were computed using the Interpolation over the Brillouin Zone method (circular data points connected by solid red lines) and the Optimized Tetrahedron method (triangular data points connected by blue dashed lines), respectively.}\label{fig5}
\end{figure*}

We initiate our study by substituting half of the strontium (Sr) atoms in \ch{SrB3C3} \cite{zhu2020} with ammonium (\ch{NH4}) units, as \ch{SrB3C3} serves as an experimentally verified superconductor, thereby offering a solid foundation for our investigation. 
Introducing new units into a crystal structure typically results in symmetry breaking, which can result in a larger primitive cell and necessitate more expensive calculations.  To circumvent this, we opted for half substitution of cations with \ch{NH4} units, which preserves much of the original high symmetry and offers an efficient approach to constructing our model system.
In the resulting \ch{SrNH4B6C6},  boron (B) and carbon (C) atoms form a type-VII cage-like host framework with Sr atoms and \ch{NH4} units located in the center of cages, and the \ch{NH4} units are isomorphic to the \ch{CH4} units in  $I$$\bar4$3$m$--\ch{CSH7} compound \cite{sun2020csh7}. 
To investigate the stability of \ch{SrNH4B6C6} at ambient pressure, we conducted \textit{ab initio} molecular dynamics (MD) simulations at 100 K and 0.001 kbar using the NPT ensemble \cite{Parrinello1980,Parrinello1981}, where N represents the number of particles, P the pressure, and T the temperature. These simulations were carried out using a supercell approach \cite{Parlinski1997} with 144 atoms and a time step of 2 \textit{fs}, spanning a total duration of 40 \textit{ps}.
The results, which include energy versus simulation time (Supplemental Material 
, Fig. S1) and atomic movement trajectories (Fig. 1a), suggest that \ch{SrNH4B6C6} is stable under these conditions. Particularly, our observations indicate that the \ch{NH4} units can rotate at a small angle with the N atom at the center, which is highlighted by the larger trajectory range displayed by the hydrogen atoms compared to other constituent atoms.

To evaluate the influence of \ch{NH4} rotation on the physical properties, drawing parallels with  $I$$\bar4$3$m$--\ch{CSH7} \cite{sun2020csh7}, $Fm\bar3m$--\ch{KB2H8} \cite{gao2021kb2h8}, and $R\bar3m$--\ch{CSH7} \cite{cui2020csh7}, we constructed three model structures: one where each \ch{NH4} unit is oriented uniquely, resulting in four distinct directions ($I$23, Fig. 1b), another with \ch{NH4} units in two different directions ($Fmmm$, Fig. 1c), and a third model where all \ch{NH4} units are oriented uniformly in a single direction ($P$1, Fig. 1d). 
Phonon calculations reveal that the $I$23 and $Fmmm$ phases are dynamically stable at ambient pressure, while the $P$1 phase is dynamically unstable (Fig. S2a--c). 
The relative enthalpy values of the $I$23 and $Fmmm$ phases compared to elemental substances within the pressure range of 0 to 50 GPa are shown in Fig. S2d, and the results reveal that the $I$23 and $Fmmm$ phases have almost identical enthalpy values and become thermally stable at pressures higher than $\sim$20.8 GPa.
The small enthalpy difference between the $I$23 and $Fmmm$ phases indicates that the orientation of \ch{NH4} may be random at finite temperatures because of the entropic term of the free energy. Ideally, the crystalline structure representing the randomly oriented case would be isotropic. However, accounting for such disordered structures in first-principles calculations presents significant challenges and is beyond the scope of this work.
It is worth noting that \ch{NH4} does not exist at ambient pressure, but is predicted to be thermodynamically stable at pressures above $\sim$50 GPa \cite{Qian2016}. 
The numerical discrepancy between the pressures at which \ch{SrNH4B6C6} and \ch{NH4} become thermodynamically stable supports our hypothesis that a B--C cage-like framework can aid in stabilizing high-pressure hydrides under reduced pressure conditions.

Moreover, we analyzed the phase stability relationships of Sr--N--H--B--C systems (Fig. S3e) using the Inverse Hull Webs method \cite{evans2021visualizing}, which offers information-dense 2D representations forgoing barycentric composition axes in favor of two energy axes: a formation-energy axis and a inverse-hull-energy axis. In this method, stable compounds are located in the negative range of the inverse-hull-energy axis, while metastable ones locate in the positive range. To aid in future experimental synthesis, we have compiled a comprehensive list of potential ambient-pressure synthesis paths for \ch{SrNH4B6C6} in Table S3, based on the stable elemental and binary compounds identified in the Inverse Hull Webs figures. It is important to note, however, that the actual feasible reactions may extend beyond those listed.

Similar to the parent compound \ch{SrB3C3} \cite{zhu2020}, both the newly proposed $I$23 and $Fmmm$ phases of \ch{SrNH4B6C6} exhibit metallic behavior at ambient pressure, with several bands crossing the Fermi level, as confirmed by electronic band structure calculations (Fig. 2). 
Boron and carbon atoms make a substantial contribution to the electronic density of states (DOS) at the Fermi level ($N(E_f)$), while the contributions from Sr, N, and H atoms appear to be minimal (Fig. 2a--b, right panel). This minimal contribution from Sr, N, and H species is further confirmed by the projected band structure in the left panel of Fig. 2. 
Interestingly, while the DOS of \ch{SrB3C3} exhibits a noticeable slope near the Fermi energy \cite{zhu2023}, the DOS of the two new \ch{SrNH4B6C6} phases display a distinct van Hove singularity,
suggesting that \ch{SrBH4B6C6} may exhibit $T_\text{c}$ higher than the parent \ch{SrB3C3}.

Therefore, we further investigated their superconductive properties. 
The computed electron-phonon coupling constant ($\lambda$) for the $I$23 and $Fmmm$ phases of \ch{SrNH4B6C6} are 1.79 and 2.01, respectively (Table S1). These values significantly exceed that of \ch{SrB3C3} (0.95) \cite{zhu2023}.
The high $\lambda$ values in \ch{SrNH4B6C6} are primarily contributed by the low-frequency vibrational modes, while the high-frequency modes dominated by H atoms have no contribution (Fig. 3).
More specifically, the low-frequency vibrational modes for the $I$23 phase, ranging between 0--842 cm$^{-1}$ can be broken down into two segments: those dominated by Sr and the B--C framework, contributing 27\% and 73\% to $\lambda$, respectively (Fig. 3b). In contrast, for the $Fmmm$ phase, the low-frequency vibrational modes (0--833 cm$^{-1}$) can be divided into three parts. In addition to the modes dominated by Sr, and B--C framework, contributing 34\% and 45\% to $\lambda$, respectively, there is a third segment dominated by H, which is a departure from the $I$23 phase. Arising from the rotational vibration modes of the \ch{NH4} complex (as shown in the inside view of Fig. 3a), the H-dominated segment contributes 21\% to $\lambda$, and, coupled with the notable phonon softening observed at the Y point, it results in an estimated $T_\text{c}$ for the $Fmmm$ phase (73 K) marginally superior to that of the $I$23 phase (70 K). 
In addition to using the interpolation over the Brillouin Zone method \cite{wierzbowska2006interpolated}, we also employed the optimized tetrahedron method \cite{mitsuaki2014opttetra} to calculate electron-phonon coupling properties. This choice was motivated by the fact that the mode-dependent electron-phonon coupling constant exactly diverges at the $\Gamma$ point. Calculations using the optimized tetrahedron method, which excludes the $\Gamma$ point, indicated that estimated $T_\text{c}$ values for $Fmmm$ and $I$23 phases reach 83--85 K (Fig. 3),  such $T_\text{c}$ values are more than double that of its precursor, \ch{SrB3C3} (31 K, \cite{zhu2023}). 
These results underscore the critical role that the orientation of \ch{NH4} units plays in shaping the superconducting properties of B--C clathrates. A higher $T_\text{c}$ in \ch{SrNH4B6C6} could potentially be achieved by exploring various orientation scenarios of the \ch{NH4} units.

Our findings validate the strategy of integrating hydride units into B--C clathrate superconductors, a method that stabilizes hydrides at reduced pressures and enhances the $T_\text{c}$ of B--C clathrate compounds, which motivates us to employ this technique on a broader array of isomorphic compounds.

High-throughput calculations were then carried out to investigate B--C clathrate materials containing \ch{NH4} and electron-positive elements, by substituting strontium atoms in various phases of \ch{SrNH4B6C6} with other metal atoms.
All metallic elements (M) in the periodic table, excluding transuranic elements, were taken into consideration.
Among the 68 \ch{MNH4B6C6} compounds (a total of 204 crystal structures) studied, 24 compounds (comprising 55 crystal structures, with structural parameters listed in Table S2) were confirmed to be dynamically stable at ambient pressure. These compounds stabilized in the $I$23, $I$222, $Fmmm$, or $P$1 phases (Fig. 4), with the $I$222 phase obtained by a slight rotation of \ch{NH4} units in the $I$23 phase.
The ambient-pressure dynamic stability of \ch{MNH4B6C6} was restricted to metal cations with ionic radii similar to \ch{NH4} ($\sim$1.0 \AA). The distribution of stable compounds in the periodic table follows the diagonal rule, highlighting the importance of a good size match between the B--C framework, \ch{NH4} units, and the metal cations. 
Thermal stability of dynamically stable phases has also been studied by calculating the inverse hull web and the relative enthalpy values of \ch{MNH4B6C6} phases compared to the constituent elements at 0--200 GPa, as shown in Fig. S3--25.  Notably, \ch{ThNH4B6C6} and \ch{PaNH4B6C6} become thermally stable at pressures down to $\sim$6.4 and $\sim$9.8 GPa, respectively, which is yet another strong evidence that B--C cage-like framework can aid in stabilizing high-pressure hydrides at lower pressure conditions.

We next explored the superconductivity of the dynamically stable \ch{MNH4B6C6} compounds, excluding those containing elements with partially filled \textit{f}-shells. Recent research has indicated that B--C clathrates containing elements with partially filled \textit{f}-shells, such as \ch{CeB3C3} and \ch{SmB3C3}, could display heavy-fermion behavior \cite{Rao2023}, which are expected to exhibit unconventional superconductivity. 
Through this study, we identified 10 stable \ch{MNH4B6C6} compounds as ambient-pressure superconductors, with $T_\text{c}$ values reaching up to 115 K in $P$1--\ch{PbNH4B6C6}, exceeding the boiling point (77 K) of liquid nitrogen. Although these \ch{MNH4B6C6} compounds possess identical stoichiometry, and some even share the same phases with the same space group, their superconducting attributes differ noticeably. This variation stems from their characteristic vibrational frequencies, electron-phonon coupling constants, and electronic density of states at the Fermi level, as listed in Table S1. 

Moreover, we compared the total number of valence electrons, $N(E_f)$, $\lambda$, and $T_\text{c}$ across \ch{SrB3C3} and a series of \ch{MNH4B6C6} compounds. Our findings, as illustrated in Fig. 5, show a direct correlation between $\lambda$ and $T_\text{c}$ with ($N(E_f)$), and an inverse relationship between the total number of valence electrons in the compounds and their $T_\text{c}$, indicating that fewer valence electrons in the metal elements correlate with higher critical temperatures. 
Table S1 shows the $T_\text{c}$ values obtained by solving both the Migdal-Eliashberg equation and the AD-McMillan equation.  The combined use of these equations provides a comprehensive prediction range for $T_\text{c}$ values and has proven to be informative. For instance, the material \ch{(La,Y)H10}, which has been experimentally verified as a high-temperature superconductor with a measured $T_\text{c}$ of up to 253 K at 183 GPa, shows that \ch{LaYH20} exhibits a $\lambda$ value as high as 3.87 at 180 GPa. Using the Migdal-Eliashberg and AD-McMillan equations, the predicted $T_\text{c}$ values range from 281 K to 300 K and from 232 K to 266 K, respectively, consistent with the measured $T_\text{c}$ value \cite{semenok2021superconductivity}. However, a recent study showed that the Migdal-Eliashberg theory loses validity for $\lambda$ values between 3.0 and 3.7, regardless of the underlying model Hamiltonian \cite{yuzbashyan2022breakdown}. Therefore, these predicted $T_\text{c}$ values may be questionable and suggest a need for refining theoretical models and conducting further experimental verifications to ensure accuracy at high $\lambda$ values.

Anharmonicity can influence both the dynamic stability and superconductivity of hydrides \cite{Belli2022,Lucrezi2023}. In our work, we explored anharmonic effects in \ch{NH4}-doped boron-carbon clathrate compounds by examining  $Fmmm$--\ch{SrNH4B6C6} at 0 K and ambient pressure. The comparison between anharmonic and harmonic phonon dispersions (Fig. S26) shows that $Fmmm$--\ch{SrNH4B6C6} displays only a mild degree of anharmonicity. This slight anharmonicity leads to the softening of high-frequency phonon modes, yet it does not adversely affect the structural stability of the compound.  In addition, as mentioned above, the high-frequency modes dominated by H atoms do not contribute to superconductivity. Therefore, anharmonic effects are not expected to have a significant impact on the superconductivity of \ch{NH4}-doped boron-carbon clathrate compounds.

\section*{Conclusion}
\label{conclusion}

In summary, we proposed a novel approach of incorporating hydride units into B--C clathrate superconductors, heralding a new frontier in the search for high-temperature superconductors at ambient pressure. By uniting the structural robustness of $sp^3$-bonded B--C frameworks with the superior $T_\text{c}$ intrinsic to hydrides, we have unveiled a series of new \ch{MNH4B6C6} superconductors, which exhibit dynamic stability and promising superconducting properties at ambient pressure. The predicted values of $T_\text{c}$ substantially higher than those of conventional B--C compounds, particularly a $T_\text{c}$
soaring to 85 K in \ch{SrNH4B6C6} and up to 115 K in \ch{PbNH4B6C6}, marks a significant leap toward the vision of practical high-temperature superconductivity. 
Our research further reveals that the orientation of the \ch{NH4} units plays a critical role in shaping the superconducting properties of B--C clathrate frameworks. While our proposed materials are pending experimental synthesis and confirmation, the calculated negative formation enthalpies fortify our anticipation for successful synthesis endeavors.
Looking forward,  as the pursuit of ambient superconductors continues, our work could serve as a foundational stone, laying down the principles and strategies that future investigations can build upon.
By combining diverse hydride units---such as \ch{SH6} in \ch{H3S} \cite{duan2014}, \ch{ReH9} in \ch{BaReH9} \cite{Muramatsu2015}, and \ch{TeH12} and \ch{TeH14} in \ch{H4Te} \cite{Zhong2016}---with various clathrate compounds (including but not limited to B--C, B--N, and B--Si clathrate materials structured in different types of clathrate frameworks), we could potentially discover new ambient-pressure high-temperature superconductors.

\section*{Methods}

Geometry optimizations, the isothermal-isobaric ensemble (NPT ensemble) molecular dynamics (MD) simulations, as well as electronic structure calculations including band structures and densities of states (DOS), were performed using density functional theory (DFT) in the Vienna ab-initio Simulation Package (VASP) \cite{Kresse1993,Kresse1999}.
The exchange-correlation functional was chosen to be the Perdew-Burke-Ernzerhof generalized gradient approximation \cite{Perdew1996} with the Semiempirical Grimme's DFT-D2 type of Van der Waals correction \cite{Grimme2006}.
The electron-ion interaction is described by projector-augmented-wave potentials \cite{Blchl1994}, and the official recommended potentials for DFT calculations \cite{vasppaw} were used.
A kinetic cutoff energy of 800 eV and Monkhorst-Pack \textit{k} meshes with a grid spacing of 0.15 \AA $^{-1}$ were adopted to ensure the enthalpy convergence to better than 1 meV/atom. Calculations were carried out at 0.001 kbar to approximate 1 \textit{atm.} pressure.
The NPT MD simulations were performed at 100 K and 0.001 kbar using a supercell (144 atoms), with temperature and pressure controlled using the Parinello-Rahman algorithm \cite{Parrinello1980,Parrinello1981} method.

For all the \ch{MNH4B6C6} compounds, phonon calculations were carried out using a finite displacement method through the supercell approach \cite{Parlinski1997} (144 atoms) using the VASP package \cite{Kresse1993,Kresse1999} coupled to the PHONOPY code \cite{phonopy-phono3py-JPCM,phonopy-phono3py-JPSJ}.

Additionally, for $Fmmm$--\ch{SrNH4B6C6}, a harmonic phonon calculation was also preformed using a finite displacement approach with a 144-atom supercell in PHONOPY \cite{phonopy-phono3py-JPCM,phonopy-phono3py-JPSJ} and Quantum-ESPRESSO \cite{Giannozzi2009} software. Then, an anharmonic calculation in the NVT ensemble at 0 K and ambient pressure was preformed using the Quantum-ESPRESSO \cite{Giannozzi2009} software within the SSCHA \cite{monacelli2021stochastic} method.

Electron-phonon coupling (EPC) calculations of dynamically stable non-\textit{f}-electron compounds were carried out using the Quantum ESPRESSO (QE) program \cite{Giannozzi2009}. 
The displacement vectors were visualized by using the ``Interactive phonon visualizer'' tool in the QE program.
The pseudopotentials employed were obtained from the standard solid-state pseudopotentials (SSSP) library \cite{ssspv130} optimized for efficiency, with a kinetic energy cutoff for wave functions of 80 Ry and a kinetic energy cutoff for charge density and potential of 640 Ry.
The EPC properties were computed using the Interpolation over the Brillouin Zone method \cite{wierzbowska2006interpolated} and the optimized tetrahedron method \cite{mitsuaki2014opttetra}, respectively. For calculations using the interpolation over the Brillouin Zone method, 
the Brillouin zone sampling $\Gamma$-centered Monkhorst-Pack \cite{Monkhorst1976} scheme was applied using Methfessel-Paxton \cite{Methfessel1989} smearing with a broadening width of 0.030 Ry.
Different \textit{\textbf{q}} meshes were used for different phases to compute the EPC matrix elements: 3 $\times$ 3 $\times$ 3 \textit{\textbf{q}} meshes and 12 $\times$ 12 $\times$ 12 \textit{\textbf{k}} meshes were used for $Fmmm$ (36 atoms) and $P$1 (18 atoms) phases, while 2 $\times$ 2 $\times$ 2 \textit{\textbf{q}} meshes  and 8 $\times$ 8 $\times$ 8 \textit{\textbf{k}} meshes were used for $I$23 (72 atoms) and $I$222 (72 atoms) phases.
The EPC constant $\lambda$ and $\omega_{log}$ were calculated using a set of Gaussian broadenings with an increment of 0.005 Ry from 0.0 to 0.050 Ry and converged to 0.030 Ry.
For calculations using the optimized tetrahedron method, 3 $\times$ 3 $\times$ 3 \textit{\textbf{q}} meshes and 12 $\times$ 12 $\times$ 12 \textit{\textbf{k}} meshes were used for $Fmmm$ (36 atoms) and $P$1 (18 atoms) phases, while 3 $\times$ 3 $\times$ 3 \textit{\textbf{q}} meshes and 6 $\times$ 6 $\times$ 6 \textit{\textbf{k}} meshes were used for $I$23 (72 atoms) and $I$222 (72 atoms) phases.

The $T_\text{c}$s were estimated using the Allen-Dynes-corrected McMillan (AD-M) equation \cite{Allen1975}
\begin{equation}
	T_c = f_1f_2\frac{\omega_{log}}{1.2}exp\left[-\frac{1.04(1+\lambda)}{\lambda-\mu^{*}(1+0.62\lambda)}\right ] \label{eq:ADM}
\end{equation}
with renormalized Coulomb potential $\mu$$^\star$ values of 0.10 and 0.13.
$f_{1}$ and $f_{2}$ are the strong coupling factor and shape factor given by
\begin{equation}
	f_ {1} =\left[1+\left[\frac {\lambda }{2.46(1+3.8\mu^{*})}\right]^{3/2}\right]^{1/3} \label{eq:f1}
\end{equation}
\begin{equation}
	f_{2}=1+\frac{(\omega_{2}/\omega_{log}-1)\lambda^{2}}{\lambda^{2}+\left[1.82(1+6.3\mu^{*})(\omega_{2}/\omega_{\log})\right]^{2}}.\label{eq:f2}
\end{equation}
The $T_\text{c}$s were also calculated by numerically solving the isotropic Migdal-Eliashberg equations \cite{eliashberg1960} by using the Elk Code \cite{elk}.

\section*{Data availability}

All data are available in the paper and from the author	upon request.

\section*{References}

\bibliography{M-NH4-B6C6}

\section*{acknowledgments}

This work was supported by the startup funds of the office of the Dean of SASN of Rutgers University-Newark. The authors acknowledge the Office of Advanced Research Computing (OARC) at Rutgers for providing access to the Amarel cluster and associated research computing resources.

\section*{AUTHOR CONTRIBUTIONS}

Y.S. performed simulations and analyzed the data.
L.Z. proposed the initial idea and supervised the project.
Y.S. and L.Z. co-wrote the manuscript.

\section*{COMPETING INTERESTS}

The authors declare no competing interests.

\section*{ADDITIONAL INFORMATION}

\textbf{Supplementary information} The online version contains supplementary material.
\newpage
\begin{widetext}
\begin{appendices}


\section*{Supplementary material}

# Supplemental Material of "Hydride Units Filled B-C Clathrate: A New Pathway for High-Temperature Superconductivity at Ambient Pressure"
Ying Sun and Li Zhu*

*Department of Physics, Rutgers University, Newark, New Jersey 07102, United States*

## CONTENTS

## 1.1 SrNH$_4$B$_6$C$_6$

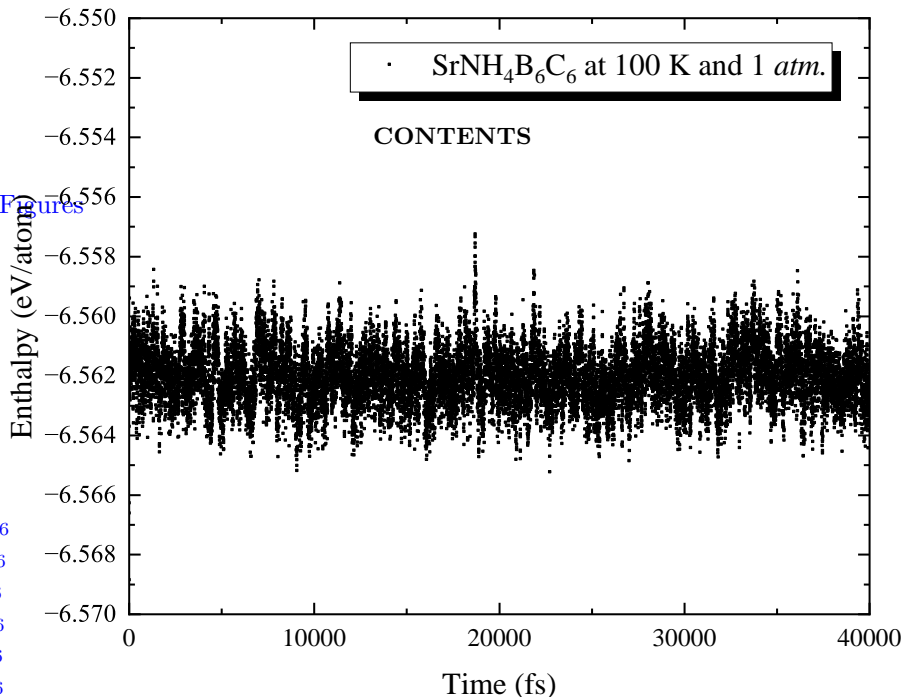

FIG. S1. Enthalpy evolution of SrNH$_4$B$_6$C$_6$ during the 100 K metadynamics simulations at ambient pressure.

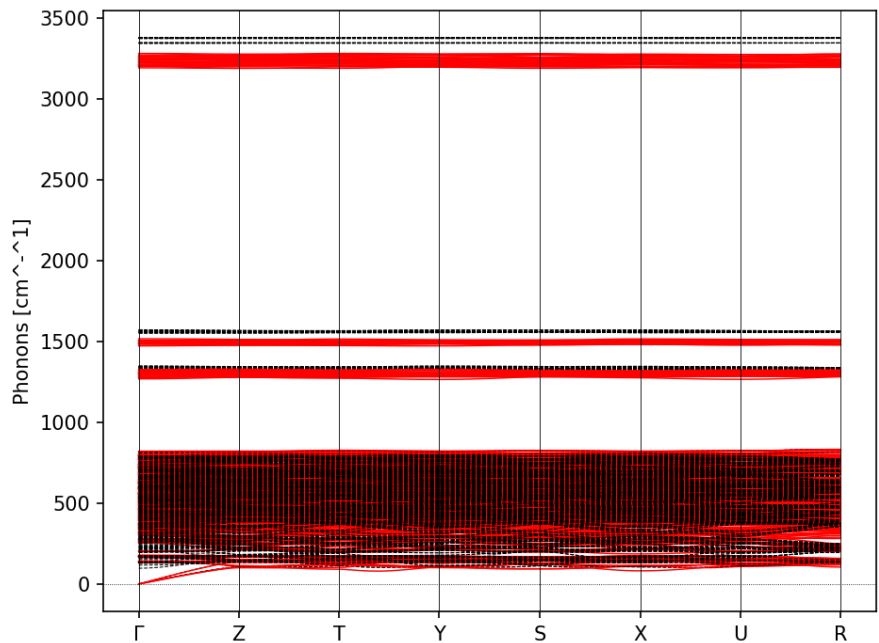

FIG. S2. Comparison between the SSCHA anharmonic (red lines) and the harmonic (black dash lines) phonon dispersion of $Fmmm$ – SrNH$_4$B$_6$C$_6$ at 0 K and ambient pressure.

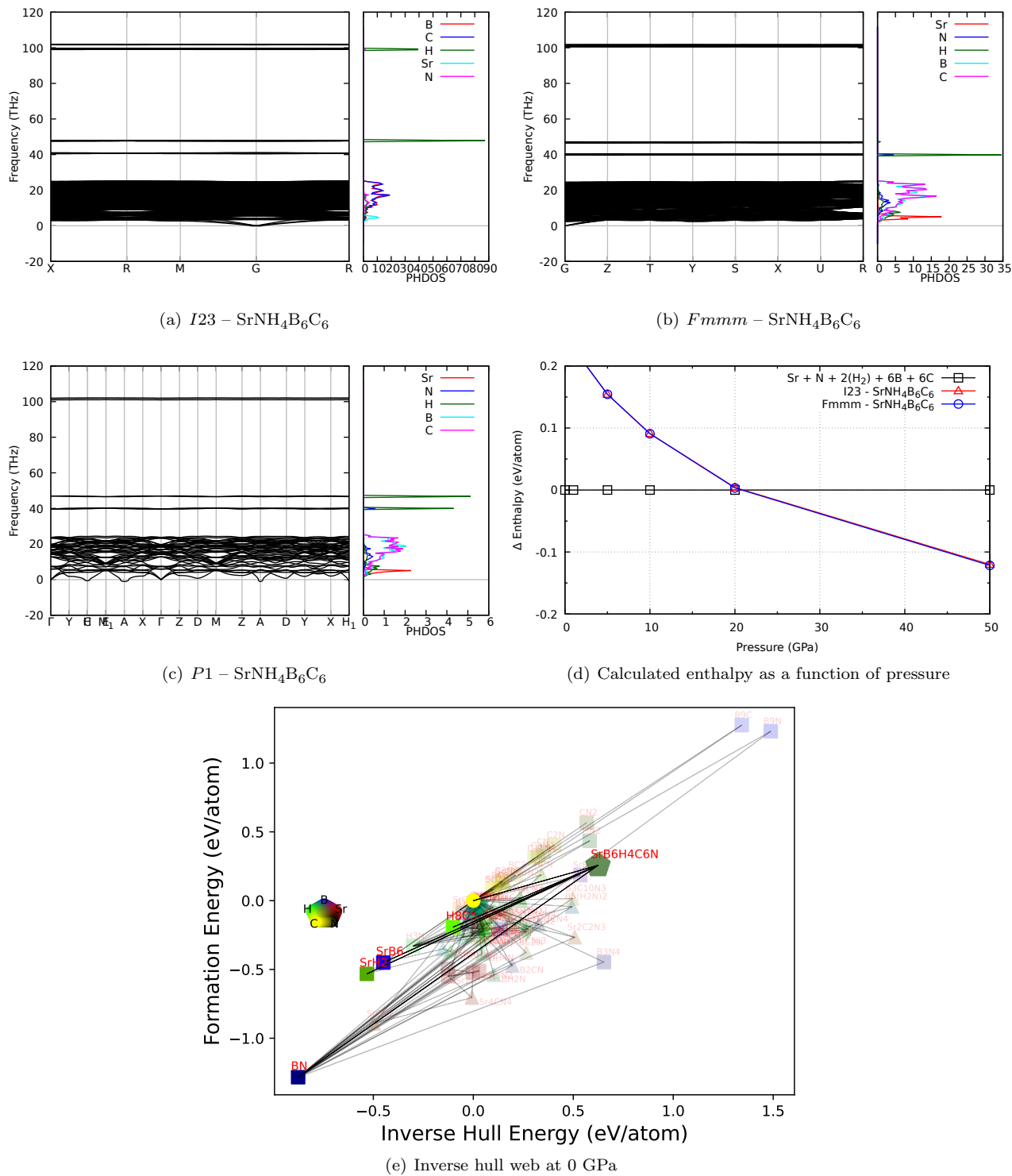

FIG. S3. (a–c) Phonon dispersion (left panel) and projected phonon densities of states (PHDOS, right panel) of different $SrNH_4B_6C_6$ phases at ambient pressure. The absence of an imaginary frequency in the phonon spectra indicates that the structures are dynamically stable at ambient pressure. (d) Calculated enthalpy as a function of pressure for $SrNH_4B_6C_6$ relative to simple elemental substance, and the phase transition pressure is ∼20.8 GPa. (e) Inverse hull web of Sr-N-H-B-C system at 0 K and ambient pressure. Compounds located in the positive range of the inverse-hull-energy axis indicates that they are metastable, while the others located in the negative range of the inverse-hull-energy axis are stable ones. We leave the $SrNH_4B_6C_6$ compound and its hull reactants opaque, while we make all other phases partially transparent. Ground state structures of simple substances and other compounds were collected from the Inorganic Crystal Structure Database (ICSD) [1].

## 1.2 NaNH$_4$B$_6$C$_6$

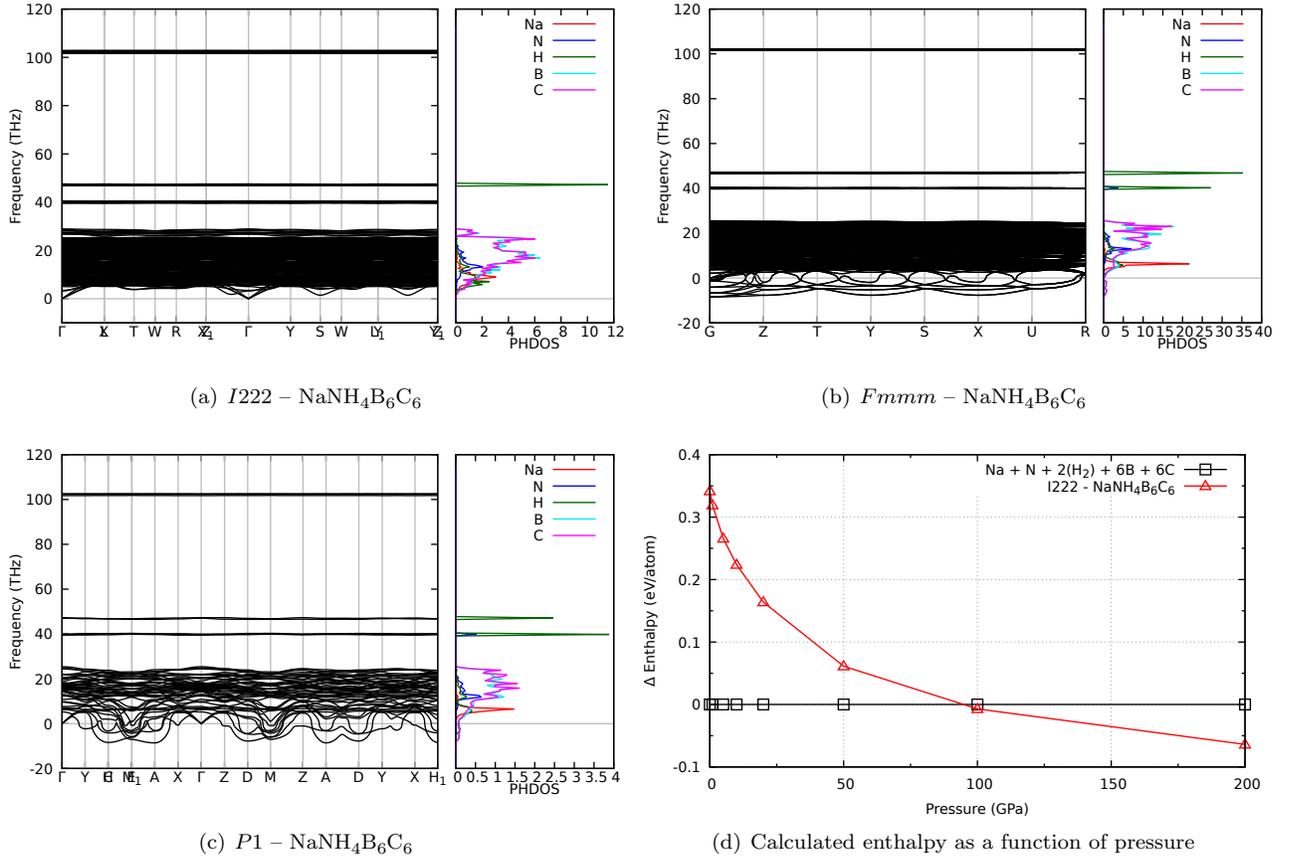

(a) $I222$ – NaNH$_4$B$_6$C$_6$

(b) $Fmmm$ – NaNH$_4$B$_6$C$_6$

(c) $P1$ – NaNH$_4$B$_6$C$_6$

(d) Calculated enthalpy as a function of pressure

(e) Inverse hull web at 0 GPa

FIG. S4. (a–c) Phonon dispersion (left panel) and projected phonon densities of states (PHDOS, right panel) of different NaNH$_4$B$_6$C$_6$ phases at ambient pressure. The absence of an imaginary frequency in the phonon spectra indicates that the structures are dynamically stable at ambient pressure. (d) Calculated enthalpy as a function of pressure for NaNH$_4$B$_6$C$_6$ relative to simple elemental substance, and the phase transition pressure is ∼95 GPa. (e) Inverse hull web of Na-N-H-B-C system at 0 K and ambient pressure. Compounds located in the positive range of the inverse-hull-energy axis indicates that they are metastable, while the others located in the negative range of the inverse-hull-energy axis are stable ones. We leave the NaNH$_4$B$_6$C$_6$ compound and its hull reactants opaque, while we make all other phases partially transparent. Ground state structures of simple substances and other compounds were collected from the Inorganic Crystal Structure Database (ICSD) [1].

## 1.3 KNH$_4$B$_6$C$_6$

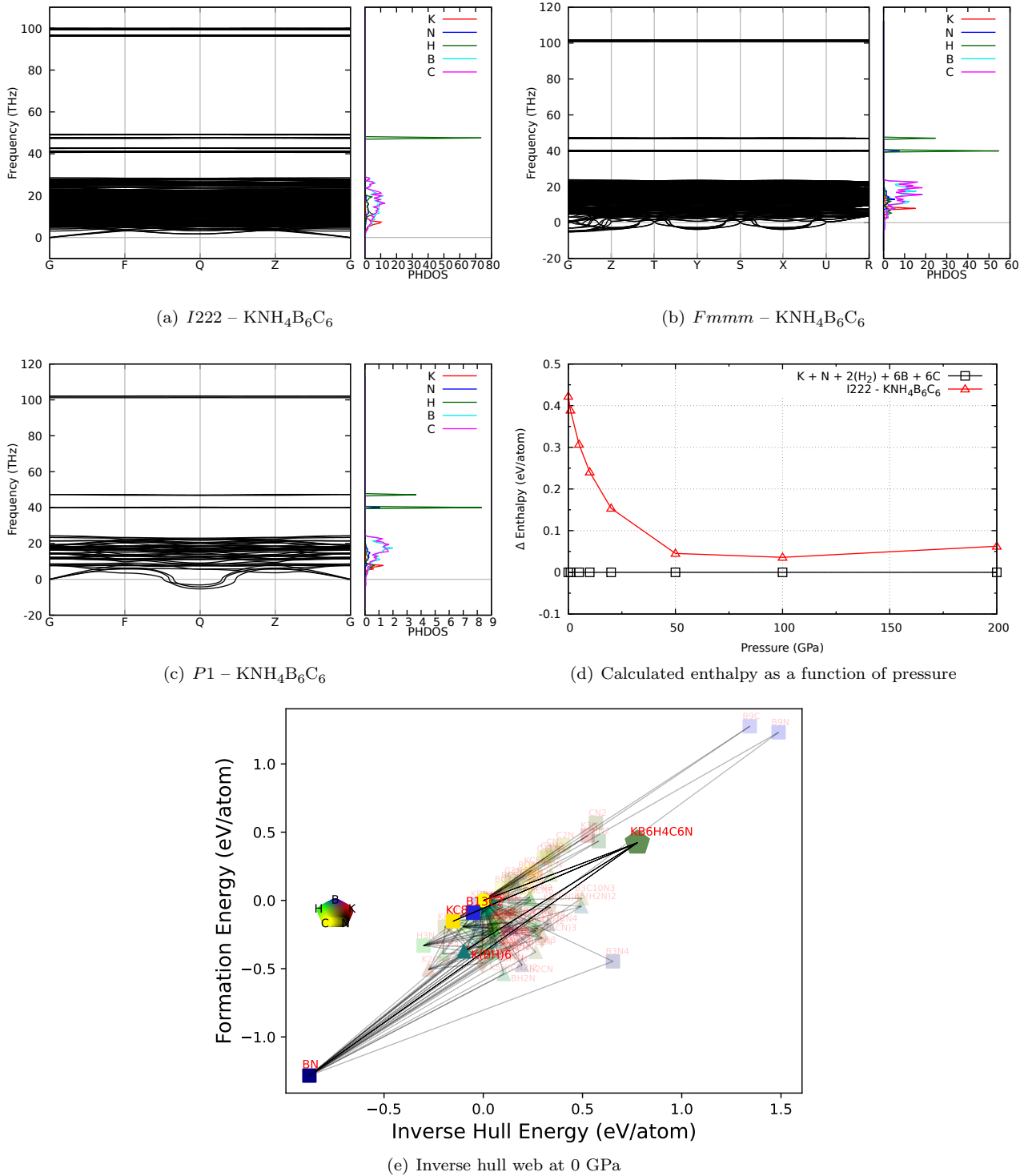

FIG. S5. (a–c) Phonon dispersion (left panel) and projected phonon densities of states (PHDOS, right panel) of different KNH$_4$B$_6$C$_6$ phases at ambient pressure. The absence of an imaginary frequency in the phonon spectra indicates that the structures are dynamically stable at ambient pressure. (d) Calculated enthalpy as a function of pressure for KNH$_4$B$_6$C$_6$ relative to simple elemental substance. (e) Inverse hull web of K-N-H-B-C system at 0 K and ambient pressure. Compounds located in the positive range of the inverse-hull-energy axis indicates that they are metastable, while the others located in the negative range of the inverse-hull-energy axis are stable ones. We leave the KNH$_4$B$_6$C$_6$ compound and its hull reactants opaque, while we make all other phases partially transparent. Ground state structures of simple substances and other compounds were collected from the Inorganic Crystal Structure Database (ICSD) [1].

## 1.4 $CaNH_4B_6C_6$

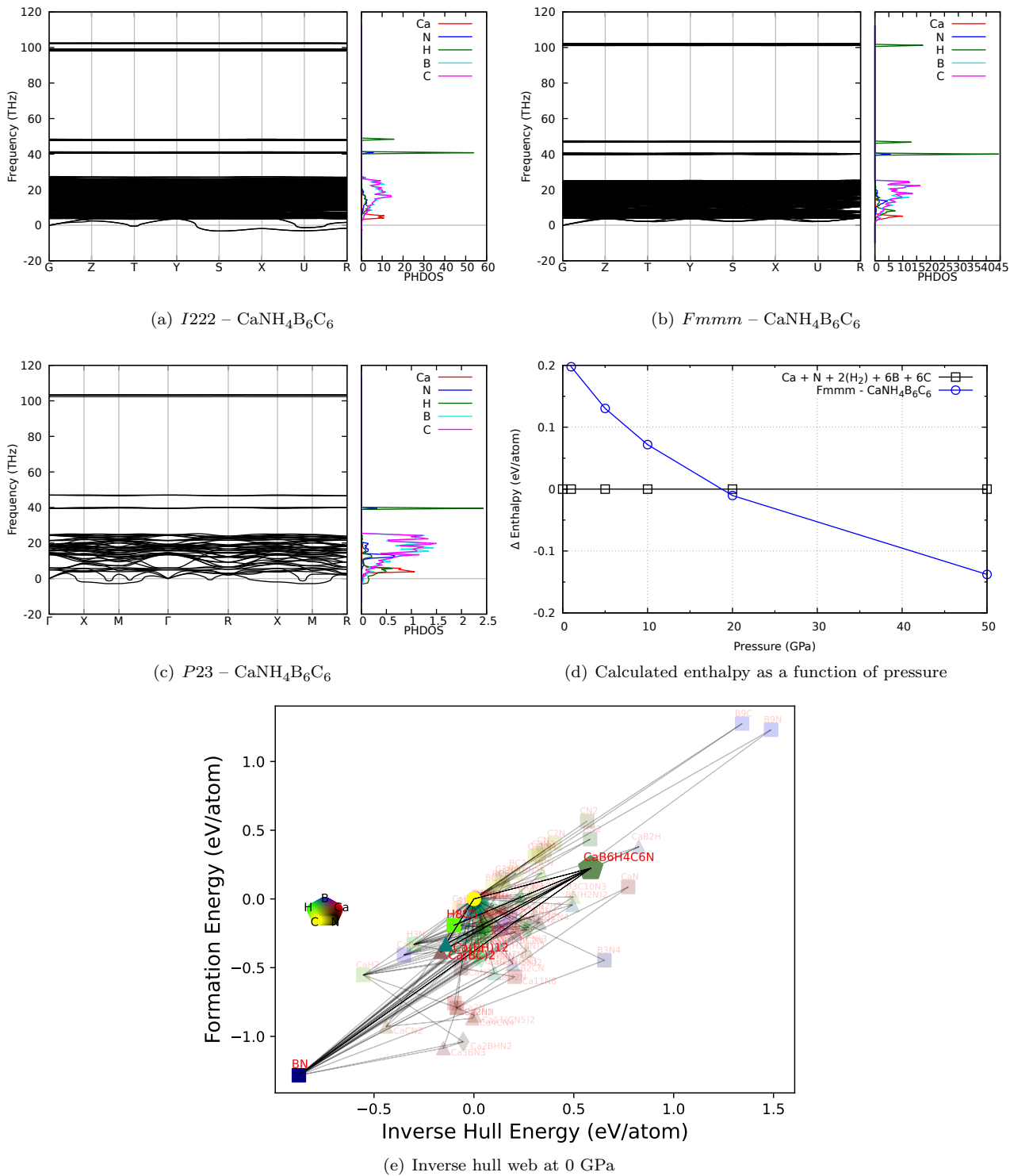

FIG. S6. (a–c) Phonon dispersion (left panel) and projected phonon densities of states (PHDOS, right panel) of different $CaNH_4B_6C_6$ phases at ambient pressure. The absence of an imaginary frequency in the phonon spectra indicates that the structures are dynamically stable at ambient pressure. (d) Calculated enthalpy as a function of pressure for $CaNH_4B_6C_6$ relative to simple elemental substance, and the phase transition pressure is ∼18.7 GPa. (e) Inverse hull web of Ca-N-H-B-C system at 0 K and ambient pressure. Compounds located in the positive range of the inverse-hull-energy axis indicates that they are metastable, while the others located in the negative range of the inverse-hull-energy axis are stable ones. We leave the $CaNH_4B_6C_6$ compound and its hull reactants opaque, while we make all other phases partially transparent. Ground state structures of simple substances and other compounds were collected from the Inorganic Crystal Structure Database (ICSD) [1].

## 1.5 YNH$_4$B$_6$C$_6$

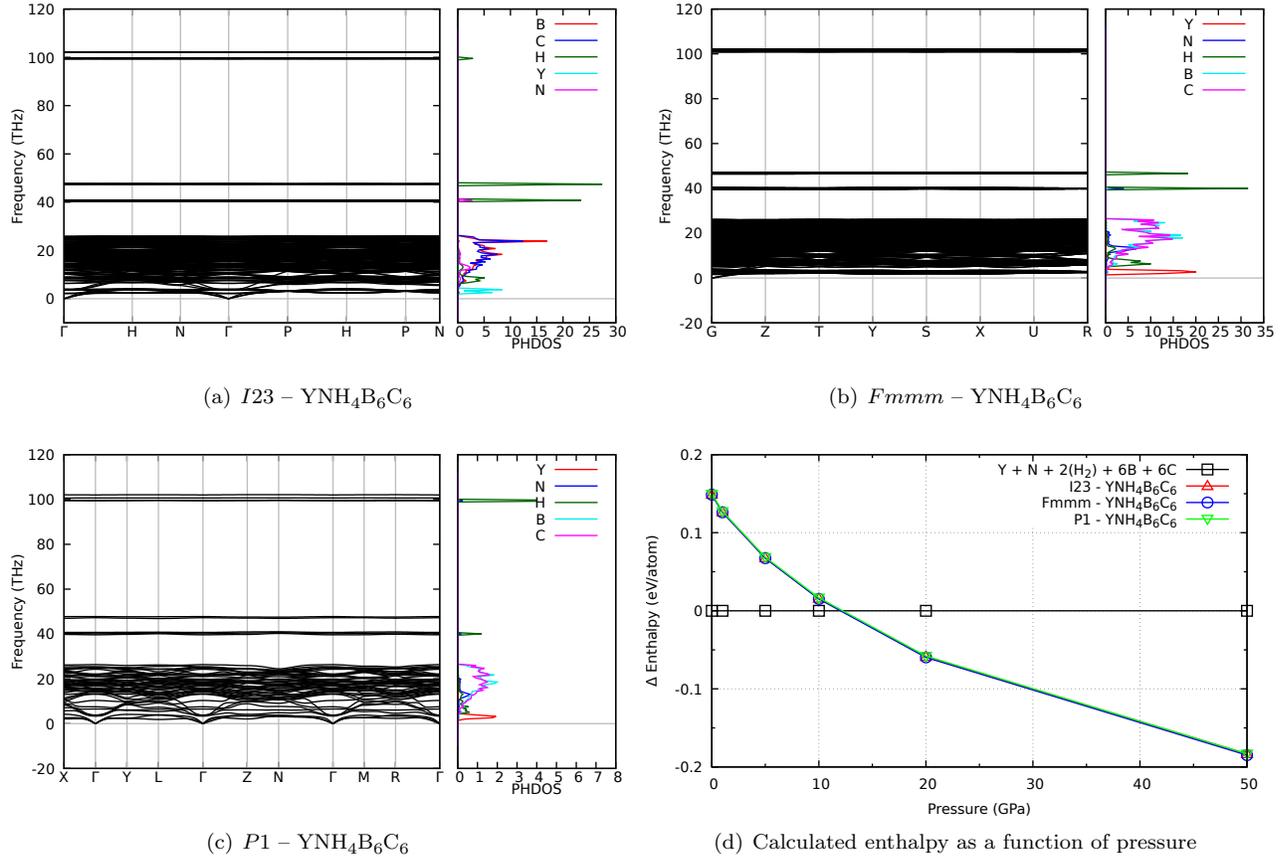

(a) $I23$ – YNH$_4$B$_6$C$_6$

(b) $Fmmm$ – YNH$_4$B$_6$C$_6$

(c) $P1$ – YNH$_4$B$_6$C$_6$

(d) Calculated enthalpy as a function of pressure

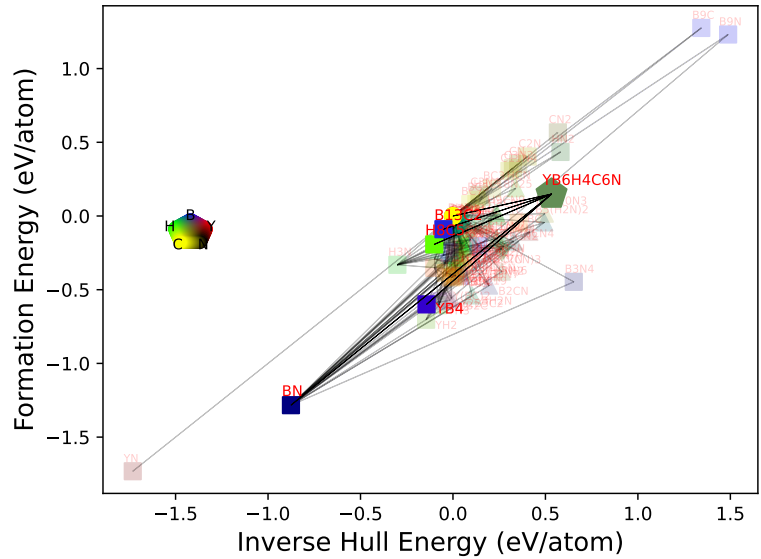

(e) Inverse hull web at 0 GPa

FIG. S7. (a–c) Phonon dispersion (left panel) and projected phonon densities of states (PHDOS, right panel) of different YNH$_4$B$_6$C$_6$ phases at ambient pressure. The absence of an imaginary frequency in the phonon spectra indicates that the structures are dynamically stable at ambient pressure. (d) Calculated enthalpy as a function of pressure for YNH$_4$B$_6$C$_6$ relative to simple elemental substance, and the phase transition pressure is ∼12.1 GPa. (e) Inverse hull web of Y-N-H-B-C system at 0 K and ambient pressure. Compounds located in the positive range of the inverse-hull-energy axis indicates that they are metastable, while the others located in the negative range of the inverse-hull-energy axis are stable ones. We leave the YNH$_4$B$_6$C$_6$ compound and its hull reactants opaque, while we make all other phases partially transparent. Ground state structures of simple substances and other compounds were collected from the Inorganic Crystal Structure Database (ICSD) [1].

### 1.6 $PbNH_4B_6C_6$

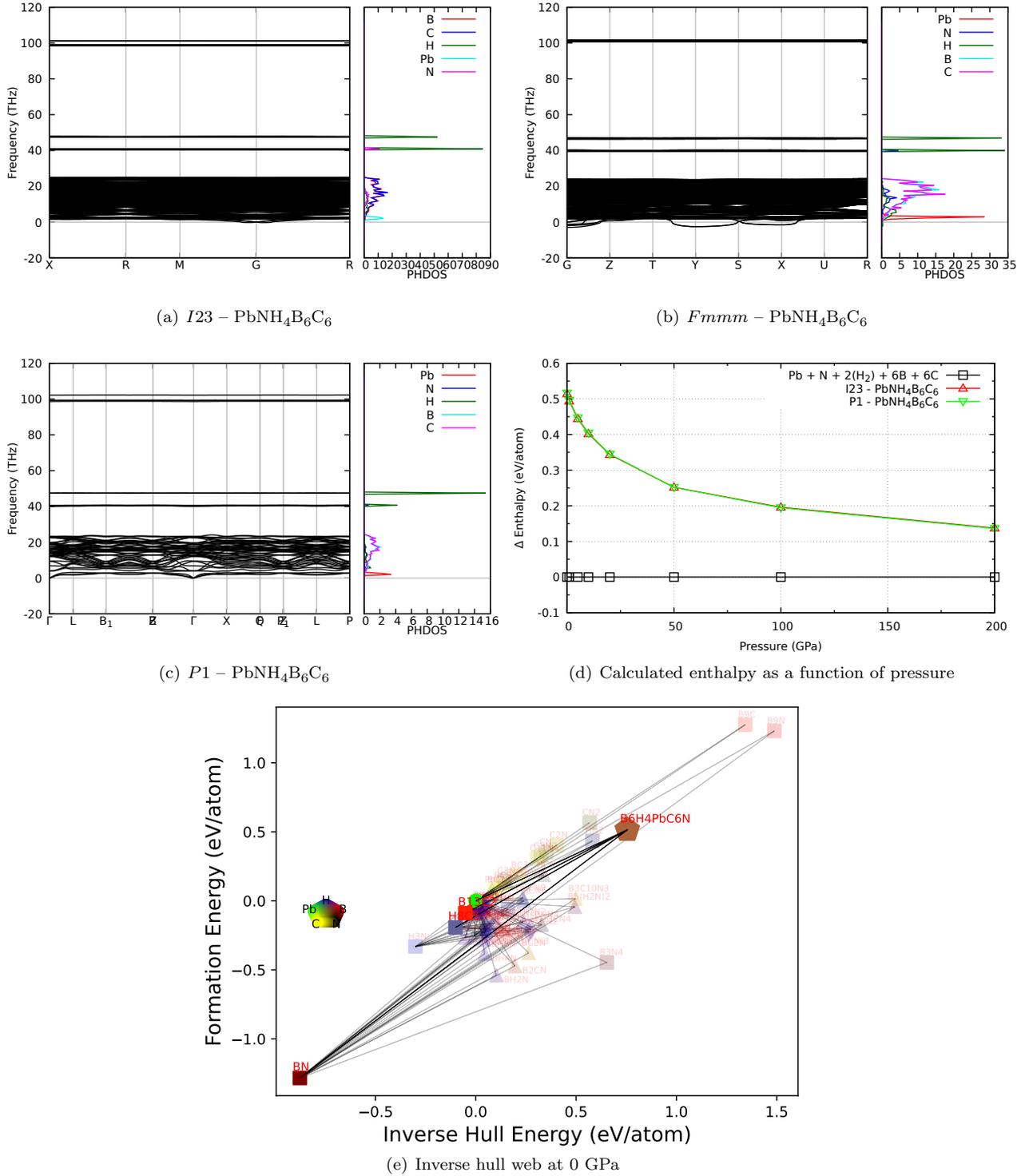

FIG. S8. (a–c) Phonon dispersion (left panel) and projected phonon densities of states (PHDOS, right panel) of different $PbNH_4B_6C_6$ phases at ambient pressure. The absence of an imaginary frequency in the phonon spectra indicates that the structures are dynamically stable at ambient pressure. (d) Calculated enthalpy as a function of pressure for $PbNH_4B_6C_6$ relative to simple elemental substance. (e) Inverse hull web of Pb-N-H-B-C system at 0 K and ambient pressure. Compounds located in the positive range of the inverse-hull-energy axis indicates that they are metastable, while the others located in the negative range of the inverse-hull-energy axis are stable ones. We leave the $PbNH_4B_6C_6$ compound and its hull reactants opaque, while we make all other phases partially transparent. Ground state structures of simple substances and other compounds were collected from the Inorganic Crystal Structure Database (ICSD) [1].

## 1.7 LaNH$_4$B$_6$C$_6$

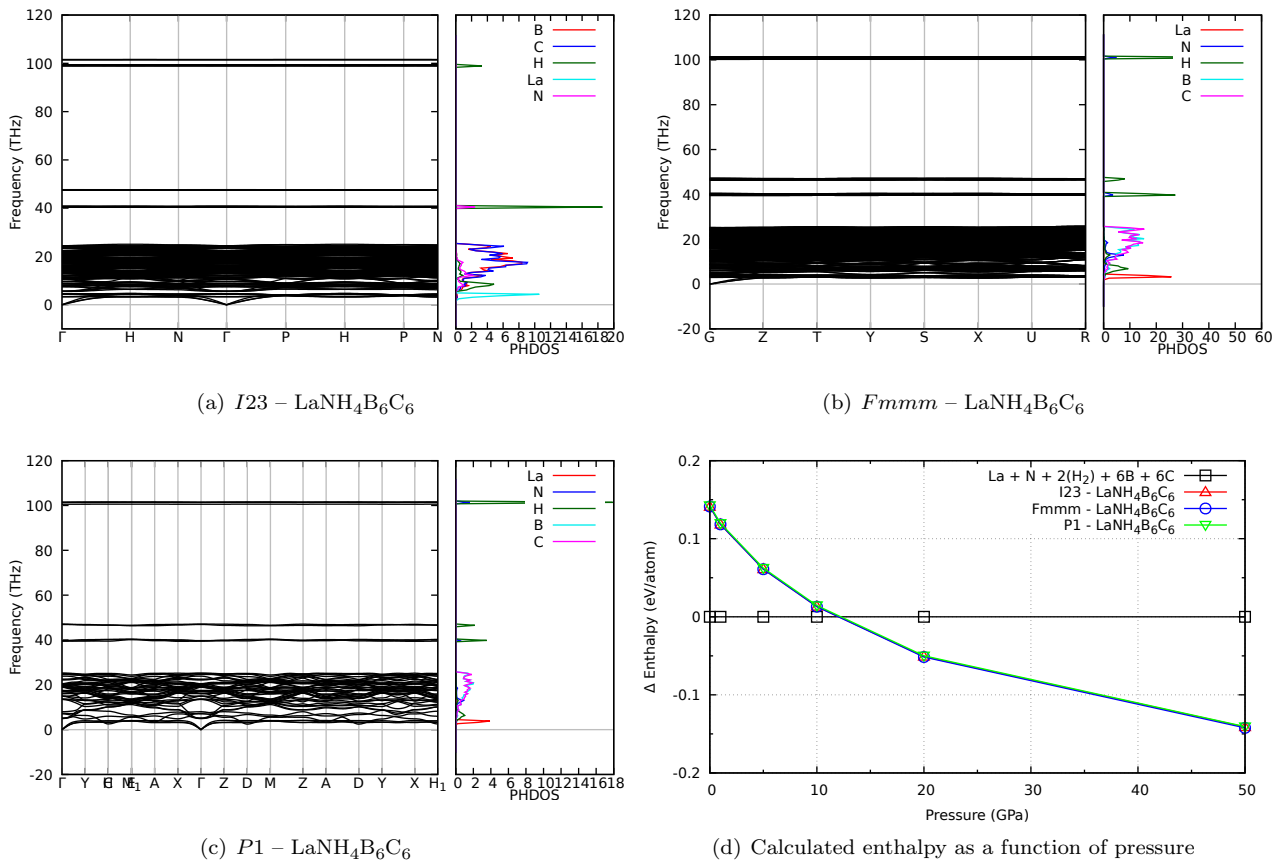

(a) $I23$ – LaNH$_4$B$_6$C$_6$

(b) $Fmmm$ – LaNH$_4$B$_6$C$_6$

(c) $P1$ – LaNH$_4$B$_6$C$_6$

(d) Calculated enthalpy as a function of pressure

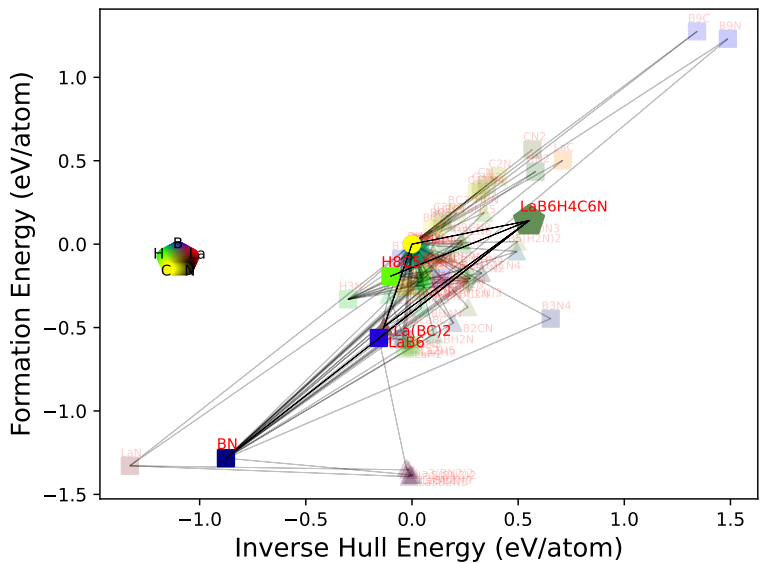

(e) Inverse hull web at 0 GPa

FIG. S9. (a–c) Phonon dispersion (left panel) and projected phonon densities of states (PHDOS, right panel) of different LaNH$_4$B$_6$C$_6$ phases at ambient pressure. The absence of an imaginary frequency in the phonon spectra indicates that the structures are dynamically stable at ambient pressure. (d) Calculated enthalpy as a function of pressure for LaNH$_4$B$_6$C$_6$ relative to simple elemental substance, and the phase transition pressure is ∼12.0 GPa. (e) Inverse hull web of La-N-H-B-C system at 0 K and ambient pressure. Compounds located in the positive range of the inverse-hull-energy axis indicates that they are metastable, while the others located in the negative range of the inverse-hull-energy axis are stable ones. We leave the LaNH$_4$B$_6$C$_6$ compound and its hull reactants opaque, while we make all other phases partially transparent. Ground state structures of simple substances and other compounds were collected from the Inorganic Crystal Structure Database (ICSD) [1].

## 1.8 PrNH$_4$B$_6$C$_6$

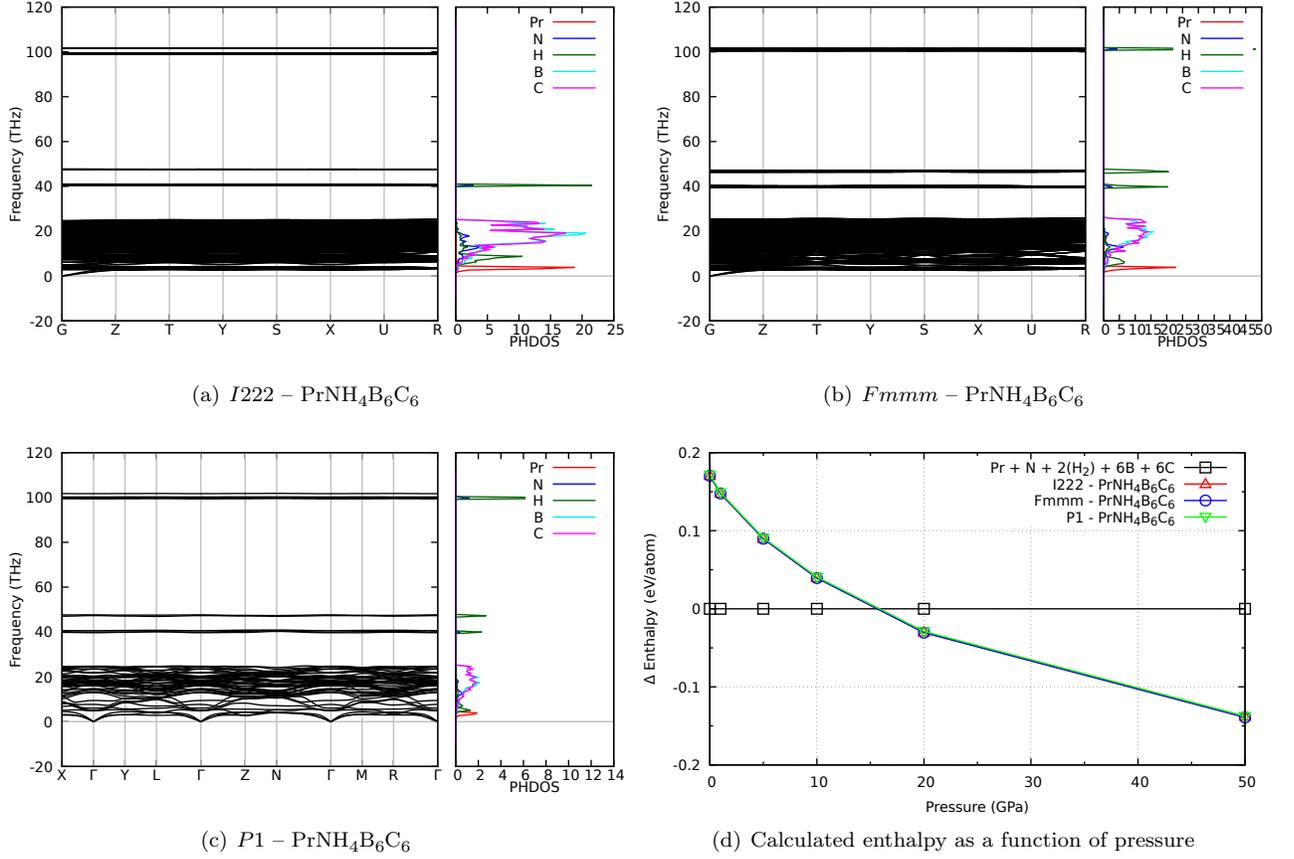

FIG. S10. (a–c) Phonon dispersion (left panel) and projected phonon densities of states (PHDOS, right panel) of different PrNH$_4$B$_6$C$_6$ phases at ambient pressure. The absence of an imaginary frequency in the phonon spectra indicates that the structures are dynamically stable at ambient pressure. (d) Calculated enthalpy as a function of pressure for PrNH$_4$B$_6$C$_6$ relative to simple elemental substance, and the phase transition pressure is ∼15.7 GPa. Ground state structures of simple substances were collected from the Inorganic Crystal Structure Database (ICSD) [1].

## 1.9 NdNH$_4$B$_6$C$_6$

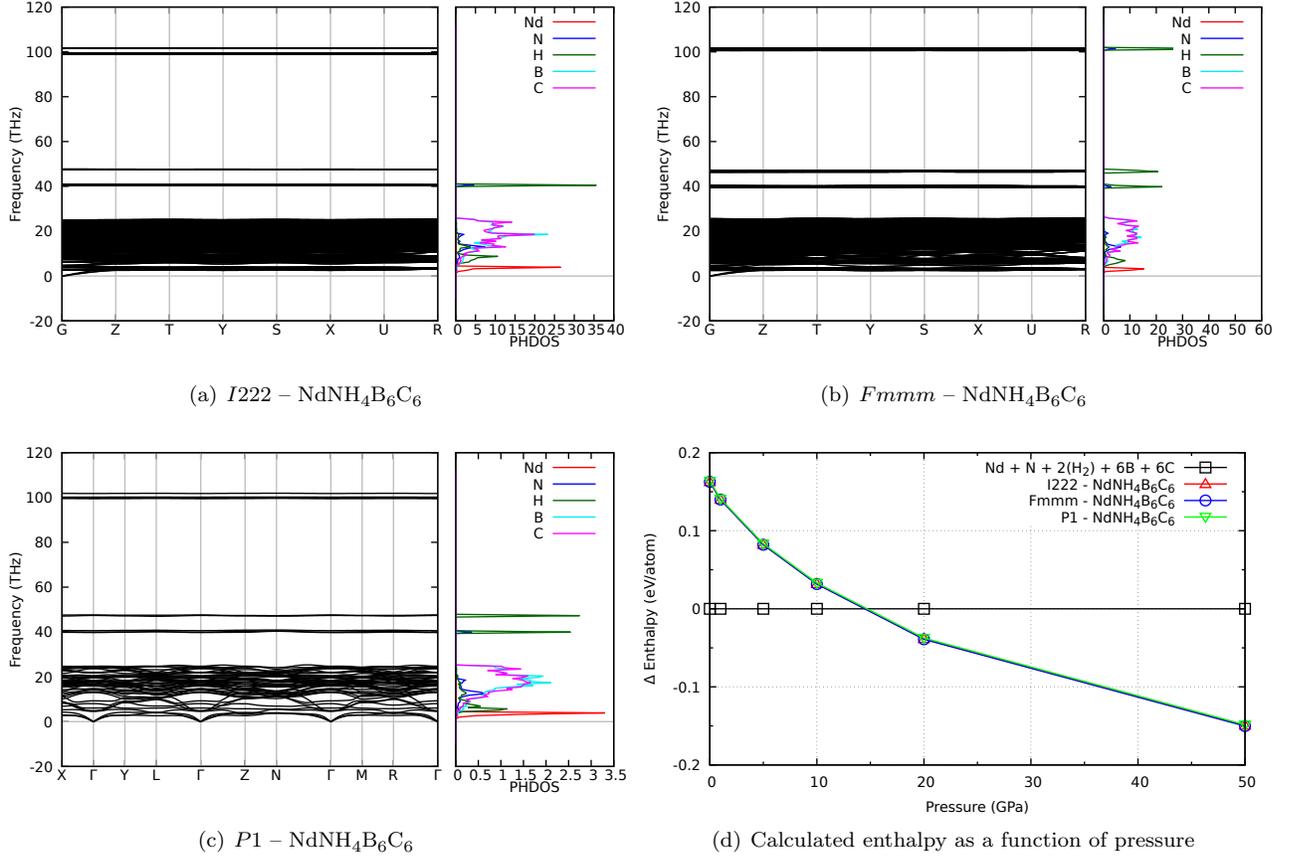

(a) $I222$ – NdNH$_4$B$_6$C$_6$

(b) $Fmmm$ – NdNH$_4$B$_6$C$_6$

(c) $P1$ – NdNH$_4$B$_6$C$_6$

(d) Calculated enthalpy as a function of pressure

FIG. S11. (a–c) Phonon dispersion (left panel) and projected phonon densities of states (PHDOS, right panel) of different NdNH$_4$B$_6$C$_6$ phases at ambient pressure. The absence of an imaginary frequency in the phonon spectra indicates that the structures are dynamically stable at ambient pressure. (d) Calculated enthalpy as a function of pressure for NdNH$_4$B$_6$C$_6$ relative to simple elemental substance, and the phase transition pressure is ∼14.6 GPa. Ground state structures of simple substances were collected from the Inorganic Crystal Structure Database (ICSD) [1].

## 1.10 PmNH$_4$B$_6$C$_6$

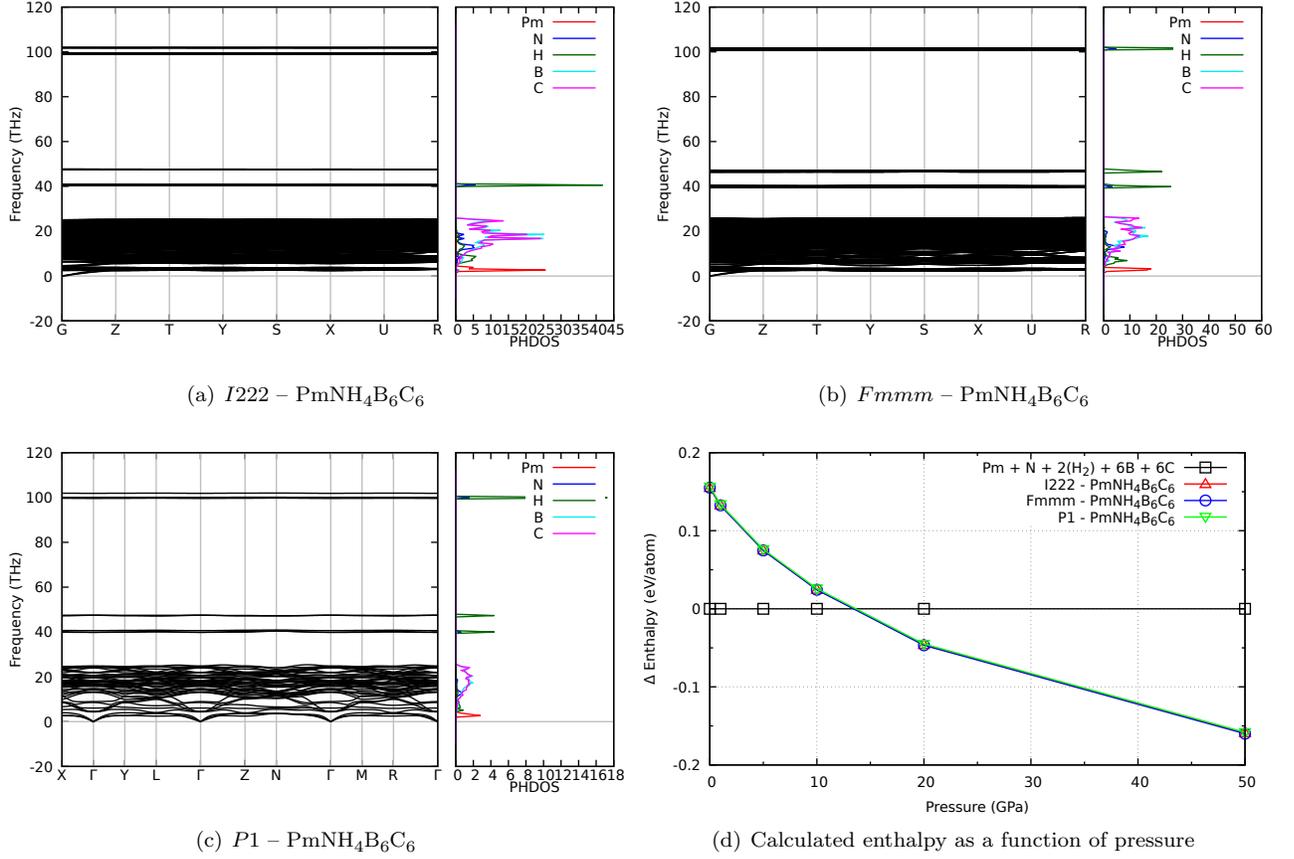

(a) $I222$ – PmNH$_4$B$_6$C$_6$

(b) $Fmmm$ – PmNH$_4$B$_6$C$_6$

(c) $P1$ – PmNH$_4$B$_6$C$_6$

(d) Calculated enthalpy as a function of pressure

FIG. S12. (a–c) Phonon dispersion (left panel) and projected phonon densities of states (PHDOS, right panel) of different PmNH$_4$B$_6$C$_6$ phases at ambient pressure. The absence of an imaginary frequency in the phonon spectra indicates that the structures are dynamically stable at ambient pressure. (d) Calculated enthalpy as a function of pressure for PmNH$_4$B$_6$C$_6$ relative to simple elemental substance, and the phase transition pressure is ∼13.6 GPa. Ground state structures of simple substances were collected from the Inorganic Crystal Structure Database (ICSD) [1].

## 1.11 SmNH$_4$B$_6$C$_6$

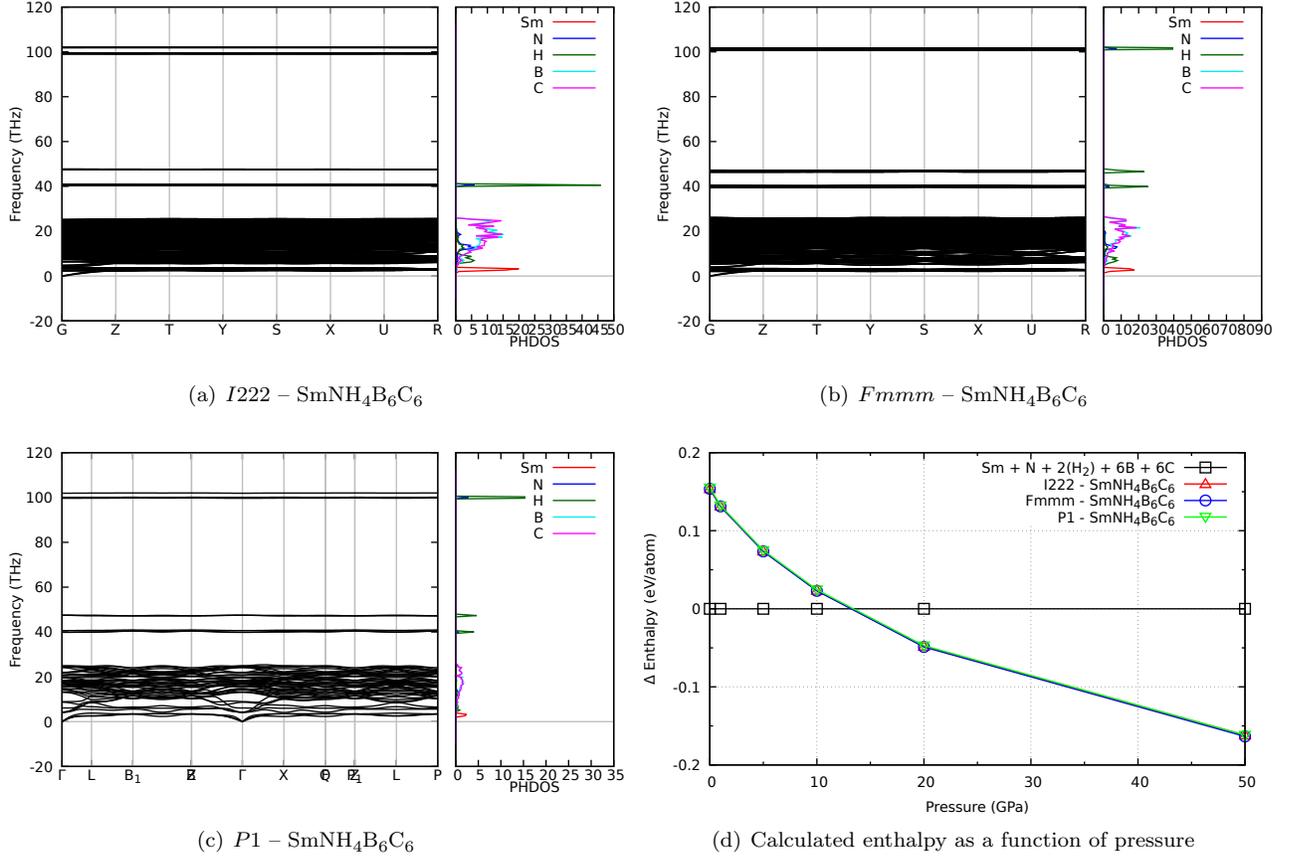

(a) $I222$ – SmNH$_4$B$_6$C$_6$

(b) $Fmmm$ – SmNH$_4$B$_6$C$_6$

(c) $P1$ – SmNH$_4$B$_6$C$_6$

(d) Calculated enthalpy as a function of pressure

FIG. S13. (a–c) Phonon dispersion (left panel) and projected phonon densities of states (PHDOS, right panel) of different SmNH$_4$B$_6$C$_6$ phases at ambient pressure. The absence of an imaginary frequency in the phonon spectra indicates that the structures are dynamically stable at ambient pressure. (d) Calculated enthalpy as a function of pressure for SmNH$_4$B$_6$C$_6$ relative to simple elemental substance, and the phase transition pressure is ∼13.3 GPa. Ground state structures of simple substances were collected from the Inorganic Crystal Structure Database (ICSD) [1].

### 1.12 EuNH$_4$B$_6$C$_6$

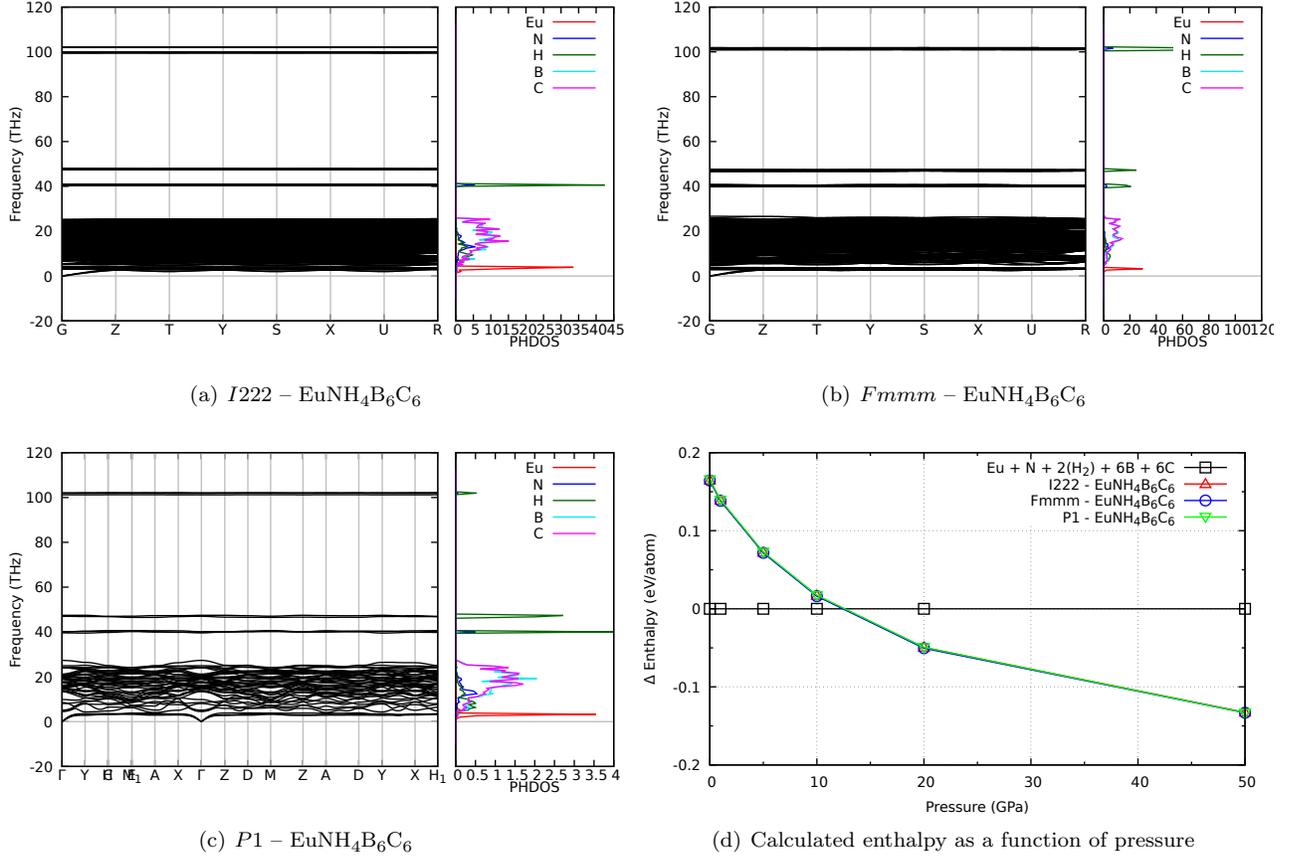

(a) $I222$ – EuNH$_4$B$_6$C$_6$

(b) $Fmmm$ – EuNH$_4$B$_6$C$_6$

(c) $P1$ – EuNH$_4$B$_6$C$_6$

(d) Calculated enthalpy as a function of pressure

FIG. S14. (a–c) Phonon dispersion (left panel) and projected phonon densities of states (PHDOS, right panel) of different EuNH$_4$B$_6$C$_6$ phases at ambient pressure. The absence of an imaginary frequency in the phonon spectra indicates that the structures are dynamically stable at ambient pressure. (d) Calculated enthalpy as a function of pressure for EuNH$_4$B$_6$C$_6$ relative to simple elemental substance, and the phase transition pressure is ∼12.5 GPa. Ground state structures of simple substances were collected from the Inorganic Crystal Structure Database (ICSD) [1].

## 1.13 GdNH$_4$B$_6$C$_6$

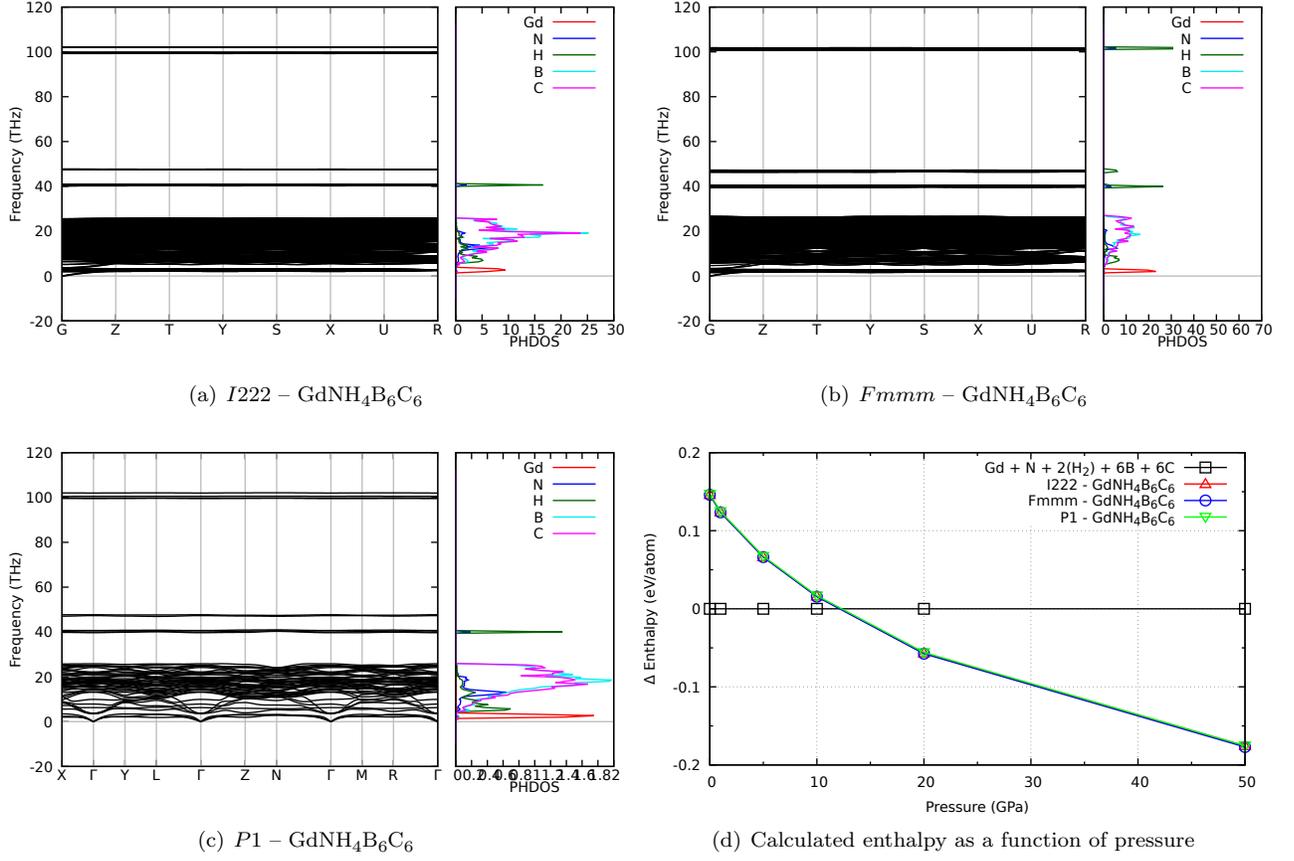

(a) $I222$ – GdNH$_4$B$_6$C$_6$

(b) $Fmmm$ – GdNH$_4$B$_6$C$_6$

(c) $P1$ – GdNH$_4$B$_6$C$_6$

(d) Calculated enthalpy as a function of pressure

FIG. S15. (a–c) Phonon dispersion (left panel) and projected phonon densities of states (PHDOS, right panel) of different GdNH$_4$B$_6$C$_6$ phases at ambient pressure. The absence of an imaginary frequency in the phonon spectra indicates that the structures are dynamically stable at ambient pressure. (d) Calculated enthalpy as a function of pressure for GdNH$_4$B$_6$C$_6$ relative to simple elemental substance, and the phase transition pressure is ∼12.3 GPa. Ground state structures of simple substances were collected from the Inorganic Crystal Structure Database (ICSD) [1].

## 1.14 TbNH$_4$B$_6$C$_6$

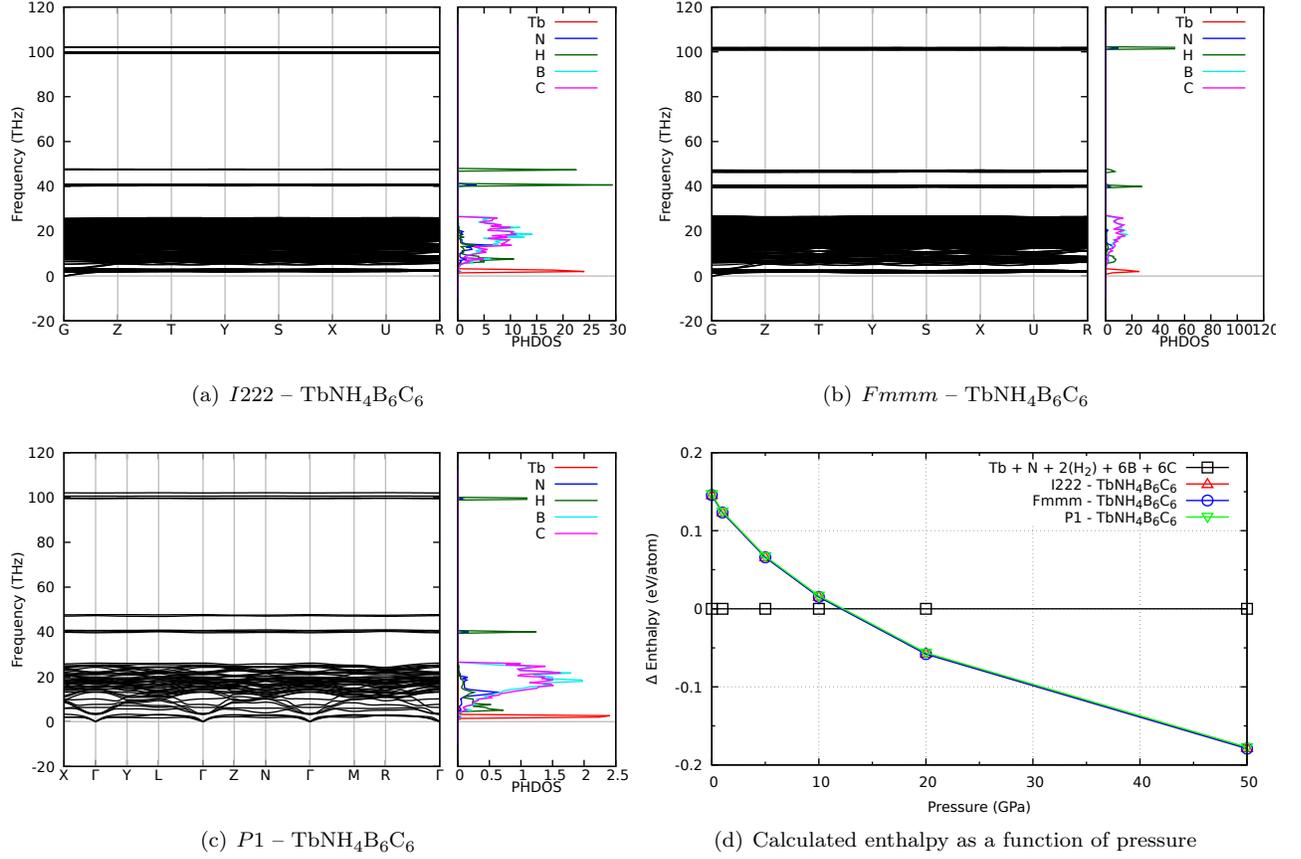

(a) $I222$ – TbNH$_4$B$_6$C$_6$

(b) $Fmmm$ – TbNH$_4$B$_6$C$_6$

(c) $P1$ – TbNH$_4$B$_6$C$_6$

(d) Calculated enthalpy as a function of pressure

FIG. S16. (a–c) Phonon dispersion (left panel) and projected phonon densities of states (PHDOS, right panel) of different TbNH$_4$B$_6$C$_6$ phases at ambient pressure. The absence of an imaginary frequency in the phonon spectra indicates that the structures are dynamically stable at ambient pressure. (d) Calculated enthalpy as a function of pressure for TbNH$_4$B$_6$C$_6$ relative to simple elemental substance, and the phase transition pressure is ∼12.2 GPa. Ground state structures of simple substances were collected from the Inorganic Crystal Structure Database (ICSD) [1].

### 1.15 DyNH$_4$B$_6$C$_6$

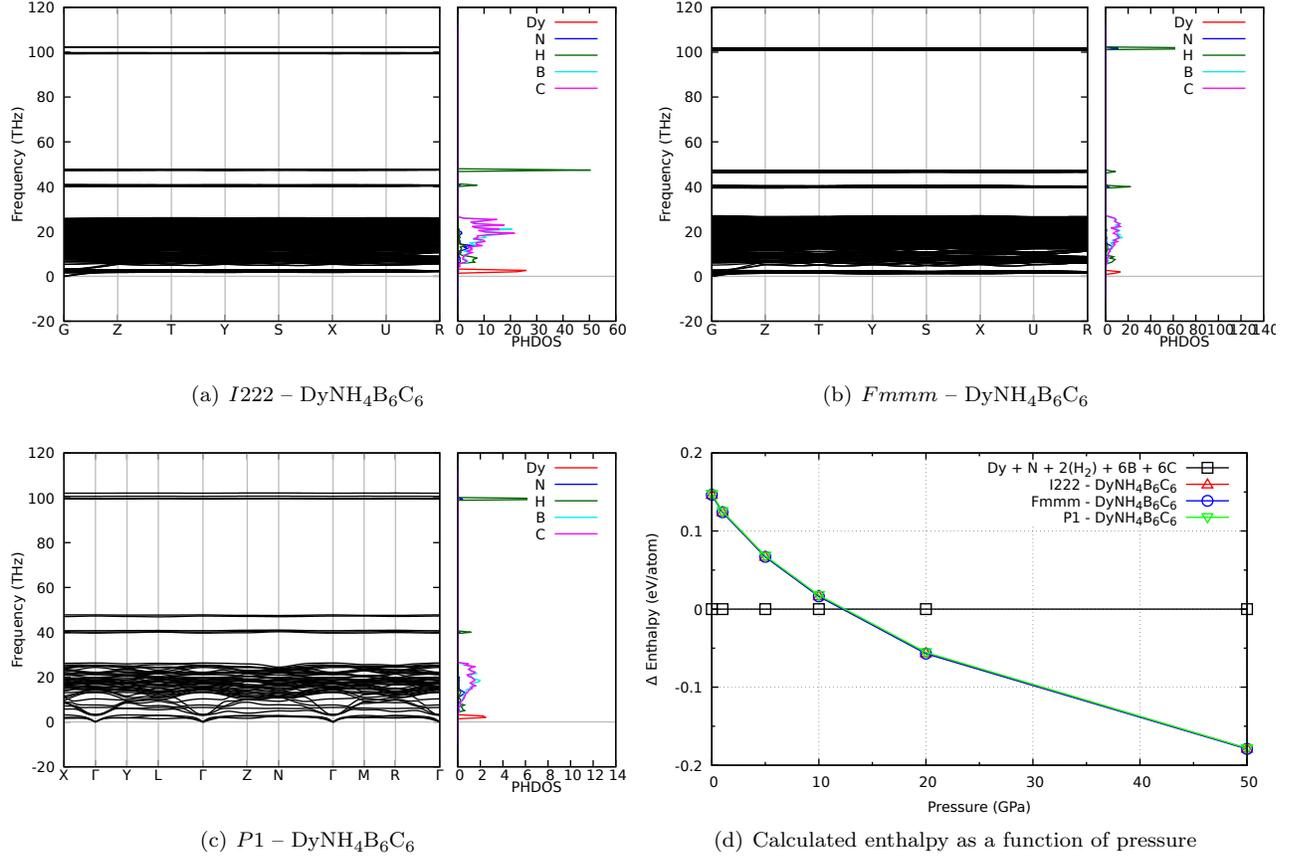

(a) $I222$ – DyNH$_4$B$_6$C$_6$

(b) $Fmmm$ – DyNH$_4$B$_6$C$_6$

(c) $P1$ – DyNH$_4$B$_6$C$_6$

(d) Calculated enthalpy as a function of pressure

FIG. S17. (a–c) Phonon dispersion (left panel) and projected phonon densities of states (PHDOS, right panel) of different DyNH$_4$B$_6$C$_6$ phases at ambient pressure. The absence of an imaginary frequency in the phonon spectra indicates that the structures are dynamically stable at ambient pressure. (d) Calculated enthalpy as a function of pressure for DyNH$_4$B$_6$C$_6$ relative to simple elemental substance, and the phase transition pressure is ∼12.3 GPa. Ground state structures of simple substances were collected from the Inorganic Crystal Structure Database (ICSD) [1].

## 1.16 $HoNH_4B_6C_6$

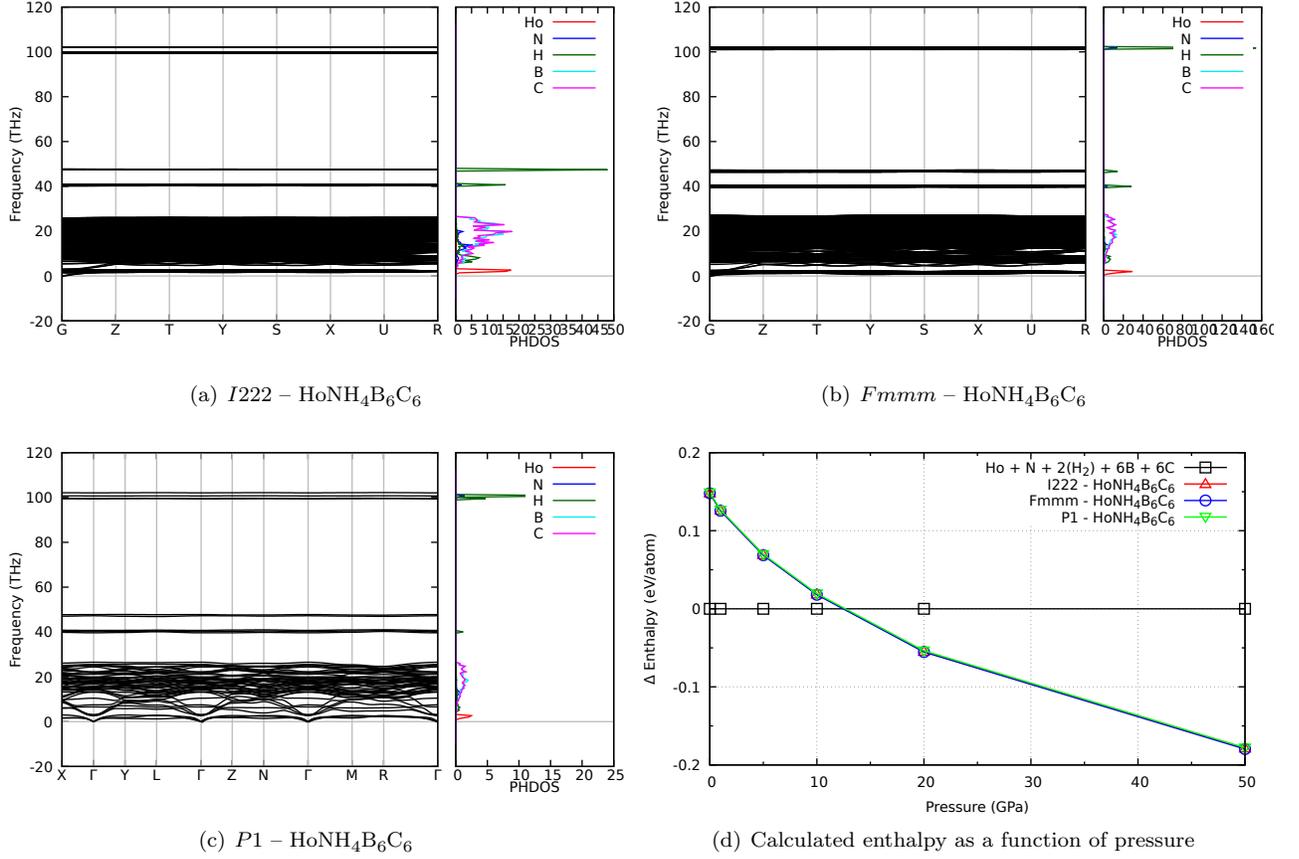

(a) $I222$ – $HoNH_4B_6C_6$

(b) $Fmmm$ – $HoNH_4B_6C_6$

(c) $P1$ – $HoNH_4B_6C_6$

(d) Calculated enthalpy as a function of pressure

FIG. S18. (a–c) Phonon dispersion (left panel) and projected phonon densities of states (PHDOS, right panel) of different $HoNH_4B_6C_6$ phases at ambient pressure. The absence of an imaginary frequency in the phonon spectra indicates that the structures are dynamically stable at ambient pressure. (d) Calculated enthalpy as a function of pressure for $HoNH_4B_6C_6$ relative to simple elemental substance, and the phase transition pressure is ∼12.5 GPa. Ground state structures of simple substances were collected from the Inorganic Crystal Structure Database (ICSD) [1].

### 1.17 ErNH$_4$B$_6$C$_6$

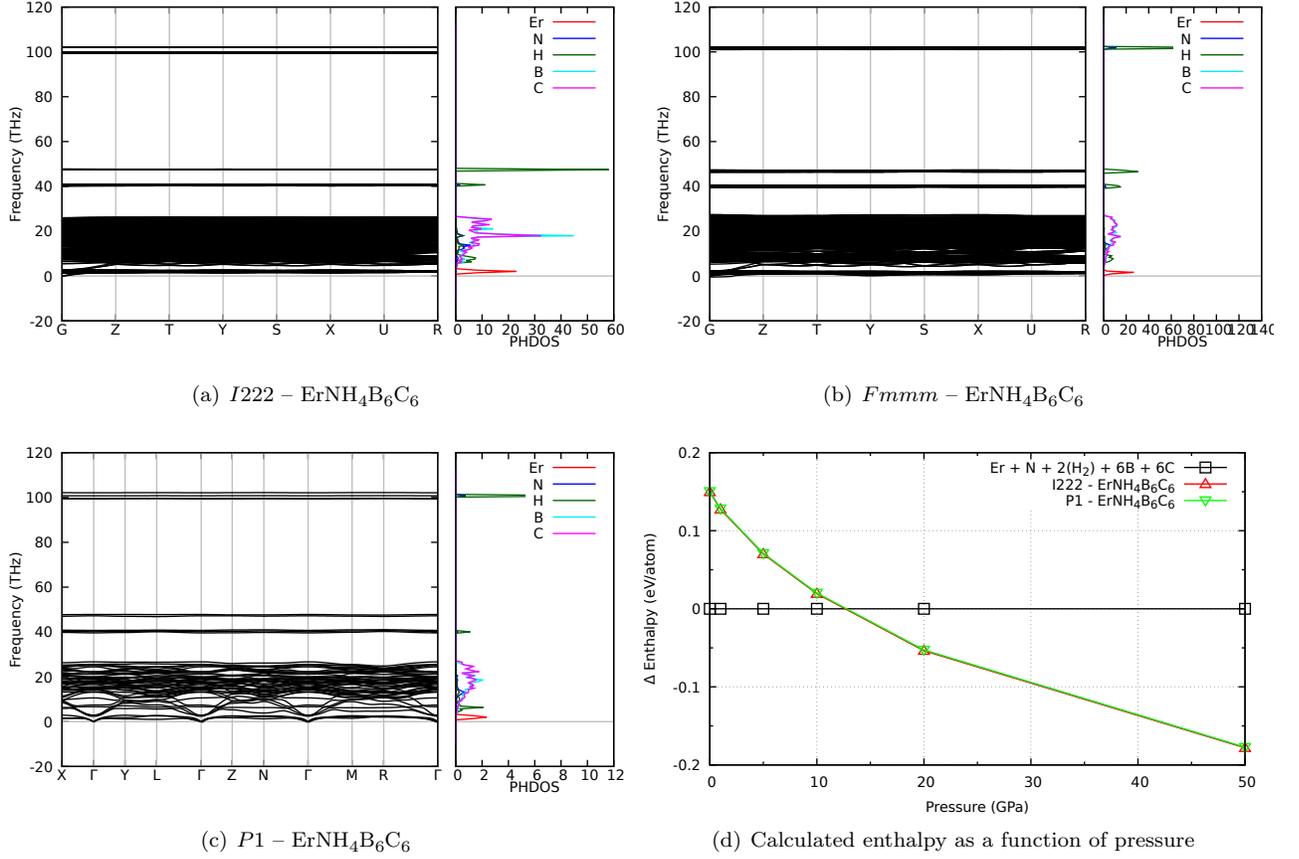

(a) $I222$ – ErNH$_4$B$_6$C$_6$

(b) $Fmmm$ – ErNH$_4$B$_6$C$_6$

(c) $P1$ – ErNH$_4$B$_6$C$_6$

(d) Calculated enthalpy as a function of pressure

FIG. S19. (a–c) Phonon dispersion (left panel) and projected phonon densities of states (PHDOS, right panel) of different ErNH$_4$B$_6$C$_6$ phases at ambient pressure. The absence of an imaginary frequency in the phonon spectra indicates that the structures are dynamically stable at ambient pressure. (d) Calculated enthalpy as a function of pressure for ErNH$_4$B$_6$C$_6$ relative to simple elemental substance, and the phase transition pressure is ∼12.7 GPa. Ground state structures of simple substances were collected from the Inorganic Crystal Structure Database (ICSD) [1].

## 1.18 TmNH$_4$B$_6$C$_6$

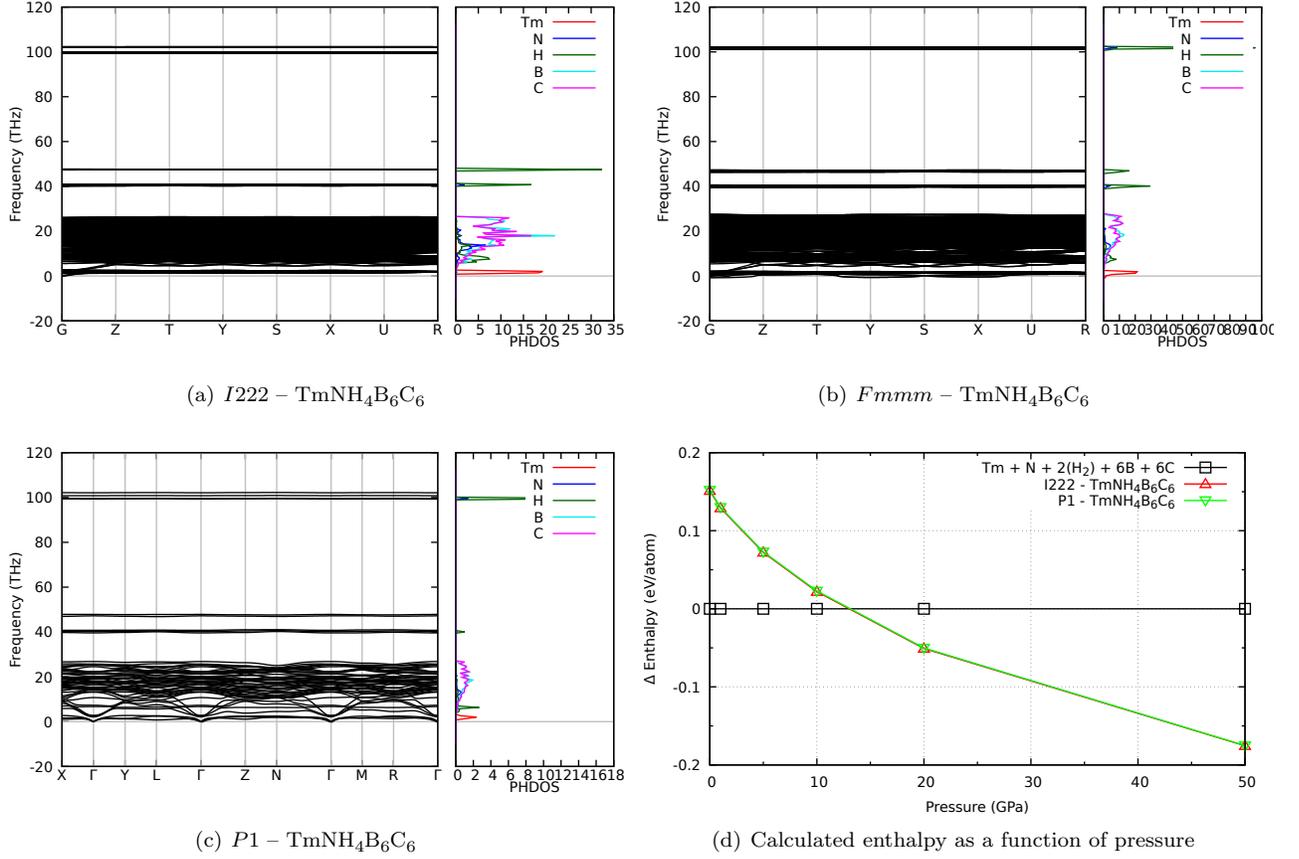

(a) $I222$ – TmNH$_4$B$_6$C$_6$

(b) $Fmmm$ – TmNH$_4$B$_6$C$_6$

(c) $P1$ – TmNH$_4$B$_6$C$_6$

(d) Calculated enthalpy as a function of pressure

FIG. S20. (a–c) Phonon dispersion (left panel) and projected phonon densities of states (PHDOS, right panel) of different TmNH$_4$B$_6$C$_6$ phases at ambient pressure. The absence of an imaginary frequency in the phonon spectra indicates that the structures are dynamically stable at ambient pressure. (d) Calculated enthalpy as a function of pressure for TmNH$_4$B$_6$C$_6$ relative to simple elemental substance, and the phase transition pressure is ∼13.0 GPa. Ground state structures of simple substances were collected from the Inorganic Crystal Structure Database (ICSD) [1].

### 1.19 YbNH$_4$B$_6$C$_6$

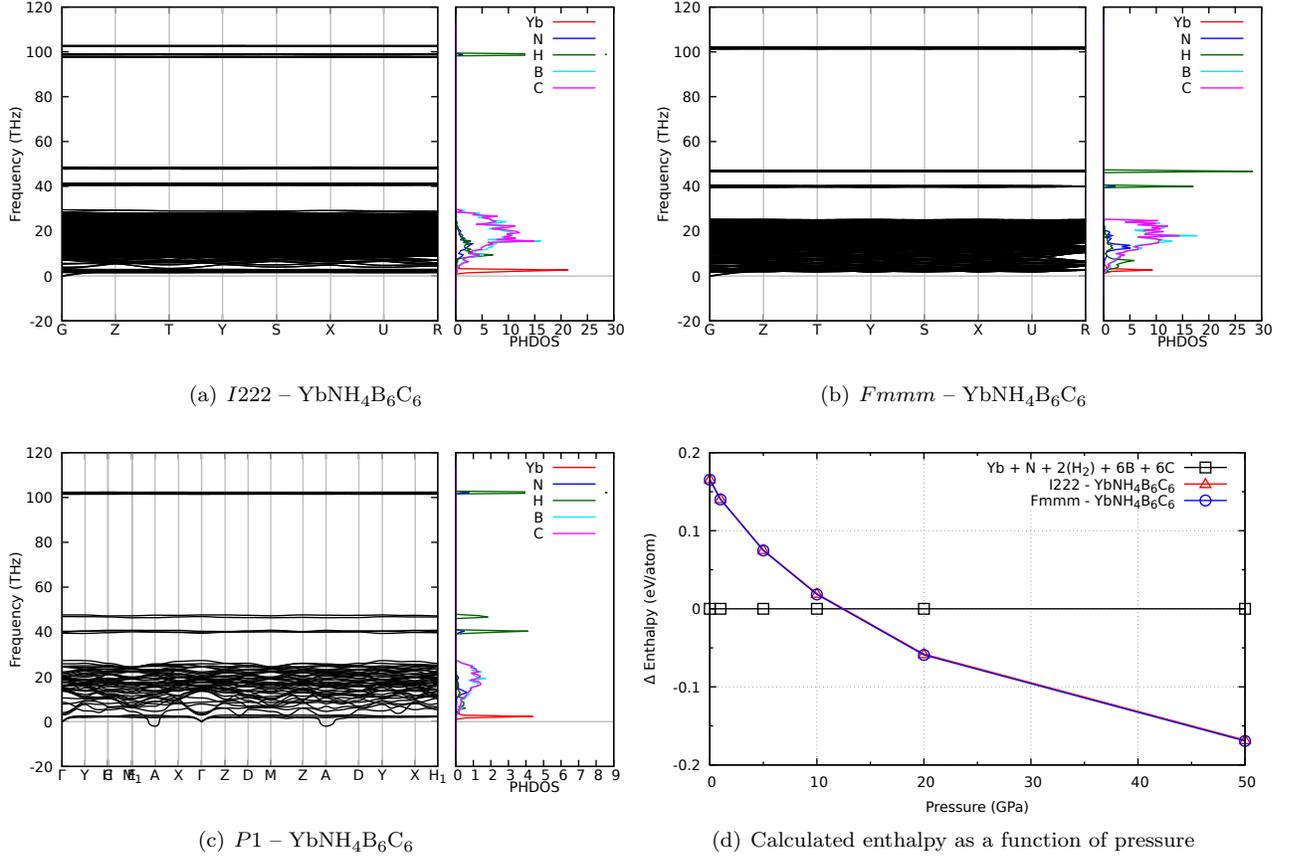

(a) $I222$ – YbNH$_4$B$_6$C$_6$

(b) $Fmmm$ – YbNH$_4$B$_6$C$_6$

(c) $P1$ – YbNH$_4$B$_6$C$_6$

(d) Calculated enthalpy as a function of pressure

FIG. S21. (a–c) Phonon dispersion (left panel) and projected phonon densities of states (PHDOS, right panel) of different YbNH$_4$B$_6$C$_6$ phases at ambient pressure. The absence of an imaginary frequency in the phonon spectra indicates that the structures are dynamically stable at ambient pressure. (d) Calculated enthalpy as a function of pressure for YbNH$_4$B$_6$C$_6$ relative to simple elemental substance, and the phase transition pressure is ∼12.4 GPa. Ground state structures of simple substances were collected from the Inorganic Crystal Structure Database (ICSD) [1].

## 1.20 LuNH$_4$B$_6$C$_6$

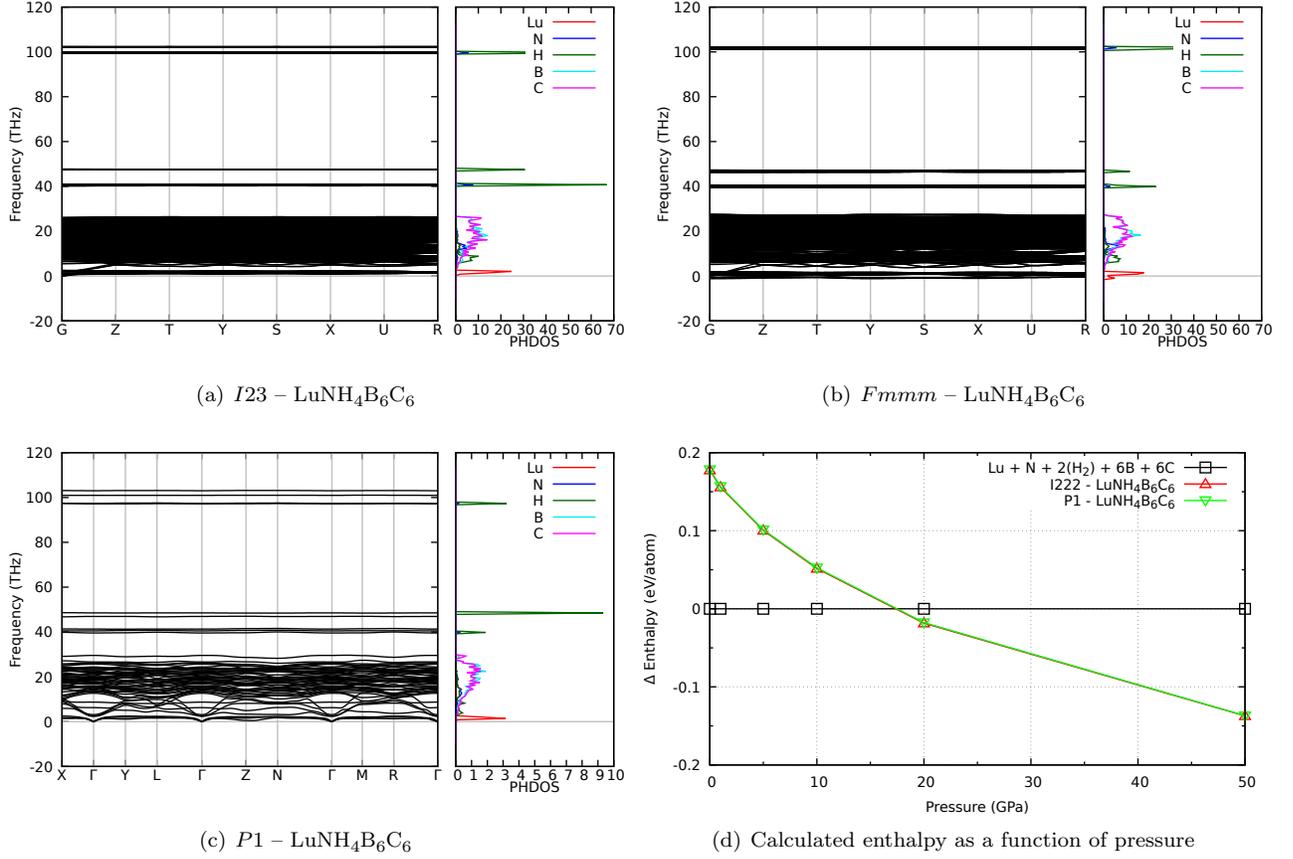

(a) $I23$ – LuNH$_4$B$_6$C$_6$

(b) $Fmmm$ – LuNH$_4$B$_6$C$_6$

(c) $P1$ – LuNH$_4$B$_6$C$_6$

(d) Calculated enthalpy as a function of pressure

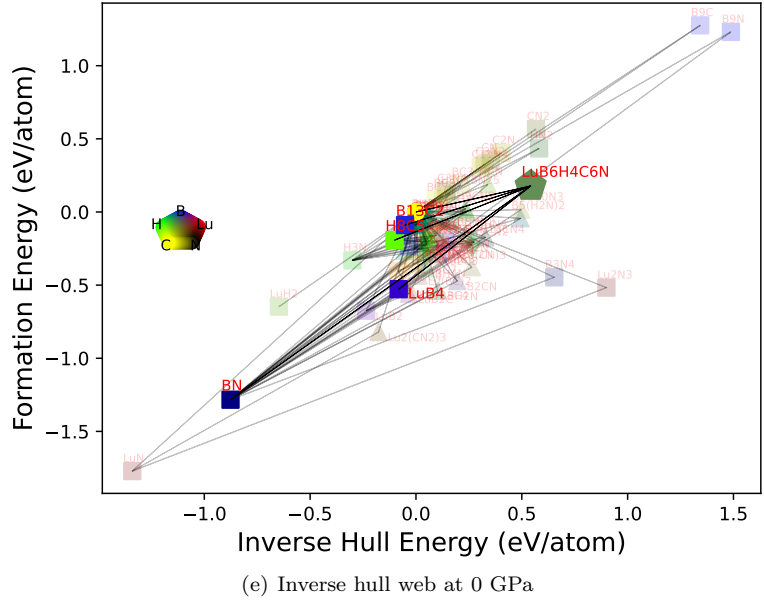

(e) Inverse hull web at 0 GPa

FIG. S22. (a–c) Phonon dispersion (left panel) and projected phonon densities of states (PHDOS, right panel) of different LuNH$_4$B$_6$C$_6$ phases at ambient pressure. The absence of an imaginary frequency in the phonon spectra indicates that the structures are dynamically stable at ambient pressure. (d) Calculated enthalpy as a function of pressure for LuNH$_4$B$_6$C$_6$ relative to simple elemental substance, and the phase transition pressure is ∼17.4 GPa. (e) Inverse hull web of Lu-N-H-B-C system at 0 K and ambient pressure. Compounds located in the positive range of the inverse-hull-energy axis indicates that they are metastable, while the others located in the negative range of the inverse-hull-energy axis are stable ones. We leave the LuNH$_4$B$_6$C$_6$ compound and its hull reactants opaque, while we make all other phases partially transparent. Ground state structures of simple substances and other compounds were collected from the Inorganic Crystal Structure Database (ICSD) [1].


### 1.21 AcNH$_4$B$_6$C$_6$

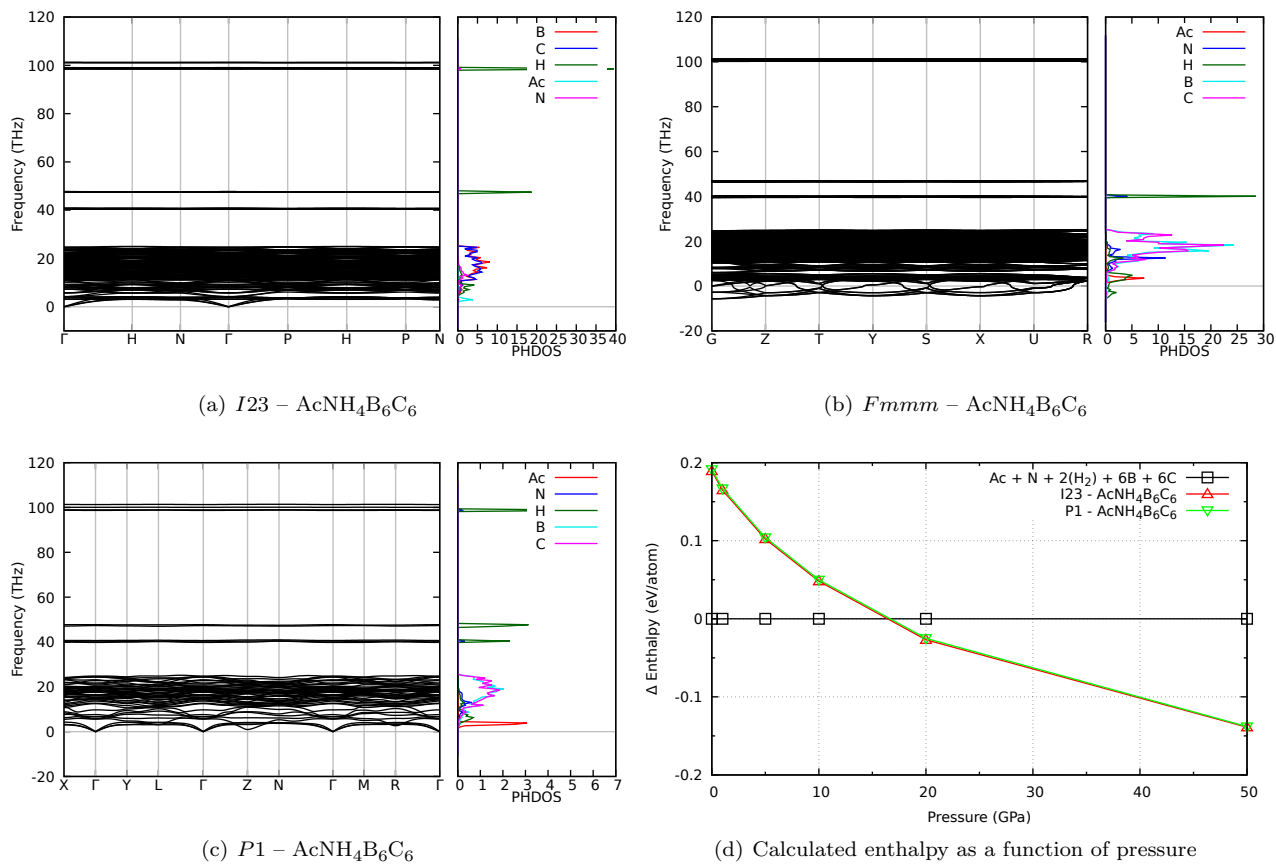

(a) $I23$ – AcNH$_4$B$_6$C$_6$

(b) $Fmmm$ – AcNH$_4$B$_6$C$_6$

(c) $P1$ – AcNH$_4$B$_6$C$_6$

(d) Calculated enthalpy as a function of pressure

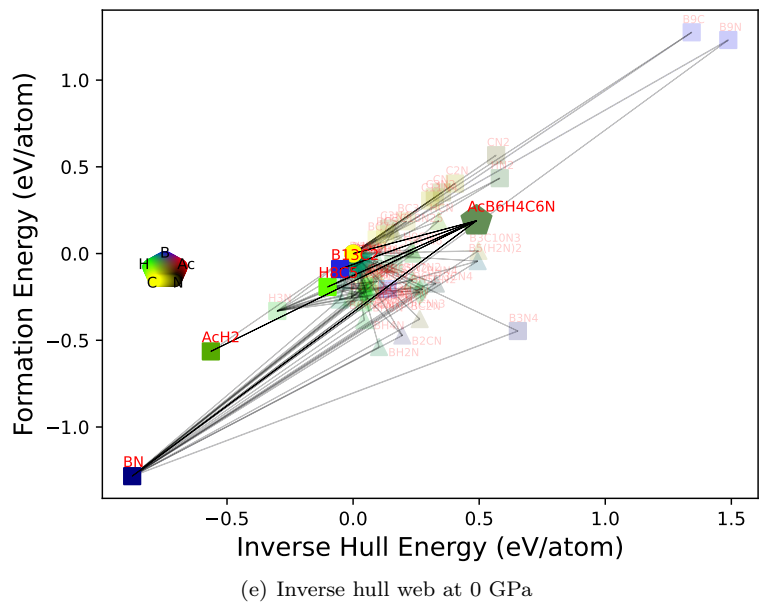

(e) Inverse hull web at 0 GPa

FIG. S23. (a–c) Phonon dispersion (left panel) and projected phonon densities of states (PHDOS, right panel) of different AcNH$_4$B$_6$C$_6$ phases at ambient pressure. The absence of an imaginary frequency in the phonon spectra indicates that the structures are dynamically stable at ambient pressure. (d) Calculated enthalpy as a function of pressure for AcNH$_4$B$_6$C$_6$ relative to simple elemental substance, and the phase transition pressure is ∼16.4 GPa. (e) Inverse hull web of Ac-N-H-B-C system at 0 K and ambient pressure. Compounds located in the positive range of the inverse-hull-energy axis indicates that they are metastable, while the others located in the negative range of the inverse-hull-energy axis are stable ones. We leave the AcNH$_4$B$_6$C$_6$ compound and its hull reactants opaque, while we make all other phases partially transparent. Ground state structures of simple substances and other compounds were collected from the Inorganic Crystal Structure Database (ICSD) [1].

## 1.22 ThNH$_4$B$_6$C$_6$

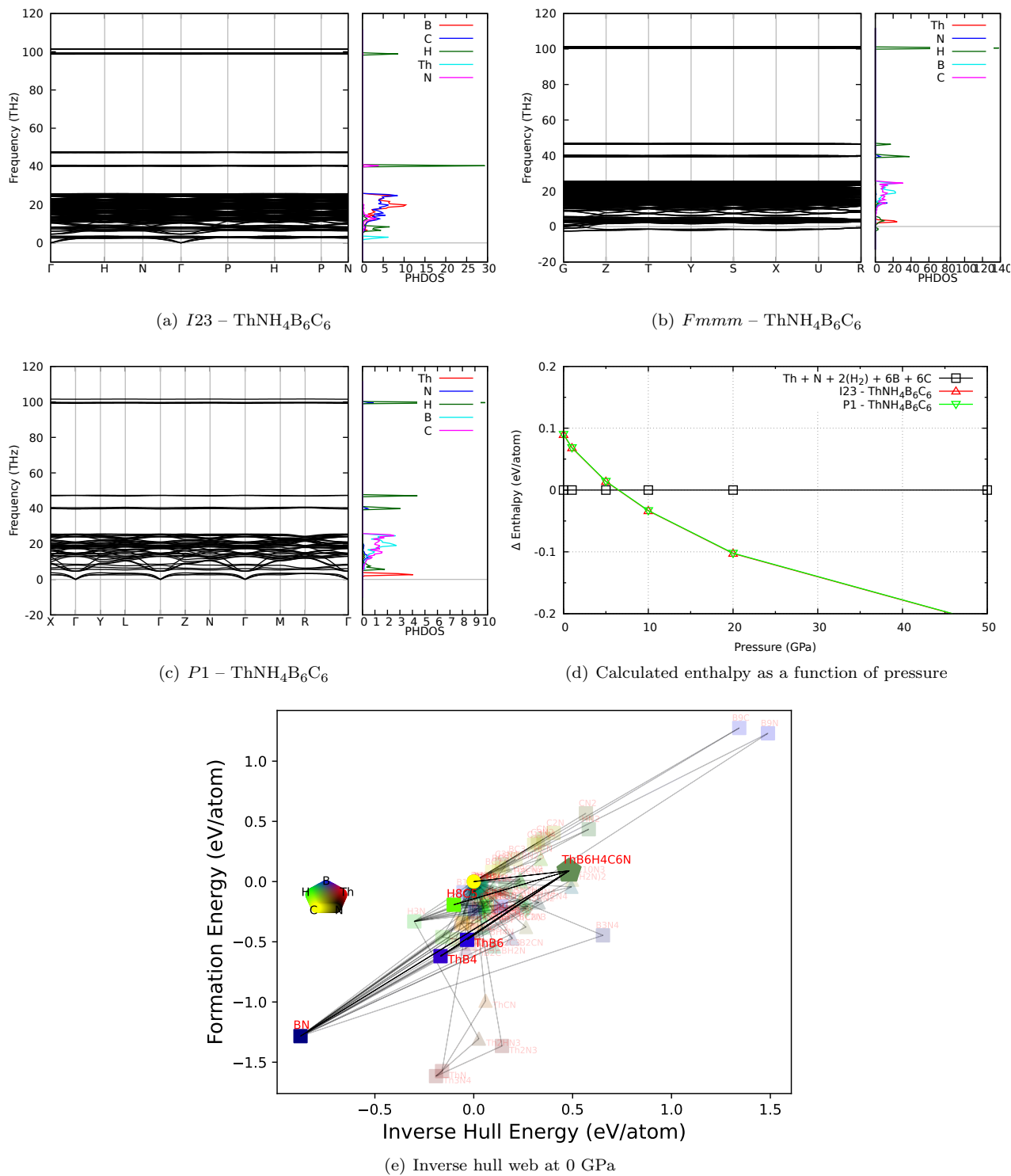

(a) $I23$ – ThNH$_4$B$_6$C$_6$

(b) $Fmmm$ – ThNH$_4$B$_6$C$_6$

(c) $P1$ – ThNH$_4$B$_6$C$_6$

(d) Calculated enthalpy as a function of pressure

(e) Inverse hull web at 0 GPa

FIG. S24. (a–c) Phonon dispersion (left panel) and projected phonon densities of states (PHDOS, right panel) of different ThNH$_4$B$_6$C$_6$ phases at ambient pressure. The absence of an imaginary frequency in the phonon spectra indicates that the structures are dynamically stable at ambient pressure. (d) Calculated enthalpy as a function of pressure for ThNH$_4$B$_6$C$_6$ relative to simple elemental substance, and the phase transition pressure is ∼6.4 GPa. (e) Inverse hull web of Th-N-H-B-C system at 0 K and ambient pressure. Compounds located in the positive range of the inverse-hull-energy axis indicates that they are metastable, while the others located in the negative range of the inverse-hull-energy axis are stable ones. We leave the ThNH$_4$B$_6$C$_6$ compound and its hull reactants opaque, while we make all other phases partially transparent. Ground state structures of simple substances and other compounds were collected from the Inorganic Crystal Structure Database (ICSD) [1].

### 1.23 PaNH$_4$B$_6$C$_6$

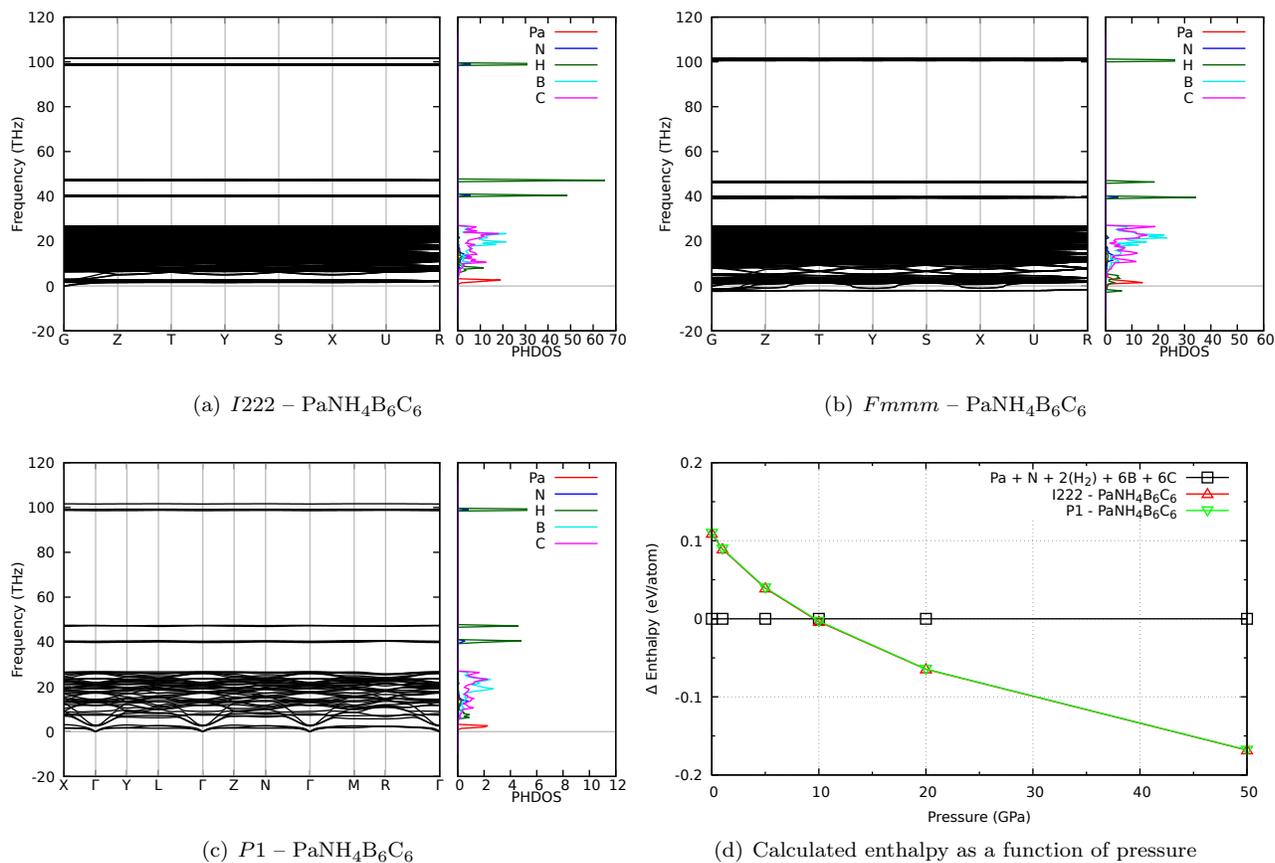

(a) $I222$ – PaNH$_4$B$_6$C$_6$

(b) $Fmmm$ – PaNH$_4$B$_6$C$_6$

(c) $P1$ – PaNH$_4$B$_6$C$_6$

(d) Calculated enthalpy as a function of pressure

FIG. S25. (a–c) Phonon dispersion (left panel) and projected phonon densities of states (PHDOS, right panel) of different PaNH$_4$B$_6$C$_6$ phases at ambient pressure. The absence of an imaginary frequency in the phonon spectra indicates that the structures are dynamically stable at ambient pressure. (d) Calculated enthalpy as a function of pressure for PaNH$_4$B$_6$C$_6$ relative to simple elemental substance, and the phase transition pressure is ∼9.8 GPa. Ground state structures of simple substances were collected from the Inorganic Crystal Structure Database (ICSD) [1].

### 1.24 UNH$_4$B$_6$C$_6$

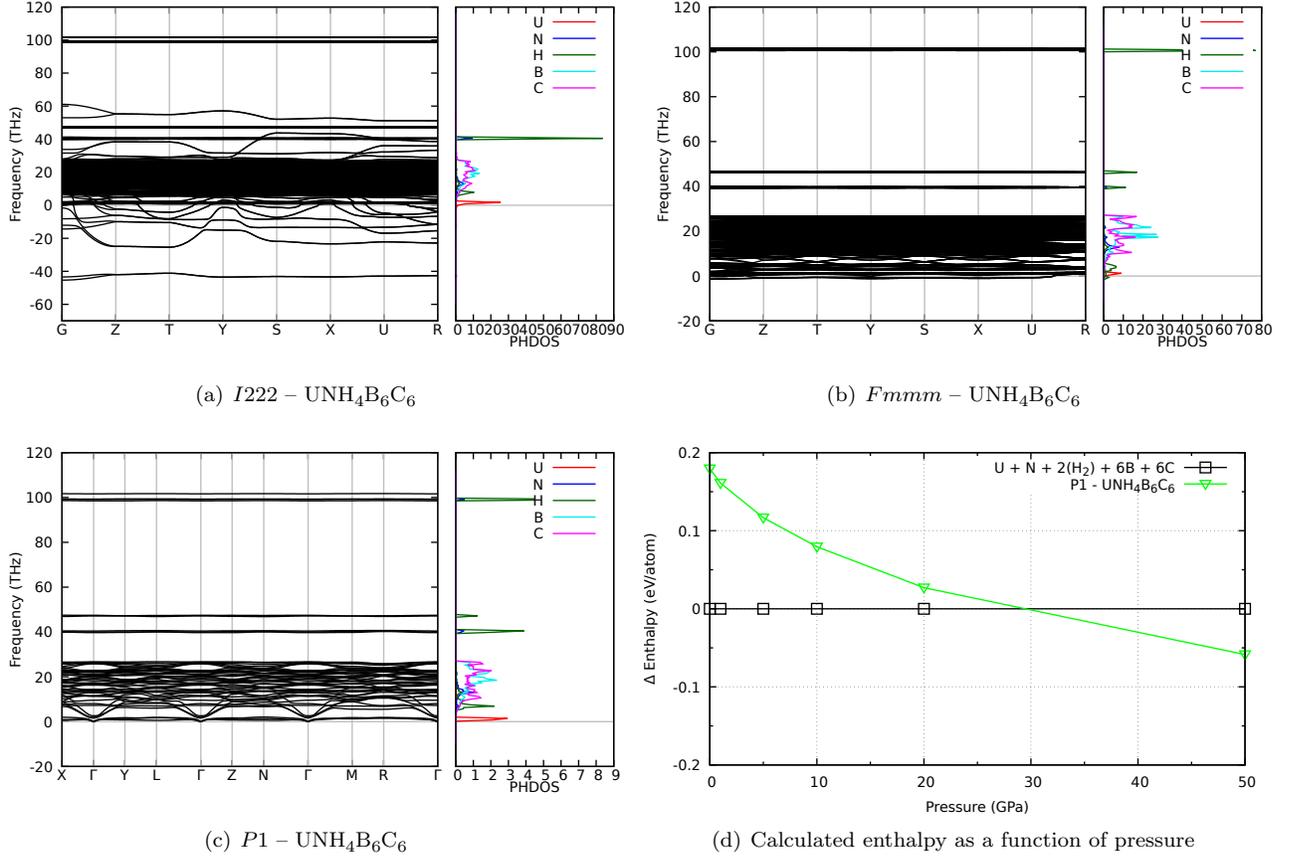

(a) $I222$ – UNH$_4$B$_6$C$_6$

(b) $Fmmm$ – UNH$_4$B$_6$C$_6$

(c) $P1$ – UNH$_4$B$_6$C$_6$

(d) Calculated enthalpy as a function of pressure

FIG. S26. (a–c) Phonon dispersion (left panel) and projected phonon densities of states (PHDOS, right panel) of different UNH$_4$B$_6$C$_6$ phases at ambient pressure. The absence of an imaginary frequency in the phonon spectra indicates that the structures are dynamically stable at ambient pressure. (d) Calculated enthalpy as a function of pressure for UNH$_4$B$_6$C$_6$ relative to simple elemental substance, and the phase transition pressure is ∼29.5 GPa. Ground state structures of simple substances were collected from the Inorganic Crystal Structure Database (ICSD) [1].

# II. SUPPLEMENTARY TABLES

## 2.1 Electron-phonon interaction properties of $MNH_4B_6C_6$ compounds.

TABLE S1: Summary of electron-phonon interaction properties for different phases of $MNH_4B_6C_6$ at ambient pressure. The properties were computed using the Interpolation over the Brillouin Zone method and the Optimized Tetrahedron method, respectively. The phases are categorized by their compound and space group (S.G.). Listed values include the electronic density of states at the Fermi level ($N(E_f)$ in states/Ry/$f.u.$), electron-phonon coupling constant ($\lambda$), logarithmically averaged phonon frequency ($\omega_{log}$ in K), and the critical temperature ($T_c$ in K) estimated using Coulomb pseudopotential values of $\mu^\star = 0.10$–0.13, derived from solving the AD-McMillan (AD-M) equation and Eliashberg (E) equations. The computational demand was extraordinarily high, but in particular with $I222$–$LuNH_4B_6C_6$ and $P1$–$LaNH_4B_6C_6$. Attempts to compute the electron-phonon coupling properties of $I222$–$LuNH_4B_6C_6$ and $P1$–$LaNH_4B_6C_6$ phases using 64-core nodes over 14 days (the maximum allowable computation time) for some single q-points did not yield results, leading us to abandon using the tetrahedron method for these phases.

| Compound | S.G. | Interpolation over the Brillouin Zone | | | | | Optimized tetrahedron method | | | | |
|---|---|---|---|---|---|---|---|---|---|---|---|
| | | $N(E_f)$ | $\lambda$ | $\omega_{log}$ | $T_c$ (AD-M) | $T_c$ (E) | $N(E_f)$ | $\lambda$ | $\omega_{log}$ | $T_c$ (AD-M) | $T_c$ (E) |
| $NaNH_4B_6C_6$ | $I222$ | 16.97 | 0.95 | 481.60 | 28–33 | 29–34 | 19.69 | 1.43 | 446.16 | 49–55 | 53–58 |
| $KNH_4B_6C_6$ | $I222$ | 19.52 | 2.34 | 272.37 | 52–56 | 59–63 | 23.84 | 3.71 | 191.22 | 56–62 | 71–76 |
| $SrNH_4B_6C_6$ | $I23$ | 18.25 | 1.79 | 422.67 | 60–66 | 65–70 | 19.51 | 3.02 | 301.52 | 70–77 | 80–85 |
| $YNH_4B_6C_6$ | $I23$ | 13.85 | 2.10 | 192.57 | 33–36 | 42–47 | 12.18 | 3.12 | 148.07 | 37–41 | 51–56 |
| $LaNH_4B_6C_6$ | $I23$ | 14.34 | 1.01 | 534.06 | 35–41 | 37–42 | 13.11 | 1.07 | 513.51 | 37–43 | 39–44 |
| $LuNH_4B_6C_6$ | $I222$ | 13.90 | 1.54 | 252.71 | 31–34 | 39–44 | – | – | – | – | – |
| $PbNH_4B_6C_6$ | $I23$ | 18.82 | 2.31 | 318.72 | 59–64 | 67–72 | 18.78 | 3.76 | 355.95 | 96–104 | 109–115 |
| $AcNH_4B_6C_6$ | $I23$ | 14.84 | 1.10 | 515.00 | 38–44 | 41–47 | 12.64 | 1.42 | 481.26 | 52–58 | 56–62 |
| $ThNH_4B_6C_6$ | $I23$ | 12.14 | 0.81 | 635.46 | 26–32 | 28–33 | 11.42 | 0.99 | 594.10 | 37–43 | 40–46 |
| $CaNH_4B_6C_6$ | $Fmmm$ | 17.85 | 2.52 | 250.19 | 51–56 | 68–73 | 21.47 | 3.42 | 187.6 | 51–57 | 73–79 |
| $SrNH_4B_6C_6$ | $Fmmm$ | 18.44 | 2.01 | 377.43 | 61–66 | 68–73 | 22.58 | 3.53 | 203.86 | 57–63 | 77–83 |
| $YNH_4B_6C_6$ | $Fmmm$ | 13.68 | 1.98 | 162.90 | 26–29 | 42–47 | 12.68 | 3.6 | 109.85 | 31–35 | 44–49 |
| $LaNH_4B_6C_6$ | $Fmmm$ | 14.10 | 1.02 | 525.15 | 35–41 | 37–42 | 13.22 | 1.16 | 408.1 | 34–39 | 36–40 |
| $YNH_4B_6C_6$ | $P1$ | 13.85 | 2.55 | 221.40 | 46–51 | 55–60 | 14.18 | 1.17 | 333.78 | 28–32 | 29–33 |
| $LaNH_4B_6C_6$ | $P1$ | 14.13 | 1.23 | 480.22 | 43–49 | 46–51 | – | – | – | – | – |
| $LuNH_4B_6C_6$ | $P1$ | 12.60 | 1.26 | 351.16 | 33–38 | 37–42 | 10.69 | 1.26 | 375.04 | 35–40 | 38–43 |
| $PbNH_4B_6C_6$ | $P1$ | 20.13 | 3.73 | 223.81 | 65–72 | 80–86 | 22.53 | 3.91 | 180.57 | 56–62 | 78–84 |
| $AcNH_4B_6C_6$ | $P1$ | 14.88 | 1.22 | 508.39 | 45–51 | 48–54 | 14.50 | 0.94 | 468.24 | 27–32 | 28–33 |
| $ThNH_4B_6C_6$ | $P1$ | 12.21 | 0.68 | 666.47 | 17–22 | 18–24 | 12.96 | 0.82 | 567.11 | 24–29 | 25–31 |

# III. SUPPLEMENTARY INFORMATION

## 3.1 Structure parameters of $MNH_4B_6C_6$ compounds.

TABLE S2: Structural parameters of proposed $MNH_4B_6C_6$ compounds at ambient pressure.

| Compound S.G.(Setting) | Lattice (Å,°) | Atomic coordinates (fractional) | | | |
|---|---|---|---|---|---|
| | | Atoms | X | Y | Z |
| $AcNH_4B_6C_6$ | a = 9.9078 | Ac(2a) | 0.0000 | 0.0000 | 0.0000 |
| $I23$ (1) | b = 9.9078 | Ac(6b) | 0.0000 | 0.5000 | 0.5000 |
| | c = 9.9078 | B(24f) | 0.1279 | 0.0004 | 0.7503 |
| | $\alpha = 90$ | B(24f) | 0.3710 | 0.9960 | 0.7505 |
| | $\beta = 90$ | C(24f) | 0.1229 | 0.7499 | 0.9989 |
| | $\gamma = 90$ | C(24f) | 0.8763 | 0.2498 | 0.5027 |
| | | H(24f) | 0.2213 | 0.8304 | 0.6933 |
| | | H(8c) | 0.3107 | 0.3107 | 0.3107 |
| | | N(8c) | 0.2513 | 0.2513 | 0.2513 |
| $AcNH_4B_6C_6$ | a = 4.9198 | Ac(1a) | 0.1948 | 0.8082 | 0.1951 |
| $P1$ (1) | b = 4.9205 | N(1a) | 0.6918 | 0.3057 | 0.6918 |
| | c = 4.9838 | H(1a) | 0.8536 | 0.1923 | 0.7592 |
| | $\alpha = 89.9$ | H(1a) | 0.5711 | 0.1838 | 0.5797 |
| | $\beta = 89.9$ | H(1a) | 0.7567 | 0.4616 | 0.5732 |
| | $\gamma = 89.1$ | H(1a) | 0.5817 | 0.3806 | 0.8496 |
| | | B(1a) | 0.9339 | 0.8031 | 0.6956 |
| | | B(1a) | 0.1925 | 0.3088 | 0.9385 |
| | | B(1a) | 0.6937 | 0.0617 | 0.1921 |
| | | B(1a) | 0.4546 | 0.8063 | 0.6958 |
| | | B(1a) | 0.1921 | 0.3088 | 0.4510 |
| | | B(1a) | 0.6971 | 0.5544 | 0.1920 |
| | | C(1a) | 0.1955 | 0.0545 | 0.6947 |
| | | C(1a) | 0.6945 | 0.8073 | 0.9481 |
| | | C(1a) | 0.9483 | 0.3080 | 0.1940 |
| | | C(1a) | 0.1934 | 0.5600 | 0.6947 |
| | | C(1a) | 0.6945 | 0.8078 | 0.4403 |
| | | C(1a) | 0.4400 | 0.3074 | 0.1945 |
| $CaNH_4B_6C_6$ | a = 9.7007 | B(16m) | 0.0000 | 0.6255 | 0.7518 |
| $Fmmm$ (1) | b = 9.6567 | B(16n) | 0.7478 | 0.0000 | 0.1260 |
| | c = 9.6568 | B(16o) | 0.8743 | 0.2519 | 0.0000 |
| | $\alpha = 90$ | C(16o) | 0.7488 | 0.1218 | 0.0000 |
| | $\beta = 90$ | C(16m) | 0.0000 | 0.2510 | 0.1214 |
| | $\gamma = 90$ | C(16n) | 0.3784 | 0.0000 | 0.7513 |
| | | Ca(4a) | 0.0000 | 0.0000 | 0.0000 |
| | | Ca(4b) | 0.0000 | 0.0000 | 0.5000 |
| | | H(32p) | 0.6895 | 0.8107 | 0.3107 |
| | | N(8f) | 0.2500 | 0.2500 | 0.2500 |
| $DyNH_4B_6C_6$ | a = 9.6964 | B(16m) | 0.0000 | 0.2515 | 0.1258 |
| $Fmmm$ (1) | b = 9.5719 | B(16o) | 0.7516 | 0.6284 | 0.0000 |
| | c = 9.5761 | B(16n) | 0.6268 | 0.0000 | 0.7517 |
| | $\alpha = 90$ | C(16n) | 0.2510 | 0.0000 | 0.1205 |
| | $\beta = 90$ | C(16m) | 0.0000 | 0.6198 | 0.7510 |
| | $\gamma = 90$ | C(16o) | 0.6202 | 0.7507 | 0.0000 |
| | | Dy(4a) | 0.0000 | 0.0000 | 0.0000 |
| | | Dy(4b) | 0.0000 | 0.0000 | 0.5000 |
| | | H(32p) | 0.8107 | 0.8111 | 0.8113 |
| | | N(8f) | 0.2500 | 0.2500 | 0.2500 |
| $DyNH_4B_6C_6$ | a = 9.6986 | B(8k) | 0.2497 | 0.9997 | 0.8740 |
| $I222$ (1) | b = 9.7007 | B(8k) | 0.9998 | 0.1260 | 0.7503 |
| | c = 9.7008 | B(8k) | 0.1260 | 0.2497 | 0.0002 |
| | $\alpha = 90$ | B(8k) | 0.2493 | 0.0025 | 0.6265 |
| | $\beta = 90$ | B(8k) | 0.0025 | 0.3735 | 0.7507 |
| | $\gamma = 90$ | B(8k) | 0.3735 | 0.2493 | 0.9975 |
| | | C(8k) | 0.0009 | 0.2502 | 0.8801 |

| Compound | Lattice | Atomic coordinates (fractional) | | | |
|---|---|---|---|---|---|
| S.G.(Setting) | (Å,°) | Atoms | X | Y | Z |
| | | C(8k) | 0.2502 | 0.1199 | 0.9991 |
| | | C(8k) | 0.1199 | 0.0009 | 0.7498 |
| | | C(8k) | 0.9985 | 0.2504 | 0.6205 |
| | | C(8k) | 0.2504 | 0.3795 | 0.0015 |
| | | C(8k) | 0.3795 | 0.9985 | 0.7496 |
| | | Dy(2a) | 0.0000 | 0.0000 | 0.0000 |
| | | Dy(2c) | 0.0000 | 0.0000 | 0.5000 |
| | | Dy(2d) | 0.0000 | 0.5000 | 0.0000 |
| | | Dy(2b) | 0.5000 | 0.0000 | 0.0000 |
| | | H(8k) | 0.3080 | 0.1687 | 0.7806 |
| | | H(8k) | 0.1687 | 0.2194 | 0.6919 |
| | | H(8k) | 0.2194 | 0.3081 | 0.8313 |
| | | H(8k) | 0.3118 | 0.3117 | 0.6883 |
| | | N(8k) | 0.2512 | 0.2512 | 0.7488 |
| DyNH$_4$B$_6$C$_6$ | a = 4.8228 | Dy(1a) | 0.8094 | 0.8098 | 0.8095 |
| P1 (1) | b = 4.8460 | N(1a) | 0.3053 | 0.3055 | 0.3049 |
| | c = 4.8924 | H(1a) | 0.1789 | 0.1830 | 0.1924 |
| | $\alpha$ = 89.7 | H(1a) | 0.3736 | 0.4634 | 0.1839 |
| | $\beta$ = 89.9 | H(1a) | 0.4680 | 0.1893 | 0.3723 |
| | $\gamma$ = 89.9 | H(1a) | 0.1959 | 0.3813 | 0.4662 |
| | | B(1a) | 0.3091 | 0.0564 | 0.8056 |
| | | B(1a) | 0.0637 | 0.8046 | 0.3095 |
| | | B(1a) | 0.8051 | 0.3092 | 0.0606 |
| | | B(1a) | 0.3090 | 0.5595 | 0.8057 |
| | | B(1a) | 0.5517 | 0.8047 | 0.3091 |
| | | B(1a) | 0.8057 | 0.3095 | 0.5551 |
| | | C(1a) | 0.3080 | 0.8077 | 0.0472 |
| | | C(1a) | 0.0463 | 0.3078 | 0.8078 |
| | | C(1a) | 0.8078 | 0.0474 | 0.3082 |
| | | C(1a) | 0.3077 | 0.8076 | 0.5668 |
| | | C(1a) | 0.5674 | 0.3077 | 0.8075 |
| | | C(1a) | 0.8074 | 0.5655 | 0.3076 |
| ErNH$_4$B$_6$C$_6$ | a = 9.6816 | B(8k) | 0.2497 | 0.9997 | 0.8741 |
| I222 (1) | b = 9.6837 | B(8k) | 0.9998 | 0.1259 | 0.7503 |
| | c = 9.6838 | B(8k) | 0.1259 | 0.2497 | 0.0002 |
| | $\alpha$ = 90 | B(8k) | 0.2493 | 0.0025 | 0.6263 |
| | $\beta$ = 90 | B(8k) | 0.0025 | 0.3736 | 0.7507 |
| | $\gamma$ = 90 | B(8k) | 0.3737 | 0.2493 | 0.9975 |
| | | C(8k) | 0.0009 | 0.2502 | 0.8803 |
| | | C(8k) | 0.2502 | 0.1197 | 0.9991 |
| | | C(8k) | 0.1197 | 0.0009 | 0.7498 |
| | | C(8k) | 0.9985 | 0.2504 | 0.6204 |
| | | C(8k) | 0.2504 | 0.3797 | 0.0015 |
| | | C(8k) | 0.3796 | 0.9985 | 0.7496 |
| | | Er(2a) | 0.0000 | 0.0000 | 0.0000 |
| | | Er(2c) | 0.0000 | 0.0000 | 0.5000 |
| | | Er(2d) | 0.0000 | 0.5000 | 0.0000 |
| | | Er(2b) | 0.5000 | 0.0000 | 0.0000 |
| | | H(8k) | 0.3081 | 0.1687 | 0.7808 |
| | | H(8k) | 0.1687 | 0.2192 | 0.6918 |
| | | H(8k) | 0.2192 | 0.3082 | 0.8313 |
| | | H(8k) | 0.3118 | 0.3118 | 0.6882 |
| | | N(8k) | 0.2512 | 0.2512 | 0.7488 |
| ErNH$_4$B$_6$C$_6$ | a = 4.8093 | Er(1a) | 0.8099 | 0.8106 | 0.8102 |
| P1 (1) | b = 4.8387 | N(1a) | 0.3053 | 0.3054 | 0.3048 |
| | c = 4.8922 | H(1a) | 0.1775 | 0.1828 | 0.1938 |
| | $\alpha$ = 89.6 | H(1a) | 0.3727 | 0.4635 | 0.1828 |
| | $\beta$ = 89.8 | H(1a) | 0.4690 | 0.1889 | 0.3710 |
| | $\gamma$ = 89.9 | H(1a) | 0.1971 | 0.3816 | 0.4668 |

| Compound | Lattice | Atomic coordinates (fractional) | | | |
|---|---|---|---|---|---|
| S.G.(Setting) | (Å,°) | Atoms | X | Y | Z |
| | | B(1a) | 0.3092 | 0.0557 | 0.8055 |
| | | B(1a) | 0.0639 | 0.8045 | 0.3097 |
| | | B(1a) | 0.8051 | 0.3093 | 0.0603 |
| | | B(1a) | 0.3091 | 0.5602 | 0.8057 |
| | | B(1a) | 0.5514 | 0.8045 | 0.3091 |
| | | B(1a) | 0.8057 | 0.3098 | 0.5553 |
| | | C(1a) | 0.3080 | 0.8076 | 0.0468 |
| | | C(1a) | 0.0458 | 0.3078 | 0.8078 |
| | | C(1a) | 0.8078 | 0.0470 | 0.3083 |
| | | C(1a) | 0.3077 | 0.8076 | 0.5672 |
| | | C(1a) | 0.5677 | 0.3077 | 0.8075 |
| | | C(1a) | 0.8074 | 0.5655 | 0.3074 |
| $EuNH_4B_6C_6$ | a = 9.7160 | B(16m) | 0.0000 | 0.2515 | 0.1258 |
| $Fmmm$ (1) | b = 9.5980 | B(16o) | 0.7520 | 0.6272 | 0.0000 |
| | c = 9.5987 | B(16n) | 0.6263 | 0.0000 | 0.7516 |
| | $\alpha = 90$ | C(16n) | 0.2510 | 0.0000 | 0.1226 |
| | $\beta = 90$ | C(16m) | 0.0000 | 0.6217 | 0.7508 |
| | $\gamma = 90$ | C(16o) | 0.6220 | 0.7511 | 0.0000 |
| | | Eu(4a) | 0.0000 | 0.0000 | 0.0000 |
| | | Eu(4b) | 0.0000 | 0.0000 | 0.5000 |
| | | H(32p) | 0.8105 | 0.8110 | 0.8111 |
| | | N(8f) | 0.2500 | 0.2500 | 0.2500 |
| $EuNH_4B_6C_6$ | a = 9.7223 | B(8k) | 0.6254 | 0.4993 | 0.7495 |
| $I222$ (1) | b = 9.7223 | B(8k) | 0.2495 | 0.1254 | 0.9993 |
| | c = 9.7198 | B(8k) | 0.9993 | 0.2495 | 0.1254 |
| | $\alpha = 90$ | B(8k) | 0.3735 | 0.0045 | 0.2497 |
| | $\beta = 90$ | B(8k) | 0.7497 | 0.8735 | 0.5045 |
| | $\gamma = 90$ | B(8k) | 0.0045 | 0.2497 | 0.3735 |
| | | C(8k) | 0.6219 | 0.7500 | 0.5007 |
| | | C(8k) | 0.0007 | 0.1219 | 0.2500 |
| | | C(8k) | 0.2500 | 0.0007 | 0.1219 |
| | | C(8k) | 0.3774 | 0.2504 | 0.9971 |
| | | C(8k) | 0.9971 | 0.3774 | 0.2504 |
| | | C(8k) | 0.7504 | 0.4971 | 0.8774 |
| | | Eu(2a) | 0.0000 | 0.0000 | 0.0000 |
| | | Eu(2b) | 0.5000 | 0.0000 | 0.0000 |
| | | Eu(2d) | 0.0000 | 0.5000 | 0.0000 |
| | | Eu(2c) | 0.0000 | 0.0000 | 0.5000 |
| | | H(8k) | 0.7183 | 0.6693 | 0.8080 |
| | | H(8k) | 0.3080 | 0.2183 | 0.1693 |
| | | H(8k) | 0.1693 | 0.3080 | 0.2183 |
| | | H(8k) | 0.3114 | 0.3114 | 0.3114 |
| | | N(8k) | 0.2510 | 0.2510 | 0.2510 |
| $EuNH_4B_6C_6$ | a = 6.8172 | Eu(1a) | 0.8067 | 0.8069 | 0.6134 |
| $P1$ (1) | b = 4.7757 | N(1a) | 0.3067 | 0.3070 | 0.6134 |
| | c = 4.8523 | H(1a) | 0.1841 | 0.1844 | 0.3730 |
| | $\alpha = 90$ | H(1a) | 0.1874 | 0.4289 | 0.6189 |
| | $\beta = 135$ | H(1a) | 0.4258 | 0.4291 | 0.6077 |
| | $\gamma = 90$ | H(1a) | 0.4294 | 0.1844 | 0.8538 |
| | | B(1a) | 0.8070 | 0.3067 | 0.8643 |
| | | B(1a) | 0.3067 | 0.0598 | 0.1133 |
| | | B(1a) | 0.0614 | 0.8063 | 0.3672 |
| | | B(1a) | 0.8063 | 0.3068 | 0.3623 |
| | | B(1a) | 0.3067 | 0.5536 | 0.1133 |
| | | B(1a) | 0.5519 | 0.8062 | 0.8594 |
| | | C(1a) | 0.0498 | 0.3067 | 0.8562 |
| | | C(1a) | 0.8067 | 0.0510 | 0.1133 |
| | | C(1a) | 0.3072 | 0.8067 | 0.3593 |
| | | C(1a) | 0.5635 | 0.3067 | 0.3705 |

31Continued

| Compound | Lattice | Atomic coordinates (fractional) | | | |
|---|---|---|---|---|---|
| S.G.(Setting) | (Å,°) | Atoms | X | Y | Z |
| | | C(1a) | 0.8067 | 0.5623 | 0.1133 |
| | | C(1a) | 0.3062 | 0.8067 | 0.8674 |
| $GdNH_4B_6C_6$ | a = 9.7162 | B(16m) | 0.0000 | 0.2515 | 0.1260 |
| $Fmmm$ (1) | b = 9.5961 | B(16o) | 0.7516 | 0.6284 | 0.0000 |
| | c = 9.5997 | B(16n) | 0.6270 | 0.0000 | 0.7517 |
| | $\alpha = 90$ | C(16n) | 0.2510 | 0.0000 | 0.1207 |
| | $\beta = 90$ | C(16m) | 0.0000 | 0.6201 | 0.7509 |
| | $\gamma = 90$ | C(16o) | 0.6204 | 0.7507 | 0.0000 |
| | | Gd(4a) | 0.0000 | 0.0000 | 0.0000 |
| | | Gd(4b) | 0.0000 | 0.0000 | 0.5000 |
| | | H(32p) | 0.8105 | 0.8110 | 0.8112 |
| | | N(8f) | 0.2500 | 0.2500 | 0.2500 |
| $GdNH_4B_6C_6$ | a = 9.7207 | B(8k) | 0.1261 | 0.9998 | 0.2497 |
| $I222$ (1) | b = 9.7207 | B(8k) | 0.2497 | 0.1261 | 0.9998 |
| | c = 9.7186 | B(8k) | 0.9998 | 0.2497 | 0.1261 |
| | $\alpha = 90$ | B(8k) | 0.3734 | 0.0026 | 0.2493 |
| | $\beta = 90$ | B(8k) | 0.2493 | 0.3734 | 0.0026 |
| | $\gamma = 90$ | B(8k) | 0.0026 | 0.2493 | 0.3734 |
| | | C(8k) | 0.1201 | 0.2502 | 0.0009 |
| | | C(8k) | 0.0009 | 0.1201 | 0.2502 |
| | | C(8k) | 0.2502 | 0.0009 | 0.1201 |
| | | C(8k) | 0.3793 | 0.2503 | 0.9984 |
| | | C(8k) | 0.9984 | 0.3793 | 0.2503 |
| | | C(8k) | 0.2503 | 0.9984 | 0.3793 |
| | | Gd(2a) | 0.0000 | 0.0000 | 0.0000 |
| | | Gd(2b) | 0.5000 | 0.0000 | 0.0000 |
| | | Gd(2d) | 0.0000 | 0.5000 | 0.0000 |
| | | Gd(2c) | 0.0000 | 0.0000 | 0.5000 |
| | | H(8k) | 0.2196 | 0.1688 | 0.3079 |
| | | H(8k) | 0.3079 | 0.2196 | 0.1688 |
| | | H(8k) | 0.1688 | 0.3079 | 0.2196 |
| | | H(8k) | 0.3116 | 0.3116 | 0.3116 |
| | | N(8k) | 0.2512 | 0.2512 | 0.2512 |
| $GdNH_4B_6C_6$ | a = 4.8393 | Gd(1a) | 0.8090 | 0.8093 | 0.8091 |
| $P1$ (1) | b = 4.8546 | N(1a) | 0.3053 | 0.3055 | 0.3050 |
| | c = 4.8920 | H(1a) | 0.1807 | 0.1833 | 0.1907 |
| | $\alpha = 89.8$ | H(1a) | 0.3747 | 0.4633 | 0.1853 |
| | $\beta = 89.9$ | H(1a) | 0.4669 | 0.1897 | 0.3740 |
| | $\gamma = 89.9$ | H(1a) | 0.1944 | 0.3809 | 0.4654 |
| | | B(1a) | 0.3091 | 0.0574 | 0.8055 |
| | | B(1a) | 0.0634 | 0.8048 | 0.3094 |
| | | B(1a) | 0.8052 | 0.3092 | 0.0608 |
| | | B(1a) | 0.3089 | 0.5585 | 0.8057 |
| | | B(1a) | 0.5521 | 0.8049 | 0.3090 |
| | | B(1a) | 0.8057 | 0.3092 | 0.5550 |
| | | C(1a) | 0.3079 | 0.8077 | 0.0477 |
| | | C(1a) | 0.0468 | 0.3078 | 0.8078 |
| | | C(1a) | 0.8078 | 0.0478 | 0.3081 |
| | | C(1a) | 0.3077 | 0.8076 | 0.5664 |
| | | C(1a) | 0.5669 | 0.3078 | 0.8076 |
| | | C(1a) | 0.8075 | 0.5655 | 0.3078 |
| $HoNH_4B_6C_6$ | a = 9.6877 | B(16m) | 0.0000 | 0.2516 | 0.1257 |
| $Fmmm$ (1) | b = 9.5613 | B(16o) | 0.7516 | 0.6284 | 0.0000 |
| | c = 9.5663 | B(16n) | 0.6268 | 0.0000 | 0.7517 |
| | $\alpha = 90$ | C(16n) | 0.2510 | 0.0000 | 0.1205 |
| | $\beta = 90$ | C(16m) | 0.0000 | 0.6197 | 0.7510 |
| | $\gamma = 90$ | C(16o) | 0.6201 | 0.7507 | 0.0000 |
| | | H(32p) | 0.8107 | 0.8111 | 0.8114 |
| | | Ho(4a) | 0.0000 | 0.0000 | 0.0000 |

Continued on next page

| Compound | Lattice | Atomic coordinates (fractional) | | | |
|---|---|---|---|---|---|
| S.G.(Setting) | (Å,°) | Atoms | X | Y | Z |
| | | Ho(4b) | 0.0000 | 0.0000 | 0.5000 |
| | | N(8f) | 0.2500 | 0.2500 | 0.2500 |
| HoNH$_4$B$_6$C$_6$ | a = 9.6900 | B(8k) | 0.2497 | 0.9997 | 0.8741 |
| I222 (1) | b = 9.6922 | B(8k) | 0.9998 | 0.1260 | 0.7503 |
| | c = 9.6922 | B(8k) | 0.1259 | 0.2497 | 0.0002 |
| | $\alpha$ = 90 | B(8k) | 0.2493 | 0.0025 | 0.6264 |
| | $\beta$ = 90 | B(8k) | 0.0025 | 0.3736 | 0.7507 |
| | $\gamma$ = 90 | B(8k) | 0.3736 | 0.2493 | 0.9975 |
| | | C(8k) | 0.0009 | 0.2502 | 0.8802 |
| | | C(8k) | 0.2502 | 0.1198 | 0.9991 |
| | | C(8k) | 0.1198 | 0.0009 | 0.7498 |
| | | C(8k) | 0.9985 | 0.2504 | 0.6204 |
| | | C(8k) | 0.2504 | 0.3796 | 0.0015 |
| | | C(8k) | 0.3796 | 0.9985 | 0.7496 |
| | | H(8k) | 0.3081 | 0.1687 | 0.7807 |
| | | H(8k) | 0.1687 | 0.2194 | 0.6918 |
| | | H(8k) | 0.2193 | 0.3081 | 0.8313 |
| | | H(8k) | 0.3118 | 0.3117 | 0.6883 |
| | | Ho(2a) | 0.0000 | 0.0000 | 0.0000 |
| | | Ho(2c) | 0.0000 | 0.0000 | 0.5000 |
| | | Ho(2d) | 0.0000 | 0.5000 | 0.0000 |
| | | Ho(2b) | 0.5000 | 0.0000 | 0.0000 |
| | | N(8k) | 0.2512 | 0.2512 | 0.7488 |
| HoNH$_4$B$_6$C$_6$ | a = 4.8159 | Ho(1a) | 0.8096 | 0.8102 | 0.8098 |
| P1 (1) | b = 4.8423 | N(1a) | 0.3053 | 0.3054 | 0.3048 |
| | c = 4.8924 | H(1a) | 0.1782 | 0.1829 | 0.1931 |
| | $\alpha$ = 89.6 | H(1a) | 0.3731 | 0.4635 | 0.1834 |
| | $\beta$ = 89.9 | H(1a) | 0.4685 | 0.1891 | 0.3716 |
| | $\gamma$ = 89.9 | H(1a) | 0.1966 | 0.3815 | 0.4665 |
| | | B(1a) | 0.3092 | 0.0560 | 0.8056 |
| | | B(1a) | 0.0638 | 0.8046 | 0.3096 |
| | | B(1a) | 0.8051 | 0.3093 | 0.0605 |
| | | B(1a) | 0.3090 | 0.5599 | 0.8057 |
| | | B(1a) | 0.5515 | 0.8046 | 0.3091 |
| | | B(1a) | 0.8057 | 0.3096 | 0.5552 |
| | | C(1a) | 0.3080 | 0.8076 | 0.0470 |
| | | C(1a) | 0.0460 | 0.3078 | 0.8078 |
| | | C(1a) | 0.8078 | 0.0472 | 0.3083 |
| | | C(1a) | 0.3077 | 0.8076 | 0.5670 |
| | | C(1a) | 0.5676 | 0.3077 | 0.8075 |
| | | C(1a) | 0.8074 | 0.5655 | 0.3075 |
| KNH$_4$B$_6$C$_6$ | a = 9.6857 | B(8k) | 0.0000 | 0.2500 | 0.1348 |
| I222 (1) | b = 9.7888 | B(8k) | 0.1273 | 0.0085 | 0.2539 |
| | c = 9.7939 | B(8k) | 0.7451 | 0.6269 | 0.4868 |
| | $\alpha$ = 90 | B(8k) | 0.5000 | 0.7500 | 0.8864 |
| | $\beta$ = 90 | B(8k) | 0.3728 | 0.0085 | 0.2461 |
| | $\gamma$ = 90 | B(8k) | 0.2451 | 0.3731 | 0.0132 |
| | | C(8k) | 0.2476 | 0.9941 | 0.1260 |
| | | C(8k) | 0.1112 | 0.2500 | 0.0000 |
| | | C(8k) | 0.5072 | 0.6256 | 0.7556 |
| | | C(8k) | 0.2524 | 0.9941 | 0.3740 |
| | | C(8k) | 0.8663 | 0.7500 | 0.5000 |
| | | C(8k) | 0.9928 | 0.3744 | 0.2556 |
| | | H(8k) | 0.1642 | 0.3104 | 0.2525 |
| | | H(8k) | 0.2514 | 0.1869 | 0.3332 |
| | | H(8k) | 0.7486 | 0.6869 | 0.6668 |
| | | H(8k) | 0.8358 | 0.8104 | 0.7475 |
| | | K(2a) | 0.0000 | 0.0000 | 0.0000 |
| | | K(2c) | 0.0000 | 0.0000 | 0.5000 |

| Compound | Lattice | Atomic coordinates (fractional) | | | |
|---|---|---|---|---|---|
| S.G.(Setting) | (Å,°) | Atoms | X | Y | Z |
| | | K(2b) | 0.5000 | 0.0000 | 0.0000 |
| | | K(2d) | 0.0000 | 0.5000 | 0.0000 |
| | | N(8k) | 0.2500 | 0.2494 | 0.2500 |
| LaNH$_4$B$_6$C$_6$ | a = 9.8075 | B(16m) | 0.0000 | 0.2513 | 0.1272 |
| Fmmm (1) | b = 9.6993 | B(16o) | 0.7514 | 0.6290 | 0.0000 |
| | c = 9.7018 | B(16n) | 0.6278 | 0.0000 | 0.7514 |
| | $\alpha = 90$ | C(16n) | 0.2508 | 0.0000 | 0.1223 |
| | $\beta = 90$ | C(16m) | 0.0000 | 0.6218 | 0.7507 |
| | $\gamma = 90$ | C(16o) | 0.6219 | 0.7507 | 0.0000 |
| | | H(32p) | 0.8100 | 0.8105 | 0.8106 |
| | | La(4a) | 0.0000 | 0.0000 | 0.0000 |
| | | La(4b) | 0.0000 | 0.0000 | 0.5000 |
| | | N(8f) | 0.2500 | 0.2500 | 0.2500 |
| LaNH$_4$B$_6$C$_6$ | a = 9.8112 | B(24f) | 0.7497 | 0.5003 | 0.6268 |
| I23 (1) | b = 9.8112 | B(24f) | 0.2495 | 0.9969 | 0.3724 |
| | c = 9.8112 | C(24f) | 0.5009 | 0.2499 | 0.6216 |
| | $\alpha = 90$ | C(24f) | 0.4980 | 0.2497 | 0.8778 |
| | $\beta = 90$ | H(24f) | 0.8074 | 0.3305 | 0.7197 |
| | $\gamma = 90$ | H(8c) | 0.1889 | 0.1889 | 0.1889 |
| | | La(2a) | 0.0000 | 0.0000 | 0.0000 |
| | | La(6b) | 0.0000 | 0.5000 | 0.5000 |
| | | N(8c) | 0.2488 | 0.2488 | 0.2488 |
| LaNH$_4$B$_6$C$_6$ | a = 6.8830 | La(1a) | 0.1933 | 0.1932 | 0.3866 |
| P1 (1) | b = 4.8444 | N(1a) | 0.6933 | 0.6933 | 0.3867 |
| | c = 4.8762 | H(1a) | 0.8130 | 0.8145 | 0.6273 |
| | $\alpha = 90$ | H(1a) | 0.5715 | 0.5725 | 0.3847 |
| | $\beta = 135$ | H(1a) | 0.8151 | 0.5722 | 0.3887 |
| | $\gamma = 90$ | H(1a) | 0.5737 | 0.8144 | 0.1459 |
| | | B(1a) | 0.9391 | 0.6933 | 0.1323 |
| | | B(1a) | 0.1933 | 0.9368 | 0.8867 |
| | | B(1a) | 0.6936 | 0.1934 | 0.6295 |
| | | B(1a) | 0.4475 | 0.6933 | 0.6410 |
| | | B(1a) | 0.1933 | 0.4498 | 0.8867 |
| | | B(1a) | 0.6931 | 0.1935 | 0.1438 |
| | | C(1a) | 0.1933 | 0.6933 | 0.1430 |
| | | C(1a) | 0.6933 | 0.9493 | 0.8867 |
| | | C(1a) | 0.9491 | 0.1933 | 0.6423 |
| | | C(1a) | 0.1933 | 0.6933 | 0.6303 |
| | | C(1a) | 0.6933 | 0.4374 | 0.8867 |
| | | C(1a) | 0.4376 | 0.1933 | 0.1310 |
| LuNH$_4$B$_6$C$_6$ | a = 9.6804 | B(8k) | 0.4998 | 0.3743 | 0.7497 |
| I222 (1) | b = 9.6804 | B(8k) | 0.6257 | 0.2503 | 0.4998 |
| | c = 9.6780 | B(8k) | 0.7497 | 0.5002 | 0.6257 |
| | $\alpha = 90$ | B(8k) | 0.5025 | 0.1261 | 0.7492 |
| | $\beta = 90$ | B(8k) | 0.8739 | 0.2508 | 0.5025 |
| | $\gamma = 90$ | B(8k) | 0.7492 | 0.4975 | 0.8739 |
| | | C(8k) | 0.7502 | 0.3804 | 0.5009 |
| | | C(8k) | 0.6196 | 0.4991 | 0.7502 |
| | | C(8k) | 0.5009 | 0.2498 | 0.6196 |
| | | C(8k) | 0.7504 | 0.1203 | 0.4985 |
| | | C(8k) | 0.8797 | 0.5015 | 0.7504 |
| | | C(8k) | 0.4985 | 0.2496 | 0.8797 |
| | | H(8k) | 0.6683 | 0.2802 | 0.8081 |
| | | H(8k) | 0.7198 | 0.1919 | 0.6683 |
| | | H(8k) | 0.8081 | 0.3317 | 0.7198 |
| | | H(8k) | 0.8118 | 0.1882 | 0.8118 |
| | | Lu(2a) | 0.0000 | 0.0000 | 0.0000 |
| | | Lu(2d) | 0.0000 | 0.5000 | 0.0000 |
| | | Lu(2b) | 0.5000 | 0.0000 | 0.0000 |

| Compound | Lattice | Atomic coordinates (fractional) | | | |
|---|---|---|---|---|---|
| S.G.(Setting) | (Å,°) | Atoms | X | Y | Z |
| | | Lu(2c) | 0.0000 | 0.0000 | 0.5000 |
| | | N(8k) | 0.7512 | 0.2488 | 0.7512 |
| $LuNH_4B_6C_6$ | a = 4.7214 | Lu(1a) | 0.8136 | 0.8176 | 0.8625 |
| $P1$ (1) | b = 4.7727 | N(1a) | 0.3045 | 0.3055 | 0.2983 |
| | c = 4.9437 | H(1a) | 0.1694 | 0.1778 | 0.2002 |
| | $\alpha = 89.4$ | H(1a) | 0.3501 | 0.4704 | 0.1724 |
| | $\beta = 89.7$ | H(1a) | 0.4812 | 0.1912 | 0.3438 |
| | $\gamma = 89.8$ | H(1a) | 0.2087 | 0.3737 | 0.4702 |
| | | B(1a) | 0.3107 | 0.0497 | 0.8019 |
| | | B(1a) | 0.0707 | 0.8042 | 0.3165 |
| | | B(1a) | 0.8059 | 0.3121 | 0.0567 |
| | | B(1a) | 0.3105 | 0.5651 | 0.8021 |
| | | B(1a) | 0.5429 | 0.8019 | 0.3141 |
| | | B(1a) | 0.8055 | 0.3099 | 0.5509 |
| | | C(1a) | 0.3092 | 0.8069 | 0.0459 |
| | | C(1a) | 0.0444 | 0.3080 | 0.8032 |
| | | C(1a) | 0.8081 | 0.0495 | 0.3077 |
| | | C(1a) | 0.3073 | 0.8078 | 0.5651 |
| | | C(1a) | 0.5699 | 0.3073 | 0.8041 |
| | | C(1a) | 0.8075 | 0.5615 | 0.3044 |
| $NaNH_4B_6C_6$ | a = 9.6463 | B(8k) | 0.2478 | 0.3886 | 0.4993 |
| $I222$ (1) | b = 9.5936 | B(8k) | 0.9999 | 0.7504 | 0.1106 |
| | c = 9.5825 | B(8k) | 0.8838 | 0.0003 | 0.2482 |
| | $\alpha = 90$ | B(8k) | 0.7404 | 0.6368 | 0.0001 |
| | $\beta = 90$ | B(8k) | 0.5003 | 0.2418 | 0.8623 |
| | $\gamma = 90$ | B(8k) | 0.6333 | 0.9996 | 0.2617 |
| | | C(8k) | 0.4999 | 0.3711 | 0.7465 |
| | | C(8k) | 0.7508 | 0.0003 | 0.1336 |
| | | C(8k) | 0.8663 | 0.7547 | 0.0001 |
| | | C(8k) | 0.0006 | 0.6119 | 0.2400 |
| | | C(8k) | 0.7546 | 0.0003 | 0.3932 |
| | | C(8k) | 0.1058 | 0.2575 | 0.4997 |
| | | H(8k) | 0.6927 | 0.8153 | 0.1896 |
| | | H(8k) | 0.8086 | 0.6891 | 0.1867 |
| | | H(8k) | 0.3150 | 0.3091 | 0.8087 |
| | | H(8k) | 0.6879 | 0.6910 | 0.3109 |
| | | N(8k) | 0.7501 | 0.7502 | 0.2498 |
| | | Na(2a) | 0.0000 | 0.0000 | 0.0000 |
| | | Na(2d) | 0.0000 | 0.5000 | 0.0000 |
| | | Na(2c) | 0.0000 | 0.0000 | 0.5000 |
| | | Na(2b) | 0.5000 | 0.0000 | 0.0000 |
| $NdNH_4B_6C_6$ | a = 9.7764 | B(16m) | 0.0000 | 0.2514 | 0.1266 |
| $Fmmm$ (1) | b = 9.6671 | B(16o) | 0.7515 | 0.6286 | 0.0000 |
| | c = 9.6692 | B(16n) | 0.6274 | 0.0000 | 0.7515 |
| | $\alpha = 90$ | C(16n) | 0.2509 | 0.0000 | 0.1214 |
| | $\beta = 90$ | C(16m) | 0.0000 | 0.6208 | 0.7508 |
| | $\gamma = 90$ | C(16o) | 0.6210 | 0.7507 | 0.0000 |
| | | H(32p) | 0.8102 | 0.8106 | 0.8108 |
| | | N(8f) | 0.2500 | 0.2500 | 0.2500 |
| | | Nd(4a) | 0.0000 | 0.0000 | 0.0000 |
| | | Nd(4b) | 0.0000 | 0.0000 | 0.5000 |
| $NdNH_4B_6C_6$ | a = 9.7787 | B(8k) | 0.2497 | 0.1264 | 0.9997 |
| $I222$ (1) | b = 9.7807 | B(8k) | 0.4998 | 0.7497 | 0.6265 |
| | c = 9.7807 | B(8k) | 0.1265 | 0.9998 | 0.2497 |
| | $\alpha = 90$ | B(8k) | 0.2494 | 0.3729 | 0.0029 |
| | $\beta = 90$ | B(8k) | 0.0029 | 0.2494 | 0.3729 |
| | $\gamma = 90$ | B(8k) | 0.8729 | 0.5029 | 0.7494 |
| | | C(8k) | 0.0009 | 0.1207 | 0.2502 |
| | | C(8k) | 0.7502 | 0.5009 | 0.6207 |

| Compound | Lattice | Atomic coordinates (fractional) | | | |
| --- | --- | --- | --- | --- | --- |
| S.G.(Setting) | (Å,°) | Atoms | X | Y | Z |
| | | C(8k) | 0.1207 | 0.2502 | 0.0009 |
| | | C(8k) | 0.4982 | 0.8786 | 0.7503 |
| | | C(8k) | 0.2503 | 0.9982 | 0.3786 |
| | | C(8k) | 0.3786 | 0.2503 | 0.9982 |
| | | H(8k) | 0.3074 | 0.2203 | 0.1689 |
| | | H(8k) | 0.6689 | 0.8075 | 0.7204 |
| | | H(8k) | 0.2203 | 0.1690 | 0.3075 |
| | | H(8k) | 0.8114 | 0.8113 | 0.8113 |
| | | N(8k) | 0.2513 | 0.2513 | 0.2513 |
| | | Nd(2a) | 0.0000 | 0.0000 | 0.0000 |
| | | Nd(2d) | 0.0000 | 0.5000 | 0.0000 |
| | | Nd(2c) | 0.0000 | 0.0000 | 0.5000 |
| | | Nd(2b) | 0.5000 | 0.0000 | 0.0000 |
| NdNH$_4$B$_6$C$_6$ | a = 4.8788 | Nd(1a) | 0.8085 | 0.8087 | 0.8086 |
| $P1$ (1) | b = 4.8814 | N(1a) | 0.3053 | 0.3054 | 0.3051 |
| | c = 4.9073 | H(1a) | 0.1833 | 0.1839 | 0.1887 |
| | $\alpha = 89.9$ | H(1a) | 0.4650 | 0.1908 | 0.3757 |
| | $\beta = 89.9$ | H(1a) | 0.1928 | 0.3794 | 0.4641 |
| | $\gamma = 89.9$ | H(1a) | 0.3757 | 0.4631 | 0.1876 |
| | | B(1a) | 0.0633 | 0.8050 | 0.3092 |
| | | B(1a) | 0.8053 | 0.3091 | 0.0614 |
| | | B(1a) | 0.3091 | 0.0597 | 0.8054 |
| | | B(1a) | 0.5524 | 0.8052 | 0.3089 |
| | | B(1a) | 0.8057 | 0.3089 | 0.5545 |
| | | B(1a) | 0.3087 | 0.5562 | 0.8056 |
| | | C(1a) | 0.8078 | 0.0491 | 0.3080 |
| | | C(1a) | 0.3079 | 0.8078 | 0.0490 |
| | | C(1a) | 0.0485 | 0.3078 | 0.8078 |
| | | C(1a) | 0.8075 | 0.5647 | 0.3079 |
| | | C(1a) | 0.3078 | 0.8075 | 0.5652 |
| | | C(1a) | 0.5656 | 0.3078 | 0.8075 |
| PaNH$_4$B$_6$C$_6$ | a = 9.7037 | B(8k) | 0.2497 | 0.9999 | 0.8741 |
| $I222$ (1) | b = 9.7061 | B(8k) | 0.9999 | 0.1259 | 0.7503 |
| | c = 9.7062 | B(8k) | 0.1259 | 0.2497 | 0.0001 |
| | $\alpha = 90$ | B(8k) | 0.2493 | 0.0016 | 0.6265 |
| | $\beta = 90$ | B(8k) | 0.0016 | 0.3734 | 0.7507 |
| | $\gamma = 90$ | B(8k) | 0.3734 | 0.2493 | 0.9984 |
| | | C(8k) | 0.0007 | 0.2501 | 0.8802 |
| | | C(8k) | 0.2501 | 0.1198 | 0.9993 |
| | | C(8k) | 0.1198 | 0.0007 | 0.7499 |
| | | C(8k) | 0.9993 | 0.2506 | 0.6202 |
| | | C(8k) | 0.2506 | 0.3799 | 0.0007 |
| | | C(8k) | 0.3798 | 0.9993 | 0.7494 |
| | | H(8k) | 0.3078 | 0.1683 | 0.7804 |
| | | H(8k) | 0.1683 | 0.2197 | 0.6921 |
| | | H(8k) | 0.2196 | 0.3078 | 0.8316 |
| | | H(8k) | 0.3118 | 0.3117 | 0.6883 |
| | | N(8k) | 0.2512 | 0.2512 | 0.7488 |
| | | Pa(2a) | 0.0000 | 0.0000 | 0.0000 |
| | | Pa(2c) | 0.0000 | 0.0000 | 0.5000 |
| | | Pa(2d) | 0.0000 | 0.5000 | 0.0000 |
| | | Pa(2b) | 0.5000 | 0.0000 | 0.0000 |
| PaNH$_4$B$_6$C$_6$ | a = 4.8393 | Pa(1a) | 0.1907 | 0.1907 | 0.1903 |
| $P1$ (1) | b = 4.8653 | N(1a) | 0.6947 | 0.6950 | 0.6950 |
| | c = 4.8666 | H(1a) | 0.8183 | 0.8144 | 0.8135 |
| | $\alpha = 89.6$ | H(1a) | 0.6283 | 0.8094 | 0.5312 |
| | $\beta = 89.9$ | H(1a) | 0.5305 | 0.6312 | 0.8109 |
| | $\gamma = 89.9$ | H(1a) | 0.8072 | 0.5302 | 0.6292 |
| | | B(1a) | 0.6907 | 0.1946 | 0.9406 |

| Compound | Lattice | Atomic coordinates (fractional) | | | |
|---|---|---|---|---|---|
| S.G.(Setting) | (Å,°) | Atoms | X | Y | Z |
| | | B(1a) | 0.9393 | 0.6906 | 0.1948 |
| | | B(1a) | 0.1948 | 0.9391 | 0.6908 |
| | | B(1a) | 0.6911 | 0.1941 | 0.4435 |
| | | B(1a) | 0.4449 | 0.6910 | 0.1945 |
| | | B(1a) | 0.1945 | 0.4451 | 0.6906 |
| | | C(1a) | 0.6920 | 0.9533 | 0.1921 |
| | | C(1a) | 0.9529 | 0.1922 | 0.6919 |
| | | C(1a) | 0.1922 | 0.6914 | 0.9528 |
| | | C(1a) | 0.6925 | 0.4323 | 0.1924 |
| | | C(1a) | 0.4330 | 0.1924 | 0.6925 |
| | | C(1a) | 0.1924 | 0.6930 | 0.4332 |
| PbNH$_4$B$_6$C$_6$ | a = 9.8705 | B(24f) | 0.2487 | 0.9985 | 0.8768 |
| I23 (1) | b = 9.8705 | B(24f) | 0.7513 | 0.5080 | 0.1287 |
| | c = 9.8705 | C(24f) | 0.0006 | 0.2490 | 0.8756 |
| | $\alpha$ = 90 | C(24f) | 0.9942 | 0.2508 | 0.6235 |
| | $\beta$ = 90 | H(24f) | 0.3059 | 0.1682 | 0.7788 |
| | $\gamma$ = 90 | H(8c) | 0.6897 | 0.6897 | 0.6897 |
| | | N(8c) | 0.7493 | 0.7493 | 0.7493 |
| | | Pb(2a) | 0.0000 | 0.0000 | 0.0000 |
| | | Pb(6b) | 0.0000 | 0.5000 | 0.5000 |
| PbNH$_4$B$_6$C$_6$ | a = 4.8946 | B(1a) | 0.6962 | 0.9433 | 0.8045 |
| P1 (1) | b = 4.8946 | B(1a) | 0.9436 | 0.1913 | 0.3095 |
| | c = 4.8946 | B(1a) | 0.1935 | 0.6970 | 0.0590 |
| | $\alpha$ = 89.8 | B(1a) | 0.6972 | 0.4460 | 0.8070 |
| | $\beta$ = 89.8 | B(1a) | 0.4458 | 0.1935 | 0.3105 |
| | $\gamma$ = 89.8 | B(1a) | 0.1913 | 0.6964 | 0.5570 |
| | | C(1a) | 0.6953 | 0.1953 | 0.0548 |
| | | C(1a) | 0.9473 | 0.6951 | 0.8080 |
| | | C(1a) | 0.1947 | 0.9474 | 0.3084 |
| | | C(1a) | 0.6951 | 0.1947 | 0.5604 |
| | | C(1a) | 0.4414 | 0.6953 | 0.8086 |
| | | C(1a) | 0.1952 | 0.4413 | 0.3086 |
| | | H(1a) | 0.7469 | 0.8585 | 0.1932 |
| | | H(1a) | 0.5720 | 0.5715 | 0.1846 |
| | | H(1a) | 0.5794 | 0.7465 | 0.4715 |
| | | H(1a) | 0.8581 | 0.5799 | 0.3607 |
| | | N(1a) | 0.6910 | 0.6910 | 0.3044 |
| | | Pb(1a) | 0.1959 | 0.1959 | 0.8092 |
| PmNH$_4$B$_6$C$_6$ | a = 9.7573 | B(16m) | 0.0000 | 0.2514 | 0.1264 |
| Fmmm (1) | b = 9.6450 | B(16o) | 0.7515 | 0.6286 | 0.0000 |
| | c = 9.6476 | B(16n) | 0.6272 | 0.0000 | 0.7516 |
| | $\alpha$ = 90 | C(16n) | 0.2509 | 0.0000 | 0.1212 |
| | $\beta$ = 90 | C(16m) | 0.0000 | 0.6206 | 0.7509 |
| | $\gamma$ = 90 | C(16o) | 0.6208 | 0.7507 | 0.0000 |
| | | H(32p) | 0.8103 | 0.8107 | 0.8109 |
| | | N(8f) | 0.2500 | 0.2500 | 0.2500 |
| | | Pm(4a) | 0.0000 | 0.0000 | 0.0000 |
| | | Pm(4b) | 0.0000 | 0.0000 | 0.5000 |
| PmNH$_4$B$_6$C$_6$ | a = 9.7621 | B(8k) | 0.1263 | 0.9998 | 0.2497 |
| I222 (1) | b = 9.7621 | B(8k) | 0.2497 | 0.1263 | 0.9998 |
| | c = 9.7597 | B(8k) | 0.4998 | 0.7497 | 0.6263 |
| | $\alpha$ = 90 | B(8k) | 0.8730 | 0.5028 | 0.7494 |
| | $\beta$ = 90 | B(8k) | 0.2494 | 0.3730 | 0.0028 |
| | $\gamma$ = 90 | B(8k) | 0.0028 | 0.2494 | 0.3730 |
| | | C(8k) | 0.1205 | 0.2502 | 0.0009 |
| | | C(8k) | 0.0009 | 0.1205 | 0.2502 |
| | | C(8k) | 0.7502 | 0.5009 | 0.6205 |
| | | C(8k) | 0.3789 | 0.2503 | 0.9982 |
| | | C(8k) | 0.4982 | 0.8789 | 0.7503 |

| Compound | Lattice | Atomic coordinates (fractional) | | | |
| S.G.(Setting) | (Å,°) | Atoms | X | Y | Z |
| --- | --- | --- | --- | --- | --- |
| | | C(8k) | 0.2503 | 0.9982 | 0.3789 |
| | | H(8k) | 0.2201 | 0.1689 | 0.3076 |
| | | H(8k) | 0.3076 | 0.2201 | 0.1689 |
| | | H(8k) | 0.6689 | 0.8076 | 0.7201 |
| | | H(8k) | 0.3114 | 0.3114 | 0.3114 |
| | | N(8k) | 0.2513 | 0.2513 | 0.2513 |
| | | Pm(2a) | 0.0000 | 0.0000 | 0.0000 |
| | | Pm(2b) | 0.5000 | 0.0000 | 0.0000 |
| | | Pm(2d) | 0.0000 | 0.5000 | 0.0000 |
| | | Pm(2c) | 0.0000 | 0.0000 | 0.5000 |
| $PmNH_4B_6C_6$ | a = 4.8771 | Pm(1a) | 0.8087 | 0.8087 | 0.8087 |
| $P1$ (1) | b = 4.8780 | N(1a) | 0.3052 | 0.3053 | 0.3052 |
| | c = 4.8870 | H(1a) | 0.1846 | 0.1846 | 0.1863 |
| | $\alpha = 89.9$ | H(1a) | 0.4647 | 0.1904 | 0.3766 |
| | $\beta = 89.9$ | H(1a) | 0.1909 | 0.3777 | 0.4643 |
| | $\gamma = 89.9$ | H(1a) | 0.3764 | 0.4641 | 0.1892 |
| | | B(1a) | 0.0618 | 0.8051 | 0.3092 |
| | | B(1a) | 0.8052 | 0.3091 | 0.0611 |
| | | B(1a) | 0.3091 | 0.0605 | 0.8053 |
| | | B(1a) | 0.5539 | 0.8054 | 0.3089 |
| | | B(1a) | 0.8055 | 0.3089 | 0.5547 |
| | | B(1a) | 0.3089 | 0.5553 | 0.8055 |
| | | C(1a) | 0.8078 | 0.0485 | 0.3079 |
| | | C(1a) | 0.3079 | 0.8078 | 0.0485 |
| | | C(1a) | 0.0483 | 0.3079 | 0.8078 |
| | | C(1a) | 0.8075 | 0.5654 | 0.3078 |
| | | C(1a) | 0.3078 | 0.8075 | 0.5655 |
| | | C(1a) | 0.5657 | 0.3078 | 0.8075 |
| $PrNH_4B_6C_6$ | a = 9.7965 | B(16m) | 0.0000 | 0.2514 | 0.1268 |
| $Fmmm$ (1) | b = 9.6901 | B(16o) | 0.7515 | 0.6287 | 0.0000 |
| | c = 9.6919 | B(16n) | 0.6275 | 0.0000 | 0.7515 |
| | $\alpha = 90$ | C(16n) | 0.2509 | 0.0000 | 0.1216 |
| | $\beta = 90$ | C(16m) | 0.0000 | 0.6211 | 0.7508 |
| | $\gamma = 90$ | C(16o) | 0.6212 | 0.7507 | 0.0000 |
| | | H(32p) | 0.8101 | 0.8105 | 0.8107 |
| | | N(8f) | 0.2500 | 0.2500 | 0.2500 |
| | | Pr(4a) | 0.0000 | 0.0000 | 0.0000 |
| | | Pr(4b) | 0.0000 | 0.0000 | 0.5000 |
| $PrNH_4B_6C_6$ | a = 9.8010 | B(8k) | 0.1266 | 0.9998 | 0.2497 |
| $I222$ (1) | b = 9.8010 | B(8k) | 0.2497 | 0.1266 | 0.9998 |
| | c = 9.7989 | B(8k) | 0.4998 | 0.7497 | 0.6266 |
| | $\alpha = 90$ | B(8k) | 0.8727 | 0.5030 | 0.7494 |
| | $\beta = 90$ | B(8k) | 0.2494 | 0.3727 | 0.0030 |
| | $\gamma = 90$ | B(8k) | 0.0030 | 0.2494 | 0.3727 |
| | | C(8k) | 0.1210 | 0.2501 | 0.0010 |
| | | C(8k) | 0.0010 | 0.1210 | 0.2501 |
| | | C(8k) | 0.7501 | 0.5010 | 0.6210 |
| | | C(8k) | 0.3784 | 0.2503 | 0.9981 |
| | | C(8k) | 0.4981 | 0.8784 | 0.7503 |
| | | C(8k) | 0.2503 | 0.9981 | 0.3784 |
| | | H(8k) | 0.2206 | 0.1690 | 0.3073 |
| | | H(8k) | 0.3073 | 0.2206 | 0.1690 |
| | | H(8k) | 0.6690 | 0.8073 | 0.7206 |
| | | H(8k) | 0.3113 | 0.3113 | 0.3113 |
| | | N(8k) | 0.2513 | 0.2513 | 0.2513 |
| | | Pr(2a) | 0.0000 | 0.0000 | 0.0000 |
| | | Pr(2b) | 0.5000 | 0.0000 | 0.0000 |
| | | Pr(2d) | 0.0000 | 0.5000 | 0.0000 |
| | | Pr(2c) | 0.0000 | 0.0000 | 0.5000 |

| Compound | Lattice | Atomic coordinates (fractional) | | | |
|---|---|---|---|---|---|
| S.G.(Setting) | (Å,°) | Atoms | X | Y | Z |
| PrNH$_4$B$_6$C$_6$ | a = 4.8829 | Pr(1a) | 0.1916 | 0.1914 | 0.8085 |
| P1 (1) | b = 4.8861 | N(1a) | 0.6947 | 0.6945 | 0.3050 |
| | c = 4.9232 | H(1a) | 0.8177 | 0.8164 | 0.1904 |
| | $\alpha = 90.1$ | H(1a) | 0.5352 | 0.8090 | 0.3753 |
| | $\beta = 90.1$ | H(1a) | 0.8058 | 0.6191 | 0.4636 |
| | $\gamma = 90$ | H(1a) | 0.6243 | 0.5379 | 0.1866 |
| | | B(1a) | 0.9357 | 0.1953 | 0.3091 |
| | | B(1a) | 0.1947 | 0.6909 | 0.0617 |
| | | B(1a) | 0.6908 | 0.9406 | 0.8055 |
| | | B(1a) | 0.4487 | 0.1948 | 0.3089 |
| | | B(1a) | 0.1942 | 0.6912 | 0.5542 |
| | | B(1a) | 0.6916 | 0.4435 | 0.8057 |
| | | C(1a) | 0.1923 | 0.9503 | 0.3080 |
| | | C(1a) | 0.6921 | 0.1923 | 0.0495 |
| | | C(1a) | 0.9510 | 0.6922 | 0.8078 |
| | | C(1a) | 0.1926 | 0.4359 | 0.3079 |
| | | C(1a) | 0.6922 | 0.1926 | 0.5648 |
| | | C(1a) | 0.4347 | 0.6922 | 0.8075 |
| SmNH$_4$B$_6$C$_6$ | a = 9.7453 | B(16m) | 0.0000 | 0.2514 | 0.1263 |
| Fmmm (1) | b = 9.6303 | B(16o) | 0.7515 | 0.6285 | 0.0000 |
| | c = 9.6332 | B(16n) | 0.6271 | 0.0000 | 0.7516 |
| | $\alpha = 90$ | C(16n) | 0.2509 | 0.0000 | 0.1210 |
| | $\beta = 90$ | C(16m) | 0.0000 | 0.6204 | 0.7509 |
| | $\gamma = 90$ | C(16o) | 0.6207 | 0.7507 | 0.0000 |
| | | H(32p) | 0.8104 | 0.8108 | 0.8110 |
| | | N(8f) | 0.2500 | 0.2500 | 0.2500 |
| | | Sm(4a) | 0.0000 | 0.0000 | 0.0000 |
| | | Sm(4b) | 0.0000 | 0.0000 | 0.5000 |
| SmNH$_4$B$_6$C$_6$ | a = 9.7498 | B(8k) | 0.1262 | 0.9998 | 0.2497 |
| I222 (1) | b = 9.7498 | B(8k) | 0.2497 | 0.1262 | 0.9998 |
| | c = 9.7474 | B(8k) | 0.9998 | 0.2497 | 0.1262 |
| | $\alpha = 90$ | B(8k) | 0.3731 | 0.0027 | 0.2494 |
| | $\beta = 90$ | B(8k) | 0.2494 | 0.3731 | 0.0027 |
| | $\gamma = 90$ | B(8k) | 0.0027 | 0.2494 | 0.3731 |
| | | C(8k) | 0.1204 | 0.2502 | 0.0009 |
| | | C(8k) | 0.0009 | 0.1204 | 0.2502 |
| | | C(8k) | 0.2502 | 0.0009 | 0.1204 |
| | | C(8k) | 0.3790 | 0.2503 | 0.9983 |
| | | C(8k) | 0.9983 | 0.3790 | 0.2503 |
| | | C(8k) | 0.2503 | 0.9983 | 0.3790 |
| | | H(8k) | 0.2199 | 0.1689 | 0.3077 |
| | | H(8k) | 0.3077 | 0.2199 | 0.1689 |
| | | H(8k) | 0.1689 | 0.3077 | 0.2199 |
| | | H(8k) | 0.3115 | 0.3115 | 0.3115 |
| | | N(8k) | 0.2513 | 0.2513 | 0.2513 |
| | | Sm(2a) | 0.0000 | 0.0000 | 0.0000 |
| | | Sm(2b) | 0.5000 | 0.0000 | 0.0000 |
| | | Sm(2d) | 0.0000 | 0.5000 | 0.0000 |
| | | Sm(2c) | 0.0000 | 0.0000 | 0.5000 |
| SmNH$_4$B$_6$C$_6$ | a = 4.8706 | Sm(1a) | 0.8088 | 0.8088 | 0.8088 |
| P1 (1) | b = 4.8769 | N(1a) | 0.3053 | 0.3052 | 0.3052 |
| | c = 4.8769 | H(1a) | 0.1843 | 0.1853 | 0.1855 |
| | $\alpha = 89.9$ | H(1a) | 0.3770 | 0.4644 | 0.1898 |
| | $\beta = 89.9$ | H(1a) | 0.4647 | 0.1896 | 0.3763 |
| | $\gamma = 89.9$ | H(1a) | 0.1907 | 0.3772 | 0.4648 |
| | | B(1a) | 0.3091 | 0.0605 | 0.8053 |
| | | B(1a) | 0.0610 | 0.8052 | 0.3092 |
| | | B(1a) | 0.8052 | 0.3092 | 0.0613 |
| | | B(1a) | 0.3089 | 0.5553 | 0.8055 |

| Compound | Lattice | Atomic coordinates (fractional) | | | |
| S.G.(Setting) | (Å,°) | Atoms | X | Y | Z |
|---|---|---|---|---|---|
| | | B(1a) | 0.5548 | 0.8054 | 0.3089 |
| | | B(1a) | 0.8055 | 0.3089 | 0.5544 |
| | | C(1a) | 0.3079 | 0.8078 | 0.0481 |
| | | C(1a) | 0.0480 | 0.3079 | 0.8078 |
| | | C(1a) | 0.8078 | 0.0482 | 0.3079 |
| | | C(1a) | 0.3078 | 0.8075 | 0.5659 |
| | | C(1a) | 0.5658 | 0.3078 | 0.8075 |
| | | C(1a) | 0.8075 | 0.5657 | 0.3078 |
| SrNH$_4$B$_6$C$_6$ | a = 9.7937 | B(16m) | 0.0000 | 0.2516 | 0.1263 |
| Fmmm (1) | b = 9.7372 | B(16o) | 0.7519 | 0.6269 | 0.0000 |
| | c = 9.7374 | B(16n) | 0.6265 | 0.0000 | 0.7517 |
| | $\alpha = 90$ | C(16n) | 0.2510 | 0.0000 | 0.1230 |
| | $\beta = 90$ | C(16m) | 0.0000 | 0.6225 | 0.7509 |
| | $\gamma = 90$ | C(16o) | 0.6227 | 0.7511 | 0.0000 |
| | | H(32p) | 0.8100 | 0.8102 | 0.8103 |
| | | N(8f) | 0.2500 | 0.2500 | 0.2500 |
| | | Sr(4a) | 0.0000 | 0.0000 | 0.0000 |
| | | Sr(4b) | 0.0000 | 0.0000 | 0.5000 |
| SrBH$_4$B$_6$C$_6$ | a = 9.8013 | B(24f) | 0.2493 | 0.9990 | 0.8747 |
| I23 (1) | b = 9.8013 | B(24f) | 0.7501 | 0.5061 | 0.1276 |
| | c = 9.8013 | C(24f) | 0.0008 | 0.2497 | 0.8772 |
| | $\alpha = 90$ | C(24f) | 0.9958 | 0.2505 | 0.6234 |
| | $\beta = 90$ | H(24f) | 0.3069 | 0.1685 | 0.7792 |
| | $\gamma = 90$ | H(8c) | 0.6890 | 0.6890 | 0.6890 |
| | | N(8c) | 0.7490 | 0.7490 | 0.7490 |
| | | Sr(2a) | 0.0000 | 0.0000 | 0.0000 |
| | | Sr(6b) | 0.0000 | 0.5000 | 0.5000 |
| TbNH$_4$B$_6$C$_6$ | a = 9.7057 | B(16m) | 0.0000 | 0.2515 | 0.1259 |
| Fmmm (1) | b = 9.5834 | B(16o) | 0.7516 | 0.6284 | 0.0000 |
| | c = 9.5873 | B(16n) | 0.6269 | 0.0000 | 0.7517 |
| | $\alpha = 90$ | C(16n) | 0.2510 | 0.0000 | 0.1206 |
| | $\beta = 90$ | C(16m) | 0.0000 | 0.6200 | 0.7510 |
| | $\gamma = 90$ | C(16o) | 0.6203 | 0.7507 | 0.0000 |
| | | H(32p) | 0.8106 | 0.8110 | 0.8112 |
| | | N(8f) | 0.2500 | 0.2500 | 0.2500 |
| | | Tb(4a) | 0.0000 | 0.0000 | 0.0000 |
| | | Tb(4b) | 0.0000 | 0.0000 | 0.5000 |
| TbNH$_4$B$_6$C$_6$ | a = 9.7078 | B(8k) | 0.2497 | 0.9997 | 0.8740 |
| I222 (1) | b = 9.7101 | B(8k) | 0.9998 | 0.1260 | 0.7503 |
| | c = 9.7102 | B(8k) | 0.1260 | 0.2497 | 0.0002 |
| | $\alpha = 90$ | B(8k) | 0.2493 | 0.0026 | 0.6265 |
| | $\beta = 90$ | B(8k) | 0.0026 | 0.3734 | 0.7507 |
| | $\gamma = 90$ | B(8k) | 0.3734 | 0.2493 | 0.9974 |
| | | C(8k) | 0.0009 | 0.2502 | 0.8800 |
| | | C(8k) | 0.2502 | 0.1200 | 0.9991 |
| | | C(8k) | 0.1200 | 0.0009 | 0.7498 |
| | | C(8k) | 0.9984 | 0.2503 | 0.6206 |
| | | C(8k) | 0.2503 | 0.3794 | 0.0016 |
| | | C(8k) | 0.3794 | 0.9984 | 0.7497 |
| | | H(8k) | 0.3079 | 0.1687 | 0.7806 |
| | | H(8k) | 0.1688 | 0.2195 | 0.6920 |
| | | H(8k) | 0.2194 | 0.3080 | 0.8312 |
| | | H(8k) | 0.3117 | 0.3116 | 0.6884 |
| | | N(8k) | 0.2512 | 0.2512 | 0.7488 |
| | | Tb(2a) | 0.0000 | 0.0000 | 0.0000 |
| | | Tb(2c) | 0.0000 | 0.0000 | 0.5000 |
| | | Tb(2d) | 0.0000 | 0.5000 | 0.0000 |
| | | Tb(2b) | 0.5000 | 0.0000 | 0.0000 |
| TbNH$_4$B$_6$C$_6$ | a = 4.8304 | Tb(1a) | 0.8092 | 0.8095 | 0.8093 |

| Compound | Lattice | Atomic coordinates (fractional) | | | |
|---|---|---|---|---|---|
| S.G.(Setting) | (Å,°) | Atoms | X | Y | Z |
| $P1$ (1) | b = 4.8500 | N(1a) | 0.3053 | 0.3055 | 0.3049 |
| | c = 4.8923 | H(1a) | 0.1797 | 0.1831 | 0.1917 |
| | $\alpha = 89.7$ | H(1a) | 0.3742 | 0.4633 | 0.1845 |
| | $\beta = 89.9$ | H(1a) | 0.4675 | 0.1895 | 0.3732 |
| | $\gamma = 89.9$ | H(1a) | 0.1952 | 0.3812 | 0.4658 |
| | | B(1a) | 0.3091 | 0.0569 | 0.8056 |
| | | B(1a) | 0.0635 | 0.8047 | 0.3094 |
| | | B(1a) | 0.8052 | 0.3092 | 0.0607 |
| | | B(1a) | 0.3090 | 0.5590 | 0.8057 |
| | | B(1a) | 0.5519 | 0.8048 | 0.3090 |
| | | B(1a) | 0.8057 | 0.3093 | 0.5550 |
| | | C(1a) | 0.3080 | 0.8077 | 0.0474 |
| | | C(1a) | 0.0465 | 0.3078 | 0.8078 |
| | | C(1a) | 0.8078 | 0.0476 | 0.3081 |
| | | C(1a) | 0.3077 | 0.8076 | 0.5666 |
| | | C(1a) | 0.5672 | 0.3077 | 0.8076 |
| | | C(1a) | 0.8075 | 0.5655 | 0.3077 |
| ThNH$_4$B$_6$C$_6$ | a = 9.7868 | B(24f) | 0.7497 | 0.3728 | 0.5001 |
| $I23$ (1) | b = 9.7868 | B(24f) | 0.2495 | 0.6279 | 0.9980 |
| | c = 9.7868 | C(24f) | 0.5009 | 0.3791 | 0.2499 |
| | $\alpha = 90$ | C(24f) | 0.9988 | 0.6213 | 0.7497 |
| | $\beta = 90$ | H(24f) | 0.8076 | 0.2805 | 0.3307 |
| | $\gamma = 90$ | H(8c) | 0.8113 | 0.8113 | 0.8113 |
| | | N(8c) | 0.7513 | 0.7513 | 0.7513 |
| | | Th(2a) | 0.0000 | 0.0000 | 0.0000 |
| | | Th(6b) | 0.0000 | 0.5000 | 0.5000 |
| ThNH$_4$B$_6$C$_6$ | a = 4.8855 | Th(1a) | 0.1919 | 0.1918 | 0.8081 |
| $P1$ (1) | b = 4.8860 | N(1a) | 0.6946 | 0.6947 | 0.3052 |
| | c = 4.8989 | H(1a) | 0.8160 | 0.8152 | 0.1868 |
| | $\alpha = 90.1$ | H(1a) | 0.5352 | 0.8093 | 0.3758 |
| | $\beta = 90.1$ | H(1a) | 0.8082 | 0.6226 | 0.4646 |
| | $\gamma = 90.1$ | H(1a) | 0.6235 | 0.5357 | 0.1896 |
| | | B(1a) | 0.9366 | 0.1948 | 0.3089 |
| | | B(1a) | 0.1947 | 0.6911 | 0.0629 |
| | | B(1a) | 0.6909 | 0.9373 | 0.8054 |
| | | B(1a) | 0.4476 | 0.1943 | 0.3087 |
| | | B(1a) | 0.1942 | 0.6913 | 0.5529 |
| | | B(1a) | 0.6917 | 0.4469 | 0.8058 |
| | | C(1a) | 0.1923 | 0.9507 | 0.3078 |
| | | C(1a) | 0.6922 | 0.1922 | 0.0492 |
| | | C(1a) | 0.9509 | 0.6923 | 0.8078 |
| | | C(1a) | 0.1925 | 0.4351 | 0.3077 |
| | | C(1a) | 0.6923 | 0.1925 | 0.5652 |
| | | C(1a) | 0.4348 | 0.6922 | 0.8076 |
| TmNH$_4$B$_6$C$_6$ | a = 9.6759 | B(8k) | 0.3741 | 0.4997 | 0.2503 |
| $I222$ (1) | b = 9.6759 | B(8k) | 0.7503 | 0.1259 | 0.0003 |
| | c = 9.6734 | B(8k) | 0.0003 | 0.2497 | 0.8741 |
| | $\alpha = 90$ | B(8k) | 0.6263 | 0.0025 | 0.7507 |
| | $\beta = 90$ | B(8k) | 0.2507 | 0.8737 | 0.4975 |
| | $\gamma = 90$ | B(8k) | 0.9975 | 0.2493 | 0.6263 |
| | | C(8k) | 0.3804 | 0.7502 | 0.4991 |
| | | C(8k) | 0.9991 | 0.1196 | 0.7498 |
| | | C(8k) | 0.7498 | 0.0009 | 0.8804 |
| | | C(8k) | 0.6203 | 0.2504 | 0.0015 |
| | | C(8k) | 0.0015 | 0.3797 | 0.7496 |
| | | C(8k) | 0.2496 | 0.4985 | 0.1203 |
| | | H(8k) | 0.2809 | 0.6687 | 0.1917 |
| | | H(8k) | 0.6917 | 0.2191 | 0.8313 |
| | | H(8k) | 0.8313 | 0.3083 | 0.7809 |

| Compound | Lattice | Atomic coordinates (fractional) | | | |
|---|---|---|---|---|---|
| S.G.(Setting) | (Å,°) | Atoms | X | Y | Z |
| | | H(8k) | 0.6882 | 0.3118 | 0.6882 |
| | | N(8k) | 0.7488 | 0.2512 | 0.7488 |
| | | Tm(2a) | 0.0000 | 0.0000 | 0.0000 |
| | | Tm(2b) | 0.5000 | 0.0000 | 0.0000 |
| | | Tm(2d) | 0.0000 | 0.5000 | 0.0000 |
| | | Tm(2c) | 0.0000 | 0.0000 | 0.5000 |
| TmNH$_4$B$_6$C$_6$ | a = 4.8029 | Tm(1a) | 0.8102 | 0.8112 | 0.8108 |
| $P1$ (1) | b = 4.8351 | N(1a) | 0.3053 | 0.3054 | 0.3047 |
| | c = 4.8921 | H(1a) | 0.1768 | 0.1826 | 0.1945 |
| | $\alpha = 89.6$ | H(1a) | 0.3722 | 0.4635 | 0.1822 |
| | $\beta = 89.8$ | H(1a) | 0.4694 | 0.1888 | 0.3704 |
| | $\gamma = 89.9$ | H(1a) | 0.1977 | 0.3819 | 0.4670 |
| | | B(1a) | 0.3093 | 0.0553 | 0.8055 |
| | | B(1a) | 0.0639 | 0.8044 | 0.3097 |
| | | B(1a) | 0.8050 | 0.3094 | 0.0602 |
| | | B(1a) | 0.3091 | 0.5605 | 0.8057 |
| | | B(1a) | 0.5512 | 0.8044 | 0.3092 |
| | | B(1a) | 0.8057 | 0.3099 | 0.5554 |
| | | C(1a) | 0.3080 | 0.8075 | 0.0466 |
| | | C(1a) | 0.0455 | 0.3078 | 0.8077 |
| | | C(1a) | 0.8078 | 0.0468 | 0.3084 |
| | | C(1a) | 0.3076 | 0.8076 | 0.5673 |
| | | C(1a) | 0.5679 | 0.3076 | 0.8075 |
| | | C(1a) | 0.8074 | 0.5655 | 0.3073 |
| UNH$_4$B$_6$C$_6$ | a = 4.8159 | U(1a) | 0.1897 | 0.1888 | 0.1877 |
| $P1$ (1) | b = 4.8604 | N(1a) | 0.6947 | 0.6952 | 0.6951 |
| | c = 4.8618 | H(1a) | 0.8202 | 0.8135 | 0.8131 |
| | $\alpha = 89.5$ | H(1a) | 0.6268 | 0.8106 | 0.5322 |
| | $\beta = 89.9$ | H(1a) | 0.5305 | 0.6313 | 0.8125 |
| | $\gamma = 89.8$ | H(1a) | 0.8064 | 0.5302 | 0.6276 |
| | | B(1a) | 0.6906 | 0.1949 | 0.9427 |
| | | B(1a) | 0.9403 | 0.6904 | 0.1952 |
| | | B(1a) | 0.1951 | 0.9399 | 0.6905 |
| | | B(1a) | 0.6910 | 0.1942 | 0.4417 |
| | | B(1a) | 0.4441 | 0.6908 | 0.1950 |
| | | B(1a) | 0.1947 | 0.4447 | 0.6903 |
| | | C(1a) | 0.6920 | 0.9544 | 0.1924 |
| | | C(1a) | 0.9539 | 0.1924 | 0.6919 |
| | | C(1a) | 0.1923 | 0.6910 | 0.9537 |
| | | C(1a) | 0.6926 | 0.4317 | 0.1926 |
| | | C(1a) | 0.4326 | 0.1926 | 0.6926 |
| | | C(1a) | 0.1926 | 0.6935 | 0.4332 |
| YbNH$_4$B$_6$C$_6$ | a = 9.6618 | B(16o) | 0.1255 | 0.2520 | 0.0000 |
| $Fmmm$ (1) | b = 9.6618 | B(16m) | 0.0000 | 0.1255 | 0.7480 |
| | c = 9.6588 | B(16n) | 0.7520 | 0.0000 | 0.3745 |
| | $\alpha = 90$ | C(16n) | 0.1213 | 0.0000 | 0.7488 |
| | $\beta = 90$ | C(16o) | 0.2512 | 0.1213 | 0.0000 |
| | $\gamma = 90$ | C(16m) | 0.0000 | 0.2512 | 0.8787 |
| | | H(32p) | 0.8107 | 0.3107 | 0.6893 |
| | | N(8f) | 0.2500 | 0.2500 | 0.2500 |
| | | Yb(4a) | 0.0000 | 0.0000 | 0.0000 |
| | | Yb(4b) | 0.0000 | 0.0000 | 0.5000 |
| YbNH$_4$B$_6$C$_6$ | a = 9.5213 | B(8k) | 0.1347 | 0.0008 | 0.2531 |
| $I222$ (1) | b = 9.5304 | B(8k) | 0.2497 | 0.1245 | 0.0001 |
| | c = 9.6621 | B(8k) | 0.4996 | 0.7455 | 0.6333 |
| | $\alpha = 90$ | B(8k) | 0.8795 | 0.5025 | 0.7464 |
| | $\beta = 90$ | B(8k) | 0.2429 | 0.3747 | 0.0034 |
| | $\gamma = 90$ | B(8k) | 0.0025 | 0.2470 | 0.3810 |
| | | C(8k) | 0.1112 | 0.2492 | 0.0008 |

| Compound S.G.(Setting) | Lattice (Å,°) | Atomic coordinates (fractional) | | | |
|---|---|---|---|---|---|
| | | Atoms | X | Y | Z |
| | | C(8k) | 0.0011 | 0.1123 | 0.2529 |
| | | C(8k) | 0.7522 | 0.5015 | 0.6210 |
| | | C(8k) | 0.3717 | 0.2524 | 0.9983 |
| | | C(8k) | 0.4982 | 0.8712 | 0.7545 |
| | | C(8k) | 0.2558 | 0.9977 | 0.3782 |
| | | H(8k) | 0.2314 | 0.1655 | 0.3111 |
| | | H(8k) | 0.3080 | 0.2271 | 0.1660 |
| | | H(8k) | 0.6630 | 0.8056 | 0.7295 |
| | | H(8k) | 0.8115 | 0.8170 | 0.8119 |
| | | N(8k) | 0.2526 | 0.2519 | 0.2528 |
| | | Yb(2a) | 0.0000 | 0.0000 | 0.0000 |
| | | Yb(2b) | 0.5000 | 0.0000 | 0.0000 |
| | | Yb(2d) | 0.0000 | 0.5000 | 0.0000 |
| | | Yb(2c) | 0.0000 | 0.0000 | 0.5000 |
| $YNH_4B_6C_6$ $Fmmm$ (1) | a = 9.6985 b = 9.6332 c = 9.6341 $\alpha$ = 90 $\beta$ = 90 $\gamma$ = 90 | B(16m) | 0.0000 | 0.2516 | 0.1261 |
| | | B(16o) | 0.7517 | 0.6273 | 0.0000 |
| | | B(16n) | 0.6266 | 0.0000 | 0.7517 |
| | | C(16n) | 0.2510 | 0.0000 | 0.1203 |
| | | C(16m) | 0.0000 | 0.6199 | 0.7510 |
| | | C(16o) | 0.6201 | 0.7509 | 0.0000 |
| | | H(32p) | 0.8106 | 0.8108 | 0.8109 |
| | | N(8f) | 0.2500 | 0.2500 | 0.2500 |
| | | Y(4a) | 0.0000 | 0.0000 | 0.0000 |
| | | Y(4b) | 0.0000 | 0.0000 | 0.5000 |
| $YNH_4B_6C_6$ $I23$ (1) | a = 9.7023 b = 9.7023 c = 9.7023 $\alpha$ = 90 $\beta$ = 90 $\gamma$ = 90 | B(24f) | 0.7497 | 0.5002 | 0.6260 |
| | | B(24f) | 0.2493 | 0.9975 | 0.3736 |
| | | C(24f) | 0.5009 | 0.2498 | 0.6199 |
| | | C(24f) | 0.4984 | 0.2496 | 0.8795 |
| | | H(24f) | 0.8080 | 0.3313 | 0.7195 |
| | | H(8c) | 0.6883 | 0.6883 | 0.6883 |
| | | N(8c) | 0.7488 | 0.7488 | 0.7488 |
| | | Y(2a) | 0.0000 | 0.0000 | 0.0000 |
| | | Y(6b) | 0.0000 | 0.5000 | 0.5000 |
| $YNH_4B_6C_6$ $P1$ (1) | a = 4.8234 b = 4.8451 c = 4.8948 $\alpha$ = 89.7 $\beta$ = 89.9 $\gamma$ = 89.9 | Y(1a) | 0.8095 | 0.8099 | 0.8096 |
| | | N(1a) | 0.3053 | 0.3055 | 0.3048 |
| | | H(1a) | 0.1787 | 0.1828 | 0.1928 |
| | | H(1a) | 0.3731 | 0.4634 | 0.1837 |
| | | H(1a) | 0.4683 | 0.1895 | 0.3719 |
| | | H(1a) | 0.1962 | 0.3814 | 0.4663 |
| | | B(1a) | 0.3091 | 0.0563 | 0.8056 |
| | | B(1a) | 0.0638 | 0.8046 | 0.3095 |
| | | B(1a) | 0.8051 | 0.3093 | 0.0605 |
| | | B(1a) | 0.3090 | 0.5596 | 0.8057 |
| | | B(1a) | 0.5515 | 0.8047 | 0.3091 |
| | | B(1a) | 0.8057 | 0.3095 | 0.5553 |
| | | C(1a) | 0.3080 | 0.8077 | 0.0472 |
| | | C(1a) | 0.0462 | 0.3078 | 0.8078 |
| | | C(1a) | 0.8078 | 0.0474 | 0.3082 |
| | | C(1a) | 0.3077 | 0.8076 | 0.5668 |
| | | C(1a) | 0.5675 | 0.3077 | 0.8075 |
| | | C(1a) | 0.8074 | 0.5655 | 0.3076 |

### 3.2 Possible synthesis paths of MNH$_4$B$_6$C$_6$ materials.

TABLE S3: Possible synthesis paths of MNH$_4$B$_6$C$_6$ materials at ambient pressure when the reactants are elemental and binary compounds and the calculated reaction energy values.

| Compound | Chemical reaction | Reaction energy(eV) |
|---|---|---:|
| KNH$_4$B$_6$C$_6$ | $15\,B_{13}C_2 + 95\,N_2 + 4\,KH \longrightarrow 3\,KC_8 + 189\,BN + KB_6H_4C_6N$ | -459.9543 |
| KNH$_4$B$_6$C$_6$ | $3\,B_{13}C_2 + 16.71\,KN_3 + 1.792\,B_9H_{11} \longrightarrow 49.12\,BN + 15.71\,KH + KB_6H_4C_6N$ | -107.1112 |
| KNH$_4$B$_6$C$_6$ | $3\,B_{13}C_2 + 15.81\,KN_3 + 2.688\,B_5H_7 \longrightarrow 46.44\,BN + 14.81\,KH + KB_6H_4C_6N$ | -100.8526 |
| KNH$_4$B$_6$C$_6$ | $3\,B_{13}C_2 + 0.3636\,B_9H_{11} + KB_6 + 21.64\,N_2 \longrightarrow 42.27\,BN + KB_6H_4C_6N$ | -96.0256 |
| KNH$_4$B$_6$C$_6$ | $3\,B_{13}C_2 + 0.5714\,B_5H_7 + KB_6 + 21.43\,N_2 \longrightarrow 41.86\,BN + KB_6H_4C_6N$ | -95.0212 |
| KNH$_4$B$_6$C$_6$ | $3\,B_{13}C_2 + 2\,H_2 + KB_6 + 20\,N_2 \longrightarrow 39\,BN + KB_6H_4C_6N$ | -88.2848 |
| KNH$_4$B$_6$C$_6$ | $3\,B_{13}C_2 + 1.333\,H_3N + KB_6 + 19.33\,N_2 \longrightarrow 39\,BN + KB_6H_4C_6N$ | -86.5193 |
| KNH$_4$B$_6$C$_6$ | $3\,B_{13}C_2 + 0.3636\,B_9H_{11} + K + 18.64\,N_2 \longrightarrow 36.27\,BN + KB_6H_4C_6N$ | -80.8122 |
| KNH$_4$B$_6$C$_6$ | $3\,B_{13}C_2 + KN_3 + 0.3636\,B_9H_{11} + 17.14\,N_2 \longrightarrow 36.27\,BN + KB_6H_4C_6N$ | -80.2030 |
| KNH$_4$B$_6$C$_6$ | $3\,B_{13}C_2 + 0.5714\,B_5H_7 + K + 18.43\,N_2 \longrightarrow 35.86\,BN + KB_6H_4C_6N$ | -79.8078 |
| KNH$_4$B$_6$C$_6$ | $3\,B_{13}C_2 + KN_3 + 0.5714\,B_5H_7 + 16.93\,N_2 \longrightarrow 35.86\,BN + KB_6H_4C_6N$ | -79.1985 |
| KNH$_4$B$_6$C$_6$ | $3\,B_{13}C_2 + 0.2727\,B_9H_{11} + 18.23\,N_2 + KH \longrightarrow 35.45\,BN + KB_6H_4C_6N$ | -78.4900 |
| KNH$_4$B$_6$C$_6$ | $3\,B_{13}C_2 + 0.4286\,B_5H_7 + 18.07\,N_2 + KH \longrightarrow 35.14\,BN + KB_6H_4C_6N$ | -77.7367 |
| KNH$_4$B$_6$C$_6$ | $1.2\,H_8C_5 + 6.6\,KB_6 + 17.3\,N_2 \longrightarrow 33.6\,BN + 5.6\,KH + KB_6H_4C_6N$ | -76.5682 |
| KNH$_4$B$_6$C$_6$ | $3\,B_{13}C_2 + 21.5\,N_2 + 4\,KH \longrightarrow 3\,KN_3 + 33\,BN + KB_6H_4C_6N$ | -73.3510 |
| KNH$_4$B$_6$C$_6$ | $3\,B_{13}C_2 + 12.42\,KN_3 + 0.3636\,B_9H_{11} \longrightarrow 11.42\,K + 36.27\,BN + KB_6H_4C_6N$ | -73.2432 |
| KNH$_4$B$_6$C$_6$ | $3\,B_{13}C_2 + 2\,H_2 + K + 17\,N_2 \longrightarrow 33\,BN + KB_6H_4C_6N$ | -73.0713 |
| KNH$_4$B$_6$C$_6$ | $3\,B_{13}C_2 + 1.5\,H_2 + 17\,N_2 + KH \longrightarrow 33\,BN + KB_6H_4C_6N$ | -72.6843 |
| KNH$_4$B$_6$C$_6$ | $3\,B_{13}C_2 + KN_3 + 2\,H_2 + 15.5\,N_2 \longrightarrow 33\,BN + KB_6H_4C_6N$ | -72.4621 |
| KNH$_4$B$_6$C$_6$ | $3\,B_{13}C_2 + 12.29\,KN_3 + 0.5714\,B_5H_7 \longrightarrow 11.29\,K + 35.86\,BN + KB_6H_4C_6N$ | -72.3232 |
| KNH$_4$B$_6$C$_6$ | $3\,B_{13}C_2 + 17\,N_2 + 4\,KH \longrightarrow 3\,K + 33\,BN + KB_6H_4C_6N$ | -71.5234 |
| KNH$_4$B$_6$C$_6$ | $3\,B_{13}C_2 + H_3N + 16.5\,N_2 + KH \longrightarrow 33\,BN + KB_6H_4C_6N$ | -71.3603 |
| KNH$_4$B$_6$C$_6$ | $3\,B_{13}C_2 + 1.333\,H_3N + K + 16.33\,N_2 \longrightarrow 33\,BN + KB_6H_4C_6N$ | -71.3059 |
| KNH$_4$B$_6$C$_6$ | $3\,B_{13}C_2 + KN_3 + 1.333\,H_3N + 14.83\,N_2 \longrightarrow 33\,BN + KB_6H_4C_6N$ | -70.6967 |
| KNH$_4$B$_6$C$_6$ | $3\,B_{13}C_2 + 11.33\,KN_3 + 7.167\,H_2 \longrightarrow 33\,BN + 10.33\,KH + KB_6H_4C_6N$ | -70.1657 |
| KNH$_4$B$_6$C$_6$ | $3\,B_{13}C_2 + 11.33\,KN_3 + 2\,H_2 \longrightarrow 10.33\,K + 33\,BN + KB_6H_4C_6N$ | -66.1669 |
| KNH$_4$B$_6$C$_6$ | $3\,B_{13}C_2 + 34\,H_3N + 99\,K \longrightarrow 33\,BN + 98\,KH + KB_6H_4C_6N$ | -65.9770 |
| KNH$_4$B$_6$C$_6$ | $3\,B_{13}C_2 + 9.9\,KN_3 + 4.3\,H_3N \longrightarrow 33\,BN + 8.9\,KH + KB_6H_4C_6N$ | -64.7908 |
| KNH$_4$B$_6$C$_6$ | $3\,B_{13}C_2 + 10.89\,KN_3 + 1.333\,H_3N \longrightarrow 9.889\,K + 33\,BN + KB_6H_4C_6N$ | -64.6723 |
| KNH$_4$B$_6$C$_6$ | $3\,B_{13}C_2 + 11.33\,KN_3 + 4\,KH \longrightarrow 14.33\,K + 33\,BN + KB_6H_4C_6N$ | -64.6190 |
| KNH$_4$B$_6$C$_6$ | $1.75\,B_{13}C_2 + 0.5\,H_8C_5 + KB_6 + 11.88\,N_2 \longrightarrow 22.75\,BN + KB_6H_4C_6N$ | -47.0019 |
| KNH$_4$B$_6$C$_6$ | $2.062\,B_{13}C_2 + 0.375\,H_8C_5 + 10.91\,N_2 + KH \longrightarrow 20.81\,BN + KB_6H_4C_6N$ | -41.7222 |
| KNH$_4$B$_6$C$_6$ | $6.6\,KN_3 + 1.2\,H_8C_5 + 24.8\,B \longrightarrow 18.8\,BN + 5.6\,KH + KB_6H_4C_6N$ | -35.7943 |
| KNH$_4$B$_6$C$_6$ | $3\,B_{13}C_2 + 40\,H_3N + KB_6 \longrightarrow 58\,H_2 + 39\,BN + KB_6H_4C_6N$ | -35.3220 |
| KNH$_4$B$_6$C$_6$ | $1.75\,B_{13}C_2 + 0.5\,H_8C_5 + K + 8.875\,N_2 \longrightarrow 16.75\,BN + KB_6H_4C_6N$ | -31.7884 |
| KNH$_4$B$_6$C$_6$ | $3\,B_{13}C_2 + KN_3 + 31\,H_3N \longrightarrow 44.5\,H_2 + 33\,BN + KB_6H_4C_6N$ | -31.4159 |
| KNH$_4$B$_6$C$_6$ | $1.75\,B_{13}C_2 + KN_3 + 0.5\,H_8C_5 + 7.375\,N_2 \longrightarrow 16.75\,BN + KB_6H_4C_6N$ | -31.1792 |
| KNH$_4$B$_6$C$_6$ | $1.75\,B_{13}C_2 + 5.917\,KN_3 + 0.5\,H_8C_5 \longrightarrow 4.917\,K + 16.75\,BN + KB_6H_4C_6N$ | -28.1839 |
| KNH$_4$B$_6$C$_6$ | $3\,B_{13}C_2 + 34\,H_3N + K \longrightarrow 49\,H_2 + 33\,BN + KB_6H_4C_6N$ | -28.0529 |
| KNH$_4$B$_6$C$_6$ | $3\,B_{13}C_2 + 34\,H_3N + KH \longrightarrow 49.5\,H_2 + 33\,BN + KB_6H_4C_6N$ | -27.6660 |
| KNH$_4$B$_6$C$_6$ | $3\,B_{13}C_2 + 8\,N_2 + 4\,KH \longrightarrow 3\,KB_6 + 15\,BN + KB_6H_4C_6N$ | -25.8831 |
| KNH$_4$B$_6$C$_6$ | $14.33\,KN_3 + 1.2\,H_8C_5 + 17.2\,B_5H_7 \longrightarrow 42\,H_3N + 13.33\,KB_6 + KB_6H_4C_6N$ | -20.7923 |
| KNH$_4$B$_6$C$_6$ | $3\,B_{13}C_2 + 4.808\,KN_3 + 0.3636\,B_9H_{11} \longrightarrow 3.808\,KB_6 + 13.42\,BN + KB_6H_4C_6N$ | -19.9491 |
| KNH$_4$B$_6$C$_6$ | $3\,B_{13}C_2 + 4.762\,KN_3 + 0.5714\,B_5H_7 \longrightarrow 3.762\,KB_6 + 13.29\,BN + KB_6H_4C_6N$ | -19.6752 |
| KNH$_4$B$_6$C$_6$ | $3\,B_{13}C_2 + 4.296\,KN_3 + 1.333\,H_3N \longrightarrow 3.296\,KB_6 + 13.22\,BN + KB_6H_4C_6N$ | -18.5405 |
| KNH$_4$B$_6$C$_6$ | $3\,B_{13}C_2 + 4.444\,KN_3 + 2\,H_2 \longrightarrow 3.444\,KB_6 + 12.33\,BN + KB_6H_4C_6N$ | -17.9619 |
| KNH$_4$B$_6$C$_6$ | $3.844\,KN_3 + 1.2\,H_8C_5 + 2.756\,KB_6 \longrightarrow 10.53\,BN + 5.6\,KH + KB_6H_4C_6N$ | -15.7389 |
| KNH$_4$B$_6$C$_6$ | $3\,B_{13}C_2 + KN_3 + 10.77\,H_3N \longrightarrow 4.045\,B_5H_7 + 12.77\,BN + KB_6H_4C_6N$ | -10.5073 |
| KNH$_4$B$_6$C$_6$ | $3\,B_{13}C_2 + 6.5\,KN_3 + 29.75\,H_2 \longrightarrow 18.5\,H_3N + 5.5\,KB_6 + KB_6H_4C_6N$ | -9.9328 |
| KNH$_4$B$_6$C$_6$ | $3\,B_{13}C_2 + KN_3 + 9.921\,H_3N \longrightarrow 2.342\,B_9H_{11} + 11.92\,BN + KB_6H_4C_6N$ | -9.4686 |
| KNH$_4$B$_6$C$_6$ | $8.833\,KN_3 + 11.5\,B_5H_7 + 0.75\,KC_8 \longrightarrow 25.5\,H_3N + 8.583\,KB_6 + KB_6H_4C_6N$ | -9.3349 |
| KNH$_4$B$_6$C$_6$ | $1.097\,B_{13}C_2 + 3.088\,KN_3 + 0.7611\,H_8C_5 \longrightarrow 8.265\,BN + 2.088\,KH + KB_6H_4C_6N$ | -9.1603 |
| KNH$_4$B$_6$C$_6$ | $3\,B_{13}C_2 + 13.64\,H_3N + KB_6 \longrightarrow 5.273\,B_5H_7 + 12.64\,BN + KB_6H_4C_6N$ | -8.0703 |
| KNH$_4$B$_6$C$_6$ | $3\,B_{13}C_2 + 12.53\,H_3N + KB_6 \longrightarrow 3.053\,B_9H_{11} + 11.53\,BN + KB_6H_4C_6N$ | -6.7164 |
| KNH$_4$B$_6$C$_6$ | $3\,B_{13}C_2 + 74.3\,N_2 + 50.2\,KH \longrightarrow 49.2\,KN_3 + 6.6\,B_5H_7 + KB_6H_4C_6N$ | -5.8122 |
| KNH$_4$B$_6$C$_6$ | $1.75\,B_{13}C_2 + 2.639\,KN_3 + 0.5\,H_8C_5 \longrightarrow 1.639\,KB_6 + 6.917\,BN + KB_6H_4C_6N$ | -5.2476 |

| Compound | Chemical reaction | Reaction energy(eV) |
|---|---|---|
| $KNH_4B_6C_6$ | $3\,B_{13}C_2 + 11.73\,H_3N + K \longrightarrow 4.455\,B_5H_7 + 10.73\,BN + KB_6H_4C_6N$ | -5.0300 |
| $KNH_4B_6C_6$ | $3\,B_{13}C_2 + 11.5\,H_3N + KH \longrightarrow 4.5\,B_5H_7 + 10.5\,BN + KB_6H_4C_6N$ | -4.4081 |
| $KNH_4B_6C_6$ | $3\,B_{13}C_2 + 65.5\,N_2 + 44.33\,KH \longrightarrow 43.33\,KN_3 + 3.667\,B_9H_{11} + KB_6H_4C_6N$ | -4.2605 |
| $KNH_4B_6C_6$ | $3\,B_{13}C_2 + 10.79\,H_3N + K \longrightarrow 2.579\,B_9H_{11} + 9.789\,BN + KB_6H_4C_6N$ | -3.8862 |
| $KNH_4B_6C_6$ | $3\,B_{13}C_2 + 10.55\,H_3N + KH \longrightarrow 2.605\,B_9H_{11} + 9.553\,BN + KB_6H_4C_6N$ | -3.2526 |
| $KNH_4B_6C_6$ | $150\,C + 6\,BN + 19\,KH \longrightarrow 5\,H_3N + 18\,KC_8 + KB_6H_4C_6N$ | -1.0066 |
| $YNH_4B_6C_6$ | $3\,B_{13}C_2 + 251\,H_3N + 251\,YB_2 \longrightarrow 107\,B_5H_7 + 250\,YN + YB_6H_4C_6N$ | -202.7686 |
| $YNH_4B_6C_6$ | $3\,B_{13}C_2 + 50\,Y + 34\,H_3N \longrightarrow 49\,YH_2 + 33\,BN + YB_6H_4C_6N$ | -136.4286 |
| $YNH_4B_6C_6$ | $0.3636\,B_9H_{11} + 3\,B_{13}C_2 + YB_{12} + 24.64\,N_2 \longrightarrow 48.27\,BN + YB_6H_4C_6N$ | -113.3094 |
| $YNH_4B_6C_6$ | $3\,B_{13}C_2 + YB_{12} + 0.5714\,B_5H_7 + 24.43\,N_2 \longrightarrow 47.86\,BN + YB_6H_4C_6N$ | -112.3050 |
| $YNH_4B_6C_6$ | $3\,B_{13}C_2 + 33.67\,Y + 34\,H_3N \longrightarrow 32.67\,YH_3 + 33\,BN + YB_6H_4C_6N$ | -112.0542 |
| $YNH_4B_6C_6$ | $3\,B_{13}C_2 + YB_{12} + 23\,N_2 + 2\,H_2 \longrightarrow 45\,BN + YB_6H_4C_6N$ | -105.5686 |
| $YNH_4B_6C_6$ | $56\,H_3N + 15.5\,Y_4C_7 + 6\,BN \longrightarrow 20.5\,H_8C_5 + 61\,YN + YB_6H_4C_6N$ | -104.3911 |
| $YNH_4B_6C_6$ | $3\,B_{13}C_2 + 1.333\,H_3N + YB_{12} + 22.33\,N_2 \longrightarrow 45\,BN + YB_6H_4C_6N$ | -103.8031 |
| $YNH_4B_6C_6$ | $2\,YH_2 + 3.875\,B_{13}C_2 + 22.69\,N_2 \longrightarrow 0.25\,Y_4C_7 + 44.37\,BN + YB_6H_4C_6N$ | -102.8619 |
| $YNH_4B_6C_6$ | $2\,YH_2 + 3.625\,B_{13}C_2 + 21.06\,N_2 \longrightarrow 0.25\,Y_4C_5 + 41.12\,BN + YB_6H_4C_6N$ | -94.7323 |
| $YNH_4B_6C_6$ | $0.3636\,B_9H_{11} + 3\,B_{13}C_2 + 20.64\,N_2 + YB_4 \longrightarrow 40.27\,BN + YB_6H_4C_6N$ | -93.0087 |
| $YNH_4B_6C_6$ | $3\,B_{13}C_2 + 0.5714\,B_5H_7 + 20.43\,N_2 + YB_4 \longrightarrow 39.86\,BN + YB_6H_4C_6N$ | -92.0043 |
| $YNH_4B_6C_6$ | $3.8\,YB_{12} + 1.2\,H_8C_5 + 20.3\,N_2 \longrightarrow 2.8\,YH_2 + 39.6\,BN + YB_6H_4C_6N$ | -89.5966 |
| $YNH_4B_6C_6$ | $0.3636\,B_9H_{11} + 3\,B_{13}C_2 + 19.64\,N_2 + YB_2 \longrightarrow 38.27\,BN + YB_6H_4C_6N$ | -89.1369 |
| $YNH_4B_6C_6$ | $3\,B_{13}C_2 + 0.5714\,B_5H_7 + 19.43\,N_2 + YB_2 \longrightarrow 37.86\,BN + YB_6H_4C_6N$ | -88.1325 |
| $YNH_4B_6C_6$ | $0.3636\,B_9H_{11} + 3\,B_{13}C_2 + Y + 18.64\,N_2 \longrightarrow 36.27\,BN + YB_6H_4C_6N$ | -85.7395 |
| $YNH_4B_6C_6$ | $3\,B_{13}C_2 + 19\,N_2 + 2\,H_2 + YB_4 \longrightarrow 37\,BN + YB_6H_4C_6N$ | -85.2679 |
| $YNH_4B_6C_6$ | $3\,B_{13}C_2 + Y + 0.5714\,B_5H_7 + 18.43\,N_2 \longrightarrow 35.86\,BN + YB_6H_4C_6N$ | -84.7351 |
| $YNH_4B_6C_6$ | $1.333\,YH_3 + 3.292\,B_{13}C_2 + 18.9\,N_2 \longrightarrow 0.08333\,Y_4C_7 + 36.79\,BN + YB_6H_4C_6N$ | -84.4664 |
| $YNH_4B_6C_6$ | $3\,B_{13}C_2 + 1.333\,H_3N + 18.33\,N_2 + YB_4 \longrightarrow 37\,BN + YB_6H_4C_6N$ | -83.5024 |
| $YNH_4B_6C_6$ | $2\,YH_2 + 3.25\,B_{13}C_2 + 18.62\,N_2 \longrightarrow 0.5\,Y_2C + 36.25\,BN + YB_6H_4C_6N$ | -82.3105 |
| $YNH_4B_6C_6$ | $0.3636\,B_9H_{11} + 3\,B_{13}C_2 + 18.14\,N_2 + YN \longrightarrow 36.27\,BN + YB_6H_4C_6N$ | -82.2765 |
| $YNH_4B_6C_6$ | $1.333\,YH_3 + 3.208\,B_{13}C_2 + 18.35\,N_2 \longrightarrow 0.08333\,Y_4C_5 + 35.71\,BN + YB_6H_4C_6N$ | -81.7565 |
| $YNH_4B_6C_6$ | $3\,B_{13}C_2 + 18\,N_2 + 2\,H_2 + YB_2 \longrightarrow 35\,BN + YB_6H_4C_6N$ | -81.3961 |
| $YNH_4B_6C_6$ | $3\,B_{13}C_2 + 0.5714\,B_5H_7 + 17.93\,N_2 + YN \longrightarrow 35.86\,BN + YB_6H_4C_6N$ | -81.2721 |
| $YNH_4B_6C_6$ | $6\,Y_2C + 27.6\,B_5H_7 + 32.03\,N_2 \longrightarrow 63.07\,H_3N + 11\,YB_{12} + YB_6H_4C_6N$ | -81.1942 |
| $YNH_4B_6C_6$ | $0.1818\,B_9H_{11} + YH_2 + 3\,B_{13}C_2 + 17.82\,N_2 \longrightarrow 34.64\,BN + YB_6H_4C_6N$ | -79.7579 |
| $YNH_4B_6C_6$ | $3\,B_{13}C_2 + 1.333\,H_3N + 17.33\,N_2 + YB_2 \longrightarrow 35\,BN + YB_6H_4C_6N$ | -79.6306 |
| $YNH_4B_6C_6$ | $YH_2 + 3\,B_{13}C_2 + 0.2857\,B_5H_7 + 17.71\,N_2 \longrightarrow 34.43\,BN + YB_6H_4C_6N$ | -79.2557 |
| $YNH_4B_6C_6$ | $3\,B_{13}C_2 + Y + 17\,N_2 + 2\,H_2 \longrightarrow 33\,BN + YB_6H_4C_6N$ | -77.9987 |
| $YNH_4B_6C_6$ | $1.333\,YH_3 + 3.083\,B_{13}C_2 + 17.54\,N_2 \longrightarrow 0.1667\,Y_2C + 34.08\,BN + YB_6H_4C_6N$ | -77.6159 |
| $YNH_4B_6C_6$ | $YH_3 + 0.09091\,B_9H_{11} + 3\,B_{13}C_2 + 17.41\,N_2 \longrightarrow 33.82\,BN + YB_6H_4C_6N$ | -77.5133 |
| $YNH_4B_6C_6$ | $YH_3 + 3\,B_{13}C_2 + 0.1429\,B_5H_7 + 17.36\,N_2 \longrightarrow 33.71\,BN + YB_6H_4C_6N$ | -77.2622 |
| $YNH_4B_6C_6$ | $2\,YH_2 + 3\,B_{13}C_2 + 17.5\,N_2 \longrightarrow YN + 33\,BN + YB_6H_4C_6N$ | -77.2393 |
| $YNH_4B_6C_6$ | $0.3636\,B_9H_{11} + 2.75\,B_{13}C_2 + 0.5\,Y_2C + 17.01\,N_2 \longrightarrow 33.02\,BN + YB_6H_4C_6N$ | -77.2053 |
| $YNH_4B_6C_6$ | $3\,B_{13}C_2 + Y + 1.333\,H_3N + 16.33\,N_2 \longrightarrow 33\,BN + YB_6H_4C_6N$ | -76.2332 |
| $YNH_4B_6C_6$ | $2.75\,B_{13}C_2 + 0.5\,Y_2C + 0.5714\,B_5H_7 + 16.8\,N_2 \longrightarrow 32.61\,BN + YB_6H_4C_6N$ | -76.2009 |
| $YNH_4B_6C_6$ | $1.333\,YH_3 + 3\,B_{13}C_2 + 17.17\,N_2 \longrightarrow 0.3333\,YN + 33\,BN + YB_6H_4C_6N$ | -75.9255 |
| $YNH_4B_6C_6$ | $YH_2 + 3\,B_{13}C_2 + 17\,N_2 + H_2 \longrightarrow 33\,BN + YB_6H_4C_6N$ | -75.8875 |
| $YNH_4B_6C_6$ | $YH_3 + 3\,B_{13}C_2 + 17\,N_2 + 0.5\,H_2 \longrightarrow 33\,BN + YB_6H_4C_6N$ | -75.5780 |
| $YNH_4B_6C_6$ | $2\,YH_3 + 3\,B_{13}C_2 + 17\,N_2 \longrightarrow YH_2 + 33\,BN + YB_6H_4C_6N$ | -75.2686 |
| $YNH_4B_6C_6$ | $YH_3 + 3\,B_{13}C_2 + 0.3333\,H_3N + 16.83\,N_2 \longrightarrow 33\,BN + YB_6H_4C_6N$ | -75.1367 |
| $YNH_4B_6C_6$ | $YH_2 + 3\,B_{13}C_2 + 0.6667\,H_3N + 16.67\,N_2 \longrightarrow 33\,BN + YB_6H_4C_6N$ | -75.0048 |
| $YNH_4B_6C_6$ | $1.333\,YH_3 + 3\,B_{13}C_2 + 17\,N_2 \longrightarrow 0.3333\,Y + 33\,BN + YB_6H_4C_6N$ | -74.7712 |
| $YNH_4B_6C_6$ | $3\,B_{13}C_2 + 16.5\,N_2 + YN + 2\,H_2 \longrightarrow 33\,BN + YB_6H_4C_6N$ | -74.5356 |
| $YNH_4B_6C_6$ | $3.364\,B_9H_{11} + 6\,Y_2C + 12.64\,N_2 \longrightarrow 11\,YH_3 + 24.27\,BN + YB_6H_4C_6N$ | -73.8183 |
| $YNH_4B_6C_6$ | $2\,YH_2 + 3\,B_{13}C_2 + 17\,N_2 \longrightarrow Y + 33\,BN + YB_6H_4C_6N$ | -73.7763 |
| $YNH_4B_6C_6$ | $1.333\,YH_3 + 3\,B_{13}C_2 + 16.67\,N_2 \longrightarrow 32.33\,BN + 0.3333\,YB_2 + YB_6H_4C_6N$ | -73.6387 |
| $YNH_4B_6C_6$ | $3\,B_{13}C_2 + 1.333\,H_3N + 15.83\,N_2 + YN \longrightarrow 33\,BN + YB_6H_4C_6N$ | -72.7702 |
| $YNH_4B_6C_6$ | $1.333\,YH_3 + 3\,B_{13}C_2 + 16.33\,N_2 \longrightarrow 31.67\,BN + 0.3333\,YB_4 + YB_6H_4C_6N$ | -72.3481 |
| $YNH_4B_6C_6$ | $15.33\,B_9H_{11} + 6\,Y_2C + 27.94\,N_2 \longrightarrow 54.89\,H_3N + 11\,YB_{12} + YB_6H_4C_6N$ | -71.4032 |
| $YNH_4B_6C_6$ | $2\,YH_2 + 3\,B_{13}C_2 + 16\,N_2 \longrightarrow 31\,BN + YB_2 + YB_6H_4C_6N$ | -70.3789 |
| $YNH_4B_6C_6$ | $2.75\,B_{13}C_2 + 0.5\,Y_2C + 15.37\,N_2 + 2\,H_2 \longrightarrow 29.75\,BN + YB_6H_4C_6N$ | -69.4645 |
| $YNH_4B_6C_6$ | $12.8\,B_9H_{11} + 22.8\,Y_2C + YN \longrightarrow 45.6\,YH_3 + 8.4\,B_{13}C_2 + YB_6H_4C_6N$ | -68.2358 |

| Compound | Chemical reaction | Reaction energy(eV) |
|---|---|---|
| $YNH_4B_6C_6$ | $2.75\,B_{13}C_2 + 0.5\,Y_2C + 1.333\,H_3N + 14.71\,N_2 \longrightarrow 29.75\,BN + YB_6H_4C_6N$ | -67.6990 |
| $YNH_4B_6C_6$ | $2\,YH_2 + 3\,B_{13}C_2 + 15\,N_2 \longrightarrow 29\,BN + YB_4 + YB_6H_4C_6N$ | -66.5071 |
| $YNH_4B_6C_6$ | $1.333\,YH_3 + 3\,B_{13}C_2 + 15\,N_2 \longrightarrow 0.3333\,YB_{12} + 29\,BN + YB_6H_4C_6N$ | -65.5812 |
| $YNH_4B_6C_6$ | $YH_3 + 2.687\,B_{13}C_2 + 0.125\,H_8C_5 + 14.97\,N_2 \longrightarrow 28.94\,BN + YB_6H_4C_6N$ | -65.2573 |
| $YNH_4B_6C_6$ | $0.3636\,B_9H_{11} + 2.375\,B_{13}C_2 + 0.25\,Y_4C_5 + 14.57\,N_2 \longrightarrow 28.15\,BN + YB_6H_4C_6N$ | -64.7835 |
| $YNH_4B_6C_6$ | $6\,Y_2C + 5.286\,B_5H_7 + 10.71\,N_2 \longrightarrow 11\,YH_3 + 20.43\,BN + YB_6H_4C_6N$ | -64.5273 |
| $YNH_4B_6C_6$ | $11.69\,B_9H_{11} + 21.26\,Y_2C + 0.5\,N_2 \longrightarrow 41.51\,YH_3 + 7.629\,B_{13}C_2 + YB_6H_4C_6N$ | -64.4227 |
| $YNH_4B_6C_6$ | $1.75\,B_{13}C_2 + YB_{12} + 0.5\,H_8C_5 + 14.88\,N_2 \longrightarrow 28.75\,BN + YB_6H_4C_6N$ | -64.2857 |
| $YNH_4B_6C_6$ | $12.03\,B_9H_{11} + 21.89\,Y_2C + BN \longrightarrow 42.77\,YH_3 + 7.943\,B_{13}C_2 + YB_6H_4C_6N$ | -64.0356 |
| $YNH_4B_6C_6$ | $6\,Y_2C + 29.67\,H_3N + 17.33\,YB_2 \longrightarrow 28.33\,YH_3 + 28.67\,BN + YB_6H_4C_6N$ | -63.7802 |
| $YNH_4B_6C_6$ | $2.375\,B_{13}C_2 + 0.25\,Y_4C_5 + 0.5714\,B_5H_7 + 14.37\,N_2 \longrightarrow 27.73\,BN + YB_6H_4C_6N$ | -63.7791 |
| $YNH_4B_6C_6$ | $2.867\,YB_{12} + 1.2\,H_8C_5 + 14.7\,N_2 \longrightarrow 1.867\,YH_3 + 28.4\,BN + YB_6H_4C_6N$ | -62.4719 |
| $YNH_4B_6C_6$ | $101\,YH_2 + 3\,B_{13}C_2 + 34\,H_3N \longrightarrow 100\,YH_3 + 33\,BN + YB_6H_4C_6N$ | -61.8131 |
| $YNH_4B_6C_6$ | $3\,B_{13}C_2 + 79.8\,H_3N + 79.8\,YB_2 \longrightarrow 21.4\,B_9H_{11} + 78.8\,YN + YB_6H_4C_6N$ | -60.7882 |
| $YNH_4B_6C_6$ | $10.57\,B_9H_{11} + 19.71\,Y_2C + H_3N \longrightarrow 38.43\,YH_3 + 6.857\,B_{13}C_2 + YB_6H_4C_6N$ | -58.2431 |
| $YNH_4B_6C_6$ | $2.375\,B_{13}C_2 + 0.25\,Y_4C_5 + 12.94\,N_2 + 2\,H_2 \longrightarrow 24.88\,BN + YB_6H_4C_6N$ | -57.0426 |
| $YNH_4B_6C_6$ | $0.3636\,B_9H_{11} + 2.125\,B_{13}C_2 + 12.95\,N_2 + 0.25\,Y_4C_7 \longrightarrow 24.9\,BN + YB_6H_4C_6N$ | -56.6540 |
| $YNH_4B_6C_6$ | $2.125\,B_{13}C_2 + 0.5714\,B_5H_7 + 12.74\,N_2 + 0.25\,Y_4C_7 \longrightarrow 24.48\,BN + YB_6H_4C_6N$ | -55.6495 |
| $YNH_4B_6C_6$ | $2.375\,B_{13}C_2 + 1.333\,H_3N + 0.25\,Y_4C_5 + 12.27\,N_2 \longrightarrow 24.88\,BN + YB_6H_4C_6N$ | -55.2772 |
| $YNH_4B_6C_6$ | $YH_2 + 2.375\,B_{13}C_2 + 0.25\,H_8C_5 + 12.94\,N_2 \longrightarrow 24.88\,BN + YB_6H_4C_6N$ | -55.2460 |
| $YNH_4B_6C_6$ | $2.364\,B_9H_{11} + 6\,Y_2C + 8.136\,N_2 \longrightarrow 11\,YH_2 + 15.27\,BN + YB_6H_4C_6N$ | -49.1271 |
| $YNH_4B_6C_6$ | $2.125\,B_{13}C_2 + 11.31\,N_2 + 0.25\,Y_4C_7 + 2\,H_2 \longrightarrow 21.62\,BN + YB_6H_4C_6N$ | -48.9131 |
| $YNH_4B_6C_6$ | $2.125\,B_{13}C_2 + 1.333\,H_3N + 10.65\,N_2 + 0.25\,Y_4C_7 \longrightarrow 21.62\,BN + YB_6H_4C_6N$ | -47.1477 |
| $YNH_4B_6C_6$ | $16.19\,Y_2C + 14.45\,B_5H_7 + YN \longrightarrow 32.39\,YH_3 + 5.097\,B_{13}C_2 + YB_6H_4C_6N$ | -46.9016 |
| $YNH_4B_6C_6$ | $2\,YH_2 + 3\,B_{13}C_2 + 11\,N_2 \longrightarrow YB_{12} + 21\,BN + YB_6H_4C_6N$ | -46.2064 |
| $YNH_4B_6C_6$ | $8.75\,Y_2C + 1.2\,B_5H_7 + 8.75\,N_2 \longrightarrow 0.55\,H_8C_5 + 16.5\,YN + YB_6H_4C_6N$ | -45.3484 |
| $YNH_4B_6C_6$ | $15.23\,Y_2C + 13.19\,B_5H_7 + 0.5\,N_2 \longrightarrow 29.45\,YH_3 + 4.613\,B_{13}C_2 + YB_6H_4C_6N$ | -44.9458 |
| $YNH_4B_6C_6$ | $3\,B_{13}C_2 + 46\,H_3N + YB_{12} \longrightarrow 67\,H_2 + 45\,BN + YB_6H_4C_6N$ | -44.6613 |
| $YNH_4B_6C_6$ | $15.68\,Y_2C + 13.58\,B_5H_7 + BN \longrightarrow 30.35\,YH_3 + 4.839\,B_{13}C_2 + YB_6H_4C_6N$ | -43.9872 |
| $YNH_4B_6C_6$ | $1.75\,B_{13}C_2 + 0.5\,H_8C_5 + 10.88\,N_2 + YB_4 \longrightarrow 20.75\,BN + YB_6H_4C_6N$ | -43.9850 |
| $YNH_4B_6C_6$ | $6\,Y_2C + 10\,B_5H_7 + 11.5\,N_2 \longrightarrow 22\,H_3N + 11\,YB_4 + YB_6H_4C_6N$ | -42.6445 |
| $YNH_4B_6C_6$ | $6\,Y_2C + 3.714\,B_5H_7 + 6.786\,N_2 \longrightarrow 11\,YH_2 + 12.57\,BN + YB_6H_4C_6N$ | -42.5982 |
| $YNH_4B_6C_6$ | $0.6667\,B_9H_{11} + 8.083\,Y_2C + 8.083\,N_2 \longrightarrow 0.4167\,H_8C_5 + 15.17\,YN + YB_6H_4C_6N$ | -41.1462 |
| $YNH_4B_6C_6$ | $14.26\,Y_2C + H_3N + 11.94\,B_5H_7 \longrightarrow 27.52\,YH_3 + 4.129\,B_{13}C_2 + YB_6H_4C_6N$ | -40.6234 |
| $YNH_4B_6C_6$ | $3\,B_{13}C_2 + 34\,YN + 51.5\,H_2 \longrightarrow 33\,YH_3 + 33\,BN + YB_6H_4C_6N$ | -40.1354 |
| $YNH_4B_6C_6$ | $1.75\,B_{13}C_2 + 0.5\,H_8C_5 + 9.875\,N_2 + YB_2 \longrightarrow 18.75\,BN + YB_6H_4C_6N$ | -40.1132 |
| $YNH_4B_6C_6$ | $5.556\,B_9H_{11} + 6\,Y_2C + 10.02\,N_2 \longrightarrow 19.04\,H_3N + 11\,YB_4 + YB_6H_4C_6N$ | -39.0970 |
| $YNH_4B_6C_6$ | $2\,YH_2 + 6\,Y_2C + 8.5\,N_2 + 3\,YB_2 \longrightarrow 16\,YN + YB_6H_4C_6N$ | -36.9670 |
| $YNH_4B_6C_6$ | $1.75\,B_{13}C_2 + Y + 0.5\,H_8C_5 + 8.875\,N_2 \longrightarrow 16.75\,BN + YB_6H_4C_6N$ | -36.7158 |
| $YNH_4B_6C_6$ | $1.333\,YH_3 + 6\,Y_2C + 8.167\,N_2 + 3\,YB_2 \longrightarrow 15.33\,YN + YB_6H_4C_6N$ | -35.6532 |
| $YNH_4B_6C_6$ | $3\,B_{13}C_2 + 17.17\,H_3N + 16.83\,YN \longrightarrow 15.83\,YH_3 + 33\,BN + YB_6H_4C_6N$ | -35.3006 |
| $YNH_4B_6C_6$ | $3\,B_{13}C_2 + 38\,H_3N + YB_4 \longrightarrow 55\,H_2 + 37\,BN + YB_6H_4C_6N$ | -34.9532 |
| $YNH_4B_6C_6$ | $6\,Y_2C + 15.5\,H_3N + 10.25\,YB_2 \longrightarrow 21.25\,YH_2 + 14.5\,BN + YB_6H_4C_6N$ | -34.7512 |
| $YNH_4B_6C_6$ | $6\,Y_2C + 7.5\,N_2 + 2\,H_2 + 3\,YB_2 \longrightarrow 14\,YN + YB_6H_4C_6N$ | -34.2633 |
| $YNH_4B_6C_6$ | $6\,Y_2C + 18.11\,H_3N + 5.778\,YB_4 \longrightarrow 16.78\,YH_3 + 17.11\,BN + YB_6H_4C_6N$ | -34.2205 |
| $YNH_4B_6C_6$ | $8.875\,B_9H_{11} + H_3N + 9.484\,Y_4C_7 \longrightarrow 12.08\,H_8C_5 + 36.94\,YB_2 + YB_6H_4C_6N$ | -33.9156 |
| $YNH_4B_6C_6$ | $3\,B_{13}C_2 + 36\,H_3N + YB_2 \longrightarrow 52\,H_2 + 35\,BN + YB_6H_4C_6N$ | -33.7295 |
| $YNH_4B_6C_6$ | $1.75\,B_{13}C_2 + 0.5\,H_8C_5 + 8.375\,N_2 + YN \longrightarrow 16.75\,BN + YB_6H_4C_6N$ | -33.2527 |
| $YNH_4B_6C_6$ | $3\,B_{13}C_2 + Y + 34\,H_3N \longrightarrow 49\,H_2 + 33\,BN + YB_6H_4C_6N$ | -32.9803 |
| $YNH_4B_6C_6$ | $3\,B_{13}C_2 + 16.73\,Y + 16.73\,H_3N \longrightarrow 6.6\,B_5H_7 + 15.73\,YN + YB_6H_4C_6N$ | -32.5222 |
| $YNH_4B_6C_6$ | $6\,Y_2C + 1.333\,H_3N + 6.833\,N_2 + 3\,YB_2 \longrightarrow 14\,YN + YB_6H_4C_6N$ | -32.4979 |
| $YNH_4B_6C_6$ | $2\,YH_2 + 6\,Y_2C + 7.75\,N_2 + 1.5\,YB_4 \longrightarrow 14.5\,YN + YB_6H_4C_6N$ | -32.4841 |
| $YNH_4B_6C_6$ | $2\,YH_2 + 6\,Y_2C + 0.5\,YB_{12} + 7.25\,N_2 \longrightarrow 13.5\,YN + YB_6H_4C_6N$ | -31.9022 |
| $YNH_4B_6C_6$ | $2\,YH_2 + 6\,Y_2C + 7\,N_2 + 6\,B \longrightarrow 13\,YN + YB_6H_4C_6N$ | -31.7863 |
| $YNH_4B_6C_6$ | $8.75\,B_9H_{11} + YN + 9.094\,Y_4C_7 \longrightarrow 11.53\,H_8C_5 + 36.38\,YB_2 + YB_6H_4C_6N$ | -31.3197 |
| $YNH_4B_6C_6$ | $6\,Y_2C + 0.5714\,B_5H_7 + 6.786\,N_2 + 1.571\,YB_2 \longrightarrow 12.57\,YN + YB_6H_4C_6N$ | -31.1991 |
| $YNH_4B_6C_6$ | $1.333\,YH_3 + 6\,Y_2C + 7.417\,N_2 + 1.5\,YB_4 \longrightarrow 13.83\,YN + YB_6H_4C_6N$ | -31.1702 |
| $YNH_4B_6C_6$ | $6\,Y_2C + 13.91\,H_3N + 1.576\,YB_{12} \longrightarrow 12.58\,YH_3 + 12.91\,BN + YB_6H_4C_6N$ | -31.0564 |
| $YNH_4B_6C_6$ | $YH_2 + 3\,B_{13}C_2 + 34\,H_3N \longrightarrow 50\,H_2 + 33\,BN + YB_6H_4C_6N$ | -30.8691 |
| $YNH_4B_6C_6$ | $3\,B_{13}C_2 + 33\,H_3N + YN \longrightarrow 47.5\,H_2 + 33\,BN + YB_6H_4C_6N$ | -30.8413 |

| Compound | Chemical reaction | Reaction energy(eV) |
|---|---|---|
| $YNH_4B_6C_6$ | $0.3636\,B_9H_{11} + 6\,Y_2C + 6.682\,N_2 + 1.364\,YB_2 \longrightarrow 12.36\,YN + YB_6H_4C_6N$ | -30.7780 |
| $YNH_4B_6C_6$ | $25.1\,YH_2 + 3\,B_{13}C_2 + 12.55\,N_2 \longrightarrow 6.6\,B_5H_7 + 24.1\,YN + YB_6H_4C_6N$ | -30.6617 |
| $YNH_4B_6C_6$ | $1.333\,YH_3 + 6\,Y_2C + 0.5\,YB_{12} + 6.917\,N_2 \longrightarrow 12.83\,YN + YB_6H_4C_6N$ | -30.5883 |
| $YNH_4B_6C_6$ | $YH_3 + 3\,B_{13}C_2 + 34\,H_3N \longrightarrow 50.5\,H_2 + 33\,BN + YB_6H_4C_6N$ | -30.5597 |
| $YNH_4B_6C_6$ | $1.333\,YH_3 + 6\,Y_2C + 6.667\,N_2 + 6\,B \longrightarrow 12.33\,YN + YB_6H_4C_6N$ | -30.4724 |
| $YNH_4B_6C_6$ | $6\,Y_2C + 12.33\,H_3N + 17.33\,B \longrightarrow 11\,YH_3 + 11.33\,BN + YB_6H_4C_6N$ | -30.3755 |
| $YNH_4B_6C_6$ | $3\,B_{13}C_2 + 14\,H_3N + 20\,YN \longrightarrow 19\,YH_2 + 33\,BN + YB_6H_4C_6N$ | -30.3131 |
| $YNH_4B_6C_6$ | $3\,B_{13}C_2 + 34\,YN + 35\,H_2 \longrightarrow 33\,YH_2 + 33\,BN + YB_6H_4C_6N$ | -29.9239 |
| $YNH_4B_6C_6$ | $6\,Y_2C + 6.75\,N_2 + 2\,H_2 + 1.5\,YB_4 \longrightarrow 12.5\,YN + YB_6H_4C_6N$ | -29.7803 |
| $YNH_4B_6C_6$ | $6\,Y_2C + 1.2\,B_5H_7 + 6.733\,N_2 \longrightarrow 1.467\,H_3N + 11\,YN + YB_6H_4C_6N$ | -29.7705 |
| $YNH_4B_6C_6$ | $0.6667\,B_9H_{11} + 6\,Y_2C + 6.556\,N_2 \longrightarrow 1.111\,H_3N + 11\,YN + YB_6H_4C_6N$ | -29.3448 |
| $YNH_4B_6C_6$ | $6\,Y_2C + 0.5\,YB_{12} + 6.25\,N_2 + 2\,H_2 \longrightarrow 11.5\,YN + YB_6H_4C_6N$ | -29.1984 |
| $YNH_4B_6C_6$ | $6\,Y_2C + 6\,N_2 + 2\,H_2 + 6\,B \longrightarrow 11\,YN + YB_6H_4C_6N$ | -29.0825 |
| $YNH_4B_6C_6$ | $6\,Y_2C + 0.5714\,B_5H_7 + 6.393\,N_2 + 0.7857\,YB_4 \longrightarrow 11.79\,YN + YB_6H_4C_6N$ | -28.8509 |
| $YNH_4B_6C_6$ | $2\,YH_2 + 6\,Y_2C + 0.5\,N_2 + 162\,B \longrightarrow 13\,YB_{12} + YB_6H_4C_6N$ | -28.7725 |
| $YNH_4B_6C_6$ | $2.75\,B_{13}C_2 + 0.5\,Y_2C + 30.75\,H_3N \longrightarrow 44.12\,H_2 + 29.75\,BN + YB_6H_4C_6N$ | -28.7493 |
| $YNH_4B_6C_6$ | $0.3636\,B_9H_{11} + 6\,Y_2C + 6.341\,N_2 + 0.6818\,YB_4 \longrightarrow 11.68\,YN + YB_6H_4C_6N$ | -28.7403 |
| $YNH_4B_6C_6$ | $6\,Y_2C + 0.2619\,YB_{12} + 0.5714\,B_5H_7 + 6.131\,N_2 \longrightarrow 11.26\,YN + YB_6H_4C_6N$ | -28.5461 |
| $YNH_4B_6C_6$ | $2\,YH_2 + 6\,Y_2C + YN + 174\,B \longrightarrow 14\,YB_{12} + YB_6H_4C_6N$ | -28.5406 |
| $YNH_4B_6C_6$ | $6\,Y_2C + 0.5714\,B_5H_7 + 6\,N_2 + 3.143\,B \longrightarrow 11\,YN + YB_6H_4C_6N$ | -28.4854 |
| $YNH_4B_6C_6$ | $0.3636\,B_9H_{11} + 6\,Y_2C + 0.2273\,YB_{12} + 6.114\,N_2 \longrightarrow 11.23\,YN + YB_6H_4C_6N$ | -28.4758 |
| $YNH_4B_6C_6$ | $0.3636\,B_9H_{11} + 6\,Y_2C + 6\,N_2 + 2.727\,B \longrightarrow 11\,YN + YB_6H_4C_6N$ | -28.4231 |
| $YNH_4B_6C_6$ | $1.5\,B_{13}C_2 + 0.5\,Y_2C + 0.5\,H_8C_5 + 7.25\,N_2 \longrightarrow 13.5\,BN + YB_6H_4C_6N$ | -28.1816 |
| $YNH_4B_6C_6$ | $3\,B_{13}C_2 + 14.78\,Y + 14.78\,H_3N \longrightarrow 3.667\,B_9H_{11} + 13.78\,YN + YB_6H_4C_6N$ | -28.0913 |
| $YNH_4B_6C_6$ | $6\,Y_2C + 1.333\,H_3N + 6.083\,N_2 + 1.5\,YB_4 \longrightarrow 12.5\,YN + YB_6H_4C_6N$ | -28.0149 |
| $YNH_4B_6C_6$ | $2.75\,B_9H_{11} + 6\,Y_2C + 6\,N_2 \longrightarrow 3.75\,B_5H_7 + 11\,YN + YB_6H_4C_6N$ | -28.0145 |
| $YNH_4B_6C_6$ | $0.6667\,B_9H_{11} + 6\,Y_2C + 6\,N_2 \longrightarrow 11\,YN + 1.667\,H_2 + YB_6H_4C_6N$ | -27.8736 |
| $YNH_4B_6C_6$ | $6\,Y_2C + 1.2\,B_5H_7 + 6\,N_2 \longrightarrow 11\,YN + 2.2\,H_2 + YB_6H_4C_6N$ | -27.8285 |
| $YNH_4B_6C_6$ | $1.333\,YH_3 + 6\,Y_2C + 0.5\,N_2 + 154\,B \longrightarrow 12.33\,YB_{12} + YB_6H_4C_6N$ | -27.6132 |
| $YNH_4B_6C_6$ | $6\,Y_2C + 1.333\,H_3N + 0.5\,YB_{12} + 5.583\,N_2 \longrightarrow 11.5\,YN + YB_6H_4C_6N$ | -27.4330 |
| $YNH_4B_6C_6$ | $1.333\,YH_3 + 6\,Y_2C + YN + 166\,B \longrightarrow 13.33\,YB_{12} + YB_6H_4C_6N$ | -27.3814 |
| $YNH_4B_6C_6$ | $6\,Y_2C + 1.333\,H_3N + 5.333\,N_2 + 6\,B \longrightarrow 11\,YN + YB_6H_4C_6N$ | -27.3171 |
| $YNH_4B_6C_6$ | $0.6667\,B_9H_{11} + 6\,Y_2C + 5.444\,N_2 \longrightarrow 1.111\,YH_3 + 9.889\,YN + YB_6H_4C_6N$ | -26.7153 |
| $YNH_4B_6C_6$ | $6\,Y_2C + 0.5\,N_2 + 2\,H_2 + 138\,B \longrightarrow 11\,YB_{12} + YB_6H_4C_6N$ | -26.5324 |
| $YNH_4B_6C_6$ | $22.17\,YH_2 + 3\,B_{13}C_2 + 11.08\,N_2 \longrightarrow 3.667\,B_9H_{11} + 21.17\,YN + YB_6H_4C_6N$ | -26.4482 |
| $YNH_4B_6C_6$ | $6\,Y_2C + YN + 2\,H_2 + 150\,B \longrightarrow 12\,YB_{12} + YB_6H_4C_6N$ | -26.3006 |
| $YNH_4B_6C_6$ | $6\,Y_2C + 1.2\,B_5H_7 + 5.267\,N_2 \longrightarrow 1.467\,YH_3 + 9.533\,YN + YB_6H_4C_6N$ | -26.2996 |
| $YNH_4B_6C_6$ | $2\,YH_2 + 6\,Y_2C + BN + 161\,B \longrightarrow 13\,YB_{12} + YB_6H_4C_6N$ | -26.2057 |
| $YNH_4B_6C_6$ | $6\,Y_2C + 0.5714\,B_5H_7 + 0.5\,N_2 + 135.1\,B \longrightarrow 11\,YB_{12} + YB_6H_4C_6N$ | -25.9352 |
| $YNH_4B_6C_6$ | $0.3636\,B_9H_{11} + 6\,Y_2C + 0.5\,N_2 + 134.7\,B \longrightarrow 11\,YB_{12} + YB_6H_4C_6N$ | -25.8730 |
| $YNH_4B_6C_6$ | $0.5\,YH_2 + 6\,Y_2C + H_3N + 144\,B \longrightarrow 11.5\,YB_{12} + YB_6H_4C_6N$ | -25.7683 |
| $YNH_4B_6C_6$ | $2\,YH_2 + 0.4615\,B_{13}C_2 + 5.077\,Y_2C + 6.077\,N_2 \longrightarrow 11.15\,YN + YB_6H_4C_6N$ | -25.7478 |
| $YNH_4B_6C_6$ | $2\,YH_2 + 6\,Y_2C + 0.5\,N_2 + 58\,B \longrightarrow 13\,YB_4 + YB_6H_4C_6N$ | -25.7385 |
| $YNH_4B_6C_6$ | $6\,Y_2C + 0.5714\,B_5H_7 + YN + 147.1\,B \longrightarrow 12\,YB_{12} + YB_6H_4C_6N$ | -25.7034 |
| $YNH_4B_6C_6$ | $0.3636\,B_9H_{11} + 0.2098\,B_{13}C_2 + 5.58\,Y_2C + 5.58\,N_2 \longrightarrow 10.16\,YN + YB_6H_4C_6N$ | -25.6783 |
| $YNH_4B_6C_6$ | $0.3636\,B_9H_{11} + 6\,Y_2C + YN + 146.7\,B \longrightarrow 12\,YB_{12} + YB_6H_4C_6N$ | -25.6411 |
| $YNH_4B_6C_6$ | $6\,Y_2C + 1.333\,H_3N + 138.3\,B \longrightarrow 11\,YB_{12} + 0.3333\,BN + YB_6H_4C_6N$ | -25.6226 |
| $YNH_4B_6C_6$ | $0.6667\,B_9H_{11} + 6\,Y_2C + 5.167\,N_2 \longrightarrow 1.667\,YH_2 + 9.333\,YN + YB_6H_4C_6N$ | -25.6204 |
| $YNH_4B_6C_6$ | $7\,B_9H_{11} + 0.5\,N_2 + 7.375\,Y_4C_7 \longrightarrow 9.125\,H_8C_5 + 28.5\,YB_2 + YB_6H_4C_6N$ | -25.5736 |
| $YNH_4B_6C_6$ | $0.3333\,YH_3 + 6\,Y_2C + H_3N + 142\,B \longrightarrow 11.33\,YB_{12} + YB_6H_4C_6N$ | -25.4785 |
| $YNH_4B_6C_6$ | $15.27\,H_3N + 1.2\,Y_4C_5 + 10.13\,YB_2 \longrightarrow 13.93\,YH_3 + 14.27\,BN + YB_6H_4C_6N$ | -25.3498 |
| $YNH_4B_6C_6$ | $0.2418\,B_{13}C_2 + 5.516\,Y_2C + 0.5714\,B_5H_7 + 5.516\,N_2 \longrightarrow 10.03\,YN + YB_6H_4C_6N$ | -25.3224 |
| $YNH_4B_6C_6$ | $2\,YH_2 + 6\,Y_2C + YN + 62\,B \longrightarrow 14\,YB_4 + YB_6H_4C_6N$ | -25.2733 |
| $YNH_4B_6C_6$ | $6\,Y_2C + H_3N + 0.5\,H_2 + 138\,B \longrightarrow 11\,YB_{12} + YB_6H_4C_6N$ | -25.2083 |
| $YNH_4B_6C_6$ | $6\,Y_2C + H_3N + 0.1429\,B_5H_7 + 137.3\,B \longrightarrow 11\,YB_{12} + YB_6H_4C_6N$ | -25.0590 |
| $YNH_4B_6C_6$ | $1.333\,YH_3 + 6\,Y_2C + BN + 153\,B \longrightarrow 12.33\,YB_{12} + YB_6H_4C_6N$ | -25.0464 |
| $YNH_4B_6C_6$ | $0.09091\,B_9H_{11} + 6\,Y_2C + H_3N + 137.2\,B \longrightarrow 11\,YB_{12} + YB_6H_4C_6N$ | -25.0435 |
| $YNH_4B_6C_6$ | $4.085\,B_9H_{11} + 10.73\,Y_2C + 0.5\,N_2 \longrightarrow 20.46\,YH_2 + 2.366\,B_{13}C_2 + YB_6H_4C_6N$ | -24.9676 |
| $YNH_4B_6C_6$ | $4.451\,B_9H_{11} + 11.24\,Y_2C + YN \longrightarrow 22.48\,YH_2 + 2.62\,B_{13}C_2 + YB_6H_4C_6N$ | -24.8976 |
| $YNH_4B_6C_6$ | $6\,Y_2C + 1.2\,B_5H_7 + 4.9\,N_2 \longrightarrow 2.2\,YH_2 + 8.8\,YN + YB_6H_4C_6N$ | -24.8544 |

| Compound | Chemical reaction | Reaction energy(eV) |
|---|---|---|
| $YNH_4B_6C_6$ | $6\,Y_2C + 1.333\,H_3N + 134\,B \longrightarrow 10.67\,YB_{12} + 0.3333\,YN + YB_6H_4C_6N$ | -24.8442 |
| $YNH_4B_6C_6$ | $6\,Y_2C + 1.333\,H_3N + 138\,B \longrightarrow 11\,YB_{12} + 0.1667\,N_2 + YB_6H_4C_6N$ | -24.7670 |
| $YNH_4B_6C_6$ | $1.333\,YH_3 + 6\,Y_2C + 0.5\,N_2 + 55.33\,B \longrightarrow 12.33\,YB_4 + YB_6H_4C_6N$ | -24.7348 |
| $YNH_4B_6C_6$ | $1.333\,YH_3 + 0.4615\,B_{13}C_2 + 5.077\,Y_2C + 5.744\,N_2 \longrightarrow 10.49\,YN + YB_6H_4C_6N$ | -24.4340 |
| $YNH_4B_6C_6$ | $1.2\,Y_4C_5 + 10.32\,B_5H_7 + 11.87\,N_2 \longrightarrow 22.75\,H_3N + 3.8\,YB_{12} + YB_6H_4C_6N$ | -24.4223 |
| $YNH_4B_6C_6$ | $1.333\,YH_3 + 6\,Y_2C + YN + 59.33\,B \longrightarrow 13.33\,YB_4 + YB_6H_4C_6N$ | -24.2696 |
| $YNH_4B_6C_6$ | $0.3636\,B_9H_{11} + 3\,B_{13}C_2 + 19.14\,Y + 0.5\,N_2 \longrightarrow 18.14\,YB_2 + YB_6H_4C_6N$ | -24.1232 |
| $YNH_4B_6C_6$ | $2\,YH_2 + 6\,Y_2C + 7.25\,YB_{12} + 0.5\,N_2 \longrightarrow 20.25\,YB_4 + YB_6H_4C_6N$ | -24.0465 |
| $YNH_4B_6C_6$ | $6\,Y_2C + 2\,H_2 + BN + 137\,B \longrightarrow 11\,YB_{12} + YB_6H_4C_6N$ | -23.9656 |
| $YNH_4B_6C_6$ | $6\,Y_2C + 0.5\,N_2 + 2\,H_2 + 50\,B \longrightarrow 11\,YB_4 + YB_6H_4C_6N$ | -23.9652 |
| $YNH_4B_6C_6$ | $3\,B_{13}C_2 + 18.93\,Y + 0.5714\,B_5H_7 + 0.5\,N_2 \longrightarrow 17.93\,YB_2 + YB_6H_4C_6N$ | -23.8248 |
| $YNH_4B_6C_6$ | $6\,Y_2C + YN + 2\,H_2 + 54\,B \longrightarrow 12\,YB_4 + YB_6H_4C_6N$ | -23.5000 |
| $YNH_4B_6C_6$ | $2\,YH_2 + 6\,Y_2C + 7.75\,YB_{12} + YN \longrightarrow 21.75\,YB_4 + YB_6H_4C_6N$ | -23.4646 |
| $YNH_4B_6C_6$ | $4.197\,B_9H_{11} + 11.04\,Y_2C + BN \longrightarrow 21.08\,YH_2 + 2.521\,B_{13}C_2 + YB_6H_4C_6N$ | -23.3857 |
| $YNH_4B_6C_6$ | $6\,Y_2C + 0.5714\,B_5H_7 + BN + 134.1\,B \longrightarrow 11\,YB_{12} + YB_6H_4C_6N$ | -23.3685 |
| $YNH_4B_6C_6$ | $6\,Y_2C + 0.5714\,B_5H_7 + 0.5\,N_2 + 47.14\,B \longrightarrow 11\,YB_4 + YB_6H_4C_6N$ | -23.3680 |
| $YNH_4B_6C_6$ | $0.3636\,B_9H_{11} + 6\,Y_2C + BN + 133.7\,B \longrightarrow 11\,YB_{12} + YB_6H_4C_6N$ | -23.3062 |
| $YNH_4B_6C_6$ | $0.3636\,B_9H_{11} + 6\,Y_2C + 0.5\,N_2 + 46.73\,B \longrightarrow 11\,YB_4 + YB_6H_4C_6N$ | -23.3058 |
| $YNH_4B_6C_6$ | $2\,YH_2 + 6\,Y_2C + BN + 57\,B \longrightarrow 13\,YB_4 + YB_6H_4C_6N$ | -23.1718 |
| $YNH_4B_6C_6$ | $1.333\,YH_3 + 6\,Y_2C + 6.917\,YB_{12} + 0.5\,N_2 \longrightarrow 19.25\,YB_4 + YB_6H_4C_6N$ | -23.1206 |
| $YNH_4B_6C_6$ | $0.5\,YH_2 + 6\,Y_2C + H_3N + 52\,B \longrightarrow 11.5\,YB_4 + YB_6H_4C_6N$ | -23.0845 |
| $YNH_4B_6C_6$ | $6\,Y_2C + 1.333\,H_3N + 50.33\,B \longrightarrow 0.3333\,BN + 11\,YB_4 + YB_6H_4C_6N$ | -23.0554 |
| $YNH_4B_6C_6$ | $0.4615\,B_{13}C_2 + 5.077\,Y_2C + 5.077\,N_2 + 2\,H_2 \longrightarrow 9.154\,YN + YB_6H_4C_6N$ | -23.0441 |
| $YNH_4B_6C_6$ | $6\,Y_2C + 0.5714\,B_5H_7 + YN + 51.14\,B \longrightarrow 12\,YB_4 + YB_6H_4C_6N$ | -22.9028 |
| $YNH_4B_6C_6$ | $0.3636\,B_9H_{11} + 6\,Y_2C + YN + 50.73\,B \longrightarrow 12\,YB_4 + YB_6H_4C_6N$ | -22.8406 |
| $YNH_4B_6C_6$ | $0.3333\,YH_3 + 6\,Y_2C + H_3N + 51.33\,B \longrightarrow 11.33\,YB_4 + YB_6H_4C_6N$ | -22.8335 |
| $YNH_4B_6C_6$ | $2.375\,B_{13}C_2 + 25.88\,H_3N + 0.25\,Y_4C_5 \longrightarrow 36.81\,H_2 + 24.87\,BN + YB_6H_4C_6N$ | -22.7823 |
| $YNH_4B_6C_6$ | $6\,Y_2C + H_3N + 0.5\,H_2 + 50\,B \longrightarrow 11\,YB_4 + YB_6H_4C_6N$ | -22.6411 |
| $YNH_4B_6C_6$ | $1.333\,YH_3 + 6\,Y_2C + 7.417\,YB_{12} + YN \longrightarrow 20.75\,YB_4 + YB_6H_4C_6N$ | -22.5387 |
| $YNH_4B_6C_6$ | $6\,Y_2C + 6.25\,YB_{12} + 0.5\,N_2 + 2\,H_2 \longrightarrow 17.25\,YB_4 + YB_6H_4C_6N$ | -22.5066 |
| $YNH_4B_6C_6$ | $6\,Y_2C + H_3N + 0.1429\,B_5H_7 + 49.29\,B \longrightarrow 11\,YB_4 + YB_6H_4C_6N$ | -22.4919 |
| $YNH_4B_6C_6$ | $0.09091\,B_9H_{11} + 6\,Y_2C + H_3N + 49.18\,B \longrightarrow 11\,YB_4 + YB_6H_4C_6N$ | -22.4763 |
| $YNH_4B_6C_6$ | $0.3636\,B_9H_{11} + 3\,B_{13}C_2 + 19.64\,Y + BN \longrightarrow 18.64\,YB_2 + YB_6H_4C_6N$ | -22.4246 |
| $YNH_4B_6C_6$ | $6\,Y_2C + 8.667\,H_3N + 13.67\,B \longrightarrow 11\,YH_2 + 7.667\,BN + YB_6H_4C_6N$ | -22.4152 |
| $YNH_4B_6C_6$ | $6\,Y_2C + 1.333\,H_3N + 48.67\,B \longrightarrow 0.3333\,YN + 10.67\,YB_4 + YB_6H_4C_6N$ | -22.3549 |
| $YNH_4B_6C_6$ | $6\,Y_2C + 1.333\,H_3N + 50\,B \longrightarrow 0.1667\,N_2 + 11\,YB_4 + YB_6H_4C_6N$ | -22.1998 |
| $YNH_4B_6C_6$ | $6\,Y_2C + 11.4\,H_3N + 4.1\,YB_4 \longrightarrow 15.1\,YH_2 + 10.4\,BN + YB_6H_4C_6N$ | -22.1765 |
| $YNH_4B_6C_6$ | $1.333\,YH_3 + 6\,Y_2C + BN + 54.33\,B \longrightarrow 12.33\,YB_4 + YB_6H_4C_6N$ | -22.1681 |
| $YNH_4B_6C_6$ | $3\,B_{13}C_2 + 19.43\,Y + 0.5714\,B_5H_7 + BN \longrightarrow 18.43\,YB_2 + YB_6H_4C_6N$ | -22.1261 |
| $YNH_4B_6C_6$ | $6\,Y_2C + 9.471\,H_3N + 1.206\,YB_{12} \longrightarrow 12.21\,YH_2 + 8.471\,BN + YB_6H_4C_6N$ | -22.0636 |
| $YNH_4B_6C_6$ | $6\,Y_2C + 5.893\,YB_{12} + 0.5714\,B_5H_7 + 0.5\,N_2 \longrightarrow 16.89\,YB_4 + YB_6H_4C_6N$ | -21.9928 |
| $YNH_4B_6C_6$ | $0.3636\,B_9H_{11} + 6\,Y_2C + 5.841\,YB_{12} + 0.5\,N_2 \longrightarrow 16.84\,YB_4 + YB_6H_4C_6N$ | -21.9426 |
| $YNH_4B_6C_6$ | $3\,B_{13}C_2 + 17.5\,Y + 0.5\,N_2 + 2\,H_2 \longrightarrow 16.5\,YB_2 + YB_6H_4C_6N$ | -21.9417 |
| $YNH_4B_6C_6$ | $6\,Y_2C + 6.75\,YB_{12} + YN + 2\,H_2 \longrightarrow 18.75\,YB_4 + YB_6H_4C_6N$ | -21.9247 |
| $YNH_4B_6C_6$ | $3.535\,B_9H_{11} + 9.972\,Y_2C + H_3N \longrightarrow 18.94\,YH_2 + 1.986\,B_{13}C_2 + YB_6H_4C_6N$ | -21.7206 |
| $YNH_4B_6C_6$ | $1.368\,B_9H_{11} + 6\,Y_2C + 7.316\,H_3N \longrightarrow 11\,YH_3 + 6.316\,BN + YB_6H_4C_6N$ | -21.6588 |
| $YNH_4B_6C_6$ | $6\,Y_2C + 1.333\,H_3N + 6.292\,YB_{12} \longrightarrow 0.3333\,BN + 17.29\,YB_4 + YB_6H_4C_6N$ | -21.5870 |
| $YNH_4B_6C_6$ | $0.5\,YH_2 + 6\,Y_2C + H_3N + 6.5\,YB_{12} \longrightarrow 18\,YB_4 + YB_6H_4C_6N$ | -21.5675 |
| $YNH_4B_6C_6$ | $5.375\,Y_2C + H_3N + 0.125\,H_8C_5 + 123\,B \longrightarrow 9.75\,YB_{12} + YB_6H_4C_6N$ | -21.5163 |
| $YNH_4B_6C_6$ | $9.451\,Y_2C + 5.686\,B_5H_7 + 0.5\,N_2 \longrightarrow 17.9\,YH_2 + 1.725\,B_{13}C_2 + YB_6H_4C_6N$ | -21.5142 |
| $YNH_4B_6C_6$ | $2\,YH_2 + 6\,Y_2C + 7.125\,YB_{12} + BN \longrightarrow 20.12\,YB_4 + YB_6H_4C_6N$ | -21.5089 |
| $YNH_4B_6C_6$ | $0.3636\,B_9H_{11} + 6\,Y_2C + 4.636\,N_2 + 2.727\,BN \longrightarrow 11\,YN + YB_6H_4C_6N$ | -21.4228 |
| $YNH_4B_6C_6$ | $6\,Y_2C + 6.393\,YB_{12} + 0.5714\,B_5H_7 + YN \longrightarrow 18.39\,YB_4 + YB_6H_4C_6N$ | -21.4109 |
| $YNH_4B_6C_6$ | $6\,Y_2C + 2\,H_2 + BN + 49\,B \longrightarrow 11\,YB_4 + YB_6H_4C_6N$ | -21.3984 |
| $YNH_4B_6C_6$ | $0.3636\,B_9H_{11} + 6\,Y_2C + 6.341\,YB_{12} + YN \longrightarrow 18.34\,YB_4 + YB_6H_4C_6N$ | -21.3607 |
| $YNH_4B_6C_6$ | $0.3333\,YH_3 + 6\,Y_2C + H_3N + 6.417\,YB_{12} \longrightarrow 17.75\,YB_4 + YB_6H_4C_6N$ | -21.3360 |
| $YNH_4B_6C_6$ | $3\,B_{13}C_2 + 17.83\,Y + 1.333\,H_3N \longrightarrow 0.3333\,YN + 16.5\,YB_2 + YB_6H_4C_6N$ | -21.3307 |
| $YNH_4B_6C_6$ | $0.4615\,B_{13}C_2 + 5.077\,Y_2C + 1.333\,H_3N + 4.41\,N_2 \longrightarrow 9.154\,YN + YB_6H_4C_6N$ | -21.2786 |
| $YNH_4B_6C_6$ | $6\,Y_2C + H_3N + 6.25\,YB_{12} + 0.5\,H_2 \longrightarrow 17.25\,YB_4 + YB_6H_4C_6N$ | -21.1825 |
| $YNH_4B_6C_6$ | $0.09091\,B_9H_{11} + 3\,B_{13}C_2 + 17.91\,Y + H_3N \longrightarrow 16.91\,YB_2 + YB_6H_4C_6N$ | -21.1630 |

| Compound | Chemical reaction | Reaction energy(eV) |
|---|---|---|
| YNH$_4$B$_6$C$_6$ | 9.843 Y$_2$C + 6.196 B$_5$H$_7$ + YN ⟶ 19.69 YH$_2$ + 1.922 B$_{13}$C$_2$ + YB$_6$H$_4$C$_6$N | -21.1347 |
| YNH$_4$B$_6$C$_6$ | 3 B$_{13}$C$_2$ + 17.86 Y + H$_3$N + 0.1429 B$_5$H$_7$ ⟶ 16.86 YB$_2$ + YB$_6$H$_4$C$_6$N | -21.0884 |
| YNH$_4$B$_6$C$_6$ | 6 Y$_2$C + H$_3$N + 6.161 YB$_{12}$ + 0.1429 B$_5$H$_7$ ⟶ 17.16 YB$_4$ + YB$_6$H$_4$C$_6$N | -21.0541 |
| YNH$_4$B$_6$C$_6$ | 6 Y$_2$C + 6.818 H$_3$N + 2.364 B$_5$H$_7$ ⟶ 11 YH$_3$ + 5.818 BN + YB$_6$H$_4$C$_6$N | -21.0518 |
| YNH$_4$B$_6$C$_6$ | 0.09091 B$_9$H$_{11}$ + 6 Y$_2$C + H$_3$N + 6.148 YB$_{12}$ ⟶ 17.15 YB$_4$ + YB$_6$H$_4$C$_6$N | -21.0415 |
| YNH$_4$B$_6$C$_6$ | 6 Y$_2$C + 1.333 H$_3$N + 6.083 YB$_{12}$ ⟶ 0.3333 YN + 16.75 YB$_4$ + YB$_6$H$_4$C$_6$N | -20.9351 |
| YNH$_4$B$_6$C$_6$ | 6 Y$_2$C + 0.5714 B$_5$H$_7$ + BN + 46.14 B ⟶ 11 YB$_4$ + YB$_6$H$_4$C$_6$N | -20.8013 |
| YNH$_4$B$_6$C$_6$ | 5.733 B$_9$H$_{11}$ + 1.2 Y$_4$C$_5$ + 10.34 N$_2$ ⟶ 19.69 H$_3$N + 3.8 YB$_{12}$ + YB$_6$H$_4$C$_6$N | -20.7613 |
| YNH$_4$B$_6$C$_6$ | 3 B$_{13}$C$_2$ + 17.33 Y + 1.333 H$_3$N ⟶ 0.3333 BN + 16.33 YB$_2$ + YB$_6$H$_4$C$_6$N | -20.7425 |
| YNH$_4$B$_6$C$_6$ | 6 Y$_2$C + 1.333 H$_3$N + 6.25 YB$_{12}$ ⟶ 0.1667 N$_2$ + 17.25 YB$_4$ + YB$_6$H$_4$C$_6$N | -20.7412 |
| YNH$_4$B$_6$C$_6$ | 0.3636 B$_9$H$_{11}$ + 6 Y$_2$C + BN + 45.73 B ⟶ 11 YB$_4$ + YB$_6$H$_4$C$_6$N | -20.7390 |
| YNH$_4$B$_6$C$_6$ | 0.3636 B$_9$H$_{11}$ + 3 B$_{13}$C$_2$ + 18.14 Y + YN ⟶ 18.14 YB$_2$ + YB$_6$H$_4$C$_6$N | -20.6602 |
| YNH$_4$B$_6$C$_6$ | 3 B$_{13}$C$_2$ + 17.5 Y + H$_3$N + 0.5 H$_2$ ⟶ 16.5 YB$_2$ + YB$_6$H$_4$C$_6$N | -20.6177 |
| YNH$_4$B$_6$C$_6$ | 1.333 YH$_3$ + 6 Y$_2$C + 6.792 YB$_{12}$ + BN ⟶ 19.12 YB$_4$ + YB$_6$H$_4$C$_6$N | -20.5830 |
| YNH$_4$B$_6$C$_6$ | 6 Y$_2$C + 0.5714 B$_5$H$_7$ + 4.429 N$_2$ + 3.143 BN ⟶ 11 YN + YB$_6$H$_4$C$_6$N | -20.4184 |
| YNH$_4$B$_6$C$_6$ | 3 B$_{13}$C$_2$ + 17.93 Y + 0.5714 B$_5$H$_7$ + YN ⟶ 17.93 YB$_2$ + YB$_6$H$_4$C$_6$N | -20.3617 |
| YNH$_4$B$_6$C$_6$ | 3 B$_{13}$C$_2$ + 18 Y + 2 H$_2$ + BN ⟶ 17 YB$_2$ + YB$_6$H$_4$C$_6$N | -20.2430 |
| YNH$_4$B$_6$C$_6$ | 3 B$_{13}$C$_2$ + 17.5 Y + 1.333 H$_3$N ⟶ 0.1667 N$_2$ + 16.5 YB$_2$ + YB$_6$H$_4$C$_6$N | -20.1763 |
| YNH$_4$B$_6$C$_6$ | 6 Y$_2$C + 6.125 YB$_{12}$ + 2 H$_2$ + BN ⟶ 17.12 YB$_4$ + YB$_6$H$_4$C$_6$N | -19.9690 |
| YNH$_4$B$_6$C$_6$ | 9.725 Y$_2$C + 5.843 B$_5$H$_7$ + BN ⟶ 18.45 YH$_2$ + 1.863 B$_{13}$C$_2$ + YB$_6$H$_4$C$_6$N | -19.8371 |
| YNH$_4$B$_6$C$_6$ | 0.3636 B$_9$H$_{11}$ + 3 B$_{13}$C$_2$ + 10.07 Y + 0.5 N$_2$ ⟶ 9.068 YB$_4$ + YB$_6$H$_4$C$_6$N | -19.8212 |
| YNH$_4$B$_6$C$_6$ | 0.3333 YH$_3$ + 3 B$_{13}$C$_2$ + 17.17 Y + H$_3$N ⟶ 16.5 YB$_2$ + YB$_6$H$_4$C$_6$N | -19.8108 |
| YNH$_4$B$_6$C$_6$ | 6 Y$_2$C + 5.6 B$_5$H$_7$ + 6.367 N$_2$ ⟶ 11.73 H$_3$N + 11 YB$_2$ + YB$_6$H$_4$C$_6$N | -19.7699 |
| YNH$_4$B$_6$C$_6$ | 6 Y$_2$C + 0.5 N$_2$ + 23 H$_2$ + 3 YB$_2$ ⟶ 14 YH$_3$ + YB$_6$H$_4$C$_6$N | -19.6693 |
| YNH$_4$B$_6$C$_6$ | 3 B$_{13}$C$_2$ + 9.964 Y + 0.5714 B$_5$H$_7$ + 0.5 N$_2$ ⟶ 8.964 YB$_4$ + YB$_6$H$_4$C$_6$N | -19.5720 |
| YNH$_4$B$_6$C$_6$ | 0.5 YH$_2$ + 3 B$_{13}$C$_2$ + 17 Y + H$_3$N ⟶ 16.5 YB$_2$ + YB$_6$H$_4$C$_6$N | -19.5621 |
| YNH$_4$B$_6$C$_6$ | 6 Y$_2$C + 5.768 YB$_{12}$ + 0.5714 B$_5$H$_7$ + BN ⟶ 16.77 YB$_4$ + YB$_6$H$_4$C$_6$N | -19.4552 |
| YNH$_4$B$_6$C$_6$ | 0.3636 B$_9$H$_{11}$ + 6 Y$_2$C + 5.716 YB$_{12}$ + BN ⟶ 16.72 YB$_4$ + YB$_6$H$_4$C$_6$N | -19.4050 |
| YNH$_4$B$_6$C$_6$ | 6.125 B$_9$H$_{11}$ + 6.516 Y$_4$C$_7$ + BN ⟶ 7.922 H$_8$C$_5$ + 25.06 YB$_2$ + YB$_6$H$_4$C$_6$N | -19.2703 |
| YNH$_4$B$_6$C$_6$ | 5.375 Y$_2$C + H$_3$N + 0.125 H$_8$C$_5$ + 45 B ⟶ 9.75 YB$_4$ + YB$_6$H$_4$C$_6$N | -19.2408 |
| YNH$_4$B$_6$C$_6$ | 2.125 B$_{13}$C$_2$ + 22.62 H$_3$N + 0.25 Y$_4$C$_7$ ⟶ 31.94 H$_2$ + 21.62 BN + YB$_6$H$_4$C$_6$N | -18.9560 |
| YNH$_4$B$_6$C$_6$ | 8.863 Y$_2$C + H$_3$N + 4.922 B$_5$H$_7$ ⟶ 16.73 YH$_2$ + 1.431 B$_{13}$C$_2$ + YB$_6$H$_4$C$_6$N | -18.7317 |
| YNH$_4$B$_6$C$_6$ | 1.333 YH$_3$ + 3 B$_{13}$C$_2$ + 16.17 Y + 0.5 N$_2$ ⟶ 16.5 YB$_2$ + YB$_6$H$_4$C$_6$N | -18.7142 |
| YNH$_4$B$_6$C$_6$ | 6 Y$_2$C + YN + 24.5 H$_2$ + 3 YB$_2$ ⟶ 15 YH$_3$ + YB$_6$H$_4$C$_6$N | -18.6269 |
| YNH$_4$B$_6$C$_6$ | 12.52 H$_3$N + 0.8571 Y$_4$C$_7$ + 8.762 YB$_2$ ⟶ 11.19 YH$_3$ + 11.52 BN + YB$_6$H$_4$C$_6$N | -18.5499 |
| YNH$_4$B$_6$C$_6$ | 3 B$_{13}$C$_2$ + 16.5 Y + YN + 2 H$_2$ ⟶ 16.5 YB$_2$ + YB$_6$H$_4$C$_6$N | -18.4787 |
| YNH$_4$B$_6$C$_6$ | 6 Y$_2$C + 12 H$_3$N + 32.18 B ⟶ 2.909 B$_9$H$_{11}$ + 11 YN + YB$_6$H$_4$C$_6$N | -18.4691 |
| YNH$_4$B$_6$C$_6$ | 6 Y$_2$C + H$_3$N + 21.5 H$_2$ + 3 YB$_2$ ⟶ 14 YH$_3$ + YB$_6$H$_4$C$_6$N | -18.3452 |
| YNH$_4$B$_6$C$_6$ | 3.5 Y$_2$C + 0.5 H$_8$C$_5$ + 5 N$_2$ + 3 YB$_2$ ⟶ 9 YN + YB$_6$H$_4$C$_6$N | -18.3361 |
| YNH$_4$B$_6$C$_6$ | 3 B$_{13}$C$_2$ + 9.25 Y + 0.5 N$_2$ + 2 H$_2$ ⟶ 8.25 YB$_4$ + YB$_6$H$_4$C$_6$N | -18.0279 |
| YNH$_4$B$_6$C$_6$ | 0.3636 B$_9$H$_{11}$ + 3 B$_{13}$C$_2$ + 10.32 Y + BN ⟶ 9.318 YB$_4$ + YB$_6$H$_4$C$_6$N | -18.0039 |
| YNH$_4$B$_6$C$_6$ | 6 Y$_2$C + 12 H$_3$N + 28.86 B ⟶ 4.571 B$_5$H$_7$ + 11 YN + YB$_6$H$_4$C$_6$N | -17.9710 |
| YNH$_4$B$_6$C$_6$ | 5.375 Y$_2$C + H$_3$N + 5.625 YB$_{12}$ + 0.125 H$_8$C$_5$ ⟶ 15.37 YB$_4$ + YB$_6$H$_4$C$_6$N | -17.9281 |
| YNH$_4$B$_6$C$_6$ | 3.111 B$_9$H$_{11}$ + 6 Y$_2$C + 5.537 N$_2$ ⟶ 10.07 H$_3$N + 11 YB$_2$ + YB$_6$H$_4$C$_6$N | -17.7833 |
| YNH$_4$B$_6$C$_6$ | 3 B$_{13}$C$_2$ + 10.21 Y + 0.5714 B$_5$H$_7$ + BN ⟶ 9.214 YB$_4$ + YB$_6$H$_4$C$_6$N | -17.7547 |
| YNH$_4$B$_6$C$_6$ | 2 YH$_2$ + 3 B$_{13}$C$_2$ + 15.5 Y + 0.5 N$_2$ ⟶ 16.5 YB$_2$ + YB$_6$H$_4$C$_6$N | -17.7194 |
| YNH$_4$B$_6$C$_6$ | 6 Y$_2$C + 0.5 N$_2$ + 18.5 H$_2$ + 6 B ⟶ 11 YH$_3$ + YB$_6$H$_4$C$_6$N | -17.6158 |
| YNH$_4$B$_6$C$_6$ | 3 B$_{13}$C$_2$ + 9.583 Y + 1.333 H$_3$N ⟶ 0.3333 YN + 8.25 YB$_4$ + YB$_6$H$_4$C$_6$N | -17.4168 |
| YNH$_4$B$_6$C$_6$ | 0.3636 B$_9$H$_{11}$ + 6 C + YB$_{12}$ + 5.136 N$_2$ ⟶ 9.273 BN + YB$_6$H$_4$C$_6$N | -17.2327 |
| YNH$_4$B$_6$C$_6$ | 6 Y$_2$C + 0.5 YB$_{12}$ + 0.5 N$_2$ + 19.25 H$_2$ ⟶ 11.5 YH$_3$ + YB$_6$H$_4$C$_6$N | -17.2105 |
| YNH$_4$B$_6$C$_6$ | 2.687 B$_{13}$C$_2$ + 15.47 Y + H$_3$N + 0.125 H$_8$C$_5$ ⟶ 14.47 YB$_2$ + YB$_6$H$_4$C$_6$N | -17.1979 |
| YNH$_4$B$_6$C$_6$ | 0.09091 B$_9$H$_{11}$ + 3 B$_{13}$C$_2$ + 9.455 Y + H$_3$N ⟶ 8.455 YB$_4$ + YB$_6$H$_4$C$_6$N | -17.1521 |
| YNH$_4$B$_6$C$_6$ | 3 B$_{13}$C$_2$ + 9.429 Y + H$_3$N + 0.1429 B$_5$H$_7$ ⟶ 8.429 YB$_4$ + YB$_6$H$_4$C$_6$N | -17.0898 |
| YNH$_4$B$_6$C$_6$ | 1.333 YH$_3$ + 3 B$_{13}$C$_2$ + 16.67 Y + BN ⟶ 17 YB$_2$ + YB$_6$H$_4$C$_6$N | -17.0155 |
| YNH$_4$B$_6$C$_6$ | 3 B$_{13}$C$_2$ + 9.167 Y + 1.333 H$_3$N ⟶ 0.3333 BN + 8.167 YB$_4$ + YB$_6$H$_4$C$_6$N | -16.8682 |
| YNH$_4$B$_6$C$_6$ | 6 Y$_2$C + 22.25 H$_2$ + BN + 2.5 YB$_2$ ⟶ 13.5 YH$_3$ + YB$_6$H$_4$C$_6$N | -16.7603 |
| YNH$_4$B$_6$C$_6$ | 6 Y$_2$C + 0.5 N$_2$ + 20.75 H$_2$ + 1.5 YB$_4$ ⟶ 12.5 YH$_3$ + YB$_6$H$_4$C$_6$N | -16.7500 |
| YNH$_4$B$_6$C$_6$ | 3 B$_{13}$C$_2$ + 9.25 Y + H$_3$N + 0.5 H$_2$ ⟶ 8.25 YB$_4$ + YB$_6$H$_4$C$_6$N | -16.7038 |
| YNH$_4$B$_6$C$_6$ | 6 Y$_2$C + YN + 20 H$_2$ + 6 B ⟶ 12 YH$_3$ + YB$_6$H$_4$C$_6$N | -16.5734 |
| YNH$_4$B$_6$C$_6$ | 6 Y$_2$C + 15.37 H$_3$N + 3.371 YB$_{12}$ ⟶ 3.829 B$_9$H$_{11}$ + 14.37 YN + YB$_6$H$_4$C$_6$N | -16.4541 |

| Compound | Chemical reaction | Reaction energy(eV) |
|---|---|---|
| $YNH_4B_6C_6$ | $1.4\,B_9H_{11} + 1.2\,Y_4C_5 + 3.8\,N_2 \longrightarrow 3.8\,YH_3 + 6.6\,BN + YB_6H_4C_6N$ | -16.4104 |
| $YNH_4B_6C_6$ | $0.6667\,B_9H_{11} + 6\,Y_2C + 0.5\,N_2 + 14.83\,H_2 \longrightarrow 11\,YH_3 + YB_6H_4C_6N$ | -16.4068 |
| $YNH_4B_6C_6$ | $2\,YH_2 + 6\,Y_2C + 4\,N_2 + 6\,BN \longrightarrow 13\,YN + YB_6H_4C_6N$ | -16.3857 |
| $YNH_4B_6C_6$ | $6\,Y_2C + 1.2\,B_5H_7 + 0.5\,N_2 + 14.3\,H_2 \longrightarrow 11\,YH_3 + YB_6H_4C_6N$ | -16.3618 |
| $YNH_4B_6C_6$ | $0.3636\,B_9H_{11} + 3\,B_{13}C_2 + 9.068\,Y + YN \longrightarrow 9.068\,YB_4 + YB_6H_4C_6N$ | -16.3581 |
| $YNH_4B_6C_6$ | $6\,Y_2C + 8.167\,H_3N + 3\,YB_2 \longrightarrow 6.833\,YH_3 + 7.167\,YN + YB_6H_4C_6N$ | -16.3268 |
| $YNH_4B_6C_6$ | $41.25\,B_9H_{11} + 6\,Y_2C + 0.5\,N_2 \longrightarrow 64.25\,B_5H_7 + 11\,YB_4 + YB_6H_4C_6N$ | -16.3044 |
| $YNH_4B_6C_6$ | $6\,Y_2C + H_3N + 17\,H_2 + 6\,B \longrightarrow 11\,YH_3 + YB_6H_4C_6N$ | -16.2917 |
| $YNH_4B_6C_6$ | $3\,B_{13}C_2 + 9.25\,Y + 1.333\,H_3N \longrightarrow 0.1667\,N_2 + 8.25\,YB_4 + YB_6H_4C_6N$ | -16.2624 |
| $YNH_4B_6C_6$ | $6\,C + YB_{12} + 0.5714\,B_5H_7 + 4.929\,N_2 \longrightarrow 8.857\,BN + YB_6H_4C_6N$ | -16.2283 |
| $YNH_4B_6C_6$ | $1.02\,B_{13}C_2 + 3.961\,Y_2C + 8.255\,H_3N \longrightarrow 6.922\,YH_3 + 7.255\,BN + YB_6H_4C_6N$ | -16.2189 |
| $YNH_4B_6C_6$ | $3\,B_{13}C_2 + 9.5\,Y + 2\,H_2 + BN \longrightarrow 8.5\,YB_4 + YB_6H_4C_6N$ | -16.2106 |
| $YNH_4B_6C_6$ | $6\,Y_2C + 0.5\,YB_{12} + YN + 20.75\,H_2 \longrightarrow 12.5\,YH_3 + YB_6H_4C_6N$ | -16.1681 |
| $YNH_4B_6C_6$ | $3\,B_{13}C_2 + 8.964\,Y + 0.5714\,B_5H_7 + YN \longrightarrow 8.964\,YB_4 + YB_6H_4C_6N$ | -16.1090 |
| $YNH_4B_6C_6$ | $6\,Y_2C + 14.93\,H_3N + 2.928\,YB_{12} \longrightarrow 5.826\,B_5H_7 + 13.93\,YN + YB_6H_4C_6N$ | -16.0845 |
| $YNH_4B_6C_6$ | $2\,YH_2 + 3\,B_{13}C_2 + 16\,Y + BN \longrightarrow 17\,YB_2 + YB_6H_4C_6N$ | -16.0207 |
| $YNH_4B_6C_6$ | $0.3333\,YH_3 + 3\,B_{13}C_2 + 8.917\,Y + H_3N \longrightarrow 8.25\,YB_4 + YB_6H_4C_6N$ | -15.8969 |
| $YNH_4B_6C_6$ | $6\,Y_2C + H_3N + 0.5\,YB_{12} + 17.75\,H_2 \longrightarrow 11.5\,YH_3 + YB_6H_4C_6N$ | -15.8864 |
| $YNH_4B_6C_6$ | $1.125\,B_{13}C_2 + 0.25\,Y_4C_5 + 0.5\,H_8C_5 + 4.812\,N_2 \longrightarrow 8.625\,BN + YB_6H_4C_6N$ | -15.7597 |
| $YNH_4B_6C_6$ | $6\,Y_2C + YN + 22.25\,H_2 + 1.5\,YB_4 \longrightarrow 13.5\,YH_3 + YB_6H_4C_6N$ | -15.7075 |
| $YNH_4B_6C_6$ | $0.5\,YH_2 + 3\,B_{13}C_2 + 8.75\,Y + H_3N \longrightarrow 8.25\,YB_4 + YB_6H_4C_6N$ | -15.6482 |
| $YNH_4B_6C_6$ | $2.338\,B_9H_{11} + 6\,Y_2C + 0.5\,N_2 \longrightarrow 7.239\,YH_3 + 3.761\,YB_4 + YB_6H_4C_6N$ | -15.5466 |
| $YNH_4B_6C_6$ | $1.079\,B_9H_{11} + 6\,Y_2C + 4.711\,H_3N \longrightarrow 11\,YH_2 + 3.711\,BN + YB_6H_4C_6N$ | -15.5423 |
| $YNH_4B_6C_6$ | $6\,Y_2C + 3.86\,B_5H_7 + 0.5\,N_2 \longrightarrow 7.674\,YH_3 + 3.326\,YB_4 + YB_6H_4C_6N$ | -15.5011 |
| $YNH_4B_6C_6$ | $6\,Y_2C + H_3N + 19.25\,H_2 + 1.5\,YB_4 \longrightarrow 12.5\,YH_3 + YB_6H_4C_6N$ | -15.4259 |
| $YNH_4B_6C_6$ | $0.6667\,B_9H_{11} + 6\,Y_2C + YN + 16.33\,H_2 \longrightarrow 12\,YH_3 + YB_6H_4C_6N$ | -15.3644 |
| $YNH_4B_6C_6$ | $6\,Y_2C + 0.5\,N_2 + 16\,H_2 + 3\,YB_2 \longrightarrow 14\,YH_2 + YB_6H_4C_6N$ | -15.3371 |
| $YNH_4B_6C_6$ | $6\,Y_2C + 1.2\,B_5H_7 + YN + 15.8\,H_2 \longrightarrow 12\,YH_3 + YB_6H_4C_6N$ | -15.3193 |
| $YNH_4B_6C_6$ | $1.333\,YH_3 + 3\,B_{13}C_2 + 15.17\,Y + YN \longrightarrow 16.5\,YB_2 + YB_6H_4C_6N$ | -15.2512 |
| $YNH_4B_6C_6$ | $44.75\,B_9H_{11} + 6\,Y_2C + YN \longrightarrow 69.75\,B_5H_7 + 12\,YB_4 + YB_6H_4C_6N$ | -15.2399 |
| $YNH_4B_6C_6$ | $6\,Y_2C + 33.38\,B_5H_7 + 0.5\,N_2 \longrightarrow 11\,YH_3 + 17.88\,B_9H_{11} + YB_6H_4C_6N$ | -15.1530 |
| $YNH_4B_6C_6$ | $43.12\,B_9H_{11} + 6\,Y_2C + H_3N \longrightarrow 67.62\,B_5H_7 + 11\,YB_4 + YB_6H_4C_6N$ | -15.1071 |
| $YNH_4B_6C_6$ | $0.6667\,B_9H_{11} + 6\,Y_2C + H_3N + 13.33\,H_2 \longrightarrow 11\,YH_3 + YB_6H_4C_6N$ | -15.0828 |
| $YNH_4B_6C_6$ | $1.333\,YH_3 + 6\,Y_2C + 3.667\,N_2 + 6\,BN \longrightarrow 12.33\,YN + YB_6H_4C_6N$ | -15.0719 |
| $YNH_4B_6C_6$ | $6\,Y_2C + 4.318\,H_3N + 1.864\,B_5H_7 \longrightarrow 11\,YH_2 + 3.318\,BN + YB_6H_4C_6N$ | -15.0638 |
| $YNH_4B_6C_6$ | $6\,Y_2C + 18.5\,H_2 + BN + 5\,B \longrightarrow 11\,YH_3 + YB_6H_4C_6N$ | -15.0490 |
| $YNH_4B_6C_6$ | $0.3636\,B_9H_{11} + 3\,B_{13}C_2 + 0.5\,N_2 + 20.14\,YB_2 \longrightarrow 19.14\,YB_4 + YB_6H_4C_6N$ | -15.0448 |
| $YNH_4B_6C_6$ | $6\,Y_2C + H_3N + 1.2\,B_5H_7 + 12.8\,H_2 \longrightarrow 11\,YH_3 + YB_6H_4C_6N$ | -15.0377 |
| $YNH_4B_6C_6$ | $3\,B_{13}C_2 + 0.5714\,B_5H_7 + 0.5\,N_2 + 19.93\,YB_2 \longrightarrow 18.93\,YB_4 + YB_6H_4C_6N$ | -14.8449 |
| $YNH_4B_6C_6$ | $1.333\,YH_3 + 3\,B_{13}C_2 + 7.917\,Y + 0.5\,N_2 \longrightarrow 8.25\,YB_4 + YB_6H_4C_6N$ | -14.8003 |
| $YNH_4B_6C_6$ | $6\,Y_2C + 0.4167\,YB_{12} + 19.12\,H_2 + BN \longrightarrow 11.42\,YH_3 + YB_6H_4C_6N$ | -14.7113 |
| $YNH_4B_6C_6$ | $6\,Y_2C + 6.667\,H_3N + 6\,B \longrightarrow 5.333\,YH_3 + 5.667\,YN + YB_6H_4C_6N$ | -14.6958 |
| $YNH_4B_6C_6$ | $3\,B_{13}C_2 + 8.25\,Y + YN + 2\,H_2 \longrightarrow 8.25\,YB_4 + YB_6H_4C_6N$ | -14.5648 |
| $YNH_4B_6C_6$ | $2.507\,B_9H_{11} + 6\,Y_2C + YN \longrightarrow 7.859\,YH_3 + 4.141\,YB_4 + YB_6H_4C_6N$ | -14.4172 |
| $YNH_4B_6C_6$ | $6\,Y_2C + 15\,H_3N + 3\,YB_2 \longrightarrow 14\,YN + 20.5\,H_2 + YB_6H_4C_6N$ | -14.4023 |
| $YNH_4B_6C_6$ | $6\,Y_2C + 4.14\,B_5H_7 + YN \longrightarrow 8.326\,YH_3 + 3.674\,YB_4 + YB_6H_4C_6N$ | -14.3684 |
| $YNH_4B_6C_6$ | $6\,Y_2C + 20.38\,H_2 + BN + 1.25\,YB_4 \longrightarrow 12.25\,YH_3 + YB_6H_4C_6N$ | -14.3275 |
| $YNH_4B_6C_6$ | $2.169\,B_9H_{11} + 6\,Y_2C + H_3N \longrightarrow 7.62\,YH_3 + 3.38\,YB_4 + YB_6H_4C_6N$ | -14.3095 |
| $YNH_4B_6C_6$ | $6\,Y_2C + H_3N + 3.581\,B_5H_7 \longrightarrow 8.023\,YH_3 + 2.977\,YB_4 + YB_6H_4C_6N$ | -14.2673 |
| $YNH_4B_6C_6$ | $2\,YH_2 + 3\,B_{13}C_2 + 14.5\,Y + YN \longrightarrow 16.5\,YB_2 + YB_6H_4C_6N$ | -14.2563 |
| $YNH_4B_6C_6$ | $6\,Y_2C + 6.917\,H_3N + 0.5\,YB_{12} \longrightarrow 5.583\,YH_3 + 5.917\,YN + YB_6H_4C_6N$ | -14.2201 |
| $YNH_4B_6C_6$ | $6\,Y_2C + 0.5\,N_2 + 13\,H_2 + 6\,B \longrightarrow 11\,YH_2 + YB_6H_4C_6N$ | -14.2119 |
| $YNH_4B_6C_6$ | $6\,Y_2C + 6.8\,H_3N + 3\,YB_2 \longrightarrow 8.2\,YH_2 + 5.8\,YN + YB_6H_4C_6N$ | -14.1743 |
| $YNH_4B_6C_6$ | $16.73\,YH_3 + 3\,B_{13}C_2 + 8.367\,N_2 \longrightarrow 6.6\,B_5H_7 + 15.73\,YN + YB_6H_4C_6N$ | -14.1731 |
| $YNH_4B_6C_6$ | $7.029\,B_5H_7 + 8.033\,N_2 + 0.8571\,Y_4C_7 \longrightarrow 15.07\,H_3N + 2.429\,YB_{12} + YB_6H_4C_6N$ | -14.1288 |
| $YNH_4B_6C_6$ | $0.5556\,B_9H_{11} + 6\,Y_2C + 15.44\,H_2 + BN \longrightarrow 11\,YH_3 + YB_6H_4C_6N$ | -14.0416 |
| $YNH_4B_6C_6$ | $6\,Y_2C + H_3N + 14.5\,H_2 + 3\,YB_2 \longrightarrow 14\,YH_2 + YB_6H_4C_6N$ | -14.0131 |
| $YNH_4B_6C_6$ | $6\,Y_2C + B_5H_7 + 15\,H_2 + BN \longrightarrow 11\,YH_3 + YB_6H_4C_6N$ | -14.0040 |
| $YNH_4B_6C_6$ | $6\,Y_2C + YN + 17\,H_2 + 3\,YB_2 \longrightarrow 15\,YH_2 + YB_6H_4C_6N$ | -13.9853 |
| $YNH_4B_6C_6$ | $6\,Y_2C + 36.75\,B_5H_7 + YN \longrightarrow 12\,YH_3 + 19.75\,B_9H_{11} + YB_6H_4C_6N$ | -13.9837 |

| Compound | Chemical reaction | Reaction energy(eV) |
|---|---|---|
| $YNH_4B_6C_6$ | $6\,Y_2C + H_3N + 30\,B_5H_7 \longrightarrow 11\,YH_3 + 16\,B_9H_{11} + YB_6H_4C_6N$ | -13.9557 |
| $YNH_4B_6C_6$ | $6\,Y_2C + 4.667\,B_5H_7 + 0.5\,N_2 \longrightarrow 9.556\,YH_3 + 1.444\,YB_{12} + YB_6H_4C_6N$ | -13.9099 |
| $YNH_4B_6C_6$ | $5.556\,B_9H_{11} + 6\,Y_2C + 0.5\,N_2 \longrightarrow 28.56\,H_2 + 11\,YB_4 + YB_6H_4C_6N$ | -13.8906 |
| $YNH_4B_6C_6$ | $40.37\,B_9H_{11} + 6\,Y_2C + BN \longrightarrow 62.87\,B_5H_7 + 11\,YB_4 + YB_6H_4C_6N$ | -13.8875 |
| $YNH_4B_6C_6$ | $2\,YH_2 + 1.2\,Y_4C_5 + 4.9\,N_2 + 3\,YB_2 \longrightarrow 8.8\,YN + YB_6H_4C_6N$ | -13.8543 |
| $YNH_4B_6C_6$ | $3.5\,Y_2C + 0.5\,H_8C_5 + 4.25\,N_2 + 1.5\,YB_4 \longrightarrow 7.5\,YN + YB_6H_4C_6N$ | -13.8531 |
| $YNH_4B_6C_6$ | $6\,Y_2C + 5.267\,H_3N + 1.2\,B_5H_7 \longrightarrow 6.733\,YH_3 + 4.267\,YN + YB_6H_4C_6N$ | -13.8360 |
| $YNH_4B_6C_6$ | $0.6667\,B_9H_{11} + 6\,Y_2C + 5.444\,H_3N \longrightarrow 6.556\,YH_3 + 4.444\,YN + YB_6H_4C_6N$ | -13.8310 |
| $YNH_4B_6C_6$ | $2\,YH_2 + 3\,B_{13}C_2 + 7.25\,Y + 0.5\,N_2 \longrightarrow 8.25\,YB_4 + YB_6H_4C_6N$ | -13.8055 |
| $YNH_4B_6C_6$ | $2.687\,B_{13}C_2 + 8.234\,Y + H_3N + 0.125\,H_8C_5 \longrightarrow 7.234\,YB_4 + YB_6H_4C_6N$ | -13.7658 |
| $YNH_4B_6C_6$ | $2.906\,B_9H_{11} + 6\,Y_2C + 0.5\,N_2 \longrightarrow 9.321\,YH_3 + 1.679\,YB_{12} + YB_6H_4C_6N$ | -13.7078 |
| $YNH_4B_6C_6$ | $6\,Y_2C + 3\,N_2 + 2\,H_2 + 6\,BN \longrightarrow 11\,YN + YB_6H_4C_6N$ | -13.6820 |
| $YNH_4B_6C_6$ | $6\,Y_2C + 0.5\,YB_{12} + 0.5\,N_2 + 13.5\,H_2 \longrightarrow 11.5\,YH_2 + YB_6H_4C_6N$ | -13.6519 |
| $YNH_4B_6C_6$ | $25.25\,B_9H_{11} + 6\,Y_2C + 12\,H_3N \longrightarrow 44.25\,B_5H_7 + 11\,YN + YB_6H_4C_6N$ | -13.6472 |
| $YNH_4B_6C_6$ | $3\,B_{13}C_2 + 0.5\,N_2 + 2\,H_2 + 18.5\,YB_2 \longrightarrow 17.5\,YB_4 + YB_6H_4C_6N$ | -13.6396 |
| $YNH_4B_6C_6$ | $6\,Y_2C + 7.417\,H_3N + 1.5\,YB_4 \longrightarrow 6.083\,YH_3 + 6.417\,YN + YB_6H_4C_6N$ | -13.6187 |
| $YNH_4B_6C_6$ | $6\,Y_2C + 10\,B_5H_7 + 0.5\,N_2 \longrightarrow 33\,H_2 + 11\,YB_4 + YB_6H_4C_6N$ | -13.5149 |
| $YNH_4B_6C_6$ | $0.4615\,B_{13}C_2 + 5.077\,Y_2C + 0.5\,N_2 + 15.73\,H_2 \longrightarrow 9.154\,YH_3 + YB_6H_4C_6N$ | -13.5018 |
| $YNH_4B_6C_6$ | $3.5\,Y_2C + 0.5\,YB_{12} + 0.5\,H_8C_5 + 3.75\,N_2 \longrightarrow 6.5\,YN + YB_6H_4C_6N$ | -13.2712 |
| $YNH_4B_6C_6$ | $1.871\,B_9H_{11} + 6\,Y_2C + 0.5\,N_2 \longrightarrow 8.29\,YH_2 + 2.71\,YB_4 + YB_6H_4C_6N$ | -13.2216 |
| $YNH_4B_6C_6$ | $6\,Y_2C + 12\,H_3N + 6\,B \longrightarrow 11\,YN + 16\,H_2 + YB_6H_4C_6N$ | -13.1937 |
| $YNH_4B_6C_6$ | $3\,B_{13}C_2 + 15.55\,H_3N + YB_{12} \longrightarrow 6.091\,B_5H_7 + 14.55\,BN + YB_6H_4C_6N$ | -13.1810 |
| $YNH_4B_6C_6$ | $3.5\,Y_2C + 0.5\,H_8C_5 + 3.5\,N_2 + 6\,B \longrightarrow 6\,YN + YB_6H_4C_6N$ | -13.1553 |
| $YNH_4B_6C_6$ | $2.296\,B_9H_{11} + 6\,Y_2C + BN \longrightarrow 7.085\,YH_3 + 3.915\,YB_4 + YB_6H_4C_6N$ | -13.1459 |
| $YNH_4B_6C_6$ | $0.3636\,B_9H_{11} + 3\,B_{13}C_2 + BN + 20.64\,YB_2 \longrightarrow 19.64\,YB_4 + YB_6H_4C_6N$ | -13.1089 |
| $YNH_4B_6C_6$ | $6\,Y_2C + 3.791\,B_5H_7 + BN \longrightarrow 7.512\,YH_3 + 3.488\,YB_4 + YB_6H_4C_6N$ | -13.1012 |
| $YNH_4B_6C_6$ | $6\,Y_2C + 3.053\,B_5H_7 + 0.5\,N_2 \longrightarrow 8.684\,YH_2 + 2.316\,YB_4 + YB_6H_4C_6N$ | -13.0752 |
| $YNH_4B_6C_6$ | $1.2\,H_8C_5 + 5.867\,B_5H_7 + 24.33\,YN \longrightarrow 23.33\,YH_2 + 23.33\,BN + YB_6H_4C_6N$ | -13.0738 |
| $YNH_4B_6C_6$ | $6\,Y_2C + 5.6\,H_3N + 6\,B \longrightarrow 6.4\,YH_2 + 4.6\,YN + YB_6H_4C_6N$ | -13.0158 |
| $YNH_4B_6C_6$ | $0.6667\,B_9H_{11} + 6\,Y_2C + 0.5\,N_2 + 9.333\,H_2 \longrightarrow 11\,YH_2 + YB_6H_4C_6N$ | -13.0030 |
| $YNH_4B_6C_6$ | $1.333\,YH_3 + 3\,B_{13}C_2 + 8.167\,Y + BN \longrightarrow 8.5\,YB_4 + YB_6H_4C_6N$ | -12.9830 |
| $YNH_4B_6C_6$ | $6\,Y_2C + 1.2\,B_5H_7 + 0.5\,N_2 + 8.8\,H_2 \longrightarrow 11\,YH_2 + YB_6H_4C_6N$ | -12.9579 |
| $YNH_4B_6C_6$ | $3\,B_{13}C_2 + 0.5714\,B_5H_7 + BN + 20.43\,YB_2 \longrightarrow 19.43\,YB_4 + YB_6H_4C_6N$ | -12.9090 |
| $YNH_4B_6C_6$ | $6\,Y_2C + H_3N + 11.5\,H_2 + 6\,B \longrightarrow 11\,YH_2 + YB_6H_4C_6N$ | -12.8879 |
| $YNH_4B_6C_6$ | $6\,Y_2C + 0.5\,N_2 + 14.5\,H_2 + 1.5\,YB_4 \longrightarrow 12.5\,YH_2 + YB_6H_4C_6N$ | -12.8820 |
| $YNH_4B_6C_6$ | $3\,B_{13}C_2 + 1.333\,H_3N + 19.17\,YB_2 \longrightarrow 0.3333\,YN + 17.83\,YB_4 + YB_6H_4C_6N$ | -12.8703 |
| $YNH_4B_6C_6$ | $6\,Y_2C + YN + 14\,H_2 + 6\,B \longrightarrow 12\,YH_2 + YB_6H_4C_6N$ | -12.8601 |
| $YNH_4B_6C_6$ | $6\,Y_2C + H_3N + 4.303\,B_5H_7 \longrightarrow 9.707\,YH_3 + 1.293\,YB_{12} + YB_6H_4C_6N$ | -12.8430 |
| $YNH_4B_6C_6$ | $0.8723\,B_{13}C_2 + 4.255\,Y_2C + 6.34\,H_3N \longrightarrow 7.511\,YH_2 + 5.34\,BN + YB_6H_4C_6N$ | -12.8284 |
| $YNH_4B_6C_6$ | $6\,Y_2C + 34.75\,B_5H_7 + BN \longrightarrow 11\,YH_3 + 18.75\,B_9H_{11} + YB_6H_4C_6N$ | -12.7360 |
| $YNH_4B_6C_6$ | $0.09091\,B_9H_{11} + 3\,B_{13}C_2 + H_3N + 18.91\,YB_2 \longrightarrow 17.91\,YB_4 + YB_6H_4C_6N$ | -12.6668 |
| $YNH_4B_6C_6$ | $2.679\,B_9H_{11} + 6\,Y_2C + H_3N \longrightarrow 9.491\,YH_3 + 1.509\,YB_{12} + YB_6H_4C_6N$ | -12.6566 |
| $YNH_4B_6C_6$ | $6\,Y_2C + 12.5\,H_3N + 0.5\,YB_{12} \longrightarrow 11.5\,YN + 16.75\,H_2 + YB_6H_4C_6N$ | -12.6476 |
| $YNH_4B_6C_6$ | $6\,B_9H_{11} + 6\,Y_2C + YN \longrightarrow 31\,H_2 + 12\,YB_4 + YB_6H_4C_6N$ | -12.6194 |
| $YNH_4B_6C_6$ | $3\,B_{13}C_2 + H_3N + 0.1429\,B_5H_7 + 18.86\,YB_2 \longrightarrow 17.86\,YB_4 + YB_6H_4C_6N$ | -12.6168 |
| $YNH_4B_6C_6$ | $6\,Y_2C + 5.03\,B_5H_7 + YN \longrightarrow 10.4\,YH_3 + 1.596\,YB_{12} + YB_6H_4C_6N$ | -12.6102 |
| $YNH_4B_6C_6$ | $6\,Y_2C + 15.5\,H_2 + BN + 2.5\,YB_2 \longrightarrow 13.5\,YH_2 + YB_6H_4C_6N$ | -12.5828 |
| $YNH_4B_6C_6$ | $5.556\,B_9H_{11} + 6\,Y_2C + H_3N \longrightarrow 30.06\,H_2 + 11\,YB_4 + YB_6H_4C_6N$ | -12.5665 |
| $YNH_4B_6C_6$ | $1.2\,Y_4C_5 + 2.2\,B_5H_7 + 3\,N_2 \longrightarrow 3.8\,YH_3 + 5\,BN + YB_6H_4C_6N$ | -12.5433 |
| $YNH_4B_6C_6$ | $1.333\,YH_3 + 1.2\,Y_4C_5 + 4.567\,N_2 + 3\,YB_2 \longrightarrow 8.133\,YN + YB_6H_4C_6N$ | -12.5405 |
| $YNH_4B_6C_6$ | $3\,B_{13}C_2 + 1.333\,H_3N + 18.33\,YB_2 \longrightarrow 0.3333\,BN + 17.33\,YB_4 + YB_6H_4C_6N$ | -12.5194 |
| $YNH_4B_6C_6$ | $6\,Y_2C + 5.8\,H_3N + 0.5\,YB_{12} \longrightarrow 6.7\,YH_2 + 4.8\,YN + YB_6H_4C_6N$ | -12.4613 |
| $YNH_4B_6C_6$ | $0.4615\,B_{13}C_2 + 5.077\,Y_2C + YN + 17.23\,H_2 \longrightarrow 10.15\,YH_3 + YB_6H_4C_6N$ | -12.4594 |
| $YNH_4B_6C_6$ | $1.2\,H_8C_5 + 5.1\,N_2 + 3.8\,YB_4 \longrightarrow 2.8\,YH_2 + 9.2\,BN + YB_6H_4C_6N$ | -12.4540 |
| $YNH_4B_6C_6$ | $3.132\,B_9H_{11} + 6\,Y_2C + YN \longrightarrow 10.15\,YH_3 + 1.849\,YB_{12} + YB_6H_4C_6N$ | -12.3924 |
| $YNH_4B_6C_6$ | $6\,Y_2C + H_3N + 0.5\,YB_{12} + 12\,H_2 \longrightarrow 11.5\,YH_2 + YB_6H_4C_6N$ | -12.3279 |
| $YNH_4B_6C_6$ | $3\,B_{13}C_2 + H_3N + 0.5\,H_2 + 18.5\,YB_2 \longrightarrow 17.5\,YB_4 + YB_6H_4C_6N$ | -12.3155 |
| $YNH_4B_6C_6$ | $6\,Y_2C + 0.5\,YB_{12} + YN + 14.5\,H_2 \longrightarrow 12.5\,YH_2 + YB_6H_4C_6N$ | -12.3001 |
| $YNH_4B_6C_6$ | $6\,Y_2C + 21\,B_5H_7 + 0.5\,N_2 \longrightarrow 11\,B_9H_{11} + 11\,YH_2 + YB_6H_4C_6N$ | -12.2141 |
| $YNH_4B_6C_6$ | $6\,Y_2C + 10.8\,B_5H_7 + YN \longrightarrow 35.8\,H_2 + 12\,YB_4 + YB_6H_4C_6N$ | -12.2137 |

| Compound | Chemical reaction | Reaction energy(eV) |
|---|---|---|
| $YNH_4B_6C_6$ | $6\,Y_2C + H_3N + 10\,B_5H_7 \longrightarrow 34.5\,H_2 + 11\,YB_4 + YB_6H_4C_6N$ | -12.1908 |
| $YNH_4B_6C_6$ | $0.4615\,B_{13}C_2 + 5.077\,Y_2C + H_3N + 14.23\,H_2 \longrightarrow 9.154\,YH_3 + YB_6H_4C_6N$ | -12.1777 |
| $YNH_4B_6C_6$ | $6\,Y_2C + 5.286\,B_5H_7 + 0.5\,N_2 \longrightarrow 11\,YH_3 + 20.43\,B + YB_6H_4C_6N$ | -12.0921 |
| $YNH_4B_6C_6$ | $0.3636\,B_9H_{11} + 3\,B_{13}C_2 + YN + 18.14\,YB_2 \longrightarrow 18.14\,YB_4 + YB_6H_4C_6N$ | -12.0561 |
| $YNH_4B_6C_6$ | $2\,YH_2 + 3\,B_{13}C_2 + 7.5\,Y + BN \longrightarrow 8.5\,YB_4 + YB_6H_4C_6N$ | -11.9882 |
| $YNH_4B_6C_6$ | $0.6667\,B_9H_{11} + 6\,Y_2C + 12\,H_3N \longrightarrow 11\,YN + 19.67\,H_2 + YB_6H_4C_6N$ | -11.9847 |
| $YNH_4B_6C_6$ | $6\,Y_2C + 12\,H_3N + 1.2\,B_5H_7 \longrightarrow 11\,YN + 20.2\,H_2 + YB_6H_4C_6N$ | -11.9396 |
| $YNH_4B_6C_6$ | $6\,Y_2C + 1.333\,H_3N + 2.333\,N_2 + 6\,BN \longrightarrow 11\,YN + YB_6H_4C_6N$ | -11.9165 |
| $YNH_4B_6C_6$ | $6\,Y_2C + 13.5\,H_3N + 1.5\,YB_4 \longrightarrow 12.5\,YN + 18.25\,H_2 + YB_6H_4C_6N$ | -11.9054 |
| $YNH_4B_6C_6$ | $2\,B_9H_{11} + 6\,Y_2C + YN \longrightarrow 9\,YH_2 + 3\,YB_4 + YB_6H_4C_6N$ | -11.8932 |
| $YNH_4B_6C_6$ | $14.78\,YH_3 + 3\,B_{13}C_2 + 7.389\,N_2 \longrightarrow 3.667\,B_9H_{11} + 13.78\,YN + YB_6H_4C_6N$ | -11.8866 |
| $YNH_4B_6C_6$ | $3\,B_{13}C_2 + 1.333\,H_3N + 18.5\,YB_2 \longrightarrow 0.1667\,N_2 + 17.5\,YB_4 + YB_6H_4C_6N$ | -11.8741 |
| $YNH_4B_6C_6$ | $1.677\,B_9H_{11} + 6\,Y_2C + H_3N \longrightarrow 8.726\,YH_2 + 2.274\,YB_4 + YB_6H_4C_6N$ | -11.8624 |
| $YNH_4B_6C_6$ | $3\,B_{13}C_2 + 0.5714\,B_5H_7 + YN + 17.93\,YB_2 \longrightarrow 17.93\,YB_4 + YB_6H_4C_6N$ | -11.8562 |
| $YNH_4B_6C_6$ | $0.6667\,B_9H_{11} + 6\,Y_2C + 4.133\,H_3N \longrightarrow 7.867\,YH_2 + 3.133\,YN + YB_6H_4C_6N$ | -11.7660 |
| $YNH_4B_6C_6$ | $3.5\,Y_2C + 0.5\,H_8C_5 + 0.5\,N_2 + 78\,B \longrightarrow 6\,YB_{12} + YB_6H_4C_6N$ | -11.7643 |
| $YNH_4B_6C_6$ | $6\,Y_2C + 3.263\,B_5H_7 + YN \longrightarrow 9.421\,YH_2 + 2.579\,YB_4 + YB_6H_4C_6N$ | -11.7366 |
| $YNH_4B_6C_6$ | $6\,Y_2C + H_3N + 2.737\,B_5H_7 \longrightarrow 9.079\,YH_2 + 1.921\,YB_4 + YB_6H_4C_6N$ | -11.7311 |
| $YNH_4B_6C_6$ | $6\,Y_2C + 3.92\,H_3N + 1.2\,B_5H_7 \longrightarrow 8.08\,YH_2 + 2.92\,YN + YB_6H_4C_6N$ | -11.7150 |
| $YNH_4B_6C_6$ | $3\,B_{13}C_2 + 2\,H_2 + BN + 19\,YB_2 \longrightarrow 18\,YB_4 + YB_6H_4C_6N$ | -11.7037 |
| $YNH_4B_6C_6$ | $6\,Y_2C + 6.2\,H_3N + 1.5\,YB_4 \longrightarrow 7.3\,YH_2 + 5.2\,YN + YB_6H_4C_6N$ | -11.7024 |
| $YNH_4B_6C_6$ | $0.6667\,B_9H_{11} + 6\,Y_2C + H_3N + 7.833\,H_2 \longrightarrow 11\,YH_2 + YB_6H_4C_6N$ | -11.6789 |
| $YNH_4B_6C_6$ | $0.3333\,YH_3 + 3\,B_{13}C_2 + H_3N + 17.83\,YB_2 \longrightarrow 17.17\,YB_4 + YB_6H_4C_6N$ | -11.6668 |
| $YNH_4B_6C_6$ | $6\,Y_2C + 3.447\,B_5H_7 + 0.5\,N_2 \longrightarrow 10.06\,YH_2 + 0.9362\,YB_{12} + YB_6H_4C_6N$ | -11.6585 |
| $YNH_4B_6C_6$ | $0.6667\,B_9H_{11} + 6\,Y_2C + YN + 10.33\,H_2 \longrightarrow 12\,YH_2 + YB_6H_4C_6N$ | -11.6511 |
| $YNH_4B_6C_6$ | $6\,Y_2C + 13\,H_2 + BN + 5\,B \longrightarrow 11\,YH_2 + YB_6H_4C_6N$ | -11.6452 |
| $YNH_4B_6C_6$ | $1.75\,Y_4C_5 + 1.2\,B_5H_7 + 3.5\,N_2 \longrightarrow 0.55\,H_8C_5 + 6\,YN + YB_6H_4C_6N$ | -11.6424 |
| $YNH_4B_6C_6$ | $3.905\,B_9H_{11} + 6.992\,N_2 + 0.8571\,Y_4C_7 \longrightarrow 12.98\,H_3N + 2.429\,YB_{12} + YB_6H_4C_6N$ | -11.6354 |
| $YNH_4B_6C_6$ | $6\,Y_2C + H_3N + 1.2\,B_5H_7 + 7.3\,H_2 \longrightarrow 11\,YH_2 + YB_6H_4C_6N$ | -11.6338 |
| $YNH_4B_6C_6$ | $0.3846\,B_{13}C_2 + 5.231\,Y_2C + 16.19\,H_2 + BN \longrightarrow 9.462\,YH_3 + YB_6H_4C_6N$ | -11.6207 |
| $YNH_4B_6C_6$ | $3\,B_{13}C_2 + 14.26\,H_3N + YB_{12} \longrightarrow 3.526\,B_9H_{11} + 13.26\,BN + YB_6H_4C_6N$ | -11.6170 |
| $YNH_4B_6C_6$ | $6\,Y_2C + 1.2\,B_5H_7 + YN + 9.8\,H_2 \longrightarrow 12\,YH_2 + YB_6H_4C_6N$ | -11.6060 |
| $YNH_4B_6C_6$ | $6\,Y_2C + 27.54\,H_3N + 15.54\,YB_4 \longrightarrow 11.23\,B_5H_7 + 26.54\,YN + YB_6H_4C_6N$ | -11.5848 |
| $YNH_4B_6C_6$ | $6\,Y_2C + H_3N + 13\,H_2 + 1.5\,YB_4 \longrightarrow 12.5\,YH_2 + YB_6H_4C_6N$ | -11.5579 |
| $YNH_4B_6C_6$ | $2.16\,B_9H_{11} + 6\,Y_2C + 0.5\,N_2 \longrightarrow 9.88\,YH_2 + 1.12\,YB_{12} + YB_6H_4C_6N$ | -11.5494 |
| $YNH_4B_6C_6$ | $3.5\,Y_2C + 0.5\,H_8C_5 + YN + 90\,B \longrightarrow 7\,YB_{12} + YB_6H_4C_6N$ | -11.5325 |
| $YNH_4B_6C_6$ | $6\,Y_2C + YN + 15.5\,H_2 + 1.5\,YB_4 \longrightarrow 13.5\,YH_2 + YB_6H_4C_6N$ | -11.5301 |
| $YNH_4B_6C_6$ | $5.444\,B_9H_{11} + 6\,Y_2C + BN \longrightarrow 27.94\,H_2 + 11\,YB_4 + YB_6H_4C_6N$ | -11.5253 |
| $YNH_4B_6C_6$ | $3.364\,B_9H_{11} + 6\,Y_2C + 0.5\,N_2 \longrightarrow 11\,YH_3 + 24.27\,B + YB_6H_4C_6N$ | -11.5161 |
| $YNH_4B_6C_6$ | $0.5\,YH_2 + 3\,B_{13}C_2 + H_3N + 17.5\,YB_2 \longrightarrow 17\,YB_4 + YB_6H_4C_6N$ | -11.4971 |
| $YNH_4B_6C_6$ | $6\,Y_2C + 4.636\,B_5H_7 + BN \longrightarrow 9.485\,YH_3 + 1.515\,YB_{12} + YB_6H_4C_6N$ | -11.4321 |
| $YNH_4B_6C_6$ | $1.333\,YH_3 + 3\,B_{13}C_2 + 6.917\,Y + YN \longrightarrow 8.25\,YB_4 + YB_6H_4C_6N$ | -11.3373 |
| $YNH_4B_6C_6$ | $2.887\,B_9H_{11} + 6\,Y_2C + BN \longrightarrow 9.252\,YH_3 + 1.748\,YB_{12} + YB_6H_4C_6N$ | -11.2313 |
| $YNH_4B_6C_6$ | $6\,Y_2C + H_3N + 4.857\,B_5H_7 \longrightarrow 11\,YH_3 + 18.29\,B + YB_6H_4C_6N$ | -11.2159 |
| $YNH_4B_6C_6$ | $6\,Y_2C + 0.4167\,YB_{12} + 13.42\,H_2 + BN \longrightarrow 11.42\,YH_2 + YB_6H_4C_6N$ | -11.1785 |
| $YNH_4B_6C_6$ | $6\,Y_2C + 9.8\,B_5H_7 + BN \longrightarrow 32.3\,H_2 + 11\,YB_4 + YB_6H_4C_6N$ | -11.1572 |
| $YNH_4B_6C_6$ | $1.2\,Y_4C_5 + 3.9\,N_2 + 2\,H_2 + 3\,YB_2 \longrightarrow 6.8\,YN + YB_6H_4C_6N$ | -11.1506 |
| $YNH_4B_6C_6$ | $1.2\,Y_4C_5 + 4.24\,B_5H_7 + 4.78\,N_2 \longrightarrow 8.56\,H_3N + 3.8\,YB_4 + YB_6H_4C_6N$ | -11.1051 |
| $YNH_4B_6C_6$ | $8.3\,H_3N + 1.2\,Y_4C_5 + 6.65\,YB_2 \longrightarrow 10.45\,YH_2 + 7.3\,BN + YB_6H_4C_6N$ | -11.0744 |
| $YNH_4B_6C_6$ | $1.333\,YH_3 + 3\,B_{13}C_2 + 0.5\,N_2 + 15.83\,YB_2 \longrightarrow 16.17\,YB_4 + YB_6H_4C_6N$ | -11.0446 |
| $YNH_4B_6C_6$ | $6\,Y_2C + H_3N + 17.62\,B_5H_7 \longrightarrow 9.125\,B_9H_{11} + 11\,YH_2 + YB_6H_4C_6N$ | -11.0168 |
| $YNH_4B_6C_6$ | $6\,Y_2C + 3.172\,B_5H_7 + 0.5\,N_2 \longrightarrow 6.069\,YH_3 + 4.931\,YB_2 + YB_6H_4C_6N$ | -10.9252 |
| $YNH_4B_6C_6$ | $6\,Y_2C + 32.82\,H_3N + 20.82\,YB_4 \longrightarrow 8.588\,B_9H_{11} + 31.82\,YN + YB_6H_4C_6N$ | -10.8833 |
| $YNH_4B_6C_6$ | $1.839\,B_9H_{11} + 6\,Y_2C + BN \longrightarrow 8.113\,YH_2 + 2.887\,YB_4 + YB_6H_4C_6N$ | -10.8707 |
| $YNH_4B_6C_6$ | $0.4615\,B_{13}C_2 + 5.077\,Y_2C + 5.744\,H_3N \longrightarrow 4.41\,YH_3 + 4.744\,YN + YB_6H_4C_6N$ | -10.8418 |
| $YNH_4B_6C_6$ | $6\,Y_2C + 23.25\,B_5H_7 + YN \longrightarrow 12.25\,B_9H_{11} + 12\,YH_2 + YB_6H_4C_6N$ | -10.7777 |
| $YNH_4B_6C_6$ | $6\,Y_2C + 3\,B_5H_7 + BN \longrightarrow 8.5\,YH_2 + 2.5\,YB_4 + YB_6H_4C_6N$ | -10.7267 |
| $YNH_4B_6C_6$ | $3.091\,B_9H_{11} + 6\,Y_2C + H_3N \longrightarrow 11\,YH_3 + 21.82\,B + YB_6H_4C_6N$ | -10.6866 |
| $YNH_4B_6C_6$ | $0.4615\,B_{13}C_2 + 5.077\,Y_2C + 0.5\,N_2 + 11.15\,H_2 \longrightarrow 9.154\,YH_2 + YB_6H_4C_6N$ | -10.6693 |
| $YNH_4B_6C_6$ | $3\,B_{13}C_2 + YN + 2\,H_2 + 16.5\,YB_2 \longrightarrow 16.5\,YB_4 + YB_6H_4C_6N$ | -10.6509 |

| Compound | Chemical reaction | Reaction energy(eV) |
|---|---|---|
| $YNH_4B_6C_6$ | $0.5556\,B_9H_{11} + 6\,Y_2C + 9.944\,H_2 + BN \longrightarrow 11\,YH_2 + YB_6H_4C_6N$ | -10.6377 |
| $YNH_4B_6C_6$ | $6\,Y_2C + 5.714\,B_5H_7 + YN \longrightarrow 12\,YH_3 + 22.57\,B + YB_6H_4C_6N$ | -10.6018 |
| $YNH_4B_6C_6$ | $6\,Y_2C + B_5H_7 + 9.5\,H_2 + BN \longrightarrow 11\,YH_2 + YB_6H_4C_6N$ | -10.6002 |
| $YNH_4B_6C_6$ | $6\,Y_2C + H_3N + 3.064\,B_5H_7 \longrightarrow 10.22\,YH_2 + 0.7766\,YB_{12} + YB_6H_4C_6N$ | -10.5559 |
| $YNH_4B_6C_6$ | $6\,Y_2C + 14.25\,H_2 + BN + 1.25\,YB_4 \longrightarrow 12.25\,YH_2 + YB_6H_4C_6N$ | -10.5369 |
| $YNH_4B_6C_6$ | $1.878\,B_9H_{11} + 6\,Y_2C + 0.5\,N_2 \longrightarrow 5.551\,YH_3 + 5.449\,YB_2 + YB_6H_4C_6N$ | -10.4812 |
| $YNH_4B_6C_6$ | $1.92\,B_9H_{11} + 6\,Y_2C + H_3N \longrightarrow 10.06\,YH_2 + 0.94\,YB_{12} + YB_6H_4C_6N$ | -10.4589 |
| $YNH_4B_6C_6$ | $2\,YH_2 + 3\,B_{13}C_2 + 0.5\,N_2 + 14.5\,YB_2 \longrightarrow 15.5\,YB_4 + YB_6H_4C_6N$ | -10.3660 |
| $YNH_4B_6C_6$ | $3.5\,Y_2C + 0.5\,H_8C_5 + 0.5\,N_2 + 30\,B \longrightarrow 6\,YB_4 + YB_6H_4C_6N$ | -10.3640 |
| $YNH_4B_6C_6$ | $2\,YH_2 + 3\,B_{13}C_2 + 6.25\,Y + YN \longrightarrow 8.25\,YB_4 + YB_6H_4C_6N$ | -10.3424 |
| $YNH_4B_6C_6$ | $6\,Y_2C + 3.714\,B_5H_7 + 0.5\,N_2 \longrightarrow 11\,YH_2 + 12.57\,B + YB_6H_4C_6N$ | -10.3304 |
| $YNH_4B_6C_6$ | $6\,Y_2C + H_3N + 2.966\,B_5H_7 \longrightarrow 6.586\,YH_3 + 4.414\,YB_2 + YB_6H_4C_6N$ | -10.1714 |
| $YNH_4B_6C_6$ | $6\,Y_2C + 3.702\,B_5H_7 + YN \longrightarrow 10.96\,YH_2 + 1.043\,YB_{12} + YB_6H_4C_6N$ | -10.1589 |
| $YNH_4B_6C_6$ | $6\,Y_2C + 0.5\,N_2 + 2\,H_2 + 28\,B \longrightarrow 11\,YB_2 + YB_6H_4C_6N$ | -10.0863 |
| $YNH_4B_6C_6$ | $6\,Y_2C + 2.667\,B_5H_7 + 0.5\,N_2 \longrightarrow 7.333\,YH_2 + 3.667\,YB_2 + YB_6H_4C_6N$ | -10.0500 |
| $YNH_4B_6C_6$ | $2.32\,B_9H_{11} + 6\,Y_2C + YN \longrightarrow 10.76\,YH_2 + 1.24\,YB_{12} + YB_6H_4C_6N$ | -10.0418 |
| $YNH_4B_6C_6$ | $0.6667\,B_9H_{11} + 1.617\,Y_4C_5 + 3.233\,N_2 \longrightarrow 0.4167\,H_8C_5 + 5.467\,YN + YB_6H_4C_6N$ | -10.0083 |
| $YNH_4B_6C_6$ | $3.636\,B_9H_{11} + 6\,Y_2C + YN \longrightarrow 12\,YH_3 + 26.73\,B + YB_6H_4C_6N$ | -9.9791 |
| $YNH_4B_6C_6$ | $2\,YH_2 + 4.214\,N_2 + 0.8571\,Y_4C_7 + 3\,YB_2 \longrightarrow 7.429\,YN + YB_6H_4C_6N$ | -9.9721 |
| $YNH_4B_6C_6$ | $3\,B_{13}C_2 + Y + 11.73\,H_3N \longrightarrow 4.455\,B_5H_7 + 10.73\,BN + YB_6H_4C_6N$ | -9.9573 |
| $YNH_4B_6C_6$ | $2.364\,B_9H_{11} + 6\,Y_2C + 0.5\,N_2 \longrightarrow 11\,YH_2 + 15.27\,B + YB_6H_4C_6N$ | -9.9257 |
| $YNH_4B_6C_6$ | $3.5\,Y_2C + 0.5\,H_8C_5 + YN + 34\,B \longrightarrow 7\,YB_4 + YB_6H_4C_6N$ | -9.8988 |
| $YNH_4B_6C_6$ | $2.688\,B_{13}C_2 + H_3N + 0.125\,H_8C_5 + 16.47\,YB_2 \longrightarrow 15.47\,YB_4 + YB_6H_4C_6N$ | -9.8594 |
| $YNH_4B_6C_6$ | $6\,Y_2C + 22.37\,B_5H_7 + BN \longrightarrow 11.87\,B_9H_{11} + 11\,YH_2 + YB_6H_4C_6N$ | -9.7971 |
| $YNH_4B_6C_6$ | $1.755\,B_9H_{11} + 6\,Y_2C + H_3N \longrightarrow 6.102\,YH_3 + 4.898\,YB_2 + YB_6H_4C_6N$ | -9.7563 |
| $YNH_4B_6C_6$ | $1.6\,B_9H_{11} + 6\,Y_2C + 0.5\,N_2 \longrightarrow 6.8\,YH_2 + 4.2\,YB_2 + YB_6H_4C_6N$ | -9.7352 |
| $YNH_4B_6C_6$ | $9.5\,Y + H_3N + 1.2\,H_8C_5 + 1.2\,B_5H_7 \longrightarrow 8.5\,YH_2 + YB_6H_4C_6N$ | -9.6873 |
| $YNH_4B_6C_6$ | $2.356\,B_9H_{11} + 1.2\,Y_4C_5 + 4.152\,N_2 \longrightarrow 7.304\,H_3N + 3.8\,YB_4 + YB_6H_4C_6N$ | -9.6010 |
| $YNH_4B_6C_6$ | $0.4615\,B_{13}C_2 + 5.077\,Y_2C + 10.15\,H_3N \longrightarrow 9.154\,YN + 13.23\,H_2 + YB_6H_4C_6N$ | -9.5997 |
| $YNH_4B_6C_6$ | $6\,Y_2C + 5.286\,B_5H_7 + BN \longrightarrow 11\,YH_3 + 21.43\,B + YB_6H_4C_6N$ | -9.5253 |
| $YNH_4B_6C_6$ | $6\,C + YB_{12} + 3.5\,N_2 + 2\,H_2 \longrightarrow 6\,BN + YB_6H_4C_6N$ | -9.4919 |
| $YNH_4B_6C_6$ | $6\,Y_2C + 0.5714\,B_5H_7 + 0.5\,N_2 + 25.14\,B \longrightarrow 11\,YB_2 + YB_6H_4C_6N$ | -9.4891 |
| $YNH_4B_6C_6$ | $3.5\,Y_2C + 3.75\,YB_{12} + 0.5\,H_8C_5 + 0.5\,N_2 \longrightarrow 9.75\,YB_4 + YB_6H_4C_6N$ | -9.4889 |
| $YNH_4B_6C_6$ | $6\,Y_2C + H_3N + 3.286\,B_5H_7 \longrightarrow 11\,YH_2 + 10.43\,B + YB_6H_4C_6N$ | -9.4542 |
| $YNH_4B_6C_6$ | $0.4615\,B_{13}C_2 + 5.077\,Y_2C + 4.862\,H_3N \longrightarrow 5.292\,YH_2 + 3.862\,YN + YB_6H_4C_6N$ | -9.4525 |
| $YNH_4B_6C_6$ | $0.3636\,B_9H_{11} + 6\,Y_2C + 0.5\,N_2 + 24.73\,B \longrightarrow 11\,YB_2 + YB_6H_4C_6N$ | -9.4269 |
| $YNH_4B_6C_6$ | $1.333\,H_3N + 1.2\,Y_4C_5 + 3.233\,N_2 + 3\,YB_2 \longrightarrow 6.8\,YN + YB_6H_4C_6N$ | -9.3852 |
| $YNH_4B_6C_6$ | $2\,YH_2 + 1.2\,Y_4C_5 + 4.15\,N_2 + 1.5\,YB_4 \longrightarrow 7.3\,YN + YB_6H_4C_6N$ | -9.3714 |
| $YNH_4B_6C_6$ | $0.4615\,B_{13}C_2 + 5.077\,Y_2C + H_3N + 9.654\,H_2 \longrightarrow 9.154\,YH_2 + YB_6H_4C_6N$ | -9.3452 |
| $YNH_4B_6C_6$ | $2\,YH_2 + 6\,Y_2C + 0.5\,N_2 + 32\,B \longrightarrow 13\,YB_2 + YB_6H_4C_6N$ | -9.3362 |
| $YNH_4B_6C_6$ | $0.4615\,B_{13}C_2 + 5.077\,Y_2C + YN + 12.15\,H_2 \longrightarrow 10.15\,YH_2 + YB_6H_4C_6N$ | -9.3174 |
| $YNH_4B_6C_6$ | $6\,Y_2C + 3.379\,B_5H_7 + YN \longrightarrow 6.552\,YH_3 + 5.448\,YB_2 + YB_6H_4C_6N$ | -9.3125 |
| $YNH_4B_6C_6$ | $3\,B_{13}C_2 + 12.36\,H_3N + YB_2 \longrightarrow 4.727\,B_5H_7 + 11.36\,BN + YB_6H_4C_6N$ | -9.2970 |
| $YNH_4B_6C_6$ | $6\,Y_2C + H_3N + 2.417\,B_5H_7 \longrightarrow 7.958\,YH_2 + 3.042\,YB_2 + YB_6H_4C_6N$ | -9.2216 |
| $YNH_4B_6C_6$ | $3.5\,Y_2C + 0.5\,H_8C_5 + BN + 77\,B \longrightarrow 6\,YB_{12} + YB_6H_4C_6N$ | -9.1976 |
| $YNH_4B_6C_6$ | $6\,Y_2C + 3.426\,B_5H_7 + BN \longrightarrow 9.989\,YH_2 + 1.011\,YB_{12} + YB_6H_4C_6N$ | -9.1974 |
| $YNH_4B_6C_6$ | $6\,Y_2C + 1.333\,H_3N + 28.33\,B \longrightarrow 0.3333\,BN + 11\,YB_2 + YB_6H_4C_6N$ | -9.1764 |
| $YNH_4B_6C_6$ | $1.333\,YH_3 + 6\,Y_2C + 0.5\,N_2 + 30.67\,B \longrightarrow 12.33\,YB_2 + YB_6H_4C_6N$ | -9.1736 |
| $YNH_4B_6C_6$ | $3\,B_{13}C_2 + 13\,H_3N + YB_4 \longrightarrow 5\,B_5H_7 + 12\,BN + YB_6H_4C_6N$ | -9.1111 |
| $YNH_4B_6C_6$ | $1.333\,YH_3 + 3\,B_{13}C_2 + BN + 16.33\,YB_2 \longrightarrow 16.67\,YB_4 + YB_6H_4C_6N$ | -9.1087 |
| $YNH_4B_6C_6$ | $2.091\,B_9H_{11} + 6\,Y_2C + H_3N \longrightarrow 11\,YH_2 + 12.82\,B + YB_6H_4C_6N$ | -9.0962 |
| $YNH_4B_6C_6$ | $2.147\,B_9H_{11} + 6\,Y_2C + BN \longrightarrow 9.807\,YH_2 + 1.193\,YB_{12} + YB_6H_4C_6N$ | -9.0889 |
| $YNH_4B_6C_6$ | $YB_{12} + 1.2\,H_8C_5 + 4.433\,N_2 \longrightarrow 1.867\,H_3N + 6\,BN + YB_6H_4C_6N$ | -8.9612 |
| $YNH_4B_6C_6$ | $3.364\,B_9H_{11} + 6\,Y_2C + BN \longrightarrow 11\,YH_3 + 25.27\,B + YB_6H_4C_6N$ | -8.9493 |
| $YNH_4B_6C_6$ | $1.45\,B_9H_{11} + 6\,Y_2C + H_3N \longrightarrow 7.475\,YH_2 + 3.525\,YB_2 + YB_6H_4C_6N$ | -8.9363 |
| $YNH_4B_6C_6$ | $3.5\,Y_2C + 4.25\,YB_{12} + 0.5\,H_8C_5 + YN \longrightarrow 11.25\,YB_4 + YB_6H_4C_6N$ | -8.9070 |
| $YNH_4B_6C_6$ | $6\,Y_2C + 1.333\,H_3N + 27.33\,B \longrightarrow 0.3333\,YN + 10.67\,YB_2 + YB_6H_4C_6N$ | -8.8965 |
| $YNH_4B_6C_6$ | $2\,B_9H_{11} + 6\,Y_2C + YN \longrightarrow 6\,YH_3 + 6\,YB_2 + YB_6H_4C_6N$ | -8.8395 |
| $YNH_4B_6C_6$ | $6\,Y_2C + 26\,H_2 + 6\,BN \longrightarrow 11\,YH_3 + 5\,H_3N + YB_6H_4C_6N$ | -8.8356 |
| $YNH_4B_6C_6$ | $3\,B_{13}C_2 + Y + 10.79\,H_3N \longrightarrow 2.579\,B_9H_{11} + 9.789\,BN + YB_6H_4C_6N$ | -8.8135 |

| Compound | Chemical reaction | Reaction energy(eV) |
|---|---|---|
| $YNH_4B_6C_6$ | $2\,YH_2 + 1.2\,Y_4C_5 + 0.5\,YB_{12} + 3.65\,N_2 \longrightarrow 6.3\,YN + YB_6H_4C_6N$ | -8.7895 |
| $YNH_4B_6C_6$ | $6\,Y_2C + H_3N + 0.5\,H_2 + 28\,B \longrightarrow 11\,YB_2 + YB_6H_4C_6N$ | -8.7622 |
| $YNH_4B_6C_6$ | $0.3846\,B_{13}C_2 + 5.231\,Y_2C + 11.46\,H_2 + BN \longrightarrow 9.462\,YH_2 + YB_6H_4C_6N$ | -8.6929 |
| $YNH_4B_6C_6$ | $6\,Y_2C + 4\,B_5H_7 + YN \longrightarrow 12\,YH_2 + 14\,B + YB_6H_4C_6N$ | -8.6800 |
| $YNH_4B_6C_6$ | $2\,YH_2 + 1.2\,Y_4C_5 + 3.4\,N_2 + 6\,B \longrightarrow 5.8\,YN + YB_6H_4C_6N$ | -8.6735 |
| $YNH_4B_6C_6$ | $1.333\,YH_3 + 3.881\,N_2 + 0.8571\,Y_4C_7 + 3\,YB_2 \longrightarrow 6.762\,YN + YB_6H_4C_6N$ | -8.6583 |
| $YNH_4B_6C_6$ | $6\,Y_2C + H_3N + 0.1429\,B_5H_7 + 27.29\,B \longrightarrow 11\,YB_2 + YB_6H_4C_6N$ | -8.6129 |
| $YNH_4B_6C_6$ | $0.6667\,B_9H_{11} + 8.967\,Y + H_3N + 1.2\,H_8C_5 \longrightarrow 7.967\,YH_2 + YB_6H_4C_6N$ | -8.6064 |
| $YNH_4B_6C_6$ | $0.09091\,B_9H_{11} + 6\,Y_2C + H_3N + 27.18\,B \longrightarrow 11\,YB_2 + YB_6H_4C_6N$ | -8.5974 |
| $YNH_4B_6C_6$ | $0.5\,YH_2 + 6\,Y_2C + H_3N + 29\,B \longrightarrow 11.5\,YB_2 + YB_6H_4C_6N$ | -8.5747 |
| $YNH_4B_6C_6$ | $0.3333\,YH_3 + 6\,Y_2C + H_3N + 28.67\,B \longrightarrow 11.33\,YB_2 + YB_6H_4C_6N$ | -8.5341 |
| $YNH_4B_6C_6$ | $3\,B_{13}C_2 + 11.41\,H_3N + YN \longrightarrow 4.318\,B_5H_7 + 11.41\,BN + YB_6H_4C_6N$ | -8.5231 |
| $YNH_4B_6C_6$ | $2\,YH_2 + 3\,B_{13}C_2 + BN + 15\,YB_2 \longrightarrow 16\,YB_4 + YB_6H_4C_6N$ | -8.4301 |
| $YNH_4B_6C_6$ | $6\,Y_2C + 2.833\,B_5H_7 + YN \longrightarrow 7.917\,YH_2 + 4.083\,YB_2 + YB_6H_4C_6N$ | -8.3677 |
| $YNH_4B_6C_6$ | $6\,Y_2C + YN + 2\,H_2 + 30\,B \longrightarrow 12\,YB_2 + YB_6H_4C_6N$ | -8.3594 |
| $YNH_4B_6C_6$ | $6\,Y_2C + 1.333\,H_3N + 28\,B \longrightarrow 0.1667\,N_2 + 11\,YB_2 + YB_6H_4C_6N$ | -8.3209 |
| $YNH_4B_6C_6$ | $6\,Y_2C + 3.069\,B_5H_7 + BN \longrightarrow 5.828\,YH_3 + 5.172\,YB_2 + YB_6H_4C_6N$ | -8.3014 |
| $YNH_4B_6C_6$ | $70\,YH_3 + 3\,B_{13}C_2 + 34\,YN \longrightarrow 103\,YH_2 + 33\,BN + YB_6H_4C_6N$ | -8.2631 |
| $YNH_4B_6C_6$ | $1.75\,B_{13}C_2 + 9.375\,Y + 0.5\,H_8C_5 + 0.5\,N_2 \longrightarrow 8.375\,YB_2 + YB_6H_4C_6N$ | -8.2626 |
| $YNH_4B_6C_6$ | $2.545\,B_9H_{11} + 6\,Y_2C + YN \longrightarrow 12\,YH_2 + 16.91\,B + YB_6H_4C_6N$ | -8.2441 |
| $YNH_4B_6C_6$ | $3.5\,C + YB_{12} + 0.5\,H_8C_5 + 3.5\,N_2 \longrightarrow 6\,BN + YB_6H_4C_6N$ | -8.2409 |
| $YNH_4B_6C_6$ | $1.2\,Y_4C_5 + 0.5714\,B_5H_7 + 3.186\,N_2 + 1.571\,YB_2 \longrightarrow 5.371\,YN + YB_6H_4C_6N$ | -8.0864 |
| $YNH_4B_6C_6$ | $3\,B_{13}C_2 + 11.37\,H_3N + YB_2 \longrightarrow 2.737\,B_9H_{11} + 10.37\,BN + YB_6H_4C_6N$ | -8.0832 |
| $YNH_4B_6C_6$ | $8.511\,H_3N + 1.2\,Y_4C_5 + 3.378\,YB_4 \longrightarrow 7.178\,YH_3 + 7.511\,BN + YB_6H_4C_6N$ | -8.0688 |
| $YNH_4B_6C_6$ | $1.333\,YH_3 + 1.2\,Y_4C_5 + 3.817\,N_2 + 1.5\,YB_4 \longrightarrow 6.633\,YN + YB_6H_4C_6N$ | -8.0575 |
| $YNH_4B_6C_6$ | $1.333\,YH_3 + 3\,B_{13}C_2 + YN + 13.83\,YB_2 \longrightarrow 15.17\,YB_4 + YB_6H_4C_6N$ | -8.0560 |
| $YNH_4B_6C_6$ | $1.7\,B_9H_{11} + 6\,Y_2C + YN \longrightarrow 7.35\,YH_2 + 4.65\,YB_2 + YB_6H_4C_6N$ | -8.0332 |
| $YNH_4B_6C_6$ | $2.75\,B_{13}C_2 + 0.5\,Y_2C + 10.69\,H_3N \longrightarrow 4.011\,B_5H_7 + 9.693\,BN + YB_6H_4C_6N$ | -8.0169 |
| $YNH_4B_6C_6$ | $1.055\,B_9H_{11} + 1.2\,Y_4C_5 + 2.245\,N_2 \longrightarrow 3.8\,YH_2 + 3.491\,BN + YB_6H_4C_6N$ | -7.8807 |
| $YNH_4B_6C_6$ | $1.816\,B_9H_{11} + 6\,Y_2C + BN \longrightarrow 5.327\,YH_3 + 5.673\,YB_2 + YB_6H_4C_6N$ | -7.8718 |
| $YNH_4B_6C_6$ | $8\,Y + 1.2\,H_8C_5 + 1.2\,B_5H_7 + 0.5\,N_2 \longrightarrow 7\,YH_2 + YB_6H_4C_6N$ | -7.8446 |
| $YNH_4B_6C_6$ | $3\,B_{13}C_2 + 11.95\,H_3N + YB_4 \longrightarrow 2.895\,B_9H_{11} + 10.95\,BN + YB_6H_4C_6N$ | -7.8273 |
| $YNH_4B_6C_6$ | $3.5\,Y_2C + 0.5\,H_8C_5 + BN + 29\,B \longrightarrow 6\,YB_4 + YB_6H_4C_6N$ | -7.7973 |
| $YNH_4B_6C_6$ | $6\,Y_2C + 3.714\,B_5H_7 + BN \longrightarrow 11\,YH_2 + 13.57\,B + YB_6H_4C_6N$ | -7.7637 |
| $YNH_4B_6C_6$ | $6\,Y_2C + 0.5714\,B_5H_7 + YN + 27.14\,B \longrightarrow 12\,YB_2 + YB_6H_4C_6N$ | -7.7622 |
| $YNH_4B_6C_6$ | $6\,C + 1.333\,H_3N + YB_{12} + 2.833\,N_2 \longrightarrow 6\,BN + YB_6H_4C_6N$ | -7.7264 |
| $YNH_4B_6C_6$ | $0.3636\,B_9H_{11} + 6\,Y_2C + YN + 26.73\,B \longrightarrow 12\,YB_2 + YB_6H_4C_6N$ | -7.6999 |
| $YNH_4B_6C_6$ | $0.3636\,B_9H_{11} + 1.2\,Y_4C_5 + 3.082\,N_2 + 1.364\,YB_2 \longrightarrow 5.164\,YN + YB_6H_4C_6N$ | -7.6653 |
| $YNH_4B_6C_6$ | $0.875\,B_{13}C_2 + 0.5\,H_8C_5 + 3.187\,N_2 + 0.25\,Y_4C_7 \longrightarrow 5.375\,BN + YB_6H_4C_6N$ | -7.6302 |
| $YNH_4B_6C_6$ | $2\,YH_2 + 6\,Y_2C + YN + 34\,B \longrightarrow 14\,YB_2 + YB_6H_4C_6N$ | -7.6092 |
| $YNH_4B_6C_6$ | $6\,Y_2C + 2\,H_2 + BN + 27\,B \longrightarrow 11\,YB_2 + YB_6H_4C_6N$ | -7.5195 |
| $YNH_4B_6C_6$ | $17.6\,YH_3 + 6\,Y_2C + 1.2\,B_5H_7 + 0.5\,N_2 \longrightarrow 28.6\,YH_2 + YB_6H_4C_6N$ | -7.5118 |
| $YNH_4B_6C_6$ | $1.333\,YH_3 + 1.2\,Y_4C_5 + 0.5\,YB_{12} + 3.317\,N_2 \longrightarrow 5.633\,YN + YB_6H_4C_6N$ | -7.4756 |
| $YNH_4B_6C_6$ | $6\,Y_2C + 2.583\,B_5H_7 + BN \longrightarrow 7.042\,YH_2 + 3.958\,YB_2 + YB_6H_4C_6N$ | -7.4609 |
| $YNH_4B_6C_6$ | $1.333\,YH_3 + 6\,Y_2C + YN + 32.67\,B \longrightarrow 13.33\,YB_2 + YB_6H_4C_6N$ | -7.4467 |
| $YNH_4B_6C_6$ | $6\,Y_2C + 11\,H_2 + 6\,BN \longrightarrow 6\,YH_3 + 5\,YN + YB_6H_4C_6N$ | -7.4274 |
| $YNH_4B_6C_6$ | $3\,B_{13}C_2 + 10.5\,H_3N + YN \longrightarrow 2.5\,B_9H_{11} + 10.5\,BN + YB_6H_4C_6N$ | -7.4144 |
| $YNH_4B_6C_6$ | $2\,YH_2 + 3\,B_{13}C_2 + YN + 12.5\,YB_2 \longrightarrow 14.5\,YB_4 + YB_6H_4C_6N$ | -7.3774 |
| $YNH_4B_6C_6$ | $YH_2 + 3\,B_{13}C_2 + 11.27\,H_3N \longrightarrow 4.545\,B_5H_7 + 10.27\,BN + YB_6H_4C_6N$ | -7.3763 |
| $YNH_4B_6C_6$ | $1.333\,YH_3 + 1.2\,Y_4C_5 + 3.067\,N_2 + 6\,B \longrightarrow 5.133\,YN + YB_6H_4C_6N$ | -7.3597 |
| $YNH_4B_6C_6$ | $2.364\,B_9H_{11} + 6\,Y_2C + BN \longrightarrow 11\,YH_2 + 16.27\,B + YB_6H_4C_6N$ | -7.3589 |
| $YNH_4B_6C_6$ | $2\,YH_2 + 1.2\,Y_4C_5 + 0.5\,N_2 + 75.6\,B \longrightarrow 5.8\,YB_{12} + YB_6H_4C_6N$ | -7.3289 |
| $YNH_4B_6C_6$ | $3.214\,N_2 + 0.8571\,Y_4C_7 + 2\,H_2 + 3\,YB_2 \longrightarrow 5.429\,YN + YB_6H_4C_6N$ | -7.2684 |
| $YNH_4B_6C_6$ | $18.67\,YH_3 + 0.6667\,B_9H_{11} + 6\,Y_2C + 0.5\,N_2 \longrightarrow 29.67\,YH_2 + YB_6H_4C_6N$ | -7.2268 |
| $YNH_4B_6C_6$ | $1.55\,B_9H_{11} + 6\,Y_2C + BN \longrightarrow 6.525\,YH_2 + 4.475\,YB_2 + YB_6H_4C_6N$ | -7.1560 |
| $YNH_4B_6C_6$ | $0.4615\,B_{13}C_2 + 2.577\,Y_2C + 0.5\,H_8C_5 + 2.577\,N_2 \longrightarrow 4.154\,YN + YB_6H_4C_6N$ | -7.1168 |
| $YNH_4B_6C_6$ | $14.6\,YH_3 + 6\,Y_2C + H_3N + 1.2\,B_5H_7 \longrightarrow 25.6\,YH_2 + YB_6H_4C_6N$ | -7.1160 |
| $YNH_4B_6C_6$ | $2\,YH_2 + 1.2\,Y_4C_5 + YN + 87.6\,B \longrightarrow 6.8\,YB_{12} + YB_6H_4C_6N$ | -7.0971 |
| $YNH_4B_6C_6$ | $6.929\,H_3N + 0.8571\,Y_4C_7 + 5.964\,YB_2 \longrightarrow 8.393\,YH_2 + 5.929\,BN + YB_6H_4C_6N$ | -7.0847 |
| $YNH_4B_6C_6$ | $2.75\,B_{13}C_2 + 0.5\,Y_2C + 9.849\,H_3N \longrightarrow 2.322\,B_9H_{11} + 8.849\,BN + YB_6H_4C_6N$ | -6.9869 |

| Compound | Chemical reaction | Reaction energy(eV) |
|---|---|---|
| $YNH_4B_6C_6$ | $3.5\,Y_2C + 3.625\,YB_{12} + 0.5\,H_8C_5 + BN \longrightarrow 9.625\,YB_4 + YB_6H_4C_6N$ | -6.9513 |
| $YNH_4B_6C_6$ | $5.375\,Y_2C + H_3N + 0.125\,H_8C_5 + 25.5\,B \longrightarrow 9.75\,YB_2 + YB_6H_4C_6N$ | -6.9391 |
| $YNH_4B_6C_6$ | $6\,Y_2C + 0.5714\,B_5H_7 + BN + 24.14\,B \longrightarrow 11\,YB_2 + YB_6H_4C_6N$ | -6.9224 |
| $YNH_4B_6C_6$ | $0.3636\,B_9H_{11} + 6\,Y_2C + BN + 23.73\,B \longrightarrow 11\,YB_2 + YB_6H_4C_6N$ | -6.8601 |
| $YNH_4B_6C_6$ | $YH_3 + 3\,B_{13}C_2 + 11.05\,H_3N \longrightarrow 4.591\,B_5H_7 + 10.05\,BN + YB_6H_4C_6N$ | -6.8319 |
| $YNH_4B_6C_6$ | $15.67\,YH_3 + 0.6667\,B_9H_{11} + 6\,Y_2C + H_3N \longrightarrow 26.67\,YH_2 + YB_6H_4C_6N$ | -6.8310 |
| $YNH_4B_6C_6$ | $2\,YH_2 + 6\,Y_2C + BN + 31\,B \longrightarrow 13\,YB_2 + YB_6H_4C_6N$ | -6.7694 |
| $YNH_4B_6C_6$ | $0.6667\,B_9H_{11} + 7.467\,Y + 1.2\,H_8C_5 + 0.5\,N_2 \longrightarrow 6.467\,YH_2 + YB_6H_4C_6N$ | -6.7637 |
| $YNH_4B_6C_6$ | $1.2\,Y_4C_5 + 3.15\,N_2 + 2\,H_2 + 1.5\,YB_4 \longrightarrow 5.3\,YN + YB_6H_4C_6N$ | -6.6676 |
| $YNH_4B_6C_6$ | $1.2\,Y_4C_5 + 1.2\,B_5H_7 + 3.133\,N_2 \longrightarrow 1.467\,H_3N + 3.8\,YN + YB_6H_4C_6N$ | -6.6577 |
| $YNH_4B_6C_6$ | $1.333\,YH_3 + 6\,Y_2C + BN + 29.67\,B \longrightarrow 12.33\,YB_2 + YB_6H_4C_6N$ | -6.6069 |
| $YNH_4B_6C_6$ | $1.75\,B_{13}C_2 + 9.875\,Y + 0.5\,H_8C_5 + BN \longrightarrow 8.875\,YB_2 + YB_6H_4C_6N$ | -6.5639 |
| $YNH_4B_6C_6$ | $2.514\,B_9H_{11} + 1.2\,H_8C_5 + 17.63\,YN \longrightarrow 16.63\,YH_2 + 16.63\,BN + YB_6H_4C_6N$ | -6.4996 |
| $YNH_4B_6C_6$ | $YB_{12} + 1.2\,H_8C_5 + 3.5\,N_2 \longrightarrow 2.8\,H_2 + 6\,BN + YB_6H_4C_6N$ | -6.4896 |
| $YNH_4B_6C_6$ | $6\,Y_2C + 3.667\,H_3N + 6\,BN \longrightarrow 2.333\,YH_3 + 8.667\,YN + YB_6H_4C_6N$ | -6.3947 |
| $YNH_4B_6C_6$ | $1.75\,B_{13}C_2 + 5.187\,Y + 0.5\,H_8C_5 + 0.5\,N_2 \longrightarrow 4.187\,YB_4 + YB_6H_4C_6N$ | -6.2760 |
| $YNH_4B_6C_6$ | $0.6667\,B_9H_{11} + 1.2\,Y_4C_5 + 2.956\,N_2 \longrightarrow 1.111\,H_3N + 3.8\,YN + YB_6H_4C_6N$ | -6.2320 |
| $YNH_4B_6C_6$ | $6.055\,H_3N + 1.2\,Y_4C_5 + 0.9212\,YB_{12} \longrightarrow 4.721\,YH_3 + 5.055\,BN + YB_6H_4C_6N$ | -6.2190 |
| $YNH_4B_6C_6$ | $YH_2 + 3\,B_{13}C_2 + 10.32\,H_3N \longrightarrow 2.632\,B_9H_{11} + 9.316\,BN + YB_6H_4C_6N$ | -6.2092 |
| $YNH_4B_6C_6$ | $2\,YH_2 + 0.4615\,B_{13}C_2 + 1.015\,Y_4C_5 + 3.031\,N_2 \longrightarrow 5.062\,YN + YB_6H_4C_6N$ | -6.1909 |
| $YNH_4B_6C_6$ | $1.333\,YH_3 + 1.2\,Y_4C_5 + 0.5\,N_2 + 67.6\,B \longrightarrow 5.133\,YB_{12} + YB_6H_4C_6N$ | -6.1696 |
| $YNH_4B_6C_6$ | $26\,YH_3 + 6\,Y_2C + 0.5\,N_2 + 6\,B \longrightarrow 37\,YH_2 + YB_6H_4C_6N$ | -6.1665 |
| $YNH_4B_6C_6$ | $9.25\,YH_3 + 3\,B_{13}C_2 + 4.458\,N_2 \longrightarrow 7.917\,H_3N + 8.25\,YB_4 + YB_6H_4C_6N$ | -6.1192 |
| $YNH_4B_6C_6$ | $1.2\,Y_4C_5 + 0.5\,YB_{12} + 2.65\,N_2 + 2\,H_2 \longrightarrow 4.3\,YN + YB_6H_4C_6N$ | -6.0857 |
| $YNH_4B_6C_6$ | $6.4\,H_3N + 3.1\,Y_4C_5 + 6\,BN \longrightarrow 1.9\,H_8C_5 + 11.4\,YN + YB_6H_4C_6N$ | -6.0234 |
| $YNH_4B_6C_6$ | $1.026\,B_9H_{11} + 2.117\,N_2 + 0.8571\,Y_4C_7 \longrightarrow 2.429\,YH_3 + 3.234\,BN + YB_6H_4C_6N$ | -5.9957 |
| $YNH_4B_6C_6$ | $1.2\,B_5H_7 + 2.5\,N_2 + 1.25\,Y_4C_7 \longrightarrow 0.55\,H_8C_5 + 4\,YN + YB_6H_4C_6N$ | -5.9808 |
| $YNH_4B_6C_6$ | $2\,YH_2 + 1.2\,Y_4C_5 + 0.5\,N_2 + 29.2\,B \longrightarrow 5.8\,YB_4 + YB_6H_4C_6N$ | -5.9753 |
| $YNH_4B_6C_6$ | $1.2\,Y_4C_5 + 2.4\,N_2 + 2\,H_2 + 6\,B \longrightarrow 3.8\,YN + YB_6H_4C_6N$ | -5.9698 |
| $YNH_4B_6C_6$ | $1.333\,YH_3 + 1.2\,Y_4C_5 + YN + 79.6\,B \longrightarrow 6.133\,YB_{12} + YB_6H_4C_6N$ | -5.9378 |
| $YNH_4B_6C_6$ | $6\,Y_2C + 2.8\,YB_{12} + 0.5\,N_2 + 2\,H_2 \longrightarrow 13.8\,YB_2 + YB_6H_4C_6N$ | -5.9000 |
| $YNH_4B_6C_6$ | $5.133\,H_3N + 1.2\,Y_4C_5 + 10.13\,B \longrightarrow 3.8\,YH_3 + 4.133\,BN + YB_6H_4C_6N$ | -5.8209 |
| $YNH_4B_6C_6$ | $23\,YH_3 + 6\,Y_2C + H_3N + 6\,B \longrightarrow 34\,YH_2 + YB_6H_4C_6N$ | -5.7708 |
| $YNH_4B_6C_6$ | $1.2\,Y_4C_5 + 0.5714\,B_5H_7 + 2.793\,N_2 + 0.7857\,YB_4 \longrightarrow 4.586\,YN + YB_6H_4C_6N$ | -5.7382 |
| $YNH_4B_6C_6$ | $6\,Y_2C + 6\,H_3N + 6\,BN \longrightarrow 11\,YN + 7\,H_2 + YB_6H_4C_6N$ | -5.7375 |
| $YNH_4B_6C_6$ | $6\,Y_2C + 2.514\,YB_{12} + 0.5714\,B_5H_7 + 0.5\,N_2 \longrightarrow 13.51\,YB_2 + YB_6H_4C_6N$ | -5.7300 |
| $YNH_4B_6C_6$ | $0.3636\,B_9H_{11} + 6\,Y_2C + 2.473\,YB_{12} + 0.5\,N_2 \longrightarrow 13.47\,YB_2 + YB_6H_4C_6N$ | -5.7299 |
| $YNH_4B_6C_6$ | $22\,B_9H_{11} + 6\,Y_2C + 0.5\,N_2 \longrightarrow 34\,B_5H_7 + 11\,YB_2 + YB_6H_4C_6N$ | -5.7218 |
| $YNH_4B_6C_6$ | $118.2\,B_9H_{11} + 6\,Y_2C + 0.5\,N_2 \longrightarrow 11\,YB_{12} + 185.2\,B_5H_7 + YB_6H_4C_6N$ | -5.6861 |
| $YNH_4B_6C_6$ | $6\,Y_2C + 3.2\,H_3N + 6\,BN \longrightarrow 2.8\,YH_2 + 8.2\,YN + YB_6H_4C_6N$ | -5.6597 |
| $YNH_4B_6C_6$ | $YH_3 + 3\,B_{13}C_2 + 10.08\,H_3N \longrightarrow 2.658\,B_9H_{11} + 9.079\,BN + YB_6H_4C_6N$ | -5.6531 |
| $YNH_4B_6C_6$ | $0.3636\,B_9H_{11} + 1.2\,Y_4C_5 + 2.741\,N_2 + 0.6818\,YB_4 \longrightarrow 4.482\,YN + YB_6H_4C_6N$ | -5.6276 |
| $YNH_4B_6C_6$ | $3.143\,B_5H_7 + 3.5\,N_2 + 0.8571\,Y_4C_7 \longrightarrow 6\,H_3N + 2.429\,YB_4 + YB_6H_4C_6N$ | -5.6178 |
| $YNH_4B_6C_6$ | $6\,Y_2C + 8\,H_2 + 6\,BN \longrightarrow 6\,YH_2 + 5\,YN + YB_6H_4C_6N$ | -5.5707 |
| $YNH_4B_6C_6$ | $19.6\,YH_3 + 6\,Y_2C + 1.2\,B_5H_7 + YN \longrightarrow 31.6\,YH_2 + YB_6H_4C_6N$ | -5.5410 |
| $YNH_4B_6C_6$ | $16.63\,C + 7\,H_3N + YB_{12} \longrightarrow 2.125\,H_8C_5 + 6\,BN + YB_6H_4C_6N$ | -5.5398 |
| $YNH_4B_6C_6$ | $2\,YH_2 + 1.2\,Y_4C_5 + YN + 33.2\,B \longrightarrow 6.8\,YB_4 + YB_6H_4C_6N$ | -5.5101 |
| $YNH_4B_6C_6$ | $1.333\,H_3N + 2.548\,N_2 + 0.8571\,Y_4C_7 + 3\,YB_2 \longrightarrow 5.429\,YN + YB_6H_4C_6N$ | -5.5029 |
| $YNH_4B_6C_6$ | $2\,YH_2 + 3.464\,N_2 + 0.8571\,Y_4C_7 + 1.5\,YB_4 \longrightarrow 5.929\,YN + YB_6H_4C_6N$ | -5.4891 |
| $YNH_4B_6C_6$ | $2.375\,B_{13}C_2 + 9.142\,H_3N + 0.25\,Y_4C_5 \longrightarrow 3.347\,B_5H_7 + 8.142\,BN + YB_6H_4C_6N$ | -5.4857 |
| $YNH_4B_6C_6$ | $6.667\,Y + H_3N + 1.2\,H_8C_5 + 1.2\,B_5H_7 \longrightarrow 5.667\,YH_3 + YB_6H_4C_6N$ | -5.4591 |
| $YNH_4B_6C_6$ | $32\,YH_3 + 6\,Y_2C + 0.5\,N_2 + 3\,YB_2 \longrightarrow 46\,YH_2 + YB_6H_4C_6N$ | -5.4351 |
| $YNH_4B_6C_6$ | $1.2\,Y_4C_5 + 0.2619\,YB_{12} + 0.5714\,B_5H_7 + 2.531\,N_2 \longrightarrow 4.062\,YN + YB_6H_4C_6N$ | -5.4334 |
| $YNH_4B_6C_6$ | $6\,Y_2C + 20.5\,H_2 + 6\,BN \longrightarrow 11\,YH_2 + 5\,H_3N + YB_6H_4C_6N$ | -5.4317 |
| $YNH_4B_6C_6$ | $1.2\,Y_4C_5 + 0.5714\,B_5H_7 + 2.4\,N_2 + 3.143\,B \longrightarrow 3.8\,YN + YB_6H_4C_6N$ | -5.3726 |
| $YNH_4B_6C_6$ | $0.3636\,B_9H_{11} + 1.2\,Y_4C_5 + 0.2273\,YB_{12} + 2.514\,N_2 \longrightarrow 4.027\,YN + YB_6H_4C_6N$ | -5.3631 |
| $YNH_4B_6C_6$ | $0.3636\,B_9H_{11} + 1.2\,Y_4C_5 + 2.4\,N_2 + 2.727\,B \longrightarrow 3.8\,YN + YB_6H_4C_6N$ | -5.3104 |
| $YNH_4B_6C_6$ | $27\,YH_3 + 6\,Y_2C + 0.5\,YB_{12} + 0.5\,N_2 \longrightarrow 38.5\,YH_2 + YB_6H_4C_6N$ | -5.2971 |
| $YNH_4B_6C_6$ | $20.67\,YH_3 + 0.6667\,B_9H_{11} + 6\,Y_2C + YN \longrightarrow 32.67\,YH_2 + YB_6H_4C_6N$ | -5.2560 |
| $YNH_4B_6C_6$ | $0.3636\,B_9H_{11} + 1.611\,B_{13}C_2 + 2.777\,Y_2C + 0.5\,N_2 \longrightarrow 4.555\,YB_4 + YB_6H_4C_6N$ | -5.2231 |

| Compound | Chemical reaction | Reaction energy(eV) |
|---|---|---|
| $YNH_4B_6C_6$ | $6\,Y_2C + 5.867\,H_3N + 1.2\,B_5H_7 \longrightarrow 11\,YH_2 + 2.433\,N_2 + YB_6H_4C_6N$ | -5.1900 |
| $YNH_4B_6C_6$ | $1.338\,B_{13}C_2 + 3.325\,Y_2C + 6.649\,H_3N \longrightarrow 2.278\,B_5H_7 + 5.649\,YN + YB_6H_4C_6N$ | -5.1573 |
| $YNH_4B_6C_6$ | $1.626\,B_{13}C_2 + 2.749\,Y_2C + 0.5714\,B_5H_7 + 0.5\,N_2 \longrightarrow 4.498\,YB_4 + YB_6H_4C_6N$ | -5.1246 |
| $YNH_4B_6C_6$ | $2\,YH_2 + 1.2\,Y_4C_5 + 3.65\,YB_{12} + 0.5\,N_2 \longrightarrow 9.45\,YB_4 + YB_6H_4C_6N$ | -5.1235 |
| $YNH_4B_6C_6$ | $1.2\,Y_4C_5 + 0.5\,N_2 + 2\,H_2 + 51.6\,B \longrightarrow 3.8\,YB_{12} + YB_6H_4C_6N$ | -5.0888 |
| $YNH_4B_6C_6$ | $29\,YH_3 + 6\,Y_2C + H_3N + 3\,YB_2 \longrightarrow 43\,YH_2 + YB_6H_4C_6N$ | -5.0393 |
| $YNH_4B_6C_6$ | $1.333\,YH_3 + 1.2\,Y_4C_5 + 0.5\,N_2 + 26.53\,B \longrightarrow 5.133\,YB_4 + YB_6H_4C_6N$ | -4.9716 |
| $YNH_4B_6C_6$ | $1.2\,Y_4C_5 + 1.657\,B_5H_7 + 1.643\,N_2 \longrightarrow 3.8\,YH_2 + 2.286\,BN + YB_6H_4C_6N$ | -4.9678 |
| $YNH_4B_6C_6$ | $6\,Y_2C + 1.333\,H_3N + 2.833\,YB_{12} \longrightarrow 0.3333\,BN + 13.83\,YB_2 + YB_6H_4C_6N$ | -4.9403 |
| $YNH_4B_6C_6$ | $2\,YH_2 + 0.5\,YB_{12} + 2.964\,N_2 + 0.8571\,Y_4C_7 \longrightarrow 4.929\,YN + YB_6H_4C_6N$ | -4.9072 |
| $YNH_4B_6C_6$ | $1.333\,H_3N + 1.2\,Y_4C_5 + 2.483\,N_2 + 1.5\,YB_4 \longrightarrow 5.3\,YN + YB_6H_4C_6N$ | -4.9022 |
| $YNH_4B_6C_6$ | $2.75\,B_9H_{11} + 1.2\,Y_4C_5 + 2.4\,N_2 \longrightarrow 3.75\,B_5H_7 + 3.8\,YN + YB_6H_4C_6N$ | -4.9017 |
| $YNH_4B_6C_6$ | $24\,YH_3 + 6\,Y_2C + H_3N + 0.5\,YB_{12} \longrightarrow 35.5\,YH_2 + YB_6H_4C_6N$ | -4.9013 |
| $YNH_4B_6C_6$ | $9.25\,YH_2 + 3\,B_{13}C_2 + 2.917\,N_2 \longrightarrow 4.833\,H_3N + 8.25\,YB_4 + YB_6H_4C_6N$ | -4.8990 |
| $YNH_4B_6C_6$ | $1.333\,YH_3 + 0.4615\,B_{13}C_2 + 1.015\,Y_4C_5 + 2.697\,N_2 \longrightarrow 4.395\,YN + YB_6H_4C_6N$ | -4.8770 |
| $YNH_4B_6C_6$ | $1.2\,Y_4C_5 + YN + 2\,H_2 + 63.6\,B \longrightarrow 4.8\,YB_{12} + YB_6H_4C_6N$ | -4.8570 |
| $YNH_4B_6C_6$ | $0.7\,Y_4C_5 + 0.5\,H_8C_5 + 2.9\,N_2 + 3\,YB_2 \longrightarrow 4.8\,YN + YB_6H_4C_6N$ | -4.8537 |
| $YNH_4B_6C_6$ | $6\,Y_2C + 1.333\,H_3N + 2.733\,YB_{12} \longrightarrow 0.3333\,YN + 13.4\,YB_2 + YB_6H_4C_6N$ | -4.8099 |
| $YNH_4B_6C_6$ | $1.75\,B_{13}C_2 + 8.375\,Y + 0.5\,H_8C_5 + YN \longrightarrow 8.375\,YB_2 + YB_6H_4C_6N$ | -4.7996 |
| $YNH_4B_6C_6$ | $3\,B_{13}C_2 + 1.333\,H_3N + 7.333\,YN \longrightarrow 7.667\,BN + 6.333\,YB_4 + YB_6H_4C_6N$ | -4.7993 |
| $YNH_4B_6C_6$ | $2\,YH_2 + 2.714\,N_2 + 0.8571\,Y_4C_7 + 6\,B \longrightarrow 4.429\,YN + YB_6H_4C_6N$ | -4.7913 |
| $YNH_4B_6C_6$ | $0.6667\,B_9H_{11} + 2.31\,N_2 + 1.155\,Y_4C_7 \longrightarrow 0.4167\,H_8C_5 + 3.619\,YN + YB_6H_4C_6N$ | -4.7781 |
| $YNH_4B_6C_6$ | $0.6667\,B_9H_{11} + 6\,Y_2C + 6.222\,H_3N \longrightarrow 11\,YH_2 + 2.611\,N_2 + YB_6H_4C_6N$ | -4.7643 |
| $YNH_4B_6C_6$ | $2\,YH_2 + 1.2\,Y_4C_5 + BN + 74.6\,B \longrightarrow 5.8\,YB_{12} + YB_6H_4C_6N$ | -4.7622 |
| $YNH_4B_6C_6$ | $0.6667\,B_9H_{11} + 1.2\,Y_4C_5 + 2.4\,N_2 \longrightarrow 3.8\,YN + 1.667\,H_2 + YB_6H_4C_6N$ | -4.7608 |
| $YNH_4B_6C_6$ | $19\,YH_3 + 6\,Y_2C + B_5H_7 + BN \longrightarrow 30\,YH_2 + YB_6H_4C_6N$ | -4.7208 |
| $YNH_4B_6C_6$ | $1.2\,Y_4C_5 + 1.2\,B_5H_7 + 2.4\,N_2 \longrightarrow 3.8\,YN + 2.2\,H_2 + YB_6H_4C_6N$ | -4.7158 |
| $YNH_4B_6C_6$ | $1.41\,B_{13}C_2 + 3.179\,Y_2C + 6.359\,H_3N \longrightarrow 1.371\,B_9H_{11} + 5.359\,YN + YB_6H_4C_6N$ | -4.6965 |
| $YNH_4B_6C_6$ | $0.6667\,B_9H_{11} + 6.311\,Y + H_3N + 1.2\,H_8C_5 \longrightarrow 5.311\,YH_3 + YB_6H_4C_6N$ | -4.6435 |
| $YNH_4B_6C_6$ | $2.375\,B_{13}C_2 + 8.438\,H_3N + 0.25\,Y_4C_5 \longrightarrow 1.938\,B_9H_{11} + 7.438\,BN + YB_6H_4C_6N$ | -4.6264 |
| $YNH_4B_6C_6$ | $1.724\,B_{13}C_2 + 2.552\,Y_2C + 0.5\,N_2 + 2\,H_2 \longrightarrow 4.103\,YB_4 + YB_6H_4C_6N$ | -4.6161 |
| $YNH_4B_6C_6$ | $1.333\,YH_3 + 6\,Y_2C + 3.067\,YB_{12} + 0.5\,N_2 \longrightarrow 15.4\,YB_2 + YB_6H_4C_6N$ | -4.5887 |
| $YNH_4B_6C_6$ | $6\,Y_2C + H_3N + 2.8\,YB_{12} + 0.5\,H_2 \longrightarrow 13.8\,YB_2 + YB_6H_4C_6N$ | -4.5759 |
| $YNH_4B_6C_6$ | $2\,YH_2 + 6\,Y_2C + 3.2\,YB_{12} + 0.5\,N_2 \longrightarrow 16.2\,YB_2 + YB_6H_4C_6N$ | -4.5519 |
| $YNH_4B_6C_6$ | $2\,YH_2 + 1.2\,Y_4C_5 + 4.15\,YB_{12} + YN \longrightarrow 10.95\,YB_4 + YB_6H_4C_6N$ | -4.5416 |
| $YNH_4B_6C_6$ | $6\,Y_2C + H_3N + 2.729\,YB_{12} + 0.1429\,B_5H_7 \longrightarrow 13.73\,YB_2 + YB_6H_4C_6N$ | -4.5334 |
| $YNH_4B_6C_6$ | $0.09091\,B_9H_{11} + 6\,Y_2C + H_3N + 2.718\,YB_{12} \longrightarrow 13.72\,YB_2 + YB_6H_4C_6N$ | -4.5334 |
| $YNH_4B_6C_6$ | $23.87\,B_9H_{11} + 6\,Y_2C + H_3N \longrightarrow 37.37\,B_5H_7 + 11\,YB_2 + YB_6H_4C_6N$ | -4.5246 |
| $YNH_4B_6C_6$ | $1.333\,YH_3 + 1.2\,Y_4C_5 + YN + 30.53\,B \longrightarrow 6.133\,YB_4 + YB_6H_4C_6N$ | -4.5064 |
| $YNH_4B_6C_6$ | $1.746\,B_9H_{11} + 3.034\,N_2 + 0.8571\,Y_4C_7 \longrightarrow 5.069\,H_3N + 2.429\,YB_4 + YB_6H_4C_6N$ | -4.5029 |
| $YNH_4B_6C_6$ | $1.2\,Y_4C_5 + 0.5714\,B_5H_7 + 0.5\,N_2 + 48.74\,B \longrightarrow 3.8\,YB_{12} + YB_6H_4C_6N$ | -4.4917 |
| $YNH_4B_6C_6$ | $120.1\,B_9H_{11} + 6\,Y_2C + H_3N \longrightarrow 11\,YB_{12} + 188.6\,B_5H_7 + YB_6H_4C_6N$ | -4.4888 |
| $YNH_4B_6C_6$ | $19.89\,YH_3 + 0.5556\,B_9H_{11} + 6\,Y_2C + BN \longrightarrow 30.89\,YH_2 + YB_6H_4C_6N$ | -4.4833 |
| $YNH_4B_6C_6$ | $1.75\,B_{13}C_2 + 5.438\,Y + 0.5\,H_8C_5 + BN \longrightarrow 4.438\,YB_4 + YB_6H_4C_6N$ | -4.4587 |
| $YNH_4B_6C_6$ | $3.111\,B_9H_{11} + 6\,Y_2C + 0.5\,N_2 \longrightarrow 15.11\,H_2 + 11\,YB_2 + YB_6H_4C_6N$ | -4.4445 |
| $YNH_4B_6C_6$ | $0.3636\,B_9H_{11} + 1.2\,Y_4C_5 + 0.5\,N_2 + 48.33\,B \longrightarrow 3.8\,YB_{12} + YB_6H_4C_6N$ | -4.4294 |
| $YNH_4B_6C_6$ | $0.3636\,B_9H_{11} + 3\,B_{13}C_2 + 8.255\,YN \longrightarrow 7.255\,BN + 7.255\,YB_4 + YB_6H_4C_6N$ | -4.4189 |
| $YNH_4B_6C_6$ | $7\,Y + 1.2\,H_8C_5 + 1.2\,B_5H_7 + YN \longrightarrow 7\,YH_2 + YB_6H_4C_6N$ | -4.3815 |
| $YNH_4B_6C_6$ | $5.667\,Y + 1.2\,H_8C_5 + 1.2\,B_5H_7 + 0.5\,N_2 \longrightarrow 4.667\,YH_3 + YB_6H_4C_6N$ | -4.3625 |
| $YNH_4B_6C_6$ | $0.5\,YH_2 + H_3N + 1.2\,Y_4C_5 + 57.6\,B \longrightarrow 4.3\,YB_{12} + YB_6H_4C_6N$ | -4.3248 |
| $YNH_4B_6C_6$ | $1.333\,H_3N + 1.2\,Y_4C_5 + 0.5\,YB_{12} + 1.983\,N_2 \longrightarrow 4.3\,YN + YB_6H_4C_6N$ | -4.3203 |
| $YNH_4B_6C_6$ | $3\,B_{13}C_2 + 0.5714\,B_5H_7 + 8.171\,YN \longrightarrow 7.171\,BN + 7.171\,YB_4 + YB_6H_4C_6N$ | -4.3065 |
| $YNH_4B_6C_6$ | $1.2\,H_8C_5 + 3.233\,N_2 + 2.867\,YB_4 \longrightarrow 1.867\,YH_3 + 5.467\,BN + YB_6H_4C_6N$ | -4.2766 |
| $YNH_4B_6C_6$ | $1.2\,Y_4C_5 + 0.5714\,B_5H_7 + YN + 60.74\,B \longrightarrow 4.8\,YB_{12} + YB_6H_4C_6N$ | -4.2599 |
| $YNH_4B_6C_6$ | $0.3333\,YH_3 + 6\,Y_2C + H_3N + 2.867\,YB_{12} \longrightarrow 14.2\,YB_2 + YB_6H_4C_6N$ | -4.2481 |
| $YNH_4B_6C_6$ | $0.5\,YH_2 + 6\,Y_2C + H_3N + 2.9\,YB_{12} \longrightarrow 14.4\,YB_2 + YB_6H_4C_6N$ | -4.2389 |
| $YNH_4B_6C_6$ | $6\,Y_2C + 5.6\,B_5H_7 + 0.5\,N_2 \longrightarrow 17.6\,H_2 + 11\,YB_2 + YB_6H_4C_6N$ | -4.2341 |
| $YNH_4B_6C_6$ | $1.333\,H_3N + 1.2\,Y_4C_5 + 1.733\,N_2 + 6\,B \longrightarrow 3.8\,YN + YB_6H_4C_6N$ | -4.2044 |
| $YNH_4B_6C_6$ | $0.5714\,B_5H_7 + 2.5\,N_2 + 0.8571\,Y_4C_7 + 1.571\,YB_2 \longrightarrow 4\,YN + YB_6H_4C_6N$ | -4.2042 |
| $YNH_4B_6C_6$ | $1.2\,Y_4C_5 + 0.5\,N_2 + 2\,H_2 + 21.2\,B \longrightarrow 3.8\,YB_4 + YB_6H_4C_6N$ | -4.2020 |

| Compound | Chemical reaction | Reaction energy(eV) |
|---|---|---|
| $YNH_4B_6C_6$ | $0.3636\,B_9H_{11} + 1.2\,Y_4C_5 + YN + 60.33\,B \longrightarrow 4.8\,YB_{12} + YB_6H_4C_6N$ | -4.1976 |
| $YNH_4B_6C_6$ | $1.333\,YH_3 + 1.2\,Y_4C_5 + 3.317\,YB_{12} + 0.5\,N_2 \longrightarrow 8.45\,YB_4 + YB_6H_4C_6N$ | -4.1976 |
| $YNH_4B_6C_6$ | $28\,YH_3 + 6\,Y_2C + YN + 6\,B \longrightarrow 40\,YH_2 + YB_6H_4C_6N$ | -4.1958 |
| $YNH_4B_6C_6$ | $0.3636\,B_9H_{11} + 0.2098\,B_{13}C_2 + 1.116\,Y_4C_5 + 2.232\,N_2 \longrightarrow 3.464\,YN + YB_6H_4C_6N$ | -4.1819 |
| $YNH_4B_6C_6$ | $1.333\,H_3N + 1.2\,Y_4C_5 + 51.93\,B \longrightarrow 3.8\,YB_{12} + 0.3333\,BN + YB_6H_4C_6N$ | -4.1790 |
| $YNH_4B_6C_6$ | $1.333\,YH_3 + 3.131\,N_2 + 0.8571\,Y_4C_7 + 1.5\,YB_4 \longrightarrow 5.262\,YN + YB_6H_4C_6N$ | -4.1753 |
| $YNH_4B_6C_6$ | $6\,Y_2C + 1.333\,H_3N + 2.8\,YB_{12} \longrightarrow 0.1667\,N_2 + 13.8\,YB_2 + YB_6H_4C_6N$ | -4.1346 |
| $YNH_4B_6C_6$ | $0.2418\,B_{13}C_2 + 1.103\,Y_4C_5 + 0.5714\,B_5H_7 + 2.207\,N_2 \longrightarrow 3.413\,YN + YB_6H_4C_6N$ | -4.0722 |
| $YNH_4B_6C_6$ | $1.2\,Y_4C_5 + 0.5\,N_2 + 12.2\,H_2 + 3\,YB_2 \longrightarrow 6.8\,YH_3 + YB_6H_4C_6N$ | -4.0621 |
| $YNH_4B_6C_6$ | $0.3333\,YH_3 + H_3N + 1.2\,Y_4C_5 + 55.6\,B \longrightarrow 4.133\,YB_{12} + YB_6H_4C_6N$ | -4.0350 |
| $YNH_4B_6C_6$ | $7.3\,Y + 1.2\,H_8C_5 + B_5H_7 + BN \longrightarrow 6.3\,YH_2 + YB_6H_4C_6N$ | -4.0090 |
| $YNH_4B_6C_6$ | $2.125\,B_{13}C_2 + 8.108\,H_3N + 0.25\,Y_4C_7 \longrightarrow 2.903\,B_5H_7 + 7.108\,BN + YB_6H_4C_6N$ | -3.9500 |
| $YNH_4B_6C_6$ | $29\,YH_3 + 6\,Y_2C + 0.5\,N_2 + 1.5\,YB_4 \longrightarrow 41.5\,YH_2 + YB_6H_4C_6N$ | -3.9082 |
| $YNH_4B_6C_6$ | $6\,Y_2C + 3\,YB_{12} + YN + 2\,H_2 \longrightarrow 15\,YB_2 + YB_6H_4C_6N$ | -3.8741 |
| $YNH_4B_6C_6$ | $1.75\,B_{13}C_2 + 0.5\,H_8C_5 + 0.5\,N_2 + 10.38\,YB_2 \longrightarrow 9.375\,YB_4 + YB_6H_4C_6N$ | -3.8150 |
| $YNH_4B_6C_6$ | $0.3636\,B_9H_{11} + 2.396\,N_2 + 0.8571\,Y_4C_7 + 1.364\,YB_2 \longrightarrow 3.792\,YN + YB_6H_4C_6N$ | -3.7831 |
| $YNH_4B_6C_6$ | $22.31\,YH_3 + 0.4615\,B_{13}C_2 + 5.077\,Y_2C + 0.5\,N_2 \longrightarrow 31.46\,YH_2 + YB_6H_4C_6N$ | -3.7664 |
| $YNH_4B_6C_6$ | $H_3N + 1.2\,Y_4C_5 + 0.5\,H_2 + 51.6\,B \longrightarrow 3.8\,YB_{12} + YB_6H_4C_6N$ | -3.7648 |
| $YNH_4B_6C_6$ | $2\,YH_2 + 0.5\,N_2 + 0.8571\,Y_4C_7 + 59.14\,B \longrightarrow 4.429\,YB_{12} + YB_6H_4C_6N$ | -3.7646 |
| $YNH_4B_6C_6$ | $6\,Y_2C + 9.533\,H_3N + 1.2\,B_5H_7 \longrightarrow 11\,YH_3 + 4.267\,N_2 + YB_6H_4C_6N$ | -3.7390 |
| $YNH_4B_6C_6$ | $1.2\,Y_4C_5 + YN + 2\,H_2 + 25.2\,B \longrightarrow 4.8\,YB_4 + YB_6H_4C_6N$ | -3.7368 |
| $YNH_4B_6C_6$ | $6\,Y_2C + 2.714\,YB_{12} + 0.5714\,B_5H_7 + YN \longrightarrow 14.71\,YB_2 + YB_6H_4C_6N$ | -3.7041 |
| $YNH_4B_6C_6$ | $0.3636\,B_9H_{11} + 6\,Y_2C + 2.673\,YB_{12} + YN \longrightarrow 14.67\,YB_2 + YB_6H_4C_6N$ | -3.7039 |
| $YNH_4B_6C_6$ | $3\,B_{13}C_2 + 7.6\,YN + 2\,H_2 \longrightarrow 6.6\,BN + 6.6\,YB_4 + YB_6H_4C_6N$ | -3.7028 |
| $YNH_4B_6C_6$ | $23.75\,B_9H_{11} + 6\,Y_2C + YN \longrightarrow 36.75\,B_5H_7 + 12\,YB_2 + YB_6H_4C_6N$ | -3.6952 |
| $YNH_4B_6C_6$ | $128.7\,B_9H_{11} + 6\,Y_2C + YN \longrightarrow 12\,YB_{12} + 201.7\,B_5H_7 + YB_6H_4C_6N$ | -3.6562 |
| $YNH_4B_6C_6$ | $1.333\,YH_3 + 1.2\,Y_4C_5 + 3.817\,YB_{12} + YN \longrightarrow 9.95\,YB_4 + YB_6H_4C_6N$ | -3.6157 |
| $YNH_4B_6C_6$ | $H_3N + 1.2\,Y_4C_5 + 0.1429\,B_5H_7 + 50.89\,B \longrightarrow 3.8\,YB_{12} + YB_6H_4C_6N$ | -3.6155 |
| $YNH_4B_6C_6$ | $6.683\,H_3N + 0.8571\,Y_4C_7 + 2.921\,YB_4 \longrightarrow 5.349\,YH_3 + 5.683\,BN + YB_6H_4C_6N$ | -3.6077 |
| $YNH_4B_6C_6$ | $1.2\,Y_4C_5 + 0.5714\,B_5H_7 + 0.5\,N_2 + 18.34\,B \longrightarrow 3.8\,YB_4 + YB_6H_4C_6N$ | -3.6048 |
| $YNH_4B_6C_6$ | $1.333\,YH_3 + 1.2\,Y_4C_5 + BN + 66.6\,B \longrightarrow 5.133\,YB_{12} + YB_6H_4C_6N$ | -3.6029 |
| $YNH_4B_6C_6$ | $0.6667\,B_9H_{11} + 1.2\,Y_4C_5 + 1.844\,N_2 \longrightarrow 1.111\,YH_3 + 2.689\,YN + YB_6H_4C_6N$ | -3.6026 |
| $YNH_4B_6C_6$ | $0.09091\,B_9H_{11} + H_3N + 1.2\,Y_4C_5 + 50.78\,B \longrightarrow 3.8\,YB_{12} + YB_6H_4C_6N$ | -3.5999 |
| $YNH_4B_6C_6$ | $26\,YH_3 + 6\,Y_2C + BN + 5\,B \longrightarrow 37\,YH_2 + YB_6H_4C_6N$ | -3.5998 |
| $YNH_4B_6C_6$ | $1.333\,YH_3 + 0.5\,YB_{12} + 2.631\,N_2 + 0.8571\,Y_4C_7 \longrightarrow 4.262\,YN + YB_6H_4C_6N$ | -3.5934 |
| $YNH_4B_6C_6$ | $1.2\,Y_4C_5 + 2.65\,YB_{12} + 0.5\,N_2 + 2\,H_2 \longrightarrow 6.45\,YB_4 + YB_6H_4C_6N$ | -3.5835 |
| $YNH_4B_6C_6$ | $1.736\,B_{13}C_2 + 2.529\,Y_2C + 1.333\,H_3N \longrightarrow 0.3333\,BN + 4.057\,YB_4 + YB_6H_4C_6N$ | -3.5773 |
| $YNH_4B_6C_6$ | $0.6667\,B_9H_{11} + 5.311\,Y + 1.2\,H_8C_5 + 0.5\,N_2 \longrightarrow 4.311\,YH_3 + YB_6H_4C_6N$ | -3.5469 |
| $YNH_4B_6C_6$ | $0.3636\,B_9H_{11} + 1.2\,Y_4C_5 + 0.5\,N_2 + 17.93\,B \longrightarrow 3.8\,YB_4 + YB_6H_4C_6N$ | -3.5426 |
| $YNH_4B_6C_6$ | $2\,YH_2 + YN + 0.8571\,Y_4C_7 + 71.14\,B \longrightarrow 5.429\,YB_{12} + YB_6H_4C_6N$ | -3.5328 |
| $YNH_4B_6C_6$ | $1.678\,B_{13}C_2 + 2.644\,Y_2C + 1.333\,H_3N \longrightarrow 0.3333\,YN + 3.954\,YB_4 + YB_6H_4C_6N$ | -3.5217 |
| $YNH_4B_6C_6$ | $26\,YH_3 + 6\,Y_2C + H_3N + 1.5\,YB_4 \longrightarrow 38.5\,YH_2 + YB_6H_4C_6N$ | -3.5124 |
| $YNH_4B_6C_6$ | $0.4615\,B_{13}C_2 + 1.015\,Y_4C_5 + 2.031\,N_2 + 2\,H_2 \longrightarrow 3.062\,YN + YB_6H_4C_6N$ | -3.4871 |
| $YNH_4B_6C_6$ | $6\,Y_2C + 2.7\,YB_{12} + 2\,H_2 + BN \longrightarrow 13.7\,YB_2 + YB_6H_4C_6N$ | -3.4828 |
| $YNH_4B_6C_6$ | $1.333\,YH_3 + 2.381\,N_2 + 0.8571\,Y_4C_7 + 6\,B \longrightarrow 3.762\,YN + YB_6H_4C_6N$ | -3.4775 |
| $YNH_4B_6C_6$ | $34\,YH_3 + 6\,Y_2C + YN + 3\,YB_2 \longrightarrow 49\,YH_2 + YB_6H_4C_6N$ | -3.4643 |
| $YNH_4B_6C_6$ | $0.09091\,B_9H_{11} + 1.696\,B_{13}C_2 + 2.608\,Y_2C + H_3N \longrightarrow 4.216\,YB_4 + YB_6H_4C_6N$ | -3.4438 |
| $YNH_4B_6C_6$ | $1.7\,B_{13}C_2 + 2.601\,Y_2C + H_3N + 0.1429\,B_5H_7 \longrightarrow 4.202\,YB_4 + YB_6H_4C_6N$ | -3.4191 |
| $YNH_4B_6C_6$ | $2\,YH_2 + 1.2\,Y_4C_5 + BN + 28.2\,B \longrightarrow 5.8\,YB_4 + YB_6H_4C_6N$ | -3.4086 |
| $YNH_4B_6C_6$ | $1.333\,H_3N + 1.2\,Y_4C_5 + 47.6\,B \longrightarrow 3.467\,YB_{12} + 0.3333\,YN + YB_6H_4C_6N$ | -3.4007 |
| $YNH_4B_6C_6$ | $19.31\,YH_3 + 0.4615\,B_{13}C_2 + 5.077\,Y_2C + H_3N \longrightarrow 28.46\,YH_2 + YB_6H_4C_6N$ | -3.3706 |
| $YNH_4B_6C_6$ | $29\,YH_3 + 6\,Y_2C + 0.5\,YB_{12} + YN \longrightarrow 41.5\,YH_2 + YB_6H_4C_6N$ | -3.3263 |
| $YNH_4B_6C_6$ | $1.333\,H_3N + 1.2\,Y_4C_5 + 51.6\,B \longrightarrow 3.8\,YB_{12} + 0.1667\,N_2 + YB_6H_4C_6N$ | -3.3234 |
| $YNH_4B_6C_6$ | $1.333\,YH_3 + 1.908\,B_{13}C_2 + 2.184\,Y_2C + 0.5\,N_2 \longrightarrow 4.701\,YB_4 + YB_6H_4C_6N$ | -3.3218 |
| $YNH_4B_6C_6$ | $0.5\,YH_2 + H_3N + 1.2\,Y_4C_5 + 23.2\,B \longrightarrow 4.3\,YB_4 + YB_6H_4C_6N$ | -3.3213 |
| $YNH_4B_6C_6$ | $0.6667\,B_9H_{11} + 6\,Y_2C + 9.889\,H_3N \longrightarrow 11\,YH_3 + 4.444\,N_2 + YB_6H_4C_6N$ | -3.3133 |
| $YNH_4B_6C_6$ | $6\,Y_2C + 2.414\,YB_{12} + 0.5714\,B_5H_7 + BN \longrightarrow 13.41\,YB_2 + YB_6H_4C_6N$ | -3.3128 |
| $YNH_4B_6C_6$ | $0.3636\,B_9H_{11} + 6\,Y_2C + 2.373\,YB_{12} + BN \longrightarrow 13.37\,YB_2 + YB_6H_4C_6N$ | -3.3126 |
| $YNH_4B_6C_6$ | $21.12\,B_9H_{11} + 6\,Y_2C + BN \longrightarrow 32.62\,B_5H_7 + 11\,YB_2 + YB_6H_4C_6N$ | -3.3049 |
| $YNH_4B_6C_6$ | $0.6667\,B_9H_{11} + 6.467\,Y + 1.2\,H_8C_5 + YN \longrightarrow 6.467\,YH_2 + YB_6H_4C_6N$ | -3.3006 |

| Compound | Chemical reaction | Reaction energy(eV) |
|---|---|---|
| $YNH_4B_6C_6$ | $2\,YH_2 + 2\,B_{13}C_2 + 2\,Y_2C + 0.5\,N_2 \longrightarrow 5\,YB_4 + YB_6H_4C_6N$ | -3.2935 |
| $YNH_4B_6C_6$ | $1.333\,H_3N + 1.2\,Y_4C_5 + 21.53\,B \longrightarrow 0.3333\,BN + 3.8\,YB_4 + YB_6H_4C_6N$ | -3.2922 |
| $YNH_4B_6C_6$ | $1.724\,B_{13}C_2 + 2.552\,Y_2C + H_3N + 0.5\,H_2 \longrightarrow 4.103\,YB_4 + YB_6H_4C_6N$ | -3.2920 |
| $YNH_4B_6C_6$ | $117.4\,B_9H_{11} + 6\,Y_2C + BN \longrightarrow 11\,YB_{12} + 183.9\,B_5H_7 + YB_6H_4C_6N$ | -3.2692 |
| $YNH_4B_6C_6$ | $0.3636\,B_9H_{11} + 1.749\,B_{13}C_2 + 2.502\,Y_2C + YN \longrightarrow 5.003\,YB_4 + YB_6H_4C_6N$ | -3.2100 |
| $YNH_4B_6C_6$ | $2.125\,B_{13}C_2 + 7.497\,H_3N + 0.25\,Y_4C_7 \longrightarrow 1.681\,B_9H_{11} + 6.497\,BN + YB_6H_4C_6N$ | -3.2045 |
| $YNH_4B_6C_6$ | $1.2\,Y_4C_5 + 2.72\,B_5H_7 + 3.007\,N_2 \longrightarrow 5.013\,H_3N + 3.8\,YB_2 + YB_6H_4C_6N$ | -3.2030 |
| $YNH_4B_6C_6$ | $1.2\,Y_4C_5 + 1.2\,B_5H_7 + 1.667\,N_2 \longrightarrow 1.467\,YH_3 + 2.333\,YN + YB_6H_4C_6N$ | -3.1869 |
| $YNH_4B_6C_6$ | $1.612\,B_5H_7 + 1.531\,N_2 + 0.8571\,Y_4C_7 \longrightarrow 2.429\,YH_3 + 2.061\,BN + YB_6H_4C_6N$ | -3.1618 |
| $YNH_4B_6C_6$ | $1.2\,Y_4C_5 + 0.5714\,B_5H_7 + YN + 22.34\,B \longrightarrow 4.8\,YB_4 + YB_6H_4C_6N$ | -3.1396 |
| $YNH_4B_6C_6$ | $5.375\,Y_2C + H_3N + 2.55\,YB_{12} + 0.125\,H_8C_5 \longrightarrow 12.3\,YB_2 + YB_6H_4C_6N$ | -3.1266 |
| $YNH_4B_6C_6$ | $3.111\,B_9H_{11} + 6\,Y_2C + H_3N \longrightarrow 16.61\,H_2 + 11\,YB_2 + YB_6H_4C_6N$ | -3.1204 |
| $YNH_4B_6C_6$ | $1.764\,B_{13}C_2 + 2.473\,Y_2C + 0.5714\,B_5H_7 + YN \longrightarrow 4.946\,YB_4 + YB_6H_4C_6N$ | -3.1115 |
| $YNH_4B_6C_6$ | $0.5556\,B_9H_{11} + 6.856\,Y + 1.2\,H_8C_5 + BN \longrightarrow 5.856\,YH_2 + YB_6H_4C_6N$ | -3.1082 |
| $YNH_4B_6C_6$ | $0.3636\,B_9H_{11} + 1.2\,Y_4C_5 + YN + 21.93\,B \longrightarrow 4.8\,YB_4 + YB_6H_4C_6N$ | -3.0774 |
| $YNH_4B_6C_6$ | $3.867\,H_3N + 1.2\,Y_4C_5 + 8.867\,B \longrightarrow 3.8\,YH_2 + 2.867\,BN + YB_6H_4C_6N$ | -3.0710 |
| $YNH_4B_6C_6$ | $0.3333\,YH_3 + H_3N + 1.2\,Y_4C_5 + 22.53\,B \longrightarrow 4.133\,YB_4 + YB_6H_4C_6N$ | -3.0703 |
| $YNH_4B_6C_6$ | $1.2\,Y_4C_5 + 2.293\,YB_{12} + 0.5714\,B_5H_7 + 0.5\,N_2 \longrightarrow 6.093\,YB_4 + YB_6H_4C_6N$ | -3.0697 |
| $YNH_4B_6C_6$ | $0.3636\,B_9H_{11} + 1.577\,B_{13}C_2 + 2.846\,Y_2C + BN \longrightarrow 4.693\,YB_4 + YB_6H_4C_6N$ | -3.0434 |
| $YNH_4B_6C_6$ | $6\,C + 5.667\,H_3N + 5.333\,YB_2 \longrightarrow 4.333\,YH_3 + 4.667\,BN + YB_6H_4C_6N$ | -3.0277 |
| $YNH_4B_6C_6$ | $1.2\,Y_4C_5 + YN + 13.7\,H_2 + 3\,YB_2 \longrightarrow 7.8\,YH_3 + YB_6H_4C_6N$ | -3.0196 |
| $YNH_4B_6C_6$ | $0.3636\,B_9H_{11} + 1.2\,Y_4C_5 + 2.241\,YB_{12} + 0.5\,N_2 \longrightarrow 6.041\,YB_4 + YB_6H_4C_6N$ | -3.0196 |
| $YNH_4B_6C_6$ | $1.2\,Y_4C_5 + 3.15\,YB_{12} + YN + 2\,H_2 \longrightarrow 7.95\,YB_4 + YB_6H_4C_6N$ | -3.0016 |
| $YNH_4B_6C_6$ | $31\,YH_3 + 6\,Y_2C + BN + 2.5\,YB_2 \longrightarrow 44.5\,YH_2 + YB_6H_4C_6N$ | -2.9902 |
| $YNH_4B_6C_6$ | $23.75\,YH_2 + 3\,B_{13}C_2 + 0.5\,N_2 \longrightarrow 14.5\,YH_3 + 8.25\,YB_4 + YB_6H_4C_6N$ | -2.9862 |
| $YNH_4B_6C_6$ | $0.3333\,YH_3 + 1.77\,B_{13}C_2 + 2.46\,Y_2C + H_3N \longrightarrow 4.253\,YB_4 + YB_6H_4C_6N$ | -2.9684 |
| $YNH_4B_6C_6$ | $0.5\,YH_2 + 1.793\,B_{13}C_2 + 2.414\,Y_2C + H_3N \longrightarrow 4.328\,YB_4 + YB_6H_4C_6N$ | -2.9614 |
| $YNH_4B_6C_6$ | $1.591\,B_{13}C_2 + 2.818\,Y_2C + 0.5714\,B_5H_7 + BN \longrightarrow 4.635\,YB_4 + YB_6H_4C_6N$ | -2.9448 |
| $YNH_4B_6C_6$ | $5.64\,H_3N + 1.2\,Y_4C_5 + 2.66\,YB_4 \longrightarrow 6.46\,YH_2 + 4.64\,BN + YB_6H_4C_6N$ | -2.9162 |
| $YNH_4B_6C_6$ | $6\,Y_2C + H_3N + 5.6\,B_5H_7 \longrightarrow 19.1\,H_2 + 11\,YB_2 + YB_6H_4C_6N$ | -2.9101 |
| $YNH_4B_6C_6$ | $2\,YH_2 + 0.4615\,B_{13}C_2 + 2.451\,N_2 + 0.7253\,Y_4C_7 \longrightarrow 3.901\,YN + YB_6H_4C_6N$ | -2.9059 |
| $YNH_4B_6C_6$ | $H_3N + 1.2\,Y_4C_5 + 0.5\,H_2 + 21.2\,B \longrightarrow 3.8\,YB_4 + YB_6H_4C_6N$ | -2.8779 |
| $YNH_4B_6C_6$ | $26.83\,YH_3 + 6\,Y_2C + 0.4167\,YB_{12} + BN \longrightarrow 38.25\,YH_2 + YB_6H_4C_6N$ | -2.8752 |
| $YNH_4B_6C_6$ | $1.724\,B_{13}C_2 + 2.552\,Y_2C + 1.333\,H_3N \longrightarrow 0.1667\,N_2 + 4.103\,YB_4 + YB_6H_4C_6N$ | -2.8507 |
| $YNH_4B_6C_6$ | $4.388\,H_3N + 1.2\,Y_4C_5 + 0.7824\,YB_{12} \longrightarrow 4.582\,YH_2 + 3.388\,BN + YB_6H_4C_6N$ | -2.8429 |
| $YNH_4B_6C_6$ | $1.75\,B_{13}C_2 + 4.188\,Y + 0.5\,H_8C_5 + YN \longrightarrow 4.188\,YB_4 + YB_6H_4C_6N$ | -2.8130 |
| $YNH_4B_6C_6$ | $3.5\,Y_2C + 0.5\,H_8C_5 + 0.5\,N_2 + 18\,B \longrightarrow 6\,YB_2 + YB_6H_4C_6N$ | -2.7937 |
| $YNH_4B_6C_6$ | $2.464\,N_2 + 0.8571\,Y_4C_7 + 2\,H_2 + 1.5\,YB_4 \longrightarrow 3.929\,YN + YB_6H_4C_6N$ | -2.7854 |
| $YNH_4B_6C_6$ | $1.2\,B_5H_7 + 2.448\,N_2 + 0.8571\,Y_4C_7 \longrightarrow 1.467\,H_3N + 2.429\,YN + YB_6H_4C_6N$ | -2.7755 |
| $YNH_4B_6C_6$ | $H_3N + 1.2\,Y_4C_5 + 10.7\,H_2 + 3\,YB_2 \longrightarrow 6.8\,YH_3 + YB_6H_4C_6N$ | -2.7380 |
| $YNH_4B_6C_6$ | $6\,Y_2C + 8.667\,H_3N + 6\,B \longrightarrow 11\,YH_2 + 3.833\,N_2 + YB_6H_4C_6N$ | -2.7367 |
| $YNH_4B_6C_6$ | $2\,YH_2 + 0.5\,N_2 + 0.8571\,Y_4C_7 + 23.71\,B \longrightarrow 4.429\,YB_4 + YB_6H_4C_6N$ | -2.7311 |
| $YNH_4B_6C_6$ | $H_3N + 1.2\,Y_4C_5 + 0.1429\,B_5H_7 + 20.49\,B \longrightarrow 3.8\,YB_4 + YB_6H_4C_6N$ | -2.7286 |
| $YNH_4B_6C_6$ | $0.09091\,B_9H_{11} + H_3N + 1.2\,Y_4C_5 + 20.38\,B \longrightarrow 3.8\,YB_4 + YB_6H_4C_6N$ | -2.7131 |
| $YNH_4B_6C_6$ | $1.333\,H_3N + 1.2\,Y_4C_5 + 2.692\,YB_{12} \longrightarrow 0.3333\,BN + 6.492\,YB_4 + YB_6H_4C_6N$ | -2.6640 |
| $YNH_4B_6C_6$ | $0.5\,YH_2 + H_3N + 1.2\,Y_4C_5 + 2.9\,YB_{12} \longrightarrow 7.2\,YB_4 + YB_6H_4C_6N$ | -2.6445 |
| $YNH_4B_6C_6$ | $1.333\,YH_3 + 0.5\,N_2 + 0.8571\,Y_4C_7 + 51.14\,B \longrightarrow 3.762\,YB_{12} + YB_6H_4C_6N$ | -2.6054 |
| $YNH_4B_6C_6$ | $1.862\,B_{13}C_2 + 2.276\,Y_2C + YN + 2\,H_2 \longrightarrow 4.552\,YB_4 + YB_6H_4C_6N$ | -2.6029 |
| $YNH_4B_6C_6$ | $1.333\,H_3N + 1.2\,Y_4C_5 + 19.87\,B \longrightarrow 0.3333\,YN + 3.467\,YB_4 + YB_6H_4C_6N$ | -2.5916 |
| $YNH_4B_6C_6$ | $26.75\,YH_2 + 3\,B_{13}C_2 + H_3N \longrightarrow 17.5\,YH_3 + 8.25\,YB_4 + YB_6H_4C_6N$ | -2.5905 |
| $YNH_4B_6C_6$ | $0.5\,H_8C_5 + 2.5\,N_2 + 0.5\,Y_4C_7 + 3\,YB_2 \longrightarrow 4\,YN + YB_6H_4C_6N$ | -2.5890 |
| $YNH_4B_6C_6$ | $2\,YH_2 + 1.2\,Y_4C_5 + 3.525\,YB_{12} + BN \longrightarrow 9.325\,YB_4 + YB_6H_4C_6N$ | -2.5859 |
| $YNH_4B_6C_6$ | $1.333\,YH_3 + 6\,Y_2C + 3.267\,YB_{12} + YN \longrightarrow 16.6\,YB_2 + YB_6H_4C_6N$ | -2.5627 |
| $YNH_4B_6C_6$ | $6\,C + 4.7\,Y + H_3N + 1.2\,B_5H_7 \longrightarrow 3.7\,YH_2 + YB_6H_4C_6N$ | -2.5558 |
| $YNH_4B_6C_6$ | $2\,YH_2 + 6\,Y_2C + 3.4\,YB_{12} + YN \longrightarrow 17.4\,YB_2 + YB_6H_4C_6N$ | -2.5259 |
| $YNH_4B_6C_6$ | $1.2\,Y_4C_5 + 2\,H_2 + BN + 50.6\,B \longrightarrow 3.8\,YB_{12} + YB_6H_4C_6N$ | -2.5221 |
| $YNH_4B_6C_6$ | $0.6667\,B_9H_{11} + 1.2\,Y_4C_5 + 1.567\,N_2 \longrightarrow 1.667\,YH_2 + 2.133\,YN + YB_6H_4C_6N$ | -2.5077 |
| $YNH_4B_6C_6$ | $1.2\,Y_4C_5 + 2.793\,YB_{12} + 0.5714\,B_5H_7 + YN \longrightarrow 7.593\,YB_4 + YB_6H_4C_6N$ | -2.4878 |
| $YNH_4B_6C_6$ | $6\,Y_2C + 2.818\,H_3N + 9.182\,BN \longrightarrow 0.6364\,B_5H_7 + 11\,YN + YB_6H_4C_6N$ | -2.4485 |
| $YNH_4B_6C_6$ | $0.3636\,B_9H_{11} + 1.2\,Y_4C_5 + 2.741\,YB_{12} + YN \longrightarrow 7.541\,YB_4 + YB_6H_4C_6N$ | -2.4377 |

| Compound | Chemical reaction | Reaction energy(eV) |
|---|---|---|
| $YNH_4B_6C_6$ | $1.333\,H_3N + 1.2\,Y_4C_5 + 21.2\,B \longrightarrow 0.1667\,N_2 + 3.8\,YB_4 + YB_6H_4C_6N$ | -2.4366 |
| $YNH_4B_6C_6$ | $1.69\,B_{13}C_2 + 2.621\,Y_2C + 2\,H_2 + BN \longrightarrow 4.241\,YB_4 + YB_6H_4C_6N$ | -2.4363 |
| $YNH_4B_6C_6$ | $0.3333\,YH_3 + H_3N + 1.2\,Y_4C_5 + 2.817\,YB_{12} \longrightarrow 6.95\,YB_4 + YB_6H_4C_6N$ | -2.4130 |
| $YNH_4B_6C_6$ | $1.333\,YH_3 + 1.2\,Y_4C_5 + BN + 25.53\,B \longrightarrow 5.133\,YB_4 + YB_6H_4C_6N$ | -2.4049 |
| $YNH_4B_6C_6$ | $0.3636\,B_9H_{11} + 3\,B_{13}C_2 + 4.023\,Y + 0.5\,N_2 \longrightarrow 3.023\,YB_{12} + YB_6H_4C_6N$ | -2.4033 |
| $YNH_4B_6C_6$ | $1.333\,YH_3 + YN + 0.8571\,Y_4C_7 + 63.14\,B \longrightarrow 4.762\,YB_{12} + YB_6H_4C_6N$ | -2.3735 |
| $YNH_4B_6C_6$ | $3\,B_{13}C_2 + 3.988\,Y + 0.5714\,B_5H_7 + 0.5\,N_2 \longrightarrow 2.988\,YB_{12} + YB_6H_4C_6N$ | -2.3537 |
| $YNH_4B_6C_6$ | $0.6667\,B_9H_{11} + 2.27\,N_2 + 0.8571\,Y_4C_7 \longrightarrow 1.111\,H_3N + 2.429\,YN + YB_6H_4C_6N$ | -2.3498 |
| $YNH_4B_6C_6$ | $3.333\,B_9H_{11} + 6\,Y_2C + YN \longrightarrow 16.33\,H_2 + 12\,YB_2 + YB_6H_4C_6N$ | -2.3146 |
| $YNH_4B_6C_6$ | $H_3N + 1.075\,Y_4C_5 + 0.125\,H_8C_5 + 45.6\,B \longrightarrow 3.3\,YB_{12} + YB_6H_4C_6N$ | -2.3065 |
| $YNH_4B_6C_6$ | $6\,Y_2C + 2.684\,H_3N + 9.316\,BN \longrightarrow 0.3684\,B_9H_{11} + 11\,YN + YB_6H_4C_6N$ | -2.2851 |
| $YNH_4B_6C_6$ | $2\,YH_2 + YN + 0.8571\,Y_4C_7 + 27.71\,B \longrightarrow 5.429\,YB_4 + YB_6H_4C_6N$ | -2.2659 |
| $YNH_4B_6C_6$ | $H_3N + 1.2\,Y_4C_5 + 2.65\,YB_{12} + 0.5\,H_2 \longrightarrow 6.45\,YB_4 + YB_6H_4C_6N$ | -2.2595 |
| $YNH_4B_6C_6$ | $1.511\,B_9H_{11} + 1.2\,Y_4C_5 + 2.604\,N_2 \longrightarrow 4.207\,H_3N + 3.8\,YB_2 + YB_6H_4C_6N$ | -2.2381 |
| $YNH_4B_6C_6$ | $1.333\,YH_3 + 3\,B_{13}C_2 + 6.533\,YN \longrightarrow 5.533\,BN + 6.867\,YB_4 + YB_6H_4C_6N$ | -2.2307 |
| $YNH_4B_6C_6$ | $6\,Y_2C + 18.5\,H_2 + 6\,BN \longrightarrow 11\,YH_3 + 2.5\,N_2 + YB_6H_4C_6N$ | -2.2152 |
| $YNH_4B_6C_6$ | $0.5\,YB_{12} + 1.964\,N_2 + 0.8571\,Y_4C_7 + 2\,H_2 \longrightarrow 2.929\,YN + YB_6H_4C_6N$ | -2.2035 |
| $YNH_4B_6C_6$ | $3\,B_{13}C_2 + 3.75\,Y + 0.5\,N_2 + 2\,H_2 \longrightarrow 2.75\,YB_{12} + YB_6H_4C_6N$ | -2.1815 |
| $YNH_4B_6C_6$ | $1.333\,YH_3 + 6\,Y_2C + 2.967\,YB_{12} + BN \longrightarrow 15.3\,YB_2 + YB_6H_4C_6N$ | -2.1714 |
| $YNH_4B_6C_6$ | $0.3636\,B_9H_{11} + 1.177\,B_{13}C_2 + 3.645\,Y_2C + 0.5\,N_2 \longrightarrow 6.29\,YB_2 + YB_6H_4C_6N$ | -2.1551 |
| $YNH_4B_6C_6$ | $8.8\,Y + 1.2\,H_8C_5 + 6\,BN \longrightarrow 2.8\,YH_2 + 5\,YN + YB_6H_4C_6N$ | -2.1463 |
| $YNH_4B_6C_6$ | $2\,YH_2 + 6\,Y_2C + 3.1\,YB_{12} + BN \longrightarrow 16.1\,YB_2 + YB_6H_4C_6N$ | -2.1346 |
| $YNH_4B_6C_6$ | $H_3N + 1.2\,Y_4C_5 + 2.561\,YB_{12} + 0.1429\,B_5H_7 \longrightarrow 6.361\,YB_4 + YB_6H_4C_6N$ | -2.1310 |
| $YNH_4B_6C_6$ | $0.09091\,B_9H_{11} + H_3N + 1.2\,Y_4C_5 + 2.548\,YB_{12} \longrightarrow 6.348\,YB_4 + YB_6H_4C_6N$ | -2.1185 |
| $YNH_4B_6C_6$ | $2\,YH_2 + 3\,B_{13}C_2 + 6\,YN \longrightarrow 5\,BN + 7\,YB_4 + YB_6H_4C_6N$ | -2.1136 |
| $YNH_4B_6C_6$ | $1.197\,B_{13}C_2 + 3.605\,Y_2C + 0.5714\,B_5H_7 + 0.5\,N_2 \longrightarrow 6.211\,YB_2 + YB_6H_4C_6N$ | -2.0952 |
| $YNH_4B_6C_6$ | $6\,C + 7.333\,Y + 1.333\,H_3N + 6\,BN \longrightarrow 6.333\,YN + YB_6H_4C_6N$ | -2.0892 |
| $YNH_4B_6C_6$ | $6\,Y_2C + 6\,B_5H_7 + YN \longrightarrow 19\,H_2 + 12\,YB_2 + YB_6H_4C_6N$ | -2.0892 |
| $YNH_4B_6C_6$ | $1.714\,N_2 + 0.8571\,Y_4C_7 + 2\,H_2 + 6\,B \longrightarrow 2.429\,YN + YB_6H_4C_6N$ | -2.0876 |
| $YNH_4B_6C_6$ | $3\,B_9H_{11} + 6\,Y_2C + BN \longrightarrow 14.5\,H_2 + 11\,YB_2 + YB_6H_4C_6N$ | -2.0792 |
| $YNH_4B_6C_6$ | $5.3\,Y + H_3N + 1.2\,H_8C_5 + 6\,B \longrightarrow 4.3\,YH_2 + YB_6H_4C_6N$ | -2.0743 |
| $YNH_4B_6C_6$ | $3.393\,Y_2C + H_3N + 0.5214\,H_8C_5 + 1.2\,B_5H_7 \longrightarrow 5.786\,YH_2 + YB_6H_4C_6N$ | -2.0731 |
| $YNH_4B_6C_6$ | $2\,YH_2 + 2.964\,YB_{12} + 0.5\,N_2 + 0.8571\,Y_4C_7 \longrightarrow 7.393\,YB_4 + YB_6H_4C_6N$ | -2.0393 |
| $YNH_4B_6C_6$ | $1.333\,H_3N + 1.2\,Y_4C_5 + 2.483\,YB_{12} \longrightarrow 0.3333\,YN + 5.95\,YB_4 + YB_6H_4C_6N$ | -2.0121 |
| $YNH_4B_6C_6$ | $1.2\,Y_4C_5 + 0.5\,N_2 + 7.7\,H_2 + 6\,B \longrightarrow 3.8\,YH_3 + YB_6H_4C_6N$ | -2.0086 |
| $YNH_4B_6C_6$ | $4.558\,H_3N + 0.7965\,YB_{12} + 0.8571\,Y_4C_7 \longrightarrow 3.225\,YH_3 + 3.558\,BN + YB_6H_4C_6N$ | -2.0083 |
| $YNH_4B_6C_6$ | $1.2\,Y_4C_5 + 0.5\,N_2 + 8.8\,H_2 + 3\,YB_2 \longrightarrow 6.8\,YH_2 + YB_6H_4C_6N$ | -1.9579 |
| $YNH_4B_6C_6$ | $31\,YH_3 + 6\,Y_2C + YN + 1.5\,YB_4 \longrightarrow 44.5\,YH_2 + YB_6H_4C_6N$ | -1.9374 |
| $YNH_4B_6C_6$ | $1.2\,Y_4C_5 + 0.5714\,B_5H_7 + BN + 47.74\,B \longrightarrow 3.8\,YB_{12} + YB_6H_4C_6N$ | -1.9249 |
| $YNH_4B_6C_6$ | $6\,Y_2C + 1.333\,H_3N + 9.733\,BN \longrightarrow 10.07\,YN + 0.9333\,YB_4 + YB_6H_4C_6N$ | -1.8998 |
| $YNH_4B_6C_6$ | $1.75\,B_{13}C_2 + 0.5\,H_8C_5 + BN + 10.87\,YB_2 \longrightarrow 9.875\,YB_4 + YB_6H_4C_6N$ | -1.8791 |
| $YNH_4B_6C_6$ | $6\,Y_2C + 5.4\,B_5H_7 + BN \longrightarrow 16.9\,H_2 + 11\,YB_2 + YB_6H_4C_6N$ | -1.8764 |
| $YNH_4B_6C_6$ | $0.6941\,B_{13}C_2 + 4.023\,H_3N + 0.9224\,Y_4C_5 \longrightarrow 2.689\,YH_3 + 3.023\,BN + YB_6H_4C_6N$ | -1.8651 |
| $YNH_4B_6C_6$ | $0.3636\,B_9H_{11} + 1.2\,Y_4C_5 + BN + 47.33\,B \longrightarrow 3.8\,YB_{12} + YB_6H_4C_6N$ | -1.8627 |
| $YNH_4B_6C_6$ | $0.5714\,B_5H_7 + 2.107\,N_2 + 0.8571\,Y_4C_7 + 0.7857\,YB_4 \longrightarrow 3.214\,YN + YB_6H_4C_6N$ | -1.8559 |
| $YNH_4B_6C_6$ | $1.333\,B_{13}C_2 + 3.333\,Y_2C + 0.5\,N_2 + 2\,H_2 \longrightarrow 5.667\,YB_2 + YB_6H_4C_6N$ | -1.8521 |
| $YNH_4B_6C_6$ | $1.552\,B_{13}C_2 + 2.272\,Y_2C + H_3N + 0.125\,H_8C_5 \longrightarrow 3.543\,YB_4 + YB_6H_4C_6N$ | -1.8266 |
| $YNH_4B_6C_6$ | $1.333\,H_3N + 1.2\,Y_4C_5 + 2.65\,YB_{12} \longrightarrow 0.1667\,N_2 + 6.45\,YB_4 + YB_6H_4C_6N$ | -1.8181 |
| $YNH_4B_6C_6$ | $24.31\,YH_3 + 0.4615\,B_{13}C_2 + 5.077\,Y_2C + YN \longrightarrow 34.46\,YH_2 + YB_6H_4C_6N$ | -1.7956 |
| $YNH_4B_6C_6$ | $0.3636\,B_9H_{11} + 2.055\,N_2 + 0.8571\,Y_4C_7 + 0.6818\,YB_4 \longrightarrow 3.11\,YN + YB_6H_4C_6N$ | -1.7453 |
| $YNH_4B_6C_6$ | $2\,YH_2 + 6\,C + 2.5\,N_2 + 3\,YB_2 \longrightarrow 4\,YN + YB_6H_4C_6N$ | -1.7440 |
| $YNH_4B_6C_6$ | $1.2\,Y_4C_5 + 1.2\,B_5H_7 + 1.3\,N_2 \longrightarrow 2.2\,YH_2 + 1.6\,YN + YB_6H_4C_6N$ | -1.7416 |
| $YNH_4B_6C_6$ | $6\,Y_2C + 9\,H_3N + 0.5\,YB_{12} \longrightarrow 11.5\,YH_2 + 4\,N_2 + YB_6H_4C_6N$ | -1.7353 |
| $YNH_4B_6C_6$ | $4.567\,H_3N + 1.2\,Y_4C_5 + 3\,YB_2 \longrightarrow 3.233\,YH_3 + 3.567\,YN + YB_6H_4C_6N$ | -1.7335 |
| $YNH_4B_6C_6$ | $1.333\,YH_3 + 0.5\,N_2 + 0.8571\,Y_4C_7 + 21.05\,B \longrightarrow 3.762\,YB_4 + YB_6H_4C_6N$ | -1.7274 |
| $YNH_4B_6C_6$ | $0.4615\,B_{13}C_2 + 1.333\,H_3N + 1.015\,Y_4C_5 + 1.364\,N_2 \longrightarrow 3.062\,YN + YB_6H_4C_6N$ | -1.7217 |
| $YNH_4B_6C_6$ | $28.5\,YH_3 + 6\,Y_2C + BN + 1.25\,YB_4 \longrightarrow 40.75\,YH_2 + YB_6H_4C_6N$ | -1.7178 |
| $YNH_4B_6C_6$ | $3.762\,H_3N + 0.8571\,Y_4C_7 + 8.762\,B \longrightarrow 2.429\,YH_3 + 2.762\,BN + YB_6H_4C_6N$ | -1.6641 |
| $YNH_4B_6C_6$ | $1.333\,YH_3 + 1.2\,Y_4C_5 + 3.192\,YB_{12} + BN \longrightarrow 8.325\,YB_4 + YB_6H_4C_6N$ | -1.6600 |
| $YNH_4B_6C_6$ | $1.2\,Y_4C_5 + 2\,H_2 + BN + 20.2\,B \longrightarrow 3.8\,YB_4 + YB_6H_4C_6N$ | -1.6352 |

| Compound | Chemical reaction | Reaction energy(eV) |
|---|---|---|
| $YNH_4B_6C_6$ | $0.5\,N_2 + 0.8571\,Y_4C_7 + 10.14\,H_2 + 3\,YB_2 \longrightarrow 5.429\,YH_3 + YB_6H_4C_6N$ | -1.6095 |
| $YNH_4B_6C_6$ | $1.2\,Y_4C_5 + 0.5\,YB_{12} + 0.5\,N_2 + 8.45\,H_2 \longrightarrow 4.3\,YH_3 + YB_6H_4C_6N$ | -1.6033 |
| $YNH_4B_6C_6$ | $22.92\,YH_3 + 0.3846\,B_{13}C_2 + 5.231\,Y_2C + BN \longrightarrow 32.38\,YH_2 + YB_6H_4C_6N$ | -1.5996 |
| $YNH_4B_6C_6$ | $1.717\,B_9H_{11} + 1.491\,Y_4C_5 + 0.5\,N_2 \longrightarrow 4.964\,YH_3 + 0.7275\,B_{13}C_2 + YB_6H_4C_6N$ | -1.5939 |
| $YNH_4B_6C_6$ | $1.333\,YH_3 + 0.4615\,B_{13}C_2 + 2.117\,N_2 + 0.7253\,Y_4C_7 \longrightarrow 3.234\,YN + YB_6H_4C_6N$ | -1.5921 |
| $YNH_4B_6C_6$ | $3\,B_{13}C_2 + 4.083\,Y + 1.333\,H_3N \longrightarrow 2.75\,YB_{12} + 0.3333\,YN + YB_6H_4C_6N$ | -1.5704 |
| $YNH_4B_6C_6$ | $0.2619\,YB_{12} + 0.5714\,B_5H_7 + 1.845\,N_2 + 0.8571\,Y_4C_7 \longrightarrow 2.69\,YN + YB_6H_4C_6N$ | -1.5511 |
| $YNH_4B_6C_6$ | $H_3N + 1.075\,Y_4C_5 + 0.125\,H_8C_5 + 19.2\,B \longrightarrow 3.3\,YB_4 + YB_6H_4C_6N$ | -1.5363 |
| $YNH_4B_6C_6$ | $0.5\,N_2 + 0.8571\,Y_4C_7 + 2\,H_2 + 35.14\,B \longrightarrow 2.429\,YB_{12} + YB_6H_4C_6N$ | -1.5246 |
| $YNH_4B_6C_6$ | $0.3636\,B_9H_{11} + 6\,Y_2C + 10.15\,BN \longrightarrow 9.145\,YN + 1.855\,YB_4 + YB_6H_4C_6N$ | -1.5194 |
| $YNH_4B_6C_6$ | $0.5714\,B_5H_7 + 1.714\,N_2 + 0.8571\,Y_4C_7 + 3.143\,B \longrightarrow 2.429\,YN + YB_6H_4C_6N$ | -1.4904 |
| $YNH_4B_6C_6$ | $0.3636\,B_9H_{11} + 0.2273\,YB_{12} + 1.828\,N_2 + 0.8571\,Y_4C_7 \longrightarrow 2.656\,YN + YB_6H_4C_6N$ | -1.4808 |
| $YNH_4B_6C_6$ | $0.6667\,B_9H_{11} + 6\,C + 4.167\,Y + H_3N \longrightarrow 3.167\,YH_2 + YB_6H_4C_6N$ | -1.4749 |
| $YNH_4B_6C_6$ | $2\,YH_2 + 3.464\,YB_{12} + YN + 0.8571\,Y_4C_7 \longrightarrow 8.893\,YB_4 + YB_6H_4C_6N$ | -1.4574 |
| $YNH_4B_6C_6$ | $2.857\,Y_2C + 0.6286\,H_8C_5 + 1.2\,B_5H_7 + 0.5\,N_2 \longrightarrow 4.714\,YH_2 + YB_6H_4C_6N$ | -1.4326 |
| $YNH_4B_6C_6$ | $0.3636\,B_9H_{11} + 1.714\,N_2 + 0.8571\,Y_4C_7 + 2.727\,B \longrightarrow 2.429\,YN + YB_6H_4C_6N$ | -1.4281 |
| $YNH_4B_6C_6$ | $0.6667\,B_9H_{11} + 3.202\,Y_2C + H_3N + 0.5595\,H_8C_5 \longrightarrow 5.405\,YH_2 + YB_6H_4C_6N$ | -1.4197 |
| $YNH_4B_6C_6$ | $6\,Y_2C + 0.5714\,B_5H_7 + 10.23\,BN \longrightarrow 9.229\,YN + 1.771\,YB_4 + YB_6H_4C_6N$ | -1.4070 |
| $YNH_4B_6C_6$ | $7.037\,Y_2C + 1.333\,H_3N + 12.74\,BN \longrightarrow 0.5185\,B_{13}C_2 + 13.07\,YN + YB_6H_4C_6N$ | -1.3986 |
| $YNH_4B_6C_6$ | $4.8\,H_3N + 1.2\,Y_4C_5 + 14.51\,B \longrightarrow 0.9455\,B_9H_{11} + 3.8\,YN + YB_6H_4C_6N$ | -1.3288 |
| $YNH_4B_6C_6$ | $1.333\,YH_3 + 2.046\,B_{13}C_2 + 1.908\,Y_2C + YN \longrightarrow 5.149\,YB_4 + YB_6H_4C_6N$ | -1.3087 |
| $YNH_4B_6C_6$ | $YN + 0.8571\,Y_4C_7 + 2\,H_2 + 47.14\,B \longrightarrow 3.429\,YB_{12} + YB_6H_4C_6N$ | -1.2927 |
| $YNH_4B_6C_6$ | $6\,Y_2C + 12.33\,H_3N + 6\,B \longrightarrow 11\,YH_3 + 5.667\,N_2 + YB_6H_4C_6N$ | -1.2856 |
| $YNH_4B_6C_6$ | $2\,YH_2 + 2.138\,B_{13}C_2 + 1.724\,Y_2C + YN \longrightarrow 5.448\,YB_4 + YB_6H_4C_6N$ | -1.2804 |
| $YNH_4B_6C_6$ | $1.409\,Y_4C_5 + 2.558\,B_5H_7 + 0.5\,N_2 \longrightarrow 4.636\,YH_3 + 0.5224\,B_{13}C_2 + YB_6H_4C_6N$ | -1.2738 |
| $YNH_4B_6C_6$ | $10.58\,Y + 1.2\,H_8C_5 + 10.58\,BN \longrightarrow 0.5091\,B_9H_{11} + 9.582\,YN + YB_6H_4C_6N$ | -1.2649 |
| $YNH_4B_6C_6$ | $1.333\,YH_3 + YN + 0.8571\,Y_4C_7 + 25.05\,B \longrightarrow 4.762\,YB_4 + YB_6H_4C_6N$ | -1.2622 |
| $YNH_4B_6C_6$ | $6\,Y_2C + 1.333\,H_3N + 9.111\,BN \longrightarrow 9.444\,YN + 1.556\,YB_2 + YB_6H_4C_6N$ | -1.2447 |
| $YNH_4B_6C_6$ | $6\,Y_2C + 10.67\,H_3N + 3\,YB_2 \longrightarrow 14\,YH_2 + 4.833\,N_2 + YB_6H_4C_6N$ | -1.2137 |
| $YNH_4B_6C_6$ | $2\,YH_2 + 0.8571\,Y_4C_7 + BN + 58.14\,B \longrightarrow 4.429\,YB_{12} + YB_6H_4C_6N$ | -1.1979 |
| $YNH_4B_6C_6$ | $3\,B_{13}C_2 + 3.722\,Y + 1.333\,H_3N \longrightarrow 2.722\,YB_{12} + 0.3333\,BN + YB_6H_4C_6N$ | -1.1819 |
| $YNH_4B_6C_6$ | $4.8\,H_3N + 1.2\,Y_4C_5 + 13.43\,B \longrightarrow 1.486\,B_5H_7 + 3.8\,YN + YB_6H_4C_6N$ | -1.1669 |
| $YNH_4B_6C_6$ | $1.2\,Y_4C_5 + 11.45\,H_2 + BN + 2.5\,YB_2 \longrightarrow 6.3\,YH_3 + YB_6H_4C_6N$ | -1.1531 |
| $YNH_4B_6C_6$ | $1.2\,Y_4C_5 + 0.5\,N_2 + 9.95\,H_2 + 1.5\,YB_4 \longrightarrow 5.3\,YH_3 + YB_6H_4C_6N$ | -1.1427 |
| $YNH_4B_6C_6$ | $1.333\,YH_3 + 1.874\,B_{13}C_2 + 2.253\,Y_2C + BN \longrightarrow 4.839\,YB_4 + YB_6H_4C_6N$ | -1.1420 |
| $YNH_4B_6C_6$ | $2\,YH_2 + 1.966\,B_{13}C_2 + 2.069\,Y_2C + BN \longrightarrow 5.138\,YB_4 + YB_6H_4C_6N$ | -1.1138 |
| $YNH_4B_6C_6$ | $1.333\,YH_3 + 2.631\,YB_{12} + 0.5\,N_2 + 0.8571\,Y_4C_7 \longrightarrow 6.393\,YB_4 + YB_6H_4C_6N$ | -1.1134 |
| $YNH_4B_6C_6$ | $1.2\,H_8C_5 + 2.95\,B_5H_7 + 9.75\,YN \longrightarrow 8.75\,YH_3 + 8.75\,BN + YB_6H_4C_6N$ | -1.1123 |
| $YNH_4B_6C_6$ | $3.5\,Y_2C + 0.5\,H_8C_5 + YN + 20\,B \longrightarrow 7\,YB_2 + YB_6H_4C_6N$ | -1.0668 |
| $YNH_4B_6C_6$ | $1.2\,Y_4C_5 + 2.525\,YB_{12} + 2\,H_2 + BN \longrightarrow 6.325\,YB_4 + YB_6H_4C_6N$ | -1.0460 |
| $YNH_4B_6C_6$ | $1.2\,Y_4C_5 + 0.5714\,B_5H_7 + BN + 17.34\,B \longrightarrow 3.8\,YB_4 + YB_6H_4C_6N$ | -1.0381 |
| $YNH_4B_6C_6$ | $1.333\,H_3N + 1.798\,N_2 + 0.8571\,Y_4C_7 + 1.5\,YB_4 \longrightarrow 3.929\,YN + YB_6H_4C_6N$ | -1.0200 |
| $YNH_4B_6C_6$ | $2.75\,B_9H_{11} + 1.714\,N_2 + 0.8571\,Y_4C_7 \longrightarrow 3.75\,B_5H_7 + 2.429\,YN + YB_6H_4C_6N$ | -1.0195 |
| $YNH_4B_6C_6$ | $20.75\,YH_2 + 3\,B_{13}C_2 + YN \longrightarrow 12.5\,YH_3 + 8.25\,YB_4 + YB_6H_4C_6N$ | -1.0155 |
| $YNH_4B_6C_6$ | $17.5\,YH_3 + 3\,B_{13}C_2 + 8.583\,N_2 \longrightarrow 16.17\,H_3N + 16.5\,YB_2 + YB_6H_4C_6N$ | -0.9865 |
| $YNH_4B_6C_6$ | $H_3N + 1.075\,Y_4C_5 + 2.4\,YB_{12} + 0.125\,H_8C_5 \longrightarrow 5.7\,YB_4 + YB_6H_4C_6N$ | -0.9762 |
| $YNH_4B_6C_6$ | $0.3636\,B_9H_{11} + 1.2\,Y_4C_5 + BN + 16.93\,B \longrightarrow 3.8\,YB_4 + YB_6H_4C_6N$ | -0.9758 |
| $YNH_4B_6C_6$ | $1.2\,Y_4C_5 + YN + 9.2\,H_2 + 6\,B \longrightarrow 4.8\,YH_3 + YB_6H_4C_6N$ | -0.9661 |
| $YNH_4B_6C_6$ | $0.5\,N_2 + 0.8571\,Y_4C_7 + 2\,H_2 + 15.71\,B \longrightarrow 2.429\,YB_4 + YB_6H_4C_6N$ | -0.9578 |
| $YNH_4B_6C_6$ | $0.5714\,B_5H_7 + 0.5\,N_2 + 0.8571\,Y_4C_7 + 32.29\,B \longrightarrow 2.429\,YB_{12} + YB_6H_4C_6N$ | -0.9274 |
| $YNH_4B_6C_6$ | $1.2\,H_8C_5 + 2.645\,N_2 + 5.291\,YB_2 \longrightarrow 0.5091\,B_9H_{11} + 4.291\,YN + YB_6H_4C_6N$ | -0.9174 |
| $YNH_4B_6C_6$ | $0.09091\,B_9H_{11} + 3\,B_{13}C_2 + 3.818\,Y + H_3N \longrightarrow 2.818\,YB_{12} + YB_6H_4C_6N$ | -0.9129 |
| $YNH_4B_6C_6$ | $3\,B_{13}C_2 + 3.81\,Y + H_3N + 0.1429\,B_5H_7 \longrightarrow 2.81\,YB_{12} + YB_6H_4C_6N$ | -0.9004 |
| $YNH_4B_6C_6$ | $4.667\,Y + 1.2\,H_8C_5 + 1.2\,B_5H_7 + YN \longrightarrow 4.667\,YH_3 + YB_6H_4C_6N$ | -0.8995 |
| $YNH_4B_6C_6$ | $0.6667\,B_9H_{11} + 1.714\,N_2 + 0.8571\,Y_4C_7 \longrightarrow 2.429\,YN + 1.667\,H_2 + YB_6H_4C_6N$ | -0.8786 |
| $YNH_4B_6C_6$ | $5.2\,Y + 1.2\,H_8C_5 + B_5H_7 + BN \longrightarrow 4.2\,YH_3 + YB_6H_4C_6N$ | -0.8751 |
| $YNH_4B_6C_6$ | $0.3636\,B_9H_{11} + 0.5\,N_2 + 0.8571\,Y_4C_7 + 31.87\,B \longrightarrow 2.429\,YB_{12} + YB_6H_4C_6N$ | -0.8651 |
| $YNH_4B_6C_6$ | $1.302\,B_{13}C_2 + 3.397\,Y_2C + 1.333\,H_3N \longrightarrow 0.3333\,YN + 5.46\,YB_2 + YB_6H_4C_6N$ | -0.8584 |
| $YNH_4B_6C_6$ | $3\,B_{13}C_2 + 3.75\,Y + H_3N + 0.5\,H_2 \longrightarrow 2.75\,YB_{12} + YB_6H_4C_6N$ | -0.8574 |
| $YNH_4B_6C_6$ | $16.05\,B_9H_{11} + 1.2\,Y_4C_5 + 0.5\,N_2 \longrightarrow 24.65\,B_5H_7 + 3.8\,YB_4 + YB_6H_4C_6N$ | -0.8564 |

| Compound | Chemical reaction | Reaction energy(eV) |
| --- | --- | --- |
| $YNH_4B_6C_6$ | $2.108\,B_9H_{11} + 1.599\,Y_4C_5 + YN \longrightarrow 6.397\,YH_3 + 0.998\,B_{13}C_2 + YB_6H_4C_6N$ | -0.8468 |
| $YNH_4B_6C_6$ | $1.349\,B_{13}C_2 + 3.302\,Y_2C + 1.333\,H_3N \longrightarrow 0.3333\,BN + 5.603\,YB_2 + YB_6H_4C_6N$ | -0.8443 |
| $YNH_4B_6C_6$ | $1.2\,B_5H_7 + 1.714\,N_2 + 0.8571\,Y_4C_7 \longrightarrow 2.429\,YN + 2.2\,H_2 + YB_6H_4C_6N$ | -0.8335 |
| $YNH_4B_6C_6$ | $1.2\,Y_4C_5 + 0.5\,N_2 + 5.8\,H_2 + 6\,B \longrightarrow 3.8\,YH_2 + YB_6H_4C_6N$ | -0.8327 |
| $YNH_4B_6C_6$ | $1.75\,B_{13}C_2 + 0.5\,H_8C_5 + YN + 8.375\,YB_2 \longrightarrow 8.375\,YB_4 + YB_6H_4C_6N$ | -0.8264 |
| $YNH_4B_6C_6$ | $0.4615\,B_{13}C_2 + 5.077\,Y_2C + 7.436\,H_3N \longrightarrow 9.154\,YH_2 + 3.218\,N_2 + YB_6H_4C_6N$ | -0.8236 |
| $YNH_4B_6C_6$ | $7.8\,H_3N + 1.2\,Y_4C_5 + 3\,YB_2 \longrightarrow 6.8\,YN + 9.7\,H_2 + YB_6H_4C_6N$ | -0.8228 |
| $YNH_4B_6C_6$ | $6\,Y_2C + 2\,H_2 + 10.8\,BN \longrightarrow 9.8\,YN + 1.2\,YB_4 + YB_6H_4C_6N$ | -0.8033 |
| $YNH_4B_6C_6$ | $0.6667\,B_9H_{11} + 1.2\,Y_4C_5 + 0.5\,N_2 + 4.033\,H_2 \longrightarrow 3.8\,YH_3 + YB_6H_4C_6N$ | -0.7996 |
| $YNH_4B_6C_6$ | $24.5\,YH_2 + 3\,B_{13}C_2 + BN \longrightarrow 15\,YH_3 + 8.5\,YB_4 + YB_6H_4C_6N$ | -0.7959 |
| $YNH_4B_6C_6$ | $0.6667\,B_9H_{11} + 2.667\,Y_2C + 0.6667\,H_8C_5 + 0.5\,N_2 \longrightarrow 4.333\,YH_2 + YB_6H_4C_6N$ | -0.7792 |
| $YNH_4B_6C_6$ | $6\,Y_2C + 1.333\,H_3N + 10.31\,BN \longrightarrow 0.359\,YB_{12} + 10.64\,YN + YB_6H_4C_6N$ | -0.7765 |
| $YNH_4B_6C_6$ | $0.5\,YH_2 + H_3N + 0.8571\,Y_4C_7 + 41.14\,B \longrightarrow 2.929\,YB_{12} + YB_6H_4C_6N$ | -0.7605 |
| $YNH_4B_6C_6$ | $1.2\,Y_4C_5 + 1.2\,B_5H_7 + 0.5\,N_2 + 3.5\,H_2 \longrightarrow 3.8\,YH_3 + YB_6H_4C_6N$ | -0.7545 |
| $YNH_4B_6C_6$ | $7.867\,Y + 1.2\,H_8C_5 + 6\,BN \longrightarrow 1.867\,YH_3 + 5\,YN + YB_6H_4C_6N$ | -0.7535 |
| $YNH_4B_6C_6$ | $0.8\,B_9H_{11} + 2.2\,H_3N + 1.2\,Y_4C_5 \longrightarrow 3.8\,YH_3 + 1.2\,BN + YB_6H_4C_6N$ | -0.7250 |
| $YNH_4B_6C_6$ | $6\,C + 3.467\,Y + H_3N + 1.2\,B_5H_7 \longrightarrow 2.467\,YH_3 + YB_6H_4C_6N$ | -0.7153 |
| $YNH_4B_6C_6$ | $3.92\,H_3N + 1.2\,Y_4C_5 + 3\,YB_2 \longrightarrow 3.88\,YH_2 + 2.92\,YN + YB_6H_4C_6N$ | -0.7150 |
| $YNH_4B_6C_6$ | $6\,C + 3.2\,Y + 1.2\,B_5H_7 + 0.5\,N_2 \longrightarrow 2.2\,YH_2 + YB_6H_4C_6N$ | -0.7131 |
| $YNH_4B_6C_6$ | $0.5714\,B_5H_7 + YN + 0.8571\,Y_4C_7 + 44.29\,B \longrightarrow 3.429\,YB_{12} + YB_6H_4C_6N$ | -0.6956 |
| $YNH_4B_6C_6$ | $H_3N + 1.2\,Y_4C_5 + 6.2\,H_2 + 6\,B \longrightarrow 3.8\,YH_3 + YB_6H_4C_6N$ | -0.6845 |
| $YNH_4B_6C_6$ | $10\,Y + 1.2\,H_8C_5 + 10\,BN \longrightarrow 0.8\,B_5H_7 + 9\,YN + YB_6H_4C_6N$ | -0.6562 |
| $YNH_4B_6C_6$ | $H_3N + 1.2\,Y_4C_5 + 7.3\,H_2 + 3\,YB_2 \longrightarrow 6.8\,YH_2 + YB_6H_4C_6N$ | -0.6338 |
| $YNH_4B_6C_6$ | $0.3636\,B_9H_{11} + YN + 0.8571\,Y_4C_7 + 43.87\,B \longrightarrow 3.429\,YB_{12} + YB_6H_4C_6N$ | -0.6333 |
| $YNH_4B_6C_6$ | $16\,YH_3 + 6\,Y_2C + 6\,BN \longrightarrow 22\,YH_2 + 5\,YN + YB_6H_4C_6N$ | -0.6197 |
| $YNH_4B_6C_6$ | $1.333\,H_3N + 0.8571\,Y_4C_7 + 35.48\,B \longrightarrow 2.429\,YB_{12} + 0.3333\,BN + YB_6H_4C_6N$ | -0.6147 |
| $YNH_4B_6C_6$ | $1.2\,Y_4C_5 + YN + 9.8\,H_2 + 3\,YB_2 \longrightarrow 7.8\,YH_2 + YB_6H_4C_6N$ | -0.6060 |
| $YNH_4B_6C_6$ | $0.09091\,B_9H_{11} + 1.294\,B_{13}C_2 + 3.411\,Y_2C + H_3N \longrightarrow 5.823\,YB_2 + YB_6H_4C_6N$ | -0.6038 |
| $YNH_4B_6C_6$ | $2.632\,Y_2C + H_3N + 0.6737\,H_8C_5 + 1.2\,B_5H_7 \longrightarrow 4.263\,YH_3 + YB_6H_4C_6N$ | -0.6006 |
| $YNH_4B_6C_6$ | $1.299\,B_{13}C_2 + 3.401\,Y_2C + H_3N + 0.1429\,B_5H_7 \longrightarrow 5.803\,YB_2 + YB_6H_4C_6N$ | -0.5888 |
| $YNH_4B_6C_6$ | $0.3636\,B_9H_{11} + 0.2098\,B_{13}C_2 + 1.594\,N_2 + 0.7972\,Y_4C_7 \longrightarrow 2.189\,YN + YB_6H_4C_6N$ | -0.5712 |
| $YNH_4B_6C_6$ | $2.171\,B_5H_7 + 2.367\,N_2 + 0.8571\,Y_4C_7 \longrightarrow 3.733\,H_3N + 2.429\,YB_2 + YB_6H_4C_6N$ | -0.5676 |
| $YNH_4B_6C_6$ | $YN + 0.8571\,Y_4C_7 + 11.64\,H_2 + 3\,YB_2 \longrightarrow 6.429\,YH_3 + YB_6H_4C_6N$ | -0.5670 |
| $YNH_4B_6C_6$ | $1.121\,B_9H_{11} + 1.2\,Y_4C_5 + 0.5\,N_2 \longrightarrow 2.777\,YH_3 + 1.023\,YB_4 + YB_6H_4C_6N$ | -0.5657 |
| $YNH_4B_6C_6$ | $1.2\,Y_4C_5 + 0.5\,YB_{12} + YN + 9.95\,H_2 \longrightarrow 5.3\,YH_3 + YB_6H_4C_6N$ | -0.5608 |
| $YNH_4B_6C_6$ | $0.8052\,B_9H_{11} + 1.123\,N_2 + 0.8571\,Y_4C_7 \longrightarrow 2.429\,YH_2 + 1.247\,BN + YB_6H_4C_6N$ | -0.5444 |
| $YNH_4B_6C_6$ | $1.2\,Y_4C_5 + 1.851\,B_5H_7 + 0.5\,N_2 \longrightarrow 2.986\,YH_3 + 0.814\,YB_4 + YB_6H_4C_6N$ | -0.5439 |
| $YNH_4B_6C_6$ | $1.2\,Y_4C_5 + 2.168\,YB_{12} + 0.5714\,B_5H_7 + BN \longrightarrow 5.968\,YB_4 + YB_6H_4C_6N$ | -0.5321 |
| $YNH_4B_6C_6$ | $1.333\,YH_3 + 3.131\,YB_{12} + YN + 0.8571\,Y_4C_7 \longrightarrow 7.893\,YB_4 + YB_6H_4C_6N$ | -0.5315 |
| $YNH_4B_6C_6$ | $1.333\,B_{13}C_2 + 3.333\,Y_2C + H_3N + 0.5\,H_2 \longrightarrow 5.667\,YB_2 + YB_6H_4C_6N$ | -0.5281 |
| $YNH_4B_6C_6$ | $0.3636\,B_9H_{11} + 8.061\,Y_2C + 16.12\,BN \longrightarrow 1.03\,B_{13}C_2 + 15.12\,YN + YB_6H_4C_6N$ | -0.5236 |
| $YNH_4B_6C_6$ | $0.2418\,B_{13}C_2 + 0.5714\,B_5H_7 + 1.576\,N_2 + 0.7881\,Y_4C_7 \longrightarrow 2.152\,YN + YB_6H_4C_6N$ | -0.5028 |
| $YNH_4B_6C_6$ | $1.964\,YB_{12} + 0.5\,N_2 + 0.8571\,Y_4C_7 + 2\,H_2 \longrightarrow 4.393\,YB_4 + YB_6H_4C_6N$ | -0.4994 |
| $YNH_4B_6C_6$ | $YN + 0.8571\,Y_4C_7 + 2\,H_2 + 19.71\,B \longrightarrow 3.429\,YB_4 + YB_6H_4C_6N$ | -0.4926 |
| $YNH_4B_6C_6$ | $0.3636\,B_9H_{11} + 1.2\,Y_4C_5 + 2.116\,YB_{12} + BN \longrightarrow 5.916\,YB_4 + YB_6H_4C_6N$ | -0.4820 |
| $YNH_4B_6C_6$ | $0.3333\,YH_3 + H_3N + 0.8571\,Y_4C_7 + 39.14\,B \longrightarrow 2.762\,YB_{12} + YB_6H_4C_6N$ | -0.4707 |
| $YNH_4B_6C_6$ | $4.8\,Y + H_3N + 0.5\,YB_{12} + 1.2\,H_8C_5 \longrightarrow 4.3\,YH_2 + YB_6H_4C_6N$ | -0.4587 |
| $YNH_4B_6C_6$ | $1.2\,Y_4C_5 + 9.075\,B_5H_7 + 0.5\,N_2 \longrightarrow 3.8\,YH_3 + 4.375\,B_9H_{11} + YB_6H_4C_6N$ | -0.4587 |
| $YNH_4B_6C_6$ | $7.968\,Y_2C + 0.5714\,B_5H_7 + 15.94\,BN \longrightarrow 0.9841\,B_{13}C_2 + 14.94\,YN + YB_6H_4C_6N$ | -0.4558 |
| $YNH_4B_6C_6$ | $1.499\,Y_4C_5 + 3.14\,B_5H_7 + YN \longrightarrow 5.994\,YH_3 + 0.7463\,B_{13}C_2 + YB_6H_4C_6N$ | -0.4540 |
| $YNH_4B_6C_6$ | $1.333\,H_3N + 0.5\,YB_{12} + 1.298\,N_2 + 0.8571\,Y_4C_7 \longrightarrow 2.929\,YN + YB_6H_4C_6N$ | -0.4381 |
| $YNH_4B_6C_6$ | $1.333\,YH_3 + 6\,C + 2.167\,N_2 + 3\,YB_2 \longrightarrow 3.333\,YN + YB_6H_4C_6N$ | -0.4302 |
| $YNH_4B_6C_6$ | $6.32\,H_3N + 1.2\,Y_4C_5 + 1.52\,YB_{12} \longrightarrow 1.36\,B_9H_{11} + 5.32\,YN + YB_6H_4C_6N$ | -0.4203 |
| $YNH_4B_6C_6$ | $3\,B_{13}C_2 + 3.75\,Y + 1.333\,H_3N \longrightarrow 2.75\,YB_{12} + 0.1667\,N_2 + YB_6H_4C_6N$ | -0.4160 |
| $YNH_4B_6C_6$ | $0.7\,Y_4C_5 + 0.5\,H_8C_5 + 2.15\,N_2 + 1.5\,YB_4 \longrightarrow 3.3\,YN + YB_6H_4C_6N$ | -0.3707 |
| $YNH_4B_6C_6$ | $1.909\,H_3N + 1.2\,Y_4C_5 + 1.382\,B_5H_7 \longrightarrow 3.8\,YH_3 + 0.9091\,BN + YB_6H_4C_6N$ | -0.3702 |
| $YNH_4B_6C_6$ | $0.5714\,B_5H_7 + 0.5\,N_2 + 0.8571\,Y_4C_7 + 12.86\,B \longrightarrow 2.429\,YB_4 + YB_6H_4C_6N$ | -0.3606 |
| $YNH_4B_6C_6$ | $0.4615\,B_{13}C_2 + 4.562\,Y + H_3N + 1.015\,H_8C_5 \longrightarrow 3.562\,YH_2 + YB_6H_4C_6N$ | -0.3576 |
| $YNH_4B_6C_6$ | $3\,B_{13}C_2 + 1.333\,H_3N + 11.56\,YN \longrightarrow 11.89\,BN + 10.56\,YB_2 + YB_6H_4C_6N$ | -0.3543 |
| $YNH_4B_6C_6$ | $1.2\,H_8C_5 + 2.5\,N_2 + 5\,YB_2 \longrightarrow 0.8\,B_5H_7 + 4\,YN + YB_6H_4C_6N$ | -0.3279 |

| Compound | Chemical reaction | Reaction energy(eV) |
|---|---|---|
| YNH$_4$B$_6$C$_6$ | 1.333 H$_3$N + 1.048 N$_2$ + 0.8571 Y$_4$C$_7$ + 6 B $\longrightarrow$ 2.429 YN + YB$_6$H$_4$C$_6$N | -0.3222 |
| YNH$_4$B$_6$C$_6$ | 0.3636 B$_9$H$_{11}$ + 0.5 N$_2$ + 0.8571 Y$_4$C$_7$ + 12.44 B $\longrightarrow$ 2.429 YB$_4$ + YB$_6$H$_4$C$_6$N | -0.2984 |
| YNH$_4$B$_6$C$_6$ | 0.4615 B$_{13}$C$_2$ + 1.015 Y$_4$C$_5$ + 0.5 N$_2$ + 6.592 H$_2$ $\longrightarrow$ 3.062 YH$_3$ + YB$_6$H$_4$C$_6$N | -0.2957 |
| YNH$_4$B$_6$C$_6$ | 6.162 H$_3$N + 1.2 Y$_4$C$_5$ + 1.362 YB$_{12}$ $\longrightarrow$ 2.07 B$_5$H$_7$ + 5.162 YN + YB$_6$H$_4$C$_6$N | -0.2890 |
| YNH$_4$B$_6$C$_6$ | H$_3$N + 0.8571 Y$_4$C$_7$ + 8.643 H$_2$ + 3 YB$_2$ $\longrightarrow$ 5.429 YH$_3$ + YB$_6$H$_4$C$_6$N | -0.2854 |
| YNH$_4$B$_6$C$_6$ | H$_3$N + 1.2 Y$_4$C$_5$ + 0.5 YB$_{12}$ + 6.95 H$_2$ $\longrightarrow$ 4.3 YH$_3$ + YB$_6$H$_4$C$_6$N | -0.2792 |
| YNH$_4$B$_6$C$_6$ | 1.2 Y$_4$C$_5$ + 0.5 YB$_{12}$ + 0.5 N$_2$ + 6.3 H$_2$ $\longrightarrow$ 4.3 YH$_2$ + YB$_6$H$_4$C$_6$N | -0.2727 |
| YNH$_4$B$_6$C$_6$ | 2.237 Y$_2$C + 0.7526 H$_8$C$_5$ + 1.2 B$_5$H$_7$ + 0.5 N$_2$ $\longrightarrow$ 3.474 YH$_3$ + YB$_6$H$_4$C$_6$N | -0.2328 |
| YNH$_4$B$_6$C$_6$ | 3.8 Y + 1.2 H$_8$C$_5$ + 0.5 N$_2$ + 6 B $\longrightarrow$ 2.8 YH$_2$ + YB$_6$H$_4$C$_6$N | -0.2316 |
| YNH$_4$B$_6$C$_6$ | 3.5 Y$_2$C + 0.5 H$_8$C$_5$ + BN + 17 B $\longrightarrow$ 6 YB$_2$ + YB$_6$H$_4$C$_6$N | -0.2270 |
| YNH$_4$B$_6$C$_6$ | 6 C + 7 H$_3$N + YB$_{12}$ $\longrightarrow$ 8.5 H$_2$ + 6 BN + YB$_6$H$_4$C$_6$N | -0.2234 |
| YNH$_4$B$_6$C$_6$ | 6 Y$_2$C + 12.83 H$_3$N + 0.5 YB$_{12}$ $\longrightarrow$ 11.5 YH$_3$ + 5.917 N$_2$ + YB$_6$H$_4$C$_6$N | -0.2183 |
| YNH$_4$B$_6$C$_6$ | 0.3636 B$_9$H$_{11}$ + 6 Y$_2$C + 8.909 BN $\longrightarrow$ 7.909 YN + 3.091 YB$_2$ + YB$_6$H$_4$C$_6$N | -0.2178 |
| YNH$_4$B$_6$C$_6$ | 0.6303 B$_{13}$C$_2$ + 3.194 H$_3$N + 0.9479 Y$_4$C$_5$ $\longrightarrow$ 2.791 YH$_2$ + 2.194 BN + YB$_6$H$_4$C$_6$N | -0.2083 |
| YNH$_4$B$_6$C$_6$ | 0.4615 B$_{13}$C$_2$ + 1.451 N$_2$ + 0.7253 Y$_4$C$_7$ + 2 H$_2$ $\longrightarrow$ 1.901 YN + YB$_6$H$_4$C$_6$N | -0.2022 |
| YNH$_4$B$_6$C$_6$ | H$_3$N + 0.8571 Y$_4$C$_7$ + 0.5 H$_2$ + 35.14 B $\longrightarrow$ 2.429 YB$_{12}$ + YB$_6$H$_4$C$_6$N | -0.2005 |
| YNH$_4$B$_6$C$_6$ | 0.5556 B$_9$H$_{11}$ + 4.904 Y + 1.2 H$_8$C$_5$ + BN $\longrightarrow$ 3.904 YH$_3$ + YB$_6$H$_4$C$_6$N | -0.1955 |
| YNH$_4$B$_6$C$_6$ | 3 B$_{13}$C$_2$ + Y + 0.5 N$_2$ + 25.1 H$_2$ $\longrightarrow$ 6.6 B$_5$H$_7$ + YB$_6$H$_4$C$_6$N | -0.1928 |
| YNH$_4$B$_6$C$_6$ | 2 YH$_2$ + 0.8571 Y$_4$C$_7$ + BN + 22.71 B $\longrightarrow$ 4.429 YB$_4$ + YB$_6$H$_4$C$_6$N | -0.1643 |
| YNH$_4$B$_6$C$_6$ | 6 Y$_2$C + 0.5714 B$_5$H$_7$ + 9.048 BN $\longrightarrow$ 8.048 YN + 2.952 YB$_2$ + YB$_6$H$_4$C$_6$N | -0.1637 |
| YNH$_4$B$_6$C$_6$ | 7.333 Y$_2$C + 2 H$_2$ + 14.67 BN $\longrightarrow$ 0.6667 B$_{13}$C$_2$ + 13.67 YN + YB$_6$H$_4$C$_6$N | -0.1589 |
| YNH$_4$B$_6$C$_6$ | 1.333 YH$_3$ + 1.46 B$_{13}$C$_2$ + 3.079 Y$_2$C + 0.5 N$_2$ $\longrightarrow$ 6.492 YB$_2$ + YB$_6$H$_4$C$_6$N | -0.1553 |
| YNH$_4$B$_6$C$_6$ | 1.2 Y$_4$C$_5$ + 2.048 B$_5$H$_7$ + 0.5 N$_2$ $\longrightarrow$ 3.446 YH$_3$ + 0.3535 YB$_{12}$ + YB$_6$H$_4$C$_6$N | -0.1544 |
| YNH$_4$B$_6$C$_6$ | 0.3636 B$_9$H$_{11}$ + 3 B$_{13}$C$_2$ + 4.106 Y + BN $\longrightarrow$ 3.106 YB$_{12}$ + YB$_6$H$_4$C$_6$N | -0.1058 |
| YNH$_4$B$_6$C$_6$ | 0.3333 YH$_3$ + 1.365 B$_{13}$C$_2$ + 3.27 Y$_2$C + H$_3$N $\longrightarrow$ 5.873 YB$_2$ + YB$_6$H$_4$C$_6$N | -0.1039 |
| YNH$_4$B$_6$C$_6$ | 3.5 Y$_2$C + 1.8 YB$_{12}$ + 0.5 H$_8$C$_5$ + 0.5 N$_2$ $\longrightarrow$ 7.8 YB$_2$ + YB$_6$H$_4$C$_6$N | -0.1026 |
| YNH$_4$B$_6$C$_6$ | 3.067 H$_3$N + 1.2 Y$_4$C$_5$ + 6 B $\longrightarrow$ 1.733 YH$_3$ + 2.067 YN + YB$_6$H$_4$C$_6$N | -0.1024 |
| YNH$_4$B$_6$C$_6$ | 1.2 Y$_4$C$_5$ + YN + 11.45 H$_2$ + 1.5 YB$_4$ $\longrightarrow$ 6.3 YH$_3$ + YB$_6$H$_4$C$_6$N | -0.1003 |
| YNH$_4$B$_6$C$_6$ | 1.333 B$_{13}$C$_2$ + 3.333 Y$_2$C + 1.333 H$_3$N $\longrightarrow$ 0.1667 N$_2$ + 5.667 YB$_2$ + YB$_6$H$_4$C$_6$N | -0.0867 |
| YNH$_4$B$_6$C$_6$ | 0.6667 B$_9$H$_{11}$ + 4.311 Y + 1.2 H$_8$C$_5$ + YN $\longrightarrow$ 4.311 YH$_3$ + YB$_6$H$_4$C$_6$N | -0.0839 |
| YNH$_4$B$_6$C$_6$ | 6 Y$_2$C + 9.667 H$_3$N + 1.5 YB$_4$ $\longrightarrow$ 12.5 YH$_2$ + 4.333 N$_2$ + YB$_6$H$_4$C$_6$N | -0.0826 |
| YNH$_4$B$_6$C$_6$ | 0.5 YH$_2$ + H$_3$N + 0.8571 Y$_4$C$_7$ + 17.71 B $\longrightarrow$ 2.929 YB$_4$ + YB$_6$H$_4$C$_6$N | -0.0770 |
| YNH$_4$B$_6$C$_6$ | 3.65 C + 2.35 Y$_2$C + H$_3$N + 1.2 B$_5$H$_7$ $\longrightarrow$ 3.7 YH$_2$ + YB$_6$H$_4$C$_6$N | -0.0751 |
| YNH$_4$B$_6$C$_6$ | 1.275 B$_9$H$_{11}$ + 1.2 Y$_4$C$_5$ + 0.5 N$_2$ $\longrightarrow$ 3.343 YH$_3$ + 0.4566 YB$_{12}$ + YB$_6$H$_4$C$_6$N | -0.0657 |
| YNH$_4$B$_6$C$_6$ | 3 B$_{13}$C$_2$ + 4.071 Y + 0.5714 B$_5$H$_7$ + BN $\longrightarrow$ 3.071 YB$_{12}$ + YB$_6$H$_4$C$_6$N | -0.0562 |
| YNH$_4$B$_6$C$_6$ | H$_3$N + 0.1429 B$_5$H$_7$ + 0.8571 Y$_4$C$_7$ + 34.43 B $\longrightarrow$ 2.429 YB$_{12}$ + YB$_6$H$_4$C$_6$N | -0.0512 |
| YNH$_4$B$_6$C$_6$ | 0.3333 YH$_3$ + 3 B$_{13}$C$_2$ + 3.417 Y + H$_3$N $\longrightarrow$ 2.75 YB$_{12}$ + YB$_6$H$_4$C$_6$N | -0.0505 |
| YNH$_4$B$_6$C$_6$ | 1.333 H$_3$N + 0.8571 Y$_4$C$_7$ + 16.05 B $\longrightarrow$ 0.3333 BN + 2.429 YB$_4$ + YB$_6$H$_4$C$_6$N | -0.0479 |
| YNH$_4$B$_6$C$_6$ | 0.5 YH$_2$ + 1.381 B$_{13}$C$_2$ + 3.238 Y$_2$C + H$_3$N $\longrightarrow$ 5.976 YB$_2$ + YB$_6$H$_4$C$_6$N | -0.0465 |
| YNH$_4$B$_6$C$_6$ | 0.6667 B$_9$H$_{11}$ + 2.491 Y$_2$C + H$_3$N + 0.7018 H$_8$C$_5$ $\longrightarrow$ 3.982 YH$_3$ + YB$_6$H$_4$C$_6$N | -0.0441 |
| YNH$_4$B$_6$C$_6$ | 1.333 YH$_3$ + 0.8571 Y$_4$C$_7$ + BN + 50.14 B $\longrightarrow$ 3.762 YB$_{12}$ + YB$_6$H$_4$C$_6$N | -0.0386 |
| YNH$_4$B$_6$C$_6$ | 0.09091 B$_9$H$_{11}$ + H$_3$N + 0.8571 Y$_4$C$_7$ + 34.32 B $\longrightarrow$ 2.429 YB$_{12}$ + YB$_6$H$_4$C$_6$N | -0.0356 |
| AcNH$_4$B$_6$C$_6$ | 34 H$_3$N + 50 Ac + 3 B$_{13}$C$_2$ $\longrightarrow$ 33 BN + 49 AcH$_2$ + AcB$_6$H$_4$C$_6$N | -115.1753 |
| AcNH$_4$B$_6$C$_6$ | 0.3636 B$_9$H$_{11}$ + 18.64 N$_2$ + Ac + 3 B$_{13}$C$_2$ $\longrightarrow$ 36.27 BN + AcB$_6$H$_4$C$_6$N | -85.0094 |
| AcNH$_4$B$_6$C$_6$ | 0.5714 B$_5$H$_7$ + 18.43 N$_2$ + Ac + 3 B$_{13}$C$_2$ $\longrightarrow$ 35.86 BN + AcB$_6$H$_4$C$_6$N | -84.0050 |
| AcNH$_4$B$_6$C$_6$ | 0.1818 B$_9$H$_{11}$ + AcH$_2$ + 17.82 N$_2$ + 3 B$_{13}$C$_2$ $\longrightarrow$ 34.64 BN + AcB$_6$H$_4$C$_6$N | -79.4466 |
| AcNH$_4$B$_6$C$_6$ | AcH$_2$ + 0.2857 B$_5$H$_7$ + 17.71 N$_2$ + 3 B$_{13}$C$_2$ $\longrightarrow$ 34.43 BN + AcB$_6$H$_4$C$_6$N | -78.9444 |
| AcNH$_4$B$_6$C$_6$ | 2 H$_2$ + 17 N$_2$ + Ac + 3 B$_{13}$C$_2$ $\longrightarrow$ 33 BN + AcB$_6$H$_4$C$_6$N | -77.2685 |
| AcNH$_4$B$_6$C$_6$ | H$_2$ + AcH$_2$ + 17 N$_2$ + 3 B$_{13}$C$_2$ $\longrightarrow$ 33 BN + AcB$_6$H$_4$C$_6$N | -75.5762 |
| AcNH$_4$B$_6$C$_6$ | 1.333 H$_3$N + 16.33 N$_2$ + Ac + 3 B$_{13}$C$_2$ $\longrightarrow$ 33 BN + AcB$_6$H$_4$C$_6$N | -75.5031 |
| AcNH$_4$B$_6$C$_6$ | AcH$_2$ + 0.6667 H$_3$N + 16.67 N$_2$ + 3 B$_{13}$C$_2$ $\longrightarrow$ 33 BN + AcB$_6$H$_4$C$_6$N | -74.6935 |
| AcNH$_4$B$_6$C$_6$ | 2 AcH$_2$ + 17 N$_2$ + 3 B$_{13}$C$_2$ $\longrightarrow$ 33 BN + Ac + AcB$_6$H$_4$C$_6$N | -73.8838 |
| AcNH$_4$B$_6$C$_6$ | 0.25 H$_8$C$_5$ + AcH$_2$ + 12.94 N$_2$ + 2.375 B$_{13}$C$_2$ $\longrightarrow$ 24.88 BN + AcB$_6$H$_4$C$_6$N | -54.9347 |
| AcNH$_4$B$_6$C$_6$ | 0.5 H$_8$C$_5$ + 8.875 N$_2$ + Ac + 1.75 B$_{13}$C$_2$ $\longrightarrow$ 16.75 BN + AcB$_6$H$_4$C$_6$N | -35.9856 |
| AcNH$_4$B$_6$C$_6$ | 34 H$_3$N + Ac + 3 B$_{13}$C$_2$ $\longrightarrow$ 49 H$_2$ + 33 BN + AcB$_6$H$_4$C$_6$N | -32.2502 |
| AcNH$_4$B$_6$C$_6$ | AcH$_2$ + 34 H$_3$N + 3 B$_{13}$C$_2$ $\longrightarrow$ 50 H$_2$ + 33 BN + AcB$_6$H$_4$C$_6$N | -30.5578 |
| AcNH$_4$B$_6$C$_6$ | 11.73 H$_3$N + Ac + 3 B$_{13}$C$_2$ $\longrightarrow$ 10.73 BN + 4.455 B$_5$H$_7$ + AcB$_6$H$_4$C$_6$N | -9.2272 |
| AcNH$_4$B$_6$C$_6$ | 10.79 H$_3$N + Ac + 3 B$_{13}$C$_2$ $\longrightarrow$ 2.579 B$_9$H$_{11}$ + 9.789 BN + AcB$_6$H$_4$C$_6$N | -8.0834 |
| AcNH$_4$B$_6$C$_6$ | AcH$_2$ + 11.27 H$_3$N + 3 B$_{13}$C$_2$ $\longrightarrow$ 10.27 BN + 4.545 B$_5$H$_7$ + AcB$_6$H$_4$C$_6$N | -7.0650 |
| AcNH$_4$B$_6$C$_6$ | AcH$_2$ + 10.32 H$_3$N + 3 B$_{13}$C$_2$ $\longrightarrow$ 2.632 B$_9$H$_{11}$ + 9.316 BN + AcB$_6$H$_4$C$_6$N | -5.8979 |

| Compound | Chemical reaction | Reaction energy(eV) |
| --- | --- | --- |
| AcNH$_4$B$_6$C$_6$ | 1.2 H$_8$C$_5$ + H$_3$N + 1.2 B$_5$H$_7$ + 9.5 Ac $\longrightarrow$ 8.5 AcH$_2$ + AcB$_6$H$_4$C$_6$N | -5.3970 |
| AcNH$_4$B$_6$C$_6$ | 0.6667 B$_9$H$_{11}$ + 1.2 H$_8$C$_5$ + H$_3$N + 8.967 Ac $\longrightarrow$ 7.967 AcH$_2$ + AcB$_6$H$_4$C$_6$N | -4.5395 |
| AcNH$_4$B$_6$C$_6$ | 1.2 H$_8$C$_5$ + 1.2 B$_5$H$_7$ + 0.5 N$_2$ + 8 Ac $\longrightarrow$ 7 AcH$_2$ + AcB$_6$H$_4$C$_6$N | -4.1825 |
| AcNH$_4$B$_6$C$_6$ | 0.6667 B$_9$H$_{11}$ + 1.2 H$_8$C$_5$ + 0.5 N$_2$ + 7.467 Ac $\longrightarrow$ 6.467 AcH$_2$ + AcB$_6$H$_4$C$_6$N | -3.3250 |
| AcNH$_4$B$_6$C$_6$ | BN + 1.2 H$_8$C$_5$ + B$_5$H$_7$ + 7.3 Ac $\longrightarrow$ 6.3 AcH$_2$ + AcB$_6$H$_4$C$_6$N | -0.6402 |
| AcNH$_4$B$_6$C$_6$ | 6 C + H$_3$N + 1.2 B$_5$H$_7$ + 4.7 Ac $\longrightarrow$ 3.7 AcH$_2$ + AcB$_6$H$_4$C$_6$N | -0.2760 |
| CaNH$_4$B$_6$C$_6$ | 3 B$_{13}$C$_2$ + 98 CaN$_2$ + 18 B$_9$H$_{11}$ $\longrightarrow$ 97 CaH$_2$ + 195 BN + CaB$_6$H$_4$C$_6$N | -472.6116 |
| CaNH$_4$B$_6$C$_6$ | 3 B$_{13}$C$_2$ + 62 CaN$_2$ + 18 B$_5$H$_7$ $\longrightarrow$ 61 CaH$_2$ + 123 BN + CaB$_6$H$_4$C$_6$N | -296.3304 |
| CaNH$_4$B$_6$C$_6$ | 34 H$_3$N + 3 B$_{13}$C$_2$ + 50 Ca $\longrightarrow$ 49 CaH$_2$ + 33 BN + CaB$_6$H$_4$C$_6$N | -113.0566 |
| CaNH$_4$B$_6$C$_6$ | 34 Ca$_2$N + 69 H$_2$ + 3 B$_{13}$C$_2$ $\longrightarrow$ 67 CaH$_2$ + 33 BN + CaB$_6$H$_4$C$_6$N | -110.5865 |
| CaNH$_4$B$_6$C$_6$ | 2 CaH$_2$ + 4 B$_{13}$C$_2$ + 23.5 N$_2$ $\longrightarrow$ 46 BN + CaC$_2$ + CaB$_6$H$_4$C$_6$N | -105.5774 |
| CaNH$_4$B$_6$C$_6$ | 3 B$_{13}$C$_2$ + 36.5 CaN$_2$ + 20 B$_9$H$_{11}$ $\longrightarrow$ 35.5 CaB$_6$ + 72 H$_3$N + CaB$_6$H$_4$C$_6$N | -97.1859 |
| CaNH$_4$B$_6$C$_6$ | CaB$_6$ + 3 B$_{13}$C$_2$ + 0.3636 B$_9$H$_{11}$ + 21.64 N$_2$ $\longrightarrow$ 42.27 BN + CaB$_6$H$_4$C$_6$N | -96.9406 |
| CaNH$_4$B$_6$C$_6$ | CaB$_6$ + 3 B$_{13}$C$_2$ + 0.5714 B$_5$H$_7$ + 21.43 N$_2$ $\longrightarrow$ 41.86 BN + CaB$_6$H$_4$C$_6$N | -95.9362 |
| CaNH$_4$B$_6$C$_6$ | 52 H$_2$ + 3 B$_{13}$C$_2$ + 17 Ca$_3$N$_2$ $\longrightarrow$ 50 CaH$_2$ + 33 BN + CaB$_6$H$_4$C$_6$N | -92.2966 |
| CaNH$_4$B$_6$C$_6$ | CaB$_6$ + 2 H$_2$ + 3 B$_{13}$C$_2$ + 20 N$_2$ $\longrightarrow$ 39 BN + CaB$_6$H$_4$C$_6$N | -89.1998 |
| CaNH$_4$B$_6$C$_6$ | CaB$_6$ + 1.333 H$_3$N + 3 B$_{13}$C$_2$ + 19.33 N$_2$ $\longrightarrow$ 39 BN + CaB$_6$H$_4$C$_6$N | -87.4343 |
| CaNH$_4$B$_6$C$_6$ | 3 B$_{13}$C$_2$ + 0.3636 B$_9$H$_{11}$ + 18.64 N$_2$ + Ca $\longrightarrow$ 36.27 BN + CaB$_6$H$_4$C$_6$N | -84.3975 |
| CaNH$_4$B$_6$C$_6$ | 3 B$_{13}$C$_2$ + 0.5714 B$_5$H$_7$ + 18.43 N$_2$ + Ca $\longrightarrow$ 35.86 BN + CaB$_6$H$_4$C$_6$N | -83.3930 |
| CaNH$_4$B$_6$C$_6$ | 0.5 Ca$_2$N + 3 B$_{13}$C$_2$ + 0.3636 B$_9$H$_{11}$ + 18.39 N$_2$ $\longrightarrow$ 36.27 BN + CaB$_6$H$_4$C$_6$N | -83.2593 |
| CaNH$_4$B$_6$C$_6$ | 3 B$_{13}$C$_2$ + 0.3333 Ca$_3$N$_2$ + 0.3636 B$_9$H$_{11}$ + 18.3 N$_2$ $\longrightarrow$ 36.27 BN + CaB$_6$H$_4$C$_6$N | -83.0751 |
| CaNH$_4$B$_6$C$_6$ | 3 B$_{13}$C$_2$ + CaN$_2$ + 0.3636 B$_9$H$_{11}$ + 17.64 N$_2$ $\longrightarrow$ 36.27 BN + CaB$_6$H$_4$C$_6$N | -82.8832 |
| CaNH$_4$B$_6$C$_6$ | 0.5 Ca$_2$N + 3 B$_{13}$C$_2$ + 0.5714 B$_5$H$_7$ + 18.18 N$_2$ $\longrightarrow$ 35.86 BN + CaB$_6$H$_4$C$_6$N | -82.2548 |
| CaNH$_4$B$_6$C$_6$ | 3 B$_{13}$C$_2$ + 0.3333 Ca$_3$N$_2$ + 0.5714 B$_5$H$_7$ + 18.1 N$_2$ $\longrightarrow$ 35.86 BN + CaB$_6$H$_4$C$_6$N | -82.0707 |
| CaNH$_4$B$_6$C$_6$ | 3 B$_{13}$C$_2$ + CaN$_2$ + 0.5714 B$_5$H$_7$ + 17.43 N$_2$ $\longrightarrow$ 35.86 BN + CaB$_6$H$_4$C$_6$N | -81.8788 |
| CaNH$_4$B$_6$C$_6$ | CaH$_2$ + 3 B$_{13}$C$_2$ + 0.1818 B$_9$H$_{11}$ + 17.82 N$_2$ $\longrightarrow$ 34.64 BN + CaB$_6$H$_4$C$_6$N | -78.8654 |
| CaNH$_4$B$_6$C$_6$ | CaH$_2$ + 3 B$_{13}$C$_2$ + 0.2857 B$_5$H$_7$ + 17.71 N$_2$ $\longrightarrow$ 34.43 BN + CaB$_6$H$_4$C$_6$N | -78.3632 |
| CaNH$_4$B$_6$C$_6$ | 3 B$_{13}$C$_2$ + 27.45 CaN$_2$ + 0.3636 B$_9$H$_{11}$ $\longrightarrow$ 8.818 Ca$_3$N$_2$ + 36.27 BN + CaB$_6$H$_4$C$_6$N | -77.8075 |
| CaNH$_4$B$_6$C$_6$ | 18 H$_2$ + 3 B$_{13}$C$_2$ + 17 CaN$_2$ $\longrightarrow$ 16 CaH$_2$ + 33 BN + CaB$_6$H$_4$C$_6$N | -77.5004 |
| CaNH$_4$B$_6$C$_6$ | 3 B$_{13}$C$_2$ + 27.14 CaN$_2$ + 0.5714 B$_5$H$_7$ $\longrightarrow$ 8.714 Ca$_3$N$_2$ + 35.86 BN + CaB$_6$H$_4$C$_6$N | -76.8629 |
| CaNH$_4$B$_6$C$_6$ | 2 H$_2$ + 3 B$_{13}$C$_2$ + 17 N$_2$ + Ca $\longrightarrow$ 33 BN + CaB$_6$H$_4$C$_6$N | -76.6566 |
| CaNH$_4$B$_6$C$_6$ | 0.5 Ca$_2$N + 2 H$_2$ + 3 B$_{13}$C$_2$ + 16.75 N$_2$ $\longrightarrow$ 33 BN + CaB$_6$H$_4$C$_6$N | -75.5184 |
| CaNH$_4$B$_6$C$_6$ | 2 H$_2$ + 3 B$_{13}$C$_2$ + 0.3333 Ca$_3$N$_2$ + 16.67 N$_2$ $\longrightarrow$ 33 BN + CaB$_6$H$_4$C$_6$N | -75.3342 |
| CaNH$_4$B$_6$C$_6$ | 2 H$_2$ + 3 B$_{13}$C$_2$ + CaN$_2$ + 16 N$_2$ $\longrightarrow$ 33 BN + CaB$_6$H$_4$C$_6$N | -75.1424 |
| CaNH$_4$B$_6$C$_6$ | H$_2$ + CaH$_2$ + 3 B$_{13}$C$_2$ + 17 N$_2$ $\longrightarrow$ 33 BN + CaB$_6$H$_4$C$_6$N | -74.9950 |
| CaNH$_4$B$_6$C$_6$ | 1.333 H$_3$N + 3 B$_{13}$C$_2$ + 16.33 N$_2$ + Ca $\longrightarrow$ 33 BN + CaB$_6$H$_4$C$_6$N | -74.8912 |
| CaNH$_4$B$_6$C$_6$ | 2 CaH$_2$ + 3 B$_{13}$C$_2$ + 18 N$_2$ $\longrightarrow$ CaN$_2$ + 33 BN + CaB$_6$H$_4$C$_6$N | -74.8476 |
| CaNH$_4$B$_6$C$_6$ | 2 CaH$_2$ + 3 B$_{13}$C$_2$ + 17.33 N$_2$ $\longrightarrow$ 0.3333 Ca$_3$N$_2$ + 33 BN + CaB$_6$H$_4$C$_6$N | -74.6557 |
| CaNH$_4$B$_6$C$_6$ | 2 CaH$_2$ + 3 B$_{13}$C$_2$ + 17.25 N$_2$ $\longrightarrow$ 0.5 Ca$_2$N + 33 BN + CaB$_6$H$_4$C$_6$N | -74.4716 |
| CaNH$_4$B$_6$C$_6$ | CaH$_2$ + 0.6667 H$_3$N + 3 B$_{13}$C$_2$ + 16.67 N$_2$ $\longrightarrow$ 33 BN + CaB$_6$H$_4$C$_6$N | -74.1123 |
| CaNH$_4$B$_6$C$_6$ | 3 B$_{13}$C$_2$ + 24.52 CaN$_2$ + 0.3636 B$_9$H$_{11}$ $\longrightarrow$ 11.76 Ca$_2$N + 36.27 BN + CaB$_6$H$_4$C$_6$N | -74.0410 |
| CaNH$_4$B$_6$C$_6$ | 0.5 Ca$_2$N + 1.333 H$_3$N + 3 B$_{13}$C$_2$ + 16.08 N$_2$ $\longrightarrow$ 33 BN + CaB$_6$H$_4$C$_6$N | -73.7530 |
| CaNH$_4$B$_6$C$_6$ | 1.333 H$_3$N + 3 B$_{13}$C$_2$ + 0.3333 Ca$_3$N$_2$ + 16 N$_2$ $\longrightarrow$ 33 BN + CaB$_6$H$_4$C$_6$N | -73.5688 |
| CaNH$_4$B$_6$C$_6$ | 1.333 H$_3$N + 3 B$_{13}$C$_2$ + CaN$_2$ + 15.33 N$_2$ $\longrightarrow$ 33 BN + CaB$_6$H$_4$C$_6$N | -73.3769 |
| CaNH$_4$B$_6$C$_6$ | 2 CaH$_2$ + 3 B$_{13}$C$_2$ + 17 N$_2$ $\longrightarrow$ 33 BN + Ca + CaB$_6$H$_4$C$_6$N | -73.3334 |
| CaNH$_4$B$_6$C$_6$ | 3 B$_{13}$C$_2$ + 24.24 CaN$_2$ + 0.5714 B$_5$H$_7$ $\longrightarrow$ 11.62 Ca$_2$N + 35.86 BN + CaB$_6$H$_4$C$_6$N | -73.1407 |
| CaNH$_4$B$_6$C$_6$ | 2 H$_2$ + 3 B$_{13}$C$_2$ + 25 CaN$_2$ $\longrightarrow$ 8 Ca$_3$N$_2$ + 33 BN + CaB$_6$H$_4$C$_6$N | -70.5376 |
| CaNH$_4$B$_6$C$_6$ | 2 CaH$_2$ + 3 B$_{13}$C$_2$ + 26 CaN$_2$ $\longrightarrow$ 9 Ca$_3$N$_2$ + 33 BN + CaB$_6$H$_4$C$_6$N | -69.6672 |
| CaNH$_4$B$_6$C$_6$ | 1.333 H$_3$N + 3 B$_{13}$C$_2$ + 24 CaN$_2$ $\longrightarrow$ 7.667 Ca$_3$N$_2$ + 33 BN + CaB$_6$H$_4$C$_6$N | -68.9640 |
| CaNH$_4$B$_6$C$_6$ | 2 H$_2$ + 3 B$_{13}$C$_2$ + 22.33 CaN$_2$ $\longrightarrow$ 10.67 Ca$_2$N + 33 BN + CaB$_6$H$_4$C$_6$N | -67.1205 |
| CaNH$_4$B$_6$C$_6$ | 2 CaH$_2$ + 3 B$_{13}$C$_2$ + 23 CaN$_2$ $\longrightarrow$ 12 Ca$_2$N + 33 BN + CaB$_6$H$_4$C$_6$N | -65.8230 |
| CaNH$_4$B$_6$C$_6$ | 1.333 H$_3$N + 3 B$_{13}$C$_2$ + 21.44 CaN$_2$ $\longrightarrow$ 10.22 Ca$_2$N + 33 BN + CaB$_6$H$_4$C$_6$N | -65.6893 |
| CaNH$_4$B$_6$C$_6$ | 9 H$_3$N + 3 B$_{13}$C$_2$ + 12.5 CaN$_2$ $\longrightarrow$ 11.5 CaH$_2$ + 33 BN + CaB$_6$H$_4$C$_6$N | -64.9206 |
| CaNH$_4$B$_6$C$_6$ | 14.29 Ca$_2$N + 19.71 H$_3$N + 3 B$_{13}$C$_2$ $\longrightarrow$ 27.57 CaH$_2$ + 33 BN + CaB$_6$H$_4$C$_6$N | -63.8463 |
| CaNH$_4$B$_6$C$_6$ | 2 CaH$_2$ + 3 B$_{13}$C$_2$ + 14 N$_2$ $\longrightarrow$ CaB$_6$ + 27 BN + CaB$_6$H$_4$C$_6$N | -60.7902 |
| CaNH$_4$B$_6$C$_6$ | 17.33 H$_3$N + 3 B$_{13}$C$_2$ + 8.333 Ca$_3$N$_2$ $\longrightarrow$ 24 CaH$_2$ + 33 BN + CaB$_6$H$_4$C$_6$N | -60.5256 |
| CaNH$_4$B$_6$C$_6$ | 3 B$_{13}$C$_2$ + 23.17 CaN$_2$ + 20 B$_5$H$_7$ $\longrightarrow$ 22.17 CaB$_6$ + 45.33 H$_3$N + CaB$_6$H$_4$C$_6$N | -59.3366 |
| CaNH$_4$B$_6$C$_6$ | 3 B$_{13}$C$_2$ + 18.64 CaN$_2$ + 0.3636 B$_9$H$_{11}$ $\longrightarrow$ 36.27 BN + 17.64 Ca + CaB$_6$H$_4$C$_6$N | -56.1779 |
| CaNH$_4$B$_6$C$_6$ | 3 B$_{13}$C$_2$ + 18.43 CaN$_2$ + 0.5714 B$_5$H$_7$ $\longrightarrow$ 35.86 BN + 17.43 Ca + CaB$_6$H$_4$C$_6$N | -55.4881 |
| CaNH$_4$B$_6$C$_6$ | CaH$_2$ + 2.375 B$_{13}$C$_2$ + 12.94 N$_2$ + 0.25 H$_8$C$_5$ $\longrightarrow$ 24.88 BN + CaB$_6$H$_4$C$_6$N | -54.3535 |

| Compound | Chemical reaction | Reaction energy(eV) |
|---|---|---|
| CaNH$_4$B$_6$C$_6$ | 2 B$_{13}$C$_2$ + 0.3636 B$_9$H$_{11}$ + 12.14 N$_2$ + CaC$_2$ $\longrightarrow$ 23.27 BN + CaB$_6$H$_4$C$_6$N | -52.1534 |
| CaNH$_4$B$_6$C$_6$ | 2 B$_{13}$C$_2$ + 0.5714 B$_5$H$_7$ + 11.93 N$_2$ + CaC$_2$ $\longrightarrow$ 22.86 BN + CaB$_6$H$_4$C$_6$N | -51.1490 |
| CaNH$_4$B$_6$C$_6$ | 2 H$_2$ + 3 B$_{13}$C$_2$ + 17 CaN$_2$ $\longrightarrow$ 33 BN + 16 Ca + CaB$_6$H$_4$C$_6$N | -50.9148 |
| CaNH$_4$B$_6$C$_6$ | 1.333 H$_3$N + 3 B$_{13}$C$_2$ + 16.33 CaN$_2$ $\longrightarrow$ 33 BN + 15.33 Ca + CaB$_6$H$_4$C$_6$N | -50.1589 |
| CaNH$_4$B$_6$C$_6$ | CaB$_6$ + 1.75 B$_{13}$C$_2$ + 11.88 N$_2$ + 0.5 H$_8$C$_5$ $\longrightarrow$ 22.75 BN + CaB$_6$H$_4$C$_6$N | -47.9169 |
| CaNH$_4$B$_6$C$_6$ | 2 CaH$_2$ + 3 B$_{13}$C$_2$ + 17 CaN$_2$ $\longrightarrow$ 33 BN + 18 Ca + CaB$_6$H$_4$C$_6$N | -47.5916 |
| CaNH$_4$B$_6$C$_6$ | 2 H$_2$ + 2 B$_{13}$C$_2$ + 10.5 N$_2$ + CaC$_2$ $\longrightarrow$ 20 BN + CaB$_6$H$_4$C$_6$N | -44.4125 |
| CaNH$_4$B$_6$C$_6$ | 1.333 H$_3$N + 2 B$_{13}$C$_2$ + 9.833 N$_2$ + CaC$_2$ $\longrightarrow$ 20 BN + CaB$_6$H$_4$C$_6$N | -42.6471 |
| CaNH$_4$B$_6$C$_6$ | 3 B$_{13}$C$_2$ + 73.55 Ca$_3$N$_2$ + 0.3636 B$_9$H$_{11}$ $\longrightarrow$ 109.8 Ca$_2$N + 36.27 BN + CaB$_6$H$_4$C$_6$N | -42.6272 |
| CaNH$_4$B$_6$C$_6$ | 3 B$_{13}$C$_2$ + 72.71 Ca$_3$N$_2$ + 0.5714 B$_5$H$_7$ $\longrightarrow$ 108.6 Ca$_2$N + 35.86 BN + CaB$_6$H$_4$C$_6$N | -42.0820 |
| CaNH$_4$B$_6$C$_6$ | 2 H$_2$ + 3 B$_{13}$C$_2$ + 67 Ca$_3$N$_2$ $\longrightarrow$ 100 Ca$_2$N + 33 BN + CaB$_6$H$_4$C$_6$N | -38.5026 |
| CaNH$_4$B$_6$C$_6$ | 1.333 H$_3$N + 3 B$_{13}$C$_2$ + 64.33 Ca$_3$N$_2$ $\longrightarrow$ 96 Ca$_2$N + 33 BN + CaB$_6$H$_4$C$_6$N | -38.2104 |
| CaNH$_4$B$_6$C$_6$ | 2 CaH$_2$ + 3 B$_{13}$C$_2$ + 69 Ca$_3$N$_2$ $\longrightarrow$ 104 Ca$_2$N + 33 BN + CaB$_6$H$_4$C$_6$N | -36.3508 |
| CaNH$_4$B$_6$C$_6$ | CaB$_6$ + 40 H$_3$N + 3 B$_{13}$C$_2$ $\longrightarrow$ 58 H$_2$ + 39 BN + CaB$_6$H$_4$C$_6$N | -36.2370 |
| CaNH$_4$B$_6$C$_6$ | 1.75 B$_{13}$C$_2$ + 8.875 N$_2$ + Ca + 0.5 H$_8$C$_5$ $\longrightarrow$ 16.75 BN + CaB$_6$H$_4$C$_6$N | -35.3737 |
| CaNH$_4$B$_6$C$_6$ | 0.5 Ca$_2$N + 1.75 B$_{13}$C$_2$ + 8.625 N$_2$ + 0.5 H$_8$C$_5$ $\longrightarrow$ 16.75 BN + CaB$_6$H$_4$C$_6$N | -34.2355 |
| CaNH$_4$B$_6$C$_6$ | 1.75 B$_{13}$C$_2$ + 0.3333 Ca$_3$N$_2$ + 8.542 N$_2$ + 0.5 H$_8$C$_5$ $\longrightarrow$ 16.75 BN + CaB$_6$H$_4$C$_6$N | -34.0513 |
| CaNH$_4$B$_6$C$_6$ | 1.75 B$_{13}$C$_2$ + CaN$_2$ + 7.875 N$_2$ + 0.5 H$_8$C$_5$ $\longrightarrow$ 16.75 BN + CaB$_6$H$_4$C$_6$N | -33.8595 |
| CaNH$_4$B$_6$C$_6$ | 32 H$_3$N + 3 B$_{13}$C$_2$ + CaN$_2$ $\longrightarrow$ 46 H$_2$ + 33 BN + CaB$_6$H$_4$C$_6$N | -32.7721 |
| CaNH$_4$B$_6$C$_6$ | 34 H$_3$N + 3 B$_{13}$C$_2$ + Ca $\longrightarrow$ 49 H$_2$ + 33 BN + CaB$_6$H$_4$C$_6$N | -31.6382 |
| CaNH$_4$B$_6$C$_6$ | 1.75 B$_{13}$C$_2$ + 12.81 CaN$_2$ + 0.5 H$_8$C$_5$ $\longrightarrow$ 3.937 Ca$_3$N$_2$ + 16.75 BN + CaB$_6$H$_4$C$_6$N | -31.5930 |
| CaNH$_4$B$_6$C$_6$ | 33.33 H$_3$N + 3 B$_{13}$C$_2$ + 0.3333 Ca$_3$N$_2$ $\longrightarrow$ 48 H$_2$ + 33 BN + CaB$_6$H$_4$C$_6$N | -31.1986 |
| CaNH$_4$B$_6$C$_6$ | 0.5 Ca$_2$N + 33.5 H$_3$N + 3 B$_{13}$C$_2$ $\longrightarrow$ 48.25 H$_2$ + 33 BN + CaB$_6$H$_4$C$_6$N | -31.1620 |
| CaNH$_4$B$_6$C$_6$ | CaH$_2$ + 34 H$_3$N + 3 B$_{13}$C$_2$ $\longrightarrow$ 50 H$_2$ + 33 BN + CaB$_6$H$_4$C$_6$N | -29.9766 |
| CaNH$_4$B$_6$C$_6$ | 1.75 B$_{13}$C$_2$ + 11.5 CaN$_2$ + 0.5 H$_8$C$_5$ $\longrightarrow$ 5.25 Ca$_2$N + 16.75 BN + CaB$_6$H$_4$C$_6$N | -29.9112 |
| CaNH$_4$B$_6$C$_6$ | 3.8 CaB$_6$ + 8.9 N$_2$ + 1.2 H$_8$C$_5$ $\longrightarrow$ 2.8 CaH$_2$ + 16.8 BN + CaB$_6$H$_4$C$_6$N | -29.8942 |
| CaNH$_4$B$_6$C$_6$ | 1.75 B$_{13}$C$_2$ + 8.875 CaN$_2$ + 0.5 H$_8$C$_5$ $\longrightarrow$ 16.75 BN + 7.875 Ca + CaB$_6$H$_4$C$_6$N | -21.9350 |
| CaNH$_4$B$_6$C$_6$ | 3 B$_{13}$C$_2$ + 5.409 CaN$_2$ + 0.3636 B$_9$H$_{11}$ $\longrightarrow$ 4.409 CaB$_6$ + 9.818 BN + CaB$_6$H$_4$C$_6$N | -20.9029 |
| CaNH$_4$B$_6$C$_6$ | 3 B$_{13}$C$_2$ + 5.357 CaN$_2$ + 0.5714 B$_5$H$_7$ $\longrightarrow$ 4.357 CaB$_6$ + 9.714 BN + CaB$_6$H$_4$C$_6$N | -20.6287 |
| CaNH$_4$B$_6$C$_6$ | 1.333 H$_3$N + 3 B$_{13}$C$_2$ + 4.833 CaN$_2$ $\longrightarrow$ 3.833 CaB$_6$ + 10 BN + CaB$_6$H$_4$C$_6$N | -19.4902 |
| CaNH$_4$B$_6$C$_6$ | 2 H$_2$ + 3 B$_{13}$C$_2$ + 5 CaN$_2$ $\longrightarrow$ 4 CaB$_6$ + 9 BN + CaB$_6$H$_4$C$_6$N | -18.9128 |
| CaNH$_4$B$_6$C$_6$ | 21 H$_3$N + 2 B$_{13}$C$_2$ + CaC$_2$ $\longrightarrow$ 29.5 H$_2$ + 20 BN + CaB$_6$H$_4$C$_6$N | -16.6070 |
| CaNH$_4$B$_6$C$_6$ | 4.25 CaN$_2$ + 1.5 B$_9$H$_{11}$ + 3 CaC$_2$ $\longrightarrow$ 6.25 CaH$_2$ + 7.5 BN + CaB$_6$H$_4$C$_6$N | -15.8050 |
| CaNH$_4$B$_6$C$_6$ | 1.75 B$_{13}$C$_2$ + 34.5 Ca$_3$N$_2$ + 0.5 H$_8$C$_5$ $\longrightarrow$ 51.25 Ca$_2$N + 16.75 BN + CaB$_6$H$_4$C$_6$N | -15.1751 |
| CaNH$_4$B$_6$C$_6$ | 4.563 Ca$_3$N$_2$ + 2.825 B$_5$H$_7$ + 1.2 H$_8$C$_5$ $\longrightarrow$ 12.69 CaH$_2$ + 8.125 BN + CaB$_6$H$_4$C$_6$N | -13.8627 |
| CaNH$_4$B$_6$C$_6$ | 20 H$_2$ + 3 B$_{13}$C$_2$ + 6.5 CaN$_2$ $\longrightarrow$ 5.5 CaB$_6$ + 12 H$_3$N + CaB$_6$H$_4$C$_6$N | -13.7155 |
| CaNH$_4$B$_6$C$_6$ | 16.73 H$_3$N + 3 B$_{13}$C$_2$ + 32.47 Ca $\longrightarrow$ 15.73 Ca$_2$N + 6.6 B$_5$H$_7$ + CaB$_6$H$_4$C$_6$N | -12.5098 |
| CaNH$_4$B$_6$C$_6$ | 2 CaH$_2$ + 3 B$_{13}$C$_2$ + 3.5 CaN$_2$ $\longrightarrow$ 4.5 CaB$_6$ + 6 BN + CaB$_6$H$_4$C$_6$N | -11.5893 |
| CaNH$_4$B$_6$C$_6$ | 11.09 H$_3$N + 3 B$_{13}$C$_2$ + CaN$_2$ $\longrightarrow$ 12.09 BN + 4.182 B$_5$H$_7$ + CaB$_6$H$_4$C$_6$N | -11.1587 |
| CaNH$_4$B$_6$C$_6$ | 3 B$_{13}$C$_2$ + 18.64 Ca$_3$N$_2$ + 0.3636 B$_9$H$_{11}$ $\longrightarrow$ 36.27 BN + 54.91 Ca + CaB$_6$H$_4$C$_6$N | -10.4659 |
| CaNH$_4$B$_6$C$_6$ | 14.78 H$_3$N + 3 B$_{13}$C$_2$ + 28.56 Ca $\longrightarrow$ 13.78 Ca$_2$N + 3.667 B$_9$H$_{11}$ + CaB$_6$H$_4$C$_6$N | -10.3995 |
| CaNH$_4$B$_6$C$_6$ | 3 B$_{13}$C$_2$ + 18.43 Ca$_3$N$_2$ + 0.5714 B$_5$H$_7$ $\longrightarrow$ 35.86 BN + 54.29 Ca + CaB$_6$H$_4$C$_6$N | -10.2858 |
| CaNH$_4$B$_6$C$_6$ | 1.333 H$_3$N + 3 B$_{13}$C$_2$ + 16.33 Ca$_3$N$_2$ $\longrightarrow$ 33 BN + 48 Ca + CaB$_6$H$_4$C$_6$N | -10.0958 |
| CaNH$_4$B$_6$C$_6$ | 10.21 H$_3$N + 3 B$_{13}$C$_2$ + CaN$_2$ $\longrightarrow$ 11.21 BN + 2.421 B$_9$H$_{11}$ + CaB$_6$H$_4$C$_6$N | -10.0850 |
| CaNH$_4$B$_6$C$_6$ | 3.856 Ca$_3$N$_2$ + 1.413 B$_9$H$_{11}$ + 1.2 H$_8$C$_5$ $\longrightarrow$ 10.57 CaH$_2$ + 6.713 BN + CaB$_6$H$_4$C$_6$N | -9.9092 |
| CaNH$_4$B$_6$C$_6$ | 2 H$_2$ + 3 B$_{13}$C$_2$ + 17 Ca$_3$N$_2$ $\longrightarrow$ 33 BN + 50 Ca + CaB$_6$H$_4$C$_6$N | -9.2166 |
| CaNH$_4$B$_6$C$_6$ | CaB$_6$ + 13.64 H$_3$N + 3 B$_{13}$C$_2$ $\longrightarrow$ 12.64 BN + 5.273 B$_5$H$_7$ + CaB$_6$H$_4$C$_6$N | -8.9853 |
| CaNH$_4$B$_6$C$_6$ | 3.8 CaN$_2$ + 12.6 B + 1.2 H$_8$C$_5$ $\longrightarrow$ 2.8 CaH$_2$ + 6.6 BN + CaB$_6$H$_4$C$_6$N | -8.8172 |
| CaNH$_4$B$_6$C$_6$ | 11.52 H$_3$N + 3 B$_{13}$C$_2$ + 0.3333 Ca$_3$N$_2$ $\longrightarrow$ 11.18 BN + 4.364 B$_5$H$_7$ + CaB$_6$H$_4$C$_6$N | -8.6455 |
| CaNH$_4$B$_6$C$_6$ | 11.73 H$_3$N + 3 B$_{13}$C$_2$ + Ca $\longrightarrow$ 10.73 BN + 4.455 B$_5$H$_7$ + CaB$_6$H$_4$C$_6$N | -8.6152 |
| CaNH$_4$B$_6$C$_6$ | 3 B$_{13}$C$_2$ + 0.3636 B$_9$H$_{11}$ + 0.5 N$_2$ + 7.045 Ca $\longrightarrow$ 6.045 CaB$_6$ + CaB$_6$H$_4$C$_6$N | -8.5682 |
| CaNH$_4$B$_6$C$_6$ | 0.5 Ca$_2$N + 11.57 H$_3$N + 3 B$_{13}$C$_2$ $\longrightarrow$ 11.07 BN + 4.386 B$_5$H$_7$ + CaB$_6$H$_4$C$_6$N | -8.4915 |
| CaNH$_4$B$_6$C$_6$ | 3 B$_{13}$C$_2$ + 0.5714 B$_5$H$_7$ + 0.5 N$_2$ + 6.976 Ca $\longrightarrow$ 5.976 CaB$_6$ + CaB$_6$H$_4$C$_6$N | -8.4326 |
| CaNH$_4$B$_6$C$_6$ | 16.73 H$_3$N + 3 B$_{13}$C$_2$ + 24.6 Ca $\longrightarrow$ 7.867 Ca$_3$N$_2$ + 6.6 B$_5$H$_7$ + CaB$_6$H$_4$C$_6$N | -7.9021 |
| CaNH$_4$B$_6$C$_6$ | 3 B$_{13}$C$_2$ + 0.5 CaN$_2$ + 0.3636 B$_9$H$_{11}$ + 6.545 Ca $\longrightarrow$ 6.045 CaB$_6$ + CaB$_6$H$_4$C$_6$N | -7.8111 |
| CaNH$_4$B$_6$C$_6$ | 3 B$_{13}$C$_2$ + 0.5 CaN$_2$ + 0.5714 B$_5$H$_7$ + 6.476 Ca $\longrightarrow$ 5.976 CaB$_6$ + CaB$_6$H$_4$C$_6$N | -7.6755 |
| CaNH$_4$B$_6$C$_6$ | 2 H$_2$ + 3 B$_{13}$C$_2$ + 0.5 N$_2$ + 6.5 Ca $\longrightarrow$ 5.5 CaB$_6$ + CaB$_6$H$_4$C$_6$N | -7.6691 |
| CaNH$_4$B$_6$C$_6$ | CaB$_6$ + 12.53 H$_3$N + 3 B$_{13}$C$_2$ $\longrightarrow$ 11.53 BN + 3.053 B$_9$H$_{11}$ + CaB$_6$H$_4$C$_6$N | -7.6314 |
| CaNH$_4$B$_6$C$_6$ | 10.6 H$_3$N + 3 B$_{13}$C$_2$ + 0.3333 Ca$_3$N$_2$ $\longrightarrow$ 10.26 BN + 2.526 B$_9$H$_{11}$ + CaB$_6$H$_4$C$_6$N | -7.5250 |
| CaNH$_4$B$_6$C$_6$ | 10.79 H$_3$N + 3 B$_{13}$C$_2$ + Ca $\longrightarrow$ 9.789 BN + 2.579 B$_9$H$_{11}$ + CaB$_6$H$_4$C$_6$N | -7.4715 |

| Compound | Chemical reaction | Reaction energy(eV) |
| --- | --- | --- |
| $CaNH_4B_6C_6$ | $0.5\,Ca_2N + 10.64\,H_3N + 3\,B_{13}C_2 \longrightarrow 10.14\,BN + 2.539\,B_9H_{11} + CaB_6H_4C_6N$ | -7.3652 |
| $CaNH_4B_6C_6$ | $1.333\,H_3N + 3\,B_{13}C_2 + 1.933\,Ca_3N_2 \longrightarrow 4.8\,CaB_6 + 4.2\,BN + CaB_6H_4C_6N$ | -7.0142 |
| $CaNH_4B_6C_6$ | $3\,B_{13}C_2 + 2.164\,Ca_3N_2 + 0.3636\,B_9H_{11} \longrightarrow 5.491\,CaB_6 + 3.327\,BN + CaB_6H_4C_6N$ | -6.9407 |
| $CaNH_4B_6C_6$ | $2\,H_2 + 3\,B_{13}C_2 + 0.5\,CaN_2 + 6\,Ca \longrightarrow 5.5\,CaB_6 + CaB_6H_4C_6N$ | -6.9120 |
| $CaNH_4B_6C_6$ | $Ca_2N + 2\,CaH_2 + 42\,B + 3\,CaC_2 \longrightarrow 6\,CaB_6 + CaB_6H_4C_6N$ | -6.8696 |
| $CaNH_4B_6C_6$ | $5.615\,Ca_2N + 2.123\,B_5H_7 + 1.2\,H_8C_5 \longrightarrow 10.23\,CaH_2 + 4.615\,BN + CaB_6H_4C_6N$ | -6.8227 |
| $CaNH_4B_6C_6$ | $3\,B_{13}C_2 + 2.143\,Ca_3N_2 + 0.5714\,B_5H_7 \longrightarrow 5.429\,CaB_6 + 3.286\,BN + CaB_6H_4C_6N$ | -6.8006 |
| $CaNH_4B_6C_6$ | $1.333\,H_3N + 3\,B_{13}C_2 + 7.167\,Ca \longrightarrow 5.5\,CaB_6 + 0.3333\,Ca_2N + CaB_6H_4C_6N$ | -6.6625 |
| $CaNH_4B_6C_6$ | $3\,B_{13}C_2 + 3.042\,Ca_3N_2 + 1.75\,B_9H_{11} \longrightarrow 8.125\,CaB_6 + 5.083\,H_3N + CaB_6H_4C_6N$ | -6.6604 |
| $CaNH_4B_6C_6$ | $3.6\,B_5H_7 + 4.033\,N_2 + 3\,CaC_2 \longrightarrow 2\,CaB_6 + 7.067\,H_3N + CaB_6H_4C_6N$ | -6.6343 |
| $CaNH_4B_6C_6$ | $1.333\,H_3N + 3\,B_{13}C_2 + 6.444\,Ca \longrightarrow 5.444\,CaB_6 + 0.3333\,BN + CaB_6H_4C_6N$ | -6.6005 |
| $CaNH_4B_6C_6$ | $3\,B_{13}C_2 + 0.5\,Ca_3N_2 + 0.3636\,B_9H_{11} + 5.545\,Ca \longrightarrow 6.045\,CaB_6 + CaB_6H_4C_6N$ | -6.5847 |
| $CaNH_4B_6C_6$ | $H_3N + 3\,B_{13}C_2 + 0.09091\,B_9H_{11} + 6.636\,Ca \longrightarrow 5.636\,CaB_6 + CaB_6H_4C_6N$ | -6.5698 |
| $CaNH_4B_6C_6$ | $1.333\,H_3N + 3\,B_{13}C_2 + 7\,Ca \longrightarrow 5.5\,CaB_6 + 0.1667\,Ca_3N_2 + CaB_6H_4C_6N$ | -6.5648 |
| $CaNH_4B_6C_6$ | $H_3N + 3\,B_{13}C_2 + 0.1429\,B_5H_7 + 6.619\,Ca \longrightarrow 5.619\,CaB_6 + CaB_6H_4C_6N$ | -6.5359 |
| $CaNH_4B_6C_6$ | $CaH_2 + 11.27\,H_3N + 3\,B_{13}C_2 \longrightarrow 10.27\,BN + 4.545\,B_5H_7 + CaB_6H_4C_6N$ | -6.4838 |
| $CaNH_4B_6C_6$ | $3\,B_{13}C_2 + BN + 0.3636\,B_9H_{11} + 7.212\,Ca \longrightarrow 6.212\,CaB_6 + CaB_6H_4C_6N$ | -6.4777 |
| $CaNH_4B_6C_6$ | $3\,B_{13}C_2 + 0.5\,Ca_3N_2 + 0.5714\,B_5H_7 + 5.476\,Ca \longrightarrow 5.976\,CaB_6 + CaB_6H_4C_6N$ | -6.4491 |
| $CaNH_4B_6C_6$ | $14.78\,H_3N + 3\,B_{13}C_2 + 21.67\,Ca \longrightarrow 6.889\,Ca_3N_2 + 3.667\,B_9H_{11} + CaB_6H_4C_6N$ | -6.3646 |
| $CaNH_4B_6C_6$ | $0.5\,H_2 + H_3N + 3\,B_{13}C_2 + 6.5\,Ca \longrightarrow 5.5\,CaB_6 + CaB_6H_4C_6N$ | -6.3450 |
| $CaNH_4B_6C_6$ | $3\,B_{13}C_2 + BN + 0.5714\,B_5H_7 + 7.143\,Ca \longrightarrow 6.143\,CaB_6 + CaB_6H_4C_6N$ | -6.3421 |
| $CaNH_4B_6C_6$ | $Ca_2N + 3\,B_{13}C_2 + 0.3636\,B_9H_{11} + 5.045\,Ca \longrightarrow 6.045\,CaB_6 + CaB_6H_4C_6N$ | -6.2918 |
| $CaNH_4B_6C_6$ | $1.75\,B_{13}C_2 + 2.969\,CaN_2 + 0.5\,H_8C_5 \longrightarrow 1.969\,CaB_6 + 4.937\,BN + CaB_6H_4C_6N$ | -6.1840 |
| $CaNH_4B_6C_6$ | $Ca_2N + 3\,B_{13}C_2 + 0.5714\,B_5H_7 + 4.976\,Ca \longrightarrow 5.976\,CaB_6 + CaB_6H_4C_6N$ | -6.1562 |
| $CaNH_4B_6C_6$ | $1.333\,H_3N + 3\,B_{13}C_2 + 6.667\,Ca \longrightarrow 5.5\,CaB_6 + 0.1667\,CaN_2 + CaB_6H_4C_6N$ | -6.1560 |
| $CaNH_4B_6C_6$ | $2.974\,Ca_2N + 1.333\,H_3N + 3\,B_{13}C_2 \longrightarrow 4.949\,CaB_6 + 3.308\,BN + CaB_6H_4C_6N$ | -6.0477 |
| $CaNH_4B_6C_6$ | $3\,B_{13}C_2 + 2.896\,Ca_3N_2 + 2.625\,B_5H_7 \longrightarrow 7.687\,CaB_6 + 4.792\,H_3N + CaB_6H_4C_6N$ | -6.0330 |
| $CaNH_4B_6C_6$ | $2\,H_2 + 3\,B_{13}C_2 + 2\,Ca_3N_2 \longrightarrow 5\,CaB_6 + 3\,BN + CaB_6H_4C_6N$ | -6.0066 |
| $CaNH_4B_6C_6$ | $1.333\,H_3N + 3\,B_{13}C_2 + 6.5\,Ca \longrightarrow 5.5\,CaB_6 + 0.1667\,N_2 + CaB_6H_4C_6N$ | -5.9037 |
| $CaNH_4B_6C_6$ | $2\,CaH_2 + 3\,B_{13}C_2 + 17\,Ca_3N_2 \longrightarrow 33\,BN + 52\,Ca + CaB_6H_4C_6N$ | -5.8934 |
| $CaNH_4B_6C_6$ | $3.329\,Ca_2N + 3\,B_{13}C_2 + 0.3636\,B_9H_{11} \longrightarrow 5.657\,CaB_6 + 2.329\,BN + CaB_6H_4C_6N$ | -5.8590 |
| $CaNH_4B_6C_6$ | $2\,CaH_2 + 0.5\,Ca_3N_2 + 39\,B + 3\,CaC_2 \longrightarrow 5.5\,CaB_6 + CaB_6H_4C_6N$ | -5.7338 |
| $CaNH_4B_6C_6$ | $3.297\,Ca_2N + 3\,B_{13}C_2 + 0.5714\,B_5H_7 \longrightarrow 5.593\,CaB_6 + 2.297\,BN + CaB_6H_4C_6N$ | -5.7293 |
| $CaNH_4B_6C_6$ | $2\,H_2 + 3\,B_{13}C_2 + 0.5\,Ca_3N_2 + 5\,Ca \longrightarrow 5.5\,CaB_6 + CaB_6H_4C_6N$ | -5.6856 |
| $CaNH_4B_6C_6$ | $2\,H_2 + 3\,B_{13}C_2 + BN + 6.667\,Ca \longrightarrow 5.667\,CaB_6 + CaB_6H_4C_6N$ | -5.5786 |
| $CaNH_4B_6C_6$ | $0.5\,CaH_2 + H_3N + 3\,B_{13}C_2 + 6\,Ca \longrightarrow 5.5\,CaB_6 + CaB_6H_4C_6N$ | -5.5142 |
| $CaNH_4B_6C_6$ | $4.171\,Ca_2N + 3\,B_{13}C_2 + 1.229\,B_9H_{11} \longrightarrow 7.343\,CaB_6 + 3.171\,H_3N + CaB_6H_4C_6N$ | -5.4103 |
| $CaNH_4B_6C_6$ | $Ca_2N + 2\,H_2 + 3\,B_{13}C_2 + 4.5\,Ca \longrightarrow 5.5\,CaB_6 + CaB_6H_4C_6N$ | -5.3927 |
| $CaNH_4B_6C_6$ | $2\,B_9H_{11} + 3.5\,N_2 + 3\,CaC_2 \longrightarrow 2\,CaB_6 + 6\,H_3N + CaB_6H_4C_6N$ | -5.3573 |
| $CaNH_4B_6C_6$ | $CaH_2 + 10.32\,H_3N + 3\,B_{13}C_2 \longrightarrow 9.316\,BN + 2.632\,B_9H_{11} + CaB_6H_4C_6N$ | -5.3167 |
| $CaNH_4B_6C_6$ | $8.667\,H_3N + 1.2\,B_5H_7 + 12.5\,CaC_2 \longrightarrow 3.833\,Ca_3N_2 + 3.8\,H_8C_5 + CaB_6H_4C_6N$ | -5.2340 |
| $CaNH_4B_6C_6$ | $3.077\,Ca_2N + 2\,H_2 + 3\,B_{13}C_2 \longrightarrow 5.154\,CaB_6 + 2.077\,BN + CaB_6H_4C_6N$ | -5.0067 |
| $CaNH_4B_6C_6$ | $4.029\,Ca_2N + 3\,B_{13}C_2 + 1.87\,B_5H_7 \longrightarrow 7.058\,CaB_6 + 3.029\,H_3N + CaB_6H_4C_6N$ | -5.0061 |
| $CaNH_4B_6C_6$ | $4.936\,Ca_2N + 1.104\,B_9H_{11} + 1.2\,H_8C_5 \longrightarrow 8.872\,CaH_2 + 3.936\,BN + CaB_6H_4C_6N$ | -4.5844 |
| $CaNH_4B_6C_6$ | $H_3N + 1.2\,B_5H_7 + 9.5\,Ca + 1.2\,H_8C_5 \longrightarrow 8.5\,CaH_2 + CaB_6H_4C_6N$ | -4.5237 |
| $CaNH_4B_6C_6$ | $H_3N + 2.687\,B_{13}C_2 + 5.823\,Ca + 0.125\,H_8C_5 \longrightarrow 4.823\,CaB_6 + CaB_6H_4C_6N$ | -4.5171 |
| $CaNH_4B_6C_6$ | $Ca_2N + 2\,H_2 + 30\,B + 3\,CaC_2 \longrightarrow 4\,CaB_6 + CaB_6H_4C_6N$ | -4.4781 |
| $CaNH_4B_6C_6$ | $2\,CaH_2 + 3\,B_{13}C_2 + 0.5\,N_2 + 4.5\,Ca \longrightarrow 5.5\,CaB_6 + CaB_6H_4C_6N$ | -4.3459 |
| $CaNH_4B_6C_6$ | $8.074\,H_3N + 0.6667\,B_9H_{11} + 11.61\,CaC_2 \longrightarrow 3.537\,Ca_3N_2 + 3.444\,H_8C_5 + CaB_6H_4C_6N$ | -4.1930 |
| $CaNH_4B_6C_6$ | $2\,CaH_2 + 0.5\,CaN_2 + 33\,B + 3\,CaC_2 \longrightarrow 4.5\,CaB_6 + CaB_6H_4C_6N$ | -4.1028 |
| $CaNH_4B_6C_6$ | $Ca_2N + 27.14\,B + 0.5714\,B_5H_7 + 3\,CaC_2 \longrightarrow 4\,CaB_6 + CaB_6H_4C_6N$ | -3.8809 |
| $CaNH_4B_6C_6$ | $Ca_2N + 26.73\,B + 0.3636\,B_9H_{11} + 3\,CaC_2 \longrightarrow 4\,CaB_6 + CaB_6H_4C_6N$ | -3.8186 |
| $CaNH_4B_6C_6$ | $H_3N + 0.6667\,B_9H_{11} + 8.967\,Ca + 1.2\,H_8C_5 \longrightarrow 7.967\,CaH_2 + CaB_6H_4C_6N$ | -3.6826 |
| $CaNH_4B_6C_6$ | $2\,CaH_2 + 3\,B_{13}C_2 + 0.5\,CaN_2 + 4\,Ca \longrightarrow 5.5\,CaB_6 + CaB_6H_4C_6N$ | -3.5888 |
| $CaNH_4B_6C_6$ | $7\,H_2 + 3\,B_{13}C_2 + 2.167\,Ca_3N_2 \longrightarrow 5.5\,CaB_6 + 3.333\,H_3N + CaB_6H_4C_6N$ | -3.4874 |
| $CaNH_4B_6C_6$ | $2\,CaH_2 + 30\,B + 0.5\,N_2 + 3\,CaC_2 \longrightarrow 4\,CaB_6 + CaB_6H_4C_6N$ | -3.4312 |
| $CaNH_4B_6C_6$ | $1.2\,B_5H_7 + 0.5\,N_2 + 8\,Ca + 1.2\,H_8C_5 \longrightarrow 7\,CaH_2 + CaB_6H_4C_6N$ | -3.3553 |
| $CaNH_4B_6C_6$ | $2\,H_2 + 0.5\,Ca_3N_2 + 27\,B + 3\,CaC_2 \longrightarrow 3.5\,CaB_6 + CaB_6H_4C_6N$ | -3.3422 |
| $CaNH_4B_6C_6$ | $11.09\,Ca_2N + 3\,B_{13}C_2 + 0.3636\,B_9H_{11} \longrightarrow 6.045\,CaB_6 + 5.045\,Ca_3N_2 + CaB_6H_4C_6N$ | -3.3366 |
| $CaNH_4B_6C_6$ | $3.25\,Ca_2N + 5.375\,H_2 + 3\,B_{13}C_2 \longrightarrow 5.5\,CaB_6 + 2.25\,H_3N + CaB_6H_4C_6N$ | -3.2500 |

| Compound | Chemical reaction | Reaction energy(eV) |
| --- | --- | --- |
| $CaNH_4B_6C_6$ | $10.95\,Ca_2N + 3\,B_{13}C_2 + 0.5714\,B_5H_7 \longrightarrow 5.976\,CaB_6 + 4.976\,Ca_3N_2 + CaB_6H_4C_6N$ | -3.2415 |
| $CaNH_4B_6C_6$ | $0.75\,B_{13}C_2 + 2.375\,N_2 + CaC_2 + 0.5\,H_8C_5 \longrightarrow 3.75\,BN + CaB_6H_4C_6N$ | -3.1296 |
| $CaNH_4B_6C_6$ | $10\,Ca_2N + 2\,H_2 + 3\,B_{13}C_2 \longrightarrow 5.5\,CaB_6 + 4.5\,Ca_3N_2 + CaB_6H_4C_6N$ | -2.7570 |
| $CaNH_4B_6C_6$ | $7.591\,H_3N + 2\,B_{13}C_2 + CaC_2 \longrightarrow 6.591\,BN + 2.682\,B_5H_7 + CaB_6H_4C_6N$ | -2.7463 |
| $CaNH_4B_6C_6$ | $0.5\,Ca_3N_2 + 24.14\,B + 0.5714\,B_5H_7 + 3\,CaC_2 \longrightarrow 3.5\,CaB_6 + CaB_6H_4C_6N$ | -2.7451 |
| $CaNH_4B_6C_6$ | $0.5\,Ca_3N_2 + 23.73\,B + 0.3636\,B_9H_{11} + 3\,CaC_2 \longrightarrow 3.5\,CaB_6 + CaB_6H_4C_6N$ | -2.6828 |
| $CaNH_4B_6C_6$ | $0.5\,CaN_2 + 1.2\,B_5H_7 + 7.5\,Ca + 1.2\,H_8C_5 \longrightarrow 7\,CaH_2 + CaB_6H_4C_6N$ | -2.5982 |
| $CaNH_4B_6C_6$ | $2\,CaH_2 + 3\,B_{13}C_2 + 1.4\,Ca_3N_2 \longrightarrow 5.2\,CaB_6 + 1.8\,BN + CaB_6H_4C_6N$ | -2.5550 |
| $CaNH_4B_6C_6$ | $0.6667\,B_9H_{11} + 0.5\,N_2 + 7.467\,Ca + 1.2\,H_8C_5 \longrightarrow 6.467\,CaH_2 + CaB_6H_4C_6N$ | -2.5142 |
| $CaNH_4B_6C_6$ | $14\,Ca_2N + 1.333\,H_3N + 3\,B_{13}C_2 \longrightarrow 5.5\,CaB_6 + 7.167\,Ca_3N_2 + CaB_6H_4C_6N$ | -2.4648 |
| $CaNH_4B_6C_6$ | $2.867\,Ca_2N + 34.4\,B + 1.2\,H_8C_5 \longrightarrow 4.733\,CaB_6 + 1.867\,H_3N + CaB_6H_4C_6N$ | -2.4491 |
| $CaNH_4B_6C_6$ | $2\,CaH_2 + 3\,B_{13}C_2 + 0.5\,Ca_3N_2 + 3\,Ca \longrightarrow 5.5\,CaB_6 + CaB_6H_4C_6N$ | -2.3624 |
| $CaNH_4B_6C_6$ | $2\,CaH_2 + 3\,B_{13}C_2 + BN + 4.667\,Ca \longrightarrow 5.667\,CaB_6 + CaB_6H_4C_6N$ | -2.2554 |
| $CaNH_4B_6C_6$ | $Ca_2N + 2\,CaH_2 + 3\,B_{13}C_2 + 2.5\,Ca \longrightarrow 5.5\,CaB_6 + CaB_6H_4C_6N$ | -2.0695 |
| $CaNH_4B_6C_6$ | $7.026\,H_3N + 2\,B_{13}C_2 + CaC_2 \longrightarrow 6.026\,BN + 1.553\,B_9H_{11} + CaB_6H_4C_6N$ | -2.0577 |
| $CaNH_4B_6C_6$ | $0.7778\,B_{13}C_2 + 2.556\,CaN_2 + 0.8889\,H_8C_5 \longrightarrow 1.556\,CaH_2 + 4.111\,BN + CaB_6H_4C_6N$ | -1.9798 |
| $CaNH_4B_6C_6$ | $2.154\,Ca_2N + 2\,CaH_2 + 3\,B_{13}C_2 \longrightarrow 5.308\,CaB_6 + 1.154\,BN + CaB_6H_4C_6N$ | -1.8550 |
| $CaNH_4B_6C_6$ | $0.5\,CaN_2 + 0.6667\,B_9H_{11} + 6.967\,Ca + 1.2\,H_8C_5 \longrightarrow 6.467\,CaH_2 + CaB_6H_4C_6N$ | -1.7571 |
| $CaNH_4B_6C_6$ | $2\,H_2 + 0.5\,CaN_2 + 21\,B + 3\,CaC_2 \longrightarrow 2.5\,CaB_6 + CaB_6H_4C_6N$ | -1.7113 |
| $CaNH_4B_6C_6$ | $5.059\,H_3N + 1.2\,B_5H_7 + 9.118\,CaC_2 \longrightarrow 4.059\,Ca_2N + 2.447\,H_8C_5 + CaB_6H_4C_6N$ | -1.3976 |
| $CaNH_4B_6C_6$ | $0.5\,Ca_3N_2 + 1.2\,B_5H_7 + 6.5\,Ca + 1.2\,H_8C_5 \longrightarrow 7\,CaH_2 + CaB_6H_4C_6N$ | -1.3718 |
| $CaNH_4B_6C_6$ | $4.364\,Ca_2N + 3\,B_{13}C_2 + 0.3636\,B_9H_{11} \longrightarrow 6.045\,CaB_6 + 1.682\,CaN_2 + CaB_6H_4C_6N$ | -1.1815 |
| $CaNH_4B_6C_6$ | $4.317\,Ca_2N + 3\,B_{13}C_2 + 0.5714\,B_5H_7 \longrightarrow 5.976\,CaB_6 + 1.659\,CaN_2 + CaB_6H_4C_6N$ | -1.1160 |
| $CaNH_4B_6C_6$ | $1.25\,CaN_2 + 1.5\,B_5H_7 + 3\,CaC_2 \longrightarrow 3.25\,CaH_2 + 1.5\,BN + CaB_6H_4C_6N$ | -1.1150 |
| $CaNH_4B_6C_6$ | $0.5\,CaN_2 + 18.14\,B + 0.5714\,B_5H_7 + 3\,CaC_2 \longrightarrow 2.5\,CaB_6 + CaB_6H_4C_6N$ | -1.1141 |
| $CaNH_4B_6C_6$ | $Ca_2N + 1.2\,B_5H_7 + 6\,Ca + 1.2\,H_8C_5 \longrightarrow 7\,CaH_2 + CaB_6H_4C_6N$ | -1.0790 |
| $CaNH_4B_6C_6$ | $0.5\,CaN_2 + 17.73\,B + 0.3636\,B_9H_{11} + 3\,CaC_2 \longrightarrow 2.5\,CaB_6 + CaB_6H_4C_6N$ | -1.0518 |
| $CaNH_4B_6C_6$ | $2\,H_2 + 18\,B + 0.5\,N_2 + 3\,CaC_2 \longrightarrow 2\,CaB_6 + CaB_6H_4C_6N$ | -1.0397 |
| $CaNH_4B_6C_6$ | $2\,CaH_2 + BN + 29\,B + 3\,CaC_2 \longrightarrow 4\,CaB_6 + CaB_6H_4C_6N$ | -0.8645 |
| $CaNH_4B_6C_6$ | $CaB_6 + 6\,C + 0.3636\,B_9H_{11} + 2.136\,N_2 \longrightarrow 3.273\,BN + CaB_6H_4C_6N$ | -0.8639 |
| $CaNH_4B_6C_6$ | $6.5\,CaH_2 + 3\,B_{13}C_2 + 2\,N_2 \longrightarrow 5.5\,CaB_6 + 3\,H_3N + CaB_6H_4C_6N$ | -0.8409 |
| $CaNH_4B_6C_6$ | $4\,Ca_2N + 2\,H_2 + 3\,B_{13}C_2 \longrightarrow 5.5\,CaB_6 + 1.5\,CaN_2 + CaB_6H_4C_6N$ | -0.8349 |
| $CaNH_4B_6C_6$ | $4.745\,H_3N + 0.6667\,B_9H_{11} + 8.49\,CaC_2 \longrightarrow 3.745\,Ca_2N + 2.196\,H_8C_5 + CaB_6H_4C_6N$ | -0.6531 |
| $CaNH_4B_6C_6$ | $6\,Ca_2N + 2\,CaH_2 + 3\,B_{13}C_2 \longrightarrow 5.5\,CaB_6 + 2.5\,Ca_3N_2 + CaB_6H_4C_6N$ | -0.6052 |
| $CaNH_4B_6C_6$ | $3.523\,Ca_2N + 3\,B_{13}C_2 + 0.3636\,B_9H_{11} \longrightarrow 6.045\,CaB_6 + 1.261\,N_2 + CaB_6H_4C_6N$ | -0.5491 |
| $CaNH_4B_6C_6$ | $0.5\,Ca_3N_2 + 0.6667\,B_9H_{11} + 5.967\,Ca + 1.2\,H_8C_5 \longrightarrow 6.467\,CaH_2 + CaB_6H_4C_6N$ | -0.5307 |
| $CaNH_4B_6C_6$ | $32.67\,Ca_2N + 1.333\,H_3N + 3\,B_{13}C_2 \longrightarrow 33\,BN + 64.33\,Ca + CaB_6H_4C_6N$ | -0.5291 |
| $CaNH_4B_6C_6$ | $3.488\,Ca_2N + 3\,B_{13}C_2 + 0.5714\,B_5H_7 \longrightarrow 5.976\,CaB_6 + 1.244\,N_2 + CaB_6H_4C_6N$ | -0.4923 |
| $CaNH_4B_6C_6$ | $15.14\,B + 0.5714\,B_5H_7 + 0.5\,N_2 + 3\,CaC_2 \longrightarrow 2\,CaB_6 + CaB_6H_4C_6N$ | -0.4425 |
| $CaNH_4B_6C_6$ | $14.73\,B + 0.3636\,B_9H_{11} + 0.5\,N_2 + 3\,CaC_2 \longrightarrow 2\,CaB_6 + CaB_6H_4C_6N$ | -0.3803 |
| $CaNH_4B_6C_6$ | $1.75\,B_{13}C_2 + 0.5\,N_2 + 3.792\,Ca + 0.5\,H_8C_5 \longrightarrow 2.792\,CaB_6 + CaB_6H_4C_6N$ | -0.3573 |
| $CaNH_4B_6C_6$ | $0.5\,CaH_2 + H_3N + 21\,B + 3\,CaC_2 \longrightarrow 2.5\,CaB_6 + CaB_6H_4C_6N$ | -0.3135 |
| $CaNH_4B_6C_6$ | $3.25\,Ca_2N + 2\,H_2 + 3\,B_{13}C_2 \longrightarrow 5.5\,CaB_6 + 1.125\,N_2 + CaB_6H_4C_6N$ | -0.2708 |
| $CaNH_4B_6C_6$ | $1.2\,B_5H_7 + 3.875\,N_2 + 4.375\,CaC_2 \longrightarrow 3.375\,CaN_2 + 0.55\,H_8C_5 + CaB_6H_4C_6N$ | -0.2570 |
| $CaNH_4B_6C_6$ | $Ca_2N + 0.6667\,B_9H_{11} + 5.467\,Ca + 1.2\,H_8C_5 \longrightarrow 6.467\,CaH_2 + CaB_6H_4C_6N$ | -0.2379 |
| $CaNH_4B_6C_6$ | $1.75\,B_{13}C_2 + 8.875\,Ca_3N_2 + 0.5\,H_8C_5 \longrightarrow 16.75\,BN + 25.62\,Ca + CaB_6H_4C_6N$ | -0.1660 |
| $CaNH_4B_6C_6$ | $1.333\,H_3N + 18.33\,B + 3\,CaC_2 \longrightarrow 2\,CaB_6 + 0.3333\,BN + CaB_6H_4C_6N$ | -0.1299 |
| $LaNH_4B_6C_6$ | $50\,La + 34\,H_3N + 3\,B_{13}C_2 \longrightarrow 33\,BN + 49\,LaH_2 + LaB_6H_4C_6N$ | -125.2998 |
| $LaNH_4B_6C_6$ | $44.56\,La + 34\,H_3N + 3\,B_{13}C_2 \longrightarrow 33\,BN + 10.89\,La_4H_9 + LaB_6H_4C_6N$ | -120.4946 |
| $LaNH_4B_6C_6$ | $40.2\,La + 34\,H_3N + 3\,B_{13}C_2 \longrightarrow 33\,BN + 19.6\,La_2H_5 + LaB_6H_4C_6N$ | -116.0742 |
| $LaNH_4B_6C_6$ | $2\,LaH_2 + 23.5\,N_2 + 4\,B_{13}C_2 \longrightarrow LaC_2 + 46\,BN + LaB_6H_4C_6N$ | -107.1535 |
| $LaNH_4B_6C_6$ | $33.67\,La + 34\,H_3N + 3\,B_{13}C_2 \longrightarrow 32.67\,LaH_3 + 33\,BN + LaB_6H_4C_6N$ | -107.1373 |
| $LaNH_4B_6C_6$ | $70\,LaC_2 + 70\,H_3N + 1.2\,B_5H_7 \longrightarrow 26.8\,H_8C_5 + 69\,LaN + LaB_6H_4C_6N$ | -101.1075 |
| $LaNH_4B_6C_6$ | $0.4444\,La_4H_9 + 22.06\,N_2 + 3.778\,B_{13}C_2 \longrightarrow 0.7778\,LaC_2 + 43.11\,BN + LaB_6H_4C_6N$ | -100.0652 |
| $LaNH_4B_6C_6$ | $2\,LaH_2 + 21.87\,N_2 + 3.75\,B_{13}C_2 \longrightarrow 42.75\,BN + 0.5\,La_2C_3 + LaB_6H_4C_6N$ | -99.0157 |
| $LaNH_4B_6C_6$ | $LaB_6 + 0.3636\,B_9H_{11} + 21.64\,N_2 + 3\,B_{13}C_2 \longrightarrow 42.27\,BN + LaB_6H_4C_6N$ | -97.3404 |
| $LaNH_4B_6C_6$ | $LaB_6 + 0.5714\,B_5H_7 + 21.43\,N_2 + 3\,B_{13}C_2 \longrightarrow 41.86\,BN + LaB_6H_4C_6N$ | -96.3360 |
| $LaNH_4B_6C_6$ | $20.9\,N_2 + 3.6\,B_{13}C_2 + 0.8\,La_2H_5 \longrightarrow 0.6\,LaC_2 + 40.8\,BN + LaB_6H_4C_6N$ | -94.4180 |
| $LaNH_4B_6C_6$ | $0.4444\,La_4H_9 + 20.79\,N_2 + 3.583\,B_{13}C_2 \longrightarrow 40.58\,BN + 0.3889\,La_2C_3 + LaB_6H_4C_6N$ | -93.7358 |
| $LaNH_4B_6C_6$ | $0.3636\,B_9H_{11} + 20.64\,N_2 + 3\,B_{13}C_2 + LaB_4 \longrightarrow 40.27\,BN + LaB_6H_4C_6N$ | -93.2975 |

| Compound | Chemical reaction | Reaction energy(eV) |
| --- | --- | --- |
| $LaNH_4B_6C_6$ | $64.67\,LaC_2 + 64.67\,H_3N + 0.6667\,B_9H_{11} \longrightarrow 24.67\,H_8C_5 + 63.67\,LaN + LaB_6H_4C_6N$ | -92.7289 |
| $LaNH_4B_6C_6$ | $0.5714\,B_5H_7 + 20.43\,N_2 + 3\,B_{13}C_2 + LaB_4 \longrightarrow 39.86\,BN + LaB_6H_4C_6N$ | -92.2931 |
| $LaNH_4B_6C_6$ | $2\,H_2 + LaB_6 + 20\,N_2 + 3\,B_{13}C_2 \longrightarrow 39\,BN + LaB_6H_4C_6N$ | -89.5996 |
| $LaNH_4B_6C_6$ | $19.93\,N_2 + 3.45\,B_{13}C_2 + 0.8\,La_2H_5 \longrightarrow 38.85\,BN + 0.3\,La_2C_3 + LaB_6H_4C_6N$ | -89.5353 |
| $LaNH_4B_6C_6$ | $LaB_6 + 1.333\,H_3N + 19.33\,N_2 + 3\,B_{13}C_2 \longrightarrow 39\,BN + LaB_6H_4C_6N$ | -87.8341 |
| $LaNH_4B_6C_6$ | $1.333\,LaH_3 + 19.17\,N_2 + 3.333\,B_{13}C_2 \longrightarrow 0.3333\,LaC_2 + 37.33\,BN + LaB_6H_4C_6N$ | -86.0414 |
| $LaNH_4B_6C_6$ | $La + 0.3636\,B_9H_{11} + 18.64\,N_2 + 3\,B_{13}C_2 \longrightarrow 36.27\,BN + LaB_6H_4C_6N$ | -85.8768 |
| $LaNH_4B_6C_6$ | $2\,H_2 + 19\,N_2 + 3\,B_{13}C_2 + LaB_4 \longrightarrow 37\,BN + LaB_6H_4C_6N$ | -85.5567 |
| $LaNH_4B_6C_6$ | $La + 0.5714\,B_5H_7 + 18.43\,N_2 + 3\,B_{13}C_2 \longrightarrow 35.86\,BN + LaB_6H_4C_6N$ | -84.8724 |
| $LaNH_4B_6C_6$ | $1.333\,H_3N + 18.33\,N_2 + 3\,B_{13}C_2 + LaB_4 \longrightarrow 37\,BN + LaB_6H_4C_6N$ | -83.7912 |
| $LaNH_4B_6C_6$ | $1.333\,LaH_3 + 18.63\,N_2 + 3.25\,B_{13}C_2 \longrightarrow 36.25\,BN + 0.1667\,La_2C_3 + LaB_6H_4C_6N$ | -83.3288 |
| $LaNH_4B_6C_6$ | $LaN + 0.3636\,B_9H_{11} + 18.14\,N_2 + 3\,B_{13}C_2 \longrightarrow 36.27\,BN + LaB_6H_4C_6N$ | -83.2195 |
| $LaNH_4B_6C_6$ | $LaN + 0.5714\,B_5H_7 + 17.93\,N_2 + 3\,B_{13}C_2 \longrightarrow 35.86\,BN + LaB_6H_4C_6N$ | -82.2150 |
| $LaNH_4B_6C_6$ | $34\,H_3N + 401\,LaH_2 + 3\,B_{13}C_2 \longrightarrow 33\,BN + 100\,La_4H_9 + LaB_6H_4C_6N$ | -81.1703 |
| $LaNH_4B_6C_6$ | $0.1818\,B_9H_{11} + LaH_2 + 17.82\,N_2 + 3\,B_{13}C_2 \longrightarrow 34.64\,BN + LaB_6H_4C_6N$ | -80.1251 |
| $LaNH_4B_6C_6$ | $0.2857\,B_5H_7 + LaH_2 + 17.71\,N_2 + 3\,B_{13}C_2 \longrightarrow 34.43\,BN + LaB_6H_4C_6N$ | -79.6229 |
| $LaNH_4B_6C_6$ | $0.25\,La_4H_9 + 0.1591\,B_9H_{11} + 17.72\,N_2 + 3\,B_{13}C_2 \longrightarrow 34.43\,BN + LaB_6H_4C_6N$ | -79.5165 |
| $LaNH_4B_6C_6$ | $0.25\,La_4H_9 + 0.25\,B_5H_7 + 17.63\,N_2 + 3\,B_{13}C_2 \longrightarrow 34.25\,BN + LaB_6H_4C_6N$ | -79.0770 |
| $LaNH_4B_6C_6$ | $0.1364\,B_9H_{11} + 17.61\,N_2 + 3\,B_{13}C_2 + 0.5\,La_2H_5 \longrightarrow 34.23\,BN + LaB_6H_4C_6N$ | -78.9225 |
| $LaNH_4B_6C_6$ | $0.2143\,B_5H_7 + 17.54\,N_2 + 3\,B_{13}C_2 + 0.5\,La_2H_5 \longrightarrow 34.07\,BN + LaB_6H_4C_6N$ | -78.5459 |
| $LaNH_4B_6C_6$ | $34\,H_3N + 201\,LaH_2 + 3\,B_{13}C_2 \longrightarrow 33\,BN + 100\,La_2H_5 + LaB_6H_4C_6N$ | -78.2303 |
| $LaNH_4B_6C_6$ | $2\,H_2 + La + 17\,N_2 + 3\,B_{13}C_2 \longrightarrow 33\,BN + LaB_6H_4C_6N$ | -78.1359 |
| $LaNH_4B_6C_6$ | $LaH_3 + 0.09091\,B_9H_{11} + 17.41\,N_2 + 3\,B_{13}C_2 \longrightarrow 33.82\,BN + LaB_6H_4C_6N$ | -77.8053 |
| $LaNH_4B_6C_6$ | $LaH_3 + 0.1429\,B_5H_7 + 17.36\,N_2 + 3\,B_{13}C_2 \longrightarrow 33.71\,BN + LaB_6H_4C_6N$ | -77.5542 |
| $LaNH_4B_6C_6$ | $2\,LaH_2 + 17.5\,N_2 + 3\,B_{13}C_2 \longrightarrow 33\,BN + LaN + LaB_6H_4C_6N$ | -77.0308 |
| $LaNH_4B_6C_6$ | $0.4444\,La_4H_9 + 17.39\,N_2 + 3\,B_{13}C_2 \longrightarrow 33\,BN + 0.7778\,LaN + LaB_6H_4C_6N$ | -76.6364 |
| $LaNH_4B_6C_6$ | $La + 1.333\,H_3N + 16.33\,N_2 + 3\,B_{13}C_2 \longrightarrow 33\,BN + LaB_6H_4C_6N$ | -76.3705 |
| $LaNH_4B_6C_6$ | $17.3\,N_2 + 3\,B_{13}C_2 + 0.8\,La_2H_5 \longrightarrow 33\,BN + 0.6\,LaN + LaB_6H_4C_6N$ | -76.3444 |
| $LaNH_4B_6C_6$ | $H_2 + LaH_2 + 17\,N_2 + 3\,B_{13}C_2 \longrightarrow 33\,BN + LaB_6H_4C_6N$ | -76.2547 |
| $LaNH_4B_6C_6$ | $0.875\,H_2 + 0.25\,La_4H_9 + 17\,N_2 + 3\,B_{13}C_2 \longrightarrow 33\,BN + LaB_6H_4C_6N$ | -76.1298 |
| $LaNH_4B_6C_6$ | $0.75\,H_2 + 17\,N_2 + 3\,B_{13}C_2 + 0.5\,La_2H_5 \longrightarrow 33\,BN + LaB_6H_4C_6N$ | -76.0197 |
| $LaNH_4B_6C_6$ | $1.333\,LaH_3 + 17.17\,N_2 + 3\,B_{13}C_2 \longrightarrow 33\,BN + 0.3333\,LaN + LaB_6H_4C_6N$ | -76.0005 |
| $LaNH_4B_6C_6$ | $LaH_3 + 0.5\,H_2 + 17\,N_2 + 3\,B_{13}C_2 \longrightarrow 33\,BN + LaB_6H_4C_6N$ | -75.8700 |
| $LaNH_4B_6C_6$ | $3\,LaH_3 + 17\,N_2 + 3\,B_{13}C_2 \longrightarrow 33\,BN + La_2H_5 + LaB_6H_4C_6N$ | -75.5707 |
| $LaNH_4B_6C_6$ | $2.333\,LaH_3 + 17\,N_2 + 3\,B_{13}C_2 \longrightarrow 33\,BN + 0.3333\,La_4H_9 + LaB_6H_4C_6N$ | -75.5236 |
| $LaNH_4B_6C_6$ | $2\,LaH_3 + 17\,N_2 + 3\,B_{13}C_2 \longrightarrow 33\,BN + LaH_2 + LaB_6H_4C_6N$ | -75.4854 |
| $LaNH_4B_6C_6$ | $2\,H_2 + LaN + 16.5\,N_2 + 3\,B_{13}C_2 \longrightarrow 33\,BN + LaB_6H_4C_6N$ | -75.4786 |
| $LaNH_4B_6C_6$ | $LaH_3 + 0.3333\,H_3N + 16.83\,N_2 + 3\,B_{13}C_2 \longrightarrow 33\,BN + LaB_6H_4C_6N$ | -75.4287 |
| $LaNH_4B_6C_6$ | $0.6667\,H_3N + LaH_2 + 16.67\,N_2 + 3\,B_{13}C_2 \longrightarrow 33\,BN + LaB_6H_4C_6N$ | -75.3720 |
| $LaNH_4B_6C_6$ | $17\,N_2 + 3\,B_{13}C_2 + 3.5\,La_2H_5 \longrightarrow 33\,BN + 1.5\,La_4H_9 + LaB_6H_4C_6N$ | -75.3589 |
| $LaNH_4B_6C_6$ | $0.5\,H_3N + 16.75\,N_2 + 3\,B_{13}C_2 + 0.5\,La_2H_5 \longrightarrow 33\,BN + LaB_6H_4C_6N$ | -75.3577 |
| $LaNH_4B_6C_6$ | $0.25\,La_4H_9 + 0.5833\,H_3N + 16.71\,N_2 + 3\,B_{13}C_2 \longrightarrow 33\,BN + LaB_6H_4C_6N$ | -75.3575 |
| $LaNH_4B_6C_6$ | $17\,N_2 + 3\,B_{13}C_2 + 2\,La_2H_5 \longrightarrow 33\,BN + 3\,LaH_2 + LaB_6H_4C_6N$ | -75.3148 |
| $LaNH_4B_6C_6$ | $100.5\,La_4H_9 + 34\,H_3N + 3\,B_{13}C_2 \longrightarrow 33\,BN + 200.5\,La_2H_5 + LaB_6H_4C_6N$ | -75.2756 |
| $LaNH_4B_6C_6$ | $2\,La_4H_9 + 17\,N_2 + 3\,B_{13}C_2 \longrightarrow 33\,BN + 7\,LaH_2 + LaB_6H_4C_6N$ | -75.2560 |
| $LaNH_4B_6C_6$ | $1.333\,LaH_3 + 17\,N_2 + 3\,B_{13}C_2 \longrightarrow 0.3333\,La + 33\,BN + LaB_6H_4C_6N$ | -75.1147 |
| $LaNH_4B_6C_6$ | $17\,N_2 + 3\,B_{13}C_2 + 0.8\,La_2H_5 \longrightarrow 0.6\,La + 33\,BN + LaB_6H_4C_6N$ | -74.7500 |
| $LaNH_4B_6C_6$ | $0.4444\,La_4H_9 + 17\,N_2 + 3\,B_{13}C_2 \longrightarrow 0.7778\,La + 33\,BN + LaB_6H_4C_6N$ | -74.5695 |
| $LaNH_4B_6C_6$ | $2\,LaH_2 + 17\,N_2 + 3\,B_{13}C_2 \longrightarrow La + 33\,BN + LaB_6H_4C_6N$ | -74.3734 |
| $LaNH_4B_6C_6$ | $1.333\,H_3N + LaN + 15.83\,N_2 + 3\,B_{13}C_2 \longrightarrow 33\,BN + LaB_6H_4C_6N$ | -73.7132 |
| $LaNH_4B_6C_6$ | $1.333\,LaH_3 + 16.33\,N_2 + 3\,B_{13}C_2 \longrightarrow 31.67\,BN + 0.3333\,LaB_4 + LaB_6H_4C_6N$ | -72.6412 |
| $LaNH_4B_6C_6$ | $1.333\,LaH_3 + 16\,N_2 + 3\,B_{13}C_2 \longrightarrow 0.3333\,LaB_6 + 31\,BN + LaB_6H_4C_6N$ | -71.2935 |
| $LaNH_4B_6C_6$ | $15.8\,N_2 + 3\,B_{13}C_2 + 0.8\,La_2H_5 \longrightarrow 30.6\,BN + 0.6\,LaB_4 + LaB_6H_4C_6N$ | -70.2975 |
| $LaNH_4B_6C_6$ | $34\,H_3N + 101\,LaH_2 + 3\,B_{13}C_2 \longrightarrow 100\,LaH_3 + 33\,BN + LaB_6H_4C_6N$ | -69.7003 |
| $LaNH_4B_6C_6$ | $50.5\,LaC_2 + 52\,H_3N + 1.5\,LaB_4 \longrightarrow 19\,H_8C_5 + 51\,LaN + LaB_6H_4C_6N$ | -69.2928 |
| $LaNH_4B_6C_6$ | $0.4444\,La_4H_9 + 15.44\,N_2 + 3\,B_{13}C_2 \longrightarrow 29.89\,BN + 0.7778\,LaB_4 + LaB_6H_4C_6N$ | -68.7979 |
| $LaNH_4B_6C_6$ | $15.2\,N_2 + 3\,B_{13}C_2 + 0.8\,La_2H_5 \longrightarrow 0.6\,LaB_6 + 29.4\,BN + LaB_6H_4C_6N$ | -67.8718 |
| $LaNH_4B_6C_6$ | $2\,LaH_2 + 15\,N_2 + 3\,B_{13}C_2 \longrightarrow 29\,BN + LaB_4 + LaB_6H_4C_6N$ | -66.9527 |
| $LaNH_4B_6C_6$ | $33.67\,La_4H_9 + 34\,H_3N + 3\,B_{13}C_2 \longrightarrow 133.7\,LaH_3 + 33\,BN + LaB_6H_4C_6N$ | -65.8387 |
| $LaNH_4B_6C_6$ | $0.4444\,La_4H_9 + 14.67\,N_2 + 3\,B_{13}C_2 \longrightarrow 0.7778\,LaB_6 + 28.33\,BN + LaB_6H_4C_6N$ | -65.6534 |

| Compound | Chemical reaction | Reaction energy(eV) |
|---|---|---|
| LaNH$_4$B$_6$C$_6$ | LaH$_3$ + 0.125 H$_8$C$_5$ + 14.97 N$_2$ + 2.687 B$_{13}$C$_2$ ⟶ 28.94 BN + LaB$_6$H$_4$C$_6$N | -65.5493 |
| LaNH$_4$B$_6$C$_6$ | 2 LaH$_2$ + 14 N$_2$ + 3 B$_{13}$C$_2$ ⟶ LaB$_6$ + 27 BN + LaB$_6$H$_4$C$_6$N | -62.9098 |
| LaNH$_4$B$_6$C$_6$ | 51.5 H$_2$ + 34 LaN + 3 B$_{13}$C$_2$ ⟶ 33 LaH$_3$ + 33 BN + LaB$_6$H$_4$C$_6$N | -62.5614 |
| LaNH$_4$B$_6$C$_6$ | 0.3636 B$_9$H$_{11}$ + 0.5 La$_2$C$_3$ + 13.76 N$_2$ + 2.25 B$_{13}$C$_2$ ⟶ 26.52 BN + LaB$_6$H$_4$C$_6$N | -61.2345 |
| LaNH$_4$B$_6$C$_6$ | 34 H$_3$N + 3 B$_{13}$C$_2$ + 101 La$_2$H$_5$ ⟶ 201 LaH$_3$ + 33 BN + LaB$_6$H$_4$C$_6$N | -61.0850 |
| LaNH$_4$B$_6$C$_6$ | 0.1875 H$_8$C$_5$ + 13.95 N$_2$ + 2.531 B$_{13}$C$_2$ + 0.5 La$_2$H$_5$ ⟶ 26.91 BN + LaB$_6$H$_4$C$_6$N | -60.5386 |
| LaNH$_4$B$_6$C$_6$ | 0.5 La$_2$C$_3$ + 0.5714 B$_5$H$_7$ + 13.55 N$_2$ + 2.25 B$_{13}$C$_2$ ⟶ 26.11 BN + LaB$_6$H$_4$C$_6$N | -60.2301 |
| LaNH$_4$B$_6$C$_6$ | 0.2187 H$_8$C$_5$ + 0.25 La$_4$H$_9$ + 13.45 N$_2$ + 2.453 B$_{13}$C$_2$ ⟶ 25.89 BN + LaB$_6$H$_4$C$_6$N | -58.0686 |
| LaNH$_4$B$_6$C$_6$ | 43.25 H$_2$ + 34 LaN + 3 B$_{13}$C$_2$ ⟶ 33 BN + 16.5 La$_2$H$_5$ + LaB$_6$H$_4$C$_6$N | -57.6223 |
| LaNH$_4$B$_6$C$_6$ | 43 LaC$_2$ + LaB$_6$ + 44 H$_3$N ⟶ 16 H$_8$C$_5$ + 43 LaN + LaB$_6$H$_4$C$_6$N | -57.1128 |
| LaNH$_4$B$_6$C$_6$ | 0.25 H$_8$C$_5$ + LaH$_2$ + 12.94 N$_2$ + 2.375 B$_{13}$C$_2$ ⟶ 24.87 BN + LaB$_6$H$_4$C$_6$N | -55.6132 |
| LaNH$_4$B$_6$C$_6$ | 39.12 H$_2$ + 34 LaN + 3 B$_{13}$C$_2$ ⟶ 33 BN + 8.25 La$_4$H$_9$ + LaB$_6$H$_4$C$_6$N | -53.9879 |
| LaNH$_4$B$_6$C$_6$ | 2 H$_2$ + 0.5 La$_2$C$_3$ + 12.12 N$_2$ + 2.25 B$_{13}$C$_2$ ⟶ 23.25 BN + LaB$_6$H$_4$C$_6$N | -53.4936 |
| LaNH$_4$B$_6$C$_6$ | LaC$_2$ + 0.3636 B$_9$H$_{11}$ + 12.14 N$_2$ + 2 B$_{13}$C$_2$ ⟶ 23.27 BN + LaB$_6$H$_4$C$_6$N | -53.0967 |
| LaNH$_4$B$_6$C$_6$ | LaC$_2$ + 0.5714 B$_5$H$_7$ + 11.93 N$_2$ + 2 B$_{13}$C$_2$ ⟶ 22.86 BN + LaB$_6$H$_4$C$_6$N | -52.0923 |
| LaNH$_4$B$_6$C$_6$ | 1.333 H$_3$N + 0.5 La$_2$C$_3$ + 11.46 N$_2$ + 2.25 B$_{13}$C$_2$ ⟶ 23.25 BN + LaB$_6$H$_4$C$_6$N | -51.7282 |
| LaNH$_4$B$_6$C$_6$ | 35 H$_2$ + 34 LaN + 3 B$_{13}$C$_2$ ⟶ 33 BN + 33 LaH$_2$ + LaB$_6$H$_4$C$_6$N | -49.8683 |
| LaNH$_4$B$_6$C$_6$ | 0.5 H$_8$C$_5$ + LaB$_6$ + 11.87 N$_2$ + 1.75 B$_{13}$C$_2$ ⟶ 22.75 BN + LaB$_6$H$_4$C$_6$N | -48.3167 |
| LaNH$_4$B$_6$C$_6$ | 17.17 H$_3$N + 16.83 LaN + 3 B$_{13}$C$_2$ ⟶ 15.83 LaH$_3$ + 33 BN + LaB$_6$H$_4$C$_6$N | -46.5511 |
| LaNH$_4$B$_6$C$_6$ | LaC$_2$ + 2 H$_2$ + 10.5 N$_2$ + 2 B$_{13}$C$_2$ ⟶ 20 BN + LaB$_6$H$_4$C$_6$N | -45.3558 |
| LaNH$_4$B$_6$C$_6$ | 15.73 H$_3$N + 18.27 LaN + 3 B$_{13}$C$_2$ ⟶ 33 BN + 8.636 La$_2$H$_5$ + LaB$_6$H$_4$C$_6$N | -45.3083 |
| LaNH$_4$B$_6$C$_6$ | 0.5 H$_8$C$_5$ + 10.87 N$_2$ + 1.75 B$_{13}$C$_2$ + LaB$_4$ ⟶ 20.75 BN + LaB$_6$H$_4$C$_6$N | -44.2738 |
| LaNH$_4$B$_6$C$_6$ | 14.9 H$_3$N + 19.1 LaN + 3 B$_{13}$C$_2$ ⟶ 33 BN + 4.524 La$_4$H$_9$ + LaB$_6$H$_4$C$_6$N | -43.9594 |
| LaNH$_4$B$_6$C$_6$ | LaC$_2$ + 1.333 H$_3$N + 9.833 N$_2$ + 2 B$_{13}$C$_2$ ⟶ 20 BN + LaB$_6$H$_4$C$_6$N | -43.5904 |
| LaNH$_4$B$_6$C$_6$ | 14 H$_3$N + 20 LaN + 3 B$_{13}$C$_2$ ⟶ 33 BN + 19 LaH$_2$ + LaB$_6$H$_4$C$_6$N | -42.1963 |
| LaNH$_4$B$_6$C$_6$ | 0.5 H$_8$C$_5$ + La + 8.875 N$_2$ + 1.75 B$_{13}$C$_2$ ⟶ 16.75 BN + LaB$_6$H$_4$C$_6$N | -36.8531 |
| LaNH$_4$B$_6$C$_6$ | LaB$_6$ + 40 H$_3$N + 3 B$_{13}$C$_2$ ⟶ 58 H$_2$ + 39 BN + LaB$_6$H$_4$C$_6$N | -36.6368 |
| LaNH$_4$B$_6$C$_6$ | 28 LaC$_2$ + 6 B + 28 H$_3$N ⟶ 10 H$_8$C$_5$ + 27 LaN + LaB$_6$H$_4$C$_6$N | -36.0248 |
| LaNH$_4$B$_6$C$_6$ | 38 H$_3$N + 3 B$_{13}$C$_2$ + LaB$_4$ ⟶ 55 H$_2$ + 37 BN + LaB$_6$H$_4$C$_6$N | -35.2420 |
| LaNH$_4$B$_6$C$_6$ | 0.5 H$_8$C$_5$ + LaN + 8.375 N$_2$ + 1.75 B$_{13}$C$_2$ ⟶ 16.75 BN + LaB$_6$H$_4$C$_6$N | -34.1957 |
| LaNH$_4$B$_6$C$_6$ | La + 34 H$_3$N + 3 B$_{13}$C$_2$ ⟶ 49 H$_2$ + 33 BN + LaB$_6$H$_4$C$_6$N | -33.1176 |
| LaNH$_4$B$_6$C$_6$ | 33 H$_3$N + LaN + 3 B$_{13}$C$_2$ ⟶ 47.5 H$_2$ + 33 BN + LaB$_6$H$_4$C$_6$N | -31.7843 |
| LaNH$_4$B$_6$C$_6$ | 173 LaH$_3$ + 34 LaN + 3 B$_{13}$C$_2$ ⟶ 33 BN + 103 La$_2$H$_5$ + LaB$_6$H$_4$C$_6$N | -31.7294 |
| LaNH$_4$B$_6$C$_6$ | 34 H$_3$N + LaH$_2$ + 3 B$_{13}$C$_2$ ⟶ 50 H$_2$ + 33 BN + LaB$_6$H$_4$C$_6$N | -31.2363 |
| LaNH$_4$B$_6$C$_6$ | 0.25 La$_4$H$_9$ + 34 H$_3$N + 3 B$_{13}$C$_2$ ⟶ 50.12 H$_2$ + 33 BN + LaB$_6$H$_4$C$_6$N | -31.1115 |
| LaNH$_4$B$_6$C$_6$ | 34 H$_3$N + 3 B$_{13}$C$_2$ + 0.5 La$_2$H$_5$ ⟶ 50.25 H$_2$ + 33 BN + LaB$_6$H$_4$C$_6$N | -31.0013 |
| LaNH$_4$B$_6$C$_6$ | LaH$_3$ + 34 H$_3$N + 3 B$_{13}$C$_2$ ⟶ 50.5 H$_2$ + 33 BN + LaB$_6$H$_4$C$_6$N | -30.8517 |
| LaNH$_4$B$_6$C$_6$ | 1.2 H$_8$C$_5$ + 3.8 LaB$_6$ + 8.9 N$_2$ ⟶ 16.8 BN + 2.8 LaH$_2$ + LaB$_6$H$_4$C$_6$N | -27.8863 |
| LaNH$_4$B$_6$C$_6$ | 1.2 H$_8$C$_5$ + 24.33 LaN + 5.867 B$_5$H$_7$ ⟶ 23.33 BN + 23.33 LaH$_2$ + LaB$_6$H$_4$C$_6$N | -27.4521 |
| LaNH$_4$B$_6$C$_6$ | 104.3 LaH$_3$ + 34 LaN + 3 B$_{13}$C$_2$ ⟶ 33 BN + 34.33 La$_4$H$_9$ + LaB$_6$H$_4$C$_6$N | -26.8815 |
| LaNH$_4$B$_6$C$_6$ | 20.62 LaC$_2$ + 20.62 H$_3$N + 0.4615 B$_{13}$C$_2$ ⟶ 7.231 H$_8$C$_5$ + 19.62 LaN + LaB$_6$H$_4$C$_6$N | -24.2035 |
| LaNH$_4$B$_6$C$_6$ | 1.2 H$_8$C$_5$ + 3.489 LaB$_6$ + 7.967 N$_2$ ⟶ 14.93 BN + 0.6222 La$_4$H$_9$ + LaB$_6$H$_4$C$_6$N | -24.0452 |
| LaNH$_4$B$_6$C$_6$ | 70 LaH$_3$ + 34 LaN + 3 B$_{13}$C$_2$ ⟶ 33 BN + 103 LaH$_2$ + LaB$_6$H$_4$C$_6$N | -22.9435 |
| LaNH$_4$B$_6$C$_6$ | 24.25 H$_3$N + 0.5 La$_2$C$_3$ + 2.25 B$_{13}$C$_2$ ⟶ 34.37 H$_2$ + 23.25 BN + LaB$_6$H$_4$C$_6$N | -21.3849 |
| LaNH$_4$B$_6$C$_6$ | 1.2 H$_8$C$_5$ + 3.24 LaB$_6$ + 7.22 N$_2$ ⟶ 13.44 BN + 1.12 La$_2$H$_5$ + LaB$_6$H$_4$C$_6$N | -20.9395 |
| LaNH$_4$B$_6$C$_6$ | 16.73 La + 16.73 H$_3$N + 3 B$_{13}$C$_2$ ⟶ 15.73 LaN + 6.6 B$_5$H$_7$ + LaB$_6$H$_4$C$_6$N | -19.9828 |
| LaNH$_4$B$_6$C$_6$ | 34 LaN + 3 B$_{13}$C$_2$ + 156.5 La$_2$H$_5$ ⟶ 33 BN + 86.5 La$_4$H$_9$ + LaB$_6$H$_4$C$_6$N | -19.5156 |
| LaNH$_4$B$_6$C$_6$ | 1.2 H$_8$C$_5$ + 17.47 LaN + 4.494 B$_5$H$_7$ ⟶ 16.47 BN + 4.118 La$_4$H$_9$ + LaB$_6$H$_4$C$_6$N | -18.6536 |
| LaNH$_4$B$_6$C$_6$ | 10.07 La + 0.3636 B$_9$H$_{11}$ + 0.5 N$_2$ + 3 B$_{13}$C$_2$ ⟶ 9.068 LaB$_4$ + LaB$_6$H$_4$C$_6$N | -18.5845 |
| LaNH$_4$B$_6$C$_6$ | 9.964 La + 0.5714 B$_5$H$_7$ + 0.5 N$_2$ + 3 B$_{13}$C$_2$ ⟶ 8.964 LaB$_4$ + LaB$_6$H$_4$C$_6$N | -18.3510 |
| LaNH$_4$B$_6$C$_6$ | LaC$_2$ + 21 H$_3$N + 2 B$_{13}$C$_2$ ⟶ 29.5 H$_2$ + 20 BN + LaB$_6$H$_4$C$_6$N | -17.5503 |
| LaNH$_4$B$_6$C$_6$ | 25.1 LaH$_2$ + 12.55 N$_2$ + 3 B$_{13}$C$_2$ ⟶ 24.1 LaN + 6.6 B$_5$H$_7$ + LaB$_6$H$_4$C$_6$N | -17.1521 |
| LaNH$_4$B$_6$C$_6$ | 14.78 La + 14.78 H$_3$N + 3 B$_{13}$C$_2$ ⟶ 13.78 LaN + 3.667 B$_9$H$_{11}$ + LaB$_6$H$_4$C$_6$N | -17.1275 |
| LaNH$_4$B$_6$C$_6$ | 1.2 H$_8$C$_5$ + 17.63 LaN + 2.514 B$_9$H$_{11}$ ⟶ 16.63 BN + 16.63 LaH$_2$ + LaB$_6$H$_4$C$_6$N | -17.0174 |
| LaNH$_4$B$_6$C$_6$ | 34 LaN + 3 B$_{13}$C$_2$ + 70 La$_2$H$_5$ ⟶ 33 BN + 173 LaH$_2$ + LaB$_6$H$_4$C$_6$N | -16.9725 |
| LaNH$_4$B$_6$C$_6$ | 2 H$_2$ + 9.25 La + 0.5 N$_2$ + 3 B$_{13}$C$_2$ ⟶ 8.25 LaB$_4$ + LaB$_6$H$_4$C$_6$N | -16.9151 |
| LaNH$_4$B$_6$C$_6$ | 10.32 La + BN + 0.3636 B$_9$H$_{11}$ + 3 B$_{13}$C$_2$ ⟶ 9.318 LaB$_4$ + LaB$_6$H$_4$C$_6$N | -16.7293 |
| LaNH$_4$B$_6$C$_6$ | 7.045 La + 0.3636 B$_9$H$_{11}$ + 0.5 N$_2$ + 3 B$_{13}$C$_2$ ⟶ 6.045 LaB$_6$ + LaB$_6$H$_4$C$_6$N | -16.5741 |
| LaNH$_4$B$_6$C$_6$ | 10.21 La + BN + 0.5714 B$_5$H$_7$ + 3 B$_{13}$C$_2$ ⟶ 9.214 LaB$_4$ + LaB$_6$H$_4$C$_6$N | -16.4958 |
| LaNH$_4$B$_6$C$_6$ | 6.976 La + 0.5714 B$_5$H$_7$ + 0.5 N$_2$ + 3 B$_{13}$C$_2$ ⟶ 5.976 LaB$_6$ + LaB$_6$H$_4$C$_6$N | -16.3637 |

| Compound | Chemical reaction | Reaction energy(eV) |
|---|---|---|
| $LaNH_4B_6C_6$ | $1.2\,H_8C_5 + 2.867\,LaB_6 + 6.1\,N_2 \longrightarrow 1.867\,LaH_3 + 11.2\,BN + LaB_6H_4C_6N$ | -16.1490 |
| $LaNH_4B_6C_6$ | $9.583\,La + 1.333\,H_3N + 3\,B_{13}C_2 \longrightarrow 0.3333\,LaN + 8.25\,LaB_4 + LaB_6H_4C_6N$ | -16.0354 |
| $LaNH_4B_6C_6$ | $9.455\,La + H_3N + 0.09091\,B_9H_{11} + 3\,B_{13}C_2 \longrightarrow 8.455\,LaB_4 + LaB_6H_4C_6N$ | -16.0084 |
| $LaNH_4B_6C_6$ | $9.429\,La + H_3N + 0.1429\,B_5H_7 + 3\,B_{13}C_2 \longrightarrow 8.429\,LaB_4 + LaB_6H_4C_6N$ | -15.9500 |
| $LaNH_4B_6C_6$ | $9.068\,La + LaN + 0.3636\,B_9H_{11} + 3\,B_{13}C_2 \longrightarrow 9.068\,LaB_4 + LaB_6H_4C_6N$ | -15.9271 |
| $LaNH_4B_6C_6$ | $9.167\,La + 1.333\,H_3N + 3\,B_{13}C_2 \longrightarrow 0.3333\,BN + 8.167\,LaB_4 + LaB_6H_4C_6N$ | -15.7681 |
| $LaNH_4B_6C_6$ | $8.964\,La + LaN + 0.5714\,B_5H_7 + 3\,B_{13}C_2 \longrightarrow 8.964\,LaB_4 + LaB_6H_4C_6N$ | -15.6937 |
| $LaNH_4B_6C_6$ | $0.5\,H_2 + 9.25\,La + H_3N + 3\,B_{13}C_2 \longrightarrow 8.25\,LaB_4 + LaB_6H_4C_6N$ | -15.5910 |
| $LaNH_4B_6C_6$ | $9.25\,La + 1.333\,H_3N + 3\,B_{13}C_2 \longrightarrow 0.1667\,N_2 + 8.25\,LaB_4 + LaB_6H_4C_6N$ | -15.1497 |
| $LaNH_4B_6C_6$ | $2\,H_2 + 6.5\,La + 0.5\,N_2 + 3\,B_{13}C_2 \longrightarrow 5.5\,LaB_6 + LaB_6H_4C_6N$ | -15.0861 |
| $LaNH_4B_6C_6$ | $2\,H_2 + 9.5\,La + BN + 3\,B_{13}C_2 \longrightarrow 8.5\,LaB_4 + LaB_6H_4C_6N$ | -15.0599 |
| $LaNH_4B_6C_6$ | $70\,La_4H_9 + 34\,LaN + 3\,B_{13}C_2 \longrightarrow 33\,BN + 313\,LaH_2 + LaB_6H_4C_6N$ | -14.9145 |
| $LaNH_4B_6C_6$ | $0.3333\,LaH_3 + 8.917\,La + H_3N + 3\,B_{13}C_2 \longrightarrow 8.25\,LaB_4 + LaB_6H_4C_6N$ | -14.8357 |
| $LaNH_4B_6C_6$ | $2\,La_2C_3 + 4.8\,B_5H_7 + 5.433\,N_2 \longrightarrow 3\,LaB_6 + 9.867\,H_3N + LaB_6H_4C_6N$ | -14.8262 |
| $LaNH_4B_6C_6$ | $8.85\,La + H_3N + 3\,B_{13}C_2 + 0.2\,La_2H_5 \longrightarrow 8.25\,LaB_4 + LaB_6H_4C_6N$ | -14.7445 |
| $LaNH_4B_6C_6$ | $8.806\,La + 0.1111\,La_4H_9 + H_3N + 3\,B_{13}C_2 \longrightarrow 8.25\,LaB_4 + LaB_6H_4C_6N$ | -14.6994 |
| $LaNH_4B_6C_6$ | $7.212\,La + BN + 0.3636\,B_9H_{11} + 3\,B_{13}C_2 \longrightarrow 6.212\,LaB_6 + LaB_6H_4C_6N$ | -14.6635 |
| $LaNH_4B_6C_6$ | $8.75\,La + H_3N + 0.5\,LaH_2 + 3\,B_{13}C_2 \longrightarrow 8.25\,LaB_4 + LaB_6H_4C_6N$ | -14.6504 |
| $LaNH_4B_6C_6$ | $22.17\,LaH_2 + 11.08\,N_2 + 3\,B_{13}C_2 \longrightarrow 21.17\,LaN + 3.667\,B_9H_{11} + LaB_6H_4C_6N$ | -14.6277 |
| $LaNH_4B_6C_6$ | $6\,BN + 14.67\,H_3N + 10.33\,La_2C_3 \longrightarrow 5\,H_8C_5 + 19.67\,LaN + LaB_6H_4C_6N$ | -14.5320 |
| $LaNH_4B_6C_6$ | $7.143\,La + BN + 0.5714\,B_5H_7 + 3\,B_{13}C_2 \longrightarrow 6.143\,LaB_6 + LaB_6H_4C_6N$ | -14.4531 |
| $LaNH_4B_6C_6$ | $2\,H_2 + 8.25\,La + LaN + 3\,B_{13}C_2 \longrightarrow 8.25\,LaB_4 + LaB_6H_4C_6N$ | -14.2578 |
| $LaNH_4B_6C_6$ | $6.833\,La + 1.333\,H_3N + 3\,B_{13}C_2 \longrightarrow 5.5\,LaB_6 + 0.3333\,LaN + LaB_6H_4C_6N$ | -14.2064 |
| $LaNH_4B_6C_6$ | $6.636\,La + H_3N + 0.09091\,B_9H_{11} + 3\,B_{13}C_2 \longrightarrow 5.636\,LaB_6 + LaB_6H_4C_6N$ | -14.1340 |
| $LaNH_4B_6C_6$ | $6.619\,La + H_3N + 0.1429\,B_5H_7 + 3\,B_{13}C_2 \longrightarrow 5.619\,LaB_6 + LaB_6H_4C_6N$ | -14.0814 |
| $LaNH_4B_6C_6$ | $24.47\,LaN + 6.765\,B_9H_{11} + 3\,B_{13}C_2 \longrightarrow 23.47\,H_3N + 23.47\,LaB_4 + LaB_6H_4C_6N$ | -14.0206 |
| $LaNH_4B_6C_6$ | $6.444\,La + 1.333\,H_3N + 3\,B_{13}C_2 \longrightarrow 5.444\,LaB_6 + 0.3333\,BN + LaB_6H_4C_6N$ | -13.9575 |
| $LaNH_4B_6C_6$ | $6.045\,La + LaN + 0.3636\,B_9H_{11} + 3\,B_{13}C_2 \longrightarrow 6.045\,LaB_6 + LaB_6H_4C_6N$ | -13.9168 |
| $LaNH_4B_6C_6$ | $1.333\,LaH_3 + 7.917\,La + 0.5\,N_2 + 3\,B_{13}C_2 \longrightarrow 8.25\,LaB_4 + LaB_6H_4C_6N$ | -13.8939 |
| $LaNH_4B_6C_6$ | $0.5\,H_2 + 6.5\,La + H_3N + 3\,B_{13}C_2 \longrightarrow 5.5\,LaB_6 + LaB_6H_4C_6N$ | -13.7620 |
| $LaNH_4B_6C_6$ | $5.976\,La + LaN + 0.5714\,B_5H_7 + 3\,B_{13}C_2 \longrightarrow 5.976\,LaB_6 + LaB_6H_4C_6N$ | -13.7063 |
| $LaNH_4B_6C_6$ | $1.2\,H_8C_5 + 13.73\,LaN + 3.745\,B_5H_7 \longrightarrow 12.73\,BN + 6.364\,La_2H_5 + LaB_6H_4C_6N$ | -13.6672 |
| $LaNH_4B_6C_6$ | $7.65\,La + 0.5\,N_2 + 3\,B_{13}C_2 + 0.8\,La_2H_5 \longrightarrow 8.25\,LaB_4 + LaB_6H_4C_6N$ | -13.5291 |
| $LaNH_4B_6C_6$ | $7.472\,La + 0.4444\,La_4H_9 + 0.5\,N_2 + 3\,B_{13}C_2 \longrightarrow 8.25\,LaB_4 + LaB_6H_4C_6N$ | -13.3487 |
| $LaNH_4B_6C_6$ | $6.5\,La + 1.333\,H_3N + 3\,B_{13}C_2 \longrightarrow 5.5\,LaB_6 + 0.1667\,N_2 + LaB_6H_4C_6N$ | -13.3207 |
| $LaNH_4B_6C_6$ | $2\,H_2 + 6.667\,La + BN + 3\,B_{13}C_2 \longrightarrow 5.667\,LaB_6 + LaB_6H_4C_6N$ | -13.1755 |
| $LaNH_4B_6C_6$ | $7.25\,La + 2\,LaH_2 + 0.5\,N_2 + 3\,B_{13}C_2 \longrightarrow 8.25\,LaB_4 + LaB_6H_4C_6N$ | -13.1526 |
| $LaNH_4B_6C_6$ | $2.667\,B_9H_{11} + 2\,La_2C_3 + 4.722\,N_2 \longrightarrow 3\,LaB_6 + 8.444\,H_3N + LaB_6H_4C_6N$ | -13.1234 |
| $LaNH_4B_6C_6$ | $0.3333\,LaH_3 + 6.167\,La + H_3N + 3\,B_{13}C_2 \longrightarrow 5.5\,LaB_6 + LaB_6H_4C_6N$ | -13.0067 |
| $LaNH_4B_6C_6$ | $6.1\,La + H_3N + 3\,B_{13}C_2 + 0.2\,La_2H_5 \longrightarrow 5.5\,LaB_6 + LaB_6H_4C_6N$ | -12.9155 |
| $LaNH_4B_6C_6$ | $6.056\,La + 0.1111\,La_4H_9 + H_3N + 3\,B_{13}C_2 \longrightarrow 5.5\,LaB_6 + LaB_6H_4C_6N$ | -12.8704 |
| $LaNH_4B_6C_6$ | $6\,La + H_3N + 0.5\,LaH_2 + 3\,B_{13}C_2 \longrightarrow 5.5\,LaB_6 + LaB_6H_4C_6N$ | -12.8214 |
| $LaNH_4B_6C_6$ | $0.125\,H_8C_5 + 8.234\,La + H_3N + 2.687\,B_{13}C_2 \longrightarrow 7.234\,LaB_4 + LaB_6H_4C_6N$ | -12.8070 |
| $LaNH_4B_6C_6$ | $1.2\,H_8C_5 + 5.1\,N_2 + 3.8\,LaB_4 \longrightarrow 9.2\,BN + 2.8\,LaH_2 + LaB_6H_4C_6N$ | -12.5232 |
| $LaNH_4B_6C_6$ | $2\,H_2 + 5.5\,La + LaN + 3\,B_{13}C_2 \longrightarrow 5.5\,LaB_6 + LaB_6H_4C_6N$ | -12.4288 |
| $LaNH_4B_6C_6$ | $0.5\,H_8C_5 + 0.5\,La_2C_3 + 4\,N_2 + B_{13}C_2 \longrightarrow 7\,BN + LaB_6H_4C_6N$ | -12.2107 |
| $LaNH_4B_6C_6$ | $5.578\,La_4H_9 + 11.16\,N_2 + 3\,B_{13}C_2 \longrightarrow 21.31\,LaN + 6.6\,B_5H_7 + LaB_6H_4C_6N$ | -12.2025 |
| $LaNH_4B_6C_6$ | $42\,B + LaN + 2\,La_2C_3 + 2\,LaH_2 \longrightarrow 6\,LaB_6 + LaB_6H_4C_6N$ | -12.1691 |
| $LaNH_4B_6C_6$ | $1.2\,H_8C_5 + 13.58\,LaN + 2.065\,B_9H_{11} \longrightarrow 12.58\,BN + 3.146\,La_4H_9 + LaB_6H_4C_6N$ | -12.1603 |
| $LaNH_4B_6C_6$ | $1.333\,LaH_3 + 5.167\,La + 0.5\,N_2 + 3\,B_{13}C_2 \longrightarrow 5.5\,LaB_6 + LaB_6H_4C_6N$ | -12.0649 |
| $LaNH_4B_6C_6$ | $1.333\,LaH_3 + 8.167\,La + BN + 3\,B_{13}C_2 \longrightarrow 8.5\,LaB_4 + LaB_6H_4C_6N$ | -12.0387 |
| $LaNH_4B_6C_6$ | $0.3636\,B_9H_{11} + 0.5\,N_2 + 3\,B_{13}C_2 + 21.14\,LaB_4 \longrightarrow 20.14\,LaB_6 + LaB_6H_4C_6N$ | -11.8882 |
| $LaNH_4B_6C_6$ | $0.5714\,B_5H_7 + 0.5\,N_2 + 3\,B_{13}C_2 + 20.93\,LaB_4 \longrightarrow 19.93\,LaB_6 + LaB_6H_4C_6N$ | -11.7239 |
| $LaNH_4B_6C_6$ | $4.9\,La + 0.5\,N_2 + 3\,B_{13}C_2 + 0.8\,La_2H_5 \longrightarrow 5.5\,LaB_6 + LaB_6H_4C_6N$ | -11.7001 |
| $LaNH_4B_6C_6$ | $11.23\,LaN + 3.154\,B_9H_{11} + 3\,B_{13}C_2 \longrightarrow 10.23\,LaB_6 + 10.23\,H_3N + LaB_6H_4C_6N$ | -11.6939 |
| $LaNH_4B_6C_6$ | $7.9\,La + BN + 3\,B_{13}C_2 + 0.8\,La_2H_5 \longrightarrow 8.5\,LaB_4 + LaB_6H_4C_6N$ | -11.6739 |
| $LaNH_4B_6C_6$ | $1.182\,B_9H_{11} + 2\,La_2C_3 + 2.818\,N_2 \longrightarrow 3\,LaH_3 + 4.636\,BN + LaB_6H_4C_6N$ | -11.5223 |
| $LaNH_4B_6C_6$ | $4.722\,La + 0.4444\,La_4H_9 + 0.5\,N_2 + 3\,B_{13}C_2 \longrightarrow 5.5\,LaB_6 + LaB_6H_4C_6N$ | -11.5197 |
| $LaNH_4B_6C_6$ | $7.722\,La + BN + 0.4444\,La_4H_9 + 3\,B_{13}C_2 \longrightarrow 8.5\,LaB_4 + LaB_6H_4C_6N$ | -11.4935 |
| $LaNH_4B_6C_6$ | $40.67\,B + 0.4444\,La_4H_9 + LaN + 2\,La_2C_3 \longrightarrow 5.778\,LaB_6 + LaB_6H_4C_6N$ | -11.4904 |

| Compound | Chemical reaction | Reaction energy(eV) |
| --- | --- | --- |
| $LaNH_4B_6C_6$ | $4.5\,La + 2\,LaH_2 + 0.5\,N_2 + 3\,B_{13}C_2 \longrightarrow 5.5\,LaB_6 + LaB_6H_4C_6N$ | -11.3236 |
| $LaNH_4B_6C_6$ | $7.5\,La + BN + 2\,LaH_2 + 3\,B_{13}C_2 \longrightarrow 8.5\,LaB_4 + LaB_6H_4C_6N$ | -11.2974 |
| $LaNH_4B_6C_6$ | $1.333\,LaH_3 + 6.917\,La + LaN + 3\,B_{13}C_2 \longrightarrow 8.25\,LaB_4 + LaB_6H_4C_6N$ | -11.2365 |
| $LaNH_4B_6C_6$ | $0.125\,H_8C_5 + 5.823\,La + H_3N + 2.688\,B_{13}C_2 \longrightarrow 4.823\,LaB_6 + LaB_6H_4C_6N$ | -11.2031 |
| $LaNH_4B_6C_6$ | $39.6\,B + LaN + 2\,La_2C_3 + 0.8\,La_2H_5 \longrightarrow 5.6\,LaB_6 + LaB_6H_4C_6N$ | -10.9709 |
| $LaNH_4B_6C_6$ | $36\,B + 2\,La_2C_3 + 2\,LaH_2 + 0.5\,N_2 \longrightarrow 5\,LaB_6 + LaB_6H_4C_6N$ | -10.8895 |
| $LaNH_4B_6C_6$ | $6.65\,La + LaN + 3\,B_{13}C_2 + 0.8\,La_2H_5 \longrightarrow 8.25\,LaB_4 + LaB_6H_4C_6N$ | -10.8718 |
| $LaNH_4B_6C_6$ | $2\,H_2 + 0.5\,N_2 + 3\,B_{13}C_2 + 19.5\,LaB_4 \longrightarrow 18.5\,LaB_6 + LaB_6H_4C_6N$ | -10.7630 |
| $LaNH_4B_6C_6$ | $20.31\,LaN + 8.846\,B_5H_7 + 3\,B_{13}C_2 \longrightarrow 19.31\,H_3N + 19.31\,LaB_4 + LaB_6H_4C_6N$ | -10.7447 |
| $LaNH_4B_6C_6$ | $6.472\,La + 0.4444\,La_4H_9 + LaN + 3\,B_{13}C_2 \longrightarrow 8.25\,LaB_4 + LaB_6H_4C_6N$ | -10.6913 |
| $LaNH_4B_6C_6$ | $6.25\,La + LaN + 2\,LaH_2 + 3\,B_{13}C_2 \longrightarrow 8.25\,LaB_4 + LaB_6H_4C_6N$ | -10.4952 |
| $LaNH_4B_6C_6$ | $1.333\,LaH_3 + 38\,B + LaN + 2\,La_2C_3 \longrightarrow 5.333\,LaB_6 + LaB_6H_4C_6N$ | -10.2858 |
| $LaNH_4B_6C_6$ | $4.926\,La_4H_9 + 9.852\,N_2 + 3\,B_{13}C_2 \longrightarrow 18.7\,LaN + 3.667\,B_9H_{11} + LaB_6H_4C_6N$ | -10.2565 |
| $LaNH_4B_6C_6$ | $10.3\,LaN + 4.556\,B_5H_7 + 3\,B_{13}C_2 \longrightarrow 9.296\,LaB_6 + 9.296\,H_3N + LaB_6H_4C_6N$ | -10.2194 |
| $LaNH_4B_6C_6$ | $34.67\,B + 0.4444\,La_4H_9 + 2\,La_2C_3 + 0.5\,N_2 \longrightarrow 4.778\,LaB_6 + LaB_6H_4C_6N$ | -10.2108 |
| $LaNH_4B_6C_6$ | $1.333\,LaH_3 + 5.333\,La + BN + 3\,B_{13}C_2 \longrightarrow 5.667\,LaB_6 + LaB_6H_4C_6N$ | -10.1543 |
| $LaNH_4B_6C_6$ | $8.255\,LaN + 0.3636\,B_9H_{11} + 3\,B_{13}C_2 \longrightarrow 7.255\,BN + 7.255\,LaB_4 + LaB_6H_4C_6N$ | -10.1078 |
| $LaNH_4B_6C_6$ | $La + 11.73\,H_3N + 3\,B_{13}C_2 \longrightarrow 10.73\,BN + 4.455\,B_5H_7 + LaB_6H_4C_6N$ | -10.0946 |
| $LaNH_4B_6C_6$ | $6.182\,LaN + 0.3636\,B_9H_{11} + 3\,B_{13}C_2 \longrightarrow 5.182\,LaB_6 + 5.182\,BN + LaB_6H_4C_6N$ | -10.0473 |
| $LaNH_4B_6C_6$ | $8.171\,LaN + 0.5714\,B_5H_7 + 3\,B_{13}C_2 \longrightarrow 7.171\,BN + 7.171\,LaB_4 + LaB_6H_4C_6N$ | -9.9410 |
| $LaNH_4B_6C_6$ | $1.2\,H_8C_5 + 4.478\,N_2 + 3.489\,LaB_4 \longrightarrow 7.956\,BN + 0.6222\,La_4H_9 + LaB_6H_4C_6N$ | -9.9400 |
| $LaNH_4B_6C_6$ | $LaN + 0.3636\,B_9H_{11} + 3\,B_{13}C_2 + 18.14\,LaB_4 \longrightarrow 18.14\,LaB_6 + LaB_6H_4C_6N$ | -9.8960 |
| $LaNH_4B_6C_6$ | $1.333\,H_3N + 7.333\,LaN + 3\,B_{13}C_2 \longrightarrow 7.667\,BN + 6.333\,LaB_4 + LaB_6H_4C_6N$ | -9.8855 |
| $LaNH_4B_6C_6$ | $6.122\,LaN + 0.5714\,B_5H_7 + 3\,B_{13}C_2 \longrightarrow 5.122\,LaB_6 + 5.122\,BN + LaB_6H_4C_6N$ | -9.8812 |
| $LaNH_4B_6C_6$ | $BN + 0.3636\,B_9H_{11} + 3\,B_{13}C_2 + 21.64\,LaB_4 \longrightarrow 20.64\,LaB_6 + LaB_6H_4C_6N$ | -9.8668 |
| $LaNH_4B_6C_6$ | $1.333\,H_3N + 5.524\,LaN + 3\,B_{13}C_2 \longrightarrow 4.524\,LaB_6 + 5.857\,BN + LaB_6H_4C_6N$ | -9.8327 |
| $LaNH_4B_6C_6$ | $5.067\,La + BN + 3\,B_{13}C_2 + 0.8\,La_2H_5 \longrightarrow 5.667\,LaB_6 + LaB_6H_4C_6N$ | -9.7895 |
| $LaNH_4B_6C_6$ | $LaN + 0.5714\,B_5H_7 + 3\,B_{13}C_2 + 17.93\,LaB_4 \longrightarrow 17.93\,LaB_6 + LaB_6H_4C_6N$ | -9.7316 |
| $LaNH_4B_6C_6$ | $H_3N + 0.09091\,B_9H_{11} + 3\,B_{13}C_2 + 19.91\,LaB_4 \longrightarrow 18.91\,LaB_6 + LaB_6H_4C_6N$ | -9.7202 |
| $LaNH_4B_6C_6$ | $BN + 0.5714\,B_5H_7 + 3\,B_{13}C_2 + 21.43\,LaB_4 \longrightarrow 20.43\,LaB_6 + LaB_6H_4C_6N$ | -9.7024 |
| $LaNH_4B_6C_6$ | $33.6\,B + 2\,La_2C_3 + 0.5\,N_2 + 0.8\,La_2H_5 \longrightarrow 4.6\,LaB_6 + LaB_6H_4C_6N$ | -9.6913 |
| $LaNH_4B_6C_6$ | $H_3N + 0.1429\,B_5H_7 + 3\,B_{13}C_2 + 19.86\,LaB_4 \longrightarrow 18.86\,LaB_6 + LaB_6H_4C_6N$ | -9.6792 |
| $LaNH_4B_6C_6$ | $1.333\,H_3N + 3\,B_{13}C_2 + 19.33\,LaB_4 \longrightarrow 18.33\,LaB_6 + 0.3333\,BN + LaB_6H_4C_6N$ | -9.6714 |
| $LaNH_4B_6C_6$ | $1.333\,H_3N + 3\,B_{13}C_2 + 20.5\,LaB_4 \longrightarrow 19.17\,LaB_6 + 0.3333\,LaN + LaB_6H_4C_6N$ | -9.6617 |
| $LaNH_4B_6C_6$ | $4.889\,La + BN + 0.4444\,La_4H_9 + 3\,B_{13}C_2 \longrightarrow 5.667\,LaB_6 + LaB_6H_4C_6N$ | -9.6091 |
| $LaNH_4B_6C_6$ | $11.41\,H_3N + LaN + 3\,B_{13}C_2 \longrightarrow 11.41\,BN + 4.318\,B_5H_7 + LaB_6H_4C_6N$ | -9.4661 |
| $LaNH_4B_6C_6$ | $0.5\,H_2 + H_3N + 3\,B_{13}C_2 + 19.5\,LaB_4 \longrightarrow 18.5\,LaB_6 + LaB_6H_4C_6N$ | -9.4389 |
| $LaNH_4B_6C_6$ | $4.667\,La + BN + 2\,LaH_2 + 3\,B_{13}C_2 \longrightarrow 5.667\,LaB_6 + LaB_6H_4C_6N$ | -9.4130 |
| $LaNH_4B_6C_6$ | $1.333\,LaH_3 + 4.167\,La + LaN + 3\,B_{13}C_2 \longrightarrow 5.5\,LaB_6 + LaB_6H_4C_6N$ | -9.4075 |
| $LaNH_4B_6C_6$ | $13\,H_3N + 3\,B_{13}C_2 + LaB_4 \longrightarrow 12\,BN + 5\,B_5H_7 + LaB_6H_4C_6N$ | -9.3999 |
| $LaNH_4B_6C_6$ | $LaB_6 + 13.64\,H_3N + 3\,B_{13}C_2 \longrightarrow 12.64\,BN + 5.273\,B_5H_7 + LaB_6H_4C_6N$ | -9.3851 |
| $LaNH_4B_6C_6$ | $2\,La_2C_3 + 3.6\,B_5H_7 + 4.033\,N_2 \longrightarrow 7.067\,H_3N + 3\,LaB_4 + LaB_6H_4C_6N$ | -9.1009 |
| $LaNH_4B_6C_6$ | $1.2\,H_8C_5 + 11.12\,LaN + 1.791\,B_9H_{11} \longrightarrow 10.12\,BN + 5.061\,La_2H_5 + LaB_6H_4C_6N$ | -9.0550 |
| $LaNH_4B_6C_6$ | $3.9\,La + LaN + 3\,B_{13}C_2 + 0.8\,La_2H_5 \longrightarrow 5.5\,LaB_6 + LaB_6H_4C_6N$ | -9.0428 |
| $LaNH_4B_6C_6$ | $1.333\,LaH_3 + 32\,B + 2\,La_2C_3 + 0.5\,N_2 \longrightarrow 4.333\,LaB_6 + LaB_6H_4C_6N$ | -9.0062 |
| $LaNH_4B_6C_6$ | $1.333\,H_3N + 3\,B_{13}C_2 + 19.5\,LaB_4 \longrightarrow 18.5\,LaB_6 + 0.1667\,N_2 + LaB_6H_4C_6N$ | -8.9976 |
| $LaNH_4B_6C_6$ | $2\,H_2 + 7.6\,LaN + 3\,B_{13}C_2 \longrightarrow 6.6\,BN + 6.6\,LaB_4 + LaB_6H_4C_6N$ | -8.9635 |
| $LaNH_4B_6C_6$ | $La + 10.79\,H_3N + 3\,B_{13}C_2 \longrightarrow 9.789\,BN + 2.579\,B_9H_{11} + LaB_6H_4C_6N$ | -8.9508 |
| $LaNH_4B_6C_6$ | $2\,H_2 + 5.714\,LaN + 3\,B_{13}C_2 \longrightarrow 4.714\,LaB_6 + 4.714\,BN + LaB_6H_4C_6N$ | -8.9084 |
| $LaNH_4B_6C_6$ | $0.3333\,LaH_3 + H_3N + 3\,B_{13}C_2 + 18.5\,LaB_4 \longrightarrow 17.83\,LaB_6 + LaB_6H_4C_6N$ | -8.9053 |
| $LaNH_4B_6C_6$ | $3.722\,La + 0.4444\,La_4H_9 + LaN + 3\,B_{13}C_2 \longrightarrow 5.5\,LaB_6 + LaB_6H_4C_6N$ | -8.8623 |
| $LaNH_4B_6C_6$ | $H_3N + 3\,B_{13}C_2 + 0.2\,La_2H_5 + 18.3\,LaB_4 \longrightarrow 17.7\,LaB_6 + LaB_6H_4C_6N$ | -8.8585 |
| $LaNH_4B_6C_6$ | $0.1111\,La_4H_9 + H_3N + 3\,B_{13}C_2 + 18.17\,LaB_4 \longrightarrow 17.61\,LaB_6 + LaB_6H_4C_6N$ | -8.8429 |
| $LaNH_4B_6C_6$ | $H_3N + 0.5\,LaH_2 + 3\,B_{13}C_2 + 18\,LaB_4 \longrightarrow 17.5\,LaB_6 + LaB_6H_4C_6N$ | -8.8309 |
| $LaNH_4B_6C_6$ | $2\,H_2 + LaN + 3\,B_{13}C_2 + 16.5\,LaB_4 \longrightarrow 16.5\,LaB_6 + LaB_6H_4C_6N$ | -8.7708 |
| $LaNH_4B_6C_6$ | $2\,H_2 + BN + 3\,B_{13}C_2 + 20\,LaB_4 \longrightarrow 19\,LaB_6 + LaB_6H_4C_6N$ | -8.7416 |
| $LaNH_4B_6C_6$ | $3.5\,La + LaN + 2\,LaH_2 + 3\,B_{13}C_2 \longrightarrow 5.5\,LaB_6 + LaB_6H_4C_6N$ | -8.6662 |
| $LaNH_4B_6C_6$ | $3\,LaC_2 + 3.6\,B_5H_7 + 4.033\,N_2 \longrightarrow 2\,LaB_6 + 7.067\,H_3N + LaB_6H_4C_6N$ | -8.6646 |
| $LaNH_4B_6C_6$ | $1.333\,LaH_3 + 0.5\,N_2 + 3\,B_{13}C_2 + 15.5\,LaB_4 \longrightarrow 15.83\,LaB_6 + LaB_6H_4C_6N$ | -8.6286 |
| $LaNH_4B_6C_6$ | $10.04\,N_2 + 3\,B_{13}C_2 + 10.04\,La_2H_5 \longrightarrow 19.08\,LaN + 6.6\,B_5H_7 + LaB_6H_4C_6N$ | -8.5380 |

| Compound | Chemical reaction | Reaction energy(eV) |
|---|---|---|
| LaNH$_4$B$_6$C$_6$ | 3 LaC$_2$ + 36 B + LaN + 2 LaH$_2$ ⟶ 5 LaB$_6$ + LaB$_6$H$_4$C$_6$N | -8.4610 |
| LaNH$_4$B$_6$C$_6$ | 0.5 N$_2$ + 3 B$_{13}$C$_2$ + 0.8 La$_2$H$_5$ + 14.7 LaB$_4$ ⟶ 15.3 LaB$_6$ + LaB$_6$H$_4$C$_6$N | -8.4412 |
| LaNH$_4$B$_6$C$_6$ | 0.4444 La$_4$H$_9$ + 0.5 N$_2$ + 3 B$_{13}$C$_2$ + 14.17 LaB$_4$ ⟶ 14.94 LaB$_6$ + LaB$_6$H$_4$C$_6$N | -8.3790 |
| LaNH$_4$B$_6$C$_6$ | 10.5 H$_3$N + LaN + 3 B$_{13}$C$_2$ ⟶ 10.5 BN + 2.5 B$_9$H$_{11}$ + LaB$_6$H$_4$C$_6$N | -8.3573 |
| LaNH$_4$B$_6$C$_6$ | 2 LaH$_2$ + 0.5 N$_2$ + 3 B$_{13}$C$_2$ + 13.5 LaB$_4$ ⟶ 14.5 LaB$_6$ + LaB$_6$H$_4$C$_6$N | -8.3307 |
| LaNH$_4$B$_6$C$_6$ | 35 B + BN + 2 La$_2$C$_3$ + 2 LaH$_2$ ⟶ 5 LaB$_6$ + LaB$_6$H$_4$C$_6$N | -8.3228 |
| LaNH$_4$B$_6$C$_6$ | 2 La$_2$C$_3$ + 1.857 B$_5$H$_7$ + 2.143 N$_2$ ⟶ 3 LaH$_3$ + 3.286 BN + LaB$_6$H$_4$C$_6$N | -8.2579 |
| LaNH$_4$B$_6$C$_6$ | 1.045 B$_9$H$_{11}$ + 2 La$_2$C$_3$ + 2.205 N$_2$ ⟶ 3.409 BN + 1.5 La$_2$H$_5$ + LaB$_6$H$_4$C$_6$N | -8.1705 |
| LaNH$_4$B$_6$C$_6$ | 11.95 H$_3$N + 3 B$_{13}$C$_2$ + LaB$_4$ ⟶ 10.95 BN + 2.895 B$_9$H$_{11}$ + LaB$_6$H$_4$C$_6$N | -8.1161 |
| LaNH$_4$B$_6$C$_6$ | 2 H$_2$ + 30 B + LaN + 2 La$_2$C$_3$ ⟶ 4 LaB$_6$ + LaB$_6$H$_4$C$_6$N | -8.0578 |
| LaNH$_4$B$_6$C$_6$ | LaB$_6$ + 12.53 H$_3$N + 3 B$_{13}$C$_2$ ⟶ 11.53 BN + 3.053 B$_9$H$_{11}$ + LaB$_6$H$_4$C$_6$N | -8.0312 |
| LaNH$_4$B$_6$C$_6$ | 1.2 H$_8$C$_5$ + 9.5 La + H$_3$N + 1.2 B$_5$H$_7$ ⟶ 8.5 LaH$_2$ + LaB$_6$H$_4$C$_6$N | -7.8702 |
| LaNH$_4$B$_6$C$_6$ | 1.2 H$_8$C$_5$ + 3.98 N$_2$ + 3.24 LaB$_4$ ⟶ 6.96 BN + 1.12 La$_2$H$_5$ + LaB$_6$H$_4$C$_6$N | -7.8405 |
| LaNH$_4$B$_6$C$_6$ | 2 B$_9$H$_{11}$ + 2 La$_2$C$_3$ + 3.5 N$_2$ ⟶ 6 H$_3$N + 3 LaB$_4$ + LaB$_6$H$_4$C$_6$N | -7.8238 |
| LaNH$_4$B$_6$C$_6$ | 3 LaC$_2$ + 34.67 B + 0.4444 La$_4$H$_9$ + LaN ⟶ 4.778 LaB$_6$ + LaB$_6$H$_4$C$_6$N | -7.7822 |
| LaNH$_4$B$_6$C$_6$ | 1.2 H$_8$C$_5$ + 9.75 LaN + 2.95 B$_5$H$_7$ ⟶ 8.75 LaH$_3$ + 8.75 BN + LaB$_6$H$_4$C$_6$N | -7.7515 |
| LaNH$_4$B$_6$C$_6$ | 11.27 H$_3$N + LaH$_2$ + 3 B$_{13}$C$_2$ ⟶ 10.27 BN + 4.545 B$_5$H$_7$ + LaB$_6$H$_4$C$_6$N | -7.7435 |
| LaNH$_4$B$_6$C$_6$ | 33.67 B + BN + 0.4444 La$_4$H$_9$ + 2 La$_2$C$_3$ ⟶ 4.778 LaB$_6$ + LaB$_6$H$_4$C$_6$N | -7.6440 |
| LaNH$_4$B$_6$C$_6$ | 0.25 La$_4$H$_9$ + 11.22 H$_3$N + 3 B$_{13}$C$_2$ ⟶ 10.22 BN + 4.557 B$_5$H$_7$ + LaB$_6$H$_4$C$_6$N | -7.5599 |
| LaNH$_4$B$_6$C$_6$ | H$_3$N + 54.5 LaH$_2$ + 3 B$_{13}$C$_2$ ⟶ 5.5 LaB$_6$ + 12 La$_4$H$_9$ + LaB$_6$H$_4$C$_6$N | -7.5259 |
| LaNH$_4$B$_6$C$_6$ | 27.14 B + LaN + 2 La$_2$C$_3$ + 0.5714 B$_5$H$_7$ ⟶ 4 LaB$_6$ + LaB$_6$H$_4$C$_6$N | -7.4606 |
| LaNH$_4$B$_6$C$_6$ | 26.73 B + LaN + 0.3636 B$_9$H$_{11}$ + 2 La$_2$C$_3$ ⟶ 4 LaB$_6$ + LaB$_6$H$_4$C$_6$N | -7.3983 |
| LaNH$_4$B$_6$C$_6$ | 11.16 H$_3$N + 3 B$_{13}$C$_2$ + 0.5 La$_2$H$_5$ ⟶ 10.16 BN + 4.568 B$_5$H$_7$ + LaB$_6$H$_4$C$_6$N | -7.3910 |
| LaNH$_4$B$_6$C$_6$ | 3 LaC$_2$ + 2 B$_9$H$_{11}$ + 3.5 N$_2$ ⟶ 2 LaB$_6$ + 6 H$_3$N + LaB$_6$H$_4$C$_6$N | -7.3875 |
| LaNH$_4$B$_6$C$_6$ | 42.5 LaH$_2$ + 0.5 N$_2$ + 3 B$_{13}$C$_2$ ⟶ 5.5 LaB$_6$ + 9 La$_4$H$_9$ + LaB$_6$H$_4$C$_6$N | -7.3519 |
| LaNH$_4$B$_6$C$_6$ | 0.125 H$_8$C$_5$ + H$_3$N + 2.687 B$_{13}$C$_2$ + 17.47 LaB$_4$ ⟶ 16.47 LaB$_6$ + LaB$_6$H$_4$C$_6$N | -7.3304 |
| LaNH$_4$B$_6$C$_6$ | 3 LaC$_2$ + 33.6 B + LaN + 0.8 La$_2$H$_5$ ⟶ 4.6 LaB$_6$ + LaB$_6$H$_4$C$_6$N | -7.2627 |
| LaNH$_4$B$_6$C$_6$ | 6.5 LaH$_3$ + 3.083 N$_2$ + 3 B$_{13}$C$_2$ ⟶ 5.5 LaB$_6$ + 5.167 H$_3$N + LaB$_6$H$_4$C$_6$N | -7.1987 |
| LaNH$_4$B$_6$C$_6$ | 3 LaC$_2$ + 30 B + 2 LaH$_2$ + 0.5 N$_2$ ⟶ 4 LaB$_6$ + LaB$_6$H$_4$C$_6$N | -7.1814 |
| LaNH$_4$B$_6$C$_6$ | H$_3$N + 30.5 LaH$_2$ + 3 B$_{13}$C$_2$ ⟶ 5.5 LaB$_6$ + 12 La$_2$H$_5$ + LaB$_6$H$_4$C$_6$N | -7.1731 |
| LaNH$_4$B$_6$C$_6$ | 32.6 B + BN + 2 La$_2$C$_3$ + 0.8 La$_2$H$_5$ ⟶ 4.6 LaB$_6$ + LaB$_6$H$_4$C$_6$N | -7.1245 |
| LaNH$_4$B$_6$C$_6$ | LaH$_3$ + 11.05 H$_3$N + 3 B$_{13}$C$_2$ ⟶ 10.05 BN + 4.591 B$_5$H$_7$ + LaB$_6$H$_4$C$_6$N | -7.1239 |
| LaNH$_4$B$_6$C$_6$ | 24.5 LaH$_2$ + 0.5 N$_2$ + 3 B$_{13}$C$_2$ ⟶ 5.5 LaB$_6$ + 9 La$_2$H$_5$ + LaB$_6$H$_4$C$_6$N | -7.0873 |
| LaNH$_4$B$_6$C$_6$ | 1.2 H$_8$C$_5$ + 8.556 La + H$_3$N + 1.2 B$_5$H$_7$ ⟶ 1.889 La$_4$H$_9$ + LaB$_6$H$_4$C$_6$N | -7.0367 |
| LaNH$_4$B$_6$C$_6$ | 8.867 N$_2$ + 3 B$_{13}$C$_2$ + 8.867 La$_2$H$_5$ ⟶ 16.73 LaN + 3.667 B$_9$H$_{11}$ + LaB$_6$H$_4$C$_6$N | -7.0203 |
| LaNH$_4$B$_6$C$_6$ | H$_3$N + 79.25 LaH$_2$ + 3 B$_{13}$C$_2$ ⟶ 17.5 La$_4$H$_9$ + 8.25 LaB$_4$ + LaB$_6$H$_4$C$_6$N | -6.9277 |
| LaNH$_4$B$_6$C$_6$ | 1.2 H$_8$C$_5$ + 8.967 La + H$_3$N + 0.6667 B$_9$H$_{11}$ ⟶ 7.967 LaH$_2$ + LaB$_6$H$_4$C$_6$N | -6.9120 |
| LaNH$_4$B$_6$C$_6$ | 6.5 LaH$_2$ + 2 N$_2$ + 3 B$_{13}$C$_2$ ⟶ 5.5 LaB$_6$ + 3 H$_3$N + LaB$_6$H$_4$C$_6$N | -6.8300 |
| LaNH$_4$B$_6$C$_6$ | 1.333 LaH$_3$ + 6.533 LaN + 3 B$_{13}$C$_2$ ⟶ 5.533 BN + 6.867 LaB$_4$ + LaB$_6$H$_4$C$_6$N | -6.7979 |
| LaNH$_4$B$_6$C$_6$ | 2.917 La$_2$C$_3$ + 1.2 B$_5$H$_7$ + 2.917 N$_2$ ⟶ 0.55 H$_8$C$_5$ + 4.833 LaN + LaB$_6$H$_4$C$_6$N | -6.7907 |
| LaNH$_4$B$_6$C$_6$ | 2 H$_2$ + 24 B + 2 La$_2$C$_3$ + 0.5 N$_2$ ⟶ 3 LaB$_6$ + LaB$_6$H$_4$C$_6$N | -6.7782 |
| LaNH$_4$B$_6$C$_6$ | 67.25 LaH$_2$ + 0.5 N$_2$ + 3 B$_{13}$C$_2$ ⟶ 14.5 La$_4$H$_9$ + 8.25 LaB$_4$ + LaB$_6$H$_4$C$_6$N | -6.7538 |
| LaNH$_4$B$_6$C$_6$ | 1.333 LaH$_3$ + 4.571 LaN + 3 B$_{13}$C$_2$ ⟶ 4.905 LaB$_6$ + 3.571 BN + LaB$_6$H$_4$C$_6$N | -6.7406 |
| LaNH$_4$B$_6$C$_6$ | 2.542 N$_2$ + 3 B$_{13}$C$_2$ + 3.25 La$_2$H$_5$ ⟶ 5.5 LaB$_6$ + 4.083 H$_3$N + LaB$_6$H$_4$C$_6$N | -6.7372 |
| LaNH$_4$B$_6$C$_6$ | 1.625 La$_4$H$_9$ + 2.271 N$_2$ + 3 B$_{13}$C$_2$ ⟶ 5.5 LaB$_6$ + 3.542 H$_3$N + LaB$_6$H$_4$C$_6$N | -6.7358 |
| LaNH$_4$B$_6$C$_6$ | 12.25 La$_4$H$_9$ + 0.5 N$_2$ + 3 B$_{13}$C$_2$ ⟶ 5.5 LaB$_6$ + 21.25 La$_2$H$_5$ + LaB$_6$H$_4$C$_6$N | -6.7272 |
| LaNH$_4$B$_6$C$_6$ | 15.25 La$_4$H$_9$ + H$_3$N + 3 B$_{13}$C$_2$ ⟶ 5.5 LaB$_6$ + 27.25 La$_2$H$_5$ + LaB$_6$H$_4$C$_6$N | -6.7247 |
| LaNH$_4$B$_6$C$_6$ | 7.444 H$_3$N + 2 La$_2$C$_3$ + 3.111 LaB$_4$ ⟶ 6.111 LaH$_3$ + 6.444 BN + LaB$_6$H$_4$C$_6$N | -6.6437 |
| LaNH$_4$B$_6$C$_6$ | 1.333 LaH$_3$ + LaN + 3 B$_{13}$C$_2$ + 12.5 LaB$_4$ ⟶ 13.83 LaB$_6$ + LaB$_6$H$_4$C$_6$N | -6.6363 |
| LaNH$_4$B$_6$C$_6$ | 1.333 LaH$_3$ + BN + 3 B$_{13}$C$_2$ + 16 LaB$_4$ ⟶ 16.33 LaB$_6$ + LaB$_6$H$_4$C$_6$N | -6.6071 |
| LaNH$_4$B$_6$C$_6$ | 6.32 LaN + 3 B$_{13}$C$_2$ + 0.8 La$_2$H$_5$ ⟶ 5.32 BN + 6.92 LaB$_4$ + LaB$_6$H$_4$C$_6$N | -6.6043 |
| LaNH$_4$B$_6$C$_6$ | 3 LaC$_2$ + 1.333 LaH$_3$ + 32 B + LaN ⟶ 4.333 LaB$_6$ + LaB$_6$H$_4$C$_6$N | -6.5777 |
| LaNH$_4$B$_6$C$_6$ | 10.32 H$_3$N + LaH$_2$ + 3 B$_{13}$C$_2$ ⟶ 9.316 BN + 2.632 B$_9$H$_{11}$ + LaB$_6$H$_4$C$_6$N | -6.5764 |
| LaNH$_4$B$_6$C$_6$ | 4.343 LaN + 3 B$_{13}$C$_2$ + 0.8 La$_2$H$_5$ ⟶ 4.943 LaB$_6$ + 3.343 BN + LaB$_6$H$_4$C$_6$N | -6.5465 |
| LaNH$_4$B$_6$C$_6$ | 0.4444 La$_4$H$_9$ + 6.178 LaN + 3 B$_{13}$C$_2$ ⟶ 5.178 BN + 6.956 LaB$_4$ + LaB$_6$H$_4$C$_6$N | -6.5379 |
| LaNH$_4$B$_6$C$_6$ | 3 LaC$_2$ + 28.67 B + 0.4444 La$_4$H$_9$ + 0.5 N$_2$ ⟶ 3.778 LaB$_6$ + LaB$_6$H$_4$C$_6$N | -6.5026 |
| LaNH$_4$B$_6$C$_6$ | 6 LaN + 2 LaH$_2$ + 3 B$_{13}$C$_2$ ⟶ 5 BN + 7 LaB$_4$ + LaB$_6$H$_4$C$_6$N | -6.4844 |
| LaNH$_4$B$_6$C$_6$ | 27 B + H$_3$N + 2 La$_2$C$_3$ + 0.5 LaH$_2$ ⟶ 3.5 LaB$_6$ + LaB$_6$H$_4$C$_6$N | -6.4819 |
| LaNH$_4$B$_6$C$_6$ | 0.4444 La$_4$H$_9$ + 4.19 LaN + 3 B$_{13}$C$_2$ ⟶ 4.968 LaB$_6$ + 3.19 BN + LaB$_6$H$_4$C$_6$N | -6.4799 |
| LaNH$_4$B$_6$C$_6$ | LaN + 3 B$_{13}$C$_2$ + 0.8 La$_2$H$_5$ + 11.7 LaB$_4$ ⟶ 13.3 LaB$_6$ + LaB$_6$H$_4$C$_6$N | -6.4489 |

| Compound | Chemical reaction | Reaction energy(eV) |
|---|---|---|
| LaNH$_4$B$_6$C$_6$ | $1.333\,\text{LaH}_3 + 31\,\text{B} + \text{BN} + 2\,\text{La}_2\text{C}_3 \longrightarrow 4.333\,\text{LaB}_6 + \text{LaB}_6\text{H}_4\text{C}_6\text{N}$ | -6.4394 |
| LaNH$_4$B$_6$C$_6$ | $9.25\,\text{LaH}_3 + 4.458\,\text{N}_2 + 3\,\text{B}_{13}\text{C}_2 \longrightarrow 7.917\,\text{H}_3\text{N} + 8.25\,\text{LaB}_4 + \text{LaB}_6\text{H}_4\text{C}_6\text{N}$ | -6.4376 |
| LaNH$_4$B$_6$C$_6$ | $4\,\text{LaN} + 2\,\text{LaH}_2 + 3\,\text{B}_{13}\text{C}_2 \longrightarrow 5\,\text{LaB}_6 + 3\,\text{BN} + \text{LaB}_6\text{H}_4\text{C}_6\text{N}$ | -6.4260 |
| LaNH$_4$B$_6$C$_6$ | $\text{BN} + 3\,\text{B}_{13}\text{C}_2 + 0.8\,\text{La}_2\text{H}_5 + 15.2\,\text{LaB}_4 \longrightarrow 15.8\,\text{LaB}_6 + \text{LaB}_6\text{H}_4\text{C}_6\text{N}$ | -6.4197 |
| LaNH$_4$B$_6$C$_6$ | $\text{H}_3\text{N} + 44.25\,\text{LaH}_2 + 3\,\text{B}_{13}\text{C}_2 \longrightarrow 17.5\,\text{La}_2\text{H}_5 + 8.25\,\text{LaB}_4 + \text{LaB}_6\text{H}_4\text{C}_6\text{N}$ | -6.4132 |
| LaNH$_4$B$_6$C$_6$ | $0.25\,\text{La}_4\text{H}_9 + 10.26\,\text{H}_3\text{N} + 3\,\text{B}_{13}\text{C}_2 \longrightarrow 9.257\,\text{BN} + 2.638\,\text{B}_9\text{H}_{11} + \text{LaB}_6\text{H}_4\text{C}_6\text{N}$ | -6.3899 |
| LaNH$_4$B$_6$C$_6$ | $0.9773\,\text{B}_9\text{H}_{11} + 2\,\text{La}_2\text{C}_3 + 1.898\,\text{N}_2 \longrightarrow 2.795\,\text{BN} + 0.75\,\text{La}_4\text{H}_9 + \text{LaB}_6\text{H}_4\text{C}_6\text{N}$ | -6.3887 |
| LaNH$_4$B$_6$C$_6$ | $0.4444\,\text{La}_4\text{H}_9 + \text{LaN} + 3\,\text{B}_{13}\text{C}_2 + 11.17\,\text{LaB}_4 \longrightarrow 12.94\,\text{LaB}_6 + \text{LaB}_6\text{H}_4\text{C}_6\text{N}$ | -6.3867 |
| LaNH$_4$B$_6$C$_6$ | $1.2\,\text{H}_8\text{C}_5 + 8\,\text{La} + 1.2\,\text{B}_5\text{H}_7 + 0.5\,\text{N}_2 \longrightarrow 7\,\text{LaH}_2 + \text{LaB}_6\text{H}_4\text{C}_6\text{N}$ | -6.3724 |
| LaNH$_4$B$_6$C$_6$ | $\text{BN} + 0.4444\,\text{La}_4\text{H}_9 + 3\,\text{B}_{13}\text{C}_2 + 14.67\,\text{LaB}_4 \longrightarrow 15.44\,\text{LaB}_6 + \text{LaB}_6\text{H}_4\text{C}_6\text{N}$ | -6.3575 |
| LaNH$_4$B$_6$C$_6$ | $\text{LaN} + 2\,\text{LaH}_2 + 3\,\text{B}_{13}\text{C}_2 + 10.5\,\text{LaB}_4 \longrightarrow 12.5\,\text{LaB}_6 + \text{LaB}_6\text{H}_4\text{C}_6\text{N}$ | -6.3384 |
| LaNH$_4$B$_6$C$_6$ | $38.25\,\text{LaH}_2 + 0.5\,\text{N}_2 + 3\,\text{B}_{13}\text{C}_2 \longrightarrow 14.5\,\text{La}_2\text{H}_5 + 8.25\,\text{LaB}_4 + \text{LaB}_6\text{H}_4\text{C}_6\text{N}$ | -6.3275 |
| LaNH$_4$B$_6$C$_6$ | $15.5\,\text{LaH}_2 + 0.5\,\text{N}_2 + 3\,\text{B}_{13}\text{C}_2 \longrightarrow 9\,\text{LaH}_3 + 5.5\,\text{LaB}_6 + \text{LaB}_6\text{H}_4\text{C}_6\text{N}$ | -6.3196 |
| LaNH$_4$B$_6$C$_6$ | $26.67\,\text{B} + 0.1111\,\text{La}_4\text{H}_9 + \text{H}_3\text{N} + 2\,\text{La}_2\text{C}_3 \longrightarrow 3.444\,\text{LaB}_6 + \text{LaB}_6\text{H}_4\text{C}_6\text{N}$ | -6.3122 |
| LaNH$_4$B$_6$C$_6$ | $\text{BN} + 2\,\text{LaH}_2 + 3\,\text{B}_{13}\text{C}_2 + 14\,\text{LaB}_4 \longrightarrow 15\,\text{LaB}_6 + \text{LaB}_6\text{H}_4\text{C}_6\text{N}$ | -6.3092 |
| LaNH$_4$B$_6$C$_6$ | $1.2\,\text{H}_8\text{C}_5 + 7.8\,\text{La} + \text{H}_3\text{N} + 1.2\,\text{B}_5\text{H}_7 \longrightarrow 3.4\,\text{La}_2\text{H}_5 + \text{LaB}_6\text{H}_4\text{C}_6\text{N}$ | -6.2699 |
| LaNH$_4$B$_6$C$_6$ | $10.2\,\text{H}_3\text{N} + 3\,\text{B}_{13}\text{C}_2 + 0.5\,\text{La}_2\text{H}_5 \longrightarrow 9.197\,\text{BN} + 2.645\,\text{B}_9\text{H}_{11} + \text{LaB}_6\text{H}_4\text{C}_6\text{N}$ | -6.2181 |
| LaNH$_4$B$_6$C$_6$ | $26.4\,\text{B} + \text{H}_3\text{N} + 2\,\text{La}_2\text{C}_3 + 0.2\,\text{La}_2\text{H}_5 \longrightarrow 3.4\,\text{LaB}_6 + \text{LaB}_6\text{H}_4\text{C}_6\text{N}$ | -6.1824 |
| LaNH$_4$B$_6$C$_6$ | $21.14\,\text{B} + 2\,\text{La}_2\text{C}_3 + 0.5714\,\text{B}_5\text{H}_7 + 0.5\,\text{N}_2 \longrightarrow 3\,\text{LaB}_6 + \text{LaB}_6\text{H}_4\text{C}_6\text{N}$ | -6.1810 |
| LaNH$_4$B$_6$C$_6$ | $\text{H}_3\text{N} + 18.5\,\text{LaH}_2 + 3\,\text{B}_{13}\text{C}_2 \longrightarrow 12\,\text{LaH}_3 + 5.5\,\text{LaB}_6 + \text{LaB}_6\text{H}_4\text{C}_6\text{N}$ | -6.1495 |
| LaNH$_4$B$_6$C$_6$ | $1.2\,\text{H}_8\text{C}_5 + 8.081\,\text{La} + \text{H}_3\text{N} + 0.6667\,\text{B}_9\text{H}_{11} \longrightarrow 1.77\,\text{La}_4\text{H}_9 + \text{LaB}_6\text{H}_4\text{C}_6\text{N}$ | -6.1307 |
| LaNH$_4$B$_6$C$_6$ | $20.73\,\text{B} + 0.3636\,\text{B}_9\text{H}_{11} + 2\,\text{La}_2\text{C}_3 + 0.5\,\text{N}_2 \longrightarrow 3\,\text{LaB}_6 + \text{LaB}_6\text{H}_4\text{C}_6\text{N}$ | -6.1187 |
| LaNH$_4$B$_6$C$_6$ | $0.3333\,\text{LaH}_3 + 26\,\text{B} + \text{H}_3\text{N} + 2\,\text{La}_2\text{C}_3 \longrightarrow 3.333\,\text{LaB}_6 + \text{LaB}_6\text{H}_4\text{C}_6\text{N}$ | -6.0111 |
| LaNH$_4$B$_6$C$_6$ | $3\,\text{LaC}_2 + 27.6\,\text{B} + 0.5\,\text{N}_2 + 0.8\,\text{La}_2\text{H}_5 \longrightarrow 3.6\,\text{LaB}_6 + \text{LaB}_6\text{H}_4\text{C}_6\text{N}$ | -5.9831 |
| LaNH$_4$B$_6$C$_6$ | $\text{LaH}_3 + 10.08\,\text{H}_3\text{N} + 3\,\text{B}_{13}\text{C}_2 \longrightarrow 9.079\,\text{BN} + 2.658\,\text{B}_9\text{H}_{11} + \text{LaB}_6\text{H}_4\text{C}_6\text{N}$ | -5.9451 |
| LaNH$_4$B$_6$C$_6$ | $9.25\,\text{LaH}_2 + 2.917\,\text{N}_2 + 3\,\text{B}_{13}\text{C}_2 \longrightarrow 4.833\,\text{H}_3\text{N} + 8.25\,\text{LaB}_4 + \text{LaB}_6\text{H}_4\text{C}_6\text{N}$ | -5.9130 |
| LaNH$_4$B$_6$C$_6$ | $24.33\,\text{B} + 1.333\,\text{H}_3\text{N} + 2\,\text{La}_2\text{C}_3 \longrightarrow 3\,\text{LaB}_6 + 0.3333\,\text{BN} + \text{LaB}_6\text{H}_4\text{C}_6\text{N}$ | -5.8683 |
| LaNH$_4$B$_6$C$_6$ | $3.687\,\text{N}_2 + 3\,\text{B}_{13}\text{C}_2 + 4.625\,\text{La}_2\text{H}_5 \longrightarrow 6.375\,\text{H}_3\text{N} + 8.25\,\text{LaB}_4 + \text{LaB}_6\text{H}_4\text{C}_6\text{N}$ | -5.7808 |
| LaNH$_4$B$_6$C$_6$ | $2.312\,\text{La}_4\text{H}_9 + 3.302\,\text{N}_2 + 3\,\text{B}_{13}\text{C}_2 \longrightarrow 5.604\,\text{H}_3\text{N} + 8.25\,\text{LaB}_4 + \text{LaB}_6\text{H}_4\text{C}_6\text{N}$ | -5.7789 |
| LaNH$_4$B$_6$C$_6$ | $0.5\,\text{H}_8\text{C}_5 + 5.187\,\text{La} + 0.5\,\text{N}_2 + 1.75\,\text{B}_{13}\text{C}_2 \longrightarrow 4.187\,\text{LaB}_4 + \text{LaB}_6\text{H}_4\text{C}_6\text{N}$ | -5.7788 |
| LaNH$_4$B$_6$C$_6$ | $19.12\,\text{La}_4\text{H}_9 + 0.5\,\text{N}_2 + 3\,\text{B}_{13}\text{C}_2 \longrightarrow 33.62\,\text{La}_2\text{H}_5 + 8.25\,\text{LaB}_4 + \text{LaB}_6\text{H}_4\text{C}_6\text{N}$ | -5.7652 |
| LaNH$_4$B$_6$C$_6$ | $22.13\,\text{La}_4\text{H}_9 + \text{H}_3\text{N} + 3\,\text{B}_{13}\text{C}_2 \longrightarrow 39.63\,\text{La}_2\text{H}_5 + 8.25\,\text{LaB}_4 + \text{LaB}_6\text{H}_4\text{C}_6\text{N}$ | -5.7627 |
| LaNH$_4$B$_6$C$_6$ | $5.167\,\text{La}_4\text{H}_9 + 0.5\,\text{N}_2 + 3\,\text{B}_{13}\text{C}_2 \longrightarrow 14.17\,\text{LaH}_3 + 5.5\,\text{LaB}_6 + \text{LaB}_6\text{H}_4\text{C}_6\text{N}$ | -5.7270 |
| LaNH$_4$B$_6$C$_6$ | $1.2\,\text{H}_8\text{C}_5 + 7.222\,\text{La} + 1.2\,\text{B}_5\text{H}_7 + 0.5\,\text{N}_2 \longrightarrow 1.556\,\text{La}_4\text{H}_9 + \text{LaB}_6\text{H}_4\text{C}_6\text{N}$ | -5.6860 |
| LaNH$_4$B$_6$C$_6$ | $30\,\text{B} + \text{LaN} + 2\,\text{La}_2\text{C}_3 + 2\,\text{LaH}_2 \longrightarrow 6\,\text{LaB}_4 + \text{LaB}_6\text{H}_4\text{C}_6\text{N}$ | -5.6254 |
| LaNH$_4$B$_6$C$_6$ | $0.6667\,\text{B}_9\text{H}_{11} + 2.694\,\text{La}_2\text{C}_3 + 2.694\,\text{N}_2 \longrightarrow 0.4167\,\text{H}_8\text{C}_5 + 4.389\,\text{LaN} + \text{LaB}_6\text{H}_4\text{C}_6\text{N}$ | -5.5981 |
| LaNH$_4$B$_6$C$_6$ | $\text{LaN} + 33.5\,\text{LaH}_2 + 3\,\text{B}_{13}\text{C}_2 \longrightarrow 5.5\,\text{LaB}_6 + 7\,\text{La}_4\text{H}_9 + \text{LaB}_6\text{H}_4\text{C}_6\text{N}$ | -5.5771 |
| LaNH$_4$B$_6$C$_6$ | $0.5\,\text{H}_2 + 24\,\text{B} + \text{H}_3\text{N} + 2\,\text{La}_2\text{C}_3 \longrightarrow 3\,\text{LaB}_6 + \text{LaB}_6\text{H}_4\text{C}_6\text{N}$ | -5.4541 |
| LaNH$_4$B$_6$C$_6$ | $6.167\,\text{La}_4\text{H}_9 + \text{H}_3\text{N} + 3\,\text{B}_{13}\text{C}_2 \longrightarrow 18.17\,\text{LaH}_3 + 5.5\,\text{LaB}_6 + \text{LaB}_6\text{H}_4\text{C}_6\text{N}$ | -5.4421 |
| LaNH$_4$B$_6$C$_6$ | $26\,\text{B} + 2\,\text{La}_2\text{C}_3 + 2\,\text{LaH}_2 + 0.5\,\text{N}_2 \longrightarrow 5\,\text{LaB}_4 + \text{LaB}_6\text{H}_4\text{C}_6\text{N}$ | -5.4364 |
| LaNH$_4$B$_6$C$_6$ | $1.2\,\text{H}_8\text{C}_5 + 7.467\,\text{La} + 0.6667\,\text{B}_9\text{H}_{11} + 0.5\,\text{N}_2 \longrightarrow 6.467\,\text{LaH}_2 + \text{LaB}_6\text{H}_4\text{C}_6\text{N}$ | -5.4141 |
| LaNH$_4$B$_6$C$_6$ | $1.2\,\text{H}_8\text{C}_5 + 7.373\,\text{La} + \text{H}_3\text{N} + 0.6667\,\text{B}_9\text{H}_{11} \longrightarrow 3.187\,\text{La}_2\text{H}_5 + \text{LaB}_6\text{H}_4\text{C}_6\text{N}$ | -5.4120 |
| LaNH$_4$B$_6$C$_6$ | $\text{LaN} + 19.5\,\text{LaH}_2 + 3\,\text{B}_{13}\text{C}_2 \longrightarrow 5.5\,\text{LaB}_6 + 7\,\text{La}_2\text{H}_5 + \text{LaB}_6\text{H}_4\text{C}_6\text{N}$ | -5.3714 |
| LaNH$_4$B$_6$C$_6$ | $23.29\,\text{B} + \text{H}_3\text{N} + 2\,\text{La}_2\text{C}_3 + 0.1429\,\text{B}_5\text{H}_7 \longrightarrow 3\,\text{LaB}_6 + \text{LaB}_6\text{H}_4\text{C}_6\text{N}$ | -5.3048 |
| LaNH$_4$B$_6$C$_6$ | $3\,\text{LaC}_2 + 1.333\,\text{LaH}_3 + 26\,\text{B} + 0.5\,\text{N}_2 \longrightarrow 3.333\,\text{LaB}_6 + \text{LaB}_6\text{H}_4\text{C}_6\text{N}$ | -5.2980 |
| LaNH$_4$B$_6$C$_6$ | $\text{BN} + 44\,\text{LaH}_2 + 3\,\text{B}_{13}\text{C}_2 \longrightarrow 5.667\,\text{LaB}_6 + 9.333\,\text{La}_4\text{H}_9 + \text{LaB}_6\text{H}_4\text{C}_6\text{N}$ | -5.2942 |
| LaNH$_4$B$_6$C$_6$ | $23.18\,\text{B} + \text{H}_3\text{N} + 0.09091\,\text{B}_9\text{H}_{11} + 2\,\text{La}_2\text{C}_3 \longrightarrow 3\,\text{LaB}_6 + \text{LaB}_6\text{H}_4\text{C}_6\text{N}$ | -5.2892 |
| LaNH$_4$B$_6$C$_6$ | $2\,\text{La}_2\text{C}_3 + 1.643\,\text{B}_5\text{H}_7 + 1.607\,\text{N}_2 \longrightarrow 2.214\,\text{BN} + 1.5\,\text{La}_2\text{H}_5 + \text{LaB}_6\text{H}_4\text{C}_6\text{N}$ | -5.2827 |
| LaNH$_4$B$_6$C$_6$ | $8.625\,\text{H}_3\text{N} + 0.5\,\text{La}_2\text{C}_3 + 2.25\,\text{B}_{13}\text{C}_2 \longrightarrow 7.625\,\text{BN} + 3.125\,\text{B}_5\text{H}_7 + \text{LaB}_6\text{H}_4\text{C}_6\text{N}$ | -5.2336 |
| LaNH$_4$B$_6$C$_6$ | $29.11\,\text{B} + 0.4444\,\text{La}_4\text{H}_9 + \text{LaN} + 2\,\text{La}_2\text{C}_3 \longrightarrow 5.778\,\text{LaB}_4 + \text{LaB}_6\text{H}_4\text{C}_6\text{N}$ | -5.1890 |
| LaNH$_4$B$_6$C$_6$ | $6\,\text{C} + 8.333\,\text{LaN} + 2.667\,\text{B}_5\text{H}_7 \longrightarrow 7.333\,\text{BN} + 7.333\,\text{LaH}_2 + \text{LaB}_6\text{H}_4\text{C}_6\text{N}$ | -5.1474 |
| LaNH$_4$B$_6$C$_6$ | $10.25\,\text{H}_2 + 6.5\,\text{LaN} + 3\,\text{B}_{13}\text{C}_2 \longrightarrow 5.5\,\text{LaB}_6 + 5.5\,\text{H}_3\text{N} + \text{LaB}_6\text{H}_4\text{C}_6\text{N}$ | -5.0958 |
| LaNH$_4$B$_6$C$_6$ | $23.75\,\text{LaH}_2 + 0.5\,\text{N}_2 + 3\,\text{B}_{13}\text{C}_2 \longrightarrow 14.5\,\text{LaH}_3 + 8.25\,\text{LaB}_4 + \text{LaB}_6\text{H}_4\text{C}_6\text{N}$ | -5.0906 |
| LaNH$_4$B$_6$C$_6$ | $9.75\,\text{La}_4\text{H}_9 + \text{LaN} + 3\,\text{B}_{13}\text{C}_2 \longrightarrow 5.5\,\text{LaB}_6 + 16.75\,\text{La}_2\text{H}_5 + \text{LaB}_6\text{H}_4\text{C}_6\text{N}$ | -5.0847 |
| LaNH$_4$B$_6$C$_6$ | $1.2\,\text{H}_8\text{C}_5 + 6.6\,\text{La} + 1.2\,\text{B}_5\text{H}_7 + 0.5\,\text{N}_2 \longrightarrow 2.8\,\text{La}_2\text{H}_5 + \text{LaB}_6\text{H}_4\text{C}_6\text{N}$ | -5.0545 |
| LaNH$_4$B$_6$C$_6$ | $\text{BN} + 25.33\,\text{LaH}_2 + 3\,\text{B}_{13}\text{C}_2 \longrightarrow 5.667\,\text{LaB}_6 + 9.333\,\text{La}_2\text{H}_5 + \text{LaB}_6\text{H}_4\text{C}_6\text{N}$ | -5.0198 |
| LaNH$_4$B$_6$C$_6$ | $24\,\text{B} + 1.333\,\text{H}_3\text{N} + 2\,\text{La}_2\text{C}_3 \longrightarrow 3\,\text{LaB}_6 + 0.1667\,\text{N}_2 + \text{LaB}_6\text{H}_4\text{C}_6\text{N}$ | -5.0127 |
| LaNH$_4$B$_6$C$_6$ | $25.11\,\text{B} + 0.4444\,\text{La}_4\text{H}_9 + 2\,\text{La}_2\text{C}_3 + 0.5\,\text{N}_2 \longrightarrow 4.778\,\text{LaB}_4 + \text{LaB}_6\text{H}_4\text{C}_6\text{N}$ | -5.0000 |
| LaNH$_4$B$_6$C$_6$ | $0.5\,\text{N}_2 + 3\,\text{B}_{13}\text{C}_2 + 15.5\,\text{La}_2\text{H}_5 \longrightarrow 24.5\,\text{LaH}_3 + 5.5\,\text{LaB}_6 + \text{LaB}_6\text{H}_4\text{C}_6\text{N}$ | -4.9974 |
| LaNH$_4$B$_6$C$_6$ | $\text{LaN} + 58.25\,\text{LaH}_2 + 3\,\text{B}_{13}\text{C}_2 \longrightarrow 12.5\,\text{La}_4\text{H}_9 + 8.25\,\text{LaB}_4 + \text{LaB}_6\text{H}_4\text{C}_6\text{N}$ | -4.9790 |

| Compound | Chemical reaction | Reaction energy(eV) |
|---|---|---|
| $LaNH_4B_6C_6$ | $1.2\,H_8C_5 + 8.275\,LaN + 1.475\,B_9H_{11} \longrightarrow 7.275\,LaH_3 + 7.275\,BN + LaB_6H_4C_6N$ | -4.9509 |
| $LaNH_4B_6C_6$ | $H_3N + 26.75\,LaH_2 + 3\,B_{13}C_2 \longrightarrow 17.5\,LaH_3 + 8.25\,LaB_4 + LaB_6H_4C_6N$ | -4.9205 |
| $LaNH_4B_6C_6$ | $28.4\,B + LaN + 2\,La_2C_3 + 0.8\,La_2H_5 \longrightarrow 5.6\,LaB_4 + LaB_6H_4C_6N$ | -4.8634 |
| $LaNH_4B_6C_6$ | $0.5\,H_8C_5 + 3.792\,La + 0.5\,N_2 + 1.75\,B_{13}C_2 \longrightarrow 2.792\,LaB_6 + LaB_6H_4C_6N$ | -4.8505 |
| $LaNH_4B_6C_6$ | $3\,LaC_2 + 2.8\,B_5H_7 + 3.1\,N_2 \longrightarrow 5.2\,H_3N + 2\,LaB_4 + LaB_6H_4C_6N$ | -4.8478 |
| $LaNH_4B_6C_6$ | $1.2\,H_8C_5 + 6.748\,La + 0.6667\,B_9H_{11} + 0.5\,N_2 \longrightarrow 1.437\,La_4H_9 + LaB_6H_4C_6N$ | -4.7800 |
| $LaNH_4B_6C_6$ | $LaN + 12.5\,LaH_2 + 3\,B_{13}C_2 \longrightarrow 7\,LaH_3 + 5.5\,LaB_6 + LaB_6H_4C_6N$ | -4.7743 |
| $LaNH_4B_6C_6$ | $1.2\,H_8C_5 + 6.667\,La + H_3N + 1.2\,B_5H_7 \longrightarrow 5.667\,LaH_3 + LaB_6H_4C_6N$ | -4.7196 |
| $LaNH_4B_6C_6$ | $BN + 69.5\,LaH_2 + 3\,B_{13}C_2 \longrightarrow 15\,La_4H_9 + 8.5\,LaB_4 + LaB_6H_4C_6N$ | -4.6780 |
| $LaNH_4B_6C_6$ | $24.4\,B + 2\,La_2C_3 + 0.5\,N_2 + 0.8\,La_2H_5 \longrightarrow 4.6\,LaB_4 + LaB_6H_4C_6N$ | -4.6744 |
| $LaNH_4B_6C_6$ | $BN + 12.67\,La_4H_9 + 3\,B_{13}C_2 \longrightarrow 5.667\,LaB_6 + 22\,La_2H_5 + LaB_6H_4C_6N$ | -4.6474 |
| $LaNH_4B_6C_6$ | $3\,LaC_2 + 29\,B + BN + 2\,LaH_2 \longrightarrow 4\,LaB_6 + LaB_6H_4C_6N$ | -4.6146 |
| $LaNH_4B_6C_6$ | $LaN + 33.25\,LaH_2 + 3\,B_{13}C_2 \longrightarrow 12.5\,La_2H_5 + 8.25\,LaB_4 + LaB_6H_4C_6N$ | -4.6115 |
| $LaNH_4B_6C_6$ | $22\,B + 1.333\,H_3N + 2\,La_2C_3 \longrightarrow 2.667\,LaB_6 + 0.3333\,LaN + LaB_6H_4C_6N$ | -4.5862 |
| $LaNH_4B_6C_6$ | $9.333\,B + 4.333\,H_3N + 2\,La_2C_3 \longrightarrow 3\,LaH_3 + 3.333\,BN + LaB_6H_4C_6N$ | -4.5833 |
| $LaNH_4B_6C_6$ | $H_3N + 3\,B_{13}C_2 + 18.5\,La_2H_5 \longrightarrow 30.5\,LaH_3 + 5.5\,LaB_6 + LaB_6H_4C_6N$ | -4.5714 |
| $LaNH_4B_6C_6$ | $0.9091\,B_9H_{11} + 2\,La_2C_3 + 1.591\,N_2 \longrightarrow 2.182\,BN + 3\,LaH_2 + LaB_6H_4C_6N$ | -4.5627 |
| $LaNH_4B_6C_6$ | $1.2\,H_8C_5 + 3.233\,N_2 + 2.867\,LaB_4 \longrightarrow 1.867\,LaH_3 + 5.467\,BN + LaB_6H_4C_6N$ | -4.5594 |
| $LaNH_4B_6C_6$ | $6\,B + 2\,La_2C_3 + 2\,LaH_2 + 3\,N_2 \longrightarrow 5\,LaN + LaB_6H_4C_6N$ | -4.4915 |
| $LaNH_4B_6C_6$ | $1.333\,LaH_3 + 27.33\,B + LaN + 2\,La_2C_3 \longrightarrow 5.333\,LaB_4 + LaB_6H_4C_6N$ | -4.4692 |
| $LaNH_4B_6C_6$ | $7.967\,H_3N + 0.5\,La_2C_3 + 2.25\,B_{13}C_2 \longrightarrow 6.967\,BN + 1.809\,B_9H_{11} + LaB_6H_4C_6N$ | -4.4312 |
| $LaNH_4B_6C_6$ | $6.158\,H_3N + 2\,La_2C_3 + 2.789\,LaB_4 \longrightarrow 5.158\,BN + 2.895\,La_2H_5 + LaB_6H_4C_6N$ | -4.3650 |
| $LaNH_4B_6C_6$ | $3\,LaC_2 + 2\,H_2 + 24\,B + LaN \longrightarrow 3\,LaB_6 + LaB_6H_4C_6N$ | -4.3496 |
| $LaNH_4B_6C_6$ | $4.167\,La_4H_9 + LaN + 3\,B_{13}C_2 \longrightarrow 11.17\,LaH_3 + 5.5\,LaB_6 + LaB_6H_4C_6N$ | -4.2963 |
| $LaNH_4B_6C_6$ | $1.333\,LaH_3 + 23.33\,B + 2\,La_2C_3 + 0.5\,N_2 \longrightarrow 4.333\,LaB_4 + LaB_6H_4C_6N$ | -4.2802 |
| $LaNH_4B_6C_6$ | $BN + 39.5\,LaH_2 + 3\,B_{13}C_2 \longrightarrow 15\,La_2H_5 + 8.5\,LaB_4 + LaB_6H_4C_6N$ | -4.2369 |
| $LaNH_4B_6C_6$ | $BN + 16\,LaH_2 + 3\,B_{13}C_2 \longrightarrow 9.333\,LaH_3 + 5.667\,LaB_6 + LaB_6H_4C_6N$ | -4.2237 |
| $LaNH_4B_6C_6$ | $16.73\,LaH_3 + 8.367\,N_2 + 3\,B_{13}C_2 \longrightarrow 15.73\,LaN + 6.6\,B_5H_7 + LaB_6H_4C_6N$ | -4.2227 |
| $LaNH_4B_6C_6$ | $2\,H_2 + 23\,B + BN + 2\,La_2C_3 \longrightarrow 3\,LaB_6 + LaB_6H_4C_6N$ | -4.2114 |
| $LaNH_4B_6C_6$ | $2\,La_2C_3 + 2\,LaH_2 + 3.75\,N_2 + 1.5\,LaB_4 \longrightarrow 6.5\,LaN + LaB_6H_4C_6N$ | -4.2080 |
| $LaNH_4B_6C_6$ | $1.2\,H_8C_5 + 6.173\,La + 0.6667\,B_9H_{11} + 0.5\,N_2 \longrightarrow 2.587\,La_2H_5 + LaB_6H_4C_6N$ | -4.1966 |
| $LaNH_4B_6C_6$ | $7.917\,La_4H_9 + 0.5\,N_2 + 3\,B_{13}C_2 \longrightarrow 22.42\,LaH_3 + 8.25\,LaB_4 + LaB_6H_4C_6N$ | -4.1826 |
| $LaNH_4B_6C_6$ | $16.63\,La_4H_9 + LaN + 3\,B_{13}C_2 \longrightarrow 29.13\,La_2H_5 + 8.25\,LaB_4 + LaB_6H_4C_6N$ | -4.1228 |
| $LaNH_4B_6C_6$ | $6\,B + 0.4444\,La_4H_9 + 2\,La_2C_3 + 2.889\,N_2 \longrightarrow 4.778\,LaN + LaB_6H_4C_6N$ | -4.0971 |
| $LaNH_4B_6C_6$ | $LaC_2 + 0.5\,H_8C_5 + 2.375\,N_2 + 0.75\,B_{13}C_2 \longrightarrow 3.75\,BN + LaB_6H_4C_6N$ | -4.0729 |
| $LaNH_4B_6C_6$ | $1.2\,H_8C_5 + 6.311\,La + H_3N + 0.6667\,B_9H_{11} \longrightarrow 5.311\,LaH_3 + LaB_6H_4C_6N$ | -3.9590 |
| $LaNH_4B_6C_6$ | $3\,LaC_2 + 27.67\,B + BN + 0.4444\,La_4H_9 \longrightarrow 3.778\,LaB_6 + LaB_6H_4C_6N$ | -3.9358 |
| $LaNH_4B_6C_6$ | $0.5\,H_8C_5 + 5.437\,La + BN + 1.75\,B_{13}C_2 \longrightarrow 4.437\,LaB_4 + LaB_6H_4C_6N$ | -3.9237 |
| $LaNH_4B_6C_6$ | $8.917\,La_4H_9 + H_3N + 3\,B_{13}C_2 \longrightarrow 26.42\,LaH_3 + 8.25\,LaB_4 + LaB_6H_4C_6N$ | -3.8977 |
| $LaNH_4B_6C_6$ | $3\,LaC_2 + 1.556\,B_9H_{11} + 2.685\,N_2 \longrightarrow 4.37\,H_3N + 2\,LaB_4 + LaB_6H_4C_6N$ | -3.8545 |
| $LaNH_4B_6C_6$ | $0.4444\,La_4H_9 + 2\,La_2C_3 + 3.639\,N_2 + 1.5\,LaB_4 \longrightarrow 6.278\,LaN + LaB_6H_4C_6N$ | -3.8136 |
| $LaNH_4B_6C_6$ | $6\,B + 2\,La_2C_3 + 2.8\,N_2 + 0.8\,La_2H_5 \longrightarrow 4.6\,LaN + LaB_6H_4C_6N$ | -3.8051 |
| $LaNH_4B_6C_6$ | $1.867\,LaB_6 + 6.2\,H_3N + 2\,La_2C_3 \longrightarrow 4.867\,LaH_3 + 5.2\,BN + LaB_6H_4C_6N$ | -3.7837 |
| $LaNH_4B_6C_6$ | $1.2\,H_8C_5 + 5.667\,La + 1.2\,B_5H_7 + 0.5\,N_2 \longrightarrow 4.667\,LaH_3 + LaB_6H_4C_6N$ | -3.7778 |
| $LaNH_4B_6C_6$ | $0.125\,H_8C_5 + 21.5\,B + H_3N + 1.792\,La_2C_3 \longrightarrow 2.583\,LaB_6 + LaB_6H_4C_6N$ | -3.7606 |
| $LaNH_4B_6C_6$ | $3\,LaC_2 + 21.14\,B + LaN + 0.5714\,B_5H_7 \longrightarrow 3\,LaB_6 + LaB_6H_4C_6N$ | -3.7524 |
| $LaNH_4B_6C_6$ | $1.2\,H_8C_5 + 7\,La + LaN + 1.2\,B_5H_7 \longrightarrow 7\,LaH_2 + LaB_6H_4C_6N$ | -3.7151 |
| $LaNH_4B_6C_6$ | $LaN + 3\,B_{13}C_2 + 12.5\,La_2H_5 \longrightarrow 19.5\,LaH_3 + 5.5\,LaB_6 + LaB_6H_4C_6N$ | -3.7080 |
| $LaNH_4B_6C_6$ | $2\,H_2 + 22\,B + LaN + 2\,La_2C_3 \longrightarrow 4\,LaB_4 + LaB_6H_4C_6N$ | -3.6953 |
| $LaNH_4B_6C_6$ | $3\,LaC_2 + 20.73\,B + LaN + 0.3636\,B_9H_{11} \longrightarrow 3\,LaB_6 + LaB_6H_4C_6N$ | -3.6902 |
| $LaNH_4B_6C_6$ | $LaC_2 + 7.591\,H_3N + 2\,B_{13}C_2 \longrightarrow 6.591\,BN + 2.682\,B_5H_7 + LaB_6H_4C_6N$ | -3.6896 |
| $LaNH_4B_6C_6$ | $2\,La_2C_3 + 1.536\,B_5H_7 + 1.339\,N_2 \longrightarrow 1.679\,BN + 0.75\,La_4H_9 + LaB_6H_4C_6N$ | -3.6892 |
| $LaNH_4B_6C_6$ | $3\,LaC_2 + 0.9091\,B_9H_{11} + 1.591\,N_2 \longrightarrow 2\,LaH_3 + 2.182\,BN + LaB_6H_4C_6N$ | -3.6795 |
| $LaNH_4B_6C_6$ | $BN + 19.75\,La_4H_9 + 3\,B_{13}C_2 \longrightarrow 34.75\,La_2H_5 + 8.5\,LaB_4 + LaB_6H_4C_6N$ | -3.6563 |
| $LaNH_4B_6C_6$ | $2\,La_2C_3 + 1.2\,B_5H_7 + 2.733\,N_2 \longrightarrow 1.467\,H_3N + 3\,LaN + LaB_6H_4C_6N$ | -3.6273 |
| $LaNH_4B_6C_6$ | $20.14\,B + BN + 2\,La_2C_3 + 0.5714\,B_5H_7 \longrightarrow 3\,LaB_6 + LaB_6H_4C_6N$ | -3.6142 |
| $LaNH_4B_6C_6$ | $BN + 5.333\,La_4H_9 + 3\,B_{13}C_2 \longrightarrow 14.67\,LaH_3 + 5.667\,LaB_6 + LaB_6H_4C_6N$ | -3.6119 |
| $LaNH_4B_6C_6$ | $19.73\,B + BN + 0.3636\,B_9H_{11} + 2\,La_2C_3 \longrightarrow 3\,LaB_6 + LaB_6H_4C_6N$ | -3.5520 |
| $LaNH_4B_6C_6$ | $LaN + 20.75\,LaH_2 + 3\,B_{13}C_2 \longrightarrow 12.5\,LaH_3 + 8.25\,LaB_4 + LaB_6H_4C_6N$ | -3.5453 |
| $LaNH_4B_6C_6$ | $2\,La_2C_3 + 3.55\,N_2 + 0.8\,La_2H_5 + 1.5\,LaB_4 \longrightarrow 6.1\,LaN + LaB_6H_4C_6N$ | -3.5216 |

| Compound | Chemical reaction | Reaction energy(eV) |
|---|---|---|
| LaNH$_4$B$_6$C$_6$ | 8.833 B + 3.833 H$_3$N + 2 La$_2$C$_3$ ⟶ 2.833 BN + 1.5 La$_2$H$_5$ + LaB$_6$H$_4$C$_6$N | -3.5129 |
| LaNH$_4$B$_6$C$_6$ | 2 H$_2$ + 18 B + 2 La$_2$C$_3$ + 0.5 N$_2$ ⟶ 3 LaB$_4$ + LaB$_6$H$_4$C$_6$N | -3.5063 |
| LaNH$_4$B$_6$C$_6$ | 1.333 LaH$_3$ + 6 B + 2 La$_2$C$_3$ + 2.667 N$_2$ ⟶ 4.333 LaN + LaB$_6$H$_4$C$_6$N | -3.4612 |
| LaNH$_4$B$_6$C$_6$ | 3 LaC$_2$ + 26.6 B + BN + 0.8 La$_2$H$_5$ ⟶ 3.6 LaB$_6$ + LaB$_6$H$_4$C$_6$N | -3.4164 |
| LaNH$_4$B$_6$C$_6$ | LaN + 23.75 B$_9$H$_{11}$ + 2 La$_2$C$_3$ ⟶ 4 LaB$_6$ + 36.75 B$_5$H$_7$ + LaB$_6$H$_4$C$_6$N | -3.3937 |
| LaNH$_4$B$_6$C$_6$ | 14.38 H$_2$ + 9.25 LaN + 3 B$_{13}$C$_2$ ⟶ 8.25 H$_3$N + 8.25 LaB$_4$ + LaB$_6$H$_4$C$_6$N | -3.2583 |
| LaNH$_4$B$_6$C$_6$ | 4.375 LaC$_2$ + 1.2 B$_5$H$_7$ + 2.188 N$_2$ ⟶ 0.55 H$_8$C$_5$ + 3.375 LaN + LaB$_6$H$_4$C$_6$N | -3.2491 |
| LaNH$_4$B$_6$C$_6$ | LaB$_6$ + 2 La$_2$C$_3$ + 2 LaH$_2$ + 3.5 N$_2$ ⟶ 6 LaN + LaB$_6$H$_4$C$_6$N | -3.2119 |
| LaNH$_4$B$_6$C$_6$ | 14.78 LaH$_3$ + 7.389 N$_2$ + 3 B$_{13}$C$_2$ ⟶ 13.78 LaN + 3.667 B$_9$H$_{11}$ + LaB$_6$H$_4$C$_6$N | -3.2092 |
| LaNH$_4$B$_6$C$_6$ | 0.6667 B$_9$H$_{11}$ + 2 La$_2$C$_3$ + 2.556 N$_2$ ⟶ 1.111 H$_3$N + 3 LaN + LaB$_6$H$_4$C$_6$N | -3.2016 |
| LaNH$_4$B$_6$C$_6$ | 1.333 LaH$_3$ + 2 La$_2$C$_3$ + 3.417 N$_2$ + 1.5 LaB$_4$ ⟶ 5.833 LaN + LaB$_6$H$_4$C$_6$N | -3.1777 |
| LaNH$_4$B$_6$C$_6$ | 3 LaC$_2$ + 6.111 H$_3$N + 2.778 LaB$_4$ ⟶ 4.778 LaH$_3$ + 5.111 BN + LaB$_6$H$_4$C$_6$N | -3.1431 |
| LaNH$_4$B$_6$C$_6$ | 0.5 H$_8$C$_5$ + 4.187 La + LaN + 1.75 B$_{13}$C$_2$ ⟶ 4.187 LaB$_4$ + LaB$_6$H$_4$C$_6$N | -3.1215 |
| LaNH$_4$B$_6$C$_6$ | 5.564 H$_3$N + 2 La$_2$C$_3$ + 2.641 LaB$_4$ ⟶ 4.564 BN + 1.41 La$_4$H$_9$ + LaB$_6$H$_4$C$_6$N | -3.1143 |
| LaNH$_4$B$_6$C$_6$ | 19.14 B + LaN + 2 La$_2$C$_3$ + 0.5714 B$_5$H$_7$ ⟶ 4 LaB$_4$ + LaB$_6$H$_4$C$_6$N | -3.0981 |
| LaNH$_4$B$_6$C$_6$ | 3 LaC$_2$ + 2 H$_2$ + 18 B + 0.5 N$_2$ ⟶ 2 LaB$_6$ + LaB$_6$H$_4$C$_6$N | -3.0700 |
| LaNH$_4$B$_6$C$_6$ | 0.5 N$_2$ + 3 B$_{13}$C$_2$ + 23.75 La$_2$H$_5$ ⟶ 38.25 LaH$_3$ + 8.25 LaB$_4$ + LaB$_6$H$_4$C$_6$N | -3.0647 |
| LaNH$_4$B$_6$C$_6$ | 18.73 B + LaN + 0.3636 B$_9$H$_{11}$ + 2 La$_2$C$_3$ ⟶ 4 LaB$_4$ + LaB$_6$H$_4$C$_6$N | -3.0359 |
| LaNH$_4$B$_6$C$_6$ | 1.2 H$_8$C$_5$ + 6.222 La + LaN + 1.2 B$_5$H$_7$ ⟶ 1.556 La$_4$H$_9$ + LaB$_6$H$_4$C$_6$N | -3.0286 |
| LaNH$_4$B$_6$C$_6$ | 1.2 H$_8$C$_5$ + 5.311 La + 0.6667 B$_9$H$_{11}$ + 0.5 N$_2$ ⟶ 4.311 LaH$_3$ + LaB$_6$H$_4$C$_6$N | -3.0172 |
| LaNH$_4$B$_6$C$_6$ | 18.5 B$_9$H$_{11}$ + 2 La$_2$C$_3$ + 0.5 N$_2$ ⟶ 3 LaB$_6$ + 28.5 B$_5$H$_7$ + LaB$_6$H$_4$C$_6$N | -3.0131 |
| LaNH$_4$B$_6$C$_6$ | 3 LaC$_2$ + 26 B + LaN + 2 LaH$_2$ ⟶ 5 LaB$_4$ + LaB$_6$H$_4$C$_6$N | -3.0079 |
| LaNH$_4$B$_6$C$_6$ | LaC$_2$ + 7.026 H$_3$N + 2 B$_{13}$C$_2$ ⟶ 6.026 BN + 1.553 B$_9$H$_{11}$ + LaB$_6$H$_4$C$_6$N | -3.0010 |
| LaNH$_4$B$_6$C$_6$ | BN + 24.5 LaH$_2$ + 3 B$_{13}$C$_2$ ⟶ 15 LaH$_3$ + 8.5 LaB$_4$ + LaB$_6$H$_4$C$_6$N | -2.9574 |
| LaNH$_4$B$_6$C$_6$ | 0.5 H$_8$C$_5$ + 3.958 La + BN + 1.75 B$_{13}$C$_2$ ⟶ 2.958 LaB$_6$ + LaB$_6$H$_4$C$_6$N | -2.9399 |
| LaNH$_4$B$_6$C$_6$ | 2 H$_2$ + 6 B + 2 La$_2$C$_3$ + 2 N$_2$ ⟶ 3 LaN + LaB$_6$H$_4$C$_6$N | -2.9393 |
| LaNH$_4$B$_6$C$_6$ | 15.14 B + 2 La$_2$C$_3$ + 0.5714 B$_5$H$_7$ + 0.5 N$_2$ ⟶ 3 LaB$_4$ + LaB$_6$H$_4$C$_6$N | -2.9091 |
| LaNH$_4$B$_6$C$_6$ | 8.583 B + 3.583 H$_3$N + 2 La$_2$C$_3$ ⟶ 2.583 BN + 0.75 La$_4$H$_9$ + LaB$_6$H$_4$C$_6$N | -2.8718 |
| LaNH$_4$B$_6$C$_6$ | 25 B + BN + 2 La$_2$C$_3$ + 2 LaH$_2$ ⟶ 5 LaB$_4$ + LaB$_6$H$_4$C$_6$N | -2.8696 |
| LaNH$_4$B$_6$C$_6$ | BN + 3 B$_{13}$C$_2$ + 16 La$_2$H$_5$ ⟶ 25.33 LaH$_3$ + 5.667 LaB$_6$ + LaB$_6$H$_4$C$_6$N | -2.8589 |
| LaNH$_4$B$_6$C$_6$ | 6.5 LaH$_2$ + 0.5 N$_2$ + 3 B$_{13}$C$_2$ ⟶ 4.5 H$_2$ + 5.5 LaB$_6$ + LaB$_6$H$_4$C$_6$N | -2.8578 |
| LaNH$_4$B$_6$C$_6$ | 2.4 LaN + 4.1 LaH$_2$ + 3 B$_{13}$C$_2$ ⟶ 5.5 LaB$_6$ + 1.4 H$_3$N + LaB$_6$H$_4$C$_6$N | -2.8490 |
| LaNH$_4$B$_6$C$_6$ | 14.73 B + 0.3636 B$_9$H$_{11}$ + 2 La$_2$C$_3$ + 0.5 N$_2$ ⟶ 3 LaB$_4$ + LaB$_6$H$_4$C$_6$N | -2.8469 |
| LaNH$_4$B$_6$C$_6$ | 3 LaC$_2$ + 22 B + 2 LaH$_2$ + 0.5 N$_2$ ⟶ 4 LaB$_4$ + LaB$_6$H$_4$C$_6$N | -2.8189 |
| LaNH$_4$B$_6$C$_6$ | LaB$_6$ + 0.4444 La$_4$H$_9$ + 2 La$_2$C$_3$ + 3.389 N$_2$ ⟶ 5.778 LaN + LaB$_6$H$_4$C$_6$N | -2.8175 |
| LaNH$_4$B$_6$C$_6$ | 3 LaC$_2$ + 21 B + H$_3$N + 0.5 LaH$_2$ ⟶ 2.5 LaB$_6$ + LaB$_6$H$_4$C$_6$N | -2.7738 |
| LaNH$_4$B$_6$C$_6$ | 1.2 H$_8$C$_5$ + 6.467 La + LaN + 0.6667 B$_9$H$_{11}$ ⟶ 6.467 LaH$_2$ + LaB$_6$H$_4$C$_6$N | -2.7568 |
| LaNH$_4$B$_6$C$_6$ | 6.917 La$_4$H$_9$ + LaN + 3 B$_{13}$C$_2$ ⟶ 19.42 LaH$_3$ + 8.25 LaB$_4$ + LaB$_6$H$_4$C$_6$N | -2.7519 |
| LaNH$_4$B$_6$C$_6$ | 3 LaC$_2$ + 1.333 LaH$_3$ + 25 B + BN ⟶ 3.333 LaB$_6$ + LaB$_6$H$_4$C$_6$N | -2.7313 |
| LaNH$_4$B$_6$C$_6$ | 1.2 H$_8$C$_5$ + 7.3 La + BN + B$_5$H$_7$ ⟶ 6.3 LaH$_2$ + LaB$_6$H$_4$C$_6$N | -2.6978 |
| LaNH$_4$B$_6$C$_6$ | 20 B + H$_3$N + 2 La$_2$C$_3$ + 0.5 LaH$_2$ ⟶ 3.5 LaB$_4$ + LaB$_6$H$_4$C$_6$N | -2.6648 |
| LaNH$_4$B$_6$C$_6$ | 2 H$_2$ + 2 La$_2$C$_3$ + 2.75 N$_2$ + 1.5 LaB$_4$ ⟶ 4.5 LaN + LaB$_6$H$_4$C$_6$N | -2.6558 |
| LaNH$_4$B$_6$C$_6$ | H$_3$N + 3 B$_{13}$C$_2$ + 26.75 La$_2$H$_5$ ⟶ 44.25 LaH$_3$ + 8.25 LaB$_4$ + LaB$_6$H$_4$C$_6$N | -2.6387 |
| LaNH$_4$B$_6$C$_6$ | 1.692 La$_2$C$_3$ + 2 LaH$_2$ + 2.692 N$_2$ + 0.4615 B$_{13}$C$_2$ ⟶ 4.385 LaN + LaB$_6$H$_4$C$_6$N | -2.6201 |
| LaNH$_4$B$_6$C$_6$ | 3 LaC$_2$ + 20.67 B + 0.1111 La$_4$H$_9$ + H$_3$N ⟶ 2.444 LaB$_6$ + LaB$_6$H$_4$C$_6$N | -2.6041 |
| LaNH$_4$B$_6$C$_6$ | 18.33 B + 1.333 H$_3$N + 2 La$_2$C$_3$ ⟶ 0.3333 BN + 3 LaB$_4$ + LaB$_6$H$_4$C$_6$N | -2.5964 |
| LaNH$_4$B$_6$C$_6$ | 3 LaC$_2$ + 25.11 B + 0.4444 La$_4$H$_9$ + LaN ⟶ 4.778 LaB$_4$ + LaB$_6$H$_4$C$_6$N | -2.5715 |
| LaNH$_4$B$_6$C$_6$ | 19.78 B + 0.1111 La$_4$H$_9$ + H$_3$N + 2 La$_2$C$_3$ ⟶ 3.444 LaB$_4$ + LaB$_6$H$_4$C$_6$N | -2.5557 |
| LaNH$_4$B$_6$C$_6$ | LaB$_6$ + 2 La$_2$C$_3$ + 3.3 N$_2$ + 0.8 La$_2$H$_5$ ⟶ 5.6 LaN + LaB$_6$H$_4$C$_6$N | -2.5255 |
| LaNH$_4$B$_6$C$_6$ | 19.6 B + H$_3$N + 2 La$_2$C$_3$ + 0.2 La$_2$H$_5$ ⟶ 3.4 LaB$_4$ + LaB$_6$H$_4$C$_6$N | -2.4743 |
| LaNH$_4$B$_6$C$_6$ | 3 LaC$_2$ + 20.4 B + H$_3$N + 0.2 La$_2$H$_5$ ⟶ 2.4 LaB$_6$ + LaB$_6$H$_4$C$_6$N | -2.4742 |
| LaNH$_4$B$_6$C$_6$ | 3 LaC$_2$ + 15.14 B + 0.5714 B$_5$H$_7$ + 0.5 N$_2$ ⟶ 2 LaB$_6$ + LaB$_6$H$_4$C$_6$N | -2.4728 |
| LaNH$_4$B$_6$C$_6$ | 0.9762 La$_4$H$_9$ + 2.595 LaN + 3 B$_{13}$C$_2$ ⟶ 5.5 LaB$_6$ + 1.595 H$_3$N + LaB$_6$H$_4$C$_6$N | -2.4685 |
| LaNH$_4$B$_6$C$_6$ | 24.11 B + BN + 0.4444 La$_4$H$_9$ + 2 La$_2$C$_3$ ⟶ 4.778 LaB$_4$ + LaB$_6$H$_4$C$_6$N | -2.4333 |
| LaNH$_4$B$_6$C$_6$ | 3 LaC$_2$ + 14.73 B + 0.3636 B$_9$H$_{11}$ + 0.5 N$_2$ ⟶ 2 LaB$_6$ + LaB$_6$H$_4$C$_6$N | -2.4106 |
| LaNH$_4$B$_6$C$_6$ | 1.2 H$_8$C$_5$ + 5.6 La + LaN + 1.2 B$_5$H$_7$ ⟶ 2.8 La$_2$H$_5$ + LaB$_6$H$_4$C$_6$N | -2.3971 |
| LaNH$_4$B$_6$C$_6$ | 3 LaC$_2$ + 21.11 B + 0.4444 La$_4$H$_9$ + 0.5 N$_2$ ⟶ 3.778 LaB$_4$ + LaB$_6$H$_4$C$_6$N | -2.3825 |
| LaNH$_4$B$_6$C$_6$ | 6 C + 6.176 LaN + 2.235 B$_5$H$_7$ ⟶ 5.176 BN + 1.294 La$_4$H$_9$ + LaB$_6$H$_4$C$_6$N | -2.3822 |
| LaNH$_4$B$_6$C$_6$ | 0.3333 LaH$_3$ + 19.33 B + H$_3$N + 2 La$_2$C$_3$ ⟶ 3.333 LaB$_4$ + LaB$_6$H$_4$C$_6$N | -2.3757 |
| LaNH$_4$B$_6$C$_6$ | 3.143 B + 2 La$_2$C$_3$ + 0.5714 B$_5$H$_7$ + 2 N$_2$ ⟶ 3 LaN + LaB$_6$H$_4$C$_6$N | -2.3422 |

| Compound | Chemical reaction | Reaction energy(eV) |
|---|---|---|
| LaNH$_4$B$_6$C$_6$ | 0.5 H$_8$C$_5$ + 0.5 N$_2$ + 1.75 B$_{13}$C$_2$ + 11.38 LaB$_4$ ⟶ 10.38 LaB$_6$ + LaB$_6$H$_4$C$_6$N | -2.3287 |
| LaNH$_4$B$_6$C$_6$ | 4.042 LaC$_2$ + 0.6667 B$_9$H$_{11}$ + 2.021 N$_2$ ⟶ 0.4167 H$_8$C$_5$ + 3.042 LaN + LaB$_6$H$_4$C$_6$N | -2.3263 |
| LaNH$_4$B$_6$C$_6$ | 3 LaC$_2$ + 0.3333 LaH$_3$ + 20 B + H$_3$N ⟶ 2.333 LaB$_6$ + LaB$_6$H$_4$C$_6$N | -2.3029 |
| LaNH$_4$B$_6$C$_6$ | 2.727 B + 0.3636 B$_9$H$_{11}$ + 2 La$_2$C$_3$ + 2 N$_2$ ⟶ 3 LaN + LaB$_6$H$_4$C$_6$N | -2.2799 |
| LaNH$_4$B$_6$C$_6$ | 3 LaC$_2$ + 24.4 B + LaN + 0.8 La$_2$H$_5$ ⟶ 4.6 LaB$_4$ + LaB$_6$H$_4$C$_6$N | -2.2459 |
| LaNH$_4$B$_6$C$_6$ | 0.4444 La$_4$H$_9$ + 1.692 La$_2$C$_3$ + 2.581 N$_2$ + 0.4615 B$_{13}$C$_2$ ⟶ 4.162 LaN + LaB$_6$H$_4$C$_6$N | -2.2257 |
| LaNH$_4$B$_6$C$_6$ | 2 La$_2$C$_3$ + 0.5714 B$_5$H$_7$ + 2.393 N$_2$ + 0.7857 LaB$_4$ ⟶ 3.786 LaN + LaB$_6$H$_4$C$_6$N | -2.1937 |
| LaNH$_4$B$_6$C$_6$ | 0.5 H$_8$C$_5$ + 2.792 La + LaN + 1.75 B$_{13}$C$_2$ ⟶ 2.792 LaB$_6$ + LaB$_6$H$_4$C$_6$N | -2.1931 |
| LaNH$_4$B$_6$C$_6$ | 8.333 B + 3.333 H$_3$N + 2 La$_2$C$_3$ ⟶ 2.333 BN + 3 LaH$_2$ + LaB$_6$H$_4$C$_6$N | -2.1867 |
| LaNH$_4$B$_6$C$_6$ | 0.5 H$_2$ + 18 B + H$_3$N + 2 La$_2$C$_3$ ⟶ 3 LaB$_4$ + LaB$_6$H$_4$C$_6$N | -2.1822 |
| LaNH$_4$B$_6$C$_6$ | 1.333 LaH$_3$ + LaB$_6$ + 2 La$_2$C$_3$ + 3.167 N$_2$ ⟶ 5.333 LaN + LaB$_6$H$_4$C$_6$N | -2.1816 |
| LaNH$_4$B$_6$C$_6$ | 2.773 LaN + 3 B$_{13}$C$_2$ + 1.864 La$_2$H$_5$ ⟶ 5.5 LaB$_6$ + 1.773 H$_3$N + LaB$_6$H$_4$C$_6$N | -2.1774 |
| LaNH$_4$B$_6$C$_6$ | 1.71 LaB$_6$ + 5.258 H$_3$N + 2 La$_2$C$_3$ ⟶ 4.258 BN + 2.355 La$_2$H$_5$ + LaB$_6$H$_4$C$_6$N | -2.1706 |
| LaNH$_4$B$_6$C$_6$ | 5.635 LaH$_2$ + 0.5 N$_2$ + 3 B$_{13}$C$_2$ ⟶ 4.635 LaB$_6$ + 1.038 B$_5$H$_7$ + LaB$_6$H$_4$C$_6$N | -2.1641 |
| LaNH$_4$B$_6$C$_6$ | 3 LaC$_2$ + 18.33 B + 1.333 H$_3$N ⟶ 2 LaB$_6$ + 0.3333 BN + LaB$_6$H$_4$C$_6$N | -2.1602 |
| LaNH$_4$B$_6$C$_6$ | 0.3636 B$_9$H$_{11}$ + 2 La$_2$C$_3$ + 2.341 N$_2$ + 0.6818 LaB$_4$ ⟶ 3.682 LaN + LaB$_6$H$_4$C$_6$N | -2.1510 |
| LaNH$_4$B$_6$C$_6$ | 1.2 H$_8$C$_5$ + 5.748 La + LaN + 0.6667 B$_9$H$_{11}$ ⟶ 1.437 La$_4$H$_9$ + LaB$_6$H$_4$C$_6$N | -2.1227 |
| LaNH$_4$B$_6$C$_6$ | 23.4 B + BN + 2 La$_2$C$_3$ + 0.8 La$_2$H$_5$ ⟶ 4.6 LaB$_4$ + LaB$_6$H$_4$C$_6$N | -2.1077 |
| LaNH$_4$B$_6$C$_6$ | LaN + 5.5 LaH$_2$ + 3 B$_{13}$C$_2$ ⟶ 3.5 H$_2$ + 5.5 LaB$_6$ + LaB$_6$H$_4$C$_6$N | -2.0818 |
| LaNH$_4$B$_6$C$_6$ | 1.2 H$_8$C$_5$ + 6.6 La + BN + B$_5$H$_7$ ⟶ 1.4 La$_4$H$_9$ + LaB$_6$H$_4$C$_6$N | -2.0800 |
| LaNH$_4$B$_6$C$_6$ | 3 LaC$_2$ + 6 B + 2 LaH$_2$ + 2.5 N$_2$ ⟶ 4 LaN + LaB$_6$H$_4$C$_6$N | -2.0629 |
| LaNH$_4$B$_6$C$_6$ | 3 LaC$_2$ + 20.4 B + 0.5 N$_2$ + 0.8 La$_2$H$_5$ ⟶ 3.6 LaB$_4$ + LaB$_6$H$_4$C$_6$N | -2.0569 |
| LaNH$_4$B$_6$C$_6$ | 2 La$_2$C$_3$ + 1.429 B$_5$H$_7$ + 1.071 N$_2$ ⟶ 1.143 BN + 3 LaH$_2$ + LaB$_6$H$_4$C$_6$N | -2.0516 |
| LaNH$_4$B$_6$C$_6$ | 1.625 La$_4$H$_9$ + 0.5 N$_2$ + 3 B$_{13}$C$_2$ ⟶ 5.312 H$_2$ + 5.5 LaB$_6$ + LaB$_6$H$_4$C$_6$N | -2.0464 |
| LaNH$_4$B$_6$C$_6$ | 5.536 LaH$_2$ + 0.5 N$_2$ + 3 B$_{13}$C$_2$ ⟶ 4.536 LaB$_6$ + 0.6429 B$_9$H$_{11}$ + LaB$_6$H$_4$C$_6$N | -2.0414 |
| LaNH$_4$B$_6$C$_6$ | 17.29 B + H$_3$N + 2 La$_2$C$_3$ + 0.1429 B$_5$H$_7$ ⟶ 3 LaB$_4$ + LaB$_6$H$_4$C$_6$N | -2.0329 |
| LaNH$_4$B$_6$C$_6$ | BN + 8.167 La$_4$H$_9$ + 3 B$_{13}$C$_2$ ⟶ 23.17 LaH$_3$ + 8.5 LaB$_4$ + LaB$_6$H$_4$C$_6$N | -2.0207 |
| LaNH$_4$B$_6$C$_6$ | 17.18 B + H$_3$N + 0.09091 B$_9$H$_{11}$ + 2 La$_2$C$_3$ ⟶ 3 LaB$_4$ + LaB$_6$H$_4$C$_6$N | -2.0174 |
| LaNH$_4$B$_6$C$_6$ | LaN + 3.333 B$_9$H$_{11}$ + 2 La$_2$C$_3$ ⟶ 16.33 H$_2$ + 4 LaB$_6$ + LaB$_6$H$_4$C$_6$N | -2.0130 |
| LaNH$_4$B$_6$C$_6$ | 2.667 B$_9$H$_{11}$ + 2 La$_2$C$_3$ + 0.5 N$_2$ ⟶ 12.67 H$_2$ + 3 LaB$_6$ + LaB$_6$H$_4$C$_6$N | -1.9423 |
| LaNH$_4$B$_6$C$_6$ | 1.692 La$_2$C$_3$ + 2.492 N$_2$ + 0.4615 B$_{13}$C$_2$ + 0.8 La$_2$H$_5$ ⟶ 3.985 LaN + LaB$_6$H$_4$C$_6$N | -1.9337 |
| LaNH$_4$B$_6$C$_6$ | 3.417 LaH$_3$ + 3.083 LaN + 3 B$_{13}$C$_2$ ⟶ 5.5 LaB$_6$ + 2.083 H$_3$N + LaB$_6$H$_4$C$_6$N | -1.9093 |
| LaNH$_4$B$_6$C$_6$ | 1.2 H$_8$C$_5$ + 6.856 La + BN + 0.5556 B$_9$H$_{11}$ ⟶ 5.856 LaH$_2$ + LaB$_6$H$_4$C$_6$N | -1.8992 |
| LaNH$_4$B$_6$C$_6$ | 2.75 B$_9$H$_{11}$ + 2 La$_2$C$_3$ + 2 N$_2$ ⟶ 3 LaN + 3.75 B$_5$H$_7$ + LaB$_6$H$_4$C$_6$N | -1.8712 |
| LaNH$_4$B$_6$C$_6$ | 3 LaC$_2$ + 1.333 LaH$_3$ + 23.33 B + LaN ⟶ 4.333 LaB$_4$ + LaB$_6$H$_4$C$_6$N | -1.8516 |
| LaNH$_4$B$_6$C$_6$ | 5 H$_3$N + 2 La$_2$C$_3$ + 2.5 LaB$_4$ ⟶ 4 BN + 5.5 LaH$_2$ + LaB$_6$H$_4$C$_6$N | -1.8451 |
| LaNH$_4$B$_6$C$_6$ | 6 C + 4.7 La + H$_3$N + 1.2 B$_5$H$_7$ ⟶ 3.7 LaH$_2$ + LaB$_6$H$_4$C$_6$N | -1.8424 |
| LaNH$_4$B$_6$C$_6$ | H$_3$N + 20.38 B$_9$H$_{11}$ + 2 La$_2$C$_3$ ⟶ 3 LaB$_6$ + 31.88 B$_5$H$_7$ + LaB$_6$H$_4$C$_6$N | -1.8158 |
| LaNH$_4$B$_6$C$_6$ | LaN + 2 La$_2$C$_3$ + 6 B$_5$H$_7$ ⟶ 19 H$_2$ + 4 LaB$_6$ + LaB$_6$H$_4$C$_6$N | -1.7876 |
| LaNH$_4$B$_6$C$_6$ | 3 LaC$_2$ + 2 LaH$_2$ + 3.25 N$_2$ + 1.5 LaB$_4$ ⟶ 5.5 LaN + LaB$_6$H$_4$C$_6$N | -1.7794 |
| LaNH$_4$B$_6$C$_6$ | LaN + 3 B$_{13}$C$_2$ + 20.75 La$_2$H$_5$ ⟶ 33.25 LaH$_3$ + 8.25 LaB$_4$ + LaB$_6$H$_4$C$_6$N | -1.7753 |
| LaNH$_4$B$_6$C$_6$ | 6.5 H$_2$ + 6 B + 2 La$_2$C$_3$ + 0.5 N$_2$ ⟶ 3 LaH$_3$ + LaB$_6$H$_4$C$_6$N | -1.7650 |
| LaNH$_4$B$_6$C$_6$ | 2 La$_2$C$_3$ + 4.8 B$_5$H$_7$ + 0.5 N$_2$ ⟶ 14.8 H$_2$ + 3 LaB$_6$ + LaB$_6$H$_4$C$_6$N | -1.7620 |
| LaNH$_4$B$_6$C$_6$ | 3 LaC$_2$ + 0.5 H$_2$ + 18 B + H$_3$N ⟶ 2 LaB$_6$ + LaB$_6$H$_4$C$_6$N | -1.7459 |
| LaNH$_4$B$_6$C$_6$ | 18 B + 1.333 H$_3$N + 2 La$_2$C$_3$ ⟶ 0.1667 N$_2$ + 3 LaB$_4$ + LaB$_6$H$_4$C$_6$N | -1.7409 |
| LaNH$_4$B$_6$C$_6$ | 0.6667 B$_9$H$_{11}$ + 2 La$_2$C$_3$ + 2 N$_2$ ⟶ 1.667 H$_2$ + 3 LaN + LaB$_6$H$_4$C$_6$N | -1.7304 |
| LaNH$_4$B$_6$C$_6$ | 1.333 LaH$_3$ + 22.33 B + BN + 2 La$_2$C$_3$ ⟶ 4.333 LaB$_4$ + LaB$_6$H$_4$C$_6$N | -1.7134 |
| LaNH$_4$B$_6$C$_6$ | 0.4545 LaB$_6$ + 0.3636 B$_9$H$_{11}$ + 2 La$_2$C$_3$ + 2.227 N$_2$ ⟶ 3.455 LaN + LaB$_6$H$_4$C$_6$N | -1.6982 |
| LaNH$_4$B$_6$C$_6$ | 2 La$_2$C$_3$ + 1.2 B$_5$H$_7$ + 2 N$_2$ ⟶ 2.2 H$_2$ + 3 LaN + LaB$_6$H$_4$C$_6$N | -1.6853 |
| LaNH$_4$B$_6$C$_6$ | 16.67 B + 1.333 H$_3$N + 2 La$_2$C$_3$ ⟶ 0.3333 LaN + 2.667 LaB$_4$ + LaB$_6$H$_4$C$_6$N | -1.6779 |
| LaNH$_4$B$_6$C$_6$ | 0.5238 LaB$_6$ + 2 La$_2$C$_3$ + 0.5714 B$_5$H$_7$ + 2.262 N$_2$ ⟶ 3.524 LaN + LaB$_6$H$_4$C$_6$N | -1.6719 |
| LaNH$_4$B$_6$C$_6$ | 3 LaC$_2$ + 6 B + 0.4444 La$_4$H$_9$ + 2.389 N$_2$ ⟶ 3.778 LaN + LaB$_6$H$_4$C$_6$N | -1.6685 |
| LaNH$_4$B$_6$C$_6$ | 3 LaC$_2$ + 1.333 LaH$_3$ + 19.33 B + 0.5 N$_2$ ⟶ 3.333 LaB$_4$ + LaB$_6$H$_4$C$_6$N | -1.6626 |
| LaNH$_4$B$_6$C$_6$ | 2 H$_2$ + LaB$_6$ + 2 La$_2$C$_3$ + 2.5 N$_2$ ⟶ 4 LaN + LaB$_6$H$_4$C$_6$N | -1.6597 |
| LaNH$_4$B$_6$C$_6$ | 3 LaC$_2$ + 17.29 B + H$_3$N + 0.1429 B$_5$H$_7$ ⟶ 2 LaB$_6$ + LaB$_6$H$_4$C$_6$N | -1.5966 |
| LaNH$_4$B$_6$C$_6$ | 1.333 LaH$_3$ + 1.692 La$_2$C$_3$ + 2.359 N$_2$ + 0.4615 B$_{13}$C$_2$ ⟶ 3.718 LaN + LaB$_6$H$_4$C$_6$N | -1.5899 |
| LaNH$_4$B$_6$C$_6$ | 3 LaC$_2$ + 17.18 B + H$_3$N + 0.09091 B$_9$H$_{11}$ ⟶ 2 LaB$_6$ + LaB$_6$H$_4$C$_6$N | -1.5811 |
| LaNH$_4$B$_6$C$_6$ | LaN + 4.827 LaH$_2$ + 3 B$_{13}$C$_2$ ⟶ 4.827 LaB$_6$ + 0.8077 B$_5$H$_7$ + LaB$_6$H$_4$C$_6$N | -1.5422 |
| LaNH$_4$B$_6$C$_6$ | 1.2 H$_8$C$_5$ + 5.173 La + LaN + 0.6667 B$_9$H$_{11}$ ⟶ 2.587 La$_2$H$_5$ + LaB$_6$H$_4$C$_6$N | -1.5393 |
| LaNH$_4$B$_6$C$_6$ | H$_3$N + 6.5 LaH$_2$ + 3 B$_{13}$C$_2$ ⟶ 6 H$_2$ + 5.5 LaB$_6$ + LaB$_6$H$_4$C$_6$N | -1.5338 |

| Compound | Chemical reaction | Reaction energy(eV) |
|---|---|---|
| $LaNH_4B_6C_6$ | $1.2\,H_8C_5 + 6.04\,La + BN + B_5H_7 \longrightarrow 2.52\,La_2H_5 + LaB_6H_4C_6N$ | -1.5116 |
| $LaNH_4B_6C_6$ | $6\,C + 4.289\,La + H_3N + 1.2\,B_5H_7 \longrightarrow 0.8222\,La_4H_9 + LaB_6H_4C_6N$ | -1.4796 |
| $LaNH_4B_6C_6$ | $LaN + 4.75\,LaH_2 + 3\,B_{13}C_2 \longrightarrow 4.75\,LaB_6 + 0.5\,B_9H_{11} + LaB_6H_4C_6N$ | -1.4467 |
| $LaNH_4B_6C_6$ | $3\,LaC_2 + 0.8182\,B_9H_{11} + 1.182\,N_2 \longrightarrow 1.364\,BN + La_2H_5 + LaB_6H_4C_6N$ | -1.4450 |
| $LaNH_4B_6C_6$ | $0.3636\,B_9H_{11} + 1.86\,La_2C_3 + 1.86\,N_2 + 0.2098\,B_{13}C_2 \longrightarrow 2.72\,LaN + LaB_6H_4C_6N$ | -1.4293 |
| $LaNH_4B_6C_6$ | $1.375\,La_4H_9 + LaN + 3\,B_{13}C_2 \longrightarrow 4.188\,H_2 + 5.5\,LaB_6 + LaB_6H_4C_6N$ | -1.3952 |
| $LaNH_4B_6C_6$ | $3\,LaC_2 + 0.4444\,La_4H_9 + 3.139\,N_2 + 1.5\,LaB_4 \longrightarrow 5.278\,LaN + LaB_6H_4C_6N$ | -1.3850 |
| $LaNH_4B_6C_6$ | $3\,LaC_2 + 6\,B + 2.3\,N_2 + 0.8\,La_2H_5 \longrightarrow 3.6\,LaN + LaB_6H_4C_6N$ | -1.3765 |
| $LaNH_4B_6C_6$ | $8\,H_2 + 6\,B + LaN + 2\,La_2C_3 \longrightarrow 4\,LaH_3 + LaB_6H_4C_6N$ | -1.3736 |
| $LaNH_4B_6C_6$ | $1.376\,La_4H_9 + 0.5\,N_2 + 3\,B_{13}C_2 \longrightarrow 4.502\,LaB_6 + 1.197\,B_5H_7 + LaB_6H_4C_6N$ | -1.3712 |
| $LaNH_4B_6C_6$ | $1.839\,La_2C_3 + 0.5714\,B_5H_7 + 1.839\,N_2 + 0.2418\,B_{13}C_2 \longrightarrow 2.678\,LaN + LaB_6H_4C_6N$ | -1.3619 |
| $LaNH_4B_6C_6$ | $3\,LaC_2 + 5.105\,H_3N + 2.526\,LaB_4 \longrightarrow 4.105\,BN + 2.263\,La_2H_5 + LaB_6H_4C_6N$ | -1.3616 |
| $LaNH_4B_6C_6$ | $0.5\,N_2 + 3\,B_{13}C_2 + 3.25\,La_2H_5 \longrightarrow 6.125\,H_2 + 5.5\,LaB_6 + LaB_6H_4C_6N$ | -1.3305 |
| $LaNH_4B_6C_6$ | $1.2\,H_8C_5 + 6.205\,La + BN + 0.5556\,B_9H_{11} \longrightarrow 1.301\,La_4H_9 + LaB_6H_4C_6N$ | -1.3250 |
| $LaNH_4B_6C_6$ | $5.75\,H_2 + 6\,B + 2\,La_2C_3 + 0.5\,N_2 \longrightarrow 1.5\,La_2H_5 + LaB_6H_4C_6N$ | -1.3160 |
| $LaNH_4B_6C_6$ | $3\,LaC_2 + 18\,B + 1.333\,H_3N \longrightarrow 2\,LaB_6 + 0.1667\,N_2 + LaB_6H_4C_6N$ | -1.3046 |
| $LaNH_4B_6C_6$ | $3\,LaC_2 + 8.333\,B + 3.333\,H_3N \longrightarrow 2\,LaH_3 + 2.333\,BN + LaB_6H_4C_6N$ | -1.3034 |
| $LaNH_4B_6C_6$ | $0.6667\,B_9H_{11} + 2\,La_2C_3 + 1.444\,N_2 \longrightarrow 1.111\,LaH_3 + 1.889\,LaN + LaB_6H_4C_6N$ | -1.2954 |
| $LaNH_4B_6C_6$ | $0.5\,H_8C_5 + 20\,B + LaN + 1.167\,La_2C_3 \longrightarrow 2.333\,LaB_6 + LaB_6H_4C_6N$ | -1.2838 |
| $LaNH_4B_6C_6$ | $6\,C + LaB_6 + 0.3636\,B_9H_{11} + 2.136\,N_2 \longrightarrow 3.273\,BN + LaB_6H_4C_6N$ | -1.2637 |
| $LaNH_4B_6C_6$ | $1.348\,La_4H_9 + 0.5\,N_2 + 3\,B_{13}C_2 \longrightarrow 4.391\,LaB_6 + 0.7391\,B_9H_{11} + LaB_6H_4C_6N$ | -1.2461 |
| $LaNH_4B_6C_6$ | $1.635\,LaB_6 + 4.81\,H_3N + 2\,La_2C_3 \longrightarrow 3.81\,BN + 1.159\,La_4H_9 + LaB_6H_4C_6N$ | -1.2388 |
| $LaNH_4B_6C_6$ | $1.2\,H_8C_5 + 6\,B + 5.3\,La + H_3N \longrightarrow 4.3\,LaH_2 + LaB_6H_4C_6N$ | -1.2229 |
| $LaNH_4B_6C_6$ | $3.465\,H_3N + 1.566\,La_2C_3 + 0.6512\,B_{13}C_2 \longrightarrow 2.132\,LaH_3 + 2.465\,BN + LaB_6H_4C_6N$ | -1.2040 |
| $LaNH_4B_6C_6$ | $3\,LaC_2 + 1.2\,B_5H_7 + 2.233\,N_2 \longrightarrow 1.467\,H_3N + 2\,LaN + LaB_6H_4C_6N$ | -1.1987 |
| $LaNH_4B_6C_6$ | $6\,B + 1.333\,H_3N + 2\,La_2C_3 + 1.333\,N_2 \longrightarrow 3\,LaN + LaB_6H_4C_6N$ | -1.1739 |
| $LaNH_4B_6C_6$ | $3\,LaC_2 + 1.429\,B_5H_7 + 1.071\,N_2 \longrightarrow 2\,LaH_3 + 1.143\,BN + LaB_6H_4C_6N$ | -1.1684 |
| $LaNH_4B_6C_6$ | $6\,C + 3.96\,La + H_3N + 1.2\,B_5H_7 \longrightarrow 1.48\,La_2H_5 + LaB_6H_4C_6N$ | -1.1458 |
| $LaNH_4B_6C_6$ | $1.2\,H_8C_5 + 4.667\,La + LaN + 1.2\,B_5H_7 \longrightarrow 4.667\,LaH_3 + LaB_6H_4C_6N$ | -1.1204 |
| $LaNH_4B_6C_6$ | $2\,La_2C_3 + 1.2\,B_5H_7 + 1.267\,N_2 \longrightarrow 1.467\,LaH_3 + 1.533\,LaN + LaB_6H_4C_6N$ | -1.1112 |
| $LaNH_4B_6C_6$ | $3\,LaC_2 + 3.05\,N_2 + 0.8\,La_2H_5 + 1.5\,LaB_4 \longrightarrow 5.1\,LaN + LaB_6H_4C_6N$ | -1.0930 |
| $LaNH_4B_6C_6$ | $3\,LaC_2 + 2\,H_2 + 18\,B + LaN \longrightarrow 3\,LaB_4 + LaB_6H_4C_6N$ | -1.0777 |
| $LaNH_4B_6C_6$ | $2\,H_2 + 1.692\,La_2C_3 + 1.692\,N_2 + 0.4615\,B_{13}C_2 \longrightarrow 2.385\,LaN + LaB_6H_4C_6N$ | -1.0680 |
| $LaNH_4B_6C_6$ | $3\,LaC_2 + 1.333\,LaH_3 + 6\,B + 2.167\,N_2 \longrightarrow 3.333\,LaN + LaB_6H_4C_6N$ | -1.0327 |
| $LaNH_4B_6C_6$ | $0.8571\,La_2C_3 + 2\,LaH_2 + 0.5\,N_2 + 1.714\,B_{13}C_2 \longrightarrow 2.714\,LaB_6 + LaB_6H_4C_6N$ | -1.0139 |
| $LaNH_4B_6C_6$ | $0.6667\,B_9H_{11} + 2\,La_2C_3 + 1.333\,N_2 \longrightarrow 1.667\,LaN + 0.6667\,La_2H_5 + LaB_6H_4C_6N$ | -1.0089 |
| $LaNH_4B_6C_6$ | $1.355\,B_9H_{11} + 2.318\,La_2C_3 + 0.5\,N_2 \longrightarrow 3.636\,LaH_3 + 0.4766\,B_{13}C_2 + LaB_6H_4C_6N$ | -0.9914 |
| $LaNH_4B_6C_6$ | $5.375\,H_2 + 6\,B + 2\,La_2C_3 + 0.5\,N_2 \longrightarrow 0.75\,La_4H_9 + LaB_6H_4C_6N$ | -0.9856 |
| $LaNH_4B_6C_6$ | $0.125\,H_8C_5 + 16.33\,B + H_3N + 1.792\,La_2C_3 \longrightarrow 2.583\,LaB_4 + LaB_6H_4C_6N$ | -0.9432 |
| $LaNH_4B_6C_6$ | $2\,H_2 + 17\,B + BN + 2\,La_2C_3 \longrightarrow 3\,LaB_4 + LaB_6H_4C_6N$ | -0.9395 |
| $LaNH_4B_6C_6$ | $8.75\,H_2 + 2\,La_2C_3 + 0.5\,N_2 + 1.5\,LaB_4 \longrightarrow 4.5\,LaH_3 + LaB_6H_4C_6N$ | -0.8944 |
| $LaNH_4B_6C_6$ | $1.333\,H_3N + 2\,La_2C_3 + 2.083\,N_2 + 1.5\,LaB_4 \longrightarrow 4.5\,LaN + LaB_6H_4C_6N$ | -0.8904 |
| $LaNH_4B_6C_6$ | $3\,LaC_2 + 2\,H_2 + 14\,B + 0.5\,N_2 \longrightarrow 2\,LaB_4 + LaB_6H_4C_6N$ | -0.8888 |
| $LaNH_4B_6C_6$ | $6\,C + 4.167\,La + H_3N + 0.6667\,B_9H_{11} \longrightarrow 3.167\,LaH_2 + LaB_6H_4C_6N$ | -0.8841 |
| $LaNH_4B_6C_6$ | $3\,LaC_2 + 16\,B + 1.333\,H_3N \longrightarrow 1.667\,LaB_6 + 0.3333\,LaN + LaB_6H_4C_6N$ | -0.8780 |
| $LaNH_4B_6C_6$ | $BN + 3\,B_{13}C_2 + 24.5\,La_2H_5 \longrightarrow 39.5\,LaH_3 + 8.5\,LaB_4 + LaB_6H_4C_6N$ | -0.8676 |
| $LaNH_4B_6C_6$ | $1.178\,La_4H_9 + LaN + 3\,B_{13}C_2 \longrightarrow 4.714\,LaB_6 + 0.9437\,B_5H_7 + LaB_6H_4C_6N$ | -0.8629 |
| $LaNH_4B_6C_6$ | $LaN + 1.72\,B_9H_{11} + 2.486\,La_2C_3 \longrightarrow 4.972\,LaH_3 + 0.729\,B_{13}C_2 + LaB_6H_4C_6N$ | -0.8304 |
| $LaNH_4B_6C_6$ | $6\,C + 5\,LaN + 2\,B_5H_7 \longrightarrow 4\,BN + 2\,La_2H_5 + LaB_6H_4C_6N$ | -0.8150 |
| $LaNH_4B_6C_6$ | $1.032\,B_9H_{11} + 2\,La_2C_3 + 0.5\,N_2 \longrightarrow 2.452\,LaH_3 + 0.5484\,LaB_6 + LaB_6H_4C_6N$ | -0.8095 |
| $LaNH_4B_6C_6$ | $1.2\,H_8C_5 + 6\,B + 4.822\,La + H_3N \longrightarrow 0.9556\,La_4H_9 + LaB_6H_4C_6N$ | -0.8013 |
| $LaNH_4B_6C_6$ | $1.2\,H_8C_5 + 5.684\,La + BN + 0.5556\,B_9H_{11} \longrightarrow 2.342\,La_2H_5 + LaB_6H_4C_6N$ | -0.7967 |
| $LaNH_4B_6C_6$ | $LaN + 3\,B_{13}C_2 + 2.75\,La_2H_5 \longrightarrow 4.875\,H_2 + 5.5\,LaB_6 + LaB_6H_4C_6N$ | -0.7894 |
| $LaNH_4B_6C_6$ | $3\,LaC_2 + LaB_6 + 2\,LaH_2 + 3\,N_2 \longrightarrow 5\,LaN + LaB_6H_4C_6N$ | -0.7833 |
| $LaNH_4B_6C_6$ | $2.216\,La_2C_3 + 2.042\,B_5H_7 + 0.5\,N_2 \longrightarrow 3.432\,LaH_3 + 0.3239\,B_{13}C_2 + LaB_6H_4C_6N$ | -0.7752 |
| $LaNH_4B_6C_6$ | $7\,H_2 + 6\,B + LaN + 2\,La_2C_3 \longrightarrow 2\,La_2H_5 + LaB_6H_4C_6N$ | -0.7749 |
| $LaNH_4B_6C_6$ | $3\,LaC_2 + 0.6667\,B_9H_{11} + 2.056\,N_2 \longrightarrow 1.111\,H_3N + 2\,LaN + LaB_6H_4C_6N$ | -0.7730 |
| $LaNH_4B_6C_6$ | $0.6667\,B_9H_{11} + 2\,La_2C_3 + 1.259\,N_2 \longrightarrow 0.3704\,La_4H_9 + 1.519\,LaN + LaB_6H_4C_6N$ | -0.7656 |
| $LaNH_4B_6C_6$ | $1.157\,La_4H_9 + LaN + 3\,B_{13}C_2 \longrightarrow 4.626\,LaB_6 + 0.5826\,B_9H_{11} + LaB_6H_4C_6N$ | -0.7643 |
| $LaNH_4B_6C_6$ | $3\,LaC_2 + 1.333\,LaH_3 + 2.917\,N_2 + 1.5\,LaB_4 \longrightarrow 4.833\,LaN + LaB_6H_4C_6N$ | -0.7492 |

| Compound | Chemical reaction | Reaction energy(eV) |
|---|---|---|
| $LaNH_4B_6C_6$ | $2\,La_2C_3 + 1.2\,B_5H_7 + 1.12\,N_2 \longrightarrow 1.24\,LaN + 0.88\,La_2H_5 + LaB_6H_4C_6N$ | -0.7329 |
| $LaNH_4B_6C_6$ | $1.625\,La_4H_9 + H_3N + 3\,B_{13}C_2 \longrightarrow 6.813\,H_2 + 5.5\,LaB_6 + LaB_6H_4C_6N$ | -0.7223 |
| $LaNH_4B_6C_6$ | $0.4444\,La_4H_9 + 0.8995\,La_2C_3 + 0.5\,N_2 + 1.651\,B_{13}C_2 \longrightarrow 2.577\,LaB_6 + LaB_6H_4C_6N$ | -0.7010 |
| $LaNH_4B_6C_6$ | $0.5\,N_2 + 3\,B_{13}C_2 + 2.688\,La_2H_5 \longrightarrow 4.376\,LaB_6 + 1.349\,B_5H_7 + LaB_6H_4C_6N$ | -0.6937 |
| $LaNH_4B_6C_6$ | $2\,La_2C_3 + 1.684\,B_5H_7 + 0.5\,N_2 \longrightarrow 2.596\,LaH_3 + 0.4035\,LaB_6 + LaB_6H_4C_6N$ | -0.6793 |
| $LaNH_4B_6C_6$ | $LaN + 0.6667\,La_2C_3 + 2\,LaH_2 + 2\,B_{13}C_2 \longrightarrow 3.333\,LaB_6 + LaB_6H_4C_6N$ | -0.6476 |
| $LaNH_4B_6C_6$ | $13.25\,B_9H_{11} + 2\,La_2C_3 + 0.5\,N_2 \longrightarrow 20.25\,B_5H_7 + 3\,LaB_4 + LaB_6H_4C_6N$ | -0.6402 |
| $LaNH_4B_6C_6$ | $BN + 6.667\,LaH_2 + 3\,B_{13}C_2 \longrightarrow 4.667\,H_2 + 5.667\,LaB_6 + LaB_6H_4C_6N$ | -0.6337 |
| $LaNH_4B_6C_6$ | $H_3N + 2.667\,B_9H_{11} + 2\,La_2C_3 \longrightarrow 14.17\,H_2 + 3\,LaB_6 + LaB_6H_4C_6N$ | -0.6183 |
| $LaNH_4B_6C_6$ | $5\,H_2 + 6\,B + 2\,La_2C_3 + 0.5\,N_2 \longrightarrow 3\,LaH_2 + LaB_6H_4C_6N$ | -0.6111 |
| $LaNH_4B_6C_6$ | $H_3N + 5.346\,LaH_2 + 3\,B_{13}C_2 \longrightarrow 4.346\,LaB_6 + 1.385\,B_5H_7 + LaB_6H_4C_6N$ | -0.6088 |
| $LaNH_4B_6C_6$ | $BN + 17.63\,B_9H_{11} + 2\,La_2C_3 \longrightarrow 3\,LaB_6 + 27.13\,B_5H_7 + LaB_6H_4C_6N$ | -0.5961 |
| $LaNH_4B_6C_6$ | $3\,LaC_2 + 8\,B + 3\,H_3N \longrightarrow 2\,BN + La_2H_5 + LaB_6H_4C_6N$ | -0.5899 |
| $LaNH_4B_6C_6$ | $3\,LaC_2 + 1.667\,LaB_6 + 5\,H_3N \longrightarrow 3.667\,LaH_3 + 4\,BN + LaB_6H_4C_6N$ | -0.5895 |
| $LaNH_4B_6C_6$ | $3\,LaC_2 + LaN + 18.5\,B_9H_{11} \longrightarrow 3\,LaB_6 + 28.5\,B_5H_7 + LaB_6H_4C_6N$ | -0.5845 |
| $LaNH_4B_6C_6$ | $6\,C + 3.815\,La + H_3N + 0.6667\,B_9H_{11} \longrightarrow 0.7037\,La_4H_9 + LaB_6H_4C_6N$ | -0.5736 |
| $LaNH_4B_6C_6$ | $0.5\,N_2 + 3\,B_{13}C_2 + 2.627\,La_2H_5 \longrightarrow 4.254\,LaB_6 + 0.8305\,B_9H_{11} + LaB_6H_4C_6N$ | -0.5684 |
| $LaNH_4B_6C_6$ | $2.538\,LaC_2 + 2\,LaH_2 + 2.269\,N_2 + 0.4615\,B_{13}C_2 \longrightarrow 3.538\,LaN + LaB_6H_4C_6N$ | -0.5652 |
| $LaNH_4B_6C_6$ | $3.138\,H_3N + 1.583\,La_2C_3 + 0.626\,B_{13}C_2 \longrightarrow 2.138\,BN + 1.083\,La_2H_5 + LaB_6H_4C_6N$ | -0.5621 |
| $LaNH_4B_6C_6$ | $LaN + 2.357\,La_2C_3 + 2.592\,B_5H_7 \longrightarrow 4.714\,LaH_3 + 0.5352\,B_{13}C_2 + LaB_6H_4C_6N$ | -0.5561 |
| $LaNH_4B_6C_6$ | $2.833\,H_2 + 0.6667\,B_9H_{11} + 2\,La_2C_3 + 0.5\,N_2 \longrightarrow 3\,LaH_3 + LaB_6H_4C_6N$ | -0.5561 |
| $LaNH_4B_6C_6$ | $LaN + 1.226\,B_9H_{11} + 2\,La_2C_3 \longrightarrow 3.161\,LaH_3 + 0.8387\,LaB_6 + LaB_6H_4C_6N$ | -0.5522 |
| $LaNH_4B_6C_6$ | $2.3\,H_2 + 2\,La_2C_3 + 1.2\,B_5H_7 + 0.5\,N_2 \longrightarrow 3\,LaH_3 + LaB_6H_4C_6N$ | -0.5110 |
| $LaNH_4B_6C_6$ | $3\,LaC_2 + 2\,H_2 + 6\,B + 1.5\,N_2 \longrightarrow 2\,LaN + LaB_6H_4C_6N$ | -0.5108 |
| $LaNH_4B_6C_6$ | $3\,LaC_2 + 2\,H_2 + 17\,B + BN \longrightarrow 2\,LaB_6 + LaB_6H_4C_6N$ | -0.5032 |
| $LaNH_4B_6C_6$ | $10.25\,H_2 + LaN + 2\,La_2C_3 + 1.5\,LaB_4 \longrightarrow 5.5\,LaH_3 + LaB_6H_4C_6N$ | -0.5030 |
| $LaNH_4B_6C_6$ | $3\,LaC_2 + 15.14\,B + LaN + 0.5714\,B_5H_7 \longrightarrow 3\,LaB_4 + LaB_6H_4C_6N$ | -0.4806 |
| $LaNH_4B_6C_6$ | $0.9333\,La_2C_3 + 0.5\,N_2 + 1.6\,B_{13}C_2 + 0.8\,La_2H_5 \longrightarrow 2.467\,LaB_6 + LaB_6H_4C_6N$ | -0.4741 |
| $LaNH_4B_6C_6$ | $6\,C + 3.467\,La + H_3N + 1.2\,B_5H_7 \longrightarrow 2.467\,LaH_3 + LaB_6H_4C_6N$ | -0.4709 |
| $LaNH_4B_6C_6$ | $H_3N + 5.214\,LaH_2 + 3\,B_{13}C_2 \longrightarrow 4.214\,LaB_6 + 0.8571\,B_9H_{11} + LaB_6H_4C_6N$ | -0.4451 |
| $LaNH_4B_6C_6$ | $5\,H_2 + 6\,B + H_3N + 2\,La_2C_3 \longrightarrow 3\,LaH_3 + LaB_6H_4C_6N$ | -0.4410 |
| $LaNH_4B_6C_6$ | $2.687\,LaC_2 + 0.125\,H_8C_5 + 16.12\,B + H_3N \longrightarrow 1.687\,LaB_6 + LaB_6H_4C_6N$ | -0.4387 |
| $LaNH_4B_6C_6$ | $H_3N + 2\,La_2C_3 + 4.8\,B_5H_7 \longrightarrow 16.3\,H_2 + 3\,LaB_6 + LaB_6H_4C_6N$ | -0.4379 |
| $LaNH_4B_6C_6$ | $0.6667\,B_9H_{11} + 2\,La_2C_3 + 1.167\,N_2 \longrightarrow 1.333\,LaN + 1.667\,LaH_2 + LaB_6H_4C_6N$ | -0.4369 |
| $LaNH_4B_6C_6$ | $0.5\,H_8C_5 + 4.35\,LaN + 1.75\,B_{13}C_2 \longrightarrow 3.35\,BN + 3.35\,LaB_4 + LaB_6H_4C_6N$ | -0.4342 |
| $LaNH_4B_6C_6$ | $0.3636\,B_9H_{11} + 1.342\,La_2C_3 + 0.5\,N_2 + 0.987\,B_{13}C_2 \longrightarrow 1.684\,LaB_6 + LaB_6H_4C_6N$ | -0.4328 |
| $LaNH_4B_6C_6$ | $3\,LaC_2 + 14.73\,B + LaN + 0.3636\,B_9H_{11} \longrightarrow 3\,LaB_4 + LaB_6H_4C_6N$ | -0.4183 |
| $LaNH_4B_6C_6$ | $1.2\,H_8C_5 + 6\,B + 4.44\,La + H_3N \longrightarrow 1.72\,La_2H_5 + LaB_6H_4C_6N$ | -0.4134 |
| $LaNH_4B_6C_6$ | $2\,La_2C_3 + 1.2\,B_5H_7 + 1.022\,N_2 \longrightarrow 0.4889\,La_4H_9 + 1.044\,LaN + LaB_6H_4C_6N$ | -0.4118 |
| $LaNH_4B_6C_6$ | $0.5\,H_8C_5 + 3.393\,LaN + 1.75\,B_{13}C_2 \longrightarrow 2.393\,LaB_6 + 2.393\,BN + LaB_6H_4C_6N$ | -0.4063 |
| $LaNH_4B_6C_6$ | $6\,C + 5.286\,LaN + 1.143\,B_9H_{11} \longrightarrow 4.286\,BN + 4.286\,LaH_2 + LaB_6H_4C_6N$ | -0.4044 |
| $LaNH_4B_6C_6$ | $LaN + 2\,La_2C_3 + 2\,B_5H_7 \longrightarrow 3.333\,LaH_3 + 0.6667\,LaB_6 + LaB_6H_4C_6N$ | -0.3976 |
| $LaNH_4B_6C_6$ | $0.9859\,B_9H_{11} + 2\,La_2C_3 + 0.5\,N_2 \longrightarrow 2.282\,LaH_3 + 0.7183\,LaB_4 + LaB_6H_4C_6N$ | -0.3941 |
| $LaNH_4B_6C_6$ | $3\,LaC_2 + LaB_6 + 0.4444\,La_4H_9 + 2.889\,N_2 \longrightarrow 4.778\,LaN + LaB_6H_4C_6N$ | -0.3889 |
| $LaNH_4B_6C_6$ | $3\,LaC_2 + 4.641\,H_3N + 2.41\,LaB_4 \longrightarrow 3.641\,BN + 1.103\,La_4H_9 + LaB_6H_4C_6N$ | -0.3837 |
| $LaNH_4B_6C_6$ | $1.329\,La_2C_3 + 0.5714\,B_5H_7 + 0.5\,N_2 + 1.007\,B_{13}C_2 \longrightarrow 1.658\,LaB_6 + LaB_6H_4C_6N$ | -0.3811 |
| $LaNH_4B_6C_6$ | $2\,La_2C_3 + 1.628\,B_5H_7 + 0.5\,N_2 \longrightarrow 2.465\,LaH_3 + 0.5349\,LaB_4 + LaB_6H_4C_6N$ | -0.3743 |
| $LaNH_4B_6C_6$ | $0.9492\,B_9H_{11} + 2\,La_2C_3 + 0.5\,N_2 \longrightarrow 0.4237\,LaB_6 + 1.288\,La_2H_5 + LaB_6H_4C_6N$ | -0.3663 |
| $LaNH_4B_6C_6$ | $1.2\,H_8C_5 + 5.2\,La + BN + B_5H_7 \longrightarrow 4.2\,LaH_3 + LaB_6H_4C_6N$ | -0.3626 |
| $LaNH_4B_6C_6$ | $1.2\,H_8C_5 + 4.311\,La + LaN + 0.6667\,B_9H_{11} \longrightarrow 4.311\,LaH_3 + LaB_6H_4C_6N$ | -0.3599 |
| $LaNH_4B_6C_6$ | $6.5\,LaH_3 + 0.5\,N_2 + 3\,B_{13}C_2 \longrightarrow 7.75\,H_2 + 5.5\,LaB_6 + LaB_6H_4C_6N$ | -0.3577 |
| $LaNH_4B_6C_6$ | $6\,C + 3.2\,La + 1.2\,B_5H_7 + 0.5\,N_2 \longrightarrow 2.2\,LaH_2 + LaB_6H_4C_6N$ | -0.3446 |
| $LaNH_4B_6C_6$ | $1.146\,B_9H_{11} + 2.221\,La_2C_3 + 0.5\,N_2 \longrightarrow 0.3319\,B_{13}C_2 + 1.721\,La_2H_5 + LaB_6H_4C_6N$ | -0.3439 |
| $LaNH_4B_6C_6$ | $14.14\,B + BN + 2\,La_2C_3 + 0.5714\,B_5H_7 \longrightarrow 3\,LaB_4 + LaB_6H_4C_6N$ | -0.3424 |
| $LaNH_4B_6C_6$ | $0.5\,H_8C_5 + LaN + 1.75\,B_{13}C_2 + 8.375\,LaB_4 \longrightarrow 8.375\,LaB_6 + LaB_6H_4C_6N$ | -0.3364 |
| $LaNH_4B_6C_6$ | $0.4444\,La_4H_9 + LaN + 0.709\,La_2C_3 + 1.937\,B_{13}C_2 \longrightarrow 3.196\,LaB_6 + LaB_6H_4C_6N$ | -0.3346 |
| $LaNH_4B_6C_6$ | $6.5\,H_2 + 6\,B + LaN + 2\,La_2C_3 \longrightarrow La_4H_9 + LaB_6H_4C_6N$ | -0.3344 |
| $LaNH_4B_6C_6$ | $25.1\,H_2 + La + 0.5\,N_2 + 3\,B_{13}C_2 \longrightarrow 6.6\,B_5H_7 + LaB_6H_4C_6N$ | -0.3301 |
| $LaNH_4B_6C_6$ | $2\,La_2C_3 + 6.375\,B_5H_7 + 0.5\,N_2 \longrightarrow 3\,LaH_3 + 2.875\,B_9H_{11} + LaB_6H_4C_6N$ | -0.3166 |
| $LaNH_4B_6C_6$ | $0.5\,H_8C_5 + BN + 1.75\,B_{13}C_2 + 11.88\,LaB_4 \longrightarrow 10.88\,LaB_6 + LaB_6H_4C_6N$ | -0.3072 |

| Compound | Chemical reaction | Reaction energy(eV) |
| --- | --- | --- |
| LaNH$_4$B$_6$C$_6$ | 3 LaC$_2$ + 11.14 B + 0.5714 B$_5$H$_7$ + 0.5 N$_2$ ⟶ 2 LaB$_4$ + LaB$_6$H$_4$C$_6$N | -0.2916 |
| LaNH$_4$B$_6$C$_6$ | 6 C + 3.533 La + H$_3$N + 0.6667 B$_9$H$_{11}$ ⟶ 1.267 La$_2$H$_5$ + LaB$_6$H$_4$C$_6$N | -0.2879 |
| LaNH$_4$B$_6$C$_6$ | LaN + 3 B$_{13}$C$_2$ + 2.303 La$_2$H$_5$ ⟶ 4.606 LaB$_6$ + 1.073 B$_5$H$_7$ + LaB$_6$H$_4$C$_6$N | -0.2825 |
| LaNH$_4$B$_6$C$_6$ | 13.73 B + BN + 0.3636 B$_9$H$_{11}$ + 2 La$_2$C$_3$ ⟶ 3 LaB$_4$ + LaB$_6$H$_4$C$_6$N | -0.2801 |
| LaNH$_4$B$_6$C$_6$ | 1.563 LaB$_6$ + 4.375 H$_3$N + 2 La$_2$C$_3$ ⟶ 3.375 BN + 4.563 LaH$_2$ + LaB$_6$H$_4$C$_6$N | -0.2691 |
| LaNH$_4$B$_6$C$_6$ | 6 C + LaB$_6$ + 0.5714 B$_5$H$_7$ + 1.929 N$_2$ ⟶ 2.857 BN + LaB$_6$H$_4$C$_6$N | -0.2593 |
| LaNH$_4$B$_6$C$_6$ | 3 LaC$_2$ + 0.7727 B$_9$H$_{11}$ + 0.9773 N$_2$ ⟶ 0.9545 BN + 0.5 La$_4$H$_9$ + LaB$_6$H$_4$C$_6$N | -0.2571 |
| LaNH$_4$B$_6$C$_6$ | 3 LaC$_2$ + 21 B + BN + 2 LaH$_2$ ⟶ 4 LaB$_4$ + LaB$_6$H$_4$C$_6$N | -0.2521 |
| LaNH$_4$B$_6$C$_6$ | LaN + 16.75 B$_9$H$_{11}$ + 2 La$_2$C$_3$ ⟶ 25.75 B$_5$H$_7$ + 4 LaB$_4$ + LaB$_6$H$_4$C$_6$N | -0.2299 |
| LaNH$_4$B$_6$C$_6$ | 3 LaC$_2$ + 10.73 B + 0.3636 B$_9$H$_{11}$ + 0.5 N$_2$ ⟶ 2 LaB$_4$ + LaB$_6$H$_4$C$_6$N | -0.2293 |
| LaNH$_4$B$_6$C$_6$ | 1.333 LaH$_3$ + 0.9841 La$_2$C$_3$ + 0.5 N$_2$ + 1.524 B$_{13}$C$_2$ ⟶ 2.302 LaB$_6$ + LaB$_6$H$_4$C$_6$N | -0.2279 |
| LaNH$_4$B$_6$C$_6$ | 3 LaC$_2$ + 2 H$_2$ + 2.25 N$_2$ + 1.5 LaB$_4$ ⟶ 3.5 LaN + LaB$_6$H$_4$C$_6$N | -0.2273 |
| LaNH$_4$B$_6$C$_6$ | 2 La$_2$C$_3$ + 1.541 B$_5$H$_7$ + 0.5 N$_2$ ⟶ 0.2844 LaB$_6$ + 1.358 La$_2$H$_5$ + LaB$_6$H$_4$C$_6$N | -0.2231 |
| LaNH$_4$B$_6$C$_6$ | 7.625 H$_2$ + 2 La$_2$C$_3$ + 0.5 N$_2$ + 1.5 LaB$_4$ ⟶ 2.25 La$_2$H$_5$ + LaB$_6$H$_4$C$_6$N | -0.2209 |
| LaNH$_4$B$_6$C$_6$ | 3 LaC$_2$ + 13.25 B$_9$H$_{11}$ + 0.5 N$_2$ ⟶ 2 LaB$_6$ + 20.25 B$_5$H$_7$ + LaB$_6$H$_4$C$_6$N | -0.2039 |
| LaNH$_4$B$_6$C$_6$ | 2 H$_2$ + 1.238 La$_2$C$_3$ + 0.5 N$_2$ + 1.143 B$_{13}$C$_2$ ⟶ 1.476 LaB$_6$ + LaB$_6$H$_4$C$_6$N | -0.1944 |
| LaNH$_4$B$_6$C$_6$ | 2.139 La$_2$C$_3$ + 1.742 B$_5$H$_7$ + 0.5 N$_2$ ⟶ 0.2085 B$_{13}$C$_2$ + 1.639 La$_2$H$_5$ + LaB$_6$H$_4$C$_6$N | -0.1904 |
| LaNH$_4$B$_6$C$_6$ | LaN + 3 B$_{13}$C$_2$ + 2.254 La$_2$H$_5$ ⟶ 4.508 LaB$_6$ + 0.661 B$_9$H$_{11}$ + LaB$_6$H$_4$C$_6$N | -0.1829 |
| LaNH$_4$B$_6$C$_6$ | H$_2$ + B + 0.01195 LaN + 0.03586 B$_{13}$C$_2$ ⟶ 0.2789 B$_5$H$_7$ + 0.01195 LaB$_6$H$_4$C$_6$N | -0.1812 |
| LaNH$_4$B$_6$C$_6$ | 2.538 LaC$_2$ + 0.4444 La$_4$H$_9$ + 2.158 N$_2$ + 0.4615 B$_{13}$C$_2$ ⟶ 3.316 LaN + LaB$_6$H$_4$C$_6$N | -0.1708 |
| LaNH$_4$B$_6$C$_6$ | 4.333 H$_2$ + LaN + 0.6667 B$_9$H$_{11}$ + 2 La$_2$C$_3$ ⟶ 4 LaH$_3$ + LaB$_6$H$_4$C$_6$N | -0.1646 |
| LaNH$_4$B$_6$C$_6$ | 3 LaC$_2$ + 7.833 B + 2.833 H$_3$N ⟶ 1.833 BN + 0.5 La$_4$H$_9$ + LaB$_6$H$_4$C$_6$N | -0.1625 |
| LaNH$_4$B$_6$C$_6$ | 2.97 H$_3$N + 1.591 La$_2$C$_3$ + 0.6131 B$_{13}$C$_2$ ⟶ 1.97 BN + 0.5456 La$_4$H$_9$ + LaB$_6$H$_4$C$_6$N | -0.1564 |
| LaNH$_4$B$_6$C$_6$ | H$_2$ + B + 0.01754 LaN + 0.05263 B$_{13}$C$_2$ ⟶ 0.1754 B$_9$H$_{11}$ + 0.01754 LaB$_6$H$_4$C$_6$N | -0.1563 |
| LaNH$_4$B$_6$C$_6$ | 5.577 H$_2$ + 1.692 La$_2$C$_3$ + 0.5 N$_2$ + 0.4615 B$_{13}$C$_2$ ⟶ 2.385 LaH$_3$ + LaB$_6$H$_4$C$_6$N | -0.1346 |
| LaNH$_4$B$_6$C$_6$ | 6 C + 2.956 La + 1.2 B$_5$H$_7$ + 0.5 N$_2$ ⟶ 0.4889 La$_4$H$_9$ + LaB$_6$H$_4$C$_6$N | -0.1288 |
| LaNH$_4$B$_6$C$_6$ | 3.8 H$_2$ + LaN + 2 La$_2$C$_3$ + 1.2 B$_5$H$_7$ ⟶ 4 LaH$_3$ + LaB$_6$H$_4$C$_6$N | -0.1196 |
| LaNH$_4$B$_6$C$_6$ | LaN + 0.7429 La$_2$C$_3$ + 1.886 B$_{13}$C$_2$ + 0.8 La$_2$H$_5$ ⟶ 3.086 LaB$_6$ + LaB$_6$H$_4$C$_6$N | -0.1078 |
| LaNH$_4$B$_6$C$_6$ | 3.5 LaN + 5.75 LaH$_2$ + 3 B$_{13}$C$_2$ ⟶ 2.5 H$_3$N + 8.25 LaB$_4$ + LaB$_6$H$_4$C$_6$N | -0.1073 |
| LaNH$_4$B$_6$C$_6$ | 2.083 H$_2$ + 0.6667 B$_9$H$_{11}$ + 2 La$_2$C$_3$ + 0.5 N$_2$ ⟶ 1.5 La$_2$H$_5$ + LaB$_6$H$_4$C$_6$N | -0.1071 |
| LaNH$_4$B$_6$C$_6$ | 3 LaC$_2$ + LaB$_6$ + 2.8 N$_2$ + 0.8 La$_2$H$_5$ ⟶ 4.6 LaN + LaB$_6$H$_4$C$_6$N | -0.0969 |
| LaNH$_4$B$_6$C$_6$ | 8 H$_2$ + LaB$_6$ + 2 La$_2$C$_3$ + 0.5 N$_2$ ⟶ 4 LaH$_3$ + LaB$_6$H$_4$C$_6$N | -0.0940 |
| LaNH$_4$B$_6$C$_6$ | 22.17 H$_2$ + La + 0.5 N$_2$ + 3 B$_{13}$C$_2$ ⟶ 3.667 B$_9$H$_{11}$ + LaB$_6$H$_4$C$_6$N | -0.0821 |
| LaNH$_4$B$_6$C$_6$ | LaN + 0.3636 B$_9$H$_{11}$ + 1.152 La$_2$C$_3$ + 1.273 B$_{13}$C$_2$ ⟶ 2.303 LaB$_6$ + LaB$_6$H$_4$C$_6$N | -0.0665 |
| LaNH$_4$B$_6$C$_6$ | 0.9173 B$_9$H$_{11}$ + 2 La$_2$C$_3$ + 0.5 N$_2$ ⟶ 1.218 La$_2$H$_5$ + 0.5639 LaB$_4$ + LaB$_6$H$_4$C$_6$N | -0.0643 |
| LaNH$_4$B$_6$C$_6$ | 1.55 H$_2$ + 2 La$_2$C$_3$ + 1.2 B$_5$H$_7$ + 0.5 N$_2$ ⟶ 1.5 La$_2$H$_5$ + LaB$_6$H$_4$C$_6$N | -0.0620 |
| LaNH$_4$B$_6$C$_6$ | 3 LaC$_2$ + 16 B + H$_3$N + 0.5 LaH$_2$ ⟶ 2.5 LaB$_4$ + LaB$_6$H$_4$C$_6$N | -0.0472 |
| LaNH$_4$B$_6$C$_6$ | 1.421 LaC$_2$ + 2 LaH$_2$ + 0.5 N$_2$ + 1.579 B$_{13}$C$_2$ ⟶ 2.421 LaB$_6$ + LaB$_6$H$_4$C$_6$N | -0.0371 |
| LaNH$_4$B$_6$C$_6$ | 0.9043 B$_9$H$_{11}$ + 2 La$_2$C$_3$ + 0.5 N$_2$ ⟶ 0.3565 LaB$_6$ + 0.6609 La$_4$H$_9$ + LaB$_6$H$_4$C$_6$N | -0.0340 |
| LaNH$_4$B$_6$C$_6$ | 2 La$_2$C$_3$ + 1.506 B$_5$H$_7$ + 0.5 N$_2$ ⟶ 1.309 La$_2$H$_5$ + 0.3827 LaB$_4$ + LaB$_6$H$_4$C$_6$N | -0.0214 |
| LaNH$_4$B$_6$C$_6$ | LaN + 1.138 La$_2$C$_3$ + 0.5714 B$_5$H$_7$ + 1.293 B$_{13}$C$_2$ ⟶ 2.277 LaB$_6$ + LaB$_6$H$_4$C$_6$N | -0.0147 |
| LaNH$_4$B$_6$C$_6$ | H$_3$N + 3 B$_{13}$C$_2$ + 3.25 La$_2$H$_5$ ⟶ 7.625 H$_2$ + 5.5 LaB$_6$ + LaB$_6$H$_4$C$_6$N | -0.0065 |
| LaNH$_4$B$_6$C$_6$ | 0.5 H$_8$C$_5$ + 14 B + 1.167 La$_2$C$_3$ + 0.5 N$_2$ ⟶ 1.333 LaB$_6$ + LaB$_6$H$_4$C$_6$N | -0.0042 |
| LuNH$_4$B$_6$C$_6$ | 3 B$_{13}$C$_2$ + 251 LuB$_2$ + 251 H$_3$N ⟶ 250 LuN + 107 B$_5$H$_7$ + LuB$_6$H$_4$C$_6$N | -149.6356 |
| LuNH$_4$B$_6$C$_6$ | 50 Lu + 3 B$_{13}$C$_2$ + 34 H$_3$N ⟶ 33 BN + 49 LuH$_2$ + LuB$_6$H$_4$C$_6$N | -127.6214 |
| LuNH$_4$B$_6$C$_6$ | 15.5 Lu$_4$C$_7$ + 56 H$_3$N + 6 BN ⟶ 20.5 H$_8$C$_5$ + 61 LuN + LuB$_6$H$_4$C$_6$N | -113.7139 |
| LuNH$_4$B$_6$C$_6$ | 24.64 N$_2$ + 3 B$_{13}$C$_2$ + 0.3636 B$_9$H$_{11}$ + LuB$_{12}$ ⟶ 48.27 BN + LuB$_6$H$_4$C$_6$N | -112.9722 |
| LuNH$_4$B$_6$C$_6$ | 24.43 N$_2$ + 3 B$_{13}$C$_2$ + LuB$_{12}$ + 0.5714 B$_5$H$_7$ ⟶ 47.86 BN + LuB$_6$H$_4$C$_6$N | -111.9678 |
| LuNH$_4$B$_6$C$_6$ | 23 N$_2$ + 3 B$_{13}$C$_2$ + LuB$_{12}$ + 2 H$_2$ ⟶ 45 BN + LuB$_6$H$_4$C$_6$N | -105.2314 |
| LuNH$_4$B$_6$C$_6$ | 22.33 N$_2$ + 3 B$_{13}$C$_2$ + LuB$_{12}$ + 1.333 H$_3$N ⟶ 45 BN + LuB$_6$H$_4$C$_6$N | -103.4659 |
| LuNH$_4$B$_6$C$_6$ | 22.69 N$_2$ + 3.875 B$_{13}$C$_2$ + 2 LuH$_2$ ⟶ 0.25 Lu$_4$C$_7$ + 44.38 BN + LuB$_6$H$_4$C$_6$N | -102.6011 |
| LuNH$_4$B$_6$C$_6$ | 21.33 N$_2$ + 3.667 B$_{13}$C$_2$ + 2 LuH$_2$ ⟶ 0.3333 Lu$_3$C$_4$ + 41.67 BN + LuB$_6$H$_4$C$_6$N | -95.8918 |
| LuNH$_4$B$_6$C$_6$ | 20.64 N$_2$ + 3 B$_{13}$C$_2$ + 0.3636 B$_9$H$_{11}$ + LuB$_4$ ⟶ 40.27 BN + LuB$_6$H$_4$C$_6$N | -92.8499 |
| LuNH$_4$B$_6$C$_6$ | 20.43 N$_2$ + 3 B$_{13}$C$_2$ + LuB$_4$ + 0.5714 B$_5$H$_7$ ⟶ 39.86 BN + LuB$_6$H$_4$C$_6$N | -91.8455 |
| LuNH$_4$B$_6$C$_6$ | 1.2 H$_8$C$_5$ + 20.3 N$_2$ + 3.8 LuB$_{12}$ ⟶ 39.6 BN + 2.8 LuH$_2$ + LuB$_6$H$_4$C$_6$N | -89.2944 |
| LuNH$_4$B$_6$C$_6$ | 19.64 N$_2$ + 3 B$_{13}$C$_2$ + 0.3636 B$_9$H$_{11}$ + LuB$_2$ ⟶ 38.27 BN + LuB$_6$H$_4$C$_6$N | -88.3289 |
| LuNH$_4$B$_6$C$_6$ | 19.43 N$_2$ + 3 B$_{13}$C$_2$ + LuB$_2$ + 0.5714 B$_5$H$_7$ ⟶ 37.86 BN + LuB$_6$H$_4$C$_6$N | -87.3245 |
| LuNH$_4$B$_6$C$_6$ | 18.64 N$_2$ + Lu + 3 B$_{13}$C$_2$ + 0.3636 B$_9$H$_{11}$ ⟶ 36.27 BN + LuB$_6$H$_4$C$_6$N | -85.2207 |
| LuNH$_4$B$_6$C$_6$ | 19 N$_2$ + 3 B$_{13}$C$_2$ + LuB$_4$ + 2 H$_2$ ⟶ 37 BN + LuB$_6$H$_4$C$_6$N | -85.1091 |
| LuNH$_4$B$_6$C$_6$ | 18.43 N$_2$ + Lu + 3 B$_{13}$C$_2$ + 0.5714 B$_5$H$_7$ ⟶ 35.86 BN + LuB$_6$H$_4$C$_6$N | -84.2163 |

| Compound | Chemical reaction | Reaction energy(eV) |
|---|---|---|
| $LuNH_4B_6C_6$ | $18.33\,N_2 + 3\,B_{13}C_2 + LuB_4 + 1.333\,H_3N \longrightarrow 37\,BN + LuB_6H_4C_6N$ | -83.3436 |
| $LuNH_4B_6C_6$ | $18.14\,N_2 + 3\,B_{13}C_2 + 0.3636\,B_9H_{11} + LuN \longrightarrow 36.27\,BN + LuB_6H_4C_6N$ | -81.6778 |
| $LuNH_4B_6C_6$ | $17.93\,N_2 + 3\,B_{13}C_2 + LuN + 0.5714\,B_5H_7 \longrightarrow 35.86\,BN + LuB_6H_4C_6N$ | -80.6734 |
| $LuNH_4B_6C_6$ | $18\,N_2 + 3\,B_{13}C_2 + LuB_2 + 2\,H_2 \longrightarrow 35\,BN + LuB_6H_4C_6N$ | -80.5881 |
| $LuNH_4B_6C_6$ | $17.82\,N_2 + 3\,B_{13}C_2 + 0.1818\,B_9H_{11} + LuH_2 \longrightarrow 34.64\,BN + LuB_6H_4C_6N$ | -79.4082 |
| $LuNH_4B_6C_6$ | $17.71\,N_2 + 3\,B_{13}C_2 + 0.2857\,B_5H_7 + LuH_2 \longrightarrow 34.43\,BN + LuB_6H_4C_6N$ | -78.9060 |
| $LuNH_4B_6C_6$ | $17.33\,N_2 + 3\,B_{13}C_2 + LuB_2 + 1.333\,H_3N \longrightarrow 35\,BN + LuB_6H_4C_6N$ | -78.8226 |
| $LuNH_4B_6C_6$ | $17\,N_2 + Lu + 3\,B_{13}C_2 + 2\,H_2 \longrightarrow 33\,BN + LuB_6H_4C_6N$ | -77.4798 |
| $LuNH_4B_6C_6$ | $17.5\,N_2 + 3\,B_{13}C_2 + 2\,LuH_2 \longrightarrow LuN + 33\,BN + LuB_6H_4C_6N$ | -77.1387 |
| $LuNH_4B_6C_6$ | $16.33\,N_2 + Lu + 3\,B_{13}C_2 + 1.333\,H_3N \longrightarrow 33\,BN + LuB_6H_4C_6N$ | -75.7144 |
| $LuNH_4B_6C_6$ | $17\,N_2 + 3\,B_{13}C_2 + H_2 + LuH_2 \longrightarrow 33\,BN + LuB_6H_4C_6N$ | -75.5378 |
| $LuNH_4B_6C_6$ | $16.67\,N_2 + 3\,B_{13}C_2 + 0.6667\,H_3N + LuH_2 \longrightarrow 33\,BN + LuB_6H_4C_6N$ | -74.6551 |
| $LuNH_4B_6C_6$ | $16.5\,N_2 + 3\,B_{13}C_2 + LuN + 2\,H_2 \longrightarrow 33\,BN + LuB_6H_4C_6N$ | -73.9369 |
| $LuNH_4B_6C_6$ | $17\,N_2 + 3\,B_{13}C_2 + 2\,LuH_2 \longrightarrow Lu + 33\,BN + LuB_6H_4C_6N$ | -73.5957 |
| $LuNH_4B_6C_6$ | $15.83\,N_2 + 3\,B_{13}C_2 + LuN + 1.333\,H_3N \longrightarrow 33\,BN + LuB_6H_4C_6N$ | -72.1715 |
| $LuNH_4B_6C_6$ | $16\,N_2 + 3\,B_{13}C_2 + 2\,LuH_2 \longrightarrow LuB_2 + 31\,BN + LuB_6H_4C_6N$ | -70.4875 |
| $LuNH_4B_6C_6$ | $15\,N_2 + 3\,B_{13}C_2 + 2\,LuH_2 \longrightarrow LuB_4 + 29\,BN + LuB_6H_4C_6N$ | -65.9665 |
| $LuNH_4B_6C_6$ | $0.5\,H_8C_5 + 14.88\,N_2 + 1.75\,B_{13}C_2 + LuB_{12} \longrightarrow 28.75\,BN + LuB_6H_4C_6N$ | -63.9485 |
| $LuNH_4B_6C_6$ | $14.3\,N_2 + 2.333\,B_{13}C_2 + 0.3636\,B_9H_{11} + 0.3333\,Lu_3C_4 \longrightarrow 27.61\,BN + LuB_6H_4C_6N$ | -62.9246 |
| $LuNH_4B_6C_6$ | $14.1\,N_2 + 2.333\,B_{13}C_2 + 0.3333\,Lu_3C_4 + 0.5714\,B_5H_7 \longrightarrow 27.19\,BN + LuB_6H_4C_6N$ | -61.9202 |
| $LuNH_4B_6C_6$ | $0.25\,Lu_4C_7 + 12.95\,N_2 + 2.125\,B_{13}C_2 + 0.3636\,B_9H_{11} \longrightarrow 24.9\,BN + LuB_6H_4C_6N$ | -56.2153 |
| $LuNH_4B_6C_6$ | $0.25\,Lu_4C_7 + 12.74\,N_2 + 2.125\,B_{13}C_2 + 0.5714\,B_5H_7 \longrightarrow 24.48\,BN + LuB_6H_4C_6N$ | -55.2109 |
| $LuNH_4B_6C_6$ | $12.67\,N_2 + 2.333\,B_{13}C_2 + 0.3333\,Lu_3C_4 + 2\,H_2 \longrightarrow 24.33\,BN + LuB_6H_4C_6N$ | -55.1837 |
| $LuNH_4B_6C_6$ | $0.25\,H_8C_5 + 12.94\,N_2 + 2.375\,B_{13}C_2 + LuH_2 \longrightarrow 24.88\,BN + LuB_6H_4C_6N$ | -54.8963 |
| $LuNH_4B_6C_6$ | $12\,N_2 + 2.333\,B_{13}C_2 + 0.3333\,Lu_3C_4 + 1.333\,H_3N \longrightarrow 24.33\,BN + LuB_6H_4C_6N$ | -53.4183 |
| $LuNH_4B_6C_6$ | $0.25\,Lu_4C_7 + 11.31\,N_2 + 2.125\,B_{13}C_2 + 2\,H_2 \longrightarrow 21.62\,BN + LuB_6H_4C_6N$ | -48.4744 |
| $LuNH_4B_6C_6$ | $9.484\,Lu_4C_7 + 8.875\,B_9H_{11} + H_3N \longrightarrow 12.08\,H_8C_5 + 36.94\,LuB_2 + LuB_6H_4C_6N$ | -47.1189 |
| $LuNH_4B_6C_6$ | $0.25\,Lu_4C_7 + 10.65\,N_2 + 2.125\,B_{13}C_2 + 1.333\,H_3N \longrightarrow 21.63\,BN + LuB_6H_4C_6N$ | -46.7090 |
| $LuNH_4B_6C_6$ | $11\,N_2 + 3\,B_{13}C_2 + 2\,LuH_2 \longrightarrow LuB_{12} + 21\,BN + LuB_6H_4C_6N$ | -45.8442 |
| $LuNH_4B_6C_6$ | $3\,B_{13}C_2 + LuB_{12} + 46\,H_3N \longrightarrow 67\,H_2 + 45\,BN + LuB_6H_4C_6N$ | -44.3241 |
| $LuNH_4B_6C_6$ | $9.094\,Lu_4C_7 + 8.75\,B_9H_{11} + LuN \longrightarrow 11.53\,H_8C_5 + 36.38\,LuB_2 + LuB_6H_4C_6N$ | -44.1552 |
| $LuNH_4B_6C_6$ | $0.5\,H_8C_5 + 10.87\,N_2 + 1.75\,B_{13}C_2 + LuB_4 \longrightarrow 20.75\,BN + LuB_6H_4C_6N$ | -43.8262 |
| $LuNH_4B_6C_6$ | $3\,B_{13}C_2 + 79.8\,LuB_2 + 79.8\,H_3N \longrightarrow 21.4\,B_9H_{11} + 78.8\,LuN + LuB_6H_4C_6N$ | -43.4873 |
| $LuNH_4B_6C_6$ | $0.5\,H_8C_5 + 9.875\,N_2 + 1.75\,B_{13}C_2 + LuB_2 \longrightarrow 18.75\,BN + LuB_6H_4C_6N$ | -39.3052 |
| $LuNH_4B_6C_6$ | $12.55\,N_2 + 3\,B_{13}C_2 + 25.1\,LuH_2 \longrightarrow 24.1\,LuN + 6.6\,B_5H_7 + LuB_6H_4C_6N$ | -36.3129 |
| $LuNH_4B_6C_6$ | $0.5\,H_8C_5 + 8.875\,N_2 + Lu + 1.75\,B_{13}C_2 \longrightarrow 16.75\,BN + LuB_6H_4C_6N$ | -36.1969 |
| $LuNH_4B_6C_6$ | $7.375\,Lu_4C_7 + 0.5\,N_2 + 7\,B_9H_{11} \longrightarrow 9.125\,H_8C_5 + 28.5\,LuB_2 + LuB_6H_4C_6N$ | -35.6607 |
| $LuNH_4B_6C_6$ | $3\,B_{13}C_2 + LuB_4 + 38\,H_3N \longrightarrow 55\,H_2 + 37\,BN + LuB_6H_4C_6N$ | -34.7944 |
| $LuNH_4B_6C_6$ | $16.73\,Lu + 3\,B_{13}C_2 + 16.73\,H_3N \longrightarrow 15.73\,LuN + 6.6\,B_5H_7 + LuB_6H_4C_6N$ | -33.2596 |
| $LuNH_4B_6C_6$ | $3\,B_{13}C_2 + LuB_2 + 36\,H_3N \longrightarrow 52\,H_2 + 35\,BN + LuB_6H_4C_6N$ | -32.9215 |
| $LuNH_4B_6C_6$ | $0.5\,H_8C_5 + 8.375\,N_2 + 1.75\,B_{13}C_2 + LuN \longrightarrow 16.75\,BN + LuB_6H_4C_6N$ | -32.6540 |
| $LuNH_4B_6C_6$ | $Lu + 3\,B_{13}C_2 + 34\,H_3N \longrightarrow 49\,H_2 + 33\,BN + LuB_6H_4C_6N$ | -32.4614 |
| $LuNH_4B_6C_6$ | $11.08\,N_2 + 3\,B_{13}C_2 + 22.17\,LuH_2 \longrightarrow 3.667\,B_9H_{11} + 21.17\,LuN + LuB_6H_4C_6N$ | -31.3690 |
| $LuNH_4B_6C_6$ | $3\,B_{13}C_2 + 34\,H_3N + LuH_2 \longrightarrow 50\,H_2 + 33\,BN + LuB_6H_4C_6N$ | -30.5194 |
| $LuNH_4B_6C_6$ | $3\,B_{13}C_2 + LuN + 33\,H_3N \longrightarrow 47.5\,H_2 + 33\,BN + LuB_6H_4C_6N$ | -30.2426 |
| $LuNH_4B_6C_6$ | $0.5\,N_2 + 19.14\,Lu + 3\,B_{13}C_2 + 0.3636\,B_9H_{11} \longrightarrow 18.14\,LuB_2 + LuB_6H_4C_6N$ | -28.8485 |
| $LuNH_4B_6C_6$ | $14.78\,Lu + 3\,B_{13}C_2 + 14.78\,H_3N \longrightarrow 3.667\,B_9H_{11} + 13.78\,LuN + LuB_6H_4C_6N$ | -28.6726 |
| $LuNH_4B_6C_6$ | $0.5\,N_2 + 18.93\,Lu + 3\,B_{13}C_2 + 0.5714\,B_5H_7 \longrightarrow 17.93\,LuB_2 + LuB_6H_4C_6N$ | -28.4900 |
| $LuNH_4B_6C_6$ | $6.516\,Lu_4C_7 + 6.125\,B_9H_{11} + BN \longrightarrow 7.922\,H_8C_5 + 25.06\,LuB_2 + LuB_6H_4C_6N$ | -28.0878 |
| $LuNH_4B_6C_6$ | $19.64\,Lu + 3\,B_{13}C_2 + 0.3636\,B_9H_{11} + BN \longrightarrow 18.64\,LuB_2 + LuB_6H_4C_6N$ | -27.2944 |
| $LuNH_4B_6C_6$ | $19.43\,Lu + 3\,B_{13}C_2 + BN + 0.5714\,B_5H_7 \longrightarrow 18.43\,LuB_2 + LuB_6H_4C_6N$ | -26.9358 |
| $LuNH_4B_6C_6$ | $0.5\,N_2 + 17.5\,Lu + 3\,B_{13}C_2 + 2\,H_2 \longrightarrow 16.5\,LuB_2 + LuB_6H_4C_6N$ | -26.1939 |
| $LuNH_4B_6C_6$ | $17.83\,Lu + 3\,B_{13}C_2 + 1.333\,H_3N \longrightarrow 0.3333\,LuN + 16.5\,LuB_2 + LuB_6H_4C_6N$ | -25.6094 |
| $LuNH_4B_6C_6$ | $17.91\,Lu + 3\,B_{13}C_2 + 0.09091\,B_9H_{11} + H_3N \longrightarrow 16.91\,LuB_2 + LuB_6H_4C_6N$ | -25.5335 |
| $LuNH_4B_6C_6$ | $17.86\,Lu + 3\,B_{13}C_2 + H_3N + 0.1429\,B_5H_7 \longrightarrow 16.86\,LuB_2 + LuB_6H_4C_6N$ | -25.4438 |
| $LuNH_4B_6C_6$ | $18.14\,Lu + 3\,B_{13}C_2 + 0.3636\,B_9H_{11} + LuN \longrightarrow 18.14\,LuB_2 + LuB_6H_4C_6N$ | -25.3056 |
| $LuNH_4B_6C_6$ | $3\,B_{13}C_2 + 20\,LuN + 14\,H_3N \longrightarrow 33\,BN + 19\,LuH_2 + LuB_6H_4C_6N$ | -24.9834 |
| $LuNH_4B_6C_6$ | $17.93\,Lu + 3\,B_{13}C_2 + LuN + 0.5714\,B_5H_7 \longrightarrow 17.93\,LuB_2 + LuB_6H_4C_6N$ | -24.9470 |
| $LuNH_4B_6C_6$ | $17.33\,Lu + 3\,B_{13}C_2 + 1.333\,H_3N \longrightarrow 16.33\,LuB_2 + 0.3333\,BN + LuB_6H_4C_6N$ | -24.9465 |
| $LuNH_4B_6C_6$ | $17.5\,Lu + 3\,B_{13}C_2 + 0.5\,H_2 + H_3N \longrightarrow 16.5\,LuB_2 + LuB_6H_4C_6N$ | -24.8698 |

| Compound | Chemical reaction | Reaction energy(eV) |
| --- | --- | --- |
| $LuNH_4B_6C_6$ | $18\,Lu + 3\,B_{13}C_2 + 2\,H_2 + BN \longrightarrow 17\,LuB_2 + LuB_6H_4C_6N$ | -24.6397 |
| $LuNH_4B_6C_6$ | $17.5\,Lu + 3\,B_{13}C_2 + 1.333\,H_3N \longrightarrow 0.1667\,N_2 + 16.5\,LuB_2 + LuB_6H_4C_6N$ | -24.4284 |
| $LuNH_4B_6C_6$ | $17\,Lu + 3\,B_{13}C_2 + H_3N + 0.5\,LuH_2 \longrightarrow 16.5\,LuB_2 + LuB_6H_4C_6N$ | -23.8988 |
| $LuNH_4B_6C_6$ | $16.5\,Lu + 3\,B_{13}C_2 + LuN + 2\,H_2 \longrightarrow 16.5\,LuB_2 + LuB_6H_4C_6N$ | -22.6510 |
| $LuNH_4B_6C_6$ | $0.5\,N_2 + 15.5\,Lu + 3\,B_{13}C_2 + 2\,LuH_2 \longrightarrow 16.5\,LuB_2 + LuB_6H_4C_6N$ | -22.3098 |
| $LuNH_4B_6C_6$ | $2.333\,B_{13}C_2 + 0.3333\,Lu_3C_4 + 25.33\,H_3N \longrightarrow 36\,H_2 + 24.33\,BN + LuB_6H_4C_6N$ | -21.6406 |
| $LuNH_4B_6C_6$ | $3\,B_{13}C_2 + 34\,LuN + 35\,H_2 \longrightarrow 33\,BN + 33\,LuH_2 + LuB_6H_4C_6N$ | -21.1082 |
| $LuNH_4B_6C_6$ | $11.03\,N_2 + 1.5\,Lu_3C_4 + 9.6\,B_5H_7 \longrightarrow 3.5\,LuB_{12} + 21.07\,H_3N + LuB_6H_4C_6N$ | -21.0829 |
| $LuNH_4B_6C_6$ | $0.125\,H_8C_5 + 15.47\,Lu + 2.688\,B_{13}C_2 + H_3N \longrightarrow 14.47\,LuB_2 + LuB_6H_4C_6N$ | -20.8627 |
| $LuNH_4B_6C_6$ | $16\,Lu + 3\,B_{13}C_2 + BN + 2\,LuH_2 \longrightarrow 17\,LuB_2 + LuB_6H_4C_6N$ | -20.7557 |
| $LuNH_4B_6C_6$ | $14.5\,Lu + 3\,B_{13}C_2 + LuN + 2\,LuH_2 \longrightarrow 16.5\,LuB_2 + LuB_6H_4C_6N$ | -18.7669 |
| $LuNH_4B_6C_6$ | $0.25\,Lu_4C_7 + 2.125\,B_{13}C_2 + 22.62\,H_3N \longrightarrow 31.94\,H_2 + 21.62\,BN + LuB_6H_4C_6N$ | -18.5173 |
| $LuNH_4B_6C_6$ | $9.611\,N_2 + 5.333\,B_9H_{11} + 1.5\,Lu_3C_4 \longrightarrow 3.5\,LuB_{12} + 18.22\,H_3N + LuB_6H_4C_6N$ | -17.6774 |
| $LuNH_4B_6C_6$ | $5.136\,N_2 + 0.3636\,B_9H_{11} + 6\,C + LuB_{12} \longrightarrow 9.273\,BN + LuB_6H_4C_6N$ | -16.8955 |
| $LuNH_4B_6C_6$ | $0.5\,N_2 + 10.07\,Lu + 3\,B_{13}C_2 + 0.3636\,B_9H_{11} \longrightarrow 9.068\,LuB_4 + LuB_6H_4C_6N$ | -16.0374 |
| $LuNH_4B_6C_6$ | $4.929\,N_2 + 6\,C + LuB_{12} + 0.5714\,B_5H_7 \longrightarrow 8.857\,BN + LuB_6H_4C_6N$ | -15.8911 |
| $LuNH_4B_6C_6$ | $0.5\,N_2 + 9.964\,Lu + 3\,B_{13}C_2 + 0.5714\,B_5H_7 \longrightarrow 8.964\,LuB_4 + LuB_6H_4C_6N$ | -15.8256 |
| $LuNH_4B_6C_6$ | $0.5\,N_2 + 9.25\,Lu + 3\,B_{13}C_2 + 2\,H_2 \longrightarrow 8.25\,LuB_4 + LuB_6H_4C_6N$ | -14.5386 |
| $LuNH_4B_6C_6$ | $10.32\,Lu + 3\,B_{13}C_2 + 0.3636\,B_9H_{11} + BN \longrightarrow 9.318\,LuB_4 + LuB_6H_4C_6N$ | -14.1301 |
| $LuNH_4B_6C_6$ | $9.583\,Lu + 3\,B_{13}C_2 + 1.333\,H_3N \longrightarrow 0.3333\,LuN + 8.25\,LuB_4 + LuB_6H_4C_6N$ | -13.9541 |
| $LuNH_4B_6C_6$ | $10.21\,Lu + 3\,B_{13}C_2 + BN + 0.5714\,B_5H_7 \longrightarrow 9.214\,LuB_4 + LuB_6H_4C_6N$ | -13.9183 |
| $LuNH_4B_6C_6$ | $0.5\,H_8C_5 + 4.542\,N_2 + 1.083\,B_{13}C_2 + 0.3333\,Lu_3C_4 \longrightarrow 8.083\,BN + LuB_6H_4C_6N$ | -13.9008 |
| $LuNH_4B_6C_6$ | $9.455\,Lu + 3\,B_{13}C_2 + 0.09091\,B_9H_{11} + H_3N \longrightarrow 8.455\,LuB_4 + LuB_6H_4C_6N$ | -13.5892 |
| $LuNH_4B_6C_6$ | $9.429\,Lu + 3\,B_{13}C_2 + H_3N + 0.1429\,B_5H_7 \longrightarrow 8.429\,LuB_4 + LuB_6H_4C_6N$ | -13.5363 |
| $LuNH_4B_6C_6$ | $0.8571\,Lu_4C_7 + 8.033\,N_2 + 7.029\,B_5H_7 \longrightarrow 2.429\,LuB_{12} + 15.07\,H_3N + LuB_6H_4C_6N$ | -13.4437 |
| $LuNH_4B_6C_6$ | $9.167\,Lu + 3\,B_{13}C_2 + 1.333\,H_3N \longrightarrow 8.167\,LuB_4 + 0.3333\,BN + LuB_6H_4C_6N$ | -13.4089 |
| $LuNH_4B_6C_6$ | $9.25\,Lu + 3\,B_{13}C_2 + 0.5\,H_2 + H_3N \longrightarrow 8.25\,LuB_4 + LuB_6H_4C_6N$ | -13.2145 |
| $LuNH_4B_6C_6$ | $3\,B_{13}C_2 + LuB_{12} + 15.55\,H_3N \longrightarrow 14.55\,BN + 6.091\,B_5H_7 + LuB_6H_4C_6N$ | -12.8438 |
| $LuNH_4B_6C_6$ | $1.2\,H_8C_5 + 5.1\,N_2 + 3.8\,LuB_4 \longrightarrow 9.2\,BN + 2.8\,LuH_2 + LuB_6H_4C_6N$ | -12.8297 |
| $LuNH_4B_6C_6$ | $9.25\,Lu + 3\,B_{13}C_2 + 1.333\,H_3N \longrightarrow 0.1667\,N_2 + 8.25\,LuB_4 + LuB_6H_4C_6N$ | -12.7732 |
| $LuNH_4B_6C_6$ | $4.75\,N_2 + 1.5\,Lu_3C_4 + 3\,LuB_2 + 2\,LuH_2 \longrightarrow 8.5\,LuN + LuB_6H_4C_6N$ | -12.7028 |
| $LuNH_4B_6C_6$ | $9.5\,Lu + 3\,B_{13}C_2 + 2\,H_2 + BN \longrightarrow 8.5\,LuB_4 + LuB_6H_4C_6N$ | -12.6313 |
| $LuNH_4B_6C_6$ | $9.068\,Lu + 3\,B_{13}C_2 + 0.3636\,B_9H_{11} + LuN \longrightarrow 9.068\,LuB_4 + LuB_6H_4C_6N$ | -12.4945 |
| $LuNH_4B_6C_6$ | $8.964\,Lu + 3\,B_{13}C_2 + LuN + 0.5714\,B_5H_7 \longrightarrow 8.964\,LuB_4 + LuB_6H_4C_6N$ | -12.2827 |
| $LuNH_4B_6C_6$ | $8.75\,Lu + 3\,B_{13}C_2 + H_3N + 0.5\,LuH_2 \longrightarrow 8.25\,LuB_4 + LuB_6H_4C_6N$ | -12.2435 |
| $LuNH_4B_6C_6$ | $3\,B_{13}C_2 + LuB_{12} + 14.26\,H_3N \longrightarrow 3.526\,B_9H_{11} + 13.26\,BN + LuB_6H_4C_6N$ | -11.2798 |
| $LuNH_4B_6C_6$ | $8.25\,Lu + 3\,B_{13}C_2 + LuN + 2\,H_2 \longrightarrow 8.25\,LuB_4 + LuB_6H_4C_6N$ | -10.9957 |
| $LuNH_4B_6C_6$ | $0.8571\,Lu_4C_7 + 6.992\,N_2 + 3.905\,B_9H_{11} \longrightarrow 2.429\,LuB_{12} + 12.98\,H_3N + LuB_6H_4C_6N$ | -10.9503 |
| $LuNH_4B_6C_6$ | $0.5\,N_2 + 7.25\,Lu + 3\,B_{13}C_2 + 2\,LuH_2 \longrightarrow 8.25\,LuB_4 + LuB_6H_4C_6N$ | -10.6545 |
| $LuNH_4B_6C_6$ | $0.125\,H_8C_5 + 8.234\,Lu + 2.688\,B_{13}C_2 + H_3N \longrightarrow 7.234\,LuB_4 + LuB_6H_4C_6N$ | -10.6422 |
| $LuNH_4B_6C_6$ | $4.769\,Lu_3C_4 + 8.308\,H_3N + 6\,BN \longrightarrow 2.615\,H_8C_5 + 13.31\,LuN + LuB_6H_4C_6N$ | -10.5635 |
| $LuNH_4B_6C_6$ | $3.281\,N_2 + 2.188\,Lu_3C_4 + 1.2\,B_5H_7 \longrightarrow 0.55\,H_8C_5 + 5.562\,LuN + LuB_6H_4C_6N$ | -10.4269 |
| $LuNH_4B_6C_6$ | $0.5\,H_8C_5 + 0.5\,N_2 + 9.375\,Lu + 1.75\,B_{13}C_2 \longrightarrow 8.375\,LuB_2 + LuB_6H_4C_6N$ | -10.1654 |
| $LuNH_4B_6C_6$ | $0.8571\,Lu_4C_7 + 4.214\,N_2 + 3\,LuB_2 + 2\,LuH_2 \longrightarrow 7.429\,LuN + LuB_6H_4C_6N$ | -9.7922 |
| $LuNH_4B_6C_6$ | $4\,N_2 + 1.5\,Lu_3C_4 + 1.5\,LuB_4 + 2\,LuH_2 \longrightarrow 7\,LuN + LuB_6H_4C_6N$ | -9.5076 |
| $LuNH_4B_6C_6$ | $3.75\,N_2 + 1.5\,Lu_3C_4 + 3\,LuB_2 + 2\,H_2 \longrightarrow 6.5\,LuN + LuB_6H_4C_6N$ | -9.5010 |
| $LuNH_4B_6C_6$ | $Lu + 3\,B_{13}C_2 + 11.73\,H_3N \longrightarrow 10.73\,BN + 4.455\,B_5H_7 + LuB_6H_4C_6N$ | -9.4385 |
| $LuNH_4B_6C_6$ | $3.5\,N_2 + 6\,C + LuB_{12} + 2\,H_2 \longrightarrow 6\,BN + LuB_6H_4C_6N$ | -9.1546 |
| $LuNH_4B_6C_6$ | $3\,B_{13}C_2 + LuB_4 + 13\,H_3N \longrightarrow 12\,BN + 5\,B_5H_7 + LuB_6H_4C_6N$ | -8.9523 |
| $LuNH_4B_6C_6$ | $3.031\,N_2 + 0.6667\,B_9H_{11} + 2.021\,Lu_3C_4 \longrightarrow 0.4167\,H_8C_5 + 5.063\,LuN + LuB_6H_4C_6N$ | -8.8398 |
| $LuNH_4B_6C_6$ | $7.5\,Lu + 3\,B_{13}C_2 + BN + 2\,LuH_2 \longrightarrow 8.5\,LuB_4 + LuB_6H_4C_6N$ | -8.7472 |
| $LuNH_4B_6C_6$ | $1.2\,H_8C_5 + 4.433\,N_2 + LuB_{12} \longrightarrow 1.867\,H_3N + 6\,BN + LuB_6H_4C_6N$ | -8.6240 |
| $LuNH_4B_6C_6$ | $0.5\,H_8C_5 + 9.875\,Lu + 1.75\,B_{13}C_2 + BN \longrightarrow 8.875\,LuB_2 + LuB_6H_4C_6N$ | -8.6113 |
| $LuNH_4B_6C_6$ | $3\,B_{13}C_2 + LuB_2 + 12.36\,H_3N \longrightarrow 11.36\,BN + 4.727\,B_5H_7 + LuB_6H_4C_6N$ | -8.4890 |
| $LuNH_4B_6C_6$ | $3.5\,N_2 + 1.5\,Lu_3C_4 + 0.5\,LuB_{12} + 2\,LuH_2 \longrightarrow 6\,LuN + LuB_6H_4C_6N$ | -8.3966 |
| $LuNH_4B_6C_6$ | $Lu + 3\,B_{13}C_2 + 10.79\,H_3N \longrightarrow 2.579\,B_9H_{11} + 9.789\,BN + LuB_6H_4C_6N$ | -8.2947 |
| $LuNH_4B_6C_6$ | $4.5\,N_2 + 1.5\,Lu_3C_4 + 4\,B_5H_7 \longrightarrow 3.5\,LuB_4 + 8\,H_3N + LuB_6H_4C_6N$ | -8.1927 |
| $LuNH_4B_6C_6$ | $3.25\,N_2 + 6\,B + 1.5\,Lu_3C_4 + 2\,LuH_2 \longrightarrow 5.5\,LuN + LuB_6H_4C_6N$ | -8.1499 |
| $LuNH_4B_6C_6$ | $3\,B_{13}C_2 + LuN + 11.41\,H_3N \longrightarrow 11.41\,BN + 4.318\,B_5H_7 + LuB_6H_4C_6N$ | -7.9244 |
| $LuNH_4B_6C_6$ | $0.5\,H_8C_5 + 3.5\,N_2 + 3.5\,C + LuB_{12} \longrightarrow 6\,BN + LuB_6H_4C_6N$ | -7.9037 |

| Compound | Chemical reaction | Reaction energy(eV) |
|---|---|---|
| $LuNH_4B_6C_6$ | $3.083\,N_2 + 1.5\,Lu_3C_4 + 3\,LuB_2 + 1.333\,H_3N \longrightarrow 6.5\,LuN + LuB_6H_4C_6N$ | -7.7356 |
| $LuNH_4B_6C_6$ | $1.2\,H_8C_5 + 9.5\,Lu + H_3N + 1.2\,B_5H_7 \longrightarrow 8.5\,LuH_2 + LuB_6H_4C_6N$ | -7.7307 |
| $LuNH_4B_6C_6$ | $3\,B_{13}C_2 + LuB_4 + 11.95\,H_3N \longrightarrow 2.895\,B_9H_{11} + 10.95\,BN + LuB_6H_4C_6N$ | -7.6685 |
| $LuNH_4B_6C_6$ | $2.833\,N_2 + 6\,C + LuB_{12} + 1.333\,H_3N \longrightarrow 6\,BN + LuB_6H_4C_6N$ | -7.3892 |
| $LuNH_4B_6C_6$ | $3\,B_{13}C_2 + LuB_2 + 11.37\,H_3N \longrightarrow 2.737\,B_9H_{11} + 10.37\,BN + LuB_6H_4C_6N$ | -7.2752 |
| $LuNH_4B_6C_6$ | $0.5\,H_8C_5 + 0.25\,Lu_4C_7 + 3.187\,N_2 + 0.875\,B_{13}C_2 \longrightarrow 5.375\,BN + LuB_6H_4C_6N$ | -7.1915 |
| $LuNH_4B_6C_6$ | $6.25\,Lu + 3\,B_{13}C_2 + LuN + 2\,LuH_2 \longrightarrow 8.25\,LuB_4 + LuB_6H_4C_6N$ | -7.1116 |
| $LuNH_4B_6C_6$ | $3\,B_{13}C_2 + 11.27\,H_3N + LuH_2 \longrightarrow 10.27\,BN + 4.545\,B_5H_7 + LuB_6H_4C_6N$ | -7.0266 |
| $LuNH_4B_6C_6$ | $3\,B_{13}C_2 + LuN + 10.5\,H_3N \longrightarrow 2.5\,B_9H_{11} + 10.5\,BN + LuB_6H_4C_6N$ | -6.8156 |
| $LuNH_4B_6C_6$ | $3.907\,N_2 + 2.222\,B_9H_{11} + 1.5\,Lu_3C_4 \longrightarrow 3.5\,LuB_4 + 6.815\,H_3N + LuB_6H_4C_6N$ | -6.7737 |
| $LuNH_4B_6C_6$ | $1.2\,H_8C_5 + 8.967\,Lu + 0.6667\,B_9H_{11} + H_3N \longrightarrow 7.967\,LuH_2 + LuB_6H_4C_6N$ | -6.7400 |
| $LuNH_4B_6C_6$ | $3.036\,N_2 + 1.5\,Lu_3C_4 + 1.571\,LuB_2 + 0.5714\,B_5H_7 \longrightarrow 5.071\,LuN + LuB_6H_4C_6N$ | -6.7358 |
| $LuNH_4B_6C_6$ | $1.2\,H_8C_5 + 24.33\,LuN + 5.867\,B_5H_7 \longrightarrow 23.33\,BN + 23.33\,LuH_2 + LuB_6H_4C_6N$ | -6.6651 |
| $LuNH_4B_6C_6$ | $0.5\,H_8C_5 + 8.375\,Lu + 1.75\,B_{13}C_2 + LuN \longrightarrow 8.375\,LuB_2 + LuB_6H_4C_6N$ | -6.6225 |
| $LuNH_4B_6C_6$ | $0.8571\,Lu_4C_7 + 3.464\,N_2 + 1.5\,LuB_4 + 2\,LuH_2 \longrightarrow 5.929\,LuN + LuB_6H_4C_6N$ | -6.5969 |
| $LuNH_4B_6C_6$ | $0.8571\,Lu_4C_7 + 3.214\,N_2 + 3\,LuB_2 + 2\,H_2 \longrightarrow 5.429\,LuN + LuB_6H_4C_6N$ | -6.5904 |
| $LuNH_4B_6C_6$ | $2.932\,N_2 + 0.3636\,B_9H_{11} + 1.5\,Lu_3C_4 + 1.364\,LuB_2 \longrightarrow 4.864\,LuN + LuB_6H_4C_6N$ | -6.3582 |
| $LuNH_4B_6C_6$ | $3\,N_2 + 1.5\,Lu_3C_4 + 1.5\,LuB_4 + 2\,H_2 \longrightarrow 5\,LuN + LuB_6H_4C_6N$ | -6.3058 |
| $LuNH_4B_6C_6$ | $1.25\,Lu_4C_7 + 2.5\,N_2 + 1.2\,B_5H_7 \longrightarrow 0.55\,H_8C_5 + 4\,LuN + LuB_6H_4C_6N$ | -6.1822 |
| $LuNH_4B_6C_6$ | $1.5\,Lu_3C_4 + 6.5\,LuB_2 + 8\,H_3N \longrightarrow 7\,BN + 10\,LuH_2 + LuB_6H_4C_6N$ | -6.1788 |
| $LuNH_4B_6C_6$ | $1.2\,H_8C_5 + 3.5\,N_2 + LuB_{12} \longrightarrow 2.8\,H_2 + 6\,BN + LuB_6H_4C_6N$ | -6.1524 |
| $LuNH_4B_6C_6$ | $1.2\,H_8C_5 + 0.5\,N_2 + 8\,Lu + 1.2\,B_5H_7 \longrightarrow 7\,LuH_2 + LuB_6H_4C_6N$ | -6.1417 |
| $LuNH_4B_6C_6$ | $5.667\,N_2 + 3\,B_{13}C_2 + 17.5\,LuH_2 \longrightarrow 16.5\,LuB_2 + 10.33\,H_3N + LuB_6H_4C_6N$ | -5.8902 |
| $LuNH_4B_6C_6$ | $3\,B_{13}C_2 + 10.32\,H_3N + LuH_2 \longrightarrow 2.632\,B_9H_{11} + 9.316\,BN + LuB_6H_4C_6N$ | -5.8595 |
| $LuNH_4B_6C_6$ | $2.904\,N_2 + 0.4615\,B_{13}C_2 + 1.269\,Lu_3C_4 + 2\,LuH_2 \longrightarrow 4.808\,LuN + LuB_6H_4C_6N$ | -5.7323 |
| $LuNH_4B_6C_6$ | $2.983\,N_2 + 1.5\,Lu_3C_4 + 1.2\,B_5H_7 \longrightarrow 3.5\,LuN + 1.467\,H_3N + LuB_6H_4C_6N$ | -5.6361 |
| $LuNH_4B_6C_6$ | $0.8571\,Lu_4C_7 + 2.964\,N_2 + 0.5\,LuB_{12} + 2\,LuH_2 \longrightarrow 4.929\,LuN + LuB_6H_4C_6N$ | -5.4859 |
| $LuNH_4B_6C_6$ | $0.5\,N_2 + 72\,B + 1.5\,Lu_3C_4 + 2\,LuH_2 \longrightarrow 5.5\,LuB_{12} + LuB_6H_4C_6N$ | -5.4366 |
| $LuNH_4B_6C_6$ | $0.8571\,Lu_4C_7 + 2.714\,N_2 + 6\,B + 2\,LuH_2 \longrightarrow 4.429\,LuN + LuB_6H_4C_6N$ | -5.2393 |
| $LuNH_4B_6C_6$ | $2\,N_2 + B_9H_{11} + 1.5\,Lu_3C_4 \longrightarrow 3\,BN + 3.5\,LuH_2 + LuB_6H_4C_6N$ | -5.2320 |
| $LuNH_4B_6C_6$ | $2.806\,N_2 + 0.6667\,B_9H_{11} + 1.5\,Lu_3C_4 \longrightarrow 3.5\,LuN + 1.111\,H_3N + LuB_6H_4C_6N$ | -5.2104 |
| $LuNH_4B_6C_6$ | $16.62\,C + LuB_{12} + 7\,H_3N \longrightarrow 2.125\,H_8C_5 + 6\,BN + LuB_6H_4C_6N$ | -5.2026 |
| $LuNH_4B_6C_6$ | $2.5\,N_2 + 1.5\,Lu_3C_4 + 0.5\,LuB_{12} + 2\,H_2 \longrightarrow 4\,LuN + LuB_6H_4C_6N$ | -5.1948 |
| $LuNH_4B_6C_6$ | $1.2\,H_8C_5 + 0.5\,N_2 + 7.467\,Lu + 0.6667\,B_9H_{11} \longrightarrow 6.467\,LuH_2 + LuB_6H_4C_6N$ | -5.1510 |
| $LuNH_4B_6C_6$ | $2.643\,N_2 + 1.5\,Lu_3C_4 + 0.7857\,LuB_4 + 0.5714\,B_5H_7 \longrightarrow 4.286\,LuN + LuB_6H_4C_6N$ | -5.0621 |
| $LuNH_4B_6C_6$ | $2.25\,N_2 + 6\,B + 1.5\,Lu_3C_4 + 2\,H_2 \longrightarrow 3.5\,LuN + LuB_6H_4C_6N$ | -4.9481 |
| $LuNH_4B_6C_6$ | $84\,B + LuN + 1.5\,Lu_3C_4 + 2\,LuH_2 \longrightarrow 6.5\,LuB_{12} + LuB_6H_4C_6N$ | -4.9432 |
| $LuNH_4B_6C_6$ | $1.155\,Lu_4C_7 + 2.31\,N_2 + 0.6667\,B_9H_{11} \longrightarrow 0.4167\,H_8C_5 + 3.619\,LuN + LuB_6H_4C_6N$ | -4.9185 |
| $LuNH_4B_6C_6$ | $2.591\,N_2 + 0.3636\,B_9H_{11} + 1.5\,Lu_3C_4 + 0.6818\,LuB_4 \longrightarrow 4.182\,LuN + LuB_6H_4C_6N$ | -4.9058 |
| $LuNH_4B_6C_6$ | $0.8571\,Lu_4C_7 + 2.548\,N_2 + 3\,LuB_2 + 1.333\,H_3N \longrightarrow 5.429\,LuN + LuB_6H_4C_6N$ | -4.8250 |
| $LuNH_4B_6C_6$ | $2.333\,B_{13}C_2 + 0.3333\,Lu_3C_4 + 8.97\,H_3N \longrightarrow 7.97\,BN + 3.273\,B_5H_7 + LuB_6H_4C_6N$ | -4.7258 |
| $LuNH_4B_6C_6$ | $2.333\,N_2 + 1.5\,Lu_3C_4 + 1.5\,LuB_4 + 1.333\,H_3N \longrightarrow 5\,LuN + LuB_6H_4C_6N$ | -4.5404 |
| $LuNH_4B_6C_6$ | $0.8571\,Lu_4C_7 + 3.5\,N_2 + 3.143\,B_5H_7 \longrightarrow 2.429\,LuB_4 + 6\,H_3N + LuB_6H_4C_6N$ | -4.4994 |
| $LuNH_4B_6C_6$ | $2.381\,N_2 + 1.5\,Lu_3C_4 + 0.2619\,LuB_{12} + 0.5714\,B_5H_7 \longrightarrow 3.762\,LuN + LuB_6H_4C_6N$ | -4.4802 |
| $LuNH_4B_6C_6$ | $2.364\,N_2 + 0.3636\,B_9H_{11} + 1.5\,Lu_3C_4 + 0.2273\,LuB_{12} \longrightarrow 3.727\,LuN + LuB_6H_4C_6N$ | -4.4008 |
| $LuNH_4B_6C_6$ | $2.25\,N_2 + 3.143\,B + 1.5\,Lu_3C_4 + 0.5714\,B_5H_7 \longrightarrow 3.5\,LuN + LuB_6H_4C_6N$ | -4.3510 |
| $LuNH_4B_6C_6$ | $2.25\,N_2 + 0.3636\,B_9H_{11} + 2.727\,B + 1.5\,Lu_3C_4 \longrightarrow 3.5\,LuN + LuB_6H_4C_6N$ | -4.2887 |
| $LuNH_4B_6C_6$ | $0.5\,H_8C_5 + 0.5\,N_2 + 5.187\,Lu + 1.75\,B_{13}C_2 \longrightarrow 4.187\,LuB_4 + LuB_6H_4C_6N$ | -4.2495 |
| $LuNH_4B_6C_6$ | $2.333\,B_{13}C_2 + 0.3333\,Lu_3C_4 + 8.281\,H_3N \longrightarrow 1.895\,B_9H_{11} + 7.281\,BN + LuB_6H_4C_6N$ | -3.8855 |
| $LuNH_4B_6C_6$ | $2.25\,N_2 + 2.75\,B_9H_{11} + 1.5\,Lu_3C_4 \longrightarrow 3.5\,LuN + 3.75\,B_5H_7 + LuB_6H_4C_6N$ | -3.8801 |
| $LuNH_4B_6C_6$ | $0.8571\,Lu_4C_7 + 2.5\,N_2 + 1.571\,LuB_2 + 0.5714\,B_5H_7 \longrightarrow 4\,LuN + LuB_6H_4C_6N$ | -3.8252 |
| $LuNH_4B_6C_6$ | $2.25\,N_2 + 0.6667\,B_9H_{11} + 1.5\,Lu_3C_4 \longrightarrow 3.5\,LuN + 1.667\,H_2 + LuB_6H_4C_6N$ | -3.7392 |
| $LuNH_4B_6C_6$ | $0.8571\,Lu_4C_7 + 5.964\,LuB_2 + 6.929\,H_3N \longrightarrow 5.929\,BN + 8.393\,LuH_2 + LuB_6H_4C_6N$ | -3.6965 |
| $LuNH_4B_6C_6$ | $2.25\,N_2 + 1.5\,Lu_3C_4 + 1.2\,B_5H_7 \longrightarrow 3.5\,LuN + 2.2\,H_2 + LuB_6H_4C_6N$ | -3.6941 |
| $LuNH_4B_6C_6$ | $0.25\,Lu_4C_7 + 2.125\,B_{13}C_2 + 8.108\,H_3N \longrightarrow 7.108\,BN + 2.903\,B_5H_7 + LuB_6H_4C_6N$ | -3.5113 |
| $LuNH_4B_6C_6$ | $0.8571\,Lu_4C_7 + 2.396\,N_2 + 0.3636\,B_9H_{11} + 1.364\,LuB_2 \longrightarrow 3.792\,LuN + LuB_6H_4C_6N$ | -3.4476 |
| $LuNH_4B_6C_6$ | $1.833\,N_2 + 1.5\,Lu_3C_4 + 0.5\,LuB_{12} + 1.333\,H_3N \longrightarrow 4\,LuN + LuB_6H_4C_6N$ | -3.4294 |
| $LuNH_4B_6C_6$ | $0.8571\,Lu_4C_7 + 2.464\,N_2 + 1.5\,LuB_4 + 2\,H_2 \longrightarrow 3.929\,LuN + LuB_6H_4C_6N$ | -3.3952 |
| $LuNH_4B_6C_6$ | $0.8571\,Lu_4C_7 + 3.034\,N_2 + 1.746\,B_9H_{11} \longrightarrow 2.429\,LuB_4 + 5.069\,H_3N + LuB_6H_4C_6N$ | -3.3845 |
| $LuNH_4B_6C_6$ | $0.5\,H_8C_5 + 2.813\,N_2 + 0.875\,Lu_3C_4 + 3\,LuB_2 \longrightarrow 4.625\,LuN + LuB_6H_4C_6N$ | -3.3804 |

| Compound | Chemical reaction | Reaction energy(eV) |
|---|---|---|
| $LuNH_4B_6C_6$ | $0.7253\,Lu_4C_7 + 2.451\,N_2 + 0.4615\,B_{13}C_2 + 2\,LuH_2 \longrightarrow 3.901\,LuN + LuB_6H_4C_6N$ | -3.2695 |
| $LuNH_4B_6C_6$ | $0.5\,N_2 + 48\,B + 1.5\,Lu_3C_4 + 2\,H_2 \longrightarrow 3.5\,LuB_{12} + LuB_6H_4C_6N$ | -3.2215 |
| $LuNH_4B_6C_6$ | $2.093\,N_2 + 0.2098\,B_{13}C_2 + 0.3636\,B_9H_{11} + 1.395\,Lu_3C_4 \longrightarrow 3.185\,LuN + LuB_6H_4C_6N$ | -3.1898 |
| $LuNH_4B_6C_6$ | $2.867\,N_2 + 1.5\,Lu_3C_4 + 2.6\,B_5H_7 \longrightarrow 3.5\,LuB_2 + 4.733\,H_3N + LuB_6H_4C_6N$ | -3.1866 |
| $LuNH_4B_6C_6$ | $1.583\,N_2 + 6\,B + 1.5\,Lu_3C_4 + 1.333\,H_3N \longrightarrow 3.5\,LuN + LuB_6H_4C_6N$ | -3.1827 |
| $LuNH_4B_6C_6$ | $0.5\,N_2 + 28\,B + 1.5\,Lu_3C_4 + 2\,LuH_2 \longrightarrow 5.5\,LuB_4 + LuB_6H_4C_6N$ | -3.1718 |
| $LuNH_4B_6C_6$ | $2.069\,N_2 + 0.2418\,B_{13}C_2 + 1.379\,Lu_3C_4 + 0.5714\,B_5H_7 \longrightarrow 3.137\,LuN + LuB_6H_4C_6N$ | -3.0846 |
| $LuNH_4B_6C_6$ | $0.8571\,Lu_4C_7 + 0.5\,N_2 + 59.14\,B + 2\,LuH_2 \longrightarrow 4.429\,LuB_{12} + LuB_6H_4C_6N$ | -3.0545 |
| $LuNH_4B_6C_6$ | $2.917\,N_2 + 3\,B_{13}C_2 + 9.25\,LuH_2 \longrightarrow 8.25\,LuB_4 + 4.833\,H_3N + LuB_6H_4C_6N$ | -2.9744 |
| $LuNH_4B_6C_6$ | $71\,B + 1.5\,Lu_3C_4 + BN + 2\,LuH_2 \longrightarrow 5.5\,LuB_{12} + LuB_6H_4C_6N$ | -2.8698 |
| $LuNH_4B_6C_6$ | $0.25\,Lu_4C_7 + 2.125\,B_{13}C_2 + 7.497\,H_3N \longrightarrow 1.681\,B_9H_{11} + 6.497\,BN + LuB_6H_4C_6N$ | -2.7658 |
| $LuNH_4B_6C_6$ | $60\,B + LuN + 1.5\,Lu_3C_4 + 2\,H_2 \longrightarrow 4.5\,LuB_{12} + LuB_6H_4C_6N$ | -2.7282 |
| $LuNH_4B_6C_6$ | $0.8571\,Lu_4C_7 + 2.448\,N_2 + 1.2\,B_5H_7 \longrightarrow 2.429\,LuN + 1.467\,H_3N + LuB_6H_4C_6N$ | -2.7255 |
| $LuNH_4B_6C_6$ | $0.5\,N_2 + 45.14\,B + 1.5\,Lu_3C_4 + 0.5714\,B_5H_7 \longrightarrow 3.5\,LuB_{12} + LuB_6H_4C_6N$ | -2.6243 |
| $LuNH_4B_6C_6$ | $1.2\,H_8C_5 + 7\,Lu + LuN + 1.2\,B_5H_7 \longrightarrow 7\,LuH_2 + LuB_6H_4C_6N$ | -2.5988 |
| $LuNH_4B_6C_6$ | $0.5\,N_2 + 0.3636\,B_9H_{11} + 44.73\,B + 1.5\,Lu_3C_4 \longrightarrow 3.5\,LuB_{12} + LuB_6H_4C_6N$ | -2.5621 |
| $LuNH_4B_6C_6$ | $0.8571\,Lu_4C_7 + 71.14\,B + LuN + 2\,LuH_2 \longrightarrow 5.429\,LuB_{12} + LuB_6H_4C_6N$ | -2.5612 |
| $LuNH_4B_6C_6$ | $1.904\,N_2 + 0.4615\,B_{13}C_2 + 1.269\,Lu_3C_4 + 2\,H_2 \longrightarrow 2.808\,LuN + LuB_6H_4C_6N$ | -2.5306 |
| $LuNH_4B_6C_6$ | $1.429\,N_2 + 1.5\,Lu_3C_4 + 1.571\,B_5H_7 \longrightarrow 1.857\,BN + 3.5\,LuH_2 + LuB_6H_4C_6N$ | -2.4698 |
| $LuNH_4B_6C_6$ | $54\,B + 1.5\,Lu_3C_4 + H_3N + 0.5\,LuH_2 \longrightarrow 4\,LuB_{12} + LuB_6H_4C_6N$ | -2.4512 |
| $LuNH_4B_6C_6$ | $1.2\,H_8C_5 + 7.3\,Lu + BN + B_5H_7 \longrightarrow 6.3\,LuH_2 + LuB_6H_4C_6N$ | -2.4245 |
| $LuNH_4B_6C_6$ | $0.5\,H_8C_5 + 5.437\,Lu + 1.75\,B_{13}C_2 + BN \longrightarrow 4.437\,LuB_4 + LuB_6H_4C_6N$ | -2.3422 |
| $LuNH_4B_6C_6$ | $48.33\,B + 1.5\,Lu_3C_4 + 1.333\,H_3N \longrightarrow 3.5\,LuB_{12} + 0.3333\,BN + LuB_6H_4C_6N$ | -2.3117 |
| $LuNH_4B_6C_6$ | $0.8571\,Lu_4C_7 + 2.27\,N_2 + 0.6667\,B_9H_{11} \longrightarrow 2.429\,LuN + 1.111\,H_3N + LuB_6H_4C_6N$ | -2.2998 |
| $LuNH_4B_6C_6$ | $0.8571\,Lu_4C_7 + 1.964\,N_2 + 0.5\,LuB_{12} + 2\,H_2 \longrightarrow 2.929\,LuN + LuB_6H_4C_6N$ | -2.2842 |
| $LuNH_4B_6C_6$ | $32\,B + LuN + 1.5\,Lu_3C_4 + 2\,LuH_2 \longrightarrow 6.5\,LuB_4 + LuB_6H_4C_6N$ | -2.2667 |
| $LuNH_4B_6C_6$ | $2.481\,N_2 + 1.444\,B_9H_{11} + 1.5\,Lu_3C_4 \longrightarrow 3.5\,LuB_2 + 3.963\,H_3N + LuB_6H_4C_6N$ | -2.2643 |
| $LuNH_4B_6C_6$ | $0.8571\,Lu_4C_7 + 2.107\,N_2 + 0.7857\,LuB_4 + 0.5714\,B_5H_7 \longrightarrow 3.214\,LuN + LuB_6H_4C_6N$ | -2.1515 |
| $LuNH_4B_6C_6$ | $57.14\,B + LuN + 1.5\,Lu_3C_4 + 0.5714\,B_5H_7 \longrightarrow 4.5\,LuB_{12} + LuB_6H_4C_6N$ | -2.1310 |
| $LuNH_4B_6C_6$ | $7.333\,Lu + 6\,C + 1.333\,H_3N + 6\,BN \longrightarrow 6.333\,LuN + LuB_6H_4C_6N$ | -2.0761 |
| $LuNH_4B_6C_6$ | $0.3636\,B_9H_{11} + 56.73\,B + LuN + 1.5\,Lu_3C_4 \longrightarrow 4.5\,LuB_{12} + LuB_6H_4C_6N$ | -2.0687 |
| $LuNH_4B_6C_6$ | $0.8571\,Lu_4C_7 + 1.714\,N_2 + 6\,B + 2\,H_2 \longrightarrow 2.429\,LuN + LuB_6H_4C_6N$ | -2.0375 |
| $LuNH_4B_6C_6$ | $0.8571\,Lu_4C_7 + 2.055\,N_2 + 0.3636\,B_9H_{11} + 0.6818\,LuB_4 \longrightarrow 3.11\,LuN + LuB_6H_4C_6N$ | -1.9952 |
| $LuNH_4B_6C_6$ | $3\,B_{13}C_2 + 11.56\,LuN + 1.333\,H_3N \longrightarrow 10.56\,LuB_2 + 11.89\,BN + LuB_6H_4C_6N$ | -1.9649 |
| $LuNH_4B_6C_6$ | $48\,B + 1.5\,Lu_3C_4 + 0.5\,H_2 + H_3N \longrightarrow 3.5\,LuB_{12} + LuB_6H_4C_6N$ | -1.8974 |
| $LuNH_4B_6C_6$ | $0.5\,N_2 + 3\,B_{13}C_2 + 0.3636\,B_9H_{11} + 20.14\,LuB_2 \longrightarrow 19.14\,LuB_4 + LuB_6H_4C_6N$ | -1.8134 |
| $LuNH_4B_6C_6$ | $0.5\,N_2 + 20\,B + 1.5\,Lu_3C_4 + 2\,H_2 \longrightarrow 3.5\,LuB_4 + LuB_6H_4C_6N$ | -1.7803 |
| $LuNH_4B_6C_6$ | $1.2\,H_8C_5 + 2.514\,B_9H_{11} + 17.63\,LuN \longrightarrow 16.63\,BN + 16.63\,LuH_2 + LuB_6H_4C_6N$ | -1.7604 |
| $LuNH_4B_6C_6$ | $0.5\,N_2 + 3\,B_{13}C_2 + 19.93\,LuB_2 + 0.5714\,B_5H_7 \longrightarrow 18.93\,LuB_4 + LuB_6H_4C_6N$ | -1.7484 |
| $LuNH_4B_6C_6$ | $47.29\,B + 1.5\,Lu_3C_4 + H_3N + 0.1429\,B_5H_7 \longrightarrow 3.5\,LuB_{12} + LuB_6H_4C_6N$ | -1.7481 |
| $LuNH_4B_6C_6$ | $0.09091\,B_9H_{11} + 47.18\,B + 1.5\,Lu_3C_4 + H_3N \longrightarrow 3.5\,LuB_{12} + LuB_6H_4C_6N$ | -1.7326 |
| $LuNH_4B_6C_6$ | $0.5\,N_2 + 1.5\,Lu_3C_4 + 3.5\,LuB_{12} + 2\,LuH_2 \longrightarrow 9\,LuB_4 + LuB_6H_4C_6N$ | -1.7306 |
| $LuNH_4B_6C_6$ | $0.5\,H_8C_5 + 0.5\,Lu_4C_7 + 2.5\,N_2 + 3\,LuB_2 \longrightarrow 4\,LuN + LuB_6H_4C_6N$ | -1.6825 |
| $LuNH_4B_6C_6$ | $1.5\,Lu_3C_4 + 2.6\,LuB_4 + 5.4\,H_3N \longrightarrow 4.4\,BN + 6.1\,LuH_2 + LuB_6H_4C_6N$ | -1.6799 |
| $LuNH_4B_6C_6$ | $0.8571\,Lu_4C_7 + 1.798\,N_2 + 1.5\,LuB_4 + 1.333\,H_3N \longrightarrow 3.929\,LuN + LuB_6H_4C_6N$ | -1.6298 |
| $LuNH_4B_6C_6$ | $44\,B + 1.5\,Lu_3C_4 + 1.333\,H_3N \longrightarrow 0.3333\,LuN + 3.167\,LuB_{12} + LuB_6H_4C_6N$ | -1.6205 |
| $LuNH_4B_6C_6$ | $1.2\,H_8C_5 + 6.467\,Lu + 0.6667\,B_9H_{11} + LuN \longrightarrow 6.467\,LuH_2 + LuB_6H_4C_6N$ | -1.6081 |
| $LuNH_4B_6C_6$ | $1.2\,H_8C_5 + 6.856\,Lu + 0.5556\,B_9H_{11} + BN \longrightarrow 5.856\,LuH_2 + LuB_6H_4C_6N$ | -1.5989 |
| $LuNH_4B_6C_6$ | $0.8571\,Lu_4C_7 + 1.845\,N_2 + 0.2619\,LuB_{12} + 0.5714\,B_5H_7 \longrightarrow 2.69\,LuN + LuB_6H_4C_6N$ | -1.5696 |
| $LuNH_4B_6C_6$ | $1.2\,H_8C_5 + 8.8\,Lu + 6\,BN \longrightarrow 5\,LuN + 2.8\,LuH_2 + LuB_6H_4C_6N$ | -1.5531 |
| $LuNH_4B_6C_6$ | $1.2\,H_8C_5 + 10.58\,Lu + 10.58\,BN \longrightarrow 0.5091\,B_9H_{11} + 9.582\,LuN + LuB_6H_4C_6N$ | -1.5112 |
| $LuNH_4B_6C_6$ | $0.8571\,Lu_4C_7 + 1.828\,N_2 + 0.3636\,B_9H_{11} + 0.2273\,LuB_{12} \longrightarrow 2.656\,LuN + LuB_6H_4C_6N$ | -1.4902 |
| $LuNH_4B_6C_6$ | $0.5\,N_2 + 3\,B_{13}C_2 + 18.5\,LuB_2 + 2\,H_2 \longrightarrow 17.5\,LuB_4 + LuB_6H_4C_6N$ | -1.4706 |
| $LuNH_4B_6C_6$ | $48\,B + 1.5\,Lu_3C_4 + 1.333\,H_3N \longrightarrow 0.1667\,N_2 + 3.5\,LuB_{12} + LuB_6H_4C_6N$ | -1.4561 |
| $LuNH_4B_6C_6$ | $0.8571\,Lu_4C_7 + 1.714\,N_2 + 3.143\,B + 0.5714\,B_5H_7 \longrightarrow 2.429\,LuN + LuB_6H_4C_6N$ | -1.4404 |
| $LuNH_4B_6C_6$ | $3\,B_{13}C_2 + 7.333\,LuN + 1.333\,H_3N \longrightarrow 6.333\,LuB_4 + 7.667\,BN + LuB_6H_4C_6N$ | -1.4145 |
| $LuNH_4B_6C_6$ | $4.7\,Lu + 6\,C + H_3N + 1.2\,B_5H_7 \longrightarrow 3.7\,LuH_2 + LuB_6H_4C_6N$ | -1.4111 |
| $LuNH_4B_6C_6$ | $0.8571\,Lu_4C_7 + 1.714\,N_2 + 0.3636\,B_9H_{11} + 2.727\,B \longrightarrow 2.429\,LuN + LuB_6H_4C_6N$ | -1.3781 |
| $LuNH_4B_6C_6$ | $0.5\,N_2 + 4.023\,Lu + 3\,B_{13}C_2 + 0.3636\,B_9H_{11} \longrightarrow 3.023\,LuB_{12} + LuB_6H_4C_6N$ | -1.3354 |
| $LuNH_4B_6C_6$ | $8.667\,B + 1.5\,Lu_3C_4 + 3.667\,H_3N \longrightarrow 2.667\,BN + 3.5\,LuH_2 + LuB_6H_4C_6N$ | -1.3349 |

| Compound | Chemical reaction | Reaction energy(eV) |
|---|---|---|
| LuNH$_4$B$_6$C$_6$ | 0.5 N$_2$ + 3.988 Lu + 3 B$_{13}$C$_2$ + 0.5714 B$_5$H$_7$ ⟶ 2.988 LuB$_{12}$ + LuB$_6$H$_4$C$_6$N | -1.2920 |
| LuNH$_4$B$_6$C$_6$ | 3 B$_{13}$C$_2$ + 0.3636 B$_9$H$_{11}$ + 13.09 LuN ⟶ 12.09 LuB$_2$ + 12.09 BN + LuB$_6$H$_4$C$_6$N | -1.2593 |
| LuNH$_4$B$_6$C$_6$ | 0.8571 Lu$_4$C$_7$ + 0.5 N$_2$ + 23.71 B + 2 LuH$_2$ ⟶ 4.429 LuB$_4$ + LuB$_6$H$_4$C$_6$N | -1.2309 |
| LuNH$_4$B$_6$C$_6$ | 1.5 Lu$_3$C$_4$ + 8.912 LuB$_4$ + 13.41 H$_3$N ⟶ 3.294 B$_9$H$_{11}$ + 12.41 LuN + LuB$_6$H$_4$C$_6$N | -1.2298 |
| LuNH$_4$B$_6$C$_6$ | 0.5 N$_2$ + 17.14 B + 1.5 Lu$_3$C$_4$ + 0.5714 B$_5$H$_7$ ⟶ 3.5 LuB$_4$ + LuB$_6$H$_4$C$_6$N | -1.1831 |
| LuNH$_4$B$_6$C$_6$ | 3 B$_{13}$C$_2$ + 12.95 LuN + 0.5714 B$_5$H$_7$ ⟶ 11.95 LuB$_2$ + 11.95 BN + LuB$_6$H$_4$C$_6$N | -1.1763 |
| LuNH$_4$B$_6$C$_6$ | 0.5 N$_2$ + 3.75 Lu + 3 B$_{13}$C$_2$ + 2 H$_2$ ⟶ 2.75 LuB$_{12}$ + LuB$_6$H$_4$C$_6$N | -1.1631 |
| LuNH$_4$B$_6$C$_6$ | 1.5 Lu$_3$C$_4$ + 0.7647 LuB$_{12}$ + 4.176 H$_3$N ⟶ 3.176 BN + 4.265 LuH$_2$ + LuB$_6$H$_4$C$_6$N | -1.1215 |
| LuNH$_4$B$_6$C$_6$ | 0.5 N$_2$ + 0.3636 B$_9$H$_{11}$ + 16.73 B + 1.5 Lu$_3$C$_4$ ⟶ 3.5 LuB$_4$ + LuB$_6$H$_4$C$_6$N | -1.1208 |
| LuNH$_4$B$_6$C$_6$ | 1.417 N$_2$ + 0.6667 B$_9$H$_{11}$ + 1.5 Lu$_3$C$_4$ ⟶ 1.833 LuN + 1.667 LuH$_2$ + LuB$_6$H$_4$C$_6$N | -1.0711 |
| LuNH$_4$B$_6$C$_6$ | 0.8571 Lu$_4$C$_7$ + 2.367 N$_2$ + 2.171 B$_5$H$_7$ ⟶ 2.429 LuB$_2$ + 3.733 H$_3$N + LuB$_6$H$_4$C$_6$N | -1.0258 |
| LuNH$_4$B$_6$C$_6$ | 2.5 N$_2$ + 6 C + 3 LuB$_2$ + 2 LuH$_2$ ⟶ 4 LuN + LuB$_6$H$_4$C$_6$N | -1.0154 |
| LuNH$_4$B$_6$C$_6$ | 0.8571 Lu$_4$C$_7$ + 1.714 N$_2$ + 2.75 B$_9$H$_{11}$ ⟶ 2.429 LuN + 3.75 B$_5$H$_7$ + LuB$_6$H$_4$C$_6$N | -0.9695 |
| LuNH$_4$B$_6$C$_6$ | 24 B + LuN + 1.5 Lu$_3$C$_4$ + 2 H$_2$ ⟶ 4.5 LuB$_4$ + LuB$_6$H$_4$C$_6$N | -0.8751 |
| LuNH$_4$B$_6$C$_6$ | 20.33 B + 1.5 Lu$_3$C$_4$ + 1.333 H$_3$N ⟶ 3.5 LuB$_4$ + 0.3333 BN + LuB$_6$H$_4$C$_6$N | -0.8704 |
| LuNH$_4$B$_6$C$_6$ | 1.2 H$_8$C$_5$ + 10 Lu + 10 BN ⟶ 9 LuN + 0.8 B$_5$H$_7$ + LuB$_6$H$_4$C$_6$N | -0.8560 |
| LuNH$_4$B$_6$C$_6$ | 0.8571 Lu$_4$C$_7$ + 0.5 N$_2$ + 35.14 B + 2 H$_2$ ⟶ 2.429 LuB$_{12}$ + LuB$_6$H$_4$C$_6$N | -0.8394 |
| LuNH$_4$B$_6$C$_6$ | 0.8571 Lu$_4$C$_7$ + 1.714 N$_2$ + 0.6667 B$_9$H$_{11}$ ⟶ 2.429 LuN + 1.667 H$_2$ + LuB$_6$H$_4$C$_6$N | -0.8286 |
| LuNH$_4$B$_6$C$_6$ | 1.2 H$_8$C$_5$ + 5.3 Lu + 6 B + H$_3$N ⟶ 4.3 LuH$_2$ + LuB$_6$H$_4$C$_6$N | -0.8281 |
| LuNH$_4$B$_6$C$_6$ | 22 B + 1.5 Lu$_3$C$_4$ + H$_3$N + 0.5 LuH$_2$ ⟶ 4 LuB$_4$ + LuB$_6$H$_4$C$_6$N | -0.8041 |
| LuNH$_4$B$_6$C$_6$ | 0.8571 Lu$_4$C$_7$ + 1.714 N$_2$ + 1.2 B$_5$H$_7$ ⟶ 2.429 LuN + 2.2 H$_2$ + LuB$_6$H$_4$C$_6$N | -0.7835 |
| LuNH$_4$B$_6$C$_6$ | 3 B$_{13}$C$_2$ + 12 LuN + 2 H$_2$ ⟶ 11 LuB$_2$ + 11 BN + LuB$_6$H$_4$C$_6$N | -0.7743 |
| LuNH$_4$B$_6$C$_6$ | 1.237 N$_2$ + 0.4615 B$_{13}$C$_2$ + 1.269 Lu$_3$C$_4$ + 1.333 H$_3$N ⟶ 2.808 LuN + LuB$_6$H$_4$C$_6$N | -0.7651 |
| LuNH$_4$B$_6$C$_6$ | 0.5 N$_2$ + 1.5 Lu$_3$C$_4$ + 2.5 LuB$_{12}$ + 2 H$_2$ ⟶ 6 LuB$_4$ + LuB$_6$H$_4$C$_6$N | -0.7508 |
| LuNH$_4$B$_6$C$_6$ | 0.5 H$_8$C$_5$ + 4.187 Lu + 1.75 B$_{13}$C$_2$ + LuN ⟶ 4.187 LuB$_4$ + LuB$_6$H$_4$C$_6$N | -0.7066 |
| LuNH$_4$B$_6$C$_6$ | 47 B + 1.5 Lu$_3$C$_4$ + 2 H$_2$ + BN ⟶ 3.5 LuB$_{12}$ + LuB$_6$H$_4$C$_6$N | -0.6547 |
| LuNH$_4$B$_6$C$_6$ | 3 B$_{13}$C$_2$ + 0.3636 B$_9$H$_{11}$ + 8.255 LuN ⟶ 7.255 LuB$_4$ + 7.255 BN + LuB$_6$H$_4$C$_6$N | -0.6289 |
| LuNH$_4$B$_6$C$_6$ | LuN + 1.5 Lu$_3$C$_4$ + 4 LuB$_{12}$ + 2 LuH$_2$ ⟶ 10.5 LuB$_4$ + LuB$_6$H$_4$C$_6$N | -0.6195 |
| LuNH$_4$B$_6$C$_6$ | 1.5 Lu$_3$C$_4$ + 6.885 LuB$_4$ + 11.38 H$_3$N ⟶ 10.38 LuN + 4.308 B$_5$H$_7$ + LuB$_6$H$_4$C$_6$N | -0.6071 |
| LuNH$_4$B$_6$C$_6$ | 27 B + 1.5 Lu$_3$C$_4$ + BN + 2 LuH$_2$ ⟶ 5.5 LuB$_4$ + LuB$_6$H$_4$C$_6$N | -0.6050 |
| LuNH$_4$B$_6$C$_6$ | 0.125 H$_8$C$_5$ + 42.37 B + 1.344 Lu$_3$C$_4$ + H$_3$N ⟶ 3.031 LuB$_{12}$ + LuB$_6$H$_4$C$_6$N | -0.5985 |
| LuNH$_4$B$_6$C$_6$ | 4.083 Lu + 3 B$_{13}$C$_2$ + 1.333 H$_3$N ⟶ 0.3333 LuN + 2.75 LuB$_{12}$ + LuB$_6$H$_4$C$_6$N | -0.5786 |
| LuNH$_4$B$_6$C$_6$ | 13.77 B + 1.5 Lu$_3$C$_4$ + 4.5 H$_3$N ⟶ 0.8636 B$_9$H$_{11}$ + 3.5 LuN + LuB$_6$H$_4$C$_6$N | -0.5560 |
| LuNH$_4$B$_6$C$_6$ | 3 B$_{13}$C$_2$ + 8.171 LuN + 0.5714 B$_5$H$_7$ ⟶ 7.171 LuB$_4$ + 7.171 BN + LuB$_6$H$_4$C$_6$N | -0.5531 |
| LuNH$_4$B$_6$C$_6$ | 0.8571 Lu$_4$C$_7$ + 1.298 N$_2$ + 0.5 LuB$_{12}$ + 1.333 H$_3$N ⟶ 2.929 LuN + LuB$_6$H$_4$C$_6$N | -0.5188 |
| LuNH$_4$B$_6$C$_6$ | 0.8571 Lu$_4$C$_7$ + 58.14 B + BN + 2 LuH$_2$ ⟶ 4.429 LuB$_{12}$ + LuB$_6$H$_4$C$_6$N | -0.4878 |
| LuNH$_4$B$_6$C$_6$ | 0.7972 Lu$_4$C$_7$ + 1.594 N$_2$ + 0.2098 B$_{13}$C$_2$ + 0.3636 B$_9$H$_{11}$ ⟶ 2.189 LuN + LuB$_6$H$_4$C$_6$N | -0.4827 |
| LuNH$_4$B$_6$C$_6$ | 3 B$_{13}$C$_2$ + 18.33 LuB$_2$ + 1.333 H$_3$N ⟶ 17.33 LuB$_4$ + 0.3333 BN + LuB$_6$H$_4$C$_6$N | -0.4586 |
| LuNH$_4$B$_6$C$_6$ | 20 B + 1.5 Lu$_3$C$_4$ + 0.5 H$_2$ + H$_3$N ⟶ 3.5 LuB$_4$ + LuB$_6$H$_4$C$_6$N | -0.4562 |
| LuNH$_4$B$_6$C$_6$ | 4.167 Lu + 0.6667 B$_9$H$_{11}$ + 6 C + H$_3$N ⟶ 3.167 LuH$_2$ + LuB$_6$H$_4$C$_6$N | -0.4204 |
| LuNH$_4$B$_6$C$_6$ | 3 B$_{13}$C$_2$ + 19.17 LuB$_2$ + 1.333 H$_3$N ⟶ 0.3333 LuN + 17.83 LuB$_4$ + LuB$_6$H$_4$C$_6$N | -0.4152 |
| LuNH$_4$B$_6$C$_6$ | 0.5 N$_2$ + 3 B$_{13}$C$_2$ + 14.5 LuB$_2$ + 2 LuH$_2$ ⟶ 15.5 LuB$_4$ + LuB$_6$H$_4$C$_6$N | -0.4120 |
| LuNH$_4$B$_6$C$_6$ | 0.7881 Lu$_4$C$_7$ + 1.576 N$_2$ + 0.2418 B$_{13}$C$_2$ + 0.5714 B$_5$H$_7$ ⟶ 2.152 LuN + LuB$_6$H$_4$C$_6$N | -0.4086 |
| LuNH$_4$B$_6$C$_6$ | 12.79 B + 1.5 Lu$_3$C$_4$ + 4.5 H$_3$N ⟶ 3.5 LuN + 1.357 B$_5$H$_7$ + LuB$_6$H$_4$C$_6$N | -0.4081 |
| LuNH$_4$B$_6$C$_6$ | 0.8571 Lu$_4$C$_7$ + 47.14 B + LuN + 2 H$_2$ ⟶ 3.429 LuB$_{12}$ + LuB$_6$H$_4$C$_6$N | -0.3461 |
| LuNH$_4$B$_6$C$_6$ | 0.8571 Lu$_4$C$_7$ + 27.71 B + LuN + 2 LuH$_2$ ⟶ 5.429 LuB$_4$ + LuB$_6$H$_4$C$_6$N | -0.3258 |
| LuNH$_4$B$_6$C$_6$ | 18.67 B + 1.5 Lu$_3$C$_4$ + 1.333 H$_3$N ⟶ 0.3333 LuN + 3.167 LuB$_4$ + LuB$_6$H$_4$C$_6$N | -0.3165 |
| LuNH$_4$B$_6$C$_6$ | 19.29 B + 1.5 Lu$_3$C$_4$ + H$_3$N + 0.1429 B$_5$H$_7$ ⟶ 3.5 LuB$_4$ + LuB$_6$H$_4$C$_6$N | -0.3069 |
| LuNH$_4$B$_6$C$_6$ | 0.5 N$_2$ + 1.5 Lu$_3$C$_4$ + 2.143 LuB$_{12}$ + 0.5714 B$_5$H$_7$ ⟶ 5.643 LuB$_4$ + LuB$_6$H$_4$C$_6$N | -0.3007 |
| LuNH$_4$B$_6$C$_6$ | 0.09091 B$_9$H$_{11}$ + 19.18 B + 1.5 Lu$_3$C$_4$ + H$_3$N ⟶ 3.5 LuB$_4$ + LuB$_6$H$_4$C$_6$N | -0.2913 |
| LuNH$_4$B$_6$C$_6$ | 21.14 B + LuN + 1.5 Lu$_3$C$_4$ + 0.5714 B$_5$H$_7$ ⟶ 4.5 LuB$_4$ + LuB$_6$H$_4$C$_6$N | -0.2780 |
| LuNH$_4$B$_6$C$_6$ | 0.8571 Lu$_4$C$_7$ + 1.048 N$_2$ + 6 B + 1.333 H$_3$N ⟶ 2.429 LuN + LuB$_6$H$_4$C$_6$N | -0.2721 |
| LuNH$_4$B$_6$C$_6$ | 0.5 N$_2$ + 0.3636 B$_9$H$_{11}$ + 1.5 Lu$_3$C$_4$ + 2.091 LuB$_{12}$ ⟶ 5.591 LuB$_4$ + LuB$_6$H$_4$C$_6$N | -0.2598 |
| LuNH$_4$B$_6$C$_6$ | 0.8571 Lu$_4$C$_7$ + 2.045 N$_2$ + 1.206 B$_9$H$_{11}$ ⟶ 2.429 LuB$_2$ + 3.09 H$_3$N + LuB$_6$H$_4$C$_6$N | -0.2555 |
| LuNH$_4$B$_6$C$_6$ | 0.8571 Lu$_4$C$_7$ + 0.5 N$_2$ + 32.29 B + 0.5714 B$_5$H$_7$ ⟶ 2.429 LuB$_{12}$ + LuB$_6$H$_4$C$_6$N | -0.2423 |
| LuNH$_4$B$_6$C$_6$ | 3 B$_{13}$C$_2$ + 0.09091 B$_9$H$_{11}$ + 18.91 LuB$_2$ + H$_3$N ⟶ 17.91 LuB$_4$ + LuB$_6$H$_4$C$_6$N | -0.2322 |
| LuNH$_4$B$_6$C$_6$ | 3 B$_{13}$C$_2$ + 18.86 LuB$_2$ + H$_3$N + 0.1429 B$_5$H$_7$ ⟶ 17.86 LuB$_4$ + LuB$_6$H$_4$C$_6$N | -0.2160 |
| LuNH$_4$B$_6$C$_6$ | 0.3636 B$_9$H$_{11}$ + 20.73 B + LuN + 1.5 Lu$_3$C$_4$ ⟶ 4.5 LuB$_4$ + LuB$_6$H$_4$C$_6$N | -0.2157 |
| LuNH$_4$B$_6$C$_6$ | 3 B$_{13}$C$_2$ + 7.6 LuN + 2 H$_2$ ⟶ 6.6 LuB$_4$ + 6.6 BN + LuB$_6$H$_4$C$_6$N | -0.2007 |
| LuNH$_4$B$_6$C$_6$ | 0.5 H$_8$C$_5$ + 2.063 N$_2$ + 0.875 Lu$_3$C$_4$ + 1.5 LuB$_4$ ⟶ 3.125 LuN + LuB$_6$H$_4$C$_6$N | -0.1851 |

| Compound | Chemical reaction | Reaction energy(eV) |
|---|---|---|
| LuNH$_4$B$_6$C$_6$ | 0.8571 Lu$_4$C$_7$ + 0.5 N$_2$ + 0.3636 B$_9$H$_{11}$ + 31.87 B $\longrightarrow$ 2.429 LuB$_{12}$ + LuB$_6$H$_4$C$_6$N | -0.1800 |
| LuNH$_4$B$_6$C$_6$ | 1.15 N$_2$ + 1.5 Lu$_3$C$_4$ + 1.2 B$_5$H$_7$ $\longrightarrow$ 1.3 LuN + 2.2 LuH$_2$ + LuB$_6$H$_4$C$_6$N | -0.1722 |
| LuNH$_4$B$_6$C$_6$ | 3.722 Lu + 3 B$_{13}$C$_2$ + 1.333 H$_3$N $\longrightarrow$ 2.722 LuB$_{12}$ + 0.3333 BN + LuB$_6$H$_4$C$_6$N | -0.1685 |
| LuNH$_4$B$_6$C$_6$ | 3 B$_{13}$C$_2$ + 18.5 LuB$_2$ + 0.5 H$_2$ + H$_3$N $\longrightarrow$ 17.5 LuB$_4$ + LuB$_6$H$_4$C$_6$N | -0.1465 |
| LuNH$_4$B$_6$C$_6$ | 1.5 Lu$_3$C$_4$ + 1.443 LuB$_{12}$ + 5.943 H$_3$N $\longrightarrow$ 1.257 B$_9$H$_{11}$ + 4.943 LuN + LuB$_6$H$_4$C$_6$N | -0.0709 |
| LuNH$_4$B$_6$C$_6$ | 0.8571 Lu$_4$C$_7$ + 41.14 B + H$_3$N + 0.5 LuH$_2$ $\longrightarrow$ 2.929 LuB$_{12}$ + LuB$_6$H$_4$C$_6$N | -0.0691 |
| LuNH$_4$B$_6$C$_6$ | 0.7253 Lu$_4$C$_7$ + 1.451 N$_2$ + 0.4615 B$_{13}$C$_2$ + 2 H$_2$ $\longrightarrow$ 1.901 LuN + LuB$_6$H$_4$C$_6$N | -0.0677 |
| LuNH$_4$B$_6$C$_6$ | 44.14 B + 1.5 Lu$_3$C$_4$ + BN + 0.5714 B$_5$H$_7$ $\longrightarrow$ 3.5 LuB$_{12}$ + LuB$_6$H$_4$C$_6$N | -0.0576 |
| LuNH$_4$B$_6$C$_6$ | 20 B + 1.5 Lu$_3$C$_4$ + 1.333 H$_3$N $\longrightarrow$ 0.1667 N$_2$ + 3.5 LuB$_4$ + LuB$_6$H$_4$C$_6$N | -0.0148 |
| LuNH$_4$B$_6$C$_6$ | 0.8571 Lu$_4$C$_7$ + 0.5 N$_2$ + 2.964 LuB$_{12}$ + 2 LuH$_2$ $\longrightarrow$ 7.393 LuB$_4$ + LuB$_6$H$_4$C$_6$N | -0.0103 |
| NaNH$_4$B$_6$C$_6$ | 99 B$_{13}$C$_2$ + 641 N$_2$ + 4 NaH $\longrightarrow$ 1281 BN + 3 NaC$_{64}$ + NaB$_6$H$_4$C$_6$N | -3172.0217 |
| NaNH$_4$B$_6$C$_6$ | 3 B$_{13}$C$_2$ + 218.2 B$_5$H$_7$ + 169.6 NaN$_3$ $\longrightarrow$ 507.8 H$_3$N + 56.2 Na$_3$B$_{20}$ + NaB$_6$H$_4$C$_6$N | -438.7759 |
| NaNH$_4$B$_6$C$_6$ | 47 N$_2$ + 6.6 NaB$_{15}$ + 1.2 H$_8$C$_5$ $\longrightarrow$ 5.6 NaH + 93 BN + NaB$_6$H$_4$C$_6$N | -226.3743 |
| NaNH$_4$B$_6$C$_6$ | 37.1 N$_2$ + 8.8 B$_9$H$_{11}$ + NaC$_{64}$ $\longrightarrow$ 73.2 BN + 11.6 H$_8$C$_5$ + NaB$_6$H$_4$C$_6$N | -186.6106 |
| NaNH$_4$B$_6$C$_6$ | 89.2 N$_2$ + NaC$_{64}$ + 76.6 B$_5$H$_7$ $\longrightarrow$ 177.4 H$_3$N + 29 B$_{13}$C$_2$ + NaB$_6$H$_4$C$_6$N | -179.4279 |
| NaNH$_4$B$_6$C$_6$ | 32.07 N$_2$ + NaC$_{64}$ + 13.83 B$_5$H$_7$ $\longrightarrow$ 63.14 BN + 11.6 H$_8$C$_5$ + NaB$_6$H$_4$C$_6$N | -162.3032 |
| NaNH$_4$B$_6$C$_6$ | 77.85 N$_2$ + 42.56 B$_9$H$_{11}$ + NaC$_{64}$ $\longrightarrow$ 154.7 H$_3$N + 29 B$_{13}$C$_2$ + NaB$_6$H$_4$C$_6$N | -152.2542 |
| NaNH$_4$B$_6$C$_6$ | 19 NaH + 6 BN + 1158 C $\longrightarrow$ 5 H$_3$N + 18 NaC$_{64}$ + NaB$_6$H$_4$C$_6$N | -125.2254 |
| NaNH$_4$B$_6$C$_6$ | 3 B$_{13}$C$_2$ + 26.14 N$_2$ + 0.3636 B$_9$H$_{11}$ + NaB$_{15}$ $\longrightarrow$ 51.27 BN + NaB$_6$H$_4$C$_6$N | -119.9621 |
| NaNH$_4$B$_6$C$_6$ | 3 B$_{13}$C$_2$ + 25.93 N$_2$ + NaB$_{15}$ + 0.5714 B$_5$H$_7$ $\longrightarrow$ 50.86 BN + NaB$_6$H$_4$C$_6$N | -118.9576 |
| NaNH$_4$B$_6$C$_6$ | 3 B$_{13}$C$_2$ + 24.5 N$_2$ + NaB$_{15}$ + 2 H$_2$ $\longrightarrow$ 48 BN + NaB$_6$H$_4$C$_6$N | -112.2212 |
| NaNH$_4$B$_6$C$_6$ | 3 B$_{13}$C$_2$ + 1.792 B$_9$H$_{11}$ + 16.71 NaN$_3$ $\longrightarrow$ 15.71 NaH + 49.13 BN + NaB$_6$H$_4$C$_6$N | -111.5114 |
| NaNH$_4$B$_6$C$_6$ | 1.333 H$_3$N + 3 B$_{13}$C$_2$ + 23.83 N$_2$ + NaB$_{15}$ $\longrightarrow$ 48 BN + NaB$_6$H$_4$C$_6$N | -110.4558 |
| NaNH$_4$B$_6$C$_6$ | 3 B$_{13}$C$_2$ + 2.687 B$_5$H$_7$ + 15.81 NaN$_3$ $\longrightarrow$ 14.81 NaH + 46.44 BN + NaB$_6$H$_4$C$_6$N | -105.0952 |
| NaNH$_4$B$_6$C$_6$ | 3 B$_{13}$C$_2$ + 21.97 N$_2$ + 0.3636 B$_9$H$_{11}$ + 0.3333 Na$_3$B$_{20}$ $\longrightarrow$ 42.94 BN + NaB$_6$H$_4$C$_6$N | -98.9205 |
| NaNH$_4$B$_6$C$_6$ | 3 B$_{13}$C$_2$ + 21.76 N$_2$ + 0.5714 B$_5$H$_7$ + 0.3333 Na$_3$B$_{20}$ $\longrightarrow$ 42.52 BN + NaB$_6$H$_4$C$_6$N | -97.9161 |
| NaNH$_4$B$_6$C$_6$ | 3 B$_{13}$C$_2$ + 20.33 N$_2$ + 0.3333 Na$_3$B$_{20}$ + 2 H$_2$ $\longrightarrow$ 39.67 BN + NaB$_6$H$_4$C$_6$N | -91.1796 |
| NaNH$_4$B$_6$C$_6$ | 1.333 H$_3$N + 3 B$_{13}$C$_2$ + 19.67 N$_2$ + 0.3333 Na$_3$B$_{20}$ $\longrightarrow$ 39.67 BN + NaB$_6$H$_4$C$_6$N | -89.4142 |
| NaNH$_4$B$_6$C$_6$ | 19.5 N$_2$ + 1.2 H$_8$C$_5$ + 2.2 Na$_3$B$_{20}$ $\longrightarrow$ 5.6 NaH + 38 BN + NaB$_6$H$_4$C$_6$N | -87.4999 |
| NaNH$_4$B$_6$C$_6$ | 3 B$_{13}$C$_2$ + 18.64 N$_2$ + 0.3636 B$_9$H$_{11}$ + Na $\longrightarrow$ 36.27 BN + NaB$_6$H$_4$C$_6$N | -82.2654 |
| NaNH$_4$B$_6$C$_6$ | 3 B$_{13}$C$_2$ + 17.14 N$_2$ + 0.3636 B$_9$H$_{11}$ + NaN$_3$ $\longrightarrow$ 36.27 BN + NaB$_6$H$_4$C$_6$N | -81.8387 |
| NaNH$_4$B$_6$C$_6$ | 3 B$_{13}$C$_2$ + 18.43 N$_2$ + Na + 0.5714 B$_5$H$_7$ $\longrightarrow$ 35.86 BN + NaB$_6$H$_4$C$_6$N | -81.2610 |
| NaNH$_4$B$_6$C$_6$ | 3 B$_{13}$C$_2$ + 16.93 N$_2$ + 0.5714 B$_5$H$_7$ + NaN$_3$ $\longrightarrow$ 35.86 BN + NaB$_6$H$_4$C$_6$N | -80.8342 |
| NaNH$_4$B$_6$C$_6$ | 3 B$_{13}$C$_2$ + 18.23 N$_2$ + NaH + 0.2727 B$_9$H$_{11}$ $\longrightarrow$ 35.45 BN + NaB$_6$H$_4$C$_6$N | -79.9497 |
| NaNH$_4$B$_6$C$_6$ | 3 B$_{13}$C$_2$ + 18.07 N$_2$ + NaH + 0.4286 B$_5$H$_7$ $\longrightarrow$ 35.14 BN + NaB$_6$H$_4$C$_6$N | -79.1964 |
| NaNH$_4$B$_6$C$_6$ | 3 B$_{13}$C$_2$ + 0.3636 B$_9$H$_{11}$ + 12.42 NaN$_3$ $\longrightarrow$ 36.27 BN + 11.42 Na + NaB$_6$H$_4$C$_6$N | -76.9634 |
| NaNH$_4$B$_6$C$_6$ | 3 B$_{13}$C$_2$ + 0.5714 B$_5$H$_7$ + 12.29 NaN$_3$ $\longrightarrow$ 35.86 BN + 11.29 Na + NaB$_6$H$_4$C$_6$N | -76.0181 |
| NaNH$_4$B$_6$C$_6$ | 3 B$_{13}$C$_2$ + 17 N$_2$ + Na + 2 H$_2$ $\longrightarrow$ 33 BN + NaB$_6$H$_4$C$_6$N | -74.5245 |
| NaNH$_4$B$_6$C$_6$ | 3 B$_{13}$C$_2$ + 21.5 N$_2$ + 4 NaH $\longrightarrow$ 33 BN + 3 NaN$_3$ + NaB$_6$H$_4$C$_6$N | -74.2828 |
| NaNH$_4$B$_6$C$_6$ | 3 B$_{13}$C$_2$ + 17 N$_2$ + NaH + 1.5 H$_2$ $\longrightarrow$ 33 BN + NaB$_6$H$_4$C$_6$N | -74.1440 |
| NaNH$_4$B$_6$C$_6$ | 3 B$_{13}$C$_2$ + 15.5 N$_2$ + 2 H$_2$ + NaN$_3$ $\longrightarrow$ 33 BN + NaB$_6$H$_4$C$_6$N | -74.0978 |
| NaNH$_4$B$_6$C$_6$ | 3 B$_{13}$C$_2$ + 7.167 H$_2$ + 11.33 NaN$_3$ $\longrightarrow$ 10.33 NaH + 33 BN + NaB$_6$H$_4$C$_6$N | -73.6200 |
| NaNH$_4$B$_6$C$_6$ | 3 B$_{13}$C$_2$ + 17 N$_2$ + 4 NaH $\longrightarrow$ 33 BN + 3 Na + NaB$_6$H$_4$C$_6$N | -73.0025 |
| NaNH$_4$B$_6$C$_6$ | H$_3$N + 3 B$_{13}$C$_2$ + 16.5 N$_2$ + NaH $\longrightarrow$ 33 BN + NaB$_6$H$_4$C$_6$N | -72.8200 |
| NaNH$_4$B$_6$C$_6$ | 1.333 H$_3$N + 3 B$_{13}$C$_2$ + 16.33 N$_2$ + Na $\longrightarrow$ 33 BN + NaB$_6$H$_4$C$_6$N | -72.7591 |
| NaNH$_4$B$_6$C$_6$ | 1.333 H$_3$N + 3 B$_{13}$C$_2$ + 14.83 N$_2$ + NaN$_3$ $\longrightarrow$ 33 BN + NaB$_6$H$_4$C$_6$N | -72.3324 |
| NaNH$_4$B$_6$C$_6$ | 1.75 B$_{13}$C$_2$ + 16.38 N$_2$ + NaB$_{15}$ + 0.5 H$_8$C$_5$ $\longrightarrow$ 31.75 BN + NaB$_6$H$_4$C$_6$N | -70.9383 |
| NaNH$_4$B$_6$C$_6$ | 3 B$_{13}$C$_2$ + 2 H$_2$ + 11.33 NaN$_3$ $\longrightarrow$ 33 BN + 10.33 Na + NaB$_6$H$_4$C$_6$N | -69.6880 |
| NaNH$_4$B$_6$C$_6$ | 3 B$_{13}$C$_2$ + 4 NaH + 11.33 NaN$_3$ $\longrightarrow$ 33 BN + 14.33 Na + NaB$_6$H$_4$C$_6$N | -68.1660 |
| NaNH$_4$B$_6$C$_6$ | 1.333 H$_3$N + 3 B$_{13}$C$_2$ + 10.89 NaN$_3$ $\longrightarrow$ 33 BN + 9.889 Na + NaB$_6$H$_4$C$_6$N | -68.1123 |
| NaNH$_4$B$_6$C$_6$ | 4.3 H$_3$N + 3 B$_{13}$C$_2$ + 9.9 NaN$_3$ $\longrightarrow$ 8.9 NaH + 33 BN + NaB$_6$H$_4$C$_6$N | -67.9928 |
| NaNH$_4$B$_6$C$_6$ | 34 H$_3$N + 3 B$_{13}$C$_2$ + 99 Na $\longrightarrow$ 98 NaH + 33 BN + NaB$_6$H$_4$C$_6$N | -66.7961 |
| NaNH$_4$B$_6$C$_6$ | 35.6 H$_3$N + 3.123 B$_{13}$C$_2$ + NaC$_{64}$ $\longrightarrow$ 34.6 BN + 12.85 H$_8$C$_5$ + NaB$_6$H$_4$C$_6$N | -55.2855 |
| NaNH$_4$B$_6$C$_6$ | 32.27 H$_3$N + NaC$_{64}$ + 37.27 B $\longrightarrow$ 31.27 BN + 11.6 H$_8$C$_5$ + NaB$_6$H$_4$C$_6$N | -52.2126 |
| NaNH$_4$B$_6$C$_6$ | 1.75 B$_{13}$C$_2$ + 12.21 N$_2$ + 0.5 H$_8$C$_5$ + 0.3333 Na$_3$B$_{20}$ $\longrightarrow$ 23.42 BN + NaB$_6$H$_4$C$_6$N | -49.8967 |
| NaNH$_4$B$_6$C$_6$ | 49 H$_3$N + 3 B$_{13}$C$_2$ + NaB$_{15}$ $\longrightarrow$ 48 BN + 71.5 H$_2$ + NaB$_6$H$_4$C$_6$N | -47.3418 |
| NaNH$_4$B$_6$C$_6$ | 2.063 B$_{13}$C$_2$ + 10.91 N$_2$ + NaH + 0.375 H$_8$C$_5$ $\longrightarrow$ 20.81 BN + NaB$_6$H$_4$C$_6$N | -43.1819 |
| NaNH$_4$B$_6$C$_6$ | 6 BN + 2.667 NaC$_{64}$ + 133.7 H$_2$ $\longrightarrow$ 32.93 H$_8$C$_5$ + 1.667 NaN$_3$ + NaB$_6$H$_4$C$_6$N | -39.7189 |
| NaNH$_4$B$_6$C$_6$ | 1.2 H$_8$C$_5$ + 6.6 NaN$_3$ + 24.8 B $\longrightarrow$ 5.6 NaH + 18.8 BN + NaB$_6$H$_4$C$_6$N | -38.4155 |
| NaNH$_4$B$_6$C$_6$ | 40.67 H$_3$N + 3 B$_{13}$C$_2$ + 0.3333 Na$_3$B$_{20}$ $\longrightarrow$ 39.67 BN + 59 H$_2$ + NaB$_6$H$_4$C$_6$N | -37.3341 |

| Compound | Chemical reaction | Reaction energy(eV) |
|---|---|---|
| $NaNH_4B_6C_6$ | $21.48\,H_3N + 2.942\,B_9H_{11} + NaC_{64} \longrightarrow 20.48\,BN + 11.6\,H_8C_5 + NaB_6H_4C_6N$ | -33.4715 |
| $NaNH_4B_6C_6$ | $1.75\,B_{13}C_2 + 8.875\,N_2 + Na + 0.5\,H_8C_5 \longrightarrow 16.75\,BN + NaB_6H_4C_6N$ | -33.2417 |
| $NaNH_4B_6C_6$ | $31\,H_3N + 3\,B_{13}C_2 + NaN_3 \longrightarrow 33\,BN + 44.5\,H_2 + NaB_6H_4C_6N$ | -33.0516 |
| $NaNH_4B_6C_6$ | $1.75\,B_{13}C_2 + 7.375\,N_2 + 0.5\,H_8C_5 + NaN_3 \longrightarrow 16.75\,BN + NaB_6H_4C_6N$ | -32.8149 |
| $NaNH_4B_6C_6$ | $20.41\,H_3N + NaC_{64} + 5.082\,B_5H_7 \longrightarrow 19.41\,BN + 11.6\,H_8C_5 + NaB_6H_4C_6N$ | -32.1667 |
| $NaNH_4B_6C_6$ | $1.75\,B_{13}C_2 + 0.5\,H_8C_5 + 5.917\,NaN_3 \longrightarrow 16.75\,BN + 4.917\,Na + NaB_6H_4C_6N$ | -30.7167 |
| $NaNH_4B_6C_6$ | $34\,H_3N + 3\,B_{13}C_2 + Na \longrightarrow 33\,BN + 49\,H_2 + NaB_6H_4C_6N$ | -29.5062 |
| $NaNH_4B_6C_6$ | $34\,H_3N + 3\,B_{13}C_2 + NaH \longrightarrow 33\,BN + 49.5\,H_2 + NaB_6H_4C_6N$ | -29.1257 |
| $NaNH_4B_6C_6$ | $1.378\,NaB_{15} + 1.2\,H_8C_5 + 5.222\,NaN_3 \longrightarrow 5.6\,NaH + 14.67\,BN + NaB_6H_4C_6N$ | -27.2855 |
| $NaNH_4B_6C_6$ | $1.333\,H_3N + 1.111\,NaC_{64} + 429.2\,B \longrightarrow 32.56\,B_{13}C_2 + 0.1111\,NaN_3 + NaB_6H_4C_6N$ | -26.7321 |
| $NaNH_4B_6C_6$ | $4\,NaH + 0.3\,Na_3B_{20} + 276.9\,C + 0.3333\,NaN_3 \longrightarrow 4.233\,NaC_{64} + NaB_6H_4C_6N$ | -26.4363 |
| $NaNH_4B_6C_6$ | $0.5\,N_2 + NaC_{64} + 0.5\,H_8C_5 + 399.3\,B \longrightarrow 30.25\,B_{13}C_2 + NaB_6H_4C_6N$ | -25.0142 |
| $NaNH_4B_6C_6$ | $0.5\,N_2 + NaC_{64} + 2\,H_2 + 383\,B \longrightarrow 29\,B_{13}C_2 + NaB_6H_4C_6N$ | -24.5872 |
| $NaNH_4B_6C_6$ | $0.5\,N_2 + NaC_{64} + 0.5714\,B_5H_7 + 380.1\,B \longrightarrow 29\,B_{13}C_2 + NaB_6H_4C_6N$ | -23.9901 |
| $NaNH_4B_6C_6$ | $0.5\,N_2 + 0.3636\,B_9H_{11} + NaC_{64} + 379.7\,B \longrightarrow 29\,B_{13}C_2 + NaB_6H_4C_6N$ | -23.9278 |
| $NaNH_4B_6C_6$ | $6.636\,N_2 + 0.3636\,B_9H_{11} + NaB_{15} + 6\,C \longrightarrow 12.27\,BN + NaB_6H_4C_6N$ | -23.8854 |
| $NaNH_4B_6C_6$ | $0.5\,N_2 + 4\,NaH + 0.3\,Na_3B_{20} + 255.6\,C \longrightarrow 3.9\,NaC_{64} + NaB_6H_4C_6N$ | -23.8492 |
| $NaNH_4B_6C_6$ | $1.333\,H_3N + NaC_{64} + 383.3\,B \longrightarrow 29\,B_{13}C_2 + 0.3333\,BN + NaB_6H_4C_6N$ | -23.6774 |
| $NaNH_4B_6C_6$ | $H_3N + NaC_{64} + 0.125\,H_8C_5 + 387.1\,B \longrightarrow 29.31\,B_{13}C_2 + NaB_6H_4C_6N$ | -23.3699 |
| $NaNH_4B_6C_6$ | $H_3N + NaC_{64} + 0.5\,H_2 + 383\,B \longrightarrow 29\,B_{13}C_2 + NaB_6H_4C_6N$ | -23.2632 |
| $NaNH_4B_6C_6$ | $H_3N + NaC_{64} + 0.1429\,B_5H_7 + 382.3\,B \longrightarrow 29\,B_{13}C_2 + NaB_6H_4C_6N$ | -23.1139 |
| $NaNH_4B_6C_6$ | $H_3N + 0.09091\,B_9H_{11} + NaC_{64} + 382.2\,B \longrightarrow 29\,B_{13}C_2 + NaB_6H_4C_6N$ | -23.0983 |
| $NaNH_4B_6C_6$ | $3\,B_{13}C_2 + 7\,N_2 + 4\,NaH \longrightarrow 13\,BN + Na_3B_{20} + NaB_6H_4C_6N$ | -23.0373 |
| $NaNH_4B_6C_6$ | $6.429\,N_2 + NaB_{15} + 0.5714\,B_5H_7 + 6\,C \longrightarrow 11.86\,BN + NaB_6H_4C_6N$ | -22.8809 |
| $NaNH_4B_6C_6$ | $1.333\,H_3N + NaC_{64} + 383\,B \longrightarrow 29\,B_{13}C_2 + 0.1667\,N_2 + NaB_6H_4C_6N$ | -22.8218 |
| $NaNH_4B_6C_6$ | $BN + NaC_{64} + 0.5\,H_8C_5 + 398.3\,B \longrightarrow 30.25\,B_{13}C_2 + NaB_6H_4C_6N$ | -22.4474 |
| $NaNH_4B_6C_6$ | $4\,NaH + 0.4\,NaB_{15} + 244.9\,C + 0.3333\,NaN_3 \longrightarrow 3.733\,NaC_{64} + NaB_6H_4C_6N$ | -22.4313 |
| $NaNH_4B_6C_6$ | $BN + NaC_{64} + 2\,H_2 + 382\,B \longrightarrow 29\,B_{13}C_2 + NaB_6H_4C_6N$ | -22.0205 |
| $NaNH_4B_6C_6$ | $BN + NaC_{64} + 0.5714\,B_5H_7 + 379.1\,B \longrightarrow 29\,B_{13}C_2 + NaB_6H_4C_6N$ | -21.4233 |
| $NaNH_4B_6C_6$ | $BN + 0.3636\,B_9H_{11} + NaC_{64} + 378.7\,B \longrightarrow 29\,B_{13}C_2 + NaB_6H_4C_6N$ | -21.3610 |
| $NaNH_4B_6C_6$ | $3\,B_{13}C_2 + 0.3636\,B_9H_{11} + 4.545\,NaN_3 \longrightarrow 12.64\,BN + 1.182\,Na_3B_{20} + NaB_6H_4C_6N$ | -21.2759 |
| $NaNH_4B_6C_6$ | $3\,B_{13}C_2 + 0.5714\,B_5H_7 + 4.502\,NaN_3 \longrightarrow 12.51\,BN + 1.167\,Na_3B_{20} + NaB_6H_4C_6N$ | -21.0058 |
| $NaNH_4B_6C_6$ | $4\,NaH + BN + 0.25\,Na_3B_{20} + 246\,C \longrightarrow 3.75\,NaC_{64} + NaB_6H_4C_6N$ | -20.1227 |
| $NaNH_4B_6C_6$ | $1.333\,H_3N + 3\,B_{13}C_2 + 4.069\,NaN_3 \longrightarrow 12.54\,BN + 1.023\,Na_3B_{20} + NaB_6H_4C_6N$ | -19.9089 |
| $NaNH_4B_6C_6$ | $0.5\,N_2 + 4\,NaH + 0.4\,NaB_{15} + 223.6\,C \longrightarrow 3.4\,NaC_{64} + NaB_6H_4C_6N$ | -19.8442 |
| $NaNH_4B_6C_6$ | $5.933\,N_2 + 0.3636\,B_9H_{11} + 0.09375\,NaC_{64} + 0.9063\,NaB_{15} \longrightarrow 10.87\,BN + NaB_6H_4C_6N$ | -19.5837 |
| $NaNH_4B_6C_6$ | $4\,NaH + 219.3\,C + 0.3333\,NaN_3 + 6\,B \longrightarrow 3.333\,NaC_{64} + NaB_6H_4C_6N$ | -19.4780 |
| $NaNH_4B_6C_6$ | $3\,B_{13}C_2 + 2\,H_2 + 4.207\,NaN_3 \longrightarrow 11.62\,BN + 1.069\,Na_3B_{20} + NaB_6H_4C_6N$ | -19.3182 |
| $NaNH_4B_6C_6$ | $0.4615\,B_{13}C_2 + 4\,NaH + 218.4\,C + 0.3333\,NaN_3 \longrightarrow 3.333\,NaC_{64} + NaB_6H_4C_6N$ | -18.8584 |
| $NaNH_4B_6C_6$ | $1.2\,H_8C_5 + 0.8552\,Na_3B_{20} + 4.034\,NaN_3 \longrightarrow 5.6\,NaH + 11.1\,BN + NaB_6H_4C_6N$ | -18.5836 |
| $NaNH_4B_6C_6$ | $5.725\,N_2 + 0.09375\,NaC_{64} + 0.9063\,NaB_{15} + 0.5714\,B_5H_7 \longrightarrow 10.45\,BN + NaB_6H_4C_6N$ | -18.5792 |
| $NaNH_4B_6C_6$ | $0.5\,N_2 + 4\,NaH + 198\,C + 6\,B \longrightarrow 3\,NaC_{64} + NaB_6H_4C_6N$ | -16.8908 |
| $NaNH_4B_6C_6$ | $4\,NaH + BN + 0.3333\,NaB_{15} + 219.3\,C \longrightarrow 3.333\,NaC_{64} + NaB_6H_4C_6N$ | -16.7852 |
| $NaNH_4B_6C_6$ | $0.4615\,B_{13}C_2 + 0.5\,N_2 + 4\,NaH + 197.1\,C \longrightarrow 3\,NaC_{64} + NaB_6H_4C_6N$ | -16.2713 |
| $NaNH_4B_6C_6$ | $5\,N_2 + NaB_{15} + 6\,C + 2\,H_2 \longrightarrow 9\,BN + NaB_6H_4C_6N$ | -16.1445 |
| $NaNH_4B_6C_6$ | $5.933\,N_2 + NaB_{15} + 1.2\,H_8C_5 \longrightarrow 1.867\,H_3N + 9\,BN + NaB_6H_4C_6N$ | -15.6138 |
| $NaNH_4B_6C_6$ | $5\,N_2 + NaB_{15} + 0.5\,H_8C_5 + 3.5\,C \longrightarrow 9\,BN + NaB_6H_4C_6N$ | -14.8936 |
| $NaNH_4B_6C_6$ | $0.5\,N_2 + NaC_{64} + 48.4\,H_2 + 6\,B \longrightarrow 11.6\,H_8C_5 + NaB_6H_4C_6N$ | -14.6819 |
| $NaNH_4B_6C_6$ | $0.4615\,B_{13}C_2 + 0.5\,N_2 + NaC_{64} + 49.14\,H_2 \longrightarrow 11.78\,H_8C_5 + NaB_6H_4C_6N$ | -14.5243 |
| $NaNH_4B_6C_6$ | $1.333\,H_3N + 4.333\,N_2 + NaB_{15} + 6\,C \longrightarrow 9\,BN + NaB_6H_4C_6N$ | -14.3791 |
| $NaNH_4B_6C_6$ | $4\,NaH + BN + 198\,C + 5\,B \longrightarrow 3\,NaC_{64} + NaB_6H_4C_6N$ | -14.3241 |
| $NaNH_4B_6C_6$ | $0.3846\,B_{13}C_2 + 4\,NaH + BN + 197.2\,C \longrightarrow 3\,NaC_{64} + NaB_6H_4C_6N$ | -13.8078 |
| $NaNH_4B_6C_6$ | $16.5\,H_3N + 3\,B_{13}C_2 + NaB_{15} \longrightarrow 15.5\,BN + 6.5\,B_5H_7 + NaB_6H_4C_6N$ | -13.7470 |
| $NaNH_4B_6C_6$ | $6.696\,B_9H_{11} + 1.2\,H_8C_5 + 9.139\,NaN_3 \longrightarrow 26.42\,H_3N + 2.713\,Na_3B_{20} + NaB_6H_4C_6N$ | -13.4993 |
| $NaNH_4B_6C_6$ | $0.5\,N_2 + 0.6667\,B_9H_{11} + NaC_{64} + 44.73\,H_2 \longrightarrow 11.6\,H_8C_5 + NaB_6H_4C_6N$ | -13.4730 |
| $NaNH_4B_6C_6$ | $0.5\,N_2 + NaC_{64} + 1.2\,B_5H_7 + 44.2\,H_2 \longrightarrow 11.6\,H_8C_5 + NaB_6H_4C_6N$ | -13.4279 |
| $NaNH_4B_6C_6$ | $H_3N + NaC_{64} + 46.9\,H_2 + 6\,B \longrightarrow 11.6\,H_8C_5 + NaB_6H_4C_6N$ | -13.3579 |
| $NaNH_4B_6C_6$ | $0.6667\,NaC_{64} + 0.5\,H_8C_5 + 0.3333\,NaN_3 + 260.6\,B \longrightarrow 19.58\,B_{13}C_2 + NaB_6H_4C_6N$ | -13.2833 |
| $NaNH_4B_6C_6$ | $H_3N + 0.4615\,B_{13}C_2 + NaC_{64} + 47.64\,H_2 \longrightarrow 11.78\,H_8C_5 + NaB_6H_4C_6N$ | -13.2002 |
| $NaNH_4B_6C_6$ | $5\,N_2 + NaB_{15} + 1.2\,H_8C_5 \longrightarrow 9\,BN + 2.8\,H_2 + NaB_6H_4C_6N$ | -13.1422 |

| Compound | Chemical reaction | Reaction energy(eV) |
|---|---|---|
| NaNH$_4$B$_6$C$_6$ | $0.6667\,\text{NaC}_{64} + 2\,\text{H}_2 + 0.3333\,\text{NaN}_3 + 244.3\,\text{B} \longrightarrow 18.33\,\text{B}_{13}\text{C}_2 + \text{NaB}_6\text{H}_4\text{C}_6\text{N}$ | -12.8564 |
| NaNH$_4$B$_6$C$_6$ | $4.59\,\text{N}_2 + 0.05469\,\text{NaC}_{64} + 0.9453\,\text{NaB}_{15} + 0.5\,\text{H}_8\text{C}_5 \longrightarrow 8.18\,\text{BN} + \text{NaB}_6\text{H}_4\text{C}_6\text{N}$ | -12.3842 |
| NaNH$_4$B$_6$C$_6$ | $0.6667\,\text{NaC}_{64} + 0.5714\,\text{B}_5\text{H}_7 + 0.3333\,\text{NaN}_3 + 241.5\,\text{B} \longrightarrow 18.33\,\text{B}_{13}\text{C}_2 + \text{NaB}_6\text{H}_4\text{C}_6\text{N}$ | -12.2592 |
| NaNH$_4$B$_6$C$_6$ | $0.3636\,\text{B}_9\text{H}_{11} + 0.6667\,\text{NaC}_{64} + 0.3333\,\text{NaN}_3 + 241.1\,\text{B} \longrightarrow 18.33\,\text{B}_{13}\text{C}_2 + \text{NaB}_6\text{H}_4\text{C}_6\text{N}$ | -12.1970 |
| NaNH$_4$B$_6$C$_6$ | $\text{H}_3\text{N} + 0.6667\,\text{B}_9\text{H}_{11} + \text{NaC}_{64} + 43.23\,\text{H}_2 \longrightarrow 11.6\,\text{H}_8\text{C}_5 + \text{NaB}_6\text{H}_4\text{C}_6\text{N}$ | -12.1489 |
| NaNH$_4$B$_6$C$_6$ | $10.77\,\text{H}_3\text{N} + 3\,\text{B}_{13}\text{C}_2 + \text{NaN}_3 \longrightarrow 12.77\,\text{BN} + 4.045\,\text{B}_5\text{H}_7 + \text{NaB}_6\text{H}_4\text{C}_6\text{N}$ | -12.1430 |
| NaNH$_4$B$_6$C$_6$ | $\text{BN} + \text{NaC}_{64} + 48.4\,\text{H}_2 + 5\,\text{B} \longrightarrow 11.6\,\text{H}_8\text{C}_5 + \text{NaB}_6\text{H}_4\text{C}_6\text{N}$ | -12.1152 |
| NaNH$_4$B$_6$C$_6$ | $\text{H}_3\text{N} + \text{NaC}_{64} + 1.2\,\text{B}_5\text{H}_7 + 42.7\,\text{H}_2 \longrightarrow 11.6\,\text{H}_8\text{C}_5 + \text{NaB}_6\text{H}_4\text{C}_6\text{N}$ | -12.1038 |
| NaNH$_4$B$_6$C$_6$ | $15.13\,\text{H}_3\text{N} + 3\,\text{B}_{13}\text{C}_2 + \text{NaB}_{15} \longrightarrow 14.13\,\text{BN} + 3.763\,\text{B}_9\text{H}_{11} + \text{NaB}_6\text{H}_4\text{C}_6\text{N}$ | -12.0780 |
| NaNH$_4$B$_6$C$_6$ | $0.3846\,\text{B}_{13}\text{C}_2 + \text{BN} + \text{NaC}_{64} + 49.02\,\text{H}_2 \longrightarrow 11.75\,\text{H}_8\text{C}_5 + \text{NaB}_6\text{H}_4\text{C}_6\text{N}$ | -11.9838 |
| NaNH$_4$B$_6$C$_6$ | $3\,\text{B}_{13}\text{C}_2 + 27.27\,\text{H}_2 + 5.95\,\text{NaN}_3 \longrightarrow 16.85\,\text{H}_3\text{N} + 1.65\,\text{Na}_3\text{B}_{20} + \text{NaB}_6\text{H}_4\text{C}_6\text{N}$ | -11.8534 |
| NaNH$_4$B$_6$C$_6$ | $4.297\,\text{N}_2 + 0.09375\,\text{NaC}_{64} + 0.9062\,\text{NaB}_{15} + 2\,\text{H}_2 \longrightarrow 7.594\,\text{BN} + \text{NaB}_6\text{H}_4\text{C}_6\text{N}$ | -11.8428 |
| NaNH$_4$B$_6$C$_6$ | $1.097\,\text{B}_{13}\text{C}_2 + 0.7611\,\text{H}_8\text{C}_5 + 3.088\,\text{NaN}_3 \longrightarrow 2.088\,\text{NaH} + 8.265\,\text{BN} + \text{NaB}_6\text{H}_4\text{C}_6\text{N}$ | -11.1635 |
| NaNH$_4$B$_6$C$_6$ | $\text{BN} + 0.5556\,\text{B}_9\text{H}_{11} + \text{NaC}_{64} + 45.34\,\text{H}_2 \longrightarrow 11.6\,\text{H}_8\text{C}_5 + \text{NaB}_6\text{H}_4\text{C}_6\text{N}$ | -11.1077 |
| NaNH$_4$B$_6$C$_6$ | $9.921\,\text{H}_3\text{N} + 3\,\text{B}_{13}\text{C}_2 + \text{NaN}_3 \longrightarrow 11.92\,\text{BN} + 2.342\,\text{B}_9\text{H}_{11} + \text{NaB}_6\text{H}_4\text{C}_6\text{N}$ | -11.1043 |
| NaNH$_4$B$_6$C$_6$ | $\text{BN} + \text{NaC}_{64} + \text{B}_5\text{H}_7 + 44.9\,\text{H}_2 \longrightarrow 11.6\,\text{H}_8\text{C}_5 + \text{NaB}_6\text{H}_4\text{C}_6\text{N}$ | -11.0701 |
| NaNH$_4$B$_6$C$_6$ | $10\,\text{H}_3\text{N} + \text{NaB}_{15} + 22.25\,\text{C} \longrightarrow 9\,\text{BN} + 3.25\,\text{H}_8\text{C}_5 + \text{NaB}_6\text{H}_4\text{C}_6\text{N}$ | -11.0349 |
| NaNH$_4$B$_6$C$_6$ | $1.333\,\text{H}_3\text{N} + 3.63\,\text{N}_2 + 0.09375\,\text{NaC}_{64} + 0.9062\,\text{NaB}_{15} \longrightarrow 7.594\,\text{BN} + \text{NaB}_6\text{H}_4\text{C}_6\text{N}$ | -10.0774 |
| NaNH$_4$B$_6$C$_6$ | $0.5\,\text{N}_2 + \text{NaC}_{64} + 100.6\,\text{B}_5\text{H}_7 \longrightarrow 55.25\,\text{B}_9\text{H}_{11} + 11.6\,\text{H}_8\text{C}_5 + \text{NaB}_6\text{H}_4\text{C}_6\text{N}$ | -9.6917 |
| NaNH$_4$B$_6$C$_6$ | $13.85\,\text{H}_3\text{N} + 3\,\text{B}_{13}\text{C}_2 + 0.3333\,\text{Na}_3\text{B}_{20} \longrightarrow 12.85\,\text{BN} + 5.364\,\text{B}_5\text{H}_7 + \text{NaB}_6\text{H}_4\text{C}_6\text{N}$ | -9.6126 |
| NaNH$_4$B$_6$C$_6$ | $1.333\,\text{H}_3\text{N} + 3\,\text{B}_{13}\text{C}_2 + 2.648\,\text{NaN}_3 \longrightarrow 8.278\,\text{BN} + 1.648\,\text{NaB}_{15} + \text{NaB}_6\text{H}_4\text{C}_6\text{N}$ | -9.4994 |
| NaNH$_4$B$_6$C$_6$ | $3\,\text{B}_{13}\text{C}_2 + 0.3636\,\text{B}_9\text{H}_{11} + 2.904\,\text{NaN}_3 \longrightarrow 7.712\,\text{BN} + 1.904\,\text{NaB}_{15} + \text{NaB}_6\text{H}_4\text{C}_6\text{N}$ | -9.2502 |
| NaNH$_4$B$_6$C$_6$ | $3\,\text{B}_{13}\text{C}_2 + 0.5714\,\text{B}_5\text{H}_7 + 2.881\,\text{NaN}_3 \longrightarrow 7.643\,\text{BN} + 1.881\,\text{NaB}_{15} + \text{NaB}_6\text{H}_4\text{C}_6\text{N}$ | -9.1259 |
| NaNH$_4$B$_6$C$_6$ | $\text{H}_3\text{N} + \text{NaC}_{64} + 97.27\,\text{B}_5\text{H}_7 \longrightarrow 53.37\,\text{B}_9\text{H}_{11} + 11.6\,\text{H}_8\text{C}_5 + \text{NaB}_6\text{H}_4\text{C}_6\text{N}$ | -8.4944 |
| NaNH$_4$B$_6$C$_6$ | $3\,\text{B}_{13}\text{C}_2 + 2\,\text{H}_2 + 2.722\,\text{NaN}_3 \longrightarrow 7.167\,\text{BN} + 1.722\,\text{NaB}_{15} + \text{NaB}_6\text{H}_4\text{C}_6\text{N}$ | -8.4408 |
| NaNH$_4$B$_6$C$_6$ | $12.72\,\text{H}_3\text{N} + 3\,\text{B}_{13}\text{C}_2 + 0.3333\,\text{Na}_3\text{B}_{20} \longrightarrow 11.72\,\text{BN} + 3.105\,\text{B}_9\text{H}_{11} + \text{NaB}_6\text{H}_4\text{C}_6\text{N}$ | -8.2354 |
| NaNH$_4$B$_6$C$_6$ | $7.6\,\text{B}_5\text{H}_7 + 6\,\text{C} + 5.8\,\text{NaN}_3 \longrightarrow 16.4\,\text{H}_3\text{N} + 1.6\,\text{Na}_3\text{B}_{20} + \text{NaB}_6\text{H}_4\text{C}_6\text{N}$ | -7.3378 |
| NaNH$_4$B$_6$C$_6$ | $\text{BN} + \text{NaC}_{64} + 102\,\text{B}_5\text{H}_7 \longrightarrow 56.13\,\text{B}_9\text{H}_{11} + 11.6\,\text{H}_8\text{C}_5 + \text{NaB}_6\text{H}_4\text{C}_6\text{N}$ | -7.2747 |
| NaNH$_4$B$_6$C$_6$ | $1.75\,\text{B}_{13}\text{C}_2 + 0.5\,\text{H}_8\text{C}_5 + 2.526\,\text{NaN}_3 \longrightarrow 6.578\,\text{BN} + 0.5086\,\text{Na}_3\text{B}_{20} + \text{NaB}_6\text{H}_4\text{C}_6\text{N}$ | -6.7504 |
| NaNH$_4$B$_6$C$_6$ | $0.6667\,\text{NaC}_{64} + 31.33\,\text{H}_2 + 0.3333\,\text{NaN}_3 + 6\,\text{B} \longrightarrow 7.333\,\text{H}_8\text{C}_5 + \text{NaB}_6\text{H}_4\text{C}_6\text{N}$ | -6.5944 |
| NaNH$_4$B$_6$C$_6$ | $11.73\,\text{H}_3\text{N} + 3\,\text{B}_{13}\text{C}_2 + \text{Na} \longrightarrow 10.73\,\text{BN} + 4.455\,\text{B}_5\text{H}_7 + \text{NaB}_6\text{H}_4\text{C}_6\text{N}$ | -6.4832 |
| NaNH$_4$B$_6$C$_6$ | $0.4615\,\text{B}_{13}\text{C}_2 + 0.6667\,\text{NaC}_{64} + 32.07\,\text{H}_2 + 0.3333\,\text{NaN}_3 \longrightarrow 7.518\,\text{H}_8\text{C}_5 + \text{NaB}_6\text{H}_4\text{C}_6\text{N}$ | -6.4368 |
| NaNH$_4$B$_6$C$_6$ | $6\,\text{BN} + \text{NaC}_{64} + 55.9\,\text{H}_2 \longrightarrow 5\,\text{H}_3\text{N} + 11.6\,\text{H}_8\text{C}_5 + \text{NaB}_6\text{H}_4\text{C}_6\text{N}$ | -5.9017 |
| NaNH$_4$B$_6$C$_6$ | $11.5\,\text{H}_3\text{N} + 3\,\text{B}_{13}\text{C}_2 + \text{NaH} \longrightarrow 10.5\,\text{BN} + 4.5\,\text{B}_5\text{H}_7 + \text{NaB}_6\text{H}_4\text{C}_6\text{N}$ | -5.8678 |
| NaNH$_4$B$_6$C$_6$ | $3\,\text{B}_{13}\text{C}_2 + 5.321\,\text{B}_9\text{H}_{11} + 6.393\,\text{NaN}_3 \longrightarrow 18.18\,\text{H}_3\text{N} + 5.393\,\text{NaB}_{15} + \text{NaB}_6\text{H}_4\text{C}_6\text{N}$ | -5.8529 |
| NaNH$_4$B$_6$C$_6$ | $0.6667\,\text{B}_9\text{H}_{11} + 0.6667\,\text{NaC}_{64} + 27.67\,\text{H}_2 + 0.3333\,\text{NaN}_3 \longrightarrow 7.333\,\text{H}_8\text{C}_5 + \text{NaB}_6\text{H}_4\text{C}_6\text{N}$ | -5.3855 |
| NaNH$_4$B$_6$C$_6$ | $0.6667\,\text{NaC}_{64} + 1.2\,\text{B}_5\text{H}_7 + 27.13\,\text{H}_2 + 0.3333\,\text{NaN}_3 \longrightarrow 7.333\,\text{H}_8\text{C}_5 + \text{NaB}_6\text{H}_4\text{C}_6\text{N}$ | -5.3404 |
| NaNH$_4$B$_6$C$_6$ | $10.79\,\text{H}_3\text{N} + 3\,\text{B}_{13}\text{C}_2 + \text{Na} \longrightarrow 9.789\,\text{BN} + 2.579\,\text{B}_9\text{H}_{11} + \text{NaB}_6\text{H}_4\text{C}_6\text{N}$ | -5.3394 |
| NaNH$_4$B$_6$C$_6$ | $0.5\,\text{N}_2 + 0.6\,\text{NaC}_{64} + 0.4\,\text{NaB}_{15} + 27.92\,\text{H}_2 \longrightarrow 6.48\,\text{H}_8\text{C}_5 + \text{NaB}_6\text{H}_4\text{C}_6\text{N}$ | -4.8257 |
| NaNH$_4$B$_6$C$_6$ | $10.55\,\text{H}_3\text{N} + 3\,\text{B}_{13}\text{C}_2 + \text{NaH} \longrightarrow 9.553\,\text{BN} + 2.605\,\text{B}_9\text{H}_{11} + \text{NaB}_6\text{H}_4\text{C}_6\text{N}$ | -4.7123 |
| NaNH$_4$B$_6$C$_6$ | $3\,\text{B}_{13}\text{C}_2 + 7.45\,\text{B}_5\text{H}_7 + 5.683\,\text{NaN}_3 \longrightarrow 16.05\,\text{H}_3\text{N} + 4.683\,\text{NaB}_{15} + \text{NaB}_6\text{H}_4\text{C}_6\text{N}$ | -4.6309 |
| NaNH$_4$B$_6$C$_6$ | $\text{BN} + 0.6667\,\text{NaC}_{64} + 0.3333\,\text{NaB}_{15} + 31.33\,\text{H}_2 \longrightarrow 7.333\,\text{H}_8\text{C}_5 + \text{NaB}_6\text{H}_4\text{C}_6\text{N}$ | -3.9017 |
| NaNH$_4$B$_6$C$_6$ | $3\,\text{N}_2 + \text{NaB}_{15} + 1.2\,\text{H}_8\text{C}_5 \longrightarrow 5\,\text{BN} + 0.8\,\text{B}_5\text{H}_7 + \text{NaB}_6\text{H}_4\text{C}_6\text{N}$ | -3.7112 |
| NaNH$_4$B$_6$C$_6$ | $0.5\,\text{N}_2 + \text{NaC}_{64} + 11.94\,\text{B}_5\text{H}_7 \longrightarrow 4.131\,\text{B}_{13}\text{C}_2 + 9.948\,\text{H}_8\text{C}_5 + \text{NaB}_6\text{H}_4\text{C}_6\text{N}$ | -3.6150 |
| NaNH$_4$B$_6$C$_6$ | $\text{H}_3\text{N} + 0.6\,\text{NaC}_{64} + 0.4\,\text{NaB}_{15} + 26.42\,\text{H}_2 \longrightarrow 6.48\,\text{H}_8\text{C}_5 + \text{NaB}_6\text{H}_4\text{C}_6\text{N}$ | -3.5017 |
| NaNH$_4$B$_6$C$_6$ | $0.6667\,\text{NaC}_{64} + 62.25\,\text{B}_5\text{H}_7 + 0.3333\,\text{NaN}_3 \longrightarrow 33.92\,\text{B}_9\text{H}_{11} + 7.333\,\text{H}_8\text{C}_5 + \text{NaB}_6\text{H}_4\text{C}_6\text{N}$ | -3.0468 |
| NaNH$_4$B$_6$C$_6$ | $10\,\text{H}_3\text{N} + \text{NaB}_{15} + 6\,\text{C} \longrightarrow 9\,\text{BN} + 13\,\text{H}_2 + \text{NaB}_6\text{H}_4\text{C}_6\text{N}$ | -2.9038 |
| NaNH$_4$B$_6$C$_6$ | $2.47\,\text{N}_2 + 0.3636\,\text{B}_9\text{H}_{11} + 0.3333\,\text{Na}_3\text{B}_{20} + 6\,\text{C} \longrightarrow 3.939\,\text{BN} + \text{NaB}_6\text{H}_4\text{C}_6\text{N}$ | -2.8438 |
| NaNH$_4$B$_6$C$_6$ | $0.5\,\text{N}_2 + 7.437\,\text{B}_9\text{H}_{11} + \text{NaC}_{64} \longrightarrow 4.687\,\text{B}_{13}\text{C}_2 + 9.725\,\text{H}_8\text{C}_5 + \text{NaB}_6\text{H}_4\text{C}_6\text{N}$ | -2.7971 |
| NaNH$_4$B$_6$C$_6$ | $\text{H}_3\text{N} + \text{NaC}_{64} + 11.58\,\text{B}_5\text{H}_7 \longrightarrow 3.991\,\text{B}_{13}\text{C}_2 + 10\,\text{H}_8\text{C}_5 + \text{NaB}_6\text{H}_4\text{C}_6\text{N}$ | -2.6240 |
| NaNH$_4$B$_6$C$_6$ | $2.709\,\text{N}_2 + \text{NaB}_{15} + 1.2\,\text{H}_8\text{C}_5 \longrightarrow 4.418\,\text{BN} + 0.5091\,\text{B}_9\text{H}_{11} + \text{NaB}_6\text{H}_4\text{C}_6\text{N}$ | -2.3050 |
| NaNH$_4$B$_6$C$_6$ | $2.262\,\text{N}_2 + 0.5714\,\text{B}_5\text{H}_7 + 0.3333\,\text{Na}_3\text{B}_{20} + 6\,\text{C} \longrightarrow 3.524\,\text{BN} + \text{NaB}_6\text{H}_4\text{C}_6\text{N}$ | -1.8393 |
| NaNH$_4$B$_6$C$_6$ | $\text{H}_3\text{N} + 7.21\,\text{B}_9\text{H}_{11} + \text{NaC}_{64} \longrightarrow 4.53\,\text{B}_{13}\text{C}_2 + 9.788\,\text{H}_8\text{C}_5 + \text{NaB}_6\text{H}_4\text{C}_6\text{N}$ | -1.8310 |
| NaNH$_4$B$_6$C$_6$ | $3\,\text{B}_{13}\text{C}_2 + 14.9\,\text{H}_2 + 3.2\,\text{NaN}_3 \longrightarrow 8.6\,\text{H}_3\text{N} + 2.2\,\text{NaB}_{15} + \text{NaB}_6\text{H}_4\text{C}_6\text{N}$ | -1.6133 |
| NaNH$_4$B$_6$C$_6$ | $1.75\,\text{B}_{13}\text{C}_2 + 0.5\,\text{H}_8\text{C}_5 + 1.819\,\text{NaN}_3 \longrightarrow 4.458\,\text{BN} + 0.8194\,\text{NaB}_{15} + \text{NaB}_6\text{H}_4\text{C}_6\text{N}$ | -1.5749 |
| NaNH$_4$B$_6$C$_6$ | $4\,\text{NaH} + 6\,\text{BN} + 198\,\text{C} \longrightarrow 2.5\,\text{N}_2 + 3\,\text{NaC}_{64} + \text{NaB}_6\text{H}_4\text{C}_6\text{N}$ | -1.4903 |
| NaNH$_4$B$_6$C$_6$ | $6.377\,\text{H}_3\text{N} + 0.2415\,\text{NaC}_{64} + 0.7585\,\text{NaB}_{15} \longrightarrow 5.377\,\text{BN} + 1.891\,\text{H}_8\text{C}_5 + \text{NaB}_6\text{H}_4\text{C}_6\text{N}$ | -1.3507 |
| NaNH$_4$B$_6$C$_6$ | $\text{BN} + \text{NaC}_{64} + 11.91\,\text{B}_5\text{H}_7 \longrightarrow 4.196\,\text{B}_{13}\text{C}_2 + 9.921\,\text{H}_8\text{C}_5 + \text{NaB}_6\text{H}_4\text{C}_6\text{N}$ | -1.1019 |
| NaNH$_4$B$_6$C$_6$ | $3\,\text{B}_{13}\text{C}_2 + 23.7\,\text{H}_2 + \text{NaN}_3 \longrightarrow 2\,\text{BN} + 6.2\,\text{B}_5\text{H}_7 + \text{NaB}_6\text{H}_4\text{C}_6\text{N}$ | -1.0074 |
| NaNH$_4$B$_6$C$_6$ | $3\,\text{B}_{13}\text{C}_2 + 20.94\,\text{H}_2 + \text{NaN}_3 \longrightarrow 2\,\text{BN} + 3.444\,\text{B}_9\text{H}_{11} + \text{NaB}_6\text{H}_4\text{C}_6\text{N}$ | -0.7745 |

| Compound | Chemical reaction | Reaction energy(eV) |
|---|---|---|
| $NaNH_4B_6C_6$ | $2.157\,N_2 + 0.3636\,B_9H_{11} + 0.09375\,NaC_{64} + 0.3021\,Na_3B_{20} \longrightarrow 3.314\,BN + NaB_6H_4C_6N$ | -0.5147 |
| $NaNH_4B_6C_6$ | $8.594\,H_3N + 0.09375\,NaC_{64} + 0.9062\,NaB_{15} \longrightarrow 7.594\,BN + 10.89\,H_2 + NaB_6H_4C_6N$ | -0.4641 |
| $NaNH_4B_6C_6$ | $BN + 7.418\,B_9H_{11} + NaC_{64} \longrightarrow 4.751\,B_{13}C_2 + 9.7\,H_8C_5 + NaB_6H_4C_6N$ | -0.2860 |
| $NaNH_4B_6C_6$ | $0.5\,N_2 + NaC_{64} + 13.83\,B_5H_7 \longrightarrow 11.6\,H_8C_5 + 63.14\,B + NaB_6H_4C_6N$ | -0.2307 |
| $NaNH_4B_6C_6$ | $1.333\,H_3N + 6\,C + NaN_3 + 9.333\,B \longrightarrow 3.333\,BN + NaB_6H_4C_6N$ | -0.2121 |
| $PbNH_4B_6C_6$ | $Pb + 3\,B_{13}C_2 + 0.3636\,B_9H_{11} + 18.64\,N_2 \longrightarrow 36.27\,BN + B_6H_4PbC_6N$ | -79.1596 |
| $PbNH_4B_6C_6$ | $Pb + 0.5714\,B_5H_7 + 3\,B_{13}C_2 + 18.43\,N_2 \longrightarrow 35.86\,BN + B_6H_4PbC_6N$ | -78.1552 |
| $PbNH_4B_6C_6$ | $Pb + 3\,B_{13}C_2 + 2\,H_2 + 17\,N_2 \longrightarrow 33\,BN + B_6H_4PbC_6N$ | -71.4187 |
| $PbNH_4B_6C_6$ | $Pb + 3\,B_{13}C_2 + 1.333\,H_3N + 16.33\,N_2 \longrightarrow 33\,BN + B_6H_4PbC_6N$ | -69.6533 |
| $PbNH_4B_6C_6$ | $Pb + 1.75\,B_{13}C_2 + 0.5\,H_8C_5 + 8.875\,N_2 \longrightarrow 16.75\,BN + B_6H_4PbC_6N$ | -30.1358 |
| $PbNH_4B_6C_6$ | $Pb + 3\,B_{13}C_2 + 34\,H_3N \longrightarrow 33\,BN + 49\,H_2 + B_6H_4PbC_6N$ | -26.4003 |
| $PbNH_4B_6C_6$ | $Pb + 3\,B_{13}C_2 + 11.73\,H_3N \longrightarrow 4.455\,B_5H_7 + 10.73\,BN + B_6H_4PbC_6N$ | -3.3774 |
| $PbNH_4B_6C_6$ | $Pb + 3\,B_{13}C_2 + 10.79\,H_3N \longrightarrow 2.579\,B_9H_{11} + 9.789\,BN + B_6H_4PbC_6N$ | -2.2336 |
| $SrNH_4B_6C_6$ | $3\,B_{13}C_2 + 98\,SrN_2 + 18\,B_9H_{11} \longrightarrow 195\,BN + 97\,SrH_2 + SrB_6H_4C_6N$ | -468.3416 |
| $SrNH_4B_6C_6$ | $3\,B_{13}C_2 + 62\,SrN_2 + 18\,B_5H_7 \longrightarrow 123\,BN + 61\,SrH_2 + SrB_6H_4C_6N$ | -293.4320 |
| $SrNH_4B_6C_6$ | $3\,B_{13}C_2 + 69\,H_2 + 34\,Sr_2N \longrightarrow 33\,BN + 67\,SrH_2 + SrB_6H_4C_6N$ | -126.9347 |
| $SrNH_4B_6C_6$ | $3\,B_{13}C_2 + 34\,H_3N + 50\,Sr \longrightarrow 33\,BN + 49\,SrH_2 + SrB_6H_4C_6N$ | -109.2894 |
| $SrNH_4B_6C_6$ | $3\,B_{13}C_2 + 36.5\,SrN_2 + 20\,B_9H_{11} \longrightarrow 72\,H_3N + 35.5\,SrB_6 + SrB_6H_4C_6N$ | -107.9539 |
| $SrNH_4B_6C_6$ | $4\,B_{13}C_2 + 23.5\,N_2 + 2\,SrH_2 \longrightarrow 46\,BN + SrC_2 + SrB_6H_4C_6N$ | -105.0899 |
| $SrNH_4B_6C_6$ | $3\,B_{13}C_2 + 0.3636\,B_9H_{11} + 21.64\,N_2 + SrB_6 \longrightarrow 42.27\,BN + SrB_6H_4C_6N$ | -96.0468 |
| $SrNH_4B_6C_6$ | $3\,B_{13}C_2 + 0.5714\,B_5H_7 + 21.43\,N_2 + SrB_6 \longrightarrow 41.86\,BN + SrB_6H_4C_6N$ | -95.0424 |
| $SrNH_4B_6C_6$ | $3\,B_{13}C_2 + 35\,H_2 + 34\,SrN \longrightarrow 33\,BN + 33\,SrH_2 + SrB_6H_4C_6N$ | -93.0918 |
| $SrNH_4B_6C_6$ | $3\,B_{13}C_2 + 2\,H_2 + 20\,N_2 + SrB_6 \longrightarrow 39\,BN + SrB_6H_4C_6N$ | -88.3060 |
| $SrNH_4B_6C_6$ | $3\,B_{13}C_2 + 1.333\,H_3N + 19.33\,N_2 + SrB_6 \longrightarrow 39\,BN + SrB_6H_4C_6N$ | -86.5405 |
| $SrNH_4B_6C_6$ | $3\,B_{13}C_2 + 0.3636\,B_9H_{11} + 18.64\,N_2 + Sr \longrightarrow 36.27\,BN + SrB_6H_4C_6N$ | -83.7966 |
| $SrNH_4B_6C_6$ | $3\,B_{13}C_2 + 0.3636\,B_9H_{11} + 18.39\,N_2 + 0.5\,Sr_2N \longrightarrow 36.27\,BN + SrB_6H_4C_6N$ | -82.9714 |
| $SrNH_4B_6C_6$ | $3\,B_{13}C_2 + 0.5714\,B_5H_7 + 18.43\,N_2 + Sr \longrightarrow 35.86\,BN + SrB_6H_4C_6N$ | -82.7922 |
| $SrNH_4B_6C_6$ | $3\,B_{13}C_2 + SrN + 0.3636\,B_9H_{11} + 18.14\,N_2 \longrightarrow 36.27\,BN + SrB_6H_4C_6N$ | -82.7477 |
| $SrNH_4B_6C_6$ | $3\,B_{13}C_2 + SrN_2 + 0.3636\,B_9H_{11} + 17.64\,N_2 \longrightarrow 36.27\,BN + SrB_6H_4C_6N$ | -82.3089 |
| $SrNH_4B_6C_6$ | $3\,B_{13}C_2 + 0.5714\,B_5H_7 + 18.18\,N_2 + 0.5\,Sr_2N \longrightarrow 35.86\,BN + SrB_6H_4C_6N$ | -81.9669 |
| $SrNH_4B_6C_6$ | $3\,B_{13}C_2 + SrN + 0.5714\,B_5H_7 + 17.93\,N_2 \longrightarrow 35.86\,BN + SrB_6H_4C_6N$ | -81.7433 |
| $SrNH_4B_6C_6$ | $3\,B_{13}C_2 + SrN_2 + 0.5714\,B_5H_7 + 17.43\,N_2 \longrightarrow 35.86\,BN + SrB_6H_4C_6N$ | -81.3045 |
| $SrNH_4B_6C_6$ | $3\,B_{13}C_2 + 0.1818\,B_9H_{11} + 17.82\,N_2 + SrH_2 \longrightarrow 34.64\,BN + SrB_6H_4C_6N$ | -78.3292 |
| $SrNH_4B_6C_6$ | $3\,B_{13}C_2 + 0.2857\,B_5H_7 + 17.71\,N_2 + SrH_2 \longrightarrow 34.43\,BN + SrB_6H_4C_6N$ | -77.8270 |
| $SrNH_4B_6C_6$ | $3\,B_{13}C_2 + 17\,SrN_2 + 18\,H_2 \longrightarrow 33\,BN + 16\,SrH_2 + SrB_6H_4C_6N$ | -76.3165 |
| $SrNH_4B_6C_6$ | $3\,B_{13}C_2 + 2\,H_2 + 17\,N_2 + Sr \longrightarrow 33\,BN + SrB_6H_4C_6N$ | -76.0558 |
| $SrNH_4B_6C_6$ | $3\,B_{13}C_2 + 2\,H_2 + 16.75\,N_2 + 0.5\,Sr_2N \longrightarrow 33\,BN + SrB_6H_4C_6N$ | -75.2305 |
| $SrNH_4B_6C_6$ | $3\,B_{13}C_2 + 2\,H_2 + SrN + 16.5\,N_2 \longrightarrow 33\,BN + SrB_6H_4C_6N$ | -75.0068 |
| $SrNH_4B_6C_6$ | $3\,B_{13}C_2 + SrN_2 + 2\,H_2 + 16\,N_2 \longrightarrow 33\,BN + SrB_6H_4C_6N$ | -74.5681 |
| $SrNH_4B_6C_6$ | $3\,B_{13}C_2 + H_2 + 17\,N_2 + SrH_2 \longrightarrow 33\,BN + SrB_6H_4C_6N$ | -74.4588 |
| $SrNH_4B_6C_6$ | $3\,B_{13}C_2 + 18\,N_2 + 2\,SrH_2 \longrightarrow SrN_2 + 33\,BN + SrB_6H_4C_6N$ | -74.3495 |
| $SrNH_4B_6C_6$ | $3\,B_{13}C_2 + 1.333\,H_3N + 16.33\,N_2 + Sr \longrightarrow 33\,BN + SrB_6H_4C_6N$ | -74.2903 |
| $SrNH_4B_6C_6$ | $3\,B_{13}C_2 + 17.5\,N_2 + 2\,SrH_2 \longrightarrow SrN + 33\,BN + SrB_6H_4C_6N$ | -73.9108 |
| $SrNH_4B_6C_6$ | $3\,B_{13}C_2 + 17.25\,N_2 + 2\,SrH_2 \longrightarrow 33\,BN + 0.5\,Sr_2N + SrB_6H_4C_6N$ | -73.6871 |
| $SrNH_4B_6C_6$ | $3\,B_{13}C_2 + 0.6667\,H_3N + 16.67\,N_2 + SrH_2 \longrightarrow 33\,BN + SrB_6H_4C_6N$ | -73.5761 |
| $SrNH_4B_6C_6$ | $3\,B_{13}C_2 + 1.333\,H_3N + 16.08\,N_2 + 0.5\,Sr_2N \longrightarrow 33\,BN + SrB_6H_4C_6N$ | -73.4651 |
| $SrNH_4B_6C_6$ | $3\,B_{13}C_2 + 1.333\,H_3N + SrN + 15.83\,N_2 \longrightarrow 33\,BN + SrB_6H_4C_6N$ | -73.2414 |
| $SrNH_4B_6C_6$ | $3\,B_{13}C_2 + 17\,N_2 + 2\,SrH_2 \longrightarrow 33\,BN + Sr + SrB_6H_4C_6N$ | -72.8618 |
| $SrNH_4B_6C_6$ | $3\,B_{13}C_2 + SrN_2 + 1.333\,H_3N + 15.33\,N_2 \longrightarrow 33\,BN + SrB_6H_4C_6N$ | -72.8026 |
| $SrNH_4B_6C_6$ | $3\,B_{13}C_2 + 19.71\,H_3N + 14.29\,Sr_2N \longrightarrow 33\,BN + 27.57\,SrH_2 + SrB_6H_4C_6N$ | -70.4044 |
| $SrNH_4B_6C_6$ | $3\,B_{13}C_2 + 14\,H_3N + 20\,SrN \longrightarrow 33\,BN + 19\,SrH_2 + SrB_6H_4C_6N$ | -66.8824 |
| $SrNH_4B_6C_6$ | $3\,B_{13}C_2 + 36.27\,SrN_2 + 0.3636\,B_9H_{11} \longrightarrow 35.27\,SrN + 36.27\,BN + SrB_6H_4C_6N$ | -66.8330 |
| $SrNH_4B_6C_6$ | $3\,B_{13}C_2 + 24.52\,SrN_2 + 0.3636\,B_9H_{11} \longrightarrow 36.27\,BN + 11.76\,Sr_2N + SrB_6H_4C_6N$ | -66.7319 |
| $SrNH_4B_6C_6$ | $3\,B_{13}C_2 + 73.55\,SrN + 0.3636\,B_9H_{11} \longrightarrow 36.27\,BN + 36.27\,Sr_2N + SrB_6H_4C_6N$ | -66.5211 |
| $SrNH_4B_6C_6$ | $3\,B_{13}C_2 + 35.86\,SrN_2 + 0.5714\,B_5H_7 \longrightarrow 34.86\,SrN + 35.86\,BN + SrB_6H_4C_6N$ | -66.0109 |
| $SrNH_4B_6C_6$ | $3\,B_{13}C_2 + 24.24\,SrN_2 + 0.5714\,B_5H_7 \longrightarrow 35.86\,BN + 11.62\,Sr_2N + SrB_6H_4C_6N$ | -65.9110 |
| $SrNH_4B_6C_6$ | $3\,B_{13}C_2 + 23.17\,SrN_2 + 20\,B_5H_7 \longrightarrow 45.33\,H_3N + 22.17\,SrB_6 + SrB_6H_4C_6N$ | -65.8445 |
| $SrNH_4B_6C_6$ | $3\,B_{13}C_2 + 72.71\,SrN + 0.5714\,B_5H_7 \longrightarrow 35.86\,BN + 35.86\,Sr_2N + SrB_6H_4C_6N$ | -65.7026 |
| $SrNH_4B_6C_6$ | $3\,B_{13}C_2 + 12.5\,SrN_2 + 9\,H_3N \longrightarrow 33\,BN + 11.5\,SrH_2 + SrB_6H_4C_6N$ | -63.9082 |
| $SrNH_4B_6C_6$ | $3\,B_{13}C_2 + 14\,N_2 + 2\,SrH_2 \longrightarrow 27\,BN + SrB_6 + SrB_6H_4C_6N$ | -60.6116 |

| Compound | Chemical reaction | Reaction energy(eV) |
|---|---|---|
| $SrNH_4B_6C_6$ | $3 B_{13}C_2 + 33 SrN_2 + 2 H_2 \longrightarrow 32 SrN + 33 BN + SrB_6H_4C_6N$ | -60.5281 |
| $SrNH_4B_6C_6$ | $3 B_{13}C_2 + 22.33 SrN_2 + 2 H_2 \longrightarrow 33 BN + 10.67 Sr_2N + SrB_6H_4C_6N$ | -60.4363 |
| $SrNH_4B_6C_6$ | $3 B_{13}C_2 + 2 H_2 + 67 SrN \longrightarrow 33 BN + 33 Sr_2N + SrB_6H_4C_6N$ | -60.2443 |
| $SrNH_4B_6C_6$ | $3 B_{13}C_2 + 31.67 SrN_2 + 1.333 H_3N \longrightarrow 30.67 SrN + 33 BN + SrB_6H_4C_6N$ | -59.3476 |
| $SrNH_4B_6C_6$ | $3 B_{13}C_2 + 21.44 SrN_2 + 1.333 H_3N \longrightarrow 33 BN + 10.22 Sr_2N + SrB_6H_4C_6N$ | -59.2597 |
| $SrNH_4B_6C_6$ | $3 B_{13}C_2 + 1.333 H_3N + 64.33 SrN \longrightarrow 33 BN + 31.67 Sr_2N + SrB_6H_4C_6N$ | -59.0753 |
| $SrNH_4B_6C_6$ | $3 B_{13}C_2 + 35 SrN_2 + 2 SrH_2 \longrightarrow 36 SrN + 33 BN + SrB_6H_4C_6N$ | -58.5545 |
| $SrNH_4B_6C_6$ | $3 B_{13}C_2 + 23 SrN_2 + 2 SrH_2 \longrightarrow 33 BN + 12 Sr_2N + SrB_6H_4C_6N$ | -58.4513 |
| $SrNH_4B_6C_6$ | $3 B_{13}C_2 + 69 SrN + 2 SrH_2 \longrightarrow 33 BN + 35 Sr_2N + SrB_6H_4C_6N$ | -58.2535 |
| $SrNH_4B_6C_6$ | $1.2 H_8C_5 + 24.33 SrN + 5.867 B_5H_7 \longrightarrow 23.33 BN + 23.33 SrH_2 + SrB_6H_4C_6N$ | -57.8760 |
| $SrNH_4B_6C_6$ | $3 B_{13}C_2 + 18.64 SrN_2 + 0.3636 B_9H_{11} \longrightarrow 36.27 BN + 17.64 Sr + SrB_6H_4C_6N$ | -56.0713 |
| $SrNH_4B_6C_6$ | $3 B_{13}C_2 + 18.43 SrN_2 + 0.5714 B_5H_7 \longrightarrow 35.86 BN + 17.43 Sr + SrB_6H_4C_6N$ | -55.3760 |
| $SrNH_4B_6C_6$ | $2.375 B_{13}C_2 + 0.25 H_8C_5 + 12.94 N_2 + SrH_2 \longrightarrow 24.88 BN + SrB_6H_4C_6N$ | -53.8173 |
| $SrNH_4B_6C_6$ | $2 B_{13}C_2 + 0.3636 B_9H_{11} + 12.14 N_2 + SrC_2 \longrightarrow 23.27 BN + SrB_6H_4C_6N$ | -51.5686 |
| $SrNH_4B_6C_6$ | $3 B_{13}C_2 + 17 SrN_2 + 2 H_2 \longrightarrow 33 BN + 16 Sr + SrB_6H_4C_6N$ | -50.7649 |
| $SrNH_4B_6C_6$ | $2 B_{13}C_2 + 0.5714 B_5H_7 + 11.93 N_2 + SrC_2 \longrightarrow 22.86 BN + SrB_6H_4C_6N$ | -50.5641 |
| $SrNH_4B_6C_6$ | $3 B_{13}C_2 + 16.33 SrN_2 + 1.333 H_3N \longrightarrow 33 BN + 15.33 Sr + SrB_6H_4C_6N$ | -49.9912 |
| $SrNH_4B_6C_6$ | $3 B_{13}C_2 + 17 SrN_2 + 2 SrH_2 \longrightarrow 33 BN + 18 Sr + SrB_6H_4C_6N$ | -47.5709 |
| $SrNH_4B_6C_6$ | $1.75 B_{13}C_2 + 0.5 H_8C_5 + 11.88 N_2 + SrB_6 \longrightarrow 22.75 BN + SrB_6H_4C_6N$ | -47.0231 |
| $SrNH_4B_6C_6$ | $3 B_{13}C_2 + 37.27 SrN + 0.3636 B_9H_{11} \longrightarrow 36.27 BN + 36.27 Sr + SrB_6H_4C_6N$ | -44.6994 |
| $SrNH_4B_6C_6$ | $3 B_{13}C_2 + 36.86 SrN + 0.5714 B_5H_7 \longrightarrow 35.86 BN + 35.86 Sr + SrB_6H_4C_6N$ | -44.1309 |
| $SrNH_4B_6C_6$ | $2 B_{13}C_2 + 2 H_2 + 10.5 N_2 + SrC_2 \longrightarrow 20 BN + SrB_6H_4C_6N$ | -43.8277 |
| $SrNH_4B_6C_6$ | $2 B_{13}C_2 + 1.333 H_3N + 9.833 N_2 + SrC_2 \longrightarrow 20 BN + SrB_6H_4C_6N$ | -42.0623 |
| $SrNH_4B_6C_6$ | $3 B_{13}C_2 + 2 H_2 + 34 SrN \longrightarrow 33 BN + 33 Sr + SrB_6H_4C_6N$ | -40.3915 |
| $SrNH_4B_6C_6$ | $3 B_{13}C_2 + 1.333 H_3N + 32.67 SrN \longrightarrow 33 BN + 31.67 Sr + SrB_6H_4C_6N$ | -40.0246 |
| $SrNH_4B_6C_6$ | $1.2 H_8C_5 + 17.63 SrN + 2.514 B_9H_{11} \longrightarrow 16.63 BN + 16.63 SrH_2 + SrB_6H_4C_6N$ | -38.5634 |
| $SrNH_4B_6C_6$ | $3 B_{13}C_2 + 34 SrN + 2 SrH_2 \longrightarrow 33 BN + 35 Sr + SrB_6H_4C_6N$ | -37.1975 |
| $SrNH_4B_6C_6$ | $3 B_{13}C_2 + 40 H_3N + SrB_6 \longrightarrow 58 H_2 + 39 BN + SrB_6H_4C_6N$ | -35.3432 |
| $SrNH_4B_6C_6$ | $1.75 B_{13}C_2 + 0.5 H_8C_5 + 8.875 N_2 + Sr \longrightarrow 16.75 BN + SrB_6H_4C_6N$ | -34.7729 |
| $SrNH_4B_6C_6$ | $1.75 B_{13}C_2 + 0.5 H_8C_5 + 8.625 N_2 + 0.5 Sr_2N \longrightarrow 16.75 BN + SrB_6H_4C_6N$ | -33.9476 |
| $SrNH_4B_6C_6$ | $1.75 B_{13}C_2 + 0.5 H_8C_5 + SrN + 8.375 N_2 \longrightarrow 16.75 BN + SrB_6H_4C_6N$ | -33.7239 |
| $SrNH_4B_6C_6$ | $1.75 B_{13}C_2 + SrN_2 + 0.5 H_8C_5 + 7.875 N_2 \longrightarrow 16.75 BN + SrB_6H_4C_6N$ | -33.2852 |
| $SrNH_4B_6C_6$ | $3 B_{13}C_2 + SrN_2 + 32 H_3N \longrightarrow 46 H_2 + 33 BN + SrB_6H_4C_6N$ | -32.1978 |
| $SrNH_4B_6C_6$ | $3 B_{13}C_2 + 33 H_3N + SrN \longrightarrow 47.5 H_2 + 33 BN + SrB_6H_4C_6N$ | -31.3125 |
| $SrNH_4B_6C_6$ | $3 B_{13}C_2 + 34 H_3N + Sr \longrightarrow 49 H_2 + 33 BN + SrB_6H_4C_6N$ | -31.0374 |
| $SrNH_4B_6C_6$ | $3 B_{13}C_2 + 33.5 H_3N + 0.5 Sr_2N \longrightarrow 48.25 H_2 + 33 BN + SrB_6H_4C_6N$ | -30.8741 |
| $SrNH_4B_6C_6$ | $3 B_{13}C_2 + 34 H_3N + SrH_2 \longrightarrow 50 H_2 + 33 BN + SrB_6H_4C_6N$ | -29.4404 |
| $SrNH_4B_6C_6$ | $1.2 H_8C_5 + 8.9 N_2 + 3.8 SrB_6 \longrightarrow 16.8 BN + 2.8 SrH_2 + SrB_6H_4C_6N$ | -27.9991 |
| $SrNH_4B_6C_6$ | $70 H_3N + 1.2 B_5H_7 + 70 SrC_2 \longrightarrow 26.8 H_8C_5 + 69 SrN + SrB_6H_4C_6N$ | -26.6947 |
| $SrNH_4B_6C_6$ | $1.75 B_{13}C_2 + 16.75 SrN_2 + 0.5 H_8C_5 \longrightarrow 15.75 SrN + 16.75 BN + SrB_6H_4C_6N$ | -26.3748 |
| $SrNH_4B_6C_6$ | $1.75 B_{13}C_2 + 11.5 SrN_2 + 0.5 H_8C_5 \longrightarrow 16.75 BN + 5.25 Sr_2N + SrB_6H_4C_6N$ | -26.3297 |
| $SrNH_4B_6C_6$ | $1.75 B_{13}C_2 + 0.5 H_8C_5 + 34.5 SrN \longrightarrow 16.75 BN + 16.75 Sr_2N + SrB_6H_4C_6N$ | -26.2308 |
| $SrNH_4B_6C_6$ | $64.67 H_3N + 0.6667 B_9H_{11} + 64.67 SrC_2 \longrightarrow 24.67 H_8C_5 + 63.67 SrN + SrB_6H_4C_6N$ | -23.9497 |
| $SrNH_4B_6C_6$ | $3 B_{13}C_2 + 0.3636 B_9H_{11} + 37.27 Sr_2N \longrightarrow 36.27 BN + 73.55 Sr + SrB_6H_4C_6N$ | -22.2761 |
| $SrNH_4B_6C_6$ | $3 B_{13}C_2 + 0.5714 B_5H_7 + 36.86 Sr_2N \longrightarrow 35.86 BN + 72.71 Sr + SrB_6H_4C_6N$ | -21.9576 |
| $SrNH_4B_6C_6$ | $3 B_{13}C_2 + 5.409 SrN_2 + 0.3636 B_9H_{11} \longrightarrow 9.818 BN + 4.409 SrB_6 + SrB_6H_4C_6N$ | -21.7373 |
| $SrNH_4B_6C_6$ | $1.75 B_{13}C_2 + 8.875 SrN_2 + 0.5 H_8C_5 \longrightarrow 16.75 BN + 7.875 Sr + SrB_6H_4C_6N$ | -21.5695 |
| $SrNH_4B_6C_6$ | $3 B_{13}C_2 + 5.357 SrN_2 + 0.5714 B_5H_7 \longrightarrow 9.714 BN + 4.357 SrB_6 + SrB_6H_4C_6N$ | -21.4465 |
| $SrNH_4B_6C_6$ | $3 B_{13}C_2 + 1.333 H_3N + 32.67 Sr_2N \longrightarrow 33 BN + 64.33 Sr + SrB_6H_4C_6N$ | -20.3724 |
| $SrNH_4B_6C_6$ | $3 B_{13}C_2 + 4.833 SrN_2 + 1.333 H_3N \longrightarrow 10 BN + 3.833 SrB_6 + SrB_6H_4C_6N$ | -20.1407 |
| $SrNH_4B_6C_6$ | $3 B_{13}C_2 + 2 H_2 + 34 Sr_2N \longrightarrow 33 BN + 67 Sr + SrB_6H_4C_6N$ | -19.9371 |
| $SrNH_4B_6C_6$ | $3 B_{13}C_2 + 11.23 SrN + 3.154 B_9H_{11} \longrightarrow 10.23 H_3N + 10.23 SrB_6 + SrB_6H_4C_6N$ | -19.6297 |
| $SrNH_4B_6C_6$ | $3 B_{13}C_2 + 5 SrN_2 + 2 H_2 \longrightarrow 9 BN + 4 SrB_6 + SrB_6H_4C_6N$ | -19.6165 |
| $SrNH_4B_6C_6$ | $3 B_{13}C_2 + 10.3 SrN + 4.556 B_5H_7 \longrightarrow 9.296 H_3N + 9.296 SrB_6 + SrB_6H_4C_6N$ | -17.3873 |
| $SrNH_4B_6C_6$ | $3 B_{13}C_2 + 34 Sr_2N + 2 SrH_2 \longrightarrow 33 BN + 69 Sr + SrB_6H_4C_6N$ | -16.7431 |
| $SrNH_4B_6C_6$ | $1.75 B_{13}C_2 + 0.5 H_8C_5 + 17.75 SrN \longrightarrow 16.75 BN + 16.75 Sr + SrB_6H_4C_6N$ | -16.1540 |
| $SrNH_4B_6C_6$ | $2 B_{13}C_2 + 21 H_3N + SrC_2 \longrightarrow 29.5 H_2 + 20 BN + SrB_6H_4C_6N$ | -16.0222 |
| $SrNH_4B_6C_6$ | $4.25 SrN_2 + 1.5 B_9H_{11} + 3 SrC_2 \longrightarrow 7.5 BN + 6.25 SrH_2 + SrB_6H_4C_6N$ | -14.9611 |
| $SrNH_4B_6C_6$ | $3 B_{13}C_2 + 6.5 SrN_2 + 20 H_2 \longrightarrow 12 H_3N + 5.5 SrB_6 + SrB_6H_4C_6N$ | -14.8985 |
| $SrNH_4B_6C_6$ | $6 C + 8.333 SrN + 2.667 B_5H_7 \longrightarrow 7.333 BN + 7.333 SrH_2 + SrB_6H_4C_6N$ | -14.3857 |

| Compound | Chemical reaction | Reaction energy(eV) |
| --- | --- | --- |
| $SrNH_4B_6C_6$ | $3\,B_{13}C_2 + 6.182\,SrN + 0.3636\,B_9H_{11} \longrightarrow 5.182\,BN + 5.182\,SrB_6 + SrB_6H_4C_6N$ | -13.8339 |
| $SrNH_4B_6C_6$ | $3\,B_{13}C_2 + 6.122\,SrN + 0.5714\,B_5H_7 \longrightarrow 5.122\,BN + 5.122\,SrB_6 + SrB_6H_4C_6N$ | -13.6190 |
| $SrNH_4B_6C_6$ | $3\,B_{13}C_2 + 1.333\,H_3N + 5.524\,SrN \longrightarrow 5.857\,BN + 4.524\,SrB_6 + SrB_6H_4C_6N$ | -13.0786 |
| $SrNH_4B_6C_6$ | $3\,B_{13}C_2 + 3.5\,SrN_2 + 2\,SrH_2 \longrightarrow 6\,BN + 4.5\,SrB_6 + SrB_6H_4C_6N$ | -12.5289 |
| $SrNH_4B_6C_6$ | $3\,B_{13}C_2 + 2\,H_2 + 5.714\,SrN \longrightarrow 4.714\,BN + 4.714\,SrB_6 + SrB_6H_4C_6N$ | -12.3108 |
| $SrNH_4B_6C_6$ | $3\,B_{13}C_2 + SrN_2 + 11.09\,H_3N \longrightarrow 4.182\,B_5H_7 + 12.09\,BN + SrB_6H_4C_6N$ | -10.5844 |
| $SrNH_4B_6C_6$ | $44\,H_3N + 43\,SrC_2 + SrB_6 \longrightarrow 16\,H_8C_5 + 43\,SrN + SrB_6H_4C_6N$ | -10.3983 |
| $SrNH_4B_6C_6$ | $3\,B_{13}C_2 + 0.3636\,B_9H_{11} + 0.5\,N_2 + 7.045\,Sr \longrightarrow 6.045\,SrB_6 + SrB_6H_4C_6N$ | -9.7386 |
| $SrNH_4B_6C_6$ | $3\,B_{13}C_2 + 0.5714\,B_5H_7 + 0.5\,N_2 + 6.976\,Sr \longrightarrow 5.976\,SrB_6 + SrB_6H_4C_6N$ | -9.5827 |
| $SrNH_4B_6C_6$ | $3\,B_{13}C_2 + 1.229\,B_9H_{11} + 4.171\,Sr_2N \longrightarrow 3.171\,H_3N + 7.343\,SrB_6 + SrB_6H_4C_6N$ | -9.5714 |
| $SrNH_4B_6C_6$ | $3\,B_{13}C_2 + SrN_2 + 10.21\,H_3N \longrightarrow 2.421\,B_9H_{11} + 11.21\,BN + SrB_6H_4C_6N$ | -9.5107 |
| $SrNH_4B_6C_6$ | $3\,B_{13}C_2 + 10.25\,H_2 + 6.5\,SrN \longrightarrow 5.5\,H_3N + 5.5\,SrB_6 + SrB_6H_4C_6N$ | -9.1439 |
| $SrNH_4B_6C_6$ | $1.2\,H_8C_5 + 2.123\,B_5H_7 + 5.615\,Sr_2N \longrightarrow 4.615\,BN + 10.23\,SrH_2 + SrB_6H_4C_6N$ | -9.0751 |
| $SrNH_4B_6C_6$ | $3\,B_{13}C_2 + 0.3636\,B_9H_{11} + 3.329\,Sr_2N \longrightarrow 2.329\,BN + 5.657\,SrB_6 + SrB_6H_4C_6N$ | -8.9989 |
| $SrNH_4B_6C_6$ | $3\,B_{13}C_2 + 0.5\,SrN_2 + 0.3636\,B_9H_{11} + 6.545\,Sr \longrightarrow 6.045\,SrB_6 + SrB_6H_4C_6N$ | -8.9948 |
| $SrNH_4B_6C_6$ | $3\,B_{13}C_2 + 1.87\,B_5H_7 + 4.029\,Sr_2N \longrightarrow 3.029\,H_3N + 7.058\,SrB_6 + SrB_6H_4C_6N$ | -8.9946 |
| $SrNH_4B_6C_6$ | $3\,B_{13}C_2 + 11.41\,H_3N + SrN \longrightarrow 4.318\,B_5H_7 + 11.41\,BN + SrB_6H_4C_6N$ | -8.9943 |
| $SrNH_4B_6C_6$ | $3\,B_{13}C_2 + 0.5\,SrN_2 + 0.5714\,B_5H_7 + 6.476\,Sr \longrightarrow 5.976\,SrB_6 + SrB_6H_4C_6N$ | -8.8388 |
| $SrNH_4B_6C_6$ | $3\,B_{13}C_2 + 0.5714\,B_5H_7 + 3.297\,Sr_2N \longrightarrow 2.297\,BN + 5.593\,SrB_6 + SrB_6H_4C_6N$ | -8.8305 |
| $SrNH_4B_6C_6$ | $Sr_2N + 42\,B + 3\,SrC_2 + 2\,SrH_2 \longrightarrow 6\,SrB_6 + SrB_6H_4C_6N$ | -8.8298 |
| $SrNH_4B_6C_6$ | $3\,B_{13}C_2 + 1.333\,H_3N + 2.974\,Sr_2N \longrightarrow 3.308\,BN + 4.949\,SrB_6 + SrB_6H_4C_6N$ | -8.7582 |
| $SrNH_4B_6C_6$ | $3\,B_{13}C_2 + SrN + 0.3636\,B_9H_{11} + 6.045\,Sr \longrightarrow 6.045\,SrB_6 + SrB_6H_4C_6N$ | -8.6897 |
| $SrNH_4B_6C_6$ | $3\,B_{13}C_2 + 2\,H_2 + 0.5\,N_2 + 6.5\,Sr \longrightarrow 5.5\,SrB_6 + SrB_6H_4C_6N$ | -8.6797 |
| $SrNH_4B_6C_6$ | $3\,B_{13}C_2 + SrN + 0.5714\,B_5H_7 + 5.976\,Sr \longrightarrow 5.976\,SrB_6 + SrB_6H_4C_6N$ | -8.5337 |
| $SrNH_4B_6C_6$ | $3\,B_{13}C_2 + 11.57\,H_3N + 0.5\,Sr_2N \longrightarrow 4.386\,B_5H_7 + 11.07\,BN + SrB_6H_4C_6N$ | -8.2036 |
| $SrNH_4B_6C_6$ | $3.8\,SrN_2 + 1.2\,H_8C_5 + 12.6\,B \longrightarrow 6.6\,BN + 2.8\,SrH_2 + SrB_6H_4C_6N$ | -8.1363 |
| $SrNH_4B_6C_6$ | $3\,B_{13}C_2 + 13.64\,H_3N + SrB_6 \longrightarrow 5.273\,B_5H_7 + 12.64\,BN + SrB_6H_4C_6N$ | -8.0915 |
| $SrNH_4B_6C_6$ | $3\,B_{13}C_2 + 0.3636\,B_9H_{11} + Sr_2N + 5.045\,Sr \longrightarrow 6.045\,SrB_6 + SrB_6H_4C_6N$ | -8.0881 |
| $SrNH_4B_6C_6$ | $3\,B_{13}C_2 + 11.73\,H_3N + Sr \longrightarrow 4.455\,B_5H_7 + 10.73\,BN + SrB_6H_4C_6N$ | -8.0144 |
| $SrNH_4B_6C_6$ | $3\,B_{13}C_2 + 0.5\,SrN_2 + 2\,H_2 + 6\,Sr \longrightarrow 5.5\,SrB_6 + SrB_6H_4C_6N$ | -7.9358 |
| $SrNH_4B_6C_6$ | $3\,B_{13}C_2 + 0.5714\,B_5H_7 + Sr_2N + 4.976\,Sr \longrightarrow 5.976\,SrB_6 + SrB_6H_4C_6N$ | -7.9321 |
| $SrNH_4B_6C_6$ | $3\,B_{13}C_2 + 10.5\,H_3N + SrN \longrightarrow 2.5\,B_9H_{11} + 10.5\,BN + SrB_6H_4C_6N$ | -7.8855 |
| $SrNH_4B_6C_6$ | $3\,B_{13}C_2 + 2\,H_2 + 3.077\,Sr_2N \longrightarrow 2.077\,BN + 5.154\,SrB_6 + SrB_6H_4C_6N$ | -7.8415 |
| $SrNH_4B_6C_6$ | $3\,B_{13}C_2 + 0.3636\,B_9H_{11} + BN + 7.212\,Sr \longrightarrow 6.212\,SrB_6 + SrB_6H_4C_6N$ | -7.6969 |
| $SrNH_4B_6C_6$ | $3\,B_{13}C_2 + 2\,H_2 + SrN + 5.5\,Sr \longrightarrow 5.5\,SrB_6 + SrB_6H_4C_6N$ | -7.6307 |
| $SrNH_4B_6C_6$ | $3\,B_{13}C_2 + H_3N + 0.09091\,B_9H_{11} + 6.636\,Sr \longrightarrow 5.636\,SrB_6 + SrB_6H_4C_6N$ | -7.6203 |
| $SrNH_4B_6C_6$ | $3\,B_{13}C_2 + 1.333\,H_3N + 6.444\,Sr \longrightarrow 0.3333\,BN + 5.444\,SrB_6 + SrB_6H_4C_6N$ | -7.5948 |
| $SrNH_4B_6C_6$ | $3\,B_{13}C_2 + H_3N + 0.1429\,B_5H_7 + 6.619\,Sr \longrightarrow 5.619\,SrB_6 + SrB_6H_4C_6N$ | -7.5813 |
| $SrNH_4B_6C_6$ | $3\,B_{13}C_2 + 0.5714\,B_5H_7 + BN + 7.143\,Sr \longrightarrow 6.143\,SrB_6 + SrB_6H_4C_6N$ | -7.5410 |
| $SrNH_4B_6C_6$ | $3\,B_{13}C_2 + 1.333\,H_3N + 7.167\,Sr \longrightarrow 0.3333\,Sr_2N + 5.5\,SrB_6 + SrB_6H_4C_6N$ | -7.4644 |
| $SrNH_4B_6C_6$ | $3\,B_{13}C_2 + 4\,SrN + 2\,SrH_2 \longrightarrow 3\,BN + 5\,SrB_6 + SrB_6H_4C_6N$ | -7.4150 |
| $SrNH_4B_6C_6$ | $3\,B_{13}C_2 + H_3N + 0.5\,H_2 + 6.5\,Sr \longrightarrow 5.5\,SrB_6 + SrB_6H_4C_6N$ | -7.3556 |
| $SrNH_4B_6C_6$ | $3\,B_{13}C_2 + 1.333\,H_3N + 6.833\,Sr \longrightarrow 0.3333\,SrN + 5.5\,SrB_6 + SrB_6H_4C_6N$ | -7.2639 |
| $SrNH_4B_6C_6$ | $3\,B_{13}C_2 + 1.333\,H_3N + 6.667\,Sr \longrightarrow 0.1667\,SrN_2 + 5.5\,SrB_6 + SrB_6H_4C_6N$ | -7.1622 |
| $SrNH_4B_6C_6$ | $3\,B_{13}C_2 + 10.64\,H_3N + 0.5\,Sr_2N \longrightarrow 2.539\,B_9H_{11} + 10.14\,BN + SrB_6H_4C_6N$ | -7.0773 |
| $SrNH_4B_6C_6$ | $3\,B_{13}C_2 + 2\,H_2 + Sr_2N + 4.5\,Sr \longrightarrow 5.5\,SrB_6 + SrB_6H_4C_6N$ | -7.0291 |
| $SrNH_4B_6C_6$ | $3\,B_{13}C_2 + 1.333\,H_3N + 6.5\,Sr \longrightarrow 0.1667\,N_2 + 5.5\,SrB_6 + SrB_6H_4C_6N$ | -6.9142 |
| $SrNH_4B_6C_6$ | $3\,B_{13}C_2 + 10.79\,H_3N + Sr \longrightarrow 2.579\,B_9H_{11} + 9.789\,BN + SrB_6H_4C_6N$ | -6.8706 |
| $SrNH_4B_6C_6$ | $3\,B_{13}C_2 + 12.53\,H_3N + SrB_6 \longrightarrow 3.053\,B_9H_{11} + 11.53\,BN + SrB_6H_4C_6N$ | -6.7376 |
| $SrNH_4B_6C_6$ | $3.6\,B_5H_7 + 4.033\,N_2 + 3\,SrC_2 \longrightarrow 7.067\,H_3N + 2\,SrB_6 + SrB_6H_4C_6N$ | -6.6675 |
| $SrNH_4B_6C_6$ | $3\,B_{13}C_2 + 2\,H_2 + BN + 6.667\,Sr \longrightarrow 5.667\,SrB_6 + SrB_6H_4C_6N$ | -6.6380 |
| $SrNH_4B_6C_6$ | $3\,B_{13}C_2 + H_3N + 6\,Sr + 0.5\,SrH_2 \longrightarrow 5.5\,SrB_6 + SrB_6H_4C_6N$ | -6.5571 |
| $SrNH_4B_6C_6$ | $1.2\,H_8C_5 + 1.104\,B_9H_{11} + 4.936\,Sr_2N \longrightarrow 3.936\,BN + 8.872\,SrH_2 + SrB_6H_4C_6N$ | -6.4994 |
| $SrNH_4B_6C_6$ | $3\,B_{13}C_2 + 5.375\,H_2 + 3.25\,Sr_2N \longrightarrow 2.25\,H_3N + 5.5\,SrB_6 + SrB_6H_4C_6N$ | -6.2945 |
| $SrNH_4B_6C_6$ | $SrN + 36\,B + 3\,SrC_2 + 2\,SrH_2 \longrightarrow 5\,SrB_6 + SrB_6H_4C_6N$ | -6.2811 |
| $SrNH_4B_6C_6$ | $1.75\,B_{13}C_2 + 2.969\,SrN_2 + 0.5\,H_8C_5 \longrightarrow 4.938\,BN + 1.969\,SrB_6 + SrB_6H_4C_6N$ | -6.2387 |
| $SrNH_4B_6C_6$ | $28\,H_3N + 6\,B + 28\,SrC_2 \longrightarrow 10\,H_8C_5 + 27\,SrN + SrB_6H_4C_6N$ | -5.9766 |
| $SrNH_4B_6C_6$ | $3\,B_{13}C_2 + 11.27\,H_3N + SrH_2 \longrightarrow 4.545\,B_5H_7 + 10.27\,BN + SrB_6H_4C_6N$ | -5.9476 |
| $SrNH_4B_6C_6$ | $2\,H_2 + Sr_2N + 30\,B + 3\,SrC_2 \longrightarrow 4\,SrB_6 + SrB_6H_4C_6N$ | -5.7231 |
| $SrNH_4B_6C_6$ | $6\,C + 5.286\,SrN + 1.143\,B_9H_{11} \longrightarrow 4.286\,BN + 4.286\,SrH_2 + SrB_6H_4C_6N$ | -5.6073 |

| Compound | Chemical reaction | Reaction energy(eV) |
|---|---|---|
| $SrNH_4B_6C_6$ | $3 B_{13}C_2 + 0.5 N_2 + 4.5 Sr + 2 SrH_2 \longrightarrow 5.5 SrB_6 + SrB_6H_4C_6N$ | -5.4857 |
| $SrNH_4B_6C_6$ | $1.75 B_{13}C_2 + 0.5 H_8C_5 + 17.75 Sr_2N \longrightarrow 16.75 BN + 34.5 Sr + SrB_6H_4C_6N$ | -5.4756 |
| $SrNH_4B_6C_6$ | $2 B_9H_{11} + 3.5 N_2 + 3 SrC_2 \longrightarrow 6 H_3N + 2 SrB_6 + SrB_6H_4C_6N$ | -5.3905 |
| $SrNH_4B_6C_6$ | $2.688 B_{13}C_2 + 0.125 H_8C_5 + H_3N + 5.823 Sr \longrightarrow 4.823 SrB_6 + SrB_6H_4C_6N$ | -5.3293 |
| $SrNH_4B_6C_6$ | $0.5714 B_5H_7 + Sr_2N + 27.14 B + 3 SrC_2 \longrightarrow 4 SrB_6 + SrB_6H_4C_6N$ | -5.1259 |
| $SrNH_4B_6C_6$ | $0.3636 B_9H_{11} + Sr_2N + 26.73 B + 3 SrC_2 \longrightarrow 4 SrB_6 + SrB_6H_4C_6N$ | -5.0636 |
| $SrNH_4B_6C_6$ | $3 B_{13}C_2 + 0.3636 B_9H_{11} + 6.045 Sr_2N \longrightarrow 5.045 SrN + 6.045 SrB_6 + SrB_6H_4C_6N$ | -5.0527 |
| $SrNH_4B_6C_6$ | $3 B_{13}C_2 + 0.3636 B_9H_{11} + 4.364 Sr_2N \longrightarrow 1.682 SrN_2 + 6.045 SrB_6 + SrB_6H_4C_6N$ | -5.0382 |
| $SrNH_4B_6C_6$ | $1.2 H_8C_5 + 2.867 Sr_2N + 34.4 B \longrightarrow 1.867 H_3N + 4.733 SrB_6 + SrB_6H_4C_6N$ | -5.0291 |
| $SrNH_4B_6C_6$ | $0.5 SrN_2 + 33 B + 3 SrC_2 + 2 SrH_2 \longrightarrow 4.5 SrB_6 + SrB_6H_4C_6N$ | -5.0110 |
| $SrNH_4B_6C_6$ | $3 B_{13}C_2 + 13.09 SrN + 0.3636 B_9H_{11} \longrightarrow 6.045 SrN_2 + 6.045 SrB_6 + SrB_6H_4C_6N$ | -5.0007 |
| $SrNH_4B_6C_6$ | $3 B_{13}C_2 + 0.5714 B_5H_7 + 5.976 Sr_2N \longrightarrow 4.976 SrN + 5.976 SrB_6 + SrB_6H_4C_6N$ | -4.9384 |
| $SrNH_4B_6C_6$ | $3 B_{13}C_2 + 0.5714 B_5H_7 + 4.317 Sr_2N \longrightarrow 1.659 SrN_2 + 5.976 SrB_6 + SrB_6H_4C_6N$ | -4.9242 |
| $SrNH_4B_6C_6$ | $3 B_{13}C_2 + 12.95 SrN + 0.5714 B_5H_7 \longrightarrow 5.976 SrN_2 + 5.976 SrB_6 + SrB_6H_4C_6N$ | -4.8870 |
| $SrNH_4B_6C_6$ | $3 B_{13}C_2 + 10.32 H_3N + SrH_2 \longrightarrow 2.632 B_9H_{11} + 9.316 BN + SrB_6H_4C_6N$ | -4.7805 |
| $SrNH_4B_6C_6$ | $3 B_{13}C_2 + 0.5 SrN_2 + 4 Sr + 2 SrH_2 \longrightarrow 5.5 SrB_6 + SrB_6H_4C_6N$ | -4.7418 |
| $SrNH_4B_6C_6$ | $3 B_{13}C_2 + SrN + 3.5 Sr + 2 SrH_2 \longrightarrow 5.5 SrB_6 + SrB_6H_4C_6N$ | -4.4367 |
| $SrNH_4B_6C_6$ | $3 B_{13}C_2 + 2 H_2 + 5.5 Sr_2N \longrightarrow 4.5 SrN + 5.5 SrB_6 + SrB_6H_4C_6N$ | -4.3219 |
| $SrNH_4B_6C_6$ | $3 B_{13}C_2 + 2 H_2 + 4 Sr_2N \longrightarrow 1.5 SrN_2 + 5.5 SrB_6 + SrB_6H_4C_6N$ | -4.3090 |
| $SrNH_4B_6C_6$ | $3 B_{13}C_2 + 2.154 Sr_2N + 2 SrH_2 \longrightarrow 1.154 BN + 5.308 SrB_6 + SrB_6H_4C_6N$ | -4.2865 |
| $SrNH_4B_6C_6$ | $3 B_{13}C_2 + 2 H_2 + 12 SrN \longrightarrow 5.5 SrN_2 + 5.5 SrB_6 + SrB_6H_4C_6N$ | -4.2746 |
| $SrNH_4B_6C_6$ | $0.5 N_2 + 30 B + 3 SrC_2 + 2 SrH_2 \longrightarrow 4 SrB_6 + SrB_6H_4C_6N$ | -4.1797 |
| $SrNH_4B_6C_6$ | $3 B_{13}C_2 + 0.3636 B_9H_{11} + 3.523 Sr_2N \longrightarrow 1.261 N_2 + 6.045 SrB_6 + SrB_6H_4C_6N$ | -3.9242 |
| $SrNH_4B_6C_6$ | $3 B_{13}C_2 + Sr_2N + 2.5 Sr + 2 SrH_2 \longrightarrow 5.5 SrB_6 + SrB_6H_4C_6N$ | -3.8352 |
| $SrNH_4B_6C_6$ | $3 B_{13}C_2 + 0.5714 B_5H_7 + 3.488 Sr_2N \longrightarrow 1.244 N_2 + 5.976 SrB_6 + SrB_6H_4C_6N$ | -3.8254 |
| $SrNH_4B_6C_6$ | $3 B_{13}C_2 + BN + 4.667 Sr + 2 SrH_2 \longrightarrow 5.667 SrB_6 + SrB_6H_4C_6N$ | -3.4440 |
| $SrNH_4B_6C_6$ | $1.2 H_8C_5 + H_3N + 1.2 B_5H_7 + 9.5 Sr \longrightarrow 8.5 SrH_2 + SrB_6H_4C_6N$ | -3.3736 |
| $SrNH_4B_6C_6$ | $3 B_{13}C_2 + 2 H_2 + 3.25 Sr_2N \longrightarrow 1.125 N_2 + 5.5 SrB_6 + SrB_6H_4C_6N$ | -3.3154 |
| $SrNH_4B_6C_6$ | $2 H_2 + SrN + 24 B + 3 SrC_2 \longrightarrow 3 SrB_6 + SrB_6H_4C_6N$ | -3.1743 |
| $SrNH_4B_6C_6$ | $3 B_{13}C_2 + 1.333 H_3N + 6.833 Sr_2N \longrightarrow 7.167 SrN + 5.5 SrB_6 + SrB_6H_4C_6N$ | -3.1529 |
| $SrNH_4B_6C_6$ | $3 B_{13}C_2 + 1.333 H_3N + 4.444 Sr_2N \longrightarrow 2.389 SrN_2 + 5.5 SrB_6 + SrB_6H_4C_6N$ | -3.1324 |
| $SrNH_4B_6C_6$ | $3 B_{13}C_2 + 1.333 H_3N + 13.33 SrN \longrightarrow 6.833 SrN_2 + 5.5 SrB_6 + SrB_6H_4C_6N$ | -3.0942 |
| $SrNH_4B_6C_6$ | $SrN + 0.5714 B_5H_7 + 21.14 B + 3 SrC_2 \longrightarrow 3 SrB_6 + SrB_6H_4C_6N$ | -2.5771 |
| $SrNH_4B_6C_6$ | $1.2 H_8C_5 + H_3N + 0.6667 B_9H_{11} + 8.967 Sr \longrightarrow 7.967 SrH_2 + SrB_6H_4C_6N$ | -2.5669 |
| $SrNH_4B_6C_6$ | $0.75 B_{13}C_2 + 0.5 H_8C_5 + 2.375 N_2 + SrC_2 \longrightarrow 3.75 BN + SrB_6H_4C_6N$ | -2.5448 |
| $SrNH_4B_6C_6$ | $SrN + 0.3636 B_9H_{11} + 20.73 B + 3 SrC_2 \longrightarrow 3 SrB_6 + SrB_6H_4C_6N$ | -2.5149 |
| $SrNH_4B_6C_6$ | $3 B_{13}C_2 + 7.045 SrN + 0.3636 B_9H_{11} \longrightarrow 3.023 N_2 + 6.045 SrB_6 + SrB_6H_4C_6N$ | -2.3483 |
| $SrNH_4B_6C_6$ | $3 B_{13}C_2 + 3.5 Sr_2N + 2 SrH_2 \longrightarrow 2.5 SrN + 5.5 SrB_6 + SrB_6H_4C_6N$ | -2.3311 |
| $SrNH_4B_6C_6$ | $3 B_{13}C_2 + 2.667 Sr_2N + 2 SrH_2 \longrightarrow 0.8333 SrN_2 + 5.5 SrB_6 + SrB_6H_4C_6N$ | -2.3240 |
| $SrNH_4B_6C_6$ | $1.2 H_8C_5 + 1.2 B_5H_7 + 0.5 N_2 + 8 Sr \longrightarrow 7 SrH_2 + SrB_6H_4C_6N$ | -2.3022 |
| $SrNH_4B_6C_6$ | $3 B_{13}C_2 + 8 SrN + 2 SrH_2 \longrightarrow 3.5 SrN_2 + 5.5 SrB_6 + SrB_6H_4C_6N$ | -2.3010 |
| $SrNH_4B_6C_6$ | $3 B_{13}C_2 + 2 N_2 + 6.5 SrH_2 \longrightarrow 3 H_3N + 5.5 SrB_6 + SrB_6H_4C_6N$ | -2.2715 |
| $SrNH_4B_6C_6$ | $3 B_{13}C_2 + 6.976 SrN + 0.5714 B_5H_7 \longrightarrow 2.988 N_2 + 5.976 SrB_6 + SrB_6H_4C_6N$ | -2.2650 |
| $SrNH_4B_6C_6$ | $2 B_{13}C_2 + 7.591 H_3N + SrC_2 \longrightarrow 2.682 B_5H_7 + 6.591 BN + SrB_6H_4C_6N$ | -2.1615 |
| $SrNH_4B_6C_6$ | $3 B_{13}C_2 + 16.73 H_3N + 32.47 Sr \longrightarrow 6.6 B_5H_7 + 15.73 Sr_2N + SrB_6H_4C_6N$ | -2.0624 |
| $SrNH_4B_6C_6$ | $0.4615 B_{13}C_2 + 20.62 H_3N + 20.62 SrC_2 \longrightarrow 7.231 H_8C_5 + 19.62 SrN + SrB_6H_4C_6N$ | -1.9557 |
| $SrNH_4B_6C_6$ | $0.5 SrN_2 + 2 H_2 + 21 B + 3 SrC_2 \longrightarrow 2.5 SrB_6 + SrB_6H_4C_6N$ | -1.9042 |
| $SrNH_4B_6C_6$ | $1.75 B_{13}C_2 + 0.5 H_8C_5 + 3.393 SrN \longrightarrow 2.393 BN + 2.393 SrB_6 + SrB_6H_4C_6N$ | -1.9009 |
| $SrNH_4B_6C_6$ | $3 B_{13}C_2 + 1.714 Sr_2N + 3.071 SrH_2 \longrightarrow 0.7143 H_3N + 5.5 SrB_6 + SrB_6H_4C_6N$ | -1.8909 |
| $SrNH_4B_6C_6$ | $3 B_{13}C_2 + 2 H_2 + 6.5 SrN \longrightarrow 2.75 N_2 + 5.5 SrB_6 + SrB_6H_4C_6N$ | -1.8615 |
| $SrNH_4B_6C_6$ | $3 B_{13}C_2 + 2.25 Sr_2N + 2 SrH_2 \longrightarrow 0.625 N_2 + 5.5 SrB_6 + SrB_6H_4C_6N$ | -1.7720 |
| $SrNH_4B_6C_6$ | $BN + 29 B + 3 SrC_2 + 2 SrH_2 \longrightarrow 4 SrB_6 + SrB_6H_4C_6N$ | -1.6129 |
| $SrNH_4B_6C_6$ | $0.5 SrN_2 + 1.2 H_8C_5 + 1.2 B_5H_7 + 7.5 Sr \longrightarrow 7 SrH_2 + SrB_6H_4C_6N$ | -1.5583 |
| $SrNH_4B_6C_6$ | $3 B_{13}C_2 + 1.333 H_3N + 3.25 Sr_2N \longrightarrow 1.792 N_2 + 5.5 SrB_6 + SrB_6H_4C_6N$ | -1.5499 |
| $SrNH_4B_6C_6$ | $1.2 H_8C_5 + 0.6667 B_9H_{11} + 0.5 N_2 + 7.467 Sr \longrightarrow 6.467 SrH_2 + SrB_6H_4C_6N$ | -1.4956 |
| $SrNH_4B_6C_6$ | $2 B_{13}C_2 + 7.026 H_3N + SrC_2 \longrightarrow 1.553 B_9H_{11} + 6.026 BN + SrB_6H_4C_6N$ | -1.4729 |
| $SrNH_4B_6C_6$ | $3 B_{13}C_2 + 2.4 SrN + 4.1 SrH_2 \longrightarrow 1.4 H_3N + 5.5 SrB_6 + SrB_6H_4C_6N$ | -1.4683 |
| $SrNH_4B_6C_6$ | $0.7778 B_{13}C_2 + 2.556 SrN_2 + 0.8889 H_8C_5 \longrightarrow 4.111 BN + 1.556 SrH_2 + SrB_6H_4C_6N$ | -1.3462 |
| $SrNH_4B_6C_6$ | $0.5 SrN_2 + 0.5714 B_5H_7 + 18.14 B + 3 SrC_2 \longrightarrow 2.5 SrB_6 + SrB_6H_4C_6N$ | -1.3071 |
| $SrNH_4B_6C_6$ | $1.2 H_8C_5 + SrN + 1.2 B_5H_7 + 7 Sr \longrightarrow 7 SrH_2 + SrB_6H_4C_6N$ | -1.2532 |

| Compound | Chemical reaction | Reaction energy(eV) |
|---|---|---|
| $SrNH_4B_6C_6$ | $0.5\,SrN_2 + 0.3636\,B_9H_{11} + 17.73\,B + 3\,SrC_2 \longrightarrow 2.5\,SrB_6 + SrB_6H_4C_6N$ | -1.2448 |
| $SrNH_4B_6C_6$ | $3\,B_{13}C_2 + 14.78\,H_3N + 28.56\,Sr \longrightarrow 3.667\,B_9H_{11} + 13.78\,Sr_2N + SrB_6H_4C_6N$ | -1.1760 |
| $SrNH_4B_6C_6$ | $3\,B_{13}C_2 + 1.5\,SrN_2 + 5\,SrH_2 \longrightarrow 2\,H_3N + 5.5\,SrB_6 + SrB_6H_4C_6N$ | -1.1113 |
| $SrNH_4B_6C_6$ | $2\,H_2 + 0.5\,N_2 + 18\,B + 3\,SrC_2 \longrightarrow 2\,SrB_6 + SrB_6H_4C_6N$ | -1.0729 |
| $SrNH_4B_6C_6$ | $23.75\,B_9H_{11} + Sr_2N + 3\,SrC_2 \longrightarrow 36.75\,B_5H_7 + 4\,SrB_6 + SrB_6H_4C_6N$ | -1.0590 |
| $SrNH_4B_6C_6$ | $0.5\,H_8C_5 + Sr_2N + 22.5\,B + 1.75\,SrC_2 \longrightarrow 2.75\,SrB_6 + SrB_6H_4C_6N$ | -0.7873 |
| $SrNH_4B_6C_6$ | $3\,B_{13}C_2 + 4.5\,SrN + 2\,SrH_2 \longrightarrow 1.75\,N_2 + 5.5\,SrB_6 + SrB_6H_4C_6N$ | -0.7654 |
| $SrNH_4B_6C_6$ | $0.5\,SrN_2 + 1.2\,H_8C_5 + 0.6667\,B_9H_{11} + 6.967\,Sr \longrightarrow 6.467\,SrH_2 + SrB_6H_4C_6N$ | -0.7517 |
| $SrNH_4B_6C_6$ | $1.2\,H_8C_5 + 1.2\,B_5H_7 + Sr_2N + 6\,Sr \longrightarrow 7\,SrH_2 + SrB_6H_4C_6N$ | -0.6516 |
| $SrNH_4B_6C_6$ | $1.75\,B_{13}C_2 + 0.5\,H_8C_5 + 0.5\,N_2 + 3.792\,Sr \longrightarrow 2.792\,SrB_6 + SrB_6H_4C_6N$ | -0.5744 |
| $SrNH_4B_6C_6$ | $H_3N + 21\,B + 3\,SrC_2 + 0.5\,SrH_2 \longrightarrow 2.5\,SrB_6 + SrB_6H_4C_6N$ | -0.5255 |
| $SrNH_4B_6C_6$ | $0.5714\,B_5H_7 + 0.5\,N_2 + 15.14\,B + 3\,SrC_2 \longrightarrow 2\,SrB_6 + SrB_6H_4C_6N$ | -0.4757 |
| $SrNH_4B_6C_6$ | $1.2\,H_8C_5 + SrN + 0.6667\,B_9H_{11} + 6.467\,Sr \longrightarrow 6.467\,SrH_2 + SrB_6H_4C_6N$ | -0.4466 |
| $SrNH_4B_6C_6$ | $0.3636\,B_9H_{11} + 0.5\,N_2 + 14.73\,B + 3\,SrC_2 \longrightarrow 2\,SrB_6 + SrB_6H_4C_6N$ | -0.4135 |
| $SrNH_4B_6C_6$ | $1.25\,SrN_2 + 1.5\,B_5H_7 + 3\,SrC_2 \longrightarrow 1.5\,BN + 3.25\,SrH_2 + SrB_6H_4C_6N$ | -0.3853 |
| $SrNH_4B_6C_6$ | $1.333\,H_3N + 18.33\,B + 3\,SrC_2 \longrightarrow 0.3333\,BN + 2\,SrB_6 + SrB_6H_4C_6N$ | -0.1631 |
| $SrNH_4B_6C_6$ | $3\,B_{13}C_2 + 1.333\,H_3N + 6.5\,SrN \longrightarrow 3.417\,N_2 + 5.5\,SrB_6 + SrB_6H_4C_6N$ | -0.0961 |
| $SrNH_4B_6C_6$ | $1.2\,H_8C_5 + 3.8\,SrN + 8.8\,B \longrightarrow 2.8\,BN + 2.8\,SrH_2 + SrB_6H_4C_6N$ | -0.0498 |
| $ThNH_4B_6C_6$ | $1.2\,B_5H_7 + 70\,ThC_2 + 70\,H_3N \longrightarrow 26.8\,H_8C_5 + 69\,ThN + ThB_6H_4C_6N$ | -117.0136 |
| $ThNH_4B_6C_6$ | $2\,ThH_2 + 4\,B_{13}C_2 + 23.5\,N_2 \longrightarrow 46\,BN + ThC_2 + ThB_6H_4C_6N$ | -109.2371 |
| $ThNH_4B_6C_6$ | $64.67\,ThC_2 + 0.6667\,B_9H_{11} + 64.67\,H_3N \longrightarrow 24.67\,H_8C_5 + 63.67\,ThN + ThB_6H_4C_6N$ | -107.4563 |
| $ThNH_4B_6C_6$ | $3\,B_{13}C_2 + 50\,Th + 34\,H_3N \longrightarrow 49\,ThH_2 + 33\,BN + ThB_6H_4C_6N$ | -104.4706 |
| $ThNH_4B_6C_6$ | $2\,ThH_2 + 3.75\,B_{13}C_2 + 21.88\,N_2 \longrightarrow 42.75\,BN + 0.5\,Th_2C_3 + ThB_6H_4C_6N$ | -101.0736 |
| $ThNH_4B_6C_6$ | $0.3636\,B_9H_{11} + 3\,B_{13}C_2 + 21.64\,N_2 + ThB_6 \longrightarrow 42.27\,BN + ThB_6H_4C_6N$ | -98.8171 |
| $ThNH_4B_6C_6$ | $0.5714\,B_5H_7 + 3\,B_{13}C_2 + 21.43\,N_2 + ThB_6 \longrightarrow 41.86\,BN + ThB_6H_4C_6N$ | -97.8127 |
| $ThNH_4B_6C_6$ | $ThB_4 + 0.3636\,B_9H_{11} + 3\,B_{13}C_2 + 20.64\,N_2 \longrightarrow 40.27\,BN + ThB_6H_4C_6N$ | -93.9722 |
| $ThNH_4B_6C_6$ | $ThB_4 + 0.5714\,B_5H_7 + 3\,B_{13}C_2 + 20.43\,N_2 \longrightarrow 39.86\,BN + ThB_6H_4C_6N$ | -92.9678 |
| $ThNH_4B_6C_6$ | $2\,ThH_2 + 3.5\,B_{13}C_2 + 20.25\,N_2 \longrightarrow ThC + 39.5\,BN + ThB_6H_4C_6N$ | -92.9028 |
| $ThNH_4B_6C_6$ | $3\,B_{13}C_2 + 27.13\,Th + 34\,H_3N \longrightarrow 6.533\,Th_4H_{15} + 33\,BN + ThB_6H_4C_6N$ | -91.3079 |
| $ThNH_4B_6C_6$ | $2\,H_2 + 3\,B_{13}C_2 + 20\,N_2 + ThB_6 \longrightarrow 39\,BN + ThB_6H_4C_6N$ | -91.0763 |
| $ThNH_4B_6C_6$ | $3\,B_{13}C_2 + 1.333\,H_3N + 19.33\,N_2 + ThB_6 \longrightarrow 39\,BN + ThB_6H_4C_6N$ | -89.3108 |
| $ThNH_4B_6C_6$ | $0.3636\,B_9H_{11} + 3\,B_{13}C_2 + Th + 18.64\,N_2 \longrightarrow 36.27\,BN + ThB_6H_4C_6N$ | -86.8031 |
| $ThNH_4B_6C_6$ | $ThB_4 + 2\,H_2 + 3\,B_{13}C_2 + 19\,N_2 \longrightarrow 37\,BN + ThB_6H_4C_6N$ | -86.2314 |
| $ThNH_4B_6C_6$ | $0.5714\,B_5H_7 + 3\,B_{13}C_2 + Th + 18.43\,N_2 \longrightarrow 35.86\,BN + ThB_6H_4C_6N$ | -85.7987 |
| $ThNH_4B_6C_6$ | $ThB_4 + 3\,B_{13}C_2 + 1.333\,H_3N + 18.33\,N_2 \longrightarrow 37\,BN + ThB_6H_4C_6N$ | -84.4659 |
| $ThNH_4B_6C_6$ | $0.3636\,B_9H_{11} + 3\,B_{13}C_2 + 18.14\,N_2 + ThN \longrightarrow 36.27\,BN + ThB_6H_4C_6N$ | -83.6556 |
| $ThNH_4B_6C_6$ | $0.3333\,Th_3N_4 + 0.3636\,B_9H_{11} + 3\,B_{13}C_2 + 17.97\,N_2 \longrightarrow 36.27\,BN + ThB_6H_4C_6N$ | -83.0336 |
| $ThNH_4B_6C_6$ | $0.5714\,B_5H_7 + 3\,B_{13}C_2 + 17.93\,N_2 + ThN \longrightarrow 35.86\,BN + ThB_6H_4C_6N$ | -82.6512 |
| $ThNH_4B_6C_6$ | $0.5714\,B_5H_7 + 0.3333\,Th_3N_4 + 3\,B_{13}C_2 + 17.76\,N_2 \longrightarrow 35.86\,BN + ThB_6H_4C_6N$ | -82.0292 |
| $ThNH_4B_6C_6$ | $ThH_2 + 0.1818\,B_9H_{11} + 3\,B_{13}C_2 + 17.82\,N_2 \longrightarrow 34.64\,BN + ThB_6H_4C_6N$ | -81.4954 |
| $ThNH_4B_6C_6$ | $1.5\,ThB_4 + 50.5\,ThC_2 + 52\,H_3N \longrightarrow 19\,H_8C_5 + 51\,ThN + ThB_6H_4C_6N$ | -81.2472 |
| $ThNH_4B_6C_6$ | $ThH_2 + 0.2857\,B_5H_7 + 3\,B_{13}C_2 + 17.71\,N_2 \longrightarrow 34.43\,BN + ThB_6H_4C_6N$ | -80.9932 |
| $ThNH_4B_6C_6$ | $2\,ThH_2 + 3\,B_{13}C_2 + 17.67\,N_2 \longrightarrow 0.3333\,Th_3N_4 + 33\,BN + ThB_6H_4C_6N$ | -79.9572 |
| $ThNH_4B_6C_6$ | $2\,ThH_2 + 3\,B_{13}C_2 + 17.5\,N_2 \longrightarrow 33\,BN + ThN + ThB_6H_4C_6N$ | -79.3353 |
| $ThNH_4B_6C_6$ | $30.2\,B_5H_7 + 36.5\,Th_3N_4 + 1.2\,H_8C_5 \longrightarrow 108.5\,ThH_2 + 145\,BN + ThB_6H_4C_6N$ | -79.1851 |
| $ThNH_4B_6C_6$ | $2\,H_2 + 3\,B_{13}C_2 + Th + 17\,N_2 \longrightarrow 33\,BN + ThB_6H_4C_6N$ | -79.0623 |
| $ThNH_4B_6C_6$ | $0.2667\,Th_4H_{15} + 3.067\,B_{13}C_2 + 17.43\,N_2 \longrightarrow 33.87\,BN + 0.06667\,ThC_2 + ThB_6H_4C_6N$ | -78.9282 |
| $ThNH_4B_6C_6$ | $0.2667\,Th_4H_{15} + 3.05\,B_{13}C_2 + 17.32\,N_2 \longrightarrow 33.65\,BN + 0.03333\,Th_2C_3 + ThB_6H_4C_6N$ | -78.3840 |
| $ThNH_4B_6C_6$ | $0.2667\,Th_4H_{15} + 3.033\,B_{13}C_2 + 17.22\,N_2 \longrightarrow 0.06667\,ThC + 33.43\,BN + ThB_6H_4C_6N$ | -77.8393 |
| $ThNH_4B_6C_6$ | $ThH_2 + H_2 + 3\,B_{13}C_2 + 17\,N_2 \longrightarrow 33\,BN + ThB_6H_4C_6N$ | -77.6250 |
| $ThNH_4B_6C_6$ | $0.25\,Th_4H_{15} + 0.02273\,B_9H_{11} + 3\,B_{13}C_2 + 17.1\,N_2 \longrightarrow 33.2\,BN + ThB_6H_4C_6N$ | -77.3548 |
| $ThNH_4B_6C_6$ | $3\,B_{13}C_2 + Th + 1.333\,H_3N + 16.33\,N_2 \longrightarrow 33\,BN + ThB_6H_4C_6N$ | -77.2968 |
| $ThNH_4B_6C_6$ | $0.03571\,B_5H_7 + 0.25\,Th_4H_{15} + 3\,B_{13}C_2 + 17.09\,N_2 \longrightarrow 33.18\,BN + ThB_6H_4C_6N$ | -77.2921 |
| $ThNH_4B_6C_6$ | $0.2667\,Th_4H_{15} + 3\,B_{13}C_2 + 17.04\,N_2 \longrightarrow 0.02222\,Th_3N_4 + 33\,BN + ThB_6H_4C_6N$ | -76.9763 |
| $ThNH_4B_6C_6$ | $0.2667\,Th_4H_{15} + 3\,B_{13}C_2 + 17.03\,N_2 \longrightarrow 33\,BN + 0.06667\,ThN + ThB_6H_4C_6N$ | -76.9348 |
| $ThNH_4B_6C_6$ | $0.125\,H_2 + 0.25\,Th_4H_{15} + 3\,B_{13}C_2 + 17\,N_2 \longrightarrow 33\,BN + ThB_6H_4C_6N$ | -76.8710 |
| $ThNH_4B_6C_6$ | $0.2857\,Th_4H_{15} + 3\,B_{13}C_2 + 17\,N_2 \longrightarrow 0.1429\,ThH_2 + 33\,BN + ThB_6H_4C_6N$ | -76.7633 |
| $ThNH_4B_6C_6$ | $0.25\,Th_4H_{15} + 3\,B_{13}C_2 + 0.08333\,H_3N + 16.96\,N_2 \longrightarrow 33\,BN + ThB_6H_4C_6N$ | -76.7607 |
| $ThNH_4B_6C_6$ | $ThH_2 + 3\,B_{13}C_2 + 0.6667\,H_3N + 16.67\,N_2 \longrightarrow 33\,BN + ThB_6H_4C_6N$ | -76.7423 |
| $ThNH_4B_6C_6$ | $0.2667\,Th_4H_{15} + 3\,B_{13}C_2 + 17\,N_2 \longrightarrow 33\,BN + 0.06667\,Th + ThB_6H_4C_6N$ | -76.7250 |

| Compound | Chemical reaction | Reaction energy(eV) |
|---|---|---|
| $ThNH_4B_6C_6$ | $0.2667\,Th_4H_{15} + 3\,B_{13}C_2 + 16.87\,N_2 \longrightarrow 0.06667\,ThB_4 + 32.73\,BN + ThB_6H_4C_6N$ | -76.2470 |
| $ThNH_4B_6C_6$ | $2\,ThH_2 + 3\,B_{13}C_2 + 17\,N_2 \longrightarrow 33\,BN + Th + ThB_6H_4C_6N$ | -76.1877 |
| $ThNH_4B_6C_6$ | $0.2667\,Th_4H_{15} + 3\,B_{13}C_2 + 16.8\,N_2 \longrightarrow 32.6\,BN + 0.06667\,ThB_6 + ThB_6H_4C_6N$ | -75.9240 |
| $ThNH_4B_6C_6$ | $2\,H_2 + 3\,B_{13}C_2 + 16.5\,N_2 + ThN \longrightarrow 33\,BN + ThB_6H_4C_6N$ | -75.9147 |
| $ThNH_4B_6C_6$ | $58.14\,ThH_2 + 3\,B_{13}C_2 + 34\,H_3N \longrightarrow 14.29\,Th_4H_{15} + 33\,BN + ThB_6H_4C_6N$ | -75.6892 |
| $ThNH_4B_6C_6$ | $0.3333\,Th_3N_4 + 2\,H_2 + 3\,B_{13}C_2 + 16.33\,N_2 \longrightarrow 33\,BN + ThB_6H_4C_6N$ | -75.2927 |
| $ThNH_4B_6C_6$ | $0.25\,Th_4H_{15} + 2.922\,B_{13}C_2 + 0.03125\,H_8C_5 + 16.49\,N_2 \longrightarrow 31.98\,BN + ThB_6H_4C_6N$ | -74.2909 |
| $ThNH_4B_6C_6$ | $3\,B_{13}C_2 + 1.333\,H_3N + 15.83\,N_2 + ThN \longrightarrow 33\,BN + ThB_6H_4C_6N$ | -74.1493 |
| $ThNH_4B_6C_6$ | $0.3333\,Th_3N_4 + 3\,B_{13}C_2 + 1.333\,H_3N + 15.67\,N_2 \longrightarrow 33\,BN + ThB_6H_4C_6N$ | -73.5273 |
| $ThNH_4B_6C_6$ | $ThC + 0.3636\,B_9H_{11} + 2.5\,B_{13}C_2 + 15.39\,N_2 \longrightarrow 29.77\,BN + ThB_6H_4C_6N$ | -70.0881 |
| $ThNH_4B_6C_6$ | $0.5714\,B_5H_7 + ThC + 2.5\,B_{13}C_2 + 15.18\,N_2 \longrightarrow 29.36\,BN + ThB_6H_4C_6N$ | -69.0836 |
| $ThNH_4B_6C_6$ | $2\,ThH_2 + 3\,B_{13}C_2 + 15\,N_2 \longrightarrow ThB_4 + 29\,BN + ThB_6H_4C_6N$ | -69.0186 |
| $ThNH_4B_6C_6$ | $43\,ThC_2 + 44\,H_3N + ThB_6 \longrightarrow 16\,H_8C_5 + 43\,ThN + ThB_6H_4C_6N$ | -68.0925 |
| $ThNH_4B_6C_6$ | $2\,ThH_2 + 3\,B_{13}C_2 + 14\,N_2 \longrightarrow 27\,BN + ThB_6 + ThB_6H_4C_6N$ | -64.1737 |
| $ThNH_4B_6C_6$ | $ThC + 2\,H_2 + 2.5\,B_{13}C_2 + 13.75\,N_2 \longrightarrow 26.5\,BN + ThB_6H_4C_6N$ | -62.3472 |
| $ThNH_4B_6C_6$ | $0.3636\,B_9H_{11} + 2.25\,B_{13}C_2 + 0.5\,Th_2C_3 + 13.76\,N_2 \longrightarrow 26.52\,BN + ThB_6H_4C_6N$ | -61.9173 |
| $ThNH_4B_6C_6$ | $0.5714\,B_5H_7 + 2.25\,B_{13}C_2 + 0.5\,Th_2C_3 + 13.55\,N_2 \longrightarrow 26.11\,BN + ThB_6H_4C_6N$ | -60.9128 |
| $ThNH_4B_6C_6$ | $ThC + 2.5\,B_{13}C_2 + 1.333\,H_3N + 13.08\,N_2 \longrightarrow 26.5\,BN + ThB_6H_4C_6N$ | -60.5818 |
| $ThNH_4B_6C_6$ | $ThH_2 + 2.375\,B_{13}C_2 + 0.25\,H_8C_5 + 12.94\,N_2 \longrightarrow 24.87\,BN + ThB_6H_4C_6N$ | -56.9835 |
| $ThNH_4B_6C_6$ | $25.1\,ThH_2 + 3\,B_{13}C_2 + 16.57\,N_2 \longrightarrow 6.6\,B_5H_7 + 8.033\,Th_3N_4 + ThB_6H_4C_6N$ | -56.0264 |
| $ThNH_4B_6C_6$ | $2\,H_2 + 2.25\,B_{13}C_2 + 0.5\,Th_2C_3 + 12.12\,N_2 \longrightarrow 23.25\,BN + ThB_6H_4C_6N$ | -54.1764 |
| $ThNH_4B_6C_6$ | $ThC_2 + 0.3636\,B_9H_{11} + 2\,B_{13}C_2 + 12.14\,N_2 \longrightarrow 23.27\,BN + ThB_6H_4C_6N$ | -53.7538 |
| $ThNH_4B_6C_6$ | $0.5714\,B_5H_7 + ThC_2 + 2\,B_{13}C_2 + 11.93\,N_2 \longrightarrow 22.86\,BN + ThB_6H_4C_6N$ | -52.7494 |
| $ThNH_4B_6C_6$ | $2.25\,B_{13}C_2 + 0.5\,Th_2C_3 + 1.333\,H_3N + 11.46\,N_2 \longrightarrow 23.25\,BN + ThB_6H_4C_6N$ | -52.4110 |
| $ThNH_4B_6C_6$ | $1.75\,B_{13}C_2 + 0.5\,H_8C_5 + 11.88\,N_2 + ThB_6 \longrightarrow 22.75\,BN + ThB_6H_4C_6N$ | -49.7934 |
| $ThNH_4B_6C_6$ | $22.17\,ThH_2 + 3\,B_{13}C_2 + 14.61\,N_2 \longrightarrow 7.056\,Th_3N_4 + 3.667\,B_9H_{11} + ThB_6H_4C_6N$ | -48.9372 |
| $ThNH_4B_6C_6$ | $2\,H_2 + ThC_2 + 2\,B_{13}C_2 + 10.5\,N_2 \longrightarrow 20\,BN + ThB_6H_4C_6N$ | -46.0129 |
| $ThNH_4B_6C_6$ | $ThB_4 + 1.75\,B_{13}C_2 + 0.5\,H_8C_5 + 10.87\,N_2 \longrightarrow 20.75\,BN + ThB_6H_4C_6N$ | -44.9485 |
| $ThNH_4B_6C_6$ | $63.87\,H_2 + 3\,B_{13}C_2 + 34\,ThN \longrightarrow 8.25\,Th_4H_{15} + 33\,BN + ThB_6H_4C_6N$ | -44.3560 |
| $ThNH_4B_6C_6$ | $ThC_2 + 2\,B_{13}C_2 + 1.333\,H_3N + 9.833\,N_2 \longrightarrow 20\,BN + ThB_6H_4C_6N$ | -44.2475 |
| $ThNH_4B_6C_6$ | $28\,ThC_2 + 6\,B + 28\,H_3N \longrightarrow 10\,H_8C_5 + 27\,ThN + ThB_6H_4C_6N$ | -42.6489 |
| $ThNH_4B_6C_6$ | $25.1\,ThH_2 + 3\,B_{13}C_2 + 12.55\,N_2 \longrightarrow 6.6\,B_5H_7 + 24.1\,ThN + ThB_6H_4C_6N$ | -41.0366 |
| $ThNH_4B_6C_6$ | $3\,B_{13}C_2 + 40\,H_3N + ThB_6 \longrightarrow 58\,H_2 + 39\,BN + ThB_6H_4C_6N$ | -38.1135 |
| $ThNH_4B_6C_6$ | $1.75\,B_{13}C_2 + Th + 0.5\,H_8C_5 + 8.875\,N_2 \longrightarrow 16.75\,BN + ThB_6H_4C_6N$ | -37.7794 |
| $ThNH_4B_6C_6$ | $3\,B_{13}C_2 + 18.93\,H_3N + 15.07\,ThN \longrightarrow 3.519\,Th_4H_{15} + 33\,BN + ThB_6H_4C_6N$ | -37.3961 |
| $ThNH_4B_6C_6$ | $8.5\,Th_3N_4 + 47.94\,H_2 + 3\,B_{13}C_2 \longrightarrow 6.125\,Th_4H_{15} + 33\,BN + ThB_6H_4C_6N$ | -36.6242 |
| $ThNH_4B_6C_6$ | $ThB_4 + 3\,B_{13}C_2 + 38\,H_3N \longrightarrow 55\,H_2 + 37\,BN + ThB_6H_4C_6N$ | -35.9167 |
| $ThNH_4B_6C_6$ | $22.17\,ThH_2 + 3\,B_{13}C_2 + 11.08\,N_2 \longrightarrow 3.667\,B_9H_{11} + 21.17\,ThN + ThB_6H_4C_6N$ | -35.7719 |
| $ThNH_4B_6C_6$ | $1.75\,B_{13}C_2 + 0.5\,H_8C_5 + 8.375\,N_2 + ThN \longrightarrow 16.75\,BN + ThB_6H_4C_6N$ | -34.6318 |
| $ThNH_4B_6C_6$ | $4.376\,Th_3N_4 + 3\,B_{13}C_2 + 16.49\,H_3N \longrightarrow 3.032\,Th_4H_{15} + 33\,BN + ThB_6H_4C_6N$ | -34.3093 |
| $ThNH_4B_6C_6$ | $3\,B_{13}C_2 + Th + 34\,H_3N \longrightarrow 49\,H_2 + 33\,BN + ThB_6H_4C_6N$ | -34.0439 |
| $ThNH_4B_6C_6$ | $0.3333\,Th_3N_4 + 1.75\,B_{13}C_2 + 0.5\,H_8C_5 + 8.208\,N_2 \longrightarrow 16.75\,BN + ThB_6H_4C_6N$ | -34.0098 |
| $ThNH_4B_6C_6$ | $6\,ThC + 2.068\,B_9H_{11} + 6.807\,N_2 \longrightarrow 1.25\,Th_4H_{15} + 12.61\,BN + ThB_6H_4C_6N$ | -33.7542 |
| $ThNH_4B_6C_6$ | $ThH_2 + 3\,B_{13}C_2 + 34\,H_3N \longrightarrow 50\,H_2 + 33\,BN + ThB_6H_4C_6N$ | -32.6066 |
| $ThNH_4B_6C_6$ | $3\,B_{13}C_2 + 33\,H_3N + ThN \longrightarrow 47.5\,H_2 + 33\,BN + ThB_6H_4C_6N$ | -32.2204 |
| $ThNH_4B_6C_6$ | $0.3333\,Th_3N_4 + 3\,B_{13}C_2 + 32.67\,H_3N \longrightarrow 47\,H_2 + 33\,BN + ThB_6H_4C_6N$ | -32.0398 |
| $ThNH_4B_6C_6$ | $0.25\,Th_4H_{15} + 3\,B_{13}C_2 + 34\,H_3N \longrightarrow 50.87\,H_2 + 33\,BN + ThB_6H_4C_6N$ | -31.8527 |
| $ThNH_4B_6C_6$ | $1.2\,H_8C_5 + 8.9\,N_2 + 3.8\,ThB_6 \longrightarrow 2.8\,ThH_2 + 16.8\,BN + ThB_6H_4C_6N$ | -29.6609 |
| $ThNH_4B_6C_6$ | $20.62\,ThC_2 + 0.4615\,B_{13}C_2 + 20.62\,H_3N \longrightarrow 7.231\,H_8C_5 + 19.62\,ThN + ThB_6H_4C_6N$ | -29.1956 |
| $ThNH_4B_6C_6$ | $3\,B_{13}C_2 + 16.73\,Th + 16.73\,H_3N \longrightarrow 6.6\,B_5H_7 + 15.73\,ThN + ThB_6H_4C_6N$ | -28.6217 |
| $ThNH_4B_6C_6$ | $3.25\,B_5H_7 + 6\,ThC + 5.625\,N_2 \longrightarrow 1.25\,Th_4H_{15} + 10.25\,BN + ThB_6H_4C_6N$ | -28.0414 |
| $ThNH_4B_6C_6$ | $ThC + 2.5\,B_{13}C_2 + 27.5\,H_3N \longrightarrow 39.25\,H_2 + 26.5\,BN + ThB_6H_4C_6N$ | -25.9353 |
| $ThNH_4B_6C_6$ | $3\,B_{13}C_2 + 14\,H_3N + 20\,ThN \longrightarrow 19\,ThH_2 + 33\,BN + ThB_6H_4C_6N$ | -24.8826 |
| $ThNH_4B_6C_6$ | $3\,B_{13}C_2 + 14.78\,Th + 14.78\,H_3N \longrightarrow 3.667\,B_9H_{11} + 13.78\,ThN + ThB_6H_4C_6N$ | -24.8079 |
| $ThNH_4B_6C_6$ | $7.2\,B_5H_7 + 6\,ThC + 8.233\,N_2 \longrightarrow 15.47\,H_3N + 5\,ThB_6 + ThB_6H_4C_6N$ | -24.0599 |
| $ThNH_4B_6C_6$ | $11\,B_5H_7 + 6\,C + 12.5\,Th_3N_4 \longrightarrow 36.5\,ThH_2 + 49\,BN + ThB_6H_4C_6N$ | -23.7652 |
| $ThNH_4B_6C_6$ | $3\,B_{13}C_2 + 12.8\,Th + 16.73\,H_3N \longrightarrow 6.6\,B_5H_7 + 3.933\,Th_3N_4 + ThB_6H_4C_6N$ | -23.5808 |
| $ThNH_4B_6C_6$ | $5.556\,Th_3N_4 + 3\,B_{13}C_2 + 11.78\,H_3N \longrightarrow 15.67\,ThH_2 + 33\,BN + ThB_6H_4C_6N$ | -23.1595 |
| $ThNH_4B_6C_6$ | $2.25\,B_{13}C_2 + 0.5\,Th_2C_3 + 24.25\,H_3N \longrightarrow 34.38\,H_2 + 23.25\,BN + ThB_6H_4C_6N$ | -22.0677 |
| $ThNH_4B_6C_6$ | $0.3636\,B_9H_{11} + 3\,B_{13}C_2 + 10.07\,Th + 0.5\,N_2 \longrightarrow 9.068\,ThB_4 + ThB_6H_4C_6N$ | -21.7924 |

| Compound | Chemical reaction | Reaction energy(eV) |
|---|---|---|
| $ThNH_4B_6C_6$ | $1.2 B_5H_7 + 8.75 ThC + 5.667 N_2 \longrightarrow 2.583 Th_3N_4 + 0.55 H_8C_5 + ThB_6H_4C_6N$ | -21.5769 |
| $ThNH_4B_6C_6$ | $0.5714 B_5H_7 + 3 B_{13}C_2 + 9.964 Th + 0.5 N_2 \longrightarrow 8.964 ThB_4 + ThB_6H_4C_6N$ | -21.5328 |
| $ThNH_4B_6C_6$ | $6 ThC + 4 B_9H_{11} + 7.167 N_2 \longrightarrow 13.33 H_3N + 5 ThB_6 + ThB_6H_4C_6N$ | -21.5057 |
| $ThNH_4B_6C_6$ | $ThC + 1.25 B_{13}C_2 + 0.5 H_8C_5 + 5.625 N_2 \longrightarrow 10.25 BN + ThB_6H_4C_6N$ | -21.0643 |
| $ThNH_4B_6C_6$ | $12.34 Th_3N_4 + 6.04 B_9H_{11} + 1.2 H_8C_5 \longrightarrow 36.02 ThH_2 + 48.36 BN + ThB_6H_4C_6N$ | -20.7820 |
| $ThNH_4B_6C_6$ | $3 B_{13}C_2 + 11.33 Th + 14.78 H_3N \longrightarrow 3.444 Th_3N_4 + 3.667 B_9H_{11} + ThB_6H_4C_6N$ | -20.3935 |
| $ThNH_4B_6C_6$ | $6 BN + 10.33 Th_2C_3 + 14.67 H_3N \longrightarrow 5 H_8C_5 + 19.67 ThN + ThB_6H_4C_6N$ | -20.0655 |
| $ThNH_4B_6C_6$ | $BN + 0.3636 B_9H_{11} + 3 B_{13}C_2 + 10.32 Th \longrightarrow 9.318 ThB_4 + ThB_6H_4C_6N$ | -20.0002 |
| $ThNH_4B_6C_6$ | $2 H_2 + 3 B_{13}C_2 + 9.25 Th + 0.5 N_2 \longrightarrow 8.25 ThB_4 + ThB_6H_4C_6N$ | -19.9172 |
| $ThNH_4B_6C_6$ | $0.5714 B_5H_7 + BN + 3 B_{13}C_2 + 10.21 Th \longrightarrow 9.214 ThB_4 + ThB_6H_4C_6N$ | -19.7406 |
| $ThNH_4B_6C_6$ | $35 H_2 + 3 B_{13}C_2 + 34 ThN \longrightarrow 33 ThH_2 + 33 BN + ThB_6H_4C_6N$ | -19.4758 |
| $ThNH_4B_6C_6$ | $8.083 ThC + 0.6667 B_9H_{11} + 5.222 N_2 \longrightarrow 2.361 Th_3N_4 + 0.4167 H_8C_5 + ThB_6H_4C_6N$ | -19.2436 |
| $ThNH_4B_6C_6$ | $3 B_{13}C_2 + 9.583 Th + 1.333 H_3N \longrightarrow 8.25 ThB_4 + 0.3333 ThN + ThB_6H_4C_6N$ | -19.2009 |
| $ThNH_4B_6C_6$ | $3 B_{13}C_2 + 9.5 Th + 1.333 H_3N \longrightarrow 8.25 ThB_4 + 0.08333 Th_3N_4 + ThB_6H_4C_6N$ | -19.0941 |
| $ThNH_4B_6C_6$ | $0.09091 B_9H_{11} + 3 B_{13}C_2 + 9.455 Th + H_3N \longrightarrow 8.455 ThB_4 + ThB_6H_4C_6N$ | -19.0619 |
| $ThNH_4B_6C_6$ | $0.1429 B_5H_7 + 3 B_{13}C_2 + 9.429 Th + H_3N \longrightarrow 8.429 ThB_4 + ThB_6H_4C_6N$ | -18.9970 |
| $ThNH_4B_6C_6$ | $0.25 Th_3N_4 + 0.3636 B_9H_{11} + 3 B_{13}C_2 + 9.318 Th \longrightarrow 9.068 ThB_4 + ThB_6H_4C_6N$ | -18.9653 |
| $ThNH_4B_6C_6$ | $3 B_{13}C_2 + 9.167 Th + 1.333 H_3N \longrightarrow 8.167 ThB_4 + 0.3333 BN + ThB_6H_4C_6N$ | -18.7492 |
| $ThNH_4B_6C_6$ | $0.5714 B_5H_7 + 0.25 Th_3N_4 + 3 B_{13}C_2 + 9.214 Th \longrightarrow 8.964 ThB_4 + ThB_6H_4C_6N$ | -18.7057 |
| $ThNH_4B_6C_6$ | $2 ThH_2 + 1.5 ThB_4 + 6 ThC + 6.167 N_2 \longrightarrow 2.833 Th_3N_4 + ThB_6H_4C_6N$ | -18.6919 |
| $ThNH_4B_6C_6$ | $0.3636 B_9H_{11} + 3 B_{13}C_2 + 9.068 Th + ThN \longrightarrow 9.068 ThB_4 + ThB_6H_4C_6N$ | -18.6449 |
| $ThNH_4B_6C_6$ | $3.347 Th_4H_{15} + 3 B_{13}C_2 + 8.758 N_2 \longrightarrow 6.6 B_5H_7 + 4.129 Th_3N_4 + ThB_6H_4C_6N$ | -18.6151 |
| $ThNH_4B_6C_6$ | $0.5 H_2 + 3 B_{13}C_2 + 9.25 Th + H_3N \longrightarrow 8.25 ThB_4 + ThB_6H_4C_6N$ | -18.5931 |
| $ThNH_4B_6C_6$ | $5.2 B_5H_7 + 6 ThC + 5.9 N_2 \longrightarrow 5 ThB_4 + 10.8 H_3N + ThB_6H_4C_6N$ | -18.5279 |
| $ThNH_4B_6C_6$ | $0.5714 B_5H_7 + 3 B_{13}C_2 + 8.964 Th + ThN \longrightarrow 8.964 ThB_4 + ThB_6H_4C_6N$ | -18.3853 |
| $ThNH_4B_6C_6$ | $ThC_2 + 2 B_{13}C_2 + 21 H_3N \longrightarrow 29.5 H_2 + 20 BN + ThB_6H_4C_6N$ | -18.2074 |
| $ThNH_4B_6C_6$ | $8.5 Th_3N_4 + 26.5 H_2 + 3 B_{13}C_2 \longrightarrow 24.5 ThH_2 + 33 BN + ThB_6H_4C_6N$ | -18.1525 |
| $ThNH_4B_6C_6$ | $3 B_{13}C_2 + 9.25 Th + 1.333 H_3N \longrightarrow 8.25 ThB_4 + 0.1667 N_2 + ThB_6H_4C_6N$ | -18.1518 |
| $ThNH_4B_6C_6$ | $2 H_2 + BN + 3 B_{13}C_2 + 9.5 Th \longrightarrow 8.5 ThB_4 + ThB_6H_4C_6N$ | -18.1249 |
| $ThNH_4B_6C_6$ | $2 ThH_2 + 6 ThC + 5.833 N_2 + ThB_6 \longrightarrow 2.667 Th_3N_4 + ThB_6H_4C_6N$ | -18.0675 |
| $ThNH_4B_6C_6$ | $0.06667 Th_4H_{15} + 3 B_{13}C_2 + 8.983 Th + H_3N \longrightarrow 8.25 ThB_4 + ThB_6H_4C_6N$ | -18.0088 |
| $ThNH_4B_6C_6$ | $0.5 ThH_2 + 3 B_{13}C_2 + 8.75 Th + H_3N \longrightarrow 8.25 ThB_4 + ThB_6H_4C_6N$ | -17.8745 |
| $ThNH_4B_6C_6$ | $6 BN + 16.75 ThC_2 + 16 H_3N \longrightarrow 5.25 Th_3N_4 + 5.5 H_8C_5 + ThB_6H_4C_6N$ | -17.7826 |
| $ThNH_4B_6C_6$ | $2 ThH_2 + 6 ThC + 6 B + 5.167 N_2 \longrightarrow 2.333 Th_3N_4 + ThB_6H_4C_6N$ | -17.6845 |
| $ThNH_4B_6C_6$ | $0.2667 Th_4H_{15} + 3 B_{13}C_2 + 8.183 Th + 0.5 N_2 \longrightarrow 8.25 ThB_4 + ThB_6H_4C_6N$ | -17.5799 |
| $ThNH_4B_6C_6$ | $0.25 Th_3N_4 + 2 H_2 + 3 B_{13}C_2 + 8.5 Th \longrightarrow 8.25 ThB_4 + ThB_6H_4C_6N$ | -17.0900 |
| $ThNH_4B_6C_6$ | $2 ThH_2 + 3 B_{13}C_2 + 7.25 Th + 0.5 N_2 \longrightarrow 8.25 ThB_4 + ThB_6H_4C_6N$ | -17.0426 |
| $ThNH_4B_6C_6$ | $2 H_2 + 3 B_{13}C_2 + 8.25 Th + ThN \longrightarrow 8.25 ThB_4 + ThB_6H_4C_6N$ | -16.7696 |
| $ThNH_4B_6C_6$ | $1.2 B_5H_7 + 8.75 ThC + 4.375 N_2 \longrightarrow 0.55 H_8C_5 + 7.75 ThN + ThB_6H_4C_6N$ | -16.7566 |
| $ThNH_4B_6C_6$ | $6 ThC + 2.889 B_9H_{11} + 5.13 N_2 \longrightarrow 5 ThB_4 + 9.259 H_3N + ThB_6H_4C_6N$ | -16.6832 |
| $ThNH_4B_6C_6$ | $36.27 Th_3N_4 + 0.3636 B_9H_{11} + 3 B_{13}C_2 \longrightarrow 36.27 BN + 107.8 ThN + ThB_6H_4C_6N$ | -15.9725 |
| $ThNH_4B_6C_6$ | $2.956 Th_4H_{15} + 3 B_{13}C_2 + 7.715 N_2 \longrightarrow 3.607 Th_3N_4 + 3.667 B_9H_{11} + ThB_6H_4C_6N$ | -15.8980 |
| $ThNH_4B_6C_6$ | $0.2667 Th_4H_{15} + BN + 3 B_{13}C_2 + 8.433 Th \longrightarrow 8.5 ThB_4 + ThB_6H_4C_6N$ | -15.7876 |
| $ThNH_4B_6C_6$ | $0.5714 B_5H_7 + 35.86 Th_3N_4 + 3 B_{13}C_2 \longrightarrow 35.86 BN + 106.6 ThN + ThB_6H_4C_6N$ | -15.7435 |
| $ThNH_4B_6C_6$ | $1.5 ThB_4 + 6 ThC + 0.2667 Th_4H_{15} + 5.544 N_2 \longrightarrow 2.522 Th_3N_4 + ThB_6H_4C_6N$ | -15.7110 |
| $ThNH_4B_6C_6$ | $1.386 B_9H_{11} + 2 Th_2C_3 + 3.739 N_2 \longrightarrow 0.75 Th_4H_{15} + 6.477 BN + ThB_6H_4C_6N$ | -15.6046 |
| $ThNH_4B_6C_6$ | $2.687 B_{13}C_2 + 8.234 Th + H_3N + 0.125 H_8C_5 \longrightarrow 7.234 ThB_4 + ThB_6H_4C_6N$ | -15.5535 |
| $ThNH_4B_6C_6$ | $2 ThH_2 + BN + 3 B_{13}C_2 + 7.5 Th \longrightarrow 8.5 ThB_4 + ThB_6H_4C_6N$ | -15.2503 |
| $ThNH_4B_6C_6$ | $2 ThH_2 + 6 ThC + 54 B + ThN \longrightarrow 8 ThB_6 + ThB_6H_4C_6N$ | -15.2427 |
| $ThNH_4B_6C_6$ | $6 ThC + 0.2667 Th_4H_{15} + 5.211 N_2 + ThB_6 \longrightarrow 2.356 Th_3N_4 + ThB_6H_4C_6N$ | -15.0865 |
| $ThNH_4B_6C_6$ | $31.67 Th_3N_4 + 3 B_{13}C_2 + 1.333 H_3N \longrightarrow 33 BN + 94 ThN + ThB_6H_4C_6N$ | -15.0609 |
| $ThNH_4B_6C_6$ | $2 ThH_2 + 6 ThC + 48 B + 0.5 N_2 \longrightarrow 7 ThB_6 + ThB_6H_4C_6N$ | -15.0037 |
| $ThNH_4B_6C_6$ | $4.576 ThB_4 + 6 ThC + 13.3 H_3N \longrightarrow 2.394 Th_4H_{15} + 12.3 BN + ThB_6H_4C_6N$ | -14.9444 |
| $ThNH_4B_6C_6$ | $8.083 ThC + 0.6667 B_9H_{11} + 4.042 N_2 \longrightarrow 0.4167 H_8C_5 + 7.083 ThN + ThB_6H_4C_6N$ | -14.8379 |
| $ThNH_4B_6C_6$ | $0.25 Th_3N_4 + 0.2667 Th_4H_{15} + 3 B_{13}C_2 + 7.433 Th \longrightarrow 8.25 ThB_4 + ThB_6H_4C_6N$ | -14.7527 |
| $ThNH_4B_6C_6$ | $2 ThH_2 + 0.25 Th_3N_4 + 6 ThC + 52.5 B \longrightarrow 7.75 ThB_6 + ThB_6H_4C_6N$ | -14.7165 |
| $ThNH_4B_6C_6$ | $6 ThC + 0.2667 Th_4H_{15} + 6 B + 4.544 N_2 \longrightarrow 2.022 Th_3N_4 + ThB_6H_4C_6N$ | -14.7036 |
| $ThNH_4B_6C_6$ | $0.2667 Th_4H_{15} + 3 B_{13}C_2 + 7.183 Th + ThN \longrightarrow 8.25 ThB_4 + ThB_6H_4C_6N$ | -14.4323 |
| $ThNH_4B_6C_6$ | $33 Th_3N_4 + 2 H_2 + 3 B_{13}C_2 \longrightarrow 33 BN + 98 ThN + ThB_6H_4C_6N$ | -14.3384 |
| $ThNH_4B_6C_6$ | $2 ThH_2 + 5.077 ThC + 0.4615 B_{13}C_2 + 4.551 N_2 \longrightarrow 2.026 Th_3N_4 + ThB_6H_4C_6N$ | -14.2337 |

| Compound | Chemical reaction | Reaction energy(eV) |
|---|---|---|
| $ThNH_4B_6C_6$ | $2\,ThH_2 + 0.25\,Th_3N_4 + 3\,B_{13}C_2 + 6.5\,Th \longrightarrow 8.25\,ThB_4 + ThB_6H_4C_6N$ | -14.2155 |
| $ThNH_4B_6C_6$ | $0.3636\,B_9H_{11} + 3\,B_{13}C_2 + 7.045\,Th + 0.5\,N_2 \longrightarrow 6.045\,ThB_6 + ThB_6H_4C_6N$ | -14.1730 |
| $ThNH_4B_6C_6$ | $1.5\,ThB_4 + 6\,ThC + 2\,H_2 + 4.833\,N_2 \longrightarrow 2.167\,Th_3N_4 + ThB_6H_4C_6N$ | -14.0274 |
| $ThNH_4B_6C_6$ | $2\,ThH_2 + 35\,Th_3N_4 + 3\,B_{13}C_2 \longrightarrow 33\,BN + 106\,ThN + ThB_6H_4C_6N$ | -14.0270 |
| $ThNH_4B_6C_6$ | $0.5714\,B_5H_7 + 3\,B_{13}C_2 + 6.976\,Th + 0.5\,N_2 \longrightarrow 5.976\,ThB_6 + ThB_6H_4C_6N$ | -14.0007 |
| $ThNH_4B_6C_6$ | $2\,ThH_2 + 3\,B_{13}C_2 + 6.25\,Th + ThN \longrightarrow 8.25\,ThB_4 + ThB_6H_4C_6N$ | -13.8951 |
| $ThNH_4B_6C_6$ | $1.2\,B_5H_7 + 6\,ThC + 4.567\,N_2 \longrightarrow 1.667\,Th_3N_4 + 1.467\,H_3N + ThB_6H_4C_6N$ | -13.7080 |
| $ThNH_4B_6C_6$ | $2\,ThH_2 + 1.5\,ThB_4 + 6\,ThC + 4.75\,N_2 \longrightarrow 8.5\,ThN + ThB_6H_4C_6N$ | -13.4051 |
| $ThNH_4B_6C_6$ | $6\,ThC + 2\,H_2 + 4.5\,N_2 + ThB_6 \longrightarrow 2\,Th_3N_4 + ThB_6H_4C_6N$ | -13.4030 |
| $ThNH_4B_6C_6$ | $34.07\,Th_3N_4 + 0.2667\,Th_4H_{15} + 3\,B_{13}C_2 \longrightarrow 33\,BN + 102.3\,ThN + ThB_6H_4C_6N$ | -13.3681 |
| $ThNH_4B_6C_6$ | $2\,ThH_2 + 6\,ThC + 6\,B + 4\,N_2 \longrightarrow 7\,ThN + ThB_6H_4C_6N$ | -13.3307 |
| $ThNH_4B_6C_6$ | $6\,ThC + 0.6667\,B_9H_{11} + 4.389\,N_2 \longrightarrow 1.667\,Th_3N_4 + 1.111\,H_3N + ThB_6H_4C_6N$ | -13.2823 |
| $ThNH_4B_6C_6$ | $1.2\,H_8C_5 + 4.98\,N_2 + 2.493\,ThB_6 \longrightarrow 0.3733\,Th_4H_{15} + 8.96\,BN + ThB_6H_4C_6N$ | -13.2104 |
| $ThNH_4B_6C_6$ | $2.313\,Th_4H_{15} + 3\,B_{13}C_2 + 5.615\,N_2 \longrightarrow 8.25\,ThB_4 + 10.23\,H_3N + ThB_6H_4C_6N$ | -13.1925 |
| $ThNH_4B_6C_6$ | $4.8\,B_5H_7 + 2\,Th_2C_3 + 5.433\,N_2 \longrightarrow 9.867\,H_3N + 3\,ThB_6 + ThB_6H_4C_6N$ | -13.1271 |
| $ThNH_4B_6C_6$ | $2\,ThH_2 + 6\,ThC + 4.5\,N_2 + ThB_6 \longrightarrow 8\,ThN + ThB_6H_4C_6N$ | -13.0917 |
| $ThNH_4B_6C_6$ | $6\,ThC + 1.273\,B_9H_{11} + 3.227\,N_2 \longrightarrow 5\,ThH_2 + 5.455\,BN + ThB_6H_4C_6N$ | -13.0513 |
| $ThNH_4B_6C_6$ | $9.25\,ThH_2 + 3\,B_{13}C_2 + 2.917\,N_2 \longrightarrow 8.25\,ThB_4 + 4.833\,H_3N + ThB_6H_4C_6N$ | -13.0220 |
| $ThNH_4B_6C_6$ | $6\,ThC + 2\,H_2 + 6\,B + 3.833\,N_2 \longrightarrow 1.667\,Th_3N_4 + ThB_6H_4C_6N$ | -13.0200 |
| $ThNH_4B_6C_6$ | $2\,H_2 + 3\,B_{13}C_2 + 6.5\,Th + 0.5\,N_2 \longrightarrow 5.5\,ThB_6 + ThB_6H_4C_6N$ | -12.9853 |
| $ThNH_4B_6C_6$ | $2\,ThH_2 + 6\,ThC + 34\,B + 0.5\,N_2 \longrightarrow 7\,ThB_4 + ThB_6H_4C_6N$ | -12.9834 |
| $ThNH_4B_6C_6$ | $0.7857\,ThB_4 + 0.5714\,B_5H_7 + 6\,ThC + 4.357\,N_2 \longrightarrow 1.929\,Th_3N_4 + ThB_6H_4C_6N$ | -12.9506 |
| $ThNH_4B_6C_6$ | $6\,ThC + 10.89\,H_3N + 2.649\,ThB_6 \longrightarrow 1.912\,Th_4H_{15} + 9.895\,BN + ThB_6H_4C_6N$ | -12.9339 |
| $ThNH_4B_6C_6$ | $2\,ThH_2 + 6\,ThC + 38\,B + ThN \longrightarrow 8\,ThB_4 + ThB_6H_4C_6N$ | -12.9338 |
| $ThNH_4B_6C_6$ | $B_{13}C_2 + 0.5\,Th_2C_3 + 0.5\,H_8C_5 + 4\,N_2 \longrightarrow 7\,BN + ThB_6H_4C_6N$ | -12.8935 |
| $ThNH_4B_6C_6$ | $17.54\,ThH_2 + 3\,B_{13}C_2 + 0.5\,N_2 \longrightarrow 8.25\,ThB_4 + 2.071\,Th_4H_{15} + ThB_6H_4C_6N$ | -12.8693 |
| $ThNH_4B_6C_6$ | $19.25\,ThH_2 + 3\,B_{13}C_2 + H_3N \longrightarrow 8.25\,ThB_4 + 2.5\,Th_4H_{15} + ThB_6H_4C_6N$ | -12.8377 |
| $ThNH_4B_6C_6$ | $0.6818\,ThB_4 + 6\,ThC + 0.3636\,B_9H_{11} + 4.288\,N_2 \longrightarrow 1.894\,Th_3N_4 + ThB_6H_4C_6N$ | -12.8185 |
| $ThNH_4B_6C_6$ | $0.5714\,B_5H_7 + 6\,ThC + 4.183\,N_2 + 0.5238\,ThB_6 \longrightarrow 1.841\,Th_3N_4 + ThB_6H_4C_6N$ | -12.6235 |
| $ThNH_4B_6C_6$ | $6\,ThC + 0.2667\,Th_4H_{15} + 48.4\,B + ThN \longrightarrow 7.067\,ThB_6 + ThB_6H_4C_6N$ | -12.6192 |
| $ThNH_4B_6C_6$ | $6\,ThC + 0.3636\,B_9H_{11} + 4.136\,N_2 + 0.4545\,ThB_6 \longrightarrow 1.818\,Th_3N_4 + ThB_6H_4C_6N$ | -12.5347 |
| $ThNH_4B_6C_6$ | $2\,ThH_2 + 0.25\,Th_3N_4 + 6\,ThC + 37\,B \longrightarrow 7.75\,ThB_4 + ThB_6H_4C_6N$ | -12.4797 |
| $ThNH_4B_6C_6$ | $2\,ThH_2 + 6\,ThC + BN + 47\,B \longrightarrow 7\,ThB_6 + ThB_6H_4C_6N$ | -12.4370 |
| $ThNH_4B_6C_6$ | $0.5714\,B_5H_7 + 6\,ThC + 3.143\,B + 3.833\,N_2 \longrightarrow 1.667\,Th_3N_4 + ThB_6H_4C_6N$ | -12.4229 |
| $ThNH_4B_6C_6$ | $6\,ThC + 0.2667\,Th_4H_{15} + 42.4\,B + 0.5\,N_2 \longrightarrow 6.067\,ThB_6 + ThB_6H_4C_6N$ | -12.3802 |
| $ThNH_4B_6C_6$ | $17.33\,Th_3N_4 + 19\,B_9H_{11} + 3\,B_{13}C_2 \longrightarrow 51\,ThB_4 + 68.33\,H_3N + ThB_6H_4C_6N$ | -12.3613 |
| $ThNH_4B_6C_6$ | $6\,ThC + 0.3636\,B_9H_{11} + 2.727\,B + 3.833\,N_2 \longrightarrow 1.667\,Th_3N_4 + ThB_6H_4C_6N$ | -12.3606 |
| $ThNH_4B_6C_6$ | $3\,B_{13}C_2 + 6.833\,Th + 1.333\,H_3N \longrightarrow 5.5\,ThB_6 + 0.3333\,ThN + ThB_6H_4C_6N$ | -12.2690 |
| $ThNH_4B_6C_6$ | $1.5\,ThB_4 + 6\,ThC + 1.333\,H_3N + 4.167\,N_2 \longrightarrow 2.167\,Th_3N_4 + ThB_6H_4C_6N$ | -12.2620 |
| $ThNH_4B_6C_6$ | $BN + 0.3636\,B_9H_{11} + 3\,B_{13}C_2 + 7.212\,Th \longrightarrow 6.212\,ThB_6 + ThB_6H_4C_6N$ | -12.1707 |
| $ThNH_4B_6C_6$ | $3\,B_{13}C_2 + 6.75\,Th + 1.333\,H_3N \longrightarrow 0.08333\,Th_3N_4 + 5.5\,ThB_6 + ThB_6H_4C_6N$ | -12.1622 |
| $ThNH_4B_6C_6$ | $0.25\,Th_3N_4 + 6\,ThC + 0.2667\,Th_4H_{15} + 46.9\,B \longrightarrow 6.817\,ThB_6 + ThB_6H_4C_6N$ | -12.0930 |
| $ThNH_4B_6C_6$ | $0.5714\,B_5H_7 + BN + 3\,B_{13}C_2 + 7.143\,Th \longrightarrow 6.143\,ThB_6 + ThB_6H_4C_6N$ | -11.9984 |
| $ThNH_4B_6C_6$ | $6\,ThC + 12.58\,B + 7.583\,H_3N \longrightarrow 1.25\,Th_4H_{15} + 6.583\,BN + ThB_6H_4C_6N$ | -11.9855 |
| $ThNH_4B_6C_6$ | $0.09091\,B_9H_{11} + 3\,B_{13}C_2 + 6.636\,Th + H_3N \longrightarrow 5.636\,ThB_6 + ThB_6H_4C_6N$ | -11.9581 |
| $ThNH_4B_6C_6$ | $6\,ThC + 2.75\,B_9H_{11} + 3.833\,N_2 \longrightarrow 3.75\,B_5H_7 + 1.667\,Th_3N_4 + ThB_6H_4C_6N$ | -11.9520 |
| $ThNH_4B_6C_6$ | $0.1429\,B_5H_7 + 3\,B_{13}C_2 + 6.619\,Th + H_3N \longrightarrow 5.619\,ThB_6 + ThB_6H_4C_6N$ | -11.9151 |
| $ThNH_4B_6C_6$ | $2\,ThH_2 + 1.5\,ThB_4 + 2\,Th_2C_3 + 4.833\,N_2 \longrightarrow 2.167\,Th_3N_4 + ThB_6H_4C_6N$ | -11.8999 |
| $ThNH_4B_6C_6$ | $3\,B_{13}C_2 + 6.444\,Th + 1.333\,H_3N \longrightarrow 0.3333\,BN + 5.444\,ThB_6 + ThB_6H_4C_6N$ | -11.8873 |
| $ThNH_4B_6C_6$ | $6\,ThC + 0.6667\,B_9H_{11} + 3.833\,N_2 \longrightarrow 1.667\,Th_3N_4 + 1.667\,H_2 + ThB_6H_4C_6N$ | -11.8111 |
| $ThNH_4B_6C_6$ | $2.179\,B_5H_7 + 2\,Th_2C_3 + 2.946\,N_2 \longrightarrow 0.75\,Th_4H_{15} + 4.893\,BN + ThB_6H_4C_6N$ | -11.7752 |
| $ThNH_4B_6C_6$ | $1.2\,B_5H_7 + 6\,ThC + 3.833\,N_2 \longrightarrow 1.667\,Th_3N_4 + 2.2\,H_2 + ThB_6H_4C_6N$ | -11.7660 |
| $ThNH_4B_6C_6$ | $1.2\,B_5H_7 + 2.917\,Th_2C_3 + 3.722\,N_2 \longrightarrow 1.611\,Th_3N_4 + 0.55\,H_8C_5 + ThB_6H_4C_6N$ | -11.6719 |
| $ThNH_4B_6C_6$ | $0.5\,H_2 + 3\,B_{13}C_2 + 6.5\,Th + H_3N \longrightarrow 5.5\,ThB_6 + ThB_6H_4C_6N$ | -11.6612 |
| $ThNH_4B_6C_6$ | $6\,ThC + 1.333\,H_3N + 3.833\,N_2 + ThB_6 \longrightarrow 2\,Th_3N_4 + ThB_6H_4C_6N$ | -11.6376 |
| $ThNH_4B_6C_6$ | $2.667\,B_9H_{11} + 2\,Th_2C_3 + 4.722\,N_2 \longrightarrow 8.444\,H_3N + 3\,ThB_6 + ThB_6H_4C_6N$ | -11.4243 |
| $ThNH_4B_6C_6$ | $0.25\,Th_3N_4 + 0.3636\,B_9H_{11} + 3\,B_{13}C_2 + 6.295\,Th \longrightarrow 6.045\,ThB_6 + ThB_6H_4C_6N$ | -11.3459 |
| $ThNH_4B_6C_6$ | $6\,ThC + 2\,H_2 + 42\,B + ThN \longrightarrow 6\,ThB_6 + ThB_6H_4C_6N$ | -11.3442 |
| $ThNH_4B_6C_6$ | $2\,ThH_2 + 2\,Th_2C_3 + 4.5\,N_2 + ThB_6 \longrightarrow 2\,Th_3N_4 + ThB_6H_4C_6N$ | -11.2754 |
| $ThNH_4B_6C_6$ | $6\,ThC + 6\,B + 1.333\,H_3N + 3.167\,N_2 \longrightarrow 1.667\,Th_3N_4 + ThB_6H_4C_6N$ | -11.2546 |

| Compound | Chemical reaction | Reaction energy(eV) |
|---|---|---|
| ThNH$_4$B$_6$C$_6$ | 5.077 ThC + 0.2667 Th$_4$H$_{15}$ + 0.4615 B$_{13}$C$_2$ + 3.929 N$_2$ $\longrightarrow$ 1.715 Th$_3$N$_4$ + ThB$_6$H$_4$C$_6$N | -11.2527 |
| ThNH$_4$B$_6$C$_6$ | 3.8 ThB$_4$ + 1.2 H$_8$C$_5$ + 5.1 N$_2$ $\longrightarrow$ 2.8 ThH$_2$ + 9.2 BN + ThB$_6$H$_4$C$_6$N | -11.2503 |
| ThNH$_4$B$_6$C$_6$ | 3 B$_{13}$C$_2$ + 6.5 Th + 1.333 H$_3$N $\longrightarrow$ 0.1667 N$_2$ + 5.5 ThB$_6$ + ThB$_6$H$_4$C$_6$N | -11.2198 |
| ThNH$_4$B$_6$C$_6$ | 0.5714 B$_5$H$_7$ + 0.25 Th$_3$N$_4$ + 3 B$_{13}$C$_2$ + 6.226 Th $\longrightarrow$ 5.976 ThB$_6$ + ThB$_6$H$_4$C$_6$N | -11.1736 |
| ThNH$_4$B$_6$C$_6$ | 6 ThC + 2 H$_2$ + 36 B + 0.5 N$_2$ $\longrightarrow$ 5 ThB$_6$ + ThB$_6$H$_4$C$_6$N | -11.1052 |
| ThNH$_4$B$_6$C$_6$ | 0.06667 Th$_4$H$_{15}$ + 3 B$_{13}$C$_2$ + 6.233 Th + H$_3$N $\longrightarrow$ 5.5 ThB$_6$ + ThB$_6$H$_4$C$_6$N | -11.0769 |
| ThNH$_4$B$_6$C$_6$ | 0.3636 B$_9$H$_{11}$ + 3 B$_{13}$C$_2$ + 6.045 Th + ThN $\longrightarrow$ 6.045 ThB$_6$ + ThB$_6$H$_4$C$_6$N | -11.0255 |
| ThNH$_4$B$_6$C$_6$ | 3 B$_{13}$C$_2$ + Th + 11.73 H$_3$N $\longrightarrow$ 4.455 B$_5$H$_7$ + 10.73 BN + ThB$_6$H$_4$C$_6$N | -11.0209 |
| ThNH$_4$B$_6$C$_6$ | 1.5 ThB$_4$ + 6 ThC + 0.2667 Th$_4$H$_{15}$ + 4.283 N$_2$ $\longrightarrow$ 7.567 ThN + ThB$_6$H$_4$C$_6$N | -11.0046 |
| ThNH$_4$B$_6$C$_6$ | 2 H$_2$ + BN + 3 B$_{13}$C$_2$ + 6.667 Th $\longrightarrow$ 5.667 ThB$_6$ + ThB$_6$H$_4$C$_6$N | -10.9829 |
| ThNH$_4$B$_6$C$_6$ | 0.5 ThH$_2$ + 3 B$_{13}$C$_2$ + 6 Th + H$_3$N $\longrightarrow$ 5.5 ThB$_6$ + ThB$_6$H$_4$C$_6$N | -10.9426 |
| ThNH$_4$B$_6$C$_6$ | 18.07 ThH$_2$ + BN + 3 B$_{13}$C$_2$ $\longrightarrow$ 8.5 ThB$_4$ + 2.143 Th$_4$H$_{15}$ + ThB$_6$H$_4$C$_6$N | -10.9331 |
| ThNH$_4$B$_6$C$_6$ | 6 ThC + 0.2667 Th$_4$H$_{15}$ + 6 B + 3.533 N$_2$ $\longrightarrow$ 6.067 ThN + ThB$_6$H$_4$C$_6$N | -10.9302 |
| ThNH$_4$B$_6$C$_6$ | 3.347 Th$_4$H$_{15}$ + 3 B$_{13}$C$_2$ + 6.693 N$_2$ $\longrightarrow$ 6.6 B$_5$H$_7$ + 12.39 ThN + ThB$_6$H$_4$C$_6$N | -10.9108 |
| ThNH$_4$B$_6$C$_6$ | 2 ThH$_2$ + 2 Th$_2$C$_3$ + 6 B + 3.833 N$_2$ $\longrightarrow$ 1.667 Th$_3$N$_4$ + ThB$_6$H$_4$C$_6$N | -10.8925 |
| ThNH$_4$B$_6$C$_6$ | 3 B$_{13}$C$_2$ + 13.64 H$_3$N + ThB$_6$ $\longrightarrow$ 5.273 B$_5$H$_7$ + 12.64 BN + ThB$_6$H$_4$C$_6$N | -10.8618 |
| ThNH$_4$B$_6$C$_6$ | 0.5714 B$_5$H$_7$ + 3 B$_{13}$C$_2$ + 5.976 Th + ThN $\longrightarrow$ 5.976 ThB$_6$ + ThB$_6$H$_4$C$_6$N | -10.8532 |
| ThNH$_4$B$_6$C$_6$ | 0.25 Th$_3$N$_4$ + 6 ThC + 2 H$_2$ + 40.5 B $\longrightarrow$ 5.75 ThB$_6$ + ThB$_6$H$_4$C$_6$N | -10.8179 |
| ThNH$_4$B$_6$C$_6$ | 5.58 ThC + 0.3636 B$_9$H$_{11}$ + 0.2098 B$_{13}$C$_2$ + 3.554 N$_2$ $\longrightarrow$ 1.527 Th$_3$N$_4$ + ThB$_6$H$_4$C$_6$N | -10.7920 |
| ThNH$_4$B$_6$C$_6$ | 0.5 ThH$_2$ + 6 ThC + 39 B + H$_3$N $\longrightarrow$ 5.5 ThB$_6$ + ThB$_6$H$_4$C$_6$N | -10.7557 |
| ThNH$_4$B$_6$C$_6$ | 0.5714 B$_5$H$_7$ + 6 ThC + 39.14 B + ThN $\longrightarrow$ 6 ThB$_6$ + ThB$_6$H$_4$C$_6$N | -10.7470 |
| ThNH$_4$B$_6$C$_6$ | 6 ThC + 0.2667 Th$_4$H$_{15}$ + 4.033 N$_2$ + ThB$_6$ $\longrightarrow$ 7.067 ThN + ThB$_6$H$_4$C$_6$N | -10.6912 |
| ThNH$_4$B$_6$C$_6$ | 6 ThC + 0.3636 B$_9$H$_{11}$ + 38.73 B + ThN $\longrightarrow$ 6 ThB$_6$ + ThB$_6$H$_4$C$_6$N | -10.6848 |
| ThNH$_4$B$_6$C$_6$ | 0.2667 Th$_4$H$_{15}$ + 3 B$_{13}$C$_2$ + 5.433 Th + 0.5 N$_2$ $\longrightarrow$ 5.5 ThB$_6$ + ThB$_6$H$_4$C$_6$N | -10.6480 |
| ThNH$_4$B$_6$C$_6$ | 6 ThC + 0.2667 Th$_4$H$_{15}$ + 30.27 B + 0.5 N$_2$ $\longrightarrow$ 6.067 ThB$_4$ + ThB$_6$H$_4$C$_6$N | -10.6292 |
| ThNH$_4$B$_6$C$_6$ | 0.5714 B$_5$H$_7$ + 5.516 ThC + 0.2418 B$_{13}$C$_2$ + 3.511 N$_2$ $\longrightarrow$ 1.505 Th$_3$N$_4$ + ThB$_6$H$_4$C$_6$N | -10.6153 |
| ThNH$_4$B$_6$C$_6$ | 1.2 B$_5$H$_7$ + 6 ThC + 3.733 N$_2$ $\longrightarrow$ 1.467 H$_3$N + 5 ThN + ThB$_6$H$_4$C$_6$N | -10.5981 |
| ThNH$_4$B$_6$C$_6$ | 6 ThC + 0.2667 Th$_4$H$_{15}$ + 34.27 B + ThN $\longrightarrow$ 7.067 ThB$_4$ + ThB$_6$H$_4$C$_6$N | -10.5796 |
| ThNH$_4$B$_6$C$_6$ | 0.5714 B$_5$H$_7$ + 6 ThC + 33.14 B + 0.5 N$_2$ $\longrightarrow$ 5 ThB$_6$ + ThB$_6$H$_4$C$_6$N | -10.5080 |
| ThNH$_4$B$_6$C$_6$ | 15.93 ThH$_2$ + 0.25 Th$_3$N$_4$ + 3 B$_{13}$C$_2$ $\longrightarrow$ 8.25 ThB$_4$ + 1.857 Th$_4$H$_{15}$ + ThB$_6$H$_4$C$_6$N | -10.4739 |
| ThNH$_4$B$_6$C$_6$ | 2 ThH$_2$ + 5.077 ThC + 0.4615 B$_{13}$C$_2$ + 3.538 N$_2$ $\longrightarrow$ 6.077 ThN + ThB$_6$H$_4$C$_6$N | -10.4540 |
| ThNH$_4$B$_6$C$_6$ | 6 ThC + 0.3636 B$_9$H$_{11}$ + 32.73 B + 0.5 N$_2$ $\longrightarrow$ 5 ThB$_6$ + ThB$_6$H$_4$C$_6$N | -10.4457 |
| ThNH$_4$B$_6$C$_6$ | 2 ThH$_2$ + 6 ThC + BN + 33 B $\longrightarrow$ 7 ThB$_4$ + ThB$_6$H$_4$C$_6$N | -10.4166 |
| ThNH$_4$B$_6$C$_6$ | 6 ThC + 0.6667 B$_9$H$_{11}$ + 3.241 N$_2$ $\longrightarrow$ 1.37 Th$_3$N$_4$ + 0.2222 Th$_4$H$_{15}$ + ThB$_6$H$_4$C$_6$N | -10.4081 |
| ThNH$_4$B$_6$C$_6$ | 15.39 ThH$_2$ + 3 B$_{13}$C$_2$ + ThN $\longrightarrow$ 8.25 ThB$_4$ + 1.786 Th$_4$H$_{15}$ + ThB$_6$H$_4$C$_6$N | -10.2974 |
| ThNH$_4$B$_6$C$_6$ | 0.5714 B$_5$H$_7$ + 0.25 Th$_3$N$_4$ + 6 ThC + 37.64 B $\longrightarrow$ 5.75 ThB$_6$ + ThB$_6$H$_4$C$_6$N | -10.2208 |
| ThNH$_4$B$_6$C$_6$ | 6 ThC + 36.33 B + 1.333 H$_3$N $\longrightarrow$ 0.3333 BN + 5 ThB$_6$ + ThB$_6$H$_4$C$_6$N | -10.1953 |
| ThNH$_4$B$_6$C$_6$ | 6 ThC + 0.6667 B$_9$H$_{11}$ + 3.556 N$_2$ $\longrightarrow$ 1.111 H$_3$N + 5 ThN + ThB$_6$H$_4$C$_6$N | -10.1724 |
| ThNH$_4$B$_6$C$_6$ | 0.25 Th$_3$N$_4$ + 6 ThC + 0.3636 B$_9$H$_{11}$ + 37.23 B $\longrightarrow$ 5.75 ThB$_6$ + ThB$_6$H$_4$C$_6$N | -10.1585 |
| ThNH$_4$B$_6$C$_6$ | 0.25 Th$_3$N$_4$ + 2 H$_2$ + 3 B$_{13}$C$_2$ + 5.75 Th $\longrightarrow$ 5.5 ThB$_6$ + ThB$_6$H$_4$C$_6$N | -10.1581 |
| ThNH$_4$B$_6$C$_6$ | 0.25 Th$_3$N$_4$ + 6 ThC + 0.2667 Th$_4$H$_{15}$ + 33.27 B $\longrightarrow$ 6.817 ThB$_4$ + ThB$_6$H$_4$C$_6$N | -10.1255 |
| ThNH$_4$B$_6$C$_6$ | 2 ThH$_2$ + 3 B$_{13}$C$_2$ + 4.5 Th + 0.5 N$_2$ $\longrightarrow$ 5.5 ThB$_6$ + ThB$_6$H$_4$C$_6$N | -10.1107 |
| ThNH$_4$B$_6$C$_6$ | 6 ThC + 0.06667 Th$_4$H$_{15}$ + 37.6 B + H$_3$N $\longrightarrow$ 5.267 ThB$_6$ + ThB$_6$H$_4$C$_6$N | -10.0999 |
| ThNH$_4$B$_6$C$_6$ | 0.6667 B$_9$H$_{11}$ + 2.694 Th$_2$C$_3$ + 3.426 N$_2$ $\longrightarrow$ 1.463 Th$_3$N$_4$ + 0.4167 H$_8$C$_5$ + ThB$_6$H$_4$C$_6$N | -10.0932 |
| ThNH$_4$B$_6$C$_6$ | ThB$_4$ + 3 B$_{13}$C$_2$ + 13 H$_3$N $\longrightarrow$ 5 B$_5$H$_7$ + 12 BN + ThB$_6$H$_4$C$_6$N | -10.0746 |
| ThNH$_4$B$_6$C$_6$ | 1.5 ThB$_4$ + 6 ThC + 2 H$_2$ + 3.75 N$_2$ $\longrightarrow$ 6.5 ThN + ThB$_6$H$_4$C$_6$N | -9.9845 |
| ThNH$_4$B$_6$C$_6$ | 0.3333 Th$_3$N$_4$ + 3 B$_{13}$C$_2$ + 11.3 H$_3$N $\longrightarrow$ 4.273 B$_5$H$_7$ + 11.64 BN + ThB$_6$H$_4$C$_6$N | -9.9565 |
| ThNH$_4$B$_6$C$_6$ | 1.2 B$_5$H$_7$ + 6 ThC + 3.051 N$_2$ $\longrightarrow$ 1.276 Th$_3$N$_4$ + 0.2933 Th$_4$H$_{15}$ + ThB$_6$H$_4$C$_6$N | -9.9141 |
| ThNH$_4$B$_6$C$_6$ | 6 ThC + 2 H$_2$ + 6 B + 3 N$_2$ $\longrightarrow$ 5 ThN + ThB$_6$H$_4$C$_6$N | -9.9101 |
| ThNH$_4$B$_6$C$_6$ | 3 B$_{13}$C$_2$ + 11.41 H$_3$N + ThN $\longrightarrow$ 4.318 B$_5$H$_7$ + 11.41 BN + ThB$_6$H$_4$C$_6$N | -9.9022 |
| ThNH$_4$B$_6$C$_6$ | 3 B$_{13}$C$_2$ + Th + 10.79 H$_3$N $\longrightarrow$ 9.789 BN + 2.579 B$_9$H$_{11}$ + ThB$_6$H$_4$C$_6$N | -9.8771 |
| ThNH$_4$B$_6$C$_6$ | 2 H$_2$ + 3 B$_{13}$C$_2$ + 5.5 Th + ThN $\longrightarrow$ 5.5 ThB$_6$ + ThB$_6$H$_4$C$_6$N | -9.8377 |
| ThNH$_4$B$_6$C$_6$ | 6 ThC + 0.2667 Th$_4$H$_{15}$ + BN + 41.4 B $\longrightarrow$ 6.067 ThB$_6$ + ThB$_6$H$_4$C$_6$N | -9.8134 |
| ThNH$_4$B$_6$C$_6$ | 3.6 B$_5$H$_7$ + 2 Th$_2$C$_3$ + 4.033 N$_2$ $\longrightarrow$ 3 ThB$_4$ + 7.067 H$_3$N + ThB$_6$H$_4$C$_6$N | -9.8078 |
| ThNH$_4$B$_6$C$_6$ | 6 ThC + 0.5 H$_2$ + 36 B + H$_3$N $\longrightarrow$ 5 ThB$_6$ + ThB$_6$H$_4$C$_6$N | -9.7811 |
| ThNH$_4$B$_6$C$_6$ | 6 ThC + 2 H$_2$ + 3.5 N$_2$ + ThB$_6$ $\longrightarrow$ 6 ThN + ThB$_6$H$_4$C$_6$N | -9.6711 |
| ThNH$_4$B$_6$C$_6$ | 6 ThC + 2 H$_2$ + 26 B + 0.5 N$_2$ $\longrightarrow$ 5 ThB$_4$ + ThB$_6$H$_4$C$_6$N | -9.6621 |
| ThNH$_4$B$_6$C$_6$ | 0.1429 B$_5$H$_7$ + 6 ThC + 35.29 B + H$_3$N $\longrightarrow$ 5 ThB$_6$ + ThB$_6$H$_4$C$_6$N | -9.6318 |
| ThNH$_4$B$_6$C$_6$ | 6 ThC + 0.09091 B$_9$H$_{11}$ + 35.18 B + H$_3$N $\longrightarrow$ 5 ThB$_6$ + ThB$_6$H$_4$C$_6$N | -9.6162 |
| ThNH$_4$B$_6$C$_6$ | 6 ThC + 2 H$_2$ + 30 B + ThN $\longrightarrow$ 6 ThB$_4$ + ThB$_6$H$_4$C$_6$N | -9.6125 |

| Compound | Chemical reaction | Reaction energy(eV) |
|---|---|---|
| ThNH$_4$B$_6$C$_6$ | 5.077 ThC + 2 H$_2$ + 0.4615 B$_{13}$C$_2$ + 3.218 N$_2$ $\longrightarrow$ 1.359 Th$_3$N$_4$ + ThB$_6$H$_4$C$_6$N | -9.5692 |
| ThNH$_4$B$_6$C$_6$ | 2 B$_5$H$_7$ + 6 ThC + 2.5 N$_2$ $\longrightarrow$ 5 ThH$_2$ + 4 BN + ThB$_6$H$_4$C$_6$N | -9.5358 |
| ThNH$_4$B$_6$C$_6$ | 3 B$_{13}$C$_2$ + 12.53 H$_3$N + ThB$_6$ $\longrightarrow$ 11.53 BN + 3.053 B$_9$H$_{11}$ + ThB$_6$H$_4$C$_6$N | -9.5079 |
| ThNH$_4$B$_6$C$_6$ | 2.687 B$_{13}$C$_2$ + 5.823 Th + H$_3$N + 0.125 H$_8$C$_5$ $\longrightarrow$ 4.823 ThB$_6$ + ThB$_6$H$_4$C$_6$N | -9.4749 |
| ThNH$_4$B$_6$C$_6$ | 6 ThC + 34.5 B + 1.333 H$_3$N $\longrightarrow$ 0.08333 Th$_3$N$_4$ + 4.75 ThB$_6$ + ThB$_6$H$_4$C$_6$N | -9.4355 |
| ThNH$_4$B$_6$C$_6$ | 0.7857 ThB$_4$ + 0.5714 B$_5$H$_7$ + 6 ThC + 3.393 N$_2$ $\longrightarrow$ 5.786 ThN + ThB$_6$H$_4$C$_6$N | -9.3519 |
| ThNH$_4$B$_6$C$_6$ | 6 ThC + 36 B + 1.333 H$_3$N $\longrightarrow$ 0.1667 N$_2$ + 5 ThB$_6$ + ThB$_6$H$_4$C$_6$N | -9.3397 |
| ThNH$_4$B$_6$C$_6$ | 2.58 Th$_3$N$_4$ + 0.3636 B$_9$H$_{11}$ + 3 B$_{13}$C$_2$ $\longrightarrow$ 6.739 ThB$_4$ + 9.318 BN + ThB$_6$H$_4$C$_6$N | -9.3221 |
| ThNH$_4$B$_6$C$_6$ | 0.5714 B$_5$H$_7$ + 6 ThC + 3.143 B + 3 N$_2$ $\longrightarrow$ 5 ThN + ThB$_6$H$_4$C$_6$N | -9.3130 |
| ThNH$_4$B$_6$C$_6$ | 0.6818 ThB$_4$ + 6 ThC + 0.3636 B$_9$H$_{11}$ + 3.341 N$_2$ $\longrightarrow$ 5.682 ThN + ThB$_6$H$_4$C$_6$N | -9.2845 |
| ThNH$_4$B$_6$C$_6$ | 2.292 Th$_3$N$_4$ + 3 B$_{13}$C$_2$ + 1.333 H$_3$N $\longrightarrow$ 5.875 ThB$_4$ + 9.5 BN + ThB$_6$H$_4$C$_6$N | -9.2628 |
| ThNH$_4$B$_6$C$_6$ | 6 ThC + 34 B + 1.333 H$_3$N $\longrightarrow$ 4.667 ThB$_6$ + 0.3333 ThN + ThB$_6$H$_4$C$_6$N | -9.2601 |
| ThNH$_4$B$_6$C$_6$ | 6 ThC + 0.3636 B$_9$H$_{11}$ + 2.727 B + 3 N$_2$ $\longrightarrow$ 5 ThN + ThB$_6$H$_4$C$_6$N | -9.2507 |
| ThNH$_4$B$_6$C$_6$ | 2 ThH$_2$ + 2 Th$_2$C$_3$ + 42 B + ThN $\longrightarrow$ 6 ThB$_6$ + ThB$_6$H$_4$C$_6$N | -9.2166 |
| ThNH$_4$B$_6$C$_6$ | 0.5714 B$_5$H$_7$ + 6 ThC + 3.262 N$_2$ + 0.5238 ThB$_6$ $\longrightarrow$ 5.524 ThN + ThB$_6$H$_4$C$_6$N | -9.1878 |
| ThNH$_4$B$_6$C$_6$ | 0.5714 B$_5$H$_7$ + 2.554 Th$_3$N$_4$ + 3 B$_{13}$C$_2$ $\longrightarrow$ 6.661 ThB$_4$ + 9.214 BN + ThB$_6$H$_4$C$_6$N | -9.1701 |
| ThNH$_4$B$_6$C$_6$ | 0.5 ThH$_2$ + 6 ThC + 28 B + H$_3$N $\longrightarrow$ 5.5 ThB$_4$ + ThB$_6$H$_4$C$_6$N | -9.1683 |
| ThNH$_4$B$_6$C$_6$ | 2.956 Th$_4$H$_{15}$ + 3 B$_{13}$C$_2$ + 5.911 N$_2$ $\longrightarrow$ 3.667 B$_9$H$_{11}$ + 10.82 ThN + ThB$_6$H$_4$C$_6$N | -9.1668 |
| ThNH$_4$B$_6$C$_6$ | 0.25 Th$_3$N$_4$ + 6 ThC + 2 H$_2$ + 29 B $\longrightarrow$ 5.75 ThB$_4$ + ThB$_6$H$_4$C$_6$N | -9.1584 |
| ThNH$_4$B$_6$C$_6$ | 6 ThC + 0.3636 B$_9$H$_{11}$ + 3.227 N$_2$ + 0.4545 ThB$_6$ $\longrightarrow$ 5.455 ThN + ThB$_6$H$_4$C$_6$N | -9.1421 |
| ThNH$_4$B$_6$C$_6$ | ThH$_2$ + 3 B$_{13}$C$_2$ + 11.27 H$_3$N $\longrightarrow$ 4.545 B$_5$H$_7$ + 10.27 BN + ThB$_6$H$_4$C$_6$N | -9.1138 |
| ThNH$_4$B$_6$C$_6$ | 0.5714 B$_5$H$_7$ + 6 ThC + 23.14 B + 0.5 N$_2$ $\longrightarrow$ 5 ThB$_4$ + ThB$_6$H$_4$C$_6$N | -9.0649 |
| ThNH$_4$B$_6$C$_6$ | 0.5714 B$_5$H$_7$ + 6 ThC + 27.14 B + ThN $\longrightarrow$ 6 ThB$_4$ + ThB$_6$H$_4$C$_6$N | -9.0153 |
| ThNH$_4$B$_6$C$_6$ | 6 ThC + 0.3636 B$_9$H$_{11}$ + 22.73 B + 0.5 N$_2$ $\longrightarrow$ 5 ThB$_4$ + ThB$_6$H$_4$C$_6$N | -9.0026 |
| ThNH$_4$B$_6$C$_6$ | 2 ThH$_2$ + 2 Th$_2$C$_3$ + 36 B + 0.5 N$_2$ $\longrightarrow$ 5 ThB$_6$ + ThB$_6$H$_4$C$_6$N | -8.9776 |
| ThNH$_4$B$_6$C$_6$ | 6 ThC + 0.3636 B$_9$H$_{11}$ + 26.73 B + ThN $\longrightarrow$ 6 ThB$_4$ + ThB$_6$H$_4$C$_6$N | -8.9530 |
| ThNH$_4$B$_6$C$_6$ | 1.5 ThB$_4$ + 0.2667 Th$_4$H$_{15}$ + 2 Th$_2$C$_3$ + 4.211 N$_2$ $\longrightarrow$ 1.856 Th$_3$N$_4$ + ThB$_6$H$_4$C$_6$N | -8.9189 |
| ThNH$_4$B$_6$C$_6$ | 0.3333 Th$_3$N$_4$ + 3 B$_{13}$C$_2$ + 10.4 H$_3$N $\longrightarrow$ 10.74 BN + 2.474 B$_9$H$_{11}$ + ThB$_6$H$_4$C$_6$N | -8.8594 |
| ThNH$_4$B$_6$C$_6$ | 6.765 B$_9$H$_{11}$ + 3 B$_{13}$C$_2$ + 24.47 ThN $\longrightarrow$ 23.47 ThB$_4$ + 23.47 H$_3$N + ThB$_6$H$_4$C$_6$N | -8.8566 |
| ThNH$_4$B$_6$C$_6$ | 6 ThC + 2.75 B$_9$H$_{11}$ + 3 N$_2$ $\longrightarrow$ 3.75 B$_5$H$_7$ + 5 ThN + ThB$_6$H$_4$C$_6$N | -8.8421 |
| ThNH$_4$B$_6$C$_6$ | 0.3636 B$_9$H$_{11}$ + 3 B$_{13}$C$_2$ + 8.255 ThN $\longrightarrow$ 7.255 ThB$_4$ + 7.255 BN + ThB$_6$H$_4$C$_6$N | -8.8130 |
| ThNH$_4$B$_6$C$_6$ | 3 B$_{13}$C$_2$ + 1.333 H$_3$N + 7.333 ThN $\longrightarrow$ 6.333 ThB$_4$ + 7.667 BN + ThB$_6$H$_4$C$_6$N | -8.8105 |
| ThNH$_4$B$_6$C$_6$ | 3 B$_{13}$C$_2$ + 10.5 H$_3$N + ThN $\longrightarrow$ 10.5 BN + 2.5 B$_9$H$_{11}$ + ThB$_6$H$_4$C$_6$N | -8.7935 |
| ThNH$_4$B$_6$C$_6$ | ThB$_4$ + 3 B$_{13}$C$_2$ + 11.95 H$_3$N $\longrightarrow$ 10.95 BN + 2.895 B$_9$H$_{11}$ + ThB$_6$H$_4$C$_6$N | -8.7908 |
| ThNH$_4$B$_6$C$_6$ | 6 ThC + 26.33 B + 1.333 H$_3$N $\longrightarrow$ 5 ThB$_4$ + 0.3333 BN + ThB$_6$H$_4$C$_6$N | -8.7522 |
| ThNH$_4$B$_6$C$_6$ | 6 ThC + 0.6667 B$_9$H$_{11}$ + 3 N$_2$ $\longrightarrow$ 1.667 H$_2$ + 5 ThN + ThB$_6$H$_4$C$_6$N | -8.7012 |
| ThNH$_4$B$_6$C$_6$ | 2 ThH$_2$ + 0.25 Th$_3$N$_4$ + 2 Th$_2$C$_3$ + 40.5 B $\longrightarrow$ 5.75 ThB$_6$ + ThB$_6$H$_4$C$_6$N | -8.6904 |
| ThNH$_4$B$_6$C$_6$ | 0.5714 B$_5$H$_7$ + 3 B$_{13}$C$_2$ + 8.171 ThN $\longrightarrow$ 7.171 ThB$_4$ + 7.171 BN + ThB$_6$H$_4$C$_6$N | -8.6660 |
| ThNH$_4$B$_6$C$_6$ | 1.2 B$_5$H$_7$ + 2.917 Th$_2$C$_3$ + 2.917 N$_2$ $\longrightarrow$ 0.55 H$_8$C$_5$ + 4.833 ThN + ThB$_6$H$_4$C$_6$N | -8.6656 |
| ThNH$_4$B$_6$C$_6$ | 1.2 B$_5$H$_7$ + 6 ThC + 3 N$_2$ $\longrightarrow$ 2.2 H$_2$ + 5 ThN + ThB$_6$H$_4$C$_6$N | -8.6561 |
| ThNH$_4$B$_6$C$_6$ | 0.2667 Th$_4$H$_{15}$ + BN + 3 B$_{13}$C$_2$ + 5.6 Th $\longrightarrow$ 5.667 ThB$_6$ + ThB$_6$H$_4$C$_6$N | -8.6456 |
| ThNH$_4$B$_6$C$_6$ | 6 ThC + 0.06667 Th$_4$H$_{15}$ + 27.07 B + H$_3$N $\longrightarrow$ 5.267 ThB$_4$ + ThB$_6$H$_4$C$_6$N | -8.5798 |
| ThNH$_4$B$_6$C$_6$ | 0.5714 B$_5$H$_7$ + 0.25 Th$_3$N$_4$ + 6 ThC + 26.14 B $\longrightarrow$ 5.75 ThB$_4$ + ThB$_6$H$_4$C$_6$N | -8.5612 |
| ThNH$_4$B$_6$C$_6$ | 6 ThC + 2 H$_2$ + BN + 35 B $\longrightarrow$ 5 ThB$_6$ + ThB$_6$H$_4$C$_6$N | -8.5384 |
| ThNH$_4$B$_6$C$_6$ | 2 B$_9$H$_{11}$ + 2 Th$_2$C$_3$ + 3.5 N$_2$ $\longrightarrow$ 3 ThB$_4$ + 6 H$_3$N + ThB$_6$H$_4$C$_6$N | -8.5308 |
| ThNH$_4$B$_6$C$_6$ | 2 ThH$_2$ + 1.5 ThB$_4$ + 3 ThC$_2$ + 4.167 N$_2$ $\longrightarrow$ 1.833 Th$_3$N$_4$ + ThB$_6$H$_4$C$_6$N | -8.5257 |
| ThNH$_4$B$_6$C$_6$ | 0.25 Th$_3$N$_4$ + 6 ThC + 0.3636 B$_9$H$_{11}$ + 25.73 B $\longrightarrow$ 5.75 ThB$_4$ + ThB$_6$H$_4$C$_6$N | -8.4989 |
| ThNH$_4$B$_6$C$_6$ | 2 ThH$_2$ + 0.4615 B$_{13}$C$_2$ + 1.692 Th$_2$C$_3$ + 3.423 N$_2$ $\longrightarrow$ 1.462 Th$_3$N$_4$ + ThB$_6$H$_4$C$_6$N | -8.4866 |
| ThNH$_4$B$_6$C$_6$ | 6 ThC + 0.5 H$_2$ + 26 B + H$_3$N $\longrightarrow$ 5 ThB$_4$ + ThB$_6$H$_4$C$_6$N | -8.3380 |
| ThNH$_4$B$_6$C$_6$ | 0.2667 Th$_4$H$_{15}$ + 2 Th$_2$C$_3$ + 3.878 N$_2$ + ThB$_6$ $\longrightarrow$ 1.689 Th$_3$N$_4$ + ThB$_6$H$_4$C$_6$N | -8.2945 |
| ThNH$_4$B$_6$C$_6$ | 2.375 Th$_3$N$_4$ + 2 H$_2$ + 3 B$_{13}$C$_2$ $\longrightarrow$ 6.125 ThB$_4$ + 8.5 BN + ThB$_6$H$_4$C$_6$N | -8.2936 |
| ThNH$_4$B$_6$C$_6$ | 1.5 ThB$_4$ + 6 ThC + 1.333 H$_3$N + 3.083 N$_2$ $\longrightarrow$ 6.5 ThN + ThB$_6$H$_4$C$_6$N | -8.2191 |
| ThNH$_4$B$_6$C$_6$ | 0.1429 B$_5$H$_7$ + 6 ThC + 25.29 B + H$_3$N $\longrightarrow$ 5 ThB$_4$ + ThB$_6$H$_4$C$_6$N | -8.1887 |
| ThNH$_4$B$_6$C$_6$ | 6 ThC + 0.09091 B$_9$H$_{11}$ + 25.18 B + H$_3$N $\longrightarrow$ 5 ThB$_4$ + ThB$_6$H$_4$C$_6$N | -8.1731 |
| ThNH$_4$B$_6$C$_6$ | 6 ThC + 6 B + 1.333 H$_3$N + 2.333 N$_2$ $\longrightarrow$ 5 ThN + ThB$_6$H$_4$C$_6$N | -8.1447 |
| ThNH$_4$B$_6$C$_6$ | 2 ThH$_2$ + BN + 3 B$_{13}$C$_2$ + 4.667 Th $\longrightarrow$ 5.667 ThB$_6$ + ThB$_6$H$_4$C$_6$N | -8.1084 |
| ThNH$_4$B$_6$C$_6$ | 2 ThH$_2$ + 6 ThC + 0.5 N$_2$ + 17 ThB$_6$ $\longrightarrow$ 24 ThB$_4$ + ThB$_6$H$_4$C$_6$N | -8.0769 |
| ThNH$_4$B$_6$C$_6$ | 6 ThC + 25 B + 1.333 H$_3$N $\longrightarrow$ 4.75 ThB$_4$ + 0.08333 Th$_3$N$_4$ + ThB$_6$H$_4$C$_6$N | -8.0645 |
| ThNH$_4$B$_6$C$_6$ | 6 ThC + 0.2667 Th$_4$H$_{15}$ + BN + 29.27 B $\longrightarrow$ 6.067 ThB$_4$ + ThB$_6$H$_4$C$_6$N | -8.0625 |
| ThNH$_4$B$_6$C$_6$ | 5.077 ThC + 0.2667 Th$_4$H$_{15}$ + 0.4615 B$_{13}$C$_2$ + 3.072 N$_2$ $\longrightarrow$ 5.144 ThN + ThB$_6$H$_4$C$_6$N | -8.0535 |

| Compound | Chemical reaction | Reaction energy(eV) |
|---|---|---|
| $ThNH_4B_6C_6$ | $0.25\,Th_4H_{15} + 3\,B_{13}C_2 + 10.88\,H_3N \longrightarrow 4.625\,B_5H_7 + 9.875\,BN + ThB_6H_4C_6N$ | -7.9487 |
| $ThNH_4B_6C_6$ | $ThH_2 + 3\,B_{13}C_2 + 10.32\,H_3N \longrightarrow 9.316\,BN + 2.632\,B_9H_{11} + ThB_6H_4C_6N$ | -7.9467 |
| $ThNH_4B_6C_6$ | $5.58\,ThC + 0.3636\,B_9H_{11} + 0.2098\,B_{13}C_2 + 2.79\,N_2 \longrightarrow 4.58\,ThN + ThB_6H_4C_6N$ | -7.9431 |
| $ThNH_4B_6C_6$ | $0.5714\,B_5H_7 + 6\,ThC + BN + 32.14\,B \longrightarrow 5\,ThB_6 + ThB_6H_4C_6N$ | -7.9412 |
| $ThNH_4B_6C_6$ | $6\,ThC + 0.6667\,B_9H_{11} + 2.722\,N_2 \longrightarrow 1.667\,ThH_2 + 1.111\,Th_3N_4 + ThB_6H_4C_6N$ | -7.9240 |
| $ThNH_4B_6C_6$ | $6\,ThC + 24.67\,B + 1.333\,H_3N \longrightarrow 4.667\,ThB_4 + 0.3333\,ThN + ThB_6H_4C_6N$ | -7.9132 |
| $ThNH_4B_6C_6$ | $0.2667\,Th_4H_{15} + 2\,Th_2C_3 + 6\,B + 3.211\,N_2 \longrightarrow 1.356\,Th_3N_4 + ThB_6H_4C_6N$ | -7.9115 |
| $ThNH_4B_6C_6$ | $6\,ThC + 1.333\,H_3N + 2.833\,N_2 + ThB_6 \longrightarrow 6\,ThN + ThB_6H_4C_6N$ | -7.9057 |
| $ThNH_4B_6C_6$ | $2\,ThH_2 + 3\,ThC_2 + 3.833\,N_2 + ThB_6 \longrightarrow 1.667\,Th_3N_4 + ThB_6H_4C_6N$ | -7.9013 |
| $ThNH_4B_6C_6$ | $6\,ThC + 26\,B + 1.333\,H_3N \longrightarrow 5\,ThB_4 + 0.1667\,N_2 + ThB_6H_4C_6N$ | -7.8966 |
| $ThNH_4B_6C_6$ | $6\,ThC + BN + 0.3636\,B_9H_{11} + 31.73\,B \longrightarrow 5\,ThB_6 + ThB_6H_4C_6N$ | -7.8790 |
| $ThNH_4B_6C_6$ | $2\,ThH_2 + 1.5\,ThB_4 + 2\,Th_2C_3 + 3.75\,N_2 \longrightarrow 6.5\,ThN + ThB_6H_4C_6N$ | -7.8570 |
| $ThNH_4B_6C_6$ | $6\,ThC + 0.6667\,B_9H_{11} + 2.556\,N_2 \longrightarrow 0.2222\,Th_4H_{15} + 4.111\,ThN + ThB_6H_4C_6N$ | -7.8511 |
| $ThNH_4B_6C_6$ | $2\,H_2 + 3\,B_{13}C_2 + 7.6\,ThN \longrightarrow 6.6\,ThB_4 + 6.6\,BN + ThB_6H_4C_6N$ | -7.8248 |
| $ThNH_4B_6C_6$ | $0.25\,Th_3N_4 + 0.2667\,Th_4H_{15} + 3\,B_{13}C_2 + 4.683\,Th \longrightarrow 5.5\,ThB_6 + ThB_6H_4C_6N$ | -7.8208 |
| $ThNH_4B_6C_6$ | $0.5714\,B_5H_7 + 5.516\,ThC + 0.2418\,B_{13}C_2 + 2.758\,N_2 \longrightarrow 4.516\,ThN + ThB_6H_4C_6N$ | -7.8061 |
| $ThNH_4B_6C_6$ | $5.077\,ThC + 0.4615\,B_{13}C_2 + 1.333\,H_3N + 2.551\,N_2 \longrightarrow 1.359\,Th_3N_4 + ThB_6H_4C_6N$ | -7.8038 |
| $ThNH_4B_6C_6$ | $5.375\,ThC + 32.25\,B + H_3N + 0.125\,H_8C_5 \longrightarrow 4.375\,ThB_6 + ThB_6H_4C_6N$ | -7.7907 |
| $ThNH_4B_6C_6$ | $2\,ThH_2 + 2\,Th_2C_3 + 6\,B + 3\,N_2 \longrightarrow 5\,ThN + ThB_6H_4C_6N$ | -7.7826 |
| $ThNH_4B_6C_6$ | $1.75\,B_{13}C_2 + 5.188\,Th + 0.5\,H_8C_5 + 0.5\,N_2 \longrightarrow 4.187\,ThB_4 + ThB_6H_4C_6N$ | -7.7588 |
| $ThNH_4B_6C_6$ | $1.625\,Th_4H_{15} + 3\,B_{13}C_2 + 3.896\,N_2 \longrightarrow 6.792\,H_3N + 5.5\,ThB_6 + ThB_6H_4C_6N$ | -7.7349 |
| $ThNH_4B_6C_6$ | $3.6\,B_5H_7 + 3\,ThC_2 + 4.033\,N_2 \longrightarrow 7.067\,H_3N + 2\,ThB_6 + ThB_6H_4C_6N$ | -7.6825 |
| $ThNH_4B_6C_6$ | $6.5\,ThH_2 + 3\,B_{13}C_2 + 2\,N_2 \longrightarrow 3\,H_3N + 5.5\,ThB_6 + ThB_6H_4C_6N$ | -7.6151 |
| $ThNH_4B_6C_6$ | $3.667\,ThB_4 + 2\,Th_2C_3 + 9.667\,H_3N \longrightarrow 1.667\,Th_4H_{15} + 8.667\,BN + ThB_6H_4C_6N$ | -7.6143 |
| $ThNH_4B_6C_6$ | $2\,ThH_2 + 2\,Th_2C_3 + 3.5\,N_2 + ThB_6 \longrightarrow 6\,ThN + ThB_6H_4C_6N$ | -7.5436 |
| $ThNH_4B_6C_6$ | $2\,ThH_2 + 2\,Th_2C_3 + 26\,B + 0.5\,N_2 \longrightarrow 5\,ThB_4 + ThB_6H_4C_6N$ | -7.5345 |
| $ThNH_4B_6C_6$ | $1.2\,B_5H_7 + 6\,ThC + 2.413\,N_2 \longrightarrow 0.2933\,Th_4H_{15} + 3.827\,ThN + ThB_6H_4C_6N$ | -7.5340 |
| $ThNH_4B_6C_6$ | $11.64\,ThH_2 + 3\,B_{13}C_2 + 0.5\,N_2 \longrightarrow 1.286\,Th_4H_{15} + 5.5\,ThB_6 + ThB_6H_4C_6N$ | -7.5204 |
| $ThNH_4B_6C_6$ | $2\,ThH_2 + 3\,ThC_2 + 6\,B + 3.167\,N_2 \longrightarrow 1.333\,Th_3N_4 + ThB_6H_4C_6N$ | -7.5183 |
| $ThNH_4B_6C_6$ | $0.2667\,Th_4H_{15} + 3\,B_{13}C_2 + 4.433\,Th + ThN \longrightarrow 5.5\,ThB_6 + ThB_6H_4C_6N$ | -7.5004 |
| $ThNH_4B_6C_6$ | $ThC + 2.5\,B_{13}C_2 + 9.659\,H_3N \longrightarrow 3.568\,B_5H_7 + 8.659\,BN + ThB_6H_4C_6N$ | -7.4934 |
| $ThNH_4B_6C_6$ | $2\,ThH_2 + 1.875\,Th_3N_4 + 3\,B_{13}C_2 \longrightarrow 6.625\,ThB_4 + 6.5\,BN + ThB_6H_4C_6N$ | -7.4888 |
| $ThNH_4B_6C_6$ | $13.36\,ThH_2 + 3\,B_{13}C_2 + H_3N \longrightarrow 1.714\,Th_4H_{15} + 5.5\,ThB_6 + ThB_6H_4C_6N$ | -7.4888 |
| $ThNH_4B_6C_6$ | $2\,ThH_2 + 2\,Th_2C_3 + 30\,B + ThN \longrightarrow 6\,ThB_4 + ThB_6H_4C_6N$ | -7.4849 |
| $ThNH_4B_6C_6$ | $2\,ThH_2 + 6\,ThC + 19\,ThB_6 + ThN \longrightarrow 27\,ThB_4 + ThB_6H_4C_6N$ | -7.4500 |
| $ThNH_4B_6C_6$ | $0.6667\,B_9H_{11} + 2.694\,Th_2C_3 + 2.694\,N_2 \longrightarrow 0.4167\,H_8C_5 + 4.389\,ThN + ThB_6H_4C_6N$ | -7.3634 |
| $ThNH_4B_6C_6$ | $2\,ThH_2 + 0.25\,Th_3N_4 + 3\,B_{13}C_2 + 3.75\,Th \longrightarrow 5.5\,ThB_6 + ThB_6H_4C_6N$ | -7.2836 |
| $ThNH_4B_6C_6$ | $1.5\,ThB_4 + 2\,H_2 + 2\,Th_2C_3 + 3.5\,N_2 \longrightarrow 1.5\,Th_3N_4 + ThB_6H_4C_6N$ | -7.2354 |
| $ThNH_4B_6C_6$ | $3\,B_{13}C_2 + 16.73\,H_3N + 51.2\,ThN \longrightarrow 6.6\,B_5H_7 + 16.73\,Th_3N_4 + ThB_6H_4C_6N$ | -7.1763 |
| $ThNH_4B_6C_6$ | $2\,ThH_2 + 0.25\,Th_3N_4 + 6\,ThC + 18.5\,ThB_6 \longrightarrow 26.25\,ThB_4 + ThB_6H_4C_6N$ | -7.1402 |
| $ThNH_4B_6C_6$ | $2\,ThH_2 + 3\,B_{13}C_2 + 6\,ThN \longrightarrow 7\,ThB_4 + 5\,BN + ThB_6H_4C_6N$ | -7.1187 |
| $ThNH_4B_6C_6$ | $6\,ThC + 2\,H_2 + BN + 25\,B \longrightarrow 5\,ThB_4 + ThB_6H_4C_6N$ | -7.0953 |
| $ThNH_4B_6C_6$ | $2.108\,Th_3N_4 + 0.2667\,Th_4H_{15} + 3\,B_{13}C_2 \longrightarrow 6.392\,ThB_4 + 7.433\,BN + ThB_6H_4C_6N$ | -7.0602 |
| $ThNH_4B_6C_6$ | $9.675\,ThC + 3.321\,B_9H_{11} + 0.5\,N_2 \longrightarrow 2.169\,Th_4H_{15} + 1.838\,B_{13}C_2 + ThB_6H_4C_6N$ | -7.0451 |
| $ThNH_4B_6C_6$ | $5.077\,ThC + 2\,H_2 + 0.4615\,B_{13}C_2 + 2.538\,N_2 \longrightarrow 4.077\,ThN + ThB_6H_4C_6N$ | -7.0334 |
| $ThNH_4B_6C_6$ | $2\,ThH_2 + 0.25\,Th_3N_4 + 2\,Th_2C_3 + 29\,B \longrightarrow 5.75\,ThB_4 + ThB_6H_4C_6N$ | -7.0308 |
| $ThNH_4B_6C_6$ | $0.3636\,B_9H_{11} + 3\,B_{13}C_2 + 37.27\,ThN \longrightarrow 9.068\,ThB_4 + 9.068\,Th_3N_4 + ThB_6H_4C_6N$ | -7.0231 |
| $ThNH_4B_6C_6$ | $2\,ThH_2 + 3\,B_{13}C_2 + 3.5\,Th + ThN \longrightarrow 5.5\,ThB_6 + ThB_6H_4C_6N$ | -6.9632 |
| $ThNH_4B_6C_6$ | $3\,B_{13}C_2 + 1.333\,H_3N + 38\,ThN \longrightarrow 8.25\,ThB_4 + 9.583\,Th_3N_4 + ThB_6H_4C_6N$ | -6.9189 |
| $ThNH_4B_6C_6$ | $1.2\,B_5H_7 + 2\,Th_2C_3 + 3.233\,N_2 \longrightarrow Th_3N_4 + 1.467\,H_3N + ThB_6H_4C_6N$ | -6.9159 |
| $ThNH_4B_6C_6$ | $0.5714\,B_5H_7 + 3\,B_{13}C_2 + 36.86\,ThN \longrightarrow 8.964\,ThB_4 + 8.964\,Th_3N_4 + ThB_6H_4C_6N$ | -6.8967 |
| $ThNH_4B_6C_6$ | $10.57\,ThC + 3.967\,B_9H_{11} + ThN \longrightarrow 2.642\,Th_4H_{15} + 2.285\,B_{13}C_2 + ThB_6H_4C_6N$ | -6.8491 |
| $ThNH_4B_6C_6$ | $0.25\,Th_4H_{15} + 3\,B_{13}C_2 + 9.901\,H_3N \longrightarrow 8.901\,BN + 2.678\,B_9H_{11} + ThB_6H_4C_6N$ | -6.7612 |
| $ThNH_4B_6C_6$ | $1.2\,B_5H_7 + 4.375\,ThC_2 + 2.75\,N_2 \longrightarrow 1.125\,Th_3N_4 + 0.55\,H_8C_5 + ThB_6H_4C_6N$ | -6.7512 |
| $ThNH_4B_6C_6$ | $0.2667\,Th_4H_{15} + 3\,B_{13}C_2 + 6.747\,ThN \longrightarrow 6.813\,ThB_4 + 5.747\,BN + ThB_6H_4C_6N$ | -6.6440 |
| $ThNH_4B_6C_6$ | $1.2\,B_5H_7 + 6\,ThC + 2.367\,N_2 \longrightarrow 2.2\,ThH_2 + 0.9333\,Th_3N_4 + ThB_6H_4C_6N$ | -6.6350 |
| $ThNH_4B_6C_6$ | $9.25\,ThH_2 + 3\,B_{13}C_2 + 0.5\,N_2 \longrightarrow 8.25\,ThB_4 + 7.25\,H_2 + ThB_6H_4C_6N$ | -6.6223 |
| $ThNH_4B_6C_6$ | $2\,H_2 + 2\,Th_2C_3 + 3.167\,N_2 + ThB_6 \longrightarrow 1.333\,Th_3N_4 + ThB_6H_4C_6N$ | -6.6109 |
| $ThNH_4B_6C_6$ | $0.2667\,Th_4H_{15} + 2\,Th_2C_3 + 36.4\,B + ThN \longrightarrow 5.067\,ThB_6 + ThB_6H_4C_6N$ | -6.5931 |
| $ThNH_4B_6C_6$ | $ThC + 2.5\,B_{13}C_2 + 8.908\,H_3N \longrightarrow 7.908\,BN + 2.066\,B_9H_{11} + ThB_6H_4C_6N$ | -6.5772 |

| Compound | Chemical reaction | Reaction energy(eV) |
|---|---|---|
| $ThNH_4B_6C_6$ | $8.846\,B_5H_7 + 3\,B_{13}C_2 + 20.31\,ThN \longrightarrow 19.31\,ThB_4 + 19.31\,H_3N + ThB_6H_4C_6N$ | -6.5740 |
| $ThNH_4B_6C_6$ | $3\,ThC_2 + 1.045\,B_9H_{11} + 2.205\,N_2 \longrightarrow 0.5\,Th_4H_{15} + 3.409\,BN + ThB_6H_4C_6N$ | -6.5517 |
| $ThNH_4B_6C_6$ | $5.375\,ThC + 23.5\,B + H_3N + 0.125\,H_8C_5 \longrightarrow 4.375\,ThB_4 + ThB_6H_4C_6N$ | -6.5280 |
| $ThNH_4B_6C_6$ | $0.5714\,B_5H_7 + 6\,ThC + BN + 22.14\,B \longrightarrow 5\,ThB_4 + ThB_6H_4C_6N$ | -6.4982 |
| $ThNH_4B_6C_6$ | $0.6667\,B_9H_{11} + 2\,Th_2C_3 + 3.056\,N_2 \longrightarrow Th_3N_4 + 1.111\,H_3N + ThB_6H_4C_6N$ | -6.4902 |
| $ThNH_4B_6C_6$ | $6\,ThC + BN + 0.3636\,B_9H_{11} + 21.73\,B \longrightarrow 5\,ThB_4 + ThB_6H_4C_6N$ | -6.4359 |
| $ThNH_4B_6C_6$ | $0.25\,Th_3N_4 + 10.35\,ThC + 3.805\,B_9H_{11} \longrightarrow 2.524\,Th_4H_{15} + 2.173\,B_{13}C_2 + ThB_6H_4C_6N$ | -6.4316 |
| $ThNH_4B_6C_6$ | $2\,ThH_2 + BN + 2\,Th_2C_3 + 35\,B \longrightarrow 5\,ThB_6 + ThB_6H_4C_6N$ | -6.4108 |
| $ThNH_4B_6C_6$ | $3\,ThC_2 + 2\,B_9H_{11} + 3.5\,N_2 \longrightarrow 6\,H_3N + 2\,ThB_6 + ThB_6H_4C_6N$ | -6.4054 |
| $ThNH_4B_6C_6$ | $0.2667\,Th_4H_{15} + 2\,Th_2C_3 + 30.4\,B + 0.5\,N_2 \longrightarrow 4.067\,ThB_6 + ThB_6H_4C_6N$ | -6.3541 |
| $ThNH_4B_6C_6$ | $6\,ThC + 0.2667\,Th_4H_{15} + 0.5\,N_2 + 15.13\,ThB_6 \longrightarrow 21.2\,ThB_4 + ThB_6H_4C_6N$ | -6.2614 |
| $ThNH_4B_6C_6$ | $2\,H_2 + 2\,Th_2C_3 + 6\,B + 2.5\,N_2 \longrightarrow Th_3N_4 + ThB_6H_4C_6N$ | -6.2280 |
| $ThNH_4B_6C_6$ | $2\,ThH_2 + 3\,ThC_2 + 36\,B + ThN \longrightarrow 5\,ThB_6 + ThB_6H_4C_6N$ | -6.2255 |
| $ThNH_4B_6C_6$ | $2\,H_2 + 3\,B_{13}C_2 + 34\,ThN \longrightarrow 8.25\,ThB_4 + 8.25\,Th_3N_4 + ThB_6H_4C_6N$ | -6.1964 |
| $ThNH_4B_6C_6$ | $19\,B_5H_7 + 11\,Th_3N_4 + 3\,B_{13}C_2 \longrightarrow 32\,ThB_4 + 43\,H_3N + ThB_6H_4C_6N$ | -6.1781 |
| $ThNH_4B_6C_6$ | $0.7857\,ThB_4 + 0.5714\,B_5H_7 + 2\,Th_2C_3 + 3.024\,N_2 \longrightarrow 1.262\,Th_3N_4 + ThB_6H_4C_6N$ | -6.1585 |
| $ThNH_4B_6C_6$ | $5.867\,B_5H_7 + 1.2\,H_8C_5 + 24.33\,ThN \longrightarrow 23.33\,ThH_2 + 23.33\,BN + ThB_6H_4C_6N$ | -6.0902 |
| $ThNH_4B_6C_6$ | $1.758\,Th_3N_4 + 3\,B_{13}C_2 + 1.333\,H_3N \longrightarrow 7.364\,BN + 4.273\,ThB_6 + ThB_6H_4C_6N$ | -6.0886 |
| $ThNH_4B_6C_6$ | $0.25\,Th_3N_4 + 0.2667\,Th_4H_{15} + 2\,Th_2C_3 + 34.9\,B \longrightarrow 4.817\,ThB_6 + ThB_6H_4C_6N$ | -6.0668 |
| $ThNH_4B_6C_6$ | $0.6818\,ThB_4 + 0.3636\,B_9H_{11} + 2\,Th_2C_3 + 2.955\,N_2 \longrightarrow 1.227\,Th_3N_4 + ThB_6H_4C_6N$ | -6.0264 |
| $ThNH_4B_6C_6$ | $4.687\,B_5H_7 + 8.682\,ThC + 0.5\,N_2 \longrightarrow 1.921\,Th_4H_{15} + 1.341\,B_{13}C_2 + ThB_6H_4C_6N$ | -6.0244 |
| $ThNH_4B_6C_6$ | $2\,Th_2C_3 + 7.737\,H_3N + 2.123\,ThB_6 \longrightarrow 1.281\,Th_4H_{15} + 6.737\,BN + ThB_6H_4C_6N$ | -6.0033 |
| $ThNH_4B_6C_6$ | $6\,ThC + 21.55\,B + 7.667\,H_3N \longrightarrow 1.667\,Th_3N_4 + 1.727\,B_9H_{11} + ThB_6H_4C_6N$ | -6.0011 |
| $ThNH_4B_6C_6$ | $2\,ThH_2 + 3\,ThC_2 + 30\,B + 0.5\,N_2 \longrightarrow 4\,ThB_6 + ThB_6H_4C_6N$ | -5.9864 |
| $ThNH_4B_6C_6$ | $BN + 1.75\,B_{13}C_2 + 5.437\,Th + 0.5\,H_8C_5 \longrightarrow 4.437\,ThB_4 + ThB_6H_4C_6N$ | -5.9665 |
| $ThNH_4B_6C_6$ | $4.376\,ThC + 0.8118\,B_{13}C_2 + 5.554\,H_3N \longrightarrow 0.8441\,Th_4H_{15} + 4.554\,BN + ThB_6H_4C_6N$ | -5.9561 |
| $ThNH_4B_6C_6$ | $2.25\,B_{13}C_2 + 0.5\,Th_2C_3 + 8.625\,H_3N \longrightarrow 3.125\,B_5H_7 + 7.625\,BN + ThB_6H_4C_6N$ | -5.9164 |
| $ThNH_4B_6C_6$ | $6\,ThC + 2\,H_2 + 0.5\,N_2 + 13\,ThB_6 \longrightarrow 18\,ThB_4 + ThB_6H_4C_6N$ | -5.9100 |
| $ThNH_4B_6C_6$ | $2\,ThH_2 + 3\,B_{13}C_2 + 26\,ThN \longrightarrow 8.25\,ThB_4 + 6.25\,Th_3N_4 + ThB_6H_4C_6N$ | -5.8851 |
| $ThNH_4B_6C_6$ | $3\,B_{13}C_2 + 14.78\,H_3N + 45.33\,ThN \longrightarrow 14.78\,Th_3N_4 + 3.667\,B_9H_{11} + ThB_6H_4C_6N$ | -5.8687 |
| $ThNH_4B_6C_6$ | $6\,ThC + 0.6667\,B_9H_{11} + 2.167\,N_2 \longrightarrow 1.667\,ThH_2 + 3.333\,ThN + ThB_6H_4C_6N$ | -5.8507 |
| $ThNH_4B_6C_6$ | $0.5714\,B_5H_7 + 2\,Th_2C_3 + 2.849\,N_2 + 0.5238\,ThB_6 \longrightarrow 1.175\,Th_3N_4 + ThB_6H_4C_6N$ | -5.8314 |
| $ThNH_4B_6C_6$ | $2\,ThH_2 + 0.4615\,B_{13}C_2 + 1.692\,Th_2C_3 + 2.692\,N_2 \longrightarrow 4.385\,ThN + ThB_6H_4C_6N$ | -5.7594 |
| $ThNH_4B_6C_6$ | $0.3636\,B_9H_{11} + 2\,Th_2C_3 + 2.803\,N_2 + 0.4545\,ThB_6 \longrightarrow 1.152\,Th_3N_4 + ThB_6H_4C_6N$ | -5.7426 |
| $ThNH_4B_6C_6$ | $0.5714\,B_5H_7 + 6\,ThC + 0.5\,N_2 + 11.57\,ThB_6 \longrightarrow 16.57\,ThB_4 + ThB_6H_4C_6N$ | -5.7252 |
| $ThNH_4B_6C_6$ | $6\,ThC + 0.3636\,B_9H_{11} + 0.5\,N_2 + 11.36\,ThB_6 \longrightarrow 16.36\,ThB_4 + ThB_6H_4C_6N$ | -5.7229 |
| $ThNH_4B_6C_6$ | $6\,ThC + 19.57\,B + 7.667\,H_3N \longrightarrow 2.714\,B_5H_7 + 1.667\,Th_3N_4 + ThB_6H_4C_6N$ | -5.7053 |
| $ThNH_4B_6C_6$ | $2\,ThH_2 + 0.25\,Th_3N_4 + 3\,ThC_2 + 34.5\,B \longrightarrow 4.75\,ThB_6 + ThB_6H_4C_6N$ | -5.6992 |
| $ThNH_4B_6C_6$ | $1.967\,Th_3N_4 + 0.3636\,B_9H_{11} + 3\,B_{13}C_2 \longrightarrow 6.868\,BN + 4.901\,ThB_6 + ThB_6H_4C_6N$ | -5.6812 |
| $ThNH_4B_6C_6$ | $6\,ThC + 0.9934\,B_9H_{11} + 3.941\,H_3N \longrightarrow 1.25\,Th_4H_{15} + 2.941\,BN + ThB_6H_4C_6N$ | -5.6574 |
| $ThNH_4B_6C_6$ | $2\,ThH_2 + 6\,ThC + BN + 16.5\,ThB_6 \longrightarrow 23.5\,ThB_4 + ThB_6H_4C_6N$ | -5.6544 |
| $ThNH_4B_6C_6$ | $6\,ThC + 0.2667\,Th_4H_{15} + 17.13\,ThB_6 + ThN \longrightarrow 24.2\,ThB_4 + ThB_6H_4C_6N$ | -5.6346 |
| $ThNH_4B_6C_6$ | $2\,ThH_2 + 2.538\,ThC_2 + 0.4615\,B_{13}C_2 + 2.859\,N_2 \longrightarrow 1.179\,Th_3N_4 + ThB_6H_4C_6N$ | -5.6315 |
| $ThNH_4B_6C_6$ | $0.5714\,B_5H_7 + 2\,Th_2C_3 + 3.143\,B + 2.5\,N_2 \longrightarrow Th_3N_4 + ThB_6H_4C_6N$ | -5.6308 |
| $ThNH_4B_6C_6$ | $5.598\,B_5H_7 + 9.383\,ThC + ThN \longrightarrow 2.346\,Th_4H_{15} + 1.692\,B_{13}C_2 + ThB_6H_4C_6N$ | -5.6299 |
| $ThNH_4B_6C_6$ | $6\,ThC + 20.25\,B_9H_{11} + 0.5\,N_2 \longrightarrow 5\,ThB_4 + 31.25\,B_5H_7 + ThB_6H_4C_6N$ | -5.5973 |
| $ThNH_4B_6C_6$ | $0.5714\,B_5H_7 + 1.948\,Th_3N_4 + 3\,B_{13}C_2 \longrightarrow 6.792\,BN + 4.844\,ThB_6 + ThB_6H_4C_6N$ | -5.5713 |
| $ThNH_4B_6C_6$ | $0.3636\,B_9H_{11} + 2\,Th_2C_3 + 2.727\,B + 2.5\,N_2 \longrightarrow Th_3N_4 + ThB_6H_4C_6N$ | -5.5685 |
| $ThNH_4B_6C_6$ | $3\,B_{13}C_2 + 1.333\,H_3N + 5.524\,ThN \longrightarrow 5.857\,BN + 4.524\,ThB_6 + ThB_6H_4C_6N$ | -5.5613 |
| $ThNH_4B_6C_6$ | $4.042\,ThC_2 + 0.6667\,B_9H_{11} + 2.528\,N_2 \longrightarrow 1.014\,Th_3N_4 + 0.4167\,H_8C_5 + ThB_6H_4C_6N$ | -5.5475 |
| $ThNH_4B_6C_6$ | $1.5\,ThB_4 + 0.2667\,Th_4H_{15} + 3\,ThC_2 + 3.544\,N_2 \longrightarrow 1.522\,Th_3N_4 + ThB_6H_4C_6N$ | -5.5448 |
| $ThNH_4B_6C_6$ | $6\,ThC + 29\,B_9H_{11} + 0.5\,N_2 \longrightarrow 45\,B_5H_7 + 5\,ThB_6 + ThB_6H_4C_6N$ | -5.5420 |
| $ThNH_4B_6C_6$ | $0.2667\,Th_4H_{15} + 0.4615\,B_{13}C_2 + 1.692\,Th_2C_3 + 2.801\,N_2 \longrightarrow 1.15\,Th_3N_4 + ThB_6H_4C_6N$ | -5.5056 |
| $ThNH_4B_6C_6$ | $1.5\,ThB_4 + 2\,Th_2C_3 + 1.333\,H_3N + 2.833\,N_2 \longrightarrow 1.5\,Th_3N_4 + ThB_6H_4C_6N$ | -5.4699 |
| $ThNH_4B_6C_6$ | $2.8\,B_5H_7 + 3\,ThC_2 + 3.1\,N_2 \longrightarrow 2\,ThB_4 + 5.2\,H_3N + ThB_6H_4C_6N$ | -5.4697 |
| $ThNH_4B_6C_6$ | $1.5\,ThB_4 + 0.2667\,Th_4H_{15} + 2\,Th_2C_3 + 3.283\,N_2 \longrightarrow 5.567\,ThN + ThB_6H_4C_6N$ | -5.4565 |
| $ThNH_4B_6C_6$ | $12\,ThH_2 + BN + 3\,B_{13}C_2 \longrightarrow 1.333\,Th_4H_{15} + 5.667\,ThB_6 + ThB_6H_4C_6N$ | -5.4221 |
| $ThNH_4B_6C_6$ | $0.2667\,Th_4H_{15} + 2\,Th_2C_3 + 6\,B + 2.533\,N_2 \longrightarrow 4.067\,ThN + ThB_6H_4C_6N$ | -5.3821 |
| $ThNH_4B_6C_6$ | $6\,ThC + 2.727\,BN + 0.3636\,B_9H_{11} + 2.47\,N_2 \longrightarrow 1.667\,Th_3N_4 + ThB_6H_4C_6N$ | -5.3604 |
| $ThNH_4B_6C_6$ | $0.25\,Th_3N_4 + 6\,ThC + 0.2667\,Th_4H_{15} + 16.63\,ThB_6 \longrightarrow 23.45\,ThB_4 + ThB_6H_4C_6N$ | -5.3248 |

| Compound | Chemical reaction | Reaction energy(eV) |
|---|---|---|
| $ThNH_4B_6C_6$ | $2\,H_2 + 2\,Th_2C_3 + 30\,B + ThN \longrightarrow 4\,ThB_6 + ThB_6H_4C_6N$ | -5.3181 |
| $ThNH_4B_6C_6$ | $9.25\,ThH_2 + 3\,B_{13}C_2 + H_3N \longrightarrow 8.25\,ThB_4 + 8.75\,H_2 + ThB_6H_4C_6N$ | -5.2983 |
| $ThNH_4B_6C_6$ | $6\,ThC + 2\,H_2 + 15\,ThB_6 + ThN \longrightarrow 21\,ThB_4 + ThB_6H_4C_6N$ | -5.2832 |
| $ThNH_4B_6C_6$ | $5.077\,ThC + 0.4615\,B_{13}C_2 + 1.333\,H_3N + 1.872\,N_2 \longrightarrow 4.077\,ThN + ThB_6H_4C_6N$ | -5.2680 |
| $ThNH_4B_6C_6$ | $5.37\,B_5H_7 + 0.25\,Th_3N_4 + 9.208\,ThC \longrightarrow 2.239\,Th_4H_{15} + 1.604\,B_{13}C_2 + ThB_6H_4C_6N$ | -5.2621 |
| $ThNH_4B_6C_6$ | $2\,Th_2C_3 + 10.08\,B + 5.083\,H_3N \longrightarrow 0.75\,Th_4H_{15} + 4.083\,BN + ThB_6H_4C_6N$ | -5.2433 |
| $ThNH_4B_6C_6$ | $0.2667\,Th_4H_{15} + 3\,B_{13}C_2 + 29.73\,ThN \longrightarrow 8.25\,ThB_4 + 7.183\,Th_3N_4 + ThB_6H_4C_6N$ | -5.2262 |
| $ThNH_4B_6C_6$ | $1.716\,B_5H_7 + 6\,ThC + 3.58\,H_3N \longrightarrow 1.25\,Th_4H_{15} + 2.58\,BN + ThB_6H_4C_6N$ | -5.2168 |
| $ThNH_4B_6C_6$ | $0.2667\,Th_4H_{15} + 2\,Th_2C_3 + 22.27\,B + 0.5\,N_2 \longrightarrow 4.067\,ThB_4 + ThB_6H_4C_6N$ | -5.1804 |
| $ThNH_4B_6C_6$ | $2.75\,B_9H_{11} + 2\,Th_2C_3 + 2.5\,N_2 \longrightarrow 3.75\,B_5H_7 + Th_3N_4 + ThB_6H_4C_6N$ | -5.1599 |
| $ThNH_4B_6C_6$ | $0.2667\,Th_4H_{15} + 2\,Th_2C_3 + 3.033\,N_2 + ThB_6 \longrightarrow 5.067\,ThN + ThB_6H_4C_6N$ | -5.1431 |
| $ThNH_4B_6C_6$ | $0.2667\,Th_4H_{15} + 2\,Th_2C_3 + 26.27\,B + ThN \longrightarrow 5.067\,ThB_4 + ThB_6H_4C_6N$ | -5.1307 |
| $ThNH_4B_6C_6$ | $6\,ThC + 11.37\,H_2 + 6\,B + 0.5\,N_2 \longrightarrow 1.25\,Th_4H_{15} + ThB_6H_4C_6N$ | -5.1285 |
| $ThNH_4B_6C_6$ | $0.5\,ThH_2 + 6\,ThC + H_3N + 14\,ThB_6 \longrightarrow 19.5\,ThB_4 + ThB_6H_4C_6N$ | -5.1277 |
| $ThNH_4B_6C_6$ | $10.04\,ThH_2 + 0.25\,Th_3N_4 + 3\,B_{13}C_2 \longrightarrow 1.071\,Th_4H_{15} + 5.5\,ThB_6 + ThB_6H_4C_6N$ | -5.1249 |
| $ThNH_4B_6C_6$ | $2.25\,B_{13}C_2 + 0.5\,Th_2C_3 + 7.967\,H_3N \longrightarrow 6.967\,BN + 1.809\,B_9H_{11} + ThB_6H_4C_6N$ | -5.1140 |
| $ThNH_4B_6C_6$ | $8.96\,ThC + 2.805\,B_9H_{11} + H_3N \longrightarrow 1.99\,Th_4H_{15} + 1.48\,B_{13}C_2 + ThB_6H_4C_6N$ | -5.1129 |
| $ThNH_4B_6C_6$ | $1.5\,ThB_4 + 3.5\,ThC + 0.5\,H_8C_5 + 3.167\,N_2 \longrightarrow 1.333\,Th_3N_4 + ThB_6H_4C_6N$ | -5.1084 |
| $ThNH_4B_6C_6$ | $2\,ThH_2 + 1.5\,ThB_4 + 3\,ThC_2 + 3.25\,N_2 \longrightarrow 5.5\,ThN + ThB_6H_4C_6N$ | -5.1048 |
| $ThNH_4B_6C_6$ | $0.5714\,B_5H_7 + 6\,ThC + 13.57\,ThB_6 + ThN \longrightarrow 19.57\,ThB_4 + ThB_6H_4C_6N$ | -5.0983 |
| $ThNH_4B_6C_6$ | $6\,ThC + 0.3636\,B_9H_{11} + 13.36\,ThB_6 + ThN \longrightarrow 19.36\,ThB_4 + ThB_6H_4C_6N$ | -5.0960 |
| $ThNH_4B_6C_6$ | $0.3636\,B_9H_{11} + 3\,B_{13}C_2 + 6.182\,ThN \longrightarrow 5.182\,BN + 5.182\,ThB_6 + ThB_6H_4C_6N$ | -5.0912 |
| $ThNH_4B_6C_6$ | $2\,H_2 + 2\,Th_2C_3 + 24\,B + 0.5\,N_2 \longrightarrow 3\,ThB_6 + ThB_6H_4C_6N$ | -5.0791 |
| $ThNH_4B_6C_6$ | $1.2\,B_5H_7 + 2\,Th_2C_3 + 2.733\,N_2 \longrightarrow 1.467\,H_3N + 3\,ThN + ThB_6H_4C_6N$ | -5.0500 |
| $ThNH_4B_6C_6$ | $7.342\,ThH_2 + 3\,B_{13}C_2 + 0.5\,N_2 \longrightarrow 6.342\,ThB_4 + 1.526\,B_5H_7 + ThB_6H_4C_6N$ | -5.0490 |
| $ThNH_4B_6C_6$ | $2\,ThH_2 + 3\,ThC_2 + 6\,B + 2.5\,N_2 \longrightarrow 4\,ThN + ThB_6H_4C_6N$ | -5.0304 |
| $ThNH_4B_6C_6$ | $1.2\,B_5H_7 + 9.5\,Th + H_3N + 1.2\,H_8C_5 \longrightarrow 8.5\,ThH_2 + ThB_6H_4C_6N$ | -5.0226 |
| $ThNH_4B_6C_6$ | $0.6667\,B_9H_{11} + 2\,Th_2C_3 + 2.5\,N_2 \longrightarrow Th_3N_4 + 1.667\,H_2 + ThB_6H_4C_6N$ | -5.0190 |
| $ThNH_4B_6C_6$ | $0.5714\,B_5H_7 + 3\,B_{13}C_2 + 6.122\,ThN \longrightarrow 5.122\,BN + 5.122\,ThB_6 + ThB_6H_4C_6N$ | -4.9869 |
| $ThNH_4B_6C_6$ | $1.818\,Th_3N_4 + 2\,H_2 + 3\,B_{13}C_2 \longrightarrow 6.273\,BN + 4.455\,ThB_6 + ThB_6H_4C_6N$ | -4.9843 |
| $ThNH_4B_6C_6$ | $1.2\,B_5H_7 + 2\,Th_2C_3 + 2.5\,N_2 \longrightarrow Th_3N_4 + 2.2\,H_2 + ThB_6H_4C_6N$ | -4.9739 |
| $ThNH_4B_6C_6$ | $0.25\,Th_3N_4 + 6\,ThC + 2\,H_2 + 14.5\,ThB_6 \longrightarrow 20.25\,ThB_4 + ThB_6H_4C_6N$ | -4.9734 |
| $ThNH_4B_6C_6$ | $2\,ThH_2 + BN + 2\,Th_2C_3 + 25\,B \longrightarrow 5\,ThB_4 + ThB_6H_4C_6N$ | -4.9678 |
| $ThNH_4B_6C_6$ | $6\,ThC + 1.333\,H_3N + 13.17\,ThB_6 \longrightarrow 18.17\,ThB_4 + 0.3333\,BN + ThB_6H_4C_6N$ | -4.9521 |
| $ThNH_4B_6C_6$ | $9.5\,ThH_2 + 3\,B_{13}C_2 + ThN \longrightarrow Th_4H_{15} + 5.5\,ThB_6 + ThB_6H_4C_6N$ | -4.9484 |
| $ThNH_4B_6C_6$ | $6\,ThC + 23.75\,B_9H_{11} + ThN \longrightarrow 6\,ThB_4 + 36.75\,B_5H_7 + ThB_6H_4C_6N$ | -4.9484 |
| $ThNH_4B_6C_6$ | $0.25\,Th_3N_4 + 1.75\,B_{13}C_2 + 4.437\,Th + 0.5\,H_8C_5 \longrightarrow 4.187\,ThB_4 + ThB_6H_4C_6N$ | -4.9316 |
| $ThNH_4B_6C_6$ | $9.967\,ThC + BN + 3.421\,B_9H_{11} \longrightarrow 2.242\,Th_4H_{15} + 1.983\,B_{13}C_2 + ThB_6H_4C_6N$ | -4.9277 |
| $ThNH_4B_6C_6$ | $0.2667\,Th_4H_{15} + 3\,ThC_2 + 3.211\,N_2 + ThB_6 \longrightarrow 1.356\,Th_3N_4 + ThB_6H_4C_6N$ | -4.9203 |
| $ThNH_4B_6C_6$ | $8.25\,ThH_2 + 3\,B_{13}C_2 + ThN \longrightarrow 8.25\,ThB_4 + 6.25\,H_2 + ThB_6H_4C_6N$ | -4.9121 |
| $ThNH_4B_6C_6$ | $1.2\,B_5H_7 + 6\,ThC + 1.9\,N_2 \longrightarrow 2.2\,ThH_2 + 2.8\,ThN + ThB_6H_4C_6N$ | -4.8935 |
| $ThNH_4B_6C_6$ | $6\,ThC + 34.25\,B_9H_{11} + ThN \longrightarrow 53.25\,B_5H_7 + 6\,ThB_6 + ThB_6H_4C_6N$ | -4.8820 |
| $ThNH_4B_6C_6$ | $8.5\,ThH_2 + 0.25\,Th_3N_4 + 3\,B_{13}C_2 \longrightarrow 8.25\,ThB_4 + 6.5\,H_2 + ThB_6H_4C_6N$ | -4.8732 |
| $ThNH_4B_6C_6$ | $2\,Th_2C_3 + 1.333\,H_3N + 2.5\,N_2 + ThB_6 \longrightarrow 1.333\,Th_3N_4 + ThB_6H_4C_6N$ | -4.8455 |
| $ThNH_4B_6C_6$ | $2\,ThH_2 + 3\,ThC_2 + 22\,B + 0.5\,N_2 \longrightarrow 4\,ThB_4 + ThB_6H_4C_6N$ | -4.8320 |
| $ThNH_4B_6C_6$ | $7.145\,ThH_2 + 3\,B_{13}C_2 + 0.5\,N_2 \longrightarrow 6.145\,ThB_4 + 0.9355\,B_9H_{11} + ThB_6H_4C_6N$ | -4.8234 |
| $ThNH_4B_6C_6$ | $0.25\,Th_3N_4 + 2\,H_2 + 2\,Th_2C_3 + 28.5\,B \longrightarrow 3.75\,ThB_6 + ThB_6H_4C_6N$ | -4.7918 |
| $ThNH_4B_6C_6$ | $2\,ThH_2 + 3\,ThC_2 + 3\,N_2 + ThB_6 \longrightarrow 5\,ThN + ThB_6H_4C_6N$ | -4.7914 |
| $ThNH_4B_6C_6$ | $0.5714\,B_5H_7 + 0.25\,Th_3N_4 + 6\,ThC + 13.07\,ThB_6 \longrightarrow 18.82\,ThB_4 + ThB_6H_4C_6N$ | -4.7885 |
| $ThNH_4B_6C_6$ | $0.25\,Th_3N_4 + 6\,ThC + 0.3636\,B_9H_{11} + 12.86\,ThB_6 \longrightarrow 18.61\,ThB_4 + ThB_6H_4C_6N$ | -4.7862 |
| $ThNH_4B_6C_6$ | $2\,ThH_2 + 3\,ThC_2 + 26\,B + ThN \longrightarrow 5\,ThB_4 + ThB_6H_4C_6N$ | -4.7824 |
| $ThNH_4B_6C_6$ | $ThC_2 + 0.75\,B_{13}C_2 + 0.5\,H_8C_5 + 2.375\,N_2 \longrightarrow 3.75\,BN + ThB_6H_4C_6N$ | -4.7300 |
| $ThNH_4B_6C_6$ | $0.5\,ThH_2 + 2\,Th_2C_3 + 27\,B + H_3N \longrightarrow 3.5\,ThB_6 + ThB_6H_4C_6N$ | -4.7296 |
| $ThNH_4B_6C_6$ | $0.5714\,B_5H_7 + 2\,Th_2C_3 + 27.14\,B + ThN \longrightarrow 4\,ThB_6 + ThB_6H_4C_6N$ | -4.7209 |
| $ThNH_4B_6C_6$ | $0.25\,Th_3N_4 + 0.2667\,Th_4H_{15} + 2\,Th_2C_3 + 25.27\,B \longrightarrow 4.817\,ThB_4 + ThB_6H_4C_6N$ | -4.6767 |
| $ThNH_4B_6C_6$ | $6\,ThC + 0.06667\,Th_4H_{15} + H_3N + 13.53\,ThB_6 \longrightarrow 18.8\,ThB_4 + ThB_6H_4C_6N$ | -4.6738 |
| $ThNH_4B_6C_6$ | $0.3636\,B_9H_{11} + 2\,Th_2C_3 + 26.73\,B + ThN \longrightarrow 4\,ThB_6 + ThB_6H_4C_6N$ | -4.6586 |
| $ThNH_4B_6C_6$ | $1.2\,B_5H_7 + 4.375\,ThC_2 + 2.187\,N_2 \longrightarrow 0.55\,H_8C_5 + 3.375\,ThN + ThB_6H_4C_6N$ | -4.6520 |
| $ThNH_4B_6C_6$ | $0.25\,Th_3N_4 + 6\,ThC + 22.88\,B_9H_{11} \longrightarrow 5.75\,ThB_4 + 35.38\,B_5H_7 + ThB_6H_4C_6N$ | -4.6441 |
| $ThNH_4B_6C_6$ | $0.6667\,B_9H_{11} + 2\,Th_2C_3 + 2.556\,N_2 \longrightarrow 1.111\,H_3N + 3\,ThN + ThB_6H_4C_6N$ | -4.6243 |

| Compound | Chemical reaction | Reaction energy(eV) |
|---|---|---|
| $ThNH_4B_6C_6$ | $1.75\,B_{13}C_2 + 4.187\,Th + 0.5\,H_8C_5 + ThN \longrightarrow 4.187\,ThB_4 + ThB_6H_4C_6N$ | -4.6112 |
| $ThNH_4B_6C_6$ | $6\,ThC + 9.667\,B + 4.667\,H_3N \longrightarrow 5\,ThH_2 + 3.667\,BN + ThB_6H_4C_6N$ | -4.5912 |
| $ThNH_4B_6C_6$ | $6\,ThC + 0.5\,H_2 + H_3N + 13\,ThB_6 \longrightarrow 18\,ThB_4 + ThB_6H_4C_6N$ | -4.5859 |
| $ThNH_4B_6C_6$ | $0.25\,Th_3N_4 + 6\,ThC + 32.94\,B_9H_{11} \longrightarrow 51.19\,B_5H_7 + 5.75\,ThB_6 + ThB_6H_4C_6N$ | -4.5806 |
| $ThNH_4B_6C_6$ | $0.1429\,B_5H_7 + 6\,ThC + H_3N + 12.64\,ThB_6 \longrightarrow 17.64\,ThB_4 + ThB_6H_4C_6N$ | -4.5397 |
| $ThNH_4B_6C_6$ | $6\,ThC + 0.09091\,B_9H_{11} + H_3N + 12.59\,ThB_6 \longrightarrow 17.59\,ThB_4 + ThB_6H_4C_6N$ | -4.5392 |
| $ThNH_4B_6C_6$ | $0.2667\,Th_4H_{15} + 3\,ThC_2 + 6\,B + 2.544\,N_2 \longrightarrow 1.022\,Th_3N_4 + ThB_6H_4C_6N$ | -4.5374 |
| $ThNH_4B_6C_6$ | $3.5\,ThC + 0.5\,H_8C_5 + 2.833\,N_2 + ThB_6 \longrightarrow 1.167\,Th_3N_4 + ThB_6H_4C_6N$ | -4.4840 |
| $ThNH_4B_6C_6$ | $0.5714\,B_5H_7 + 2\,Th_2C_3 + 21.14\,B + 0.5\,N_2 \longrightarrow 3\,ThB_6 + ThB_6H_4C_6N$ | -4.4819 |
| $ThNH_4B_6C_6$ | $3\,ThC_2 + 1.556\,B_9H_{11} + 2.685\,N_2 \longrightarrow 2\,ThB_4 + 4.37\,H_3N + ThB_6H_4C_6N$ | -4.4764 |
| $ThNH_4B_6C_6$ | $0.3636\,B_9H_{11} + 0.2098\,B_{13}C_2 + 1.86\,Th_2C_3 + 2.314\,N_2 \longrightarrow 0.9068\,Th_3N_4 + ThB_6H_4C_6N$ | -4.4750 |
| $ThNH_4B_6C_6$ | $9.5\,ThH_2 + BN + 3\,B_{13}C_2 \longrightarrow 8.5\,ThB_4 + 7.5\,H_2 + ThB_6H_4C_6N$ | -4.4708 |
| $ThNH_4B_6C_6$ | $2\,Th_2C_3 + 6\,B + 1.333\,H_3N + 1.833\,N_2 \longrightarrow Th_3N_4 + ThB_6H_4C_6N$ | -4.4625 |
| $ThNH_4B_6C_6$ | $6\,ThC + 1.333\,H_3N + 12.5\,ThB_6 \longrightarrow 17.25\,ThB_4 + 0.08333\,Th_3N_4 + ThB_6H_4C_6N$ | -4.4568 |
| $ThNH_4B_6C_6$ | $2\,H_2 + 3\,B_{13}C_2 + 5.714\,ThN \longrightarrow 4.714\,BN + 4.714\,ThB_6 + ThB_6H_4C_6N$ | -4.4388 |
| $ThNH_4B_6C_6$ | $1.5\,ThB_4 + 2\,H_2 + 2\,Th_2C_3 + 2.75\,N_2 \longrightarrow 4.5\,ThN + ThB_6H_4C_6N$ | -4.4364 |
| $ThNH_4B_6C_6$ | $6\,ThC + 2.889\,B_9H_{11} + 0.5\,N_2 \longrightarrow 5\,ThB_4 + 13.89\,H_2 + ThB_6H_4C_6N$ | -4.4233 |
| $ThNH_4B_6C_6$ | $0.3636\,B_9H_{11} + 2\,Th_2C_3 + 20.73\,B + 0.5\,N_2 \longrightarrow 3\,ThB_6 + ThB_6H_4C_6N$ | -4.4196 |
| $ThNH_4B_6C_6$ | $6\,ThC + 22.12\,B_9H_{11} + H_3N \longrightarrow 5\,ThB_4 + 34.62\,B_5H_7 + ThB_6H_4C_6N$ | -4.4000 |
| $ThNH_4B_6C_6$ | $0.5714\,B_5H_7 + 0.2418\,B_{13}C_2 + 1.839\,Th_2C_3 + 2.285\,N_2 \longrightarrow 0.8926\,Th_3N_4 + ThB_6H_4C_6N$ | -4.3706 |
| $ThNH_4B_6C_6$ | $2\,H_2 + 2\,Th_2C_3 + 6\,B + 2\,N_2 \longrightarrow 3\,ThN + ThB_6H_4C_6N$ | -4.3620 |
| $ThNH_4B_6C_6$ | $0.5714\,B_5H_7 + 6\,ThC + 3.143\,BN + 2.262\,N_2 \longrightarrow 1.667\,Th_3N_4 + ThB_6H_4C_6N$ | -4.3559 |
| $ThNH_4B_6C_6$ | $6\,ThC + 1.333\,H_3N + 12.33\,ThB_6 \longrightarrow 17\,ThB_4 + 0.3333\,ThN + ThB_6H_4C_6N$ | -4.3535 |
| $ThNH_4B_6C_6$ | $ThC_2 + 2\,B_{13}C_2 + 7.591\,H_3N \longrightarrow 2.682\,B_5H_7 + 6.591\,BN + ThB_6H_4C_6N$ | -4.3467 |
| $ThNH_4B_6C_6$ | $6\,ThC + 30.88\,B_9H_{11} + H_3N \longrightarrow 48.38\,B_5H_7 + 5\,ThB_6 + ThB_6H_4C_6N$ | -4.3448 |
| $ThNH_4B_6C_6$ | $2\,ThH_2 + 0.25\,Th_3N_4 + 3\,ThC_2 + 25\,B \longrightarrow 4.75\,ThB_4 + ThB_6H_4C_6N$ | -4.3283 |
| $ThNH_4B_6C_6$ | $0.6667\,B_9H_{11} + 8.967\,Th + H_3N + 1.2\,H_8C_5 \longrightarrow 7.967\,ThH_2 + ThB_6H_4C_6N$ | -4.3012 |
| $ThNH_4B_6C_6$ | $6\,ThC + 17.45\,B + 6\,H_3N \longrightarrow 1.273\,B_9H_{11} + 5\,ThN + ThB_6H_4C_6N$ | -4.2737 |
| $ThNH_4B_6C_6$ | $3.958\,B_5H_7 + 8.121\,ThC + H_3N \longrightarrow 1.78\,Th_4H_{15} + 1.061\,B_{13}C_2 + ThB_6H_4C_6N$ | -4.2509 |
| $ThNH_4B_6C_6$ | $1.75\,B_{13}C_2 + 3.792\,Th + 0.5\,H_8C_5 + 0.5\,N_2 \longrightarrow 2.792\,ThB_6 + ThB_6H_4C_6N$ | -4.2403 |
| $ThNH_4B_6C_6$ | $5.2\,B_5H_7 + 6\,ThC + 0.5\,N_2 \longrightarrow 5\,ThB_4 + 16.2\,H_2 + ThB_6H_4C_6N$ | -4.2279 |
| $ThNH_4B_6C_6$ | $2\,H_2 + 2\,Th_2C_3 + 18\,B + 0.5\,N_2 \longrightarrow 3\,ThB_4 + ThB_6H_4C_6N$ | -4.2132 |
| $ThNH_4B_6C_6$ | $0.5714\,B_5H_7 + 0.25\,Th_3N_4 + 2\,Th_2C_3 + 25.64\,B \longrightarrow 3.75\,ThB_6 + ThB_6H_4C_6N$ | -4.1947 |
| $ThNH_4B_6C_6$ | $1.2\,B_5H_7 + 8\,Th + 1.2\,H_8C_5 + 0.5\,N_2 \longrightarrow 7\,ThH_2 + ThB_6H_4C_6N$ | -4.1908 |
| $ThNH_4B_6C_6$ | $6\,ThC + 13.25\,H_2 + 6\,B + ThN \longrightarrow 1.5\,Th_4H_{15} + ThB_6H_4C_6N$ | -4.1722 |
| $ThNH_4B_6C_6$ | $2\,Th_2C_3 + 24.33\,B + 1.333\,H_3N \longrightarrow 0.3333\,BN + 3\,ThB_6 + ThB_6H_4C_6N$ | -4.1692 |
| $ThNH_4B_6C_6$ | $2\,H_2 + 2\,Th_2C_3 + 22\,B + ThN \longrightarrow 4\,ThB_4 + ThB_6H_4C_6N$ | -4.1636 |
| $ThNH_4B_6C_6$ | $6\,ThC + 1.333\,H_3N + 13\,ThB_6 \longrightarrow 18\,ThB_4 + 0.1667\,N_2 + ThB_6H_4C_6N$ | -4.1446 |
| $ThNH_4B_6C_6$ | $0.25\,Th_3N_4 + 0.3636\,B_9H_{11} + 2\,Th_2C_3 + 25.23\,B \longrightarrow 3.75\,ThB_6 + ThB_6H_4C_6N$ | -4.1324 |
| $ThNH_4B_6C_6$ | $2\,H_2 + 2\,Th_2C_3 + 2.5\,N_2 + ThB_6 \longrightarrow 4\,ThN + ThB_6H_4C_6N$ | -4.1230 |
| $ThNH_4B_6C_6$ | $3.5\,ThC + 6\,B + 0.5\,H_8C_5 + 2.167\,N_2 \longrightarrow 0.8333\,Th_3N_4 + ThB_6H_4C_6N$ | -4.1010 |
| $ThNH_4B_6C_6$ | $6\,ThC + 1.46\,B_9H_{11} + 0.5\,N_2 \longrightarrow 1.785\,ThB_4 + 0.8039\,Th_4H_{15} + ThB_6H_4C_6N$ | -4.0993 |
| $ThNH_4B_6C_6$ | $0.06667\,Th_4H_{15} + 2\,Th_2C_3 + 25.6\,B + H_3N \longrightarrow 3.267\,ThB_6 + ThB_6H_4C_6N$ | -4.0737 |
| $ThNH_4B_6C_6$ | $6\,ThC + 16\,B + 6\,H_3N \longrightarrow 2\,B_5H_7 + 5\,ThN + ThB_6H_4C_6N$ | -4.0558 |
| $ThNH_4B_6C_6$ | $2.428\,B_5H_7 + 6\,ThC + 0.5\,N_2 \longrightarrow 1.535\,ThB_4 + 0.8663\,Th_4H_{15} + ThB_6H_4C_6N$ | -3.9830 |
| $ThNH_4B_6C_6$ | $3.212\,ThB_4 + 3\,ThC_2 + 7.848\,H_3N \longrightarrow 1.303\,Th_4H_{15} + 6.848\,BN + ThB_6H_4C_6N$ | -3.9712 |
| $ThNH_4B_6C_6$ | $5.75\,ThH_2 + 3\,B_{13}C_2 + 3.5\,ThN \longrightarrow 8.25\,ThB_4 + 2.5\,H_3N + ThB_6H_4C_6N$ | -3.9466 |
| $ThNH_4B_6C_6$ | $0.25\,Th_3N_4 + 6\,ThC + 12.78\,H_2 + 6\,B \longrightarrow 1.437\,Th_4H_{15} + ThB_6H_4C_6N$ | -3.9448 |
| $ThNH_4B_6C_6$ | $6\,ThC + 13.25\,H_2 + 0.5\,N_2 + ThB_6 \longrightarrow 1.5\,Th_4H_{15} + ThB_6H_4C_6N$ | -3.9332 |
| $ThNH_4B_6C_6$ | $6\,ThC + 7.708\,H_2 + 0.6667\,B_9H_{11} + 0.5\,N_2 \longrightarrow 1.25\,Th_4H_{15} + ThB_6H_4C_6N$ | -3.9195 |
| $ThNH_4B_6C_6$ | $2\,ThH_2 + 1.273\,Th_3N_4 + 3\,B_{13}C_2 \longrightarrow 4.091\,BN + 4.818\,ThB_6 + ThB_6H_4C_6N$ | -3.9093 |
| $ThNH_4B_6C_6$ | $6\,ThC + 1.594\,B_9H_{11} + 0.5\,N_2 \longrightarrow 0.9023\,Th_4H_{15} + 1.391\,ThB_6 + ThB_6H_4C_6N$ | -3.9006 |
| $ThNH_4B_6C_6$ | $4.827\,B_5H_7 + 8.944\,ThC + BN \longrightarrow 1.986\,Th_4H_{15} + 1.472\,B_{13}C_2 + ThB_6H_4C_6N$ | -3.8764 |
| $ThNH_4B_6C_6$ | $1.2\,B_5H_7 + 6\,ThC + 7.175\,H_2 + 0.5\,N_2 \longrightarrow 1.25\,Th_4H_{15} + ThB_6H_4C_6N$ | -3.8745 |
| $ThNH_4B_6C_6$ | $1.5\,ThB_4 + 2\,H_2 + 3\,ThC_2 + 2.833\,N_2 \longrightarrow 1.167\,Th_3N_4 + ThB_6H_4C_6N$ | -3.8612 |
| $ThNH_4B_6C_6$ | $6\,ThC + 4\,B_9H_{11} + 0.5\,N_2 \longrightarrow 20\,H_2 + 5\,ThB_6 + ThB_6H_4C_6N$ | -3.8515 |
| $ThNH_4B_6C_6$ | $6\,ThC + 0.2667\,Th_4H_{15} + BN + 14.63\,ThB_6 \longrightarrow 20.7\,ThB_4 + ThB_6H_4C_6N$ | -3.8390 |
| $ThNH_4B_6C_6$ | $2\,H_2 + 0.4615\,B_{13}C_2 + 1.692\,Th_2C_3 + 2.09\,N_2 \longrightarrow 0.7949\,Th_3N_4 + ThB_6H_4C_6N$ | -3.8221 |
| $ThNH_4B_6C_6$ | $2.617\,B_5H_7 + 6\,ThC + 0.5\,N_2 \longrightarrow 0.9547\,Th_4H_{15} + 1.181\,ThB_6 + ThB_6H_4C_6N$ | -3.8051 |
| $ThNH_4B_6C_6$ | $6\,ThC + 9.875\,H_2 + 6\,B + H_3N \longrightarrow 1.25\,Th_4H_{15} + ThB_6H_4C_6N$ | -3.8044 |

| Compound | Chemical reaction | Reaction energy(eV) |
| --- | --- | --- |
| $ThNH_4B_6C_6$ | $0.7857\,ThB_4 + 0.5714\,B_5H_7 + 2\,Th_2C_3 + 2.393\,N_2 \longrightarrow 3.786\,ThN + ThB_6H_4C_6N$ | -3.8038 |
| $ThNH_4B_6C_6$ | $0.2667\,Th_4H_{15} + BN + 2\,Th_2C_3 + 29.4\,B \longrightarrow 4.067\,ThB_6 + ThB_6H_4C_6N$ | -3.7873 |
| $ThNH_4B_6C_6$ | $2\,ThH_2 + 2\,Th_2C_3 + 0.5\,N_2 + 13\,ThB_6 \longrightarrow 18\,ThB_4 + ThB_6H_4C_6N$ | -3.7825 |
| $ThNH_4B_6C_6$ | $1.5\,ThB_4 + 6\,ThC + 14.19\,H_2 + 0.5\,N_2 \longrightarrow 1.625\,Th_4H_{15} + ThB_6H_4C_6N$ | -3.7684 |
| $ThNH_4B_6C_6$ | $0.5714\,B_5H_7 + 2\,Th_2C_3 + 3.143\,B + 2\,N_2 \longrightarrow 3\,ThN + ThB_6H_4C_6N$ | -3.7649 |
| $ThNH_4B_6C_6$ | $0.5\,H_2 + 2\,Th_2C_3 + 24\,B + H_3N \longrightarrow 3\,ThB_6 + ThB_6H_4C_6N$ | -3.7550 |
| $ThNH_4B_6C_6$ | $0.6818\,ThB_4 + 0.3636\,B_9H_{11} + 2\,Th_2C_3 + 2.341\,N_2 \longrightarrow 3.682\,ThN + ThB_6H_4C_6N$ | -3.7364 |
| $ThNH_4B_6C_6$ | $0.5\,ThH_2 + 2\,Th_2C_3 + 20\,B + H_3N \longrightarrow 3.5\,ThB_4 + ThB_6H_4C_6N$ | -3.7194 |
| $ThNH_4B_6C_6$ | $0.25\,Th_3N_4 + 2\,H_2 + 2\,Th_2C_3 + 21\,B \longrightarrow 3.75\,ThB_4 + ThB_6H_4C_6N$ | -3.7095 |
| $ThNH_4B_6C_6$ | $0.3636\,B_9H_{11} + 2\,Th_2C_3 + 2.727\,B + 2\,N_2 \longrightarrow 3\,ThN + ThB_6H_4C_6N$ | -3.7026 |
| $ThNH_4B_6C_6$ | $1.643\,B_5H_7 + 3\,ThC_2 + 1.607\,N_2 \longrightarrow 0.5\,Th_4H_{15} + 2.214\,BN + ThB_6H_4C_6N$ | -3.6639 |
| $ThNH_4B_6C_6$ | $ThC_2 + 2\,B_{13}C_2 + 7.026\,H_3N \longrightarrow 6.026\,BN + 1.553\,B_9H_{11} + ThB_6H_4C_6N$ | -3.6581 |
| $ThNH_4B_6C_6$ | $4.042\,ThC_2 + 0.6667\,B_9H_{11} + 2.021\,N_2 \longrightarrow 0.4167\,H_8C_5 + 3.042\,ThN + ThB_6H_4C_6N$ | -3.6556 |
| $ThNH_4B_6C_6$ | $6.333\,ThH_2 + 0.9722\,Th_3N_4 + 3\,B_{13}C_2 \longrightarrow 8.25\,ThB_4 + 2.889\,H_3N + ThB_6H_4C_6N$ | -3.6450 |
| $ThNH_4B_6C_6$ | $6.5\,ThH_2 + 3\,B_{13}C_2 + 0.5\,N_2 \longrightarrow 4.5\,H_2 + 5.5\,ThB_6 + ThB_6H_4C_6N$ | -3.6429 |
| $ThNH_4B_6C_6$ | $0.5714\,B_5H_7 + 2\,Th_2C_3 + 2.262\,N_2 + 0.5238\,ThB_6 \longrightarrow 3.524\,ThN + ThB_6H_4C_6N$ | -3.6397 |
| $ThNH_4B_6C_6$ | $0.6667\,B_9H_{11} + 2\,Th_2C_3 + 1.907\,N_2 \longrightarrow 0.7037\,Th_3N_4 + 0.2222\,Th_4H_{15} + ThB_6H_4C_6N$ | -3.6161 |
| $ThNH_4B_6C_6$ | $0.5714\,B_5H_7 + 2\,Th_2C_3 + 15.14\,B + 0.5\,N_2 \longrightarrow 3\,ThB_4 + ThB_6H_4C_6N$ | -3.6160 |
| $ThNH_4B_6C_6$ | $1.527\,Th_3N_4 + 0.2667\,Th_4H_{15} + 3\,B_{13}C_2 \longrightarrow 5.109\,BN + 4.648\,ThB_6 + ThB_6H_4C_6N$ | -3.6067 |
| $ThNH_4B_6C_6$ | $0.1429\,B_5H_7 + 2\,Th_2C_3 + 23.29\,B + H_3N \longrightarrow 3\,ThB_6 + ThB_6H_4C_6N$ | -3.6057 |
| $ThNH_4B_6C_6$ | $0.2667\,Th_4H_{15} + 3\,ThC_2 + 30.4\,B + ThN \longrightarrow 4.067\,ThB_6 + ThB_6H_4C_6N$ | -3.6019 |
| $ThNH_4B_6C_6$ | $0.3636\,B_9H_{11} + 2\,Th_2C_3 + 2.227\,N_2 + 0.4545\,ThB_6 \longrightarrow 3.455\,ThN + ThB_6H_4C_6N$ | -3.5940 |
| $ThNH_4B_6C_6$ | $0.09091\,B_9H_{11} + 2\,Th_2C_3 + 23.18\,B + H_3N \longrightarrow 3\,ThB_6 + ThB_6H_4C_6N$ | -3.5901 |
| $ThNH_4B_6C_6$ | $7.2\,B_5H_7 + 6\,ThC + 0.5\,N_2 \longrightarrow 23.2\,H_2 + 5\,ThB_6 + ThB_6H_4C_6N$ | -3.5810 |
| $ThNH_4B_6C_6$ | $6\,ThC + 3.333\,B_9H_{11} + ThN \longrightarrow 6\,ThB_4 + 16.33\,H_2 + ThB_6H_4C_6N$ | -3.5677 |
| $ThNH_4B_6C_6$ | $0.5714\,B_5H_7 + 2\,Th_2C_3 + 19.14\,B + ThN \longrightarrow 4\,ThB_4 + ThB_6H_4C_6N$ | -3.5664 |
| $ThNH_4B_6C_6$ | $6.605\,ThH_2 + 3\,B_{13}C_2 + ThN \longrightarrow 6.605\,ThB_4 + 1.316\,B_5H_7 + ThB_6H_4C_6N$ | -3.5558 |
| $ThNH_4B_6C_6$ | $0.3636\,B_9H_{11} + 2\,Th_2C_3 + 14.73\,B + 0.5\,N_2 \longrightarrow 3\,ThB_4 + ThB_6H_4C_6N$ | -3.5538 |
| $ThNH_4B_6C_6$ | $1.2\,B_5H_7 + 3\,ThC_2 + 2.567\,N_2 \longrightarrow 0.6667\,Th_3N_4 + 1.467\,H_3N + ThB_6H_4C_6N$ | -3.5418 |
| $ThNH_4B_6C_6$ | $2\,ThH_2 + 3\,B_{13}C_2 + 4\,ThN \longrightarrow 3\,BN + 5\,ThB_6 + ThB_6H_4C_6N$ | -3.5275 |
| $ThNH_4B_6C_6$ | $3\,B_{13}C_2 + 1.333\,H_3N + 27\,ThN \longrightarrow 6.833\,Th_3N_4 + 5.5\,ThB_6 + ThB_6H_4C_6N$ | -3.5114 |
| $ThNH_4B_6C_6$ | $0.3636\,B_9H_{11} + 2\,Th_2C_3 + 18.73\,B + ThN \longrightarrow 4\,ThB_4 + ThB_6H_4C_6N$ | -3.5042 |
| $ThNH_4B_6C_6$ | $6\,ThC + 2\,H_2 + BN + 12.5\,ThB_6 \longrightarrow 17.5\,ThB_4 + ThB_6H_4C_6N$ | -3.4876 |
| $ThNH_4B_6C_6$ | $0.6667\,B_9H_{11} + 7.467\,Th + 1.2\,H_8C_5 + 0.5\,N_2 \longrightarrow 6.467\,ThH_2 + ThB_6H_4C_6N$ | -3.4693 |
| $ThNH_4B_6C_6$ | $6.789\,ThH_2 + 0.25\,Th_3N_4 + 3\,B_{13}C_2 \longrightarrow 6.539\,ThB_4 + 1.368\,B_5H_7 + ThB_6H_4C_6N$ | -3.4626 |
| $ThNH_4B_6C_6$ | $2\,ThH_2 + 2.538\,ThC_2 + 0.4615\,B_{13}C_2 + 2.269\,N_2 \longrightarrow 3.538\,ThN + ThB_6H_4C_6N$ | -3.4307 |
| $ThNH_4B_6C_6$ | $2\,ThH_2 + BN + 3\,ThC_2 + 29\,B \longrightarrow 4\,ThB_6 + ThB_6H_4C_6N$ | -3.4197 |
| $ThNH_4B_6C_6$ | $2\,Th_2C_3 + 22.5\,B + 1.333\,H_3N \longrightarrow 0.08333\,Th_3N_4 + 2.75\,ThB_6 + ThB_6H_4C_6N$ | -3.4094 |
| $ThNH_4B_6C_6$ | $6.947\,ThH_2 + 3\,B_{13}C_2 + H_3N \longrightarrow 5.947\,ThB_4 + 1.842\,B_5H_7 + ThB_6H_4C_6N$ | -3.3994 |
| $ThNH_4B_6C_6$ | $3.5\,ThC + 27\,B + 0.5\,H_8C_5 + ThN \longrightarrow 3.5\,ThB_6 + ThB_6H_4C_6N$ | -3.3826 |
| $ThNH_4B_6C_6$ | $16.75\,Th_3N_4 + 1.75\,B_{13}C_2 + 0.5\,H_8C_5 \longrightarrow 16.75\,BN + 49.25\,ThN + ThB_6H_4C_6N$ | -3.3772 |
| $ThNH_4B_6C_6$ | $0.2667\,Th_4H_{15} + 3\,ThC_2 + 24.4\,B + 0.5\,N_2 \longrightarrow 3.067\,ThB_6 + ThB_6H_4C_6N$ | -3.3629 |
| $ThNH_4B_6C_6$ | $6.435\,ThH_2 + 3\,B_{13}C_2 + ThN \longrightarrow 6.435\,ThB_4 + 0.8065\,B_9H_{11} + ThB_6H_4C_6N$ | -3.3612 |
| $ThNH_4B_6C_6$ | $0.2667\,Th_4H_{15} + 0.4615\,B_{13}C_2 + 1.692\,Th_2C_3 + 2.226\,N_2 \longrightarrow 3.451\,ThN + ThB_6H_4C_6N$ | -3.3590 |
| $ThNH_4B_6C_6$ | $6\,B_5H_7 + 6\,ThC + ThN \longrightarrow 6\,ThB_4 + 19\,H_2 + ThB_6H_4C_6N$ | -3.3423 |
| $ThNH_4B_6C_6$ | $6\,ThC + 6\,B + 4.398\,H_3N \longrightarrow 0.8495\,Th_3N_4 + 0.6129\,Th_4H_{15} + ThB_6H_4C_6N$ | -3.3276 |
| $ThNH_4B_6C_6$ | $0.25\,Th_3N_4 + 6\,ThC + 3.222\,B_9H_{11} \longrightarrow 5.75\,ThB_4 + 15.72\,H_2 + ThB_6H_4C_6N$ | -3.3151 |
| $ThNH_4B_6C_6$ | $2\,Th_2C_3 + 24\,B + 1.333\,H_3N \longrightarrow 0.1667\,N_2 + 3\,ThB_6 + ThB_6H_4C_6N$ | -3.3136 |
| $ThNH_4B_6C_6$ | $2\,Th_2C_3 + 18.33\,B + 1.333\,H_3N \longrightarrow 3\,ThB_4 + 0.3333\,BN + ThB_6H_4C_6N$ | -3.3033 |
| $ThNH_4B_6C_6$ | $0.5714\,B_5H_7 + 6\,ThC + BN + 11.07\,ThB_6 \longrightarrow 16.07\,ThB_4 + ThB_6H_4C_6N$ | -3.3027 |
| $ThNH_4B_6C_6$ | $6\,ThC + BN + 0.3636\,B_9H_{11} + 10.86\,ThB_6 \longrightarrow 15.86\,ThB_4 + ThB_6H_4C_6N$ | -3.3004 |
| $ThNH_4B_6C_6$ | $2.75\,B_9H_{11} + 2\,Th_2C_3 + 2\,N_2 \longrightarrow 3.75\,B_5H_7 + 3\,ThN + ThB_6H_4C_6N$ | -3.2940 |
| $ThNH_4B_6C_6$ | $0.3636\,B_9H_{11} + 3\,B_{13}C_2 + 25.18\,ThN \longrightarrow 6.045\,Th_3N_4 + 6.045\,ThB_6 + ThB_6H_4C_6N$ | -3.2776 |
| $ThNH_4B_6C_6$ | $17.34\,B_5H_7 + 6\,ThC + 0.5\,N_2 \longrightarrow 1.25\,Th_4H_{15} + 8.969\,B_9H_{11} + ThB_6H_4C_6N$ | -3.2680 |
| $ThNH_4B_6C_6$ | $6.613\,ThH_2 + 0.25\,Th_3N_4 + 3\,B_{13}C_2 \longrightarrow 6.363\,ThB_4 + 0.8387\,B_9H_{11} + ThB_6H_4C_6N$ | -3.2603 |
| $ThNH_4B_6C_6$ | $2\,H_2 + 3\,ThC_2 + 2.5\,N_2 + ThB_6 \longrightarrow Th_3N_4 + ThB_6H_4C_6N$ | -3.2368 |
| $ThNH_4B_6C_6$ | $2\,Th_2C_3 + 22\,B + 1.333\,H_3N \longrightarrow 2.667\,ThB_6 + 0.3333\,ThN + ThB_6H_4C_6N$ | -3.2340 |
| $ThNH_4B_6C_6$ | $0.5714\,B_5H_7 + 3\,B_{13}C_2 + 24.9\,ThN \longrightarrow 5.976\,Th_3N_4 + 5.976\,ThB_6 + ThB_6H_4C_6N$ | -3.1941 |
| $ThNH_4B_6C_6$ | $6\,ThC + 1.653\,B_9H_{11} + ThN \longrightarrow 2.219\,ThB_4 + 0.9453\,Th_4H_{15} + ThB_6H_4C_6N$ | -3.1867 |
| $ThNH_4B_6C_6$ | $0.9091\,B_9H_{11} + 2\,Th_2C_3 + 1.591\,N_2 \longrightarrow 3\,ThH_2 + 2.182\,BN + ThB_6H_4C_6N$ | -3.1828 |

| Compound | Chemical reaction | Reaction energy(eV) |
|---|---|---|
| $ThNH_4B_6C_6$ | $6\,ThC + BN + 19.37\,B_9H_{11} \longrightarrow 5\,ThB_4 + 29.87\,B_5H_7 + ThB_6H_4C_6N$ | -3.1804 |
| $ThNH_4B_6C_6$ | $3.315\,B_5H_7 + 2.894\,Th_3N_4 + 1.2\,H_8C_5 \longrightarrow 1.92\,Th_4H_{15} + 10.58\,BN + ThB_6H_4C_6N$ | -3.1704 |
| $ThNH_4B_6C_6$ | $2\,ThH_2 + 2\,Th_2C_3 + 15\,ThB_6 + ThN \longrightarrow 21\,ThB_4 + ThB_6H_4C_6N$ | -3.1556 |
| $ThNH_4B_6C_6$ | $0.6667\,B_9H_{11} + 2\,Th_2C_3 + 2\,N_2 \longrightarrow 1.667\,H_2 + 3\,ThN + ThB_6H_4C_6N$ | -3.1531 |
| $ThNH_4B_6C_6$ | $0.2667\,Th_4H_{15} + 3\,B_{13}C_2 + 4.8\,ThN \longrightarrow 3.8\,BN + 4.867\,ThB_6 + ThB_6H_4C_6N$ | -3.1486 |
| $ThNH_4B_6C_6$ | $3.5\,ThC + 21\,B + 0.5\,H_8C_5 + 0.5\,N_2 \longrightarrow 2.5\,ThB_6 + ThB_6H_4C_6N$ | -3.1436 |
| $ThNH_4B_6C_6$ | $5.375\,ThC + H_3N + 0.125\,H_8C_5 + 11.75\,ThB_6 \longrightarrow 16.12\,ThB_4 + ThB_6H_4C_6N$ | -3.1367 |
| $ThNH_4B_6C_6$ | $5.077\,ThC + 9.644\,H_2 + 0.4615\,B_{13}C_2 + 0.5\,N_2 \longrightarrow 1.019\,Th_4H_{15} + ThB_6H_4C_6N$ | -3.1346 |
| $ThNH_4B_6C_6$ | $0.06667\,Th_4H_{15} + 2\,Th_2C_3 + 19.07\,B + H_3N \longrightarrow 3.267\,ThB_4 + ThB_6H_4C_6N$ | -3.1309 |
| $ThNH_4B_6C_6$ | $6.71\,ThH_2 + 3\,B_{13}C_2 + H_3N \longrightarrow 5.71\,ThB_4 + 1.129\,B_9H_{11} + ThB_6H_4C_6N$ | -3.1271 |
| $ThNH_4B_6C_6$ | $6\,ThC + BN + 28.13\,B_9H_{11} \longrightarrow 43.63\,B_5H_7 + 5\,ThB_6 + ThB_6H_4C_6N$ | -3.1251 |
| $ThNH_4B_6C_6$ | $1.2\,B_5H_7 + 2\,Th_2C_3 + 1.718\,N_2 \longrightarrow 0.6089\,Th_3N_4 + 0.2933\,Th_4H_{15} + ThB_6H_4C_6N$ | -3.1221 |
| $ThNH_4B_6C_6$ | $3\,ThC_2 + 0.6667\,B_9H_{11} + 2.389\,N_2 \longrightarrow 0.6667\,Th_3N_4 + 1.111\,H_3N + ThB_6H_4C_6N$ | -3.1161 |
| $ThNH_4B_6C_6$ | $0.5714\,B_5H_7 + 0.25\,Th_3N_4 + 2\,Th_2C_3 + 18.14\,B \longrightarrow 3.75\,ThB_4 + ThB_6H_4C_6N$ | -3.1123 |
| $ThNH_4B_6C_6$ | $1.2\,B_5H_7 + 2\,Th_2C_3 + 2\,N_2 \longrightarrow 2.2\,H_2 + 3\,ThN + ThB_6H_4C_6N$ | -3.1080 |
| $ThNH_4B_6C_6$ | $6\,ThC + 2.889\,B_9H_{11} + H_3N \longrightarrow 5\,ThB_4 + 15.39\,H_2 + ThB_6H_4C_6N$ | -3.0992 |
| $ThNH_4B_6C_6$ | $5.8\,B_5H_7 + 0.25\,Th_3N_4 + 6\,ThC \longrightarrow 5.75\,ThB_4 + 18.3\,H_2 + ThB_6H_4C_6N$ | -3.0972 |
| $ThNH_4B_6C_6$ | $0.25\,Th_3N_4 + 0.2667\,Th_4H_{15} + 3\,ThC_2 + 28.9\,B \longrightarrow 3.817\,ThB_6 + ThB_6H_4C_6N$ | -3.0757 |
| $ThNH_4B_6C_6$ | $2.749\,B_5H_7 + 6\,ThC + ThN \longrightarrow 1.936\,ThB_4 + 1.016\,Th_4H_{15} + ThB_6H_4C_6N$ | -3.0550 |
| $ThNH_4B_6C_6$ | $2\,ThH_2 + 2.762\,ThC + 1.619\,B_{13}C_2 + 0.5\,N_2 \longrightarrow 3.762\,ThB_4 + ThB_6H_4C_6N$ | -3.0528 |
| $ThNH_4B_6C_6$ | $0.25\,Th_3N_4 + 0.3636\,B_9H_{11} + 2\,Th_2C_3 + 17.73\,B \longrightarrow 3.75\,ThB_4 + ThB_6H_4C_6N$ | -3.0501 |
| $ThNH_4B_6C_6$ | $5.635\,ThH_2 + 3\,B_{13}C_2 + 0.5\,N_2 \longrightarrow 1.038\,B_5H_7 + 4.635\,ThB_6 + ThB_6H_4C_6N$ | -3.0413 |
| $ThNH_4B_6C_6$ | $6\,ThC + 15.12\,H_2 + ThB_6 + ThN \longrightarrow 1.75\,Th_4H_{15} + ThB_6H_4C_6N$ | -2.9768 |
| $ThNH_4B_6C_6$ | $6\,ThC + 9.583\,H_2 + 0.6667\,B_9H_{11} + ThN \longrightarrow 1.5\,Th_4H_{15} + ThB_6H_4C_6N$ | -2.9632 |
| $ThNH_4B_6C_6$ | $0.25\,Th_3N_4 + 6\,ThC + 1.605\,B_9H_{11} \longrightarrow 2.11\,ThB_4 + 0.91\,Th_4H_{15} + ThB_6H_4C_6N$ | -2.9484 |
| $ThNH_4B_6C_6$ | $6\,ThC + 1.82\,B_9H_{11} + ThN \longrightarrow 1.068\,Th_4H_{15} + 1.729\,ThB_6 + ThB_6H_4C_6N$ | -2.9397 |
| $ThNH_4B_6C_6$ | $5.536\,ThH_2 + 3\,B_{13}C_2 + 0.5\,N_2 \longrightarrow 0.6429\,B_9H_{11} + 4.536\,ThB_6 + ThB_6H_4C_6N$ | -2.9291 |
| $ThNH_4B_6C_6$ | $1.2\,B_5H_7 + 6\,ThC + 9.05\,H_2 + ThN \longrightarrow 1.5\,Th_4H_{15} + ThB_6H_4C_6N$ | -2.9181 |
| $ThNH_4B_6C_6$ | $5.2\,B_5H_7 + 6\,ThC + H_3N \longrightarrow 5\,ThB_4 + 17.7\,H_2 + ThB_6H_4C_6N$ | -2.9038 |
| $ThNH_4B_6C_6$ | $0.5\,H_2 + 2\,Th_2C_3 + 18\,B + H_3N \longrightarrow 3\,ThB_4 + ThB_6H_4C_6N$ | -2.8891 |
| $ThNH_4B_6C_6$ | $6\,ThC + 4.667\,B_9H_{11} + ThN \longrightarrow 23.67\,H_2 + 6\,ThB_6 + ThB_6H_4C_6N$ | -2.8815 |
| $ThNH_4B_6C_6$ | $6\,ThC + 6\,B + 7.667\,H_3N \longrightarrow 1.667\,Th_3N_4 + 9.5\,H_2 + ThB_6H_4C_6N$ | -2.8688 |
| $ThNH_4B_6C_6$ | $0.25\,Th_3N_4 + 3.5\,ThC + 25.5\,B + 0.5\,H_8C_5 \longrightarrow 3.25\,ThB_6 + ThB_6H_4C_6N$ | -2.8563 |
| $ThNH_4B_6C_6$ | $2\,H_2 + 3\,ThC_2 + 6\,B + 1.833\,N_2 \longrightarrow 0.6667\,Th_3N_4 + ThB_6H_4C_6N$ | -2.8538 |
| $ThNH_4B_6C_6$ | $2\,ThH_2 + 0.25\,Th_3N_4 + 2\,Th_2C_3 + 14.5\,ThB_6 \longrightarrow 20.25\,ThB_4 + ThB_6H_4C_6N$ | -2.8458 |
| $ThNH_4B_6C_6$ | $7.526\,ThH_2 + BN + 3\,B_{13}C_2 \longrightarrow 6.526\,ThB_4 + 1.579\,B_5H_7 + ThB_6H_4C_6N$ | -2.8432 |
| $ThNH_4B_6C_6$ | $2.988\,B_5H_7 + 6\,ThC + ThN \longrightarrow 1.128\,Th_4H_{15} + 1.49\,ThB_6 + ThB_6H_4C_6N$ | -2.8307 |
| $ThNH_4B_6C_6$ | $2.668\,B_5H_7 + 0.25\,Th_3N_4 + 6\,ThC \longrightarrow 1.836\,ThB_4 + 0.9786\,Th_4H_{15} + ThB_6H_4C_6N$ | -2.8205 |
| $ThNH_4B_6C_6$ | $1.5\,ThB_4 + 6\,ThC + 16.06\,H_2 + ThN \longrightarrow 1.875\,Th_4H_{15} + ThB_6H_4C_6N$ | -2.8121 |
| $ThNH_4B_6C_6$ | $2\,H_2 + 3\,B_{13}C_2 + 23\,ThN \longrightarrow 5.5\,Th_3N_4 + 5.5\,ThB_6 + ThB_6H_4C_6N$ | -2.7889 |
| $ThNH_4B_6C_6$ | $0.7857\,ThB_4 + 0.5714\,B_5H_7 + 3\,ThC_2 + 2.357\,N_2 \longrightarrow 0.9286\,Th_3N_4 + ThB_6H_4C_6N$ | -2.7844 |
| $ThNH_4B_6C_6$ | $0.3636\,B_9H_{11} + 0.2098\,B_{13}C_2 + 1.86\,Th_2C_3 + 1.86\,N_2 \longrightarrow 2.72\,ThN + ThB_6H_4C_6N$ | -2.7830 |
| $ThNH_4B_6C_6$ | $6\,ThC + 17.13\,B_9H_{11} + 7.667\,H_3N \longrightarrow 29.63\,B_5H_7 + 1.667\,Th_3N_4 + ThB_6H_4C_6N$ | -2.7729 |
| $ThNH_4B_6C_6$ | $8.533\,ThH_2 + 3\,B_{13}C_2 + 11.04\,H_3N \longrightarrow 6.6\,B_5H_7 + 2.511\,Th_3N_4 + ThB_6H_4C_6N$ | -2.7651 |
| $ThNH_4B_6C_6$ | $0.25\,Th_3N_4 + 6\,ThC + 14.66\,H_2 + ThB_6 \longrightarrow 1.688\,Th_4H_{15} + ThB_6H_4C_6N$ | -2.7494 |
| $ThNH_4B_6C_6$ | $6\,C + 0.3636\,B_9H_{11} + 2.136\,N_2 + ThB_6 \longrightarrow 3.273\,BN + ThB_6H_4C_6N$ | -2.7404 |
| $ThNH_4B_6C_6$ | $6\,ThC + 1.305\,B_9H_{11} + H_3N \longrightarrow 1.437\,ThB_4 + 0.8907\,Th_4H_{15} + ThB_6H_4C_6N$ | -2.7403 |
| $ThNH_4B_6C_6$ | $0.1429\,B_5H_7 + 2\,Th_2C_3 + 17.29\,B + H_3N \longrightarrow 3\,ThB_4 + ThB_6H_4C_6N$ | -2.7398 |
| $ThNH_4B_6C_6$ | $1.2\,B_5H_7 + 5.533\,Th + H_3N + 1.2\,H_8C_5 \longrightarrow 1.133\,Th_4H_{15} + ThB_6H_4C_6N$ | -2.7393 |
| $ThNH_4B_6C_6$ | $0.25\,Th_3N_4 + 6\,ThC + 9.115\,H_2 + 0.6667\,B_9H_{11} \longrightarrow 1.438\,Th_4H_{15} + ThB_6H_4C_6N$ | -2.7358 |
| $ThNH_4B_6C_6$ | $6\,ThC + 6\,B + 3.926\,H_3N \longrightarrow 0.5185\,Th_4H_{15} + 2.926\,ThN + ThB_6H_4C_6N$ | -2.7284 |
| $ThNH_4B_6C_6$ | $0.09091\,B_9H_{11} + 2\,Th_2C_3 + 17.18\,B + H_3N \longrightarrow 3\,ThB_4 + ThB_6H_4C_6N$ | -2.7243 |
| $ThNH_4B_6C_6$ | $0.25\,Th_3N_4 + 6\,ThC + 1.763\,B_9H_{11} \longrightarrow 1.026\,Th_4H_{15} + 1.645\,ThB_6 + ThB_6H_4C_6N$ | -2.7134 |
| $ThNH_4B_6C_6$ | $0.5714\,B_5H_7 + 0.2418\,B_{13}C_2 + 1.839\,Th_2C_3 + 1.839\,N_2 \longrightarrow 2.678\,ThN + ThB_6H_4C_6N$ | -2.7051 |
| $ThNH_4B_6C_6$ | $1.5\,ThB_4 + 0.2667\,Th_4H_{15} + 3\,ThC_2 + 2.783\,N_2 \longrightarrow 4.567\,ThN + ThB_6H_4C_6N$ | -2.7044 |
| $ThNH_4B_6C_6$ | $1.2\,B_5H_7 + 0.25\,Th_3N_4 + 6\,ThC + 8.581\,H_2 \longrightarrow 1.437\,Th_4H_{15} + ThB_6H_4C_6N$ | -2.6907 |
| $ThNH_4B_6C_6$ | $1.5\,ThB_4 + 2\,Th_2C_3 + 1.333\,H_3N + 2.083\,N_2 \longrightarrow 4.5\,ThN + ThB_6H_4C_6N$ | -2.6710 |
| $ThNH_4B_6C_6$ | $0.25\,Th_3N_4 + 6\,ThC + 4.5\,B_9H_{11} \longrightarrow 22.75\,H_2 + 5.75\,ThB_6 + ThB_6H_4C_6N$ | -2.6575 |
| $ThNH_4B_6C_6$ | $0.6818\,ThB_4 + 3\,ThC_2 + 0.3636\,B_9H_{11} + 2.288\,N_2 \longrightarrow 0.8939\,Th_3N_4 + ThB_6H_4C_6N$ | -2.6523 |
| $ThNH_4B_6C_6$ | $0.2667\,Th_4H_{15} + 2.538\,ThC_2 + 0.4615\,B_{13}C_2 + 2.237\,N_2 \longrightarrow 0.8684\,Th_3N_4 + ThB_6H_4C_6N$ | -2.6506 |

| Compound | Chemical reaction | Reaction energy(eV) |
|---|---|---|
| ThNH$_4$B$_6$C$_6$ | 2.171 B$_5$H$_7$ + 6 ThC + H$_3$N $\longrightarrow$ 1.214 ThB$_4$ + 0.9465 Th$_4$H$_{15}$ + ThB$_6$H$_4$C$_6$N | -2.6362 |
| ThNH$_4$B$_6$C$_6$ | 0.2667 Th$_4$H$_{15}$ + 3 ThC$_2$ + 6 B + 2.033 N$_2$ $\longrightarrow$ 3.067 ThN + ThB$_6$H$_4$C$_6$N | -2.6299 |
| ThNH$_4$B$_6$C$_6$ | 1.5 ThB$_4$ + 3.5 ThC + 0.5 H$_8$C$_5$ + 2.5 N$_2$ $\longrightarrow$ 4 ThN + ThB$_6$H$_4$C$_6$N | -2.6204 |
| ThNH$_4$B$_6$C$_6$ | 2 Th$_2$C$_3$ + 17 B + 1.333 H$_3$N $\longrightarrow$ 2.75 ThB$_4$ + 0.08333 Th$_3$N$_4$ + ThB$_6$H$_4$C$_6$N | -2.6157 |
| ThNH$_4$B$_6$C$_6$ | 0.2667 Th$_4$H$_{15}$ + BN + 2 Th$_2$C$_3$ + 21.27 B $\longrightarrow$ 4.067 ThB$_4$ + ThB$_6$H$_4$C$_6$N | -2.6136 |
| ThNH$_4$B$_6$C$_6$ | 7.323 ThH$_2$ + BN + 3 B$_{13}$C$_2$ $\longrightarrow$ 6.323 ThB$_4$ + 0.9677 B$_9$H$_{11}$ + ThB$_6$H$_4$C$_6$N | -2.6097 |
| ThNH$_4$B$_6$C$_6$ | 6 ThC + 11.75 H$_2$ + H$_3$N + ThB$_6$ $\longrightarrow$ 1.5 Th$_4$H$_{15}$ + ThB$_6$H$_4$C$_6$N | -2.6091 |
| ThNH$_4$B$_6$C$_6$ | 2.895 B$_5$H$_7$ + 0.25 Th$_3$N$_4$ + 6 ThC $\longrightarrow$ 1.084 Th$_4$H$_{15}$ + 1.413 ThB$_6$ + ThB$_6$H$_4$C$_6$N | -2.6078 |
| ThNH$_4$B$_6$C$_6$ | 2 Th$_2$C$_3$ + 6 B + 1.333 H$_3$N + 1.333 N$_2$ $\longrightarrow$ 3 ThN + ThB$_6$H$_4$C$_6$N | -2.5966 |
| ThNH$_4$B$_6$C$_6$ | 6 ThC + 6.208 H$_2$ + 0.6667 B$_9$H$_{11}$ + H$_3$N $\longrightarrow$ 1.25 Th$_4$H$_{15}$ + ThB$_6$H$_4$C$_6$N | -2.5955 |
| ThNH$_4$B$_6$C$_6$ | 1.5 ThB$_4$ + 0.25 Th$_3$N$_4$ + 6 ThC + 15.59 H$_2$ $\longrightarrow$ 1.812 Th$_4$H$_{15}$ + ThB$_6$H$_4$C$_6$N | -2.5847 |
| ThNH$_4$B$_6$C$_6$ | 6 ThC + 1.414 B$_9$H$_{11}$ + H$_3$N $\longrightarrow$ 0.9699 Th$_4$H$_{15}$ + 1.12 ThB$_6$ + ThB$_6$H$_4$C$_6$N | -2.5802 |
| ThNH$_4$B$_6$C$_6$ | 8.4 B$_5$H$_7$ + 6 ThC + ThN $\longrightarrow$ 27.4 H$_2$ + 6 ThB$_6$ + ThB$_6$H$_4$C$_6$N | -2.5659 |
| ThNH$_4$B$_6$C$_6$ | 6 ThC + 11.38 H$_2$ + BN + 5 B $\longrightarrow$ 1.25 Th$_4$H$_{15}$ + ThB$_6$H$_4$C$_6$N | -2.5617 |
| ThNH$_4$B$_6$C$_6$ | 3 ThC$_2$ + 6.158 H$_3$N + 1.86 ThB$_6$ $\longrightarrow$ 0.9649 Th$_4$H$_{15}$ + 5.158 BN + ThB$_6$H$_4$C$_6$N | -2.5599 |
| ThNH$_4$B$_6$C$_6$ | 6 ThC + 5.875 H$_3$N + 1.813 ThB$_6$ $\longrightarrow$ 6.813 ThH$_2$ + 4.875 BN + ThB$_6$H$_4$C$_6$N | -2.5597 |
| ThNH$_4$B$_6$C$_6$ | 1.2 B$_5$H$_7$ + 6 ThC + 5.675 H$_2$ + H$_3$N $\longrightarrow$ 1.25 Th$_4$H$_{15}$ + ThB$_6$H$_4$C$_6$N | -2.5504 |
| ThNH$_4$B$_6$C$_6$ | 3.5 ThC + 6 B + 0.5 H$_8$C$_5$ + 1.75 N$_2$ $\longrightarrow$ 2.5 ThN + ThB$_6$H$_4$C$_6$N | -2.5460 |
| ThNH$_4$B$_6$C$_6$ | 6 ThC + 4 B$_9$H$_{11}$ + H$_3$N $\longrightarrow$ 21.5 H$_2$ + 5 ThB$_6$ + ThB$_6$H$_4$C$_6$N | -2.5274 |
| ThNH$_4$B$_6$C$_6$ | 2 H$_2$ + BN + 2 Th$_2$C$_3$ + 23 B $\longrightarrow$ 3 ThB$_6$ + ThB$_6$H$_4$C$_6$N | -2.5123 |
| ThNH$_4$B$_6$C$_6$ | 2.321 B$_5$H$_7$ + 6 ThC + H$_3$N $\longrightarrow$ 1.016 Th$_4$H$_{15}$ + 0.9342 ThB$_6$ + ThB$_6$H$_4$C$_6$N | -2.4956 |
| ThNH$_4$B$_6$C$_6$ | 6 C + 3.7 Th$_3$N$_4$ + 2.2 B$_9$H$_{11}$ $\longrightarrow$ 10.1 ThH$_2$ + 13.8 BN + ThB$_6$H$_4$C$_6$N | -2.4926 |
| ThNH$_4$B$_6$C$_6$ | 0.2667 Th$_4$H$_{15}$ + 3 ThC$_2$ + 18.27 B + 0.5 N$_2$ $\longrightarrow$ 3.067 ThB$_4$ + ThB$_6$H$_4$C$_6$N | -2.4778 |
| ThNH$_4$B$_6$C$_6$ | 2 ThH$_2$ + 3 B$_{13}$C$_2$ + 15 ThN $\longrightarrow$ 3.5 Th$_3$N$_4$ + 5.5 ThB$_6$ + ThB$_6$H$_4$C$_6$N | -2.4775 |
| ThNH$_4$B$_6$C$_6$ | 2 Th$_2$C$_3$ + 16.67 B + 1.333 H$_3$N $\longrightarrow$ 2.667 ThB$_4$ + 0.3333 ThN + ThB$_6$H$_4$C$_6$N | -2.4643 |
| ThNH$_4$B$_6$C$_6$ | 0.5714 B$_5$H$_7$ + 3 ThC$_2$ + 2.183 N$_2$ + 0.5238 ThB$_6$ $\longrightarrow$ 0.8413 Th$_3$N$_4$ + ThB$_6$H$_4$C$_6$N | -2.4573 |
| ThNH$_4$B$_6$C$_6$ | 2 Th$_2$C$_3$ + 18 B + 1.333 H$_3$N $\longrightarrow$ 3 ThB$_4$ + 0.1667 N$_2$ + ThB$_6$H$_4$C$_6$N | -2.4478 |
| ThNH$_4$B$_6$C$_6$ | 1.5 ThB$_4$ + 6 ThC + 12.69 H$_2$ + H$_3$N $\longrightarrow$ 1.625 Th$_4$H$_{15}$ + ThB$_6$H$_4$C$_6$N | -2.4444 |
| ThNH$_4$B$_6$C$_6$ | 0.2667 Th$_4$H$_{15}$ + 3 ThC$_2$ + 22.27 B + ThN $\longrightarrow$ 4.067 ThB$_4$ + ThB$_6$H$_4$C$_6$N | -2.4282 |
| ThNH$_4$B$_6$C$_6$ | 3.5 ThC + 16 B + 0.5 H$_8$C$_5$ + 0.5 N$_2$ $\longrightarrow$ 2.5 ThB$_4$ + ThB$_6$H$_4$C$_6$N | -2.4220 |
| ThNH$_4$B$_6$C$_6$ | 1.792 Th$_2$C$_3$ + 21.5 B + H$_3$N + 0.125 H$_8$C$_5$ $\longrightarrow$ 2.583 ThB$_6$ + ThB$_6$H$_4$C$_6$N | -2.3923 |
| ThNH$_4$B$_6$C$_6$ | 0.2667 Th$_4$H$_{15}$ + 3 ThC$_2$ + 2.533 N$_2$ + ThB$_6$ $\longrightarrow$ 4.067 ThN + ThB$_6$H$_4$C$_6$N | -2.3909 |
| ThNH$_4$B$_6$C$_6$ | 3.5 ThC + 20 B + 0.5 H$_8$C$_5$ + ThN $\longrightarrow$ 3.5 ThB$_4$ + ThB$_6$H$_4$C$_6$N | -2.3724 |
| ThNH$_4$B$_6$C$_6$ | 3 ThC$_2$ + 0.3636 B$_9$H$_{11}$ + 2.136 N$_2$ + 0.4545 ThB$_6$ $\longrightarrow$ 0.8182 Th$_3$N$_4$ + ThB$_6$H$_4$C$_6$N | -2.3685 |
| ThNH$_4$B$_6$C$_6$ | 3.835 ThC + 0.3636 B$_9$H$_{11}$ + 1.082 B$_{13}$C$_2$ + 0.5 N$_2$ $\longrightarrow$ 2.835 ThB$_4$ + ThB$_6$H$_4$C$_6$N | -2.3645 |
| ThNH$_4$B$_6$C$_6$ | 2 Th$_2$C$_3$ + 1.333 H$_3$N + 1.833 N$_2$ + ThB$_6$ $\longrightarrow$ 4 ThN + ThB$_6$H$_4$C$_6$N | -2.3576 |
| ThNH$_4$B$_6$C$_6$ | 8.1 B$_5$H$_7$ + 0.25 Th$_3$N$_4$ + 6 ThC $\longrightarrow$ 26.35 H$_2$ + 5.75 ThB$_6$ + ThB$_6$H$_4$C$_6$N | -2.3532 |
| ThNH$_4$B$_6$C$_6$ | 2 H$_2$ + 0.4615 B$_{13}$C$_2$ + 1.692 Th$_2$C$_3$ + 1.692 N$_2$ $\longrightarrow$ 2.385 ThN + ThB$_6$H$_4$C$_6$N | -2.3389 |
| ThNH$_4$B$_6$C$_6$ | 2 H$_2$ + 3 ThC$_2$ + 24 B + ThN $\longrightarrow$ 3 ThB$_6$ + ThB$_6$H$_4$C$_6$N | -2.3269 |
| ThNH$_4$B$_6$C$_6$ | 6.5 ThH$_2$ + 3 B$_{13}$C$_2$ + H$_3$N $\longrightarrow$ 6 H$_2$ + 5.5 ThB$_6$ + ThB$_6$H$_4$C$_6$N | -2.3189 |
| ThNH$_4$B$_6$C$_6$ | 1.2 B$_5$H$_7$ + 4.733 Th + 1.2 H$_8$C$_5$ + 0.5 N$_2$ $\longrightarrow$ 0.9333 Th$_4$H$_{15}$ + ThB$_6$H$_4$C$_6$N | -2.3104 |
| ThNH$_4$B$_6$C$_6$ | 3.5 ThC + 0.5 H$_8$C$_5$ + 2.25 N$_2$ + ThB$_6$ $\longrightarrow$ 3.5 ThN + ThB$_6$H$_4$C$_6$N | -2.3070 |
| ThNH$_4$B$_6$C$_6$ | 0.5714 B$_5$H$_7$ + 3.796 ThC + 1.102 B$_{13}$C$_2$ + 0.5 N$_2$ $\longrightarrow$ 2.796 ThB$_4$ + ThB$_6$H$_4$C$_6$N | -2.3054 |
| ThNH$_4$B$_6$C$_6$ | 0.6667 B$_9$H$_{11}$ + 2 Th$_2$C$_3$ + 1.556 N$_2$ $\longrightarrow$ 0.2222 Th$_4$H$_{15}$ + 2.111 ThN + ThB$_6$H$_4$C$_6$N | -2.3030 |
| ThNH$_4$B$_6$C$_6$ | 1.2 B$_5$H$_7$ + 3 ThC$_2$ + 2.233 N$_2$ $\longrightarrow$ 1.467 H$_3$N + 2 ThN + ThB$_6$H$_4$C$_6$N | -2.2978 |
| ThNH$_4$B$_6$C$_6$ | 6 ThC + 0.6667 B$_9$H$_{11}$ + 3.136 H$_3$N $\longrightarrow$ 0.5341 Th$_3$N$_4$ + 0.8495 Th$_4$H$_{15}$ + ThB$_6$H$_4$C$_6$N | -2.2957 |
| ThNH$_4$B$_6$C$_6$ | 2 ThH$_2$ + 6 ThC + 6 BN + 2.167 N$_2$ $\longrightarrow$ 2.333 Th$_3$N$_4$ + ThB$_6$H$_4$C$_6$N | -2.2840 |
| ThNH$_4$B$_6$C$_6$ | 1.2 B$_5$H$_7$ + 6 ThC + 2.953 H$_3$N $\longrightarrow$ 0.4882 Th$_3$N$_4$ + 0.8839 Th$_4$H$_{15}$ + ThB$_6$H$_4$C$_6$N | -2.2764 |
| ThNH$_4$B$_6$C$_6$ | 2 ThH$_2$ + BN + 3 ThC$_2$ + 21 B $\longrightarrow$ 4 ThB$_4$ + ThB$_6$H$_4$C$_6$N | -2.2652 |
| ThNH$_4$B$_6$C$_6$ | 7.2 B$_5$H$_7$ + 6 ThC + H$_3$N $\longrightarrow$ 24.7 H$_2$ + 5 ThB$_6$ + ThB$_6$H$_4$C$_6$N | -2.2569 |
| ThNH$_4$B$_6$C$_6$ | 0.5714 B$_5$H$_7$ + 3 ThC$_2$ + 3.143 B + 1.833 N$_2$ $\longrightarrow$ 0.6667 Th$_3$N$_4$ + ThB$_6$H$_4$C$_6$N | -2.2567 |
| ThNH$_4$B$_6$C$_6$ | 6 ThC + 2.727 BN + 0.3636 B$_9$H$_{11}$ + 1.636 N$_2$ $\longrightarrow$ 5 ThN + ThB$_6$H$_4$C$_6$N | -2.2504 |
| ThNH$_4$B$_6$C$_6$ | BN + 1.75 B$_{13}$C$_2$ + 3.958 Th + 0.5 H$_8$C$_5$ $\longrightarrow$ 2.958 ThB$_6$ + ThB$_6$H$_4$C$_6$N | -2.2379 |
| ThNH$_4$B$_6$C$_6$ | 3 ThC$_2$ + 0.3636 B$_9$H$_{11}$ + 2.727 B + 1.833 N$_2$ $\longrightarrow$ 0.6667 Th$_3$N$_4$ + ThB$_6$H$_4$C$_6$N | -2.1944 |
| ThNH$_4$B$_6$C$_6$ | 5.077 ThC + 11.52 H$_2$ + 0.4615 B$_{13}$C$_2$ + ThN $\longrightarrow$ 1.269 Th$_4$H$_{15}$ + ThB$_6$H$_4$C$_6$N | -2.1782 |
| ThNH$_4$B$_6$C$_6$ | 2.9 ThB$_4$ + 6 ThC + 6.6 H$_3$N $\longrightarrow$ 7.9 ThH$_2$ + 5.6 BN + ThB$_6$H$_4$C$_6$N | -2.1779 |
| ThNH$_4$B$_6$C$_6$ | 0.6667 B$_9$H$_{11}$ + 5.249 Th + H$_3$N + 1.2 H$_8$C$_5$ $\longrightarrow$ 1.062 Th$_4$H$_{15}$ + ThB$_6$H$_4$C$_6$N | -2.1611 |
| ThNH$_4$B$_6$C$_6$ | 21.56 B$_5$H$_7$ + 6 ThC + ThN $\longrightarrow$ 1.5 Th$_4$H$_{15}$ + 11.31 B$_9$H$_{11}$ + ThB$_6$H$_4$C$_6$N | -2.1531 |
| ThNH$_4$B$_6$C$_6$ | 2 ThH$_2$ + 2.16 ThC + 1.92 B$_{13}$C$_2$ + 0.5 N$_2$ $\longrightarrow$ 3.16 ThB$_6$ + ThB$_6$H$_4$C$_6$N | -2.1189 |
| ThNH$_4$B$_6$C$_6$ | 1.5 ThB$_4$ + 3 ThC$_2$ + 1.333 H$_3$N + 2.167 N$_2$ $\longrightarrow$ 1.167 Th$_3$N$_4$ + ThB$_6$H$_4$C$_6$N | -2.0958 |

| Compound | Chemical reaction | Reaction energy(eV) |
|---|---|---|
| $ThNH_4B_6C_6$ | $2\,H_2 + 3\,ThC_2 + 18\,B + 0.5\,N_2 \longrightarrow 2\,ThB_6 + ThB_6H_4C_6N$ | -2.0879 |
| $ThNH_4B_6C_6$ | $13.97\,B_5H_7 + 6\,ThC + H_3N \longrightarrow 1.25\,Th_4H_{15} + 7.094\,B_9H_{11} + ThB_6H_4C_6N$ | -2.0707 |
| $ThNH_4B_6C_6$ | $3.524\,ThC + 2\,H_2 + 1.238\,B_{13}C_2 + 0.5\,N_2 \longrightarrow 2.524\,ThB_4 + ThB_6H_4C_6N$ | -2.0681 |
| $ThNH_4B_6C_6$ | $6\,ThC + BN + 2.778\,B_9H_{11} \longrightarrow 5\,ThB_4 + 13.28\,H_2 + ThB_6H_4C_6N$ | -2.0580 |
| $ThNH_4B_6C_6$ | $0.4615\,B_{13}C_2 + 1.692\,Th_2C_3 + 1.333\,H_3N + 1.423\,N_2 \longrightarrow 0.7949\,Th_3N_4 + ThB_6H_4C_6N$ | -2.0566 |
| $ThNH_4B_6C_6$ | $6\,ThC + 5.043\,H_3N + ThB_6 \longrightarrow 1.011\,Th_3N_4 + 0.7419\,Th_4H_{15} + ThB_6H_4C_6N$ | -2.0417 |
| $ThNH_4B_6C_6$ | $10.04\,ThH_2 + 3\,B_{13}C_2 + 10.04\,H_3N \longrightarrow 6.6\,B_5H_7 + 9.04\,ThN + ThB_6H_4C_6N$ | -1.9863 |
| $ThNH_4B_6C_6$ | $1.2\,B_5H_7 + 2\,Th_2C_3 + 1.413\,N_2 \longrightarrow 0.2933\,Th_4H_{15} + 1.827\,ThN + ThB_6H_4C_6N$ | -1.9859 |
| $ThNH_4B_6C_6$ | $0.25\,Th_3N_4 + 0.2667\,Th_4H_{15} + 3\,ThC_2 + 21.27\,B \longrightarrow 3.817\,ThB_4 + ThB_6H_4C_6N$ | -1.9741 |
| $ThNH_4B_6C_6$ | $0.2667\,Th_4H_{15} + 2\,Th_2C_3 + 0.5\,N_2 + 11.13\,ThB_6 \longrightarrow 15.2\,ThB_4 + ThB_6H_4C_6N$ | -1.9670 |
| $ThNH_4B_6C_6$ | $6\,ThC + 6\,B + 6\,H_3N \longrightarrow 7\,H_2 + 5\,ThN + ThB_6H_4C_6N$ | -1.9657 |
| $ThNH_4B_6C_6$ | $20.51\,B_5H_7 + 0.25\,Th_3N_4 + 6\,ThC \longrightarrow 1.437\,Th_4H_{15} + 10.73\,B_9H_{11} + ThB_6H_4C_6N$ | -1.9654 |
| $ThNH_4B_6C_6$ | $7.556\,ThH_2 + 3\,B_{13}C_2 + 9.741\,H_3N \longrightarrow 2.185\,Th_3N_4 + 3.667\,B_9H_{11} + ThB_6H_4C_6N$ | -1.9629 |
| $ThNH_4B_6C_6$ | $0.25\,Th_3N_4 + 5.077\,ThC + 11.05\,H_2 + 0.4615\,B_{13}C_2 \longrightarrow 1.207\,Th_4H_{15} + ThB_6H_4C_6N$ | -1.9508 |
| $ThNH_4B_6C_6$ | $5.5\,ThH_2 + 3\,B_{13}C_2 + ThN \longrightarrow 3.5\,H_2 + 5.5\,ThB_6 + ThB_6H_4C_6N$ | -1.9327 |
| $ThNH_4B_6C_6$ | $1.2\,B_5H_7 + 6\,ThC + 2.681\,H_3N \longrightarrow 0.8296\,Th_4H_{15} + 1.681\,ThN + ThB_6H_4C_6N$ | -1.9320 |
| $ThNH_4B_6C_6$ | $2.514\,B_9H_{11} + 1.2\,H_8C_5 + 17.63\,ThN \longrightarrow 16.63\,ThH_2 + 16.63\,BN + ThB_6H_4C_6N$ | -1.9190 |
| $ThNH_4B_6C_6$ | $6\,ThC + 0.6667\,B_9H_{11} + 2.84\,H_3N \longrightarrow 0.7901\,Th_4H_{15} + 1.84\,ThN + ThB_6H_4C_6N$ | -1.9190 |
| $ThNH_4B_6C_6$ | $0.25\,Th_3N_4 + 3.5\,ThC + 19\,B + 0.5\,H_8C_5 \longrightarrow 3.25\,ThB_4 + ThB_6H_4C_6N$ | -1.9183 |
| $ThNH_4B_6C_6$ | $0.5714\,B_5H_7 + BN + 2\,Th_2C_3 + 20.14\,B \longrightarrow 3\,ThB_6 + ThB_6H_4C_6N$ | -1.9151 |
| $ThNH_4B_6C_6$ | $3\,ThC_2 + 8.833\,B + 3.833\,H_3N \longrightarrow 0.5\,Th_4H_{15} + 2.833\,BN + ThB_6H_4C_6N$ | -1.8941 |
| $ThNH_4B_6C_6$ | $5.75\,ThH_2 + 0.25\,Th_3N_4 + 3\,B_{13}C_2 \longrightarrow 3.75\,H_2 + 5.5\,ThB_6 + ThB_6H_4C_6N$ | -1.8938 |
| $ThNH_4B_6C_6$ | $3\,ThC_2 + 0.6667\,B_9H_{11} + 2.056\,N_2 \longrightarrow 1.111\,H_3N + 2\,ThN + ThB_6H_4C_6N$ | -1.8721 |
| $ThNH_4B_6C_6$ | $5\,B_5H_7 + 6\,ThC + BN \longrightarrow 5\,ThB_4 + 15.5\,H_2 + ThB_6H_4C_6N$ | -1.8702 |
| $ThNH_4B_6C_6$ | $BN + 0.3636\,B_9H_{11} + 2\,Th_2C_3 + 19.73\,B \longrightarrow 3\,ThB_6 + ThB_6H_4C_6N$ | -1.8529 |
| $ThNH_4B_6C_6$ | $2\,ThH_2 + 2.381\,ThC + 1.81\,B_{13}C_2 + ThN \longrightarrow 4.381\,ThB_4 + ThB_6H_4C_6N$ | -1.8349 |
| $ThNH_4B_6C_6$ | $1.5\,ThB_4 + 6\,ThC + 5.366\,H_3N \longrightarrow 1.091\,Th_3N_4 + 0.8065\,Th_4H_{15} + ThB_6H_4C_6N$ | -1.8317 |
| $ThNH_4B_6C_6$ | $0.2667\,Th_4H_{15} + 3\,B_{13}C_2 + 18.73\,ThN \longrightarrow 4.433\,Th_3N_4 + 5.5\,ThB_6 + ThB_6H_4C_6N$ | -1.8186 |
| $ThNH_4B_6C_6$ | $5.077\,ThC + 8.144\,H_2 + 0.4615\,B_{13}C_2 + H_3N \longrightarrow 1.019\,Th_4H_{15} + ThB_6H_4C_6N$ | -1.8105 |
| $ThNH_4B_6C_6$ | $0.25\,Th_3N_4 + 2\,H_2 + 3\,ThC_2 + 22.5\,B \longrightarrow 2.75\,ThB_6 + ThB_6H_4C_6N$ | -1.8007 |
| $ThNH_4B_6C_6$ | $3.117\,ThC + 0.2667\,Th_4H_{15} + 1.441\,B_{13}C_2 + 0.5\,N_2 \longrightarrow 3.184\,ThB_4 + ThB_6H_4C_6N$ | -1.7890 |
| $ThNH_4B_6C_6$ | $3\,ThC_2 + 2.75\,B_9H_{11} + 1.833\,N_2 \longrightarrow 3.75\,B_5H_7 + 0.6667\,Th_3N_4 + ThB_6H_4C_6N$ | -1.7858 |
| $ThNH_4B_6C_6$ | $6\,ThC + BN + 1.412\,B_9H_{11} \longrightarrow 1.926\,ThB_4 + 0.7685\,Th_4H_{15} + ThB_6H_4C_6N$ | -1.7483 |
| $ThNH_4B_6C_6$ | $0.5\,ThH_2 + 3\,ThC_2 + 21\,B + H_3N \longrightarrow 2.5\,ThB_6 + ThB_6H_4C_6N$ | -1.7385 |
| $ThNH_4B_6C_6$ | $0.5714\,B_5H_7 + 6\,C + 1.929\,N_2 + ThB_6 \longrightarrow 2.857\,BN + ThB_6H_4C_6N$ | -1.7360 |
| $ThNH_4B_6C_6$ | $0.6667\,B_9H_{11} + 4.449\,Th + 1.2\,H_8C_5 + 0.5\,N_2 \longrightarrow 0.8622\,Th_4H_{15} + ThB_6H_4C_6N$ | -1.7322 |
| $ThNH_4B_6C_6$ | $3.25\,B_5H_7 + 6\,ThC + 0.5\,N_2 \longrightarrow 1.25\,Th_4H_{15} + 10.25\,B + ThB_6H_4C_6N$ | -1.7321 |
| $ThNH_4B_6C_6$ | $0.5714\,B_5H_7 + 3\,ThC_2 + 21.14\,B + ThN \longrightarrow 3\,ThB_6 + ThB_6H_4C_6N$ | -1.7297 |
| $ThNH_4B_6C_6$ | $14.38\,H_2 + 3\,B_{13}C_2 + 9.25\,ThN \longrightarrow 8.25\,ThB_4 + 8.25\,H_3N + ThB_6H_4C_6N$ | -1.7259 |
| $ThNH_4B_6C_6$ | $1.5\,ThB_4 + 2\,H_2 + 3\,ThC_2 + 2.25\,N_2 \longrightarrow 3.5\,ThN + ThB_6H_4C_6N$ | -1.6843 |
| $ThNH_4B_6C_6$ | $0.6875\,B_{13}C_2 + 1.542\,Th_2C_3 + 3.937\,H_3N \longrightarrow 0.5208\,Th_4H_{15} + 2.937\,BN + ThB_6H_4C_6N$ | -1.6824 |
| $ThNH_4B_6C_6$ | $2\,ThH_2 + 0.25\,Th_3N_4 + 2.476\,ThC + 1.762\,B_{13}C_2 \longrightarrow 4.226\,ThB_4 + ThB_6H_4C_6N$ | -1.6729 |
| $ThNH_4B_6C_6$ | $3\,ThC_2 + 0.3636\,B_9H_{11} + 20.73\,B + ThN \longrightarrow 3\,ThB_6 + ThB_6H_4C_6N$ | -1.6675 |
| $ThNH_4B_6C_6$ | $3.382\,ThC + 0.3636\,B_9H_{11} + 1.309\,B_{13}C_2 + 0.5\,N_2 \longrightarrow 2.382\,ThB_6 + ThB_6H_4C_6N$ | -1.6606 |
| $ThNH_4B_6C_6$ | $6\,ThC + 0.6667\,B_9H_{11} + 7.667\,H_3N \longrightarrow 1.667\,Th_3N_4 + 13.17\,H_2 + ThB_6H_4C_6N$ | -1.6599 |
| $ThNH_4B_6C_6$ | $6\,ThC + 14\,B_9H_{11} + 6\,H_3N \longrightarrow 24\,B_5H_7 + 5\,ThN + ThB_6H_4C_6N$ | -1.6584 |
| $ThNH_4B_6C_6$ | $2\,ThH_2 + 3\,ThC_2 + 0.5\,N_2 + 11\,ThB_6 \longrightarrow 15\,ThB_4 + ThB_6H_4C_6N$ | -1.6571 |
| $ThNH_4B_6C_6$ | $1.792\,Th_2C_3 + 16.33\,B + H_3N + 0.125\,H_8C_5 \longrightarrow 2.583\,ThB_4 + ThB_6H_4C_6N$ | -1.6467 |
| $ThNH_4B_6C_6$ | $2\,H_2 + BN + 2\,Th_2C_3 + 17\,B \longrightarrow 3\,ThB_4 + ThB_6H_4C_6N$ | -1.6464 |
| $ThNH_4B_6C_6$ | $3\,ThC_2 + 0.6667\,B_9H_{11} + 1.833\,N_2 \longrightarrow 0.6667\,Th_3N_4 + 1.667\,H_2 + ThB_6H_4C_6N$ | -1.6449 |
| $ThNH_4B_6C_6$ | $2.348\,B_5H_7 + 6\,ThC + BN \longrightarrow 1.684\,ThB_4 + 0.8289\,Th_4H_{15} + ThB_6H_4C_6N$ | -1.6358 |
| $ThNH_4B_6C_6$ | $2\,H_2 + 2\,Th_2C_3 + 0.5\,N_2 + 9\,ThB_6 \longrightarrow 12\,ThB_4 + ThB_6H_4C_6N$ | -1.6156 |
| $ThNH_4B_6C_6$ | $1.2\,B_5H_7 + 6\,ThC + 7.667\,H_3N \longrightarrow 1.667\,Th_3N_4 + 13.7\,H_2 + ThB_6H_4C_6N$ | -1.6148 |
| $ThNH_4B_6C_6$ | $0.5714\,B_5H_7 + 3.349\,ThC + 1.326\,B_{13}C_2 + 0.5\,N_2 \longrightarrow 2.349\,ThB_6 + ThB_6H_4C_6N$ | -1.6113 |
| $ThNH_4B_6C_6$ | $2\,H_2 + 3\,ThC_2 + 6\,B + 1.5\,N_2 \longrightarrow 2\,ThN + ThB_6H_4C_6N$ | -1.6099 |
| $ThNH_4B_6C_6$ | $1.2\,B_5H_7 + 3\,ThC_2 + 1.833\,N_2 \longrightarrow 0.6667\,Th_3N_4 + 2.2\,H_2 + ThB_6H_4C_6N$ | -1.5998 |
| $ThNH_4B_6C_6$ | $4.651\,ThC + 0.6744\,B_{13}C_2 + 3.767\,H_3N \longrightarrow 3.651\,ThH_2 + 2.767\,BN + ThB_6H_4C_6N$ | -1.5771 |
| $ThNH_4B_6C_6$ | $6\,ThC + 12.94\,H_2 + BN + 0.8333\,ThB_6 \longrightarrow 1.458\,Th_4H_{15} + ThB_6H_4C_6N$ | -1.5656 |
| $ThNH_4B_6C_6$ | $6\,ThC + 8.319\,H_2 + BN + 0.5556\,B_9H_{11} \longrightarrow 1.25\,Th_4H_{15} + ThB_6H_4C_6N$ | -1.5543 |
| $ThNH_4B_6C_6$ | $6\,ThC + BN + 1.556\,B_9H_{11} \longrightarrow 0.8747\,Th_4H_{15} + 1.501\,ThB_6 + ThB_6H_4C_6N$ | -1.5338 |

| Compound | Chemical reaction | Reaction energy(eV) |
|---|---|---|
| $ThNH_4B_6C_6$ | $B_5H_7 + 6\,ThC + 7.875\,H_2 + BN \longrightarrow 1.25\,Th_4H_{15} + ThB_6H_4C_6N$ | -1.5167 |
| $ThNH_4B_6C_6$ | $5.346\,ThH_2 + 3\,B_{13}C_2 + H_3N \longrightarrow 1.385\,B_5H_7 + 4.346\,ThB_6 + ThB_6H_4C_6N$ | -1.5167 |
| $ThNH_4B_6C_6$ | $2\,H_2 + 3\,ThC_2 + 14\,B + 0.5\,N_2 \longrightarrow 2\,ThB_4 + ThB_6H_4C_6N$ | -1.5107 |
| $ThNH_4B_6C_6$ | $7.625\,H_2 + 2\,Th_2C_3 + 6\,B + 0.5\,N_2 \longrightarrow 0.75\,Th_4H_{15} + ThB_6H_4C_6N$ | -1.4930 |
| $ThNH_4B_6C_6$ | $0.5714\,B_5H_7 + 3\,ThC_2 + 15.14\,B + 0.5\,N_2 \longrightarrow 2\,ThB_6 + ThB_6H_4C_6N$ | -1.4907 |
| $ThNH_4B_6C_6$ | $6\,ThC + 9\,H_3N + ThB_6 \longrightarrow 2\,Th_3N_4 + 11.5\,H_2 + ThB_6H_4C_6N$ | -1.4864 |
| $ThNH_4B_6C_6$ | $6\,ThC + BN + 3.889\,B_9H_{11} \longrightarrow 19.39\,H_2 + 5\,ThB_6 + ThB_6H_4C_6N$ | -1.4862 |
| $ThNH_4B_6C_6$ | $3.154\,B_9H_{11} + 3\,B_{13}C_2 + 11.23\,ThN \longrightarrow 10.23\,H_3N + 10.23\,ThB_6 + ThB_6H_4C_6N$ | -1.4839 |
| $ThNH_4B_6C_6$ | $3\,ThC_2 + 1.333\,H_3N + 1.833\,N_2 + ThB_6 \longrightarrow Th_3N_4 + ThB_6H_4C_6N$ | -1.4714 |
| $ThNH_4B_6C_6$ | $4.827\,ThH_2 + 3\,B_{13}C_2 + ThN \longrightarrow 0.8077\,B_5H_7 + 4.827\,ThB_6 + ThB_6H_4C_6N$ | -1.4647 |
| $ThNH_4B_6C_6$ | $2\,H_2 + 3\,ThC_2 + 18\,B + ThN \longrightarrow 3\,ThB_4 + ThB_6H_4C_6N$ | -1.4610 |
| $ThNH_4B_6C_6$ | $3.12\,ThC + 2\,H_2 + 1.44\,B_{13}C_2 + 0.5\,N_2 \longrightarrow 2.12\,ThB_6 + ThB_6H_4C_6N$ | -1.4415 |
| $ThNH_4B_6C_6$ | $2.556\,B_5H_7 + 6\,ThC + BN \longrightarrow 0.9259\,Th_4H_{15} + 1.296\,ThB_6 + ThB_6H_4C_6N$ | -1.4406 |
| $ThNH_4B_6C_6$ | $0.5714\,B_5H_7 + 2\,Th_2C_3 + 0.5\,N_2 + 7.571\,ThB_6 \longrightarrow 10.57\,ThB_4 + ThB_6H_4C_6N$ | -1.4308 |
| $ThNH_4B_6C_6$ | $0.3636\,B_9H_{11} + 2\,Th_2C_3 + 0.5\,N_2 + 7.364\,ThB_6 \longrightarrow 10.36\,ThB_4 + ThB_6H_4C_6N$ | -1.4285 |
| $ThNH_4B_6C_6$ | $3\,ThC_2 + 0.3636\,B_9H_{11} + 14.73\,B + 0.5\,N_2 \longrightarrow 2\,ThB_6 + ThB_6H_4C_6N$ | -1.4285 |
| $ThNH_4B_6C_6$ | $1.25\,ThB_4 + 6\,ThC + 13.72\,H_2 + BN \longrightarrow 1.562\,Th_4H_{15} + ThB_6H_4C_6N$ | -1.4283 |
| $ThNH_4B_6C_6$ | $2\,Th_2C_3 + 15\,B + 5\,H_3N \longrightarrow Th_3N_4 + B_9H_{11} + ThB_6H_4C_6N$ | -1.4210 |
| $ThNH_4B_6C_6$ | $5.077\,ThC + 0.4615\,B_{13}C_2 + 3.802\,H_3N \longrightarrow 0.7006\,Th_3N_4 + 0.4938\,Th_4H_{15} + ThB_6H_4C_6N$ | -1.4172 |
| $ThNH_4B_6C_6$ | $0.25\,Th_3N_4 + 1.75\,B_{13}C_2 + 3.042\,Th + 0.5\,H_8C_5 \longrightarrow 2.792\,ThB_6 + ThB_6H_4C_6N$ | -1.4131 |
| $ThNH_4B_6C_6$ | $2\,ThH_2 + 1.418\,B_{13}C_2 + 1.055\,Th_2C_3 + 0.5\,N_2 \longrightarrow 3.109\,ThB_4 + ThB_6H_4C_6N$ | -1.4118 |
| $ThNH_4B_6C_6$ | $6.667\,ThH_2 + BN + 3\,B_{13}C_2 \longrightarrow 4.667\,H_2 + 5.667\,ThB_6 + ThB_6H_4C_6N$ | -1.4011 |
| $ThNH_4B_6C_6$ | $5.029\,ThH_2 + 0.25\,Th_3N_4 + 3\,B_{13}C_2 \longrightarrow 0.8654\,B_5H_7 + 4.779\,ThB_6 + ThB_6H_4C_6N$ | -1.3924 |
| $ThNH_4B_6C_6$ | $4.1\,ThH_2 + 3\,B_{13}C_2 + 2.4\,ThN \longrightarrow 1.4\,H_3N + 5.5\,ThB_6 + ThB_6H_4C_6N$ | -1.3920 |
| $ThNH_4B_6C_6$ | $6\,ThC + 2.068\,B_9H_{11} + 0.5\,N_2 \longrightarrow 1.25\,Th_4H_{15} + 12.61\,B + ThB_6H_4C_6N$ | -1.3780 |
| $ThNH_4B_6C_6$ | $4.75\,ThH_2 + 3\,B_{13}C_2 + ThN \longrightarrow 0.5\,B_9H_{11} + 4.75\,ThB_6 + ThB_6H_4C_6N$ | -1.3774 |
| $ThNH_4B_6C_6$ | $2\,H_2 + 3\,ThC_2 + 2\,N_2 + ThB_6 \longrightarrow 3\,ThN + ThB_6H_4C_6N$ | -1.3709 |
| $ThNH_4B_6C_6$ | $5.214\,ThH_2 + 3\,B_{13}C_2 + H_3N \longrightarrow 0.8571\,B_9H_{11} + 4.214\,ThB_6 + ThB_6H_4C_6N$ | -1.3670 |
| $ThNH_4B_6C_6$ | $1.2\,B_5H_7 + 0.25\,Th_3N_4 + 7.25\,Th + 1.2\,H_8C_5 \longrightarrow 7\,ThH_2 + ThB_6H_4C_6N$ | -1.3636 |
| $ThNH_4B_6C_6$ | $2\,ThH_2 + BN + 2\,Th_2C_3 + 12.5\,ThB_6 \longrightarrow 17.5\,ThB_4 + ThB_6H_4C_6N$ | -1.3600 |
| $ThNH_4B_6C_6$ | $6\,ThC + 7\,H_2 + 6\,B + 0.5\,N_2 \longrightarrow 5\,ThH_2 + ThB_6H_4C_6N$ | -1.3588 |
| $ThNH_4B_6C_6$ | $13.25\,B_9H_{11} + 2\,Th_2C_3 + 0.5\,N_2 \longrightarrow 3\,ThB_4 + 20.25\,B_5H_7 + ThB_6H_4C_6N$ | -1.3471 |
| $ThNH_4B_6C_6$ | $0.2667\,Th_4H_{15} + 2\,Th_2C_3 + 13.13\,ThB_6 + ThN \longrightarrow 18.2\,ThB_4 + ThB_6H_4C_6N$ | -1.3402 |
| $ThNH_4B_6C_6$ | $2.79\,ThC_2 + 0.3636\,B_9H_{11} + 0.2098\,B_{13}C_2 + 1.693\,N_2 \longrightarrow 0.5967\,Th_3N_4 + ThB_6H_4C_6N$ | -1.3368 |
| $ThNH_4B_6C_6$ | $6\,ThC + 4.481\,H_3N + ThB_6 \longrightarrow 0.6296\,Th_4H_{15} + 3.481\,ThN + ThB_6H_4C_6N$ | -1.3288 |
| $ThNH_4B_6C_6$ | $18.5\,B_9H_{11} + 2\,Th_2C_3 + 0.5\,N_2 \longrightarrow 28.5\,B_5H_7 + 3\,ThB_6 + ThB_6H_4C_6N$ | -1.3140 |
| $ThNH_4B_6C_6$ | $4.946\,ThH_2 + 0.25\,Th_3N_4 + 3\,B_{13}C_2 \longrightarrow 0.5357\,B_9H_{11} + 4.696\,ThB_6 + ThB_6H_4C_6N$ | -1.2988 |
| $ThNH_4B_6C_6$ | $8.867\,ThH_2 + 3\,B_{13}C_2 + 8.867\,H_3N \longrightarrow 3.667\,B_9H_{11} + 7.867\,ThN + ThB_6H_4C_6N$ | -1.2852 |
| $ThNH_4B_6C_6$ | $0.5714\,B_5H_7 + 2.758\,ThC_2 + 0.2418\,B_{13}C_2 + 1.672\,N_2 \longrightarrow 0.5861\,Th_3N_4 + ThB_6H_4C_6N$ | -1.2684 |
| $ThNH_4B_6C_6$ | $25.1\,H_2 + 3\,B_{13}C_2 + Th + 0.5\,N_2 \longrightarrow 6.6\,B_5H_7 + ThB_6H_4C_6N$ | -1.2564 |
| $ThNH_4B_6C_6$ | $2\,Th_2C_3 + 13.86\,B + 5\,H_3N \longrightarrow 1.571\,B_5H_7 + Th_3N_4 + ThB_6H_4C_6N$ | -1.2498 |
| $ThNH_4B_6C_6$ | $0.5714\,B_5H_7 + 6\,ThC + 3.143\,BN + 1.429\,N_2 \longrightarrow 5\,ThN + ThB_6H_4C_6N$ | -1.2460 |
| $ThNH_4B_6C_6$ | $1.5\,ThB_4 + 6\,ThC + 9.667\,H_3N \longrightarrow 2.167\,Th_3N_4 + 12.5\,H_2 + ThB_6H_4C_6N$ | -1.2281 |
| $ThNH_4B_6C_6$ | $7\,B_5H_7 + 6\,ThC + BN \longrightarrow 22.5\,H_2 + 5\,ThB_6 + ThB_6H_4C_6N$ | -1.2232 |
| $ThNH_4B_6C_6$ | $0.5714\,B_5H_7 + 0.25\,Th_3N_4 + 3\,ThC_2 + 19.64\,B \longrightarrow 2.75\,ThB_6 + ThB_6H_4C_6N$ | -1.2035 |
| $ThNH_4B_6C_6$ | $4.5\,ThH_2 + 0.6667\,Th_3N_4 + 3\,B_{13}C_2 \longrightarrow 1.667\,H_3N + 5.5\,ThB_6 + ThB_6H_4C_6N$ | -1.1852 |
| $ThNH_4B_6C_6$ | $3\,ThC_2 + 18.33\,B + 1.333\,H_3N \longrightarrow 0.3333\,BN + 2\,ThB_6 + ThB_6H_4C_6N$ | -1.1780 |
| $ThNH_4B_6C_6$ | $6\,C + 6\,BN + 7.333\,Th + 1.333\,H_3N \longrightarrow 6.333\,ThN + ThB_6H_4C_6N$ | -1.1546 |
| $ThNH_4B_6C_6$ | $3.455\,ThC + 0.3636\,B_9H_{11} + 1.273\,B_{13}C_2 + ThN \longrightarrow 3.455\,ThB_4 + ThB_6H_4C_6N$ | -1.1466 |
| $ThNH_4B_6C_6$ | $1.5\,ThB_4 + 1.167\,Th_2C_3 + 0.5\,H_8C_5 + 2.389\,N_2 \longrightarrow 0.9444\,Th_3N_4 + ThB_6H_4C_6N$ | -1.1463 |
| $ThNH_4B_6C_6$ | $0.25\,Th_3N_4 + 3\,ThC_2 + 0.3636\,B_9H_{11} + 19.23\,B \longrightarrow 2.75\,ThB_6 + ThB_6H_4C_6N$ | -1.1412 |
| $ThNH_4B_6C_6$ | $0.6667\,B_9H_{11} + 2\,Th_2C_3 + 1.389\,N_2 \longrightarrow 1.667\,ThH_2 + 0.4444\,Th_3N_4 + ThB_6H_4C_6N$ | -1.1319 |
| $ThNH_4B_6C_6$ | $2.493\,ThB_4 + 1.2\,H_8C_5 + 2.487\,N_2 \longrightarrow 0.3733\,Th_4H_{15} + 3.973\,BN + ThB_6H_4C_6N$ | -1.1305 |
| $ThNH_4B_6C_6$ | $1.2\,B_5H_7 + 6\,C + 4.7\,Th + H_3N \longrightarrow 3.7\,ThH_2 + ThB_6H_4C_6N$ | -1.1259 |
| $ThNH_4B_6C_6$ | $2.431\,Th_3N_4 + 1.636\,B_9H_{11} + 1.2\,H_8C_5 \longrightarrow 1.573\,Th_4H_{15} + 8.722\,BN + ThB_6H_4C_6N$ | -1.1056 |
| $ThNH_4B_6C_6$ | $1.75\,B_{13}C_2 + 2.792\,Th + 0.5\,H_8C_5 + ThN \longrightarrow 2.792\,ThB_6 + ThB_6H_4C_6N$ | -1.0927 |
| $ThNH_4B_6C_6$ | $3\,ThC_2 + 6\,B + 1.333\,H_3N + 1.167\,N_2 \longrightarrow 0.6667\,Th_3N_4 + ThB_6H_4C_6N$ | -1.0884 |
| $ThNH_4B_6C_6$ | $0.5714\,B_5H_7 + 3.415\,ThC + 1.293\,B_{13}C_2 + ThN \longrightarrow 3.415\,ThB_4 + ThB_6H_4C_6N$ | -1.0875 |
| $ThNH_4B_6C_6$ | $0.06667\,Th_4H_{15} + 3\,ThC_2 + 19.6\,B + H_3N \longrightarrow 2.267\,ThB_6 + ThB_6H_4C_6N$ | -1.0826 |
| $ThNH_4B_6C_6$ | $6\,ThC + 6\,B + 3.444\,H_3N \longrightarrow 3.167\,ThH_2 + 0.6111\,Th_3N_4 + ThB_6H_4C_6N$ | -1.0739 |

| Compound | Chemical reaction | Reaction energy(eV) |
|---|---|---|
| ThNH$_4$B$_6$C$_6$ | 1.5 ThB$_4$ + 6 ThC + 4.759 H$_3$N $\longrightarrow$ 0.6852 Th$_4$H$_{15}$ + 3.759 ThN + ThB$_6$H$_4$C$_6$N | -1.0619 |
| ThNH$_4$B$_6$C$_6$ | 3.492 ThC + 1.254 B$_{13}$C$_2$ + 1.333 H$_3$N $\longrightarrow$ 2.492 ThB$_4$ + 0.3333 BN + ThB$_6$H$_4$C$_6$N | -1.0609 |
| ThNH$_4$B$_6$C$_6$ | 0.7857 ThB$_4$ + 0.5714 B$_5$H$_7$ + 3 ThC$_2$ + 1.893 N$_2$ $\longrightarrow$ 2.786 ThN + ThB$_6$H$_4$C$_6$N | -1.0517 |
| ThNH$_4$B$_6$C$_6$ | 0.5714 B$_5$H$_7$ + BN + 2 Th$_2$C$_3$ + 14.14 B $\longrightarrow$ 3 ThB$_4$ + ThB$_6$H$_4$C$_6$N | -1.0493 |
| ThNH$_4$B$_6$C$_6$ | 5.077 ThC + 0.4615 B$_{13}$C$_2$ + 6.436 H$_3$N $\longrightarrow$ 1.359 Th$_3$N$_4$ + 7.654 H$_2$ + ThB$_6$H$_4$C$_6$N | -1.0476 |
| ThNH$_4$B$_6$C$_6$ | 1.2 B$_5$H$_7$ + 7 Th + 1.2 H$_8$C$_5$ + ThN $\longrightarrow$ 7 ThH$_2$ + ThB$_6$H$_4$C$_6$N | -1.0432 |
| ThNH$_4$B$_6$C$_6$ | 0.25 Th$_3$N$_4$ + 0.2667 Th$_4$H$_{15}$ + 2 Th$_2$C$_3$ + 12.63 ThB$_6$ $\longrightarrow$ 17.45 ThB$_4$ + ThB$_6$H$_4$C$_6$N | -1.0304 |
| ThNH$_4$B$_6$C$_6$ | 2 ThH$_2$ + 3 ThC$_2$ + 13 ThB$_6$ + ThN $\longrightarrow$ 18 ThB$_4$ + ThB$_6$H$_4$C$_6$N | -1.0303 |
| ThNH$_4$B$_6$C$_6$ | 0.2667 Th$_4$H$_{15}$ + 2.538 ThC$_2$ + 0.4615 B$_{13}$C$_2$ + 1.803 N$_2$ $\longrightarrow$ 2.605 ThN + ThB$_6$H$_4$C$_6$N | -1.0302 |
| ThNH$_4$B$_6$C$_6$ | 0.5 ThH$_2$ + 3 ThC$_2$ + 16 B + H$_3$N $\longrightarrow$ 2.5 ThB$_4$ + ThB$_6$H$_4$C$_6$N | -1.0169 |
| ThNH$_4$B$_6$C$_6$ | 0.5714 B$_5$H$_7$ + 3 ThC$_2$ + 3.143 B + 1.5 N$_2$ $\longrightarrow$ 2 ThN + ThB$_6$H$_4$C$_6$N | -1.0127 |
| ThNH$_4$B$_6$C$_6$ | 22.17 H$_2$ + 3 B$_{13}$C$_2$ + Th + 0.5 N$_2$ $\longrightarrow$ 3.667 B$_9$H$_{11}$ + ThB$_6$H$_4$C$_6$N | -1.0084 |
| ThNH$_4$B$_6$C$_6$ | 0.25 Th$_3$N$_4$ + 2 H$_2$ + 3 ThC$_2$ + 17 B $\longrightarrow$ 2.75 ThB$_4$ + ThB$_6$H$_4$C$_6$N | -1.0070 |
| ThNH$_4$B$_6$C$_6$ | 2.608 ThC + 0.2667 Th$_4$H$_{15}$ + 1.696 B$_{13}$C$_2$ + 0.5 N$_2$ $\longrightarrow$ 2.675 ThB$_6$ + ThB$_6$H$_4$C$_6$N | -0.9986 |
| ThNH$_4$B$_6$C$_6$ | 0.5 ThH$_2$ + 3.333 ThC + 1.333 B$_{13}$C$_2$ + H$_3$N $\longrightarrow$ 2.833 ThB$_4$ + ThB$_6$H$_4$C$_6$N | -0.9902 |
| ThNH$_4$B$_6$C$_6$ | 2 H$_2$ + 2 Th$_2$C$_3$ + 11 ThB$_6$ + ThN $\longrightarrow$ 15 ThB$_4$ + ThB$_6$H$_4$C$_6$N | -0.9888 |
| ThNH$_4$B$_6$C$_6$ | BN + 0.3636 B$_9$H$_{11}$ + 2 Th$_2$C$_3$ + 13.73 B $\longrightarrow$ 3 ThB$_4$ + ThB$_6$H$_4$C$_6$N | -0.9870 |
| ThNH$_4$B$_6$C$_6$ | 0.25 Th$_3$N$_4$ + 3.55 ThC + 0.3636 B$_9$H$_{11}$ + 1.225 B$_{13}$C$_2$ $\longrightarrow$ 3.3 ThB$_4$ + ThB$_6$H$_4$C$_6$N | -0.9846 |
| ThNH$_4$B$_6$C$_6$ | 0.6818 ThB$_4$ + 3 ThC$_2$ + 0.3636 B$_9$H$_{11}$ + 1.841 N$_2$ $\longrightarrow$ 2.682 ThN + ThB$_6$H$_4$C$_6$N | -0.9843 |
| ThNH$_4$B$_6$C$_6$ | 4.556 B$_5$H$_7$ + 3 B$_{13}$C$_2$ + 10.3 ThN $\longrightarrow$ 9.296 H$_3$N + 9.296 ThB$_6$ + ThB$_6$H$_4$C$_6$N | -0.9818 |
| ThNH$_4$B$_6$C$_6$ | 6 ThC + 1.097 B$_9$H$_{11}$ + 0.5 N$_2$ $\longrightarrow$ 4.032 ThH$_2$ + 0.9677 ThB$_4$ + ThB$_6$H$_4$C$_6$N | -0.9769 |
| ThNH$_4$B$_6$C$_6$ | 2 H$_2$ + 2.538 ThC$_2$ + 0.4615 B$_{13}$C$_2$ + 1.526 N$_2$ $\longrightarrow$ 0.5128 Th$_3$N$_4$ + ThB$_6$H$_4$C$_6$N | -0.9670 |
| ThNH$_4$B$_6$C$_6$ | 3 ThC$_2$ + 0.3636 B$_9$H$_{11}$ + 2.727 B + 1.5 N$_2$ $\longrightarrow$ 2 ThN + ThB$_6$H$_4$C$_6$N | -0.9504 |
| ThNH$_4$B$_6$C$_6$ | 0.5714 B$_5$H$_7$ + 0.25 Th$_3$N$_4$ + 3.51 ThC + 1.245 B$_{13}$C$_2$ $\longrightarrow$ 3.26 ThB$_4$ + ThB$_6$H$_4$C$_6$N | -0.9255 |
| ThNH$_4$B$_6$C$_6$ | 5.077 ThC + 0.4615 B$_{13}$C$_2$ + 3.413 H$_3$N $\longrightarrow$ 0.416 Th$_4$H$_{15}$ + 2.413 ThN + ThB$_6$H$_4$C$_6$N | -0.9231 |
| ThNH$_4$B$_6$C$_6$ | 2 ThH$_2$ + 1.714 B$_{13}$C$_2$ + 0.8571 Th$_2$C$_3$ + 0.5 N$_2$ $\longrightarrow$ 2.714 ThB$_6$ + ThB$_6$H$_4$C$_6$N | -0.9168 |
| ThNH$_4$B$_6$C$_6$ | 0.5714 B$_5$H$_7$ + 3 ThC$_2$ + 11.14 B + 0.5 N$_2$ $\longrightarrow$ 2 ThB$_4$ + ThB$_6$H$_4$C$_6$N | -0.9135 |
| ThNH$_4$B$_6$C$_6$ | 5.231 ThC + 9.933 H$_2$ + BN + 0.3846 B$_{13}$C$_2$ $\longrightarrow$ 1.058 Th$_4$H$_{15}$ + ThB$_6$H$_4$C$_6$N | -0.9001 |
| ThNH$_4$B$_6$C$_6$ | 0.5714 B$_5$H$_7$ + 3 ThC$_2$ + 1.762 N$_2$ + 0.5238 ThB$_6$ $\longrightarrow$ 2.524 ThN + ThB$_6$H$_4$C$_6$N | -0.8875 |
| ThNH$_4$B$_6$C$_6$ | 6 ThC + 6 B + 3.2 H$_3$N $\longrightarrow$ 2.8 ThH$_2$ + 2.2 ThN + ThB$_6$H$_4$C$_6$N | -0.8843 |
| ThNH$_4$B$_6$C$_6$ | 0.5714 B$_5$H$_7$ + 3 ThC$_2$ + 15.14 B + ThN $\longrightarrow$ 3 ThB$_4$ + ThB$_6$H$_4$C$_6$N | -0.8639 |
| ThNH$_4$B$_6$C$_6$ | 2.821 B$_5$H$_7$ + 6 ThC + H$_3$N $\longrightarrow$ 1.25 Th$_4$H$_{15}$ + 8.107 B + ThB$_6$H$_4$C$_6$N | -0.8559 |
| ThNH$_4$B$_6$C$_6$ | 3 ThC$_2$ + 0.3636 B$_9$H$_{11}$ + 10.73 B + 0.5 N$_2$ $\longrightarrow$ 2 ThB$_4$ + ThB$_6$H$_4$C$_6$N | -0.8512 |
| ThNH$_4$B$_6$C$_6$ | 18.72 B$_5$H$_7$ + 6 ThC + BN $\longrightarrow$ 1.25 Th$_4$H$_{15}$ + 9.844 B$_9$H$_{11}$ + ThB$_6$H$_4$C$_6$N | -0.8510 |
| ThNH$_4$B$_6$C$_6$ | 3.143 ThC + 2 H$_2$ + 1.429 B$_{13}$C$_2$ + ThN $\longrightarrow$ 3.143 ThB$_4$ + ThB$_6$H$_4$C$_6$N | -0.8502 |
| ThNH$_4$B$_6$C$_6$ | 3 ThC$_2$ + 0.3636 B$_9$H$_{11}$ + 1.727 N$_2$ + 0.4545 ThB$_6$ $\longrightarrow$ 2.455 ThN + ThB$_6$H$_4$C$_6$N | -0.8418 |
| ThNH$_4$B$_6$C$_6$ | 0.5 ThH$_2$ + 2 Th$_2$C$_3$ + H$_3$N + 10 ThB$_6$ $\longrightarrow$ 13.5 ThB$_4$ + ThB$_6$H$_4$C$_6$N | -0.8333 |
| ThNH$_4$B$_6$C$_6$ | B$_5$H$_7$ + BN + 7.3 Th + 1.2 H$_8$C$_5$ $\longrightarrow$ 6.3 ThH$_2$ + ThB$_6$H$_4$C$_6$N | -0.8269 |
| ThNH$_4$B$_6$C$_6$ | 3.602 ThC + 0.09091 B$_9$H$_{11}$ + 1.199 B$_{13}$C$_2$ + H$_3$N $\longrightarrow$ 2.602 ThB$_4$ + ThB$_6$H$_4$C$_6$N | -0.8181 |
| ThNH$_4$B$_6$C$_6$ | 6 ThC + 15.97 H$_3$N + 6.227 ThB$_6$ $\longrightarrow$ 6.273 B$_5$H$_7$ + 3.742 Th$_3$N$_4$ + ThB$_6$H$_4$C$_6$N | -0.8151 |
| ThNH$_4$B$_6$C$_6$ | 2 Th$_2$C$_3$ + 8.333 B + 3.333 H$_3$N $\longrightarrow$ 3 ThH$_2$ + 2.333 BN + ThB$_6$H$_4$C$_6$N | -0.8067 |
| ThNH$_4$B$_6$C$_6$ | 0.5714 B$_5$H$_7$ + 2 Th$_2$C$_3$ + 9.571 ThB$_6$ + ThN $\longrightarrow$ 13.57 ThB$_4$ + ThB$_6$H$_4$C$_6$N | -0.8039 |
| ThNH$_4$B$_6$C$_6$ | 0.1429 B$_5$H$_7$ + 3.592 ThC + 1.204 B$_{13}$C$_2$ + H$_3$N $\longrightarrow$ 2.592 ThB$_4$ + ThB$_6$H$_4$C$_6$N | -0.8033 |
| ThNH$_4$B$_6$C$_6$ | 0.3636 B$_9$H$_{11}$ + 2 Th$_2$C$_3$ + 9.364 ThB$_6$ + ThN $\longrightarrow$ 13.36 ThB$_4$ + ThB$_6$H$_4$C$_6$N | -0.8016 |
| ThNH$_4$B$_6$C$_6$ | 3 ThC$_2$ + 0.3636 B$_9$H$_{11}$ + 14.73 B + ThN $\longrightarrow$ 3 ThB$_4$ + ThB$_6$H$_4$C$_6$N | -0.8016 |
| ThNH$_4$B$_6$C$_6$ | 0.2667 Th$_4$H$_{15}$ + BN + 3 ThC$_2$ + 23.4 B $\longrightarrow$ 3.067 ThB$_6$ + ThB$_6$H$_4$C$_6$N | -0.7962 |
| ThNH$_4$B$_6$C$_6$ | 2 ThH$_2$ + 2.857 ThC + BN + 1.571 B$_{13}$C$_2$ $\longrightarrow$ 3.857 ThB$_4$ + ThB$_6$H$_4$C$_6$N | -0.7781 |
| ThNH$_4$B$_6$C$_6$ | 5.769 ThH$_2$ + BN + 3 B$_{13}$C$_2$ $\longrightarrow$ 1.077 B$_5$H$_7$ + 4.769 ThB$_6$ + ThB$_6$H$_4$C$_6$N | -0.7771 |
| ThNH$_4$B$_6$C$_6$ | 0.5 H$_2$ + 3 ThC$_2$ + 18 B + H$_3$N $\longrightarrow$ 2 ThB$_6$ + ThB$_6$H$_4$C$_6$N | -0.7638 |
| ThNH$_4$B$_6$C$_6$ | 3.619 ThC + 1.19 B$_{13}$C$_2$ + 1.333 H$_3$N $\longrightarrow$ 2.369 ThB$_4$ + 0.08333 Th$_3$N$_4$ + ThB$_6$H$_4$C$_6$N | -0.7626 |
| ThNH$_4$B$_6$C$_6$ | 6 ThC + 0.6667 B$_9$H$_{11}$ + 6 H$_3$N $\longrightarrow$ 10.67 H$_2$ + 5 ThN + ThB$_6$H$_4$C$_6$N | -0.7567 |
| ThNH$_4$B$_6$C$_6$ | 2 ThH$_2$ + 1.68 ThC + 2.16 B$_{13}$C$_2$ + ThN $\longrightarrow$ 3.68 ThB$_6$ + ThB$_6$H$_4$C$_6$N | -0.7473 |
| ThNH$_4$B$_6$C$_6$ | 3.524 ThC + 0.5 H$_2$ + 1.238 B$_{13}$C$_2$ + H$_3$N $\longrightarrow$ 2.524 ThB$_4$ + ThB$_6$H$_4$C$_6$N | -0.7440 |
| ThNH$_4$B$_6$C$_6$ | 2 ThH$_2$ + 0.25 Th$_3$N$_4$ + 3 ThC$_2$ + 12.5 ThB$_6$ $\longrightarrow$ 17.25 ThB$_4$ + ThB$_6$H$_4$C$_6$N | -0.7205 |
| ThNH$_4$B$_6$C$_6$ | 1.789 B$_5$H$_7$ + 6 ThC + 0.5 N$_2$ $\longrightarrow$ 4.263 ThH$_2$ + 0.7368 ThB$_4$ + ThB$_6$H$_4$C$_6$N | -0.7124 |
| ThNH$_4$B$_6$C$_6$ | 1.2 B$_5$H$_7$ + 6 ThC + 6 H$_3$N $\longrightarrow$ 11.2 H$_2$ + 5 ThN + ThB$_6$H$_4$C$_6$N | -0.7117 |
| ThNH$_4$B$_6$C$_6$ | 3.651 ThC + 1.175 B$_{13}$C$_2$ + 1.333 H$_3$N $\longrightarrow$ 2.317 ThB$_4$ + 0.3333 ThN + ThB$_6$H$_4$C$_6$N | -0.7086 |
| ThNH$_4$B$_6$C$_6$ | 16.75 B$_9$H$_{11}$ + 2 Th$_2$C$_3$ + ThN $\longrightarrow$ 4 ThB$_4$ + 25.75 B$_5$H$_7$ + ThB$_6$H$_4$C$_6$N | -0.6982 |
| ThNH$_4$B$_6$C$_6$ | 0.25 Th$_3$N$_4$ + 3.238 ThC + 2 H$_2$ + 1.381 B$_{13}$C$_2$ $\longrightarrow$ 2.988 ThB$_4$ + ThB$_6$H$_4$C$_6$N | -0.6882 |
| ThNH$_4$B$_6$C$_6$ | 0.25 Th$_3$N$_4$ + 2 H$_2$ + 2 Th$_2$C$_3$ + 10.5 ThB$_6$ $\longrightarrow$ 14.25 ThB$_4$ + ThB$_6$H$_4$C$_6$N | -0.6790 |

| Compound | Chemical reaction | Reaction energy(eV) |
|---|---|---|
| $ThNH_4B_6C_6$ | $6\,ThC + 1.143\,B_9H_{11} + 0.5\,N_2 \longrightarrow 4.286\,ThH_2 + 0.7143\,ThB_6 + ThB_6H_4C_6N$ | -0.6786 |
| $ThNH_4B_6C_6$ | $3.422\,ThC + 0.06667\,Th_4H_{15} + 1.289\,B_{13}C_2 + H_3N \longrightarrow 2.689\,ThB_4 + ThB_6H_4C_6N$ | -0.6742 |
| $ThNH_4B_6C_6$ | $1.429\,B_5H_7 + 2\,Th_2C_3 + 1.071\,N_2 \longrightarrow 3\,ThH_2 + 1.143\,BN + ThB_6H_4C_6N$ | -0.6717 |
| $ThNH_4B_6C_6$ | $5.667\,ThH_2 + BN + 3\,B_{13}C_2 \longrightarrow 0.6667\,B_9H_{11} + 4.667\,ThB_6 + ThB_6H_4C_6N$ | -0.6607 |
| $ThNH_4B_6C_6$ | $2\,Th_2C_3 + 1.333\,H_3N + 9.167\,ThB_6 \longrightarrow 12.17\,ThB_4 + 0.3333\,BN + ThB_6H_4C_6N$ | -0.6577 |
| $ThNH_4B_6C_6$ | $23.75\,B_9H_{11} + 2\,Th_2C_3 + ThN \longrightarrow 36.75\,B_5H_7 + 4\,ThB_6 + ThB_6H_4C_6N$ | -0.6540 |
| $ThNH_4B_6C_6$ | $2.577\,ThC + 0.4615\,B_{13}C_2 + 0.5\,H_8C_5 + 1.551\,N_2 \longrightarrow 0.5256\,Th_3N_4 + ThB_6H_4C_6N$ | -0.6502 |
| $ThNH_4B_6C_6$ | $0.25\,Th_3N_4 + 0.6667\,B_9H_{11} + 6.717\,Th + 1.2\,H_8C_5 \longrightarrow 6.467\,ThH_2 + ThB_6H_4C_6N$ | -0.6422 |
| $ThNH_4B_6C_6$ | $1.717\,B_9H_{11} + 2.485\,Th_2C_3 + 0.5\,N_2 \longrightarrow 0.9923\,Th_4H_{15} + 0.727\,B_{13}C_2 + ThB_6H_4C_6N$ | -0.6396 |
| $ThNH_4B_6C_6$ | $2\,ThH_2 + 0.25\,Th_3N_4 + 1.8\,ThC + 2.1\,B_{13}C_2 \longrightarrow 3.55\,ThB_6 + ThB_6H_4C_6N$ | -0.6237 |
| $ThNH_4B_6C_6$ | $0.1429\,B_5H_7 + 3\,ThC_2 + 17.29\,B + H_3N \longrightarrow 2\,ThB_6 + ThB_6H_4C_6N$ | -0.6145 |
| $ThNH_4B_6C_6$ | $3\,ThC_2 + 14.33\,B + 1.333\,H_3N \longrightarrow 2\,ThB_4 + 0.3333\,BN + ThB_6H_4C_6N$ | -0.6008 |
| $ThNH_4B_6C_6$ | $3\,ThC_2 + 0.09091\,B_9H_{11} + 17.18\,B + H_3N \longrightarrow 2\,ThB_6 + ThB_6H_4C_6N$ | -0.5990 |
| $ThNH_4B_6C_6$ | $2\,B_9H_{11} + 2\,Th_2C_3 + 0.5\,N_2 \longrightarrow 3\,ThB_4 + 9\,H_2 + ThB_6H_4C_6N$ | -0.5863 |
| $ThNH_4B_6C_6$ | $3.5\,ThC + BN + 20\,B + 0.5\,H_8C_5 \longrightarrow 2.5\,ThB_6 + ThB_6H_4C_6N$ | -0.5768 |
| $ThNH_4B_6C_6$ | $0.4615\,B_{13}C_2 + 1.692\,Th_2C_3 + 1.333\,H_3N + 1.026\,N_2 \longrightarrow 2.385\,ThN + ThB_6H_4C_6N$ | -0.5735 |
| $ThNH_4B_6C_6$ | $2.737\,ThC + 0.2667\,Th_4H_{15} + 1.632\,B_{13}C_2 + ThN \longrightarrow 3.803\,ThB_4 + ThB_6H_4C_6N$ | -0.5711 |
| $ThNH_4B_6C_6$ | $6\,ThC + 1.795\,B_9H_{11} + H_3N \longrightarrow 1.25\,Th_4H_{15} + 10.16\,B + ThB_6H_4C_6N$ | -0.5485 |
| $ThNH_4B_6C_6$ | $3\,ThC_2 + 2.75\,B_9H_{11} + 1.5\,N_2 \longrightarrow 3.75\,B_5H_7 + 2\,ThN + ThB_6H_4C_6N$ | -0.5418 |
| $ThNH_4B_6C_6$ | $9.5\,H_2 + 2\,Th_2C_3 + 6\,B + ThN \longrightarrow Th_4H_{15} + ThB_6H_4C_6N$ | -0.5367 |
| $ThNH_4B_6C_6$ | $1.167\,Th_2C_3 + 0.5\,H_8C_5 + 2.056\,N_2 + ThB_6 \longrightarrow 0.7778\,Th_3N_4 + ThB_6H_4C_6N$ | -0.5219 |
| $ThNH_4B_6C_6$ | $0.5714\,B_5H_7 + 0.25\,Th_3N_4 + 2\,Th_2C_3 + 9.071\,ThB_6 \longrightarrow 12.82\,ThB_4 + ThB_6H_4C_6N$ | -0.4941 |
| $ThNH_4B_6C_6$ | $0.25\,Th_3N_4 + 0.3636\,B_9H_{11} + 2\,Th_2C_3 + 8.864\,ThB_6 \longrightarrow 12.61\,ThB_4 + ThB_6H_4C_6N$ | -0.4918 |
| $ThNH_4B_6C_6$ | $1.846\,B_5H_7 + 6\,ThC + 0.5\,N_2 \longrightarrow 4.462\,ThH_2 + 0.5385\,ThB_6 + ThB_6H_4C_6N$ | -0.4791 |
| $ThNH_4B_6C_6$ | $3.6\,B_5H_7 + 2\,Th_2C_3 + 0.5\,N_2 \longrightarrow 3\,ThB_4 + 10.6\,H_2 + ThB_6H_4C_6N$ | -0.4511 |
| $ThNH_4B_6C_6$ | $3.093\,ThC + 1.453\,B_{13}C_2 + 1.333\,H_3N \longrightarrow 0.3333\,BN + 2.093\,ThB_6 + ThB_6H_4C_6N$ | -0.4422 |
| $ThNH_4B_6C_6$ | $2.549\,B_5H_7 + 2.346\,Th_2C_3 + 0.5\,N_2 \longrightarrow 0.923\,Th_4H_{15} + 0.5189\,B_{13}C_2 + ThB_6H_4C_6N$ | -0.4421 |
| $ThNH_4B_6C_6$ | $1.2\,B_5H_7 + 8.75\,Th_3N_4 + 8.75\,ThC \longrightarrow 0.55\,H_8C_5 + 34\,ThN + ThB_6H_4C_6N$ | -0.4295 |
| $ThNH_4B_6C_6$ | $0.06667\,Th_4H_{15} + 3\,ThC_2 + 15.07\,B + H_3N \longrightarrow 2.267\,ThB_4 + ThB_6H_4C_6N$ | -0.4284 |
| $ThNH_4B_6C_6$ | $2\,ThH_2 + 1.636\,B_{13}C_2 + 0.9091\,Th_2C_3 + ThN \longrightarrow 3.818\,ThB_4 + ThB_6H_4C_6N$ | -0.4202 |
| $ThNH_4B_6C_6$ | $3\,ThC_2 + 16.5\,B + 1.333\,H_3N \longrightarrow 0.08333\,Th_3N_4 + 1.75\,ThB_6 + ThB_6H_4C_6N$ | -0.4182 |
| $ThNH_4B_6C_6$ | $6\,BN + 4.188\,Th_2C_3 + 4.833\,H_3N \longrightarrow 2.458\,Th_3N_4 + 1.313\,H_8C_5 + ThB_6H_4C_6N$ | -0.4114 |
| $ThNH_4B_6C_6$ | $2\,ThH_2 + 1.706\,ThC_2 + 1.294\,B_{13}C_2 + 0.5\,N_2 \longrightarrow 2.706\,ThB_4 + ThB_6H_4C_6N$ | -0.4107 |
| $ThNH_4B_6C_6$ | $0.5714\,B_5H_7 + 0.25\,Th_3N_4 + 3\,ThC_2 + 14.14\,B \longrightarrow 2.75\,ThB_4 + ThB_6H_4C_6N$ | -0.4098 |
| $ThNH_4B_6C_6$ | $0.25\,Th_3N_4 + 2.832\,ThC + 0.2667\,Th_4H_{15} + 1.584\,B_{13}C_2 \longrightarrow 3.648\,ThB_4 + ThB_6H_4C_6N$ | -0.4091 |
| $ThNH_4B_6C_6$ | $6\,C + 0.6667\,B_9H_{11} + 4.167\,Th + H_3N \longrightarrow 3.167\,ThH_2 + ThB_6H_4C_6N$ | -0.4045 |
| $ThNH_4B_6C_6$ | $6\,ThC + 7\,H_3N + ThB_6 \longrightarrow 8.5\,H_2 + 6\,ThN + ThB_6H_4C_6N$ | -0.4026 |
| $ThNH_4B_6C_6$ | $3\,ThC_2 + 0.6667\,B_9H_{11} + 1.5\,N_2 \longrightarrow 1.667\,H_2 + 2\,ThN + ThB_6H_4C_6N$ | -0.4009 |
| $ThNH_4B_6C_6$ | $0.25\,Th_3N_4 + 15.87\,B_9H_{11} + 2\,Th_2C_3 \longrightarrow 3.75\,ThB_4 + 24.37\,B_5H_7 + ThB_6H_4C_6N$ | -0.3939 |
| $ThNH_4B_6C_6$ | $2\,Th_2C_3 + 12.55\,B + 4\,H_3N \longrightarrow 0.7273\,B_9H_{11} + 3\,ThN + ThB_6H_4C_6N$ | -0.3846 |
| $ThNH_4B_6C_6$ | $0.06667\,Th_4H_{15} + 2\,Th_2C_3 + H_3N + 9.533\,ThB_6 \longrightarrow 12.8\,ThB_4 + ThB_6H_4C_6N$ | -0.3794 |
| $ThNH_4B_6C_6$ | $1.074\,B_9H_{11} + 2\,Th_2C_3 + 0.5\,N_2 \longrightarrow 0.9164\,ThB_4 + 0.5209\,Th_4H_{15} + ThB_6H_4C_6N$ | -0.3764 |
| $ThNH_4B_6C_6$ | $1.2\,B_5H_7 + 3\,ThC_2 + 1.5\,N_2 \longrightarrow 2.2\,H_2 + 2\,ThN + ThB_6H_4C_6N$ | -0.3558 |
| $ThNH_4B_6C_6$ | $0.25\,Th_3N_4 + 22.44\,B_9H_{11} + 2\,Th_2C_3 \longrightarrow 34.69\,B_5H_7 + 3.75\,ThB_6 + ThB_6H_4C_6N$ | -0.3525 |
| $ThNH_4B_6C_6$ | $0.25\,Th_3N_4 + 3\,ThC_2 + 0.3636\,B_9H_{11} + 13.73\,B \longrightarrow 2.75\,ThB_4 + ThB_6H_4C_6N$ | -0.3475 |
| $ThNH_4B_6C_6$ | $1.359\,Th_3N_4 + 1.75\,B_{13}C_2 + 0.5\,H_8C_5 \longrightarrow 3.078\,ThB_4 + 4.438\,BN + ThB_6H_4C_6N$ | -0.3393 |
| $ThNH_4B_6C_6$ | $3.908\,ThC + 1.046\,B_{13}C_2 + 4.878\,H_3N \longrightarrow 1.519\,B_5H_7 + 0.9695\,Th_3N_4 + ThB_6H_4C_6N$ | -0.3296 |
| $ThNH_4B_6C_6$ | $3\,ThC_2 + 18\,B + 1.333\,H_3N \longrightarrow 0.1667\,N_2 + 2\,ThB_6 + ThB_6H_4C_6N$ | -0.3225 |
| $ThNH_4B_6C_6$ | $0.6667\,B_9H_{11} + 6.467\,Th + 1.2\,H_8C_5 + ThN \longrightarrow 6.467\,ThH_2 + ThB_6H_4C_6N$ | -0.3218 |
| $ThNH_4B_6C_6$ | $5.077\,ThC + 0.4615\,B_{13}C_2 + 5.077\,H_3N \longrightarrow 5.615\,H_2 + 4.077\,ThN + ThB_6H_4C_6N$ | -0.3112 |
| $ThNH_4B_6C_6$ | $0.25\,Th_3N_4 + 9.031\,H_2 + 2\,Th_2C_3 + 6\,B \longrightarrow 0.9375\,Th_4H_{15} + ThB_6H_4C_6N$ | -0.3093 |
| $ThNH_4B_6C_6$ | $1.2\,B_5H_7 + Th + 1.2\,H_8C_5 + 2.833\,N_2 \longrightarrow 4.667\,H_3N + ThB_6H_4C_6N$ | -0.3088 |
| $ThNH_4B_6C_6$ | $3.524\,ThC + 1.238\,B_{13}C_2 + 1.333\,H_3N \longrightarrow 2.524\,ThB_4 + 0.1667\,N_2 + ThB_6H_4C_6N$ | -0.3026 |
| $ThNH_4B_6C_6$ | $0.6667\,B_9H_{11} + 2\,Th_2C_3 + 1.167\,N_2 \longrightarrow 1.667\,ThH_2 + 1.333\,ThN + ThB_6H_4C_6N$ | -0.3026 |
| $ThNH_4B_6C_6$ | $9.5\,H_2 + 2\,Th_2C_3 + 0.5\,N_2 + ThB_6 \longrightarrow Th_4H_{15} + ThB_6H_4C_6N$ | -0.2977 |
| $ThNH_4B_6C_6$ | $1.2\,B_5H_7 + 6\,C + 3.2\,Th + 0.5\,N_2 \longrightarrow 2.2\,ThH_2 + ThB_6H_4C_6N$ | -0.2941 |
| $ThNH_4B_6C_6$ | $0.5\,H_2 + 2\,Th_2C_3 + H_3N + 9\,ThB_6 \longrightarrow 12\,ThB_4 + ThB_6H_4C_6N$ | -0.2915 |
| $ThNH_4B_6C_6$ | $1.786\,B_5H_7 + 2\,Th_2C_3 + 0.5\,N_2 \longrightarrow 0.7326\,ThB_4 + 0.5668\,Th_4H_{15} + ThB_6H_4C_6N$ | -0.2908 |
| $ThNH_4B_6C_6$ | $6\,ThC + 18.2\,H_3N + 7.9\,ThB_6 \longrightarrow 4.3\,Th_3N_4 + 4.6\,B_9H_{11} + ThB_6H_4C_6N$ | -0.2892 |
| $ThNH_4B_6C_6$ | $2.902\,ThC + 0.3636\,B_9H_{11} + 1.549\,B_{13}C_2 + ThN \longrightarrow 2.902\,ThB_6 + ThB_6H_4C_6N$ | -0.2890 |

| Compound | Chemical reaction | Reaction energy(eV) |
|---|---|---|
| $ThNH_4B_6C_6$ | $2\,ThH_2 + 1.5\,ThB_4 + 6\,C + 2.167\,N_2 \longrightarrow 0.8333\,Th_3N_4 + ThB_6H_4C_6N$ | -0.2885 |
| $ThNH_4B_6C_6$ | $0.5\,ThH_2 + 2.88\,ThC + 1.56\,B_{13}C_2 + H_3N \longrightarrow 2.38\,ThB_6 + ThB_6H_4C_6N$ | -0.2868 |
| $ThNH_4B_6C_6$ | $3.958\,H_2 + 0.6667\,B_9H_{11} + 2\,Th_2C_3 + 0.5\,N_2 \longrightarrow 0.75\,Th_4H_{15} + ThB_6H_4C_6N$ | -0.2841 |
| $ThNH_4B_6C_6$ | $1.143\,B_9H_{11} + 2\,Th_2C_3 + 0.5\,N_2 \longrightarrow 0.5714\,Th_4H_{15} + 0.7143\,ThB + ThB_6H_4C_6N$ | -0.2744 |
| $ThNH_4B_6C_6$ | $2\,Th_2C_3 + 11.71\,B + 4\,H_3N \longrightarrow 1.143\,B_5H_7 + 3\,ThN + ThB_6H_4C_6N$ | -0.2601 |
| $ThNH_4B_6C_6$ | $7.121\,ThC + 1.477\,B_9H_{11} + 0.5\,N_2 \longrightarrow 6.121\,ThH_2 + 0.5607\,B_{13}C_2 + ThB_6H_4C_6N$ | -0.2580 |
| $ThNH_4B_6C_6$ | $0.1429\,B_5H_7 + 2\,Th_2C_3 + H_3N + 8.643\,ThB_6 \longrightarrow 11.64\,ThB_4 + ThB_6H_4C_6N$ | -0.2453 |
| $ThNH_4B_6C_6$ | $0.09091\,B_9H_{11} + 2\,Th_2C_3 + H_3N + 8.591\,ThB_6 \longrightarrow 11.59\,ThB_4 + ThB_6H_4C_6N$ | -0.2448 |
| $ThNH_4B_6C_6$ | $2.667\,B_9H_{11} + 2\,Th_2C_3 + 0.5\,N_2 \longrightarrow 12.67\,H_2 + 3\,ThB_6 + ThB_6H_4C_6N$ | -0.2432 |
| $ThNH_4B_6C_6$ | $3\,ThC_2 + 16\,B + 1.333\,H_3N \longrightarrow 1.667\,ThB_6 + 0.3333\,ThN + ThB_6H_4C_6N$ | -0.2428 |
| $ThNH_4B_6C_6$ | $3\,ThC_2 + 0.6667\,B_9H_{11} + 1.241\,N_2 \longrightarrow 0.3704\,Th_3N_4 + 0.2222\,Th_4H_{15} + ThB_6H_4C_6N$ | -0.2419 |
| $ThNH_4B_6C_6$ | $6\,B + 5.3\,Th + H_3N + 1.2\,H_8C_5 \longrightarrow 4.3\,ThH_2 + ThB_6H_4C_6N$ | -0.2401 |
| $ThNH_4B_6C_6$ | $0.5714\,B_5H_7 + 2.869\,ThC + 1.566\,B_{13}C_2 + ThN \longrightarrow 2.869\,ThB_6 + ThB_6H_4C_6N$ | -0.2397 |
| $ThNH_4B_6C_6$ | $1.2\,B_5H_7 + 3.425\,H_2 + 2\,Th_2C_3 + 0.5\,N_2 \longrightarrow 0.75\,Th_4H_{15} + ThB_6H_4C_6N$ | -0.2390 |
| $ThNH_4B_6C_6$ | $BN + 0.5556\,B_9H_{11} + 6.856\,Th + 1.2\,H_8C_5 \longrightarrow 5.856\,ThH_2 + ThB_6H_4C_6N$ | -0.2257 |
| $ThNH_4B_6C_6$ | $2.79\,ThC_2 + 0.3636\,B_9H_{11} + 0.2098\,B_{13}C_2 + 1.395\,N_2 \longrightarrow 1.79\,ThN + ThB_6H_4C_6N$ | -0.2233 |
| $ThNH_4B_6C_6$ | $3.786\,B_5H_7 + 6\,ThC + ThN \longrightarrow 1.5\,Th_4H_{15} + 12.93\,B + ThB_6H_4C_6N$ | -0.2160 |
| $ThNH_4B_6C_6$ | $1.877\,B_5H_7 + 2\,Th_2C_3 + 0.5\,N_2 \longrightarrow 0.6091\,Th_4H_{15} + 0.5638\,ThB_6 + ThB_6H_4C_6N$ | -0.2059 |
| $ThNH_4B_6C_6$ | $2\,ThH_2 + 0.25\,Th_3N_4 + 1.582\,B_{13}C_2 + 0.9455\,Th_2C_3 \longrightarrow 3.641\,ThB_4 + ThB_6H_4C_6N$ | -0.2016 |
| $ThNH_4B_6C_6$ | $3.795\,ThC + 1.102\,B_{13}C_2 + 4.727\,H_3N \longrightarrow 0.9318\,Th_3N_4 + 0.9256\,B_9H_{11} + ThB_6H_4C_6N$ | -0.1976 |
| $ThNH_4B_6C_6$ | $0.5\,H_2 + 3\,ThC_2 + 14\,B + H_3N \longrightarrow 2\,ThB_4 + ThB_6H_4C_6N$ | -0.1866 |
| $ThNH_4B_6C_6$ | $0.5714\,B_5H_7 + 2.758\,ThC_2 + 0.2418\,B_{13}C_2 + 1.379\,N_2 \longrightarrow 1.758\,ThN + ThB_6H_4C_6N$ | -0.1748 |
| $ThNH_4B_6C_6$ | $3.24\,ThC + 1.38\,B_{13}C_2 + 1.333\,H_3N \longrightarrow 0.08333\,Th_3N_4 + 1.99\,ThB_6 + ThB_6H_4C_6N$ | -0.1745 |
| $ThNH_4B_6C_6$ | $0.7961\,B_9H_{11} + 2\,Th_2C_3 + 2.164\,H_3N \longrightarrow 0.75\,Th_4H_{15} + 1.164\,BN + ThB_6H_4C_6N$ | -0.1725 |
| $ThNH_4B_6C_6$ | $3.185\,ThC + 0.09091\,B_9H_{11} + 1.407\,B_{13}C_2 + H_3N \longrightarrow 2.185\,ThB_6 + ThB_6H_4C_6N$ | -0.1722 |
| $ThNH_4B_6C_6$ | $6.125\,H_2 + 2\,Th_2C_3 + 6\,B + H_3N \longrightarrow 0.75\,Th_4H_{15} + ThB_6H_4C_6N$ | -0.1690 |
| $ThNH_4B_6C_6$ | $0.25\,Th_3N_4 + 3.022\,ThC + 0.3636\,B_9H_{11} + 1.489\,B_{13}C_2 \longrightarrow 2.772\,ThB_6 + ThB_6H_4C_6N$ | -0.1654 |
| $ThNH_4B_6C_6$ | $2\,Th_2C_3 + 1.333\,H_3N + 8.5\,ThB_6 \longrightarrow 11.25\,ThB_4 + 0.08333\,Th_3N_4 + ThB_6H_4C_6N$ | -0.1624 |
| $ThNH_4B_6C_6$ | $0.1429\,B_5H_7 + 3.177\,ThC + 1.411\,B_{13}C_2 + H_3N \longrightarrow 2.177\,ThB_6 + ThB_6H_4C_6N$ | -0.1599 |
| $ThNH_4B_6C_6$ | $1.065\,Th_4H_{15} + 3\,B_{13}C_2 + 4.991\,ThN \longrightarrow 8.25\,ThB_4 + 3.991\,H_3N + ThB_6H_4C_6N$ | -0.1596 |
| $ThNH_4B_6C_6$ | $15.12\,B_9H_{11} + 2\,Th_2C_3 + H_3N \longrightarrow 3\,ThB_4 + 23.62\,B_5H_7 + ThB_6H_4C_6N$ | -0.1498 |
| $ThNH_4B_6C_6$ | $6\,ThC + 3.333\,H_2 + 0.6667\,B_9H_{11} + 0.5\,N_2 \longrightarrow 5\,ThH_2 + ThB_6H_4C_6N$ | -0.1498 |
| $ThNH_4B_6C_6$ | $2.225\,B_5H_7 + 6.789\,ThC + 0.5\,N_2 \longrightarrow 5.789\,ThH_2 + 0.3944\,B_{13}C_2 + ThB_6H_4C_6N$ | -0.1423 |
| $ThNH_4B_6C_6$ | $1.167\,Th_2C_3 + 6\,B + 0.5\,H_8C_5 + 1.389\,N_2 \longrightarrow 0.4444\,Th_3N_4 + ThB_6H_4C_6N$ | -0.1390 |
| $ThNH_4B_6C_6$ | $2\,ThH_2 + 1.421\,ThC_2 + 1.579\,B_{13}C_2 + 0.5\,N_2 \longrightarrow 2.421\,ThB_6 + ThB_6H_4C_6N$ | -0.1363 |
| $ThNH_4B_6C_6$ | $3.28\,ThC + 1.36\,B_{13}C_2 + 1.333\,H_3N \longrightarrow 1.947\,ThB_6 + 0.3333\,ThN + ThB_6H_4C_6N$ | -0.1333 |
| $ThNH_4B_6C_6$ | $1.5\,ThB_4 + 10.44\,H_2 + 2\,Th_2C_3 + 0.5\,N_2 \longrightarrow 1.125\,Th_4H_{15} + ThB_6H_4C_6N$ | -0.1330 |
| $ThNH_4B_6C_6$ | $1.2\,B_5H_7 + 6\,C + 2.973\,Th + H_3N \longrightarrow 0.4933\,Th_4H_{15} + ThB_6H_4C_6N$ | -0.1320 |
| $ThNH_4B_6C_6$ | $3.652\,B_5H_7 + 0.25\,Th_3N_4 + 6\,ThC \longrightarrow 1.438\,Th_4H_{15} + 12.26\,B + ThB_6H_4C_6N$ | -0.1285 |
| $ThNH_4B_6C_6$ | $3.12\,ThC + 0.5\,H_2 + 1.44\,B_{13}C_2 + H_3N \longrightarrow 2.12\,ThB_6 + ThB_6H_4C_6N$ | -0.1175 |
| $ThNH_4B_6C_6$ | $20.38\,B_9H_{11} + 2\,Th_2C_3 + H_3N \longrightarrow 31.88\,B_5H_7 + 3\,ThB_6 + ThB_6H_4C_6N$ | -0.1167 |
| $ThNH_4B_6C_6$ | $0.5714\,B_5H_7 + 0.25\,Th_3N_4 + 2.989\,ThC + 1.506\,B_{13}C_2 \longrightarrow 2.739\,ThB_6 + ThB_6H_4C_6N$ | -0.1162 |
| $ThNH_4B_6C_6$ | $3.5\,ThC + 0.5\,H_8C_5 + 0.5\,N_2 + 8\,ThB_6 \longrightarrow 10.5\,ThB_4 + ThB_6H_4C_6N$ | -0.1131 |
| $ThNH_4B_6C_6$ | $1.2\,B_5H_7 + 6\,ThC + 2.8\,H_2 + 0.5\,N_2 \longrightarrow 5\,ThH_2 + ThB_6H_4C_6N$ | -0.1047 |
| $ThNH_4B_6C_6$ | $3.931\,ThC + BN + 0.3636\,B_9H_{11} + 1.035\,B_{13}C_2 \longrightarrow 2.931\,ThB_4 + ThB_6H_4C_6N$ | -0.0899 |
| $ThNH_4B_6C_6$ | $0.3636\,B_9H_{11} + 0.8033\,B_{13}C_2 + 1.464\,Th_2C_3 + 0.5\,N_2 \longrightarrow 1.929\,ThB_4 + ThB_6H_4C_6N$ | -0.0856 |
| $ThNH_4B_6C_6$ | $1.75\,B_{13}C_2 + 0.5\,H_8C_5 + 4.35\,ThN \longrightarrow 3.35\,ThB_4 + 3.35\,BN + ThB_6H_4C_6N$ | -0.0710 |
| $ThNH_4B_6C_6$ | $2.64\,ThC + 2\,H_2 + 1.68\,B_{13}C_2 + ThN \longrightarrow 2.64\,ThB_6 + ThB_6H_4C_6N$ | -0.0700 |
| $ThNH_4B_6C_6$ | $4.8\,B_5H_7 + 2\,Th_2C_3 + 0.5\,N_2 \longrightarrow 14.8\,H_2 + 3\,ThB_6 + ThB_6H_4C_6N$ | -0.0629 |
| $ThNH_4B_6C_6$ | $5.077\,ThC + 6.077\,H_2 + 0.4615\,B_{13}C_2 + 0.5\,N_2 \longrightarrow 4.077\,ThH_2 + ThB_6H_4C_6N$ | -0.0608 |
| $ThNH_4B_6C_6$ | $2\,Th_2C_3 + 1.333\,H_3N + 8.333\,ThB_6 \longrightarrow 11\,ThB_4 + 0.3333\,ThN + ThB_6H_4C_6N$ | -0.0591 |
| $ThNH_4B_6C_6$ | $6.471\,H_2 + 0.4615\,B_{13}C_2 + 1.692\,Th_2C_3 + 0.5\,N_2 \longrightarrow 0.5962\,Th_4H_{15} + ThB_6H_4C_6N$ | -0.0584 |
| $ThNH_4B_6C_6$ | $3.083\,Th_3N_4 + 19\,H_2 + 3\,B_{13}C_2 \longrightarrow 8.25\,ThB_4 + 11.33\,H_3N + ThB_6H_4C_6N$ | -0.0551 |
| $ThNH_4B_6C_6$ | $1.5\,ThB_4 + 6\,ThC + 7.5\,H_3N \longrightarrow 9.25\,H_2 + 6.5\,ThN + ThB_6H_4C_6N$ | -0.0540 |
| $ThNH_4B_6C_6$ | $0.5714\,B_5H_7 + 0.826\,B_{13}C_2 + 1.449\,Th_2C_3 + 0.5\,N_2 \longrightarrow 1.899\,ThB_4 + ThB_6H_4C_6N$ | -0.0500 |
| $ThNH_4B_6C_6$ | $4.867\,Th_3N_4 + 5.4\,B_9H_{11} + 3\,B_{13}C_2 \longrightarrow 18.47\,H_3N + 13.6\,ThB_6 + ThB_6H_4C_6N$ | -0.0398 |
| $ThNH_4B_6C_6$ | $0.1429\,B_5H_7 + 3\,ThC_2 + 13.29\,B + H_3N \longrightarrow 2\,ThB_4 + ThB_6H_4C_6N$ | -0.0373 |
| $ThNH_4B_6C_6$ | $6\,ThC + 5.5\,H_2 + 6\,B + H_3N \longrightarrow 5\,ThH_2 + ThB_6H_4C_6N$ | -0.0347 |
| $ThNH_4B_6C_6$ | $0.5714\,B_5H_7 + 3.891\,ThC + BN + 1.054\,B_{13}C_2 \longrightarrow 2.891\,ThB_4 + ThB_6H_4C_6N$ | -0.0307 |
| $ThNH_4B_6C_6$ | $3\,ThC_2 + 0.09091\,B_9H_{11} + 13.18\,B + H_3N \longrightarrow 2\,ThB_4 + ThB_6H_4C_6N$ | -0.0217 |

Continued

| Compound | Chemical reaction | Reaction energy(eV) |
|---|---|---|
| $ThNH_4B_6C_6$ | $2\,H_2 + 2.538\,ThC_2 + 0.4615\,B_{13}C_2 + 1.269\,N_2 \longrightarrow 1.538\,ThN + ThB_6H_4C_6N$ | -0.0101 |
| $ThNH_4B_6C_6$ | $2.992\,ThC + 0.06667\,Th_4H_{15} + 1.504\,B_{13}C_2 + H_3N \longrightarrow 2.259\,ThB_6 + ThB_6H_4C_6N$ | -0.0067 |


\end{appendices}

\end{widetext}

\end{document}